\documentclass[ twoside]{report}

\usepackage[a4paper,margin=3cm]{geometry}
\usepackage{latexsym,amsmath,amssymb,tabularx,supertabular,bm, cite, subfigure, verbatim, fdag}

\usepackage{epsfig}
\usepackage{color}

\newcommand{\tr}{{\rm tr}}               

\newcommand{\mbf}[1]{\mbox{\boldmath $#1$}}
\newcommand{\bl}{\mbf{l}}

\newcommand{\bk}{\mbf{k}}

\usepackage{hyperref}
\usepackage{graphicx}
\usepackage{rotating}

\begin{document}


\begin{titlepage}
        \vspace*{1cm}
\thispagestyle{empty}
        \begin{center} \vspace{2cm}
                {\LARGE\bf  The high energy behavior of QCD: \\ The effective action \\  and  the triple-Pomeron-vertex\\[1cm]}
                 \vspace{1cm}
{\large Dissertation\\zur Erlangung des Doktorgrades\\
                  des Departments Physik \\
                  der Universit\"at Hamburg \\}
                \vspace{2cm}
                { \large vorgelegt von \\
                  Martin Hentschinski\\aus Deggendorf\\[2.5cm]}
                \vspace{1cm}
               { \large Hamburg 2009 \par}
                \end{center}
\end{titlepage}
\thispagestyle{empty}
\vspace*{18cm}
\begin{tabular}{p{6.5cm}l}
Erstgutachter der Dissertation: & Prof.~Dr.~J.~Bartels \\
Zweitgutachter der Dissertation: & Prof.~Dr.~V.~Schomerus \\
\\
Erstgutachter der Disputation: & Prof.~Dr.~J.~Bartels \\
Zweitgutachter der Disputation: & Prof.~Dr.~B.~A.~Kniehl \\
\\
Datum der Disputation: & 13. Juli 2009 \\
Vorsitzender des Pr\"ufungsauschusses: & Prof.~Dr.~G.~Sigl \\
Vorsitzender des Promotionsausschusses: & Prof.~Dr.~R.~Klanner \\ \\
Departmentleiter: &  Prof.~Dr.~J.~Bartels\\
Dekan der Fakult\"at f\"ur Mathematik, Informatik und Naturwissenschaften:  & Prof.~Dr.~H.~Graener \\
\end{tabular}





\newpage




%

\abstract{We study integrations over light-cone momenta in the high
  energy effective action of QCD.  After a brief review of the
  effective action, we arrive on a regularization mechanism from
  matching of effective action diagrams with QCD diagrams, which we
  apply to a re-derivation of the reggeized gluon and of the
  BFKL-equation.  We study consequences of the proposed regularization
  on the analytic structure of $2\to 3$ and $2 \to 4$ production
  amplitudes in the Multi-Regge-Kinematics. We derive a certain part
  of the 1-loop corrections to the production vertex and demonstrate
  that they yield the on-set of corrections demanded by the
  Steinmann-relations. The Reggeon-Particle-2-Reggeon vertex is
  determined and applied to the construction of various signature
  configurations of the production amplitudes.  We extend the proposed
  regularization method to states of three and four reggeized gluons
  and propose a supplement to the effective Lagrangian.  We derive
  vertices for the $1-3$ and $2-4$ reggeized-gluon-transition inside
  the elastic amplitude and verify that signature conservation is
  obeyed.  Integral equations for the state of three and four
  reggeized gluons are formulated and shown to be in accordance with a
  result by Bartels and W\"usthoff. In a second part we investigate
  the high-energy behavior of QCD for different surface topologies of
  color graphs. After a brief review of the planar limit (bootstrap
  and gluon reggeization) and of the cylinder topology (BFKL) we
  investigate the $3\to3$ scattering in the triple Regge limit which
  belongs to the pair-of-pants topology.  We re-derive the triple
  Pomeron vertex function and show that it belongs to a specific set
  of graphs in color space which we identify as the
  analog of the Mandelstam diagram. We then extend the study to the high-energy behavior of $\mathcal{N}=4$ SYM where we find a new class of color graphs not present in QCD.}   \\

\begin{center}
{ \bf Zusammenfassung}  
\end{center}

{ Wir untersuchen Integrationen \"uber Lichtkegelimpulse in der
  effektiven Wirkung der Hochenergie-QCD.  Nach einem kurzem
  \"Uberblick \"uber die effektive Wirkung erhalten wir durch Abgleich
  von Diagrammen der effektiven Wirkung und der QCD einen
  Regularisierungsmechanismus, den wir zu einer erneuten Ableitung des
  reggesierten Gluons und der BFKL-Gleichung benutzen. Wir untersuchen
  die Auswirkungen des vorgeschlagenen Regularisierungsmechanismus'
  auf die analytische Struktur von $2\to 3$ und $2\to 4$
  Produktionsamplituden in der Multi-Regge-Kinematik.  Wir bestimmen
  einen Teil der 1-Schleifen-Korrekturen zur Produktionsvertex und
  zeigen, dass sie Anzeichen f\"ur die aufgrund der
  Steinmann-Beziehungen ben\"otigten Korrekturen liefern.  Die
  Reggeon-Teilchen-2 Reggeon-Vertex wird bestimmt und zur Konstruktion
  verschiedener Signatur-Konfigurationen von Produktionsamplituden
  benutzt.  Wir erweitern den vorgeschlagenen
  Regularisierungsmechanismus zur Beschreibung von Zust\"anden
  bestehend aus drei bzw. vier reggesierten Gluonen und schlagen eine
  Erg\"anzung zum effektiven Lagrangian vor.  Vertizes die den
  \"Ubergang von $1-3$ und $2-4$ reggesierten Gluonen beschreiben
  werden f\"ur die elastische Streuamplitude abgeleitet und wir
  zeigen, dass Signaturerhaltung in der effektive Wirkung erf\"ullt
  ist. Wir formulieren Integralgleichungen f\"ur den Zustand 
  aus drei bzw. vier reggesierten Gluonen und zeigen, dass diese im
  Farbsinglet in \"Ubereinstimmung mit einem fr\"uherem Ergebnis von
  Bartels und W\"usthoff sind.  In einem zweiten Teil untersuchen wir
  das Hochenergieverhalten der QCD f\"ur verschiedene
  Oberfl\"achentopologien von Farbgraphen.  Nach einem kurzem
  \"Uberblick \"uber den planaren Limes (Boostrap und Reggesierung
  des Gluons) und der Zylindertopologie (BFKL) untersuchen wir die
  Hosentopologie. Wir reproduzieren die Triple-Pomeron-Vertex-Funktion
  und zeigen, dass sie zu einer bestimmten Gruppe von Farbgraphen
  geh\"ohrt, die wir als das Analog zum Mandelstam-Diagramm
  identifizieren. Wir erweitern die Untersuchung auf $\mathcal{N}=4$
  SYM und finden eine neue Klasse von Farbgraphen, die in der QCD
  nicht vorhanden ist. }  \newpage

\cleardoublepage
\pagenumbering{roman}
\tableofcontents
\cleardoublepage
\pagenumbering{arabic}
\chapter{Introduction}

Quantum Chromodynamics (QCD) is nowadays well established as the
microscopic theory of strong interactions. It is stated in terms of a
non-abelian gauge theory with the gauge group $SU(N_c)$ with $N_c = 3$
the number of colors and with quarks and gluons as its elementary
degrees of freedom. Due to its complicated mathematical structure, an
exact analytic determination of its correlation functions is currently
out of reach and one needs to apply approximative methods to arrive on
predictions from the theory.  Generally, the strong coupling constant
$\alpha_s$ is large and cannot be used as a small parameter to expand
correlation functions in a perturbative series.  In that case one
mainly relies on lattice models, using   simulations on a
computer, and QCD sum rules.  Fortunately, there exists a class of
processes where the strong coupling is small and a perturbative
treatment can be given meaning: Due to asymptotic freedom
\cite{Politzer:1973fx,Gross:1973id}, the strong-coupling constant
$\alpha_s$ turns out to be small in presence of a hard scale and
consequently hard scattering processes allow for a perturbative
description.

Before the advent of QCD, Regge-theory \cite{Collins:1977jy} was used
to describe the high-energy limit of scattering amplitudes of strong
interactions. The analysis was mainly based on fundamental principles
such as analyticity, unitarity and Lorentz-invariance of scattering
amplitudes.  In the high-energy- or Regge-limit, where the squared
center-of-mass energy $s$ is considerably larger than the momentum
transfer $t$ and all other mass scales, scattering amplitudes were
shown to be determined by the position of singularities of their
partial wave amplitude in the complex angular momentum plane.
Starting from the assumption that the leading singularities in the
complex angular momentum plane were pole-singularities only,
so-called Regge-poles, phenomenological successful models could be
constructed, which were used to describe scattering processes at high
energies. In particular, apart from Regge-poles with the quantum
numbers of physical particles, the Pomeron, a Regge-pole with the
quantum numbers of the vacuum and even parity, was introduced to
explain the experimentally observed asymptotic rise of the total
cross-section with center of mass energy.  By analytical continuation,
Regge-poles could be furthermore related to the spectrum of hadronic
states in the crossed channel: All strongly interacting particles
turned out to lie on remarkably straight Regge-trajectories, passing
for the mass of particles through their physical spin and it was one
of the triumphs of Regge-theory, to relate experimental results in
different physical channels by crossing \cite{Abarbanel:1975me}.

However it was soon realized that in a relativistic theory
Regge-poles require the simultaneous presence of branch-cuts,
so-called Regge-cuts, in the complex angular momentum plane, which
 could be understood to arise from the exchange of two or
more Regge-poles.  Starting with the pioneering work of Gribov
\cite{Gribov:1968fc}, a Reggeon-field-theory was developed
\cite{Abarbanel:1975me,Gribov:1968fc, Baker:1976cv}, which describes
the interactions between the various Reggeons and physical particles.
Reggeon-field-theory is formulated in two space dimensions, transverse
to the scattering axis, and one time-like dimension, given by the
rapidity variable.

As soon as QCD was identified as the correct description of strong
interactions, it was natural to ask, whether the findings of
Regge-theory could be derived directly from the new, fundamental
theory of the strong interactions.  Unfortunately, physics described
by Regge-theory is typically soft and there is not a hard scale which would justify
a perturbative treatment.   There exists,
however, a class of processes, like scattering of two highly virtual
photons at high center-of-mass energies, where the virtually of the
photon provides a hard scale $Q^2$ which allows for a perturbative
study of the Regge-limit.  Nevertheless, large logarithms in the
center-of-mass-energy $\sqrt{s}$ can compensate for the smallness of
the coupling and  therefore spoil the validity of any finite order
correction; a resummation of terms enhanced by large logarithms
becomes necessary.  The resulting Leading Logarithmic Approximation
(LLA) resums maximally enhanced perturbative terms of the order
$(\alpha_s \ln s)^n$ to all orders in the strong coupling constant,
which is achieved by the Balitsky-Fadin-Kuraev-Lipatov
(BFKL)-equation
\cite{Lipatov:1976zz,Kuraev:1976ge,Fadin:1975cb, Kuraev:1977fs,Balitsky:1978ic}.  For
the scattering of colored objects one finds as a result of the
LLA-resummation  reggeization of QCD-scattering amplitudes:
The interaction between scattering particles is mediated through the
exchange of a single Regge-pole. This Regge-pole carries the quantum
numbers of the gluon and its trajectory $j(t)$ passes for $t=0$
(corresponding to the vanishing mass of the gluon) through one
(corresponding to the gluon's physical spin). Consequently this Regge
pole is naturally identified as the reggeized gluon.  With the
BFKL-equation nowadays also known within the
Next-to-Leading-Logarithmic-Approximation (NLLA) \cite{Fadin:1998py,Ciafaloni:1998gs },
reggeization of QCD-amplitudes is now  confirmed also if sub-leading corrections are taken
into account \cite{Fadin:2006bj} . For scattering of colorless objects  the interaction is mediated at LLA and NLLA by the
famous BFKL-Pomeron. It is given as a  state of two reggeized
gluons and yields a cut rather than a pole in the complex angular
momentum plane.

The BFKL-Pomeron predicts for the total cross-section a power-like
rise with $s$, which ultimately would violate unitarity. It is
therefore believed that at some value of $s$ so-called unitarity
corrections will become sizable that will restore unitarity. In the
last 15 years a number of alternative approaches have been developed,
which are meant to supplement the BFKL-equation. Among the most
popular ones are the dipole model
\cite{Nikolaev:1990ja,Nikolaev:1991et,Mueller:1993rr,Mueller:1994gb},
the BK-equation
\cite{Balitsky:1995ub,Kovchegov:1996ty,Kovchegov:1997pc}, the
Balitsky-hierachy
\cite{Balitsky:1995ub,Balitsky:1998ya,Balitsky:2001gj} and the
JIMWLK-equation
\cite{Jalilian-Marian:1996xn,Jalilian-Marian:1997jx,Jalilian-Marian:1997gr,Jalilian-Marian:1997dw,Jalilian-Marian:1998cb,Kovner:2000pt,Weigert:2000gi,Iancu:2000hn,Ferreiro:2001qy}.
They are all based on the idea of saturation of gluon densities at
high energies. While the power-like rise of the total-cross-section
predicted by the BFKL-equation can be attributed to an infinite rise
of gluon densities, these are believed to saturate at a
certain value of $s$, which tames the growth of cross-sections.

Seen from the point of view of Regge-theory, the existence of moving
Regge-poles allows to conclude that the high energy behavior of QCD can
be reformulated in terms of a 2+1 dimensional Reggeon field theory.
Such a reformulation seems to be the most comprehensive description of
the high-energy behavior of QCD, in particular to address the issue of unitarity
corrections. Both $t$-channel and $s$-channel unitarity
can be regarded to be fulfilled automatically in such a reformulation.
Elements of this field theory are at first given by generalizations of
the BFKL-equation to the exchange of $n > 2$ reggeized gluons, which
is described by the BKP-equation \cite{Bartels:1980pe,
  Kwiecinski:1980wb}. Furthermore vertices describing transitions from
 two to four \cite{Bartels:1994jj,Bartels:1993ih,Bartels:1992ym}
and two to six reggeized gluons \cite{Bartels:1999aw} have been
derived so far.

A number of remarkable properties have been found at LLA for the known
elements of Reggeon-field-theory: They are invariant under M\"obius
transformation in two-dimensional transverse coordinate space and
 states of arbitrary many reggeized gluons are integrable in the
large $N_c$ limit \cite{Lipatov:1993yb,Lipatov:1994xy,Faddeev:1994zg}.
These properties are generally believed for QCD to be inherited from
its maximal supersymmetric extension, ${\cal N}=4$
Super-Yang-Mills-Theorie (SYM): Both theories coincide with each other
within the LLA.  ${\cal N}=4$ SYM is on the other hand known to be
conformal invariant in $3+1$ dimensions and believed to be integrable
with an enormous amount of activity in this direction in recent years
\cite{Lipatov:1997vu,Kotikov:2003fb,
  Kotikov:2004er,Minahan:2002ve,Beisert:2003yb,Kotikov:2007cy,Eden:2006rx,Beisert:2006ez,Bajnok:2008qj,Lipatov:2009nt}.
The above mentioned properties of Reggeon-field-theory gained further
interest with the advance of the AdS/CFT correspondence, which relates
${\cal N}=4$ SYM, to supergravity on five-dimensional AdS space
\cite{Maldacena:1997re,Gubser:1998bc,Witten:1998qj}.  In particular
the question arises whether a gravity dual of Reggeon-field theory in
string theory exists.

In spite of all these progresses and impressive results, understanding
of Reggeon-field-theory is  still limited and a complete
description of the high energy behavior for QCD and its supersymmetric
extensions is not yet achieved.  A systematic approach to derive for
non-abelian gauge theories  the
elements of Reggeon field theory promises to be the effective action
proposed by Lipatov in \cite{Lipatov:1995pn,Lipatov:1996ts}. The
reformulation of quantum-field-theories such as QCD as effective field
theories has nowadays become a popular and powerful method for the
analysis of multi-scale problems.  In particular the use of effective
field theories generally simplifies practical calculations and allows
to derive results which are only very hard or even impossible to be
obtained directly from QCD, due to the sheer complexity of the
underlying expressions\footnote{For a pedagogical introduction to
  effective field theories in particle physics see 
  \cite{Neubert:2005mu,Polchinski:1992ed}}.  Popular examples of such
effective theories range  nowadays from Chiral-Perturbation-Theory \cite{Weinberg:1978kz},
over Soft-Collinear-Effective Field \cite{Bauer:2000yr,Bauer:2001ct,
  Bauer:2001yt ,Chay:2002vy, Beneke:2002ph,Hill:2002vw } and
Heavy-Quark-Effective-Theory \cite{Georgi:1990um} up to
Non-Relativistic QCD \cite{Caswell:1985ui,Bodwin:1994jh} .

A first proposal for an effective field theory for QCD in the
Regge-limit has then been given in \cite{Lipatov:1991nf} which later
could be explicitly re-derived directly from the QCD-action
\cite{Kirschner:1994gd,Kirschner:1994xi}. A more general effective
action, which implicitly includes the first effective action, was then
proposed in \cite{Lipatov:1995pn,Lipatov:1996ts}. It is thought to
provide a systematic approach to the unitarization of the BFKL-Pomeron
and for the determination of sub-leading corrections beyond the LLA.

In the high-energy limit, scattering particles are close to opposite
parts of the light-cone and therefore separated by a large relative
difference in rapidity. The interaction between individual parts of
the amplitude is highly non-local in rapidity, which complicates the
analysis in terms of Feynman-diagrams considerably.  The effective
action then aims on a local description, where the interaction between
QCD-particles, quarks, gluons and ghosts is restricted to a small
interval in rapidity. Interaction between regions significantly
separated in rapidity is on the other hand mediated by reggeized
gluons.  To arrive at such a description, it is needed to integrate
out the modes of highly virtual particles, which spoil for usual QCD
such a factorization.  In the case of the effective action this is not
done in a literal sense within the QCD-path-integral, but the
effective description is obtained from the study of QCD
tree-amplitudes in the high energy limit. As a result of such an
analysis one obtains apart from the  reggeized gluon
fields an infinite number of new, induced vertices, which supplement
the QCD-Lagrangian.

In contrast to $2+1$ dimensional Reggeon-field-theory, the effective
action \cite{Lipatov:1995pn,Lipatov:1996ts} is stated in usual $3+1$
dimensional space-time. For real production processes at three-level,
the conversion into a $2 + 1$ dimensional description takes place
automatically: The kinematics of produced particles is fixed and the
mass-shall condition of produced particles reduces amplitudes of the
effective action immediately into the required $2+1$ dimensional form.
However for the determination of virtual corrections, an integration
over longitudinal momenta is needed.  While corresponding integrations
over transverse momenta can be performed by using standard techniques,
longitudinal integrations take a special role in the effective action.
They contain an implicit integration over rapidity and are therefore
in potential conflict with the desired locality in rapidity. 
Attempting a naive unrestricted integration, one makes
the unpleasant observation that some integrals turn out to be
divergent and even worse some contributions seem to be counted twice. The
emergence of new divergences in effective theories, which have not
been present in the underlying theory is natural and occurs also for
other effective field theories: Effective theories provide
simplifications in the description of amplitudes compared to the
underlying full theory, which sometimes spoils convergences of
integrals due neglecting certain parts.  For the effective action
\cite{Lipatov:1995pn,Lipatov:1996ts} this requires  to find a
regularization that ensures locality in rapidity and  absence
of over-counting.  Apart from that, another point requires
clarification, if one attempts to carry out integrals over
longitudinal momenta within the effective action: The induced vertices
that supplement for the effective action the QCD-Lagrangian yield
poles in the longitudinal momenta. Whenever those induced vertices
appear as a part of an integral, a suitable prescription how to
circumvent these singularities needs to be given.

The first part of this thesis is dedicated to the study of the following
points: We propose a regularization mechanism that ensure locality of
effective amplitudes in rapidity in accordance with the underlying
principles of the effective action. To arrive on such a prescription
we match the elastic scattering amplitude of the effective action at
low orders in the strong coupling with the corresponding QCD-
expression, which further allows to find the correct pole prescription
for the relevant induced vertex.  The obtained prescription is then
used for a re-derivation of the reggeized gluon and of the BFKL-equation
from the effective action within leading logarithmic accuracy.

A suitable testing ground for the found prescription and the
LLA-resummations within the effective action is given by the
determination of loop-corrections to production amplitudes: Due to the
Steinmann-relations \cite{Steinmann:1960}, these amplitudes have a
non-trivial structure with respect to their energy discontinuities.
The latter arise from logarithms and resumming them within the LLA in the
effective action, the question arises in which sense
production amplitudes from the effective action are in agreement with
the Steinmann-relations \cite{Steinmann:1960}.

To arrive from the effective action on a description in terms of a
field theory of reggeized gluons, a description for  states of
$n>2$ reggeized gluons is further needed. In particular the
regularization method needs to be extended and the pole structure of
higher induced vertices is to be determined.  As a reference process
we consider elastic quark-quark scattering with exchange of up to four
reggeized gluons. The description of the  state of three and four
reggeized gluons requires apart from pairwise interaction between
reggeized gluons, also vertices that describe transitions from
one-to-three and two-to-four reggeized gluons. For the state of four
reggeized gluons the obtained result is then shown to agree with the
two-to-four Reggeon transition vertex derived in
\cite{Bartels:1994jj}.

Another route for a deeper understanding of Reggeon-field-theory is
given by a systematic study of its large $N_c$ limit: In the classical
paper by 't Hooft \cite{'tHooft:1973jz} it has been shown that an
expansion in the rank of the gauge group $N_c$ organizes the color
structure of Feynman diagrams in terms of topologies of
two-dimensional surfaces which resemble the loop expansion of a closed
string theory. In particular with the advance of the AdS/CFT
correspondence where the string coupling on the gravity side could be
shown to be proportional to $N_c^{-1}$ , this observation has gained
new attention.

From the point of view of Reggeon-field-theory, it is at first the
integrability of  states of $n > 2$ of reggeized gluons which
trigger the interest in the large $N_c$ expansion. They belong, like
the BFKL-Pomeron, to the class of color factors with the topology of a
cylinder.  The leading term of the large $N_c$ expansion is on the other
hand  given by color factors that have the topology of a plane, and
one usually refers to these leading contributions as 'planar'
diagrams.  As it is the leading term in the expansion, the 'large
$N_c$ limit' and the 'planar' limit are often used interchangeable for
each other. Planar diagrams contribute, for example, to gluon-gluon
scattering amplitudes or to multi-gluon production amplitudes.

The integrable BKP-states belong then together with the BFKL-Pomeron
to the class of processes in which the $N_c$ leading color factors do
not fit onto the plane, and the whole expansion starts with a formally
sub-leading term.  An example of such a physical process is given by
the scattering of two electromagnetic currents or virtual photons in
the Regge-limit, where the $N_c$-leading color factor has the topology
of a cylinder.

\begin{figure}[htbp] \centering
   \parbox{6cm}{ \center \includegraphics[width=4.5cm]{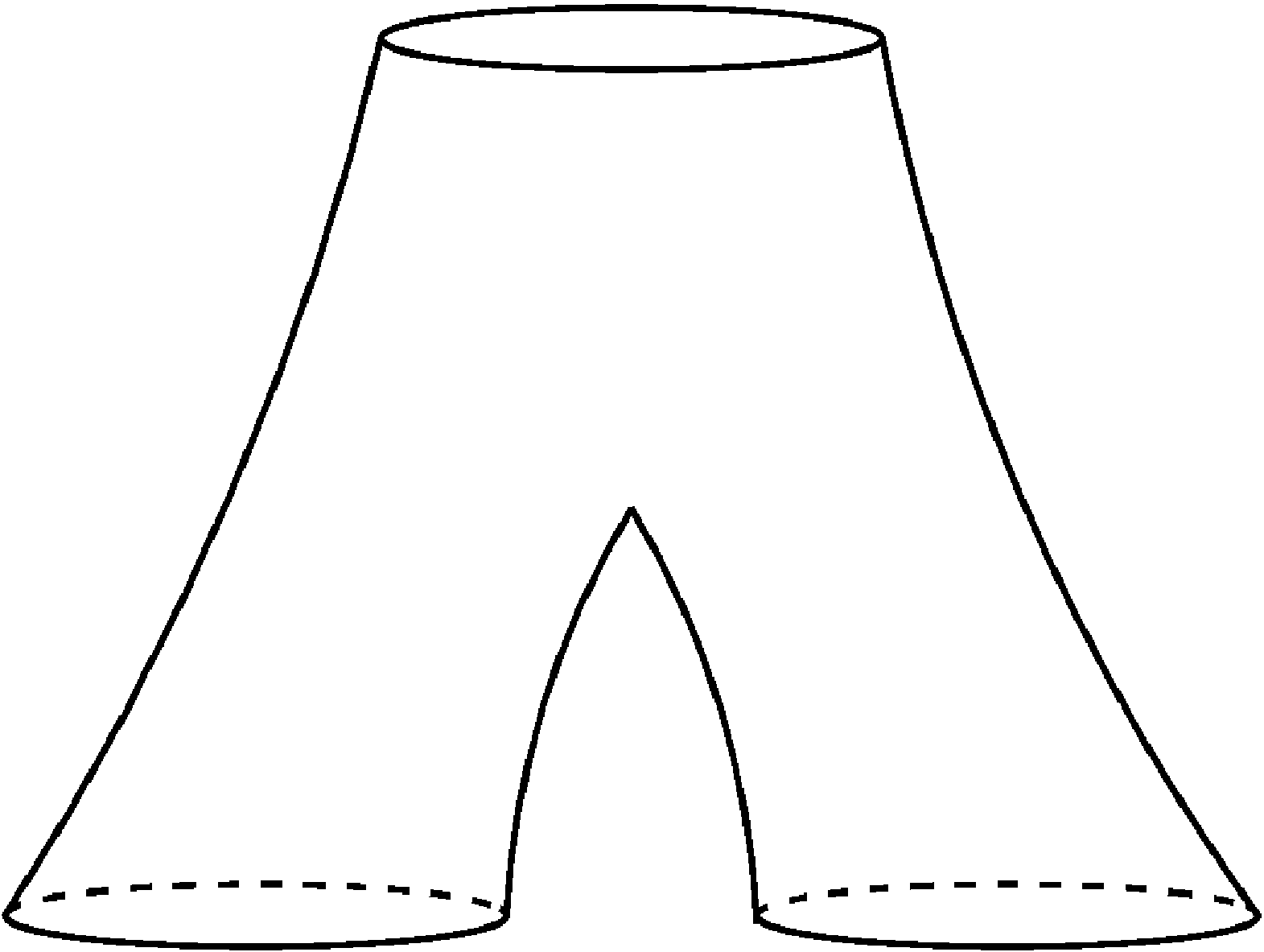}}
\parbox{4cm}{\includegraphics[height = 4.5cm]{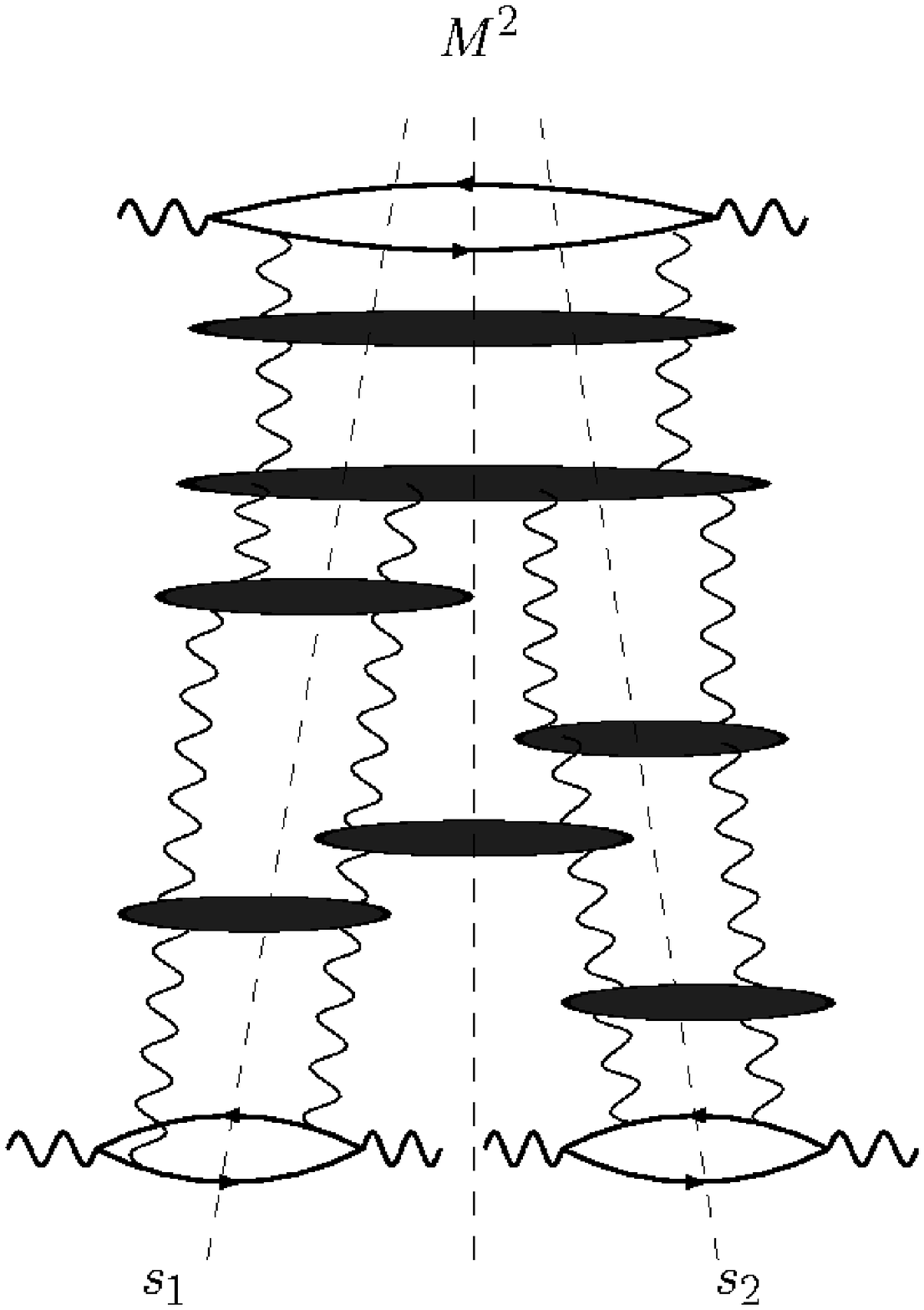}}
\parbox{4cm}{\includegraphics[height = 4.5cm]{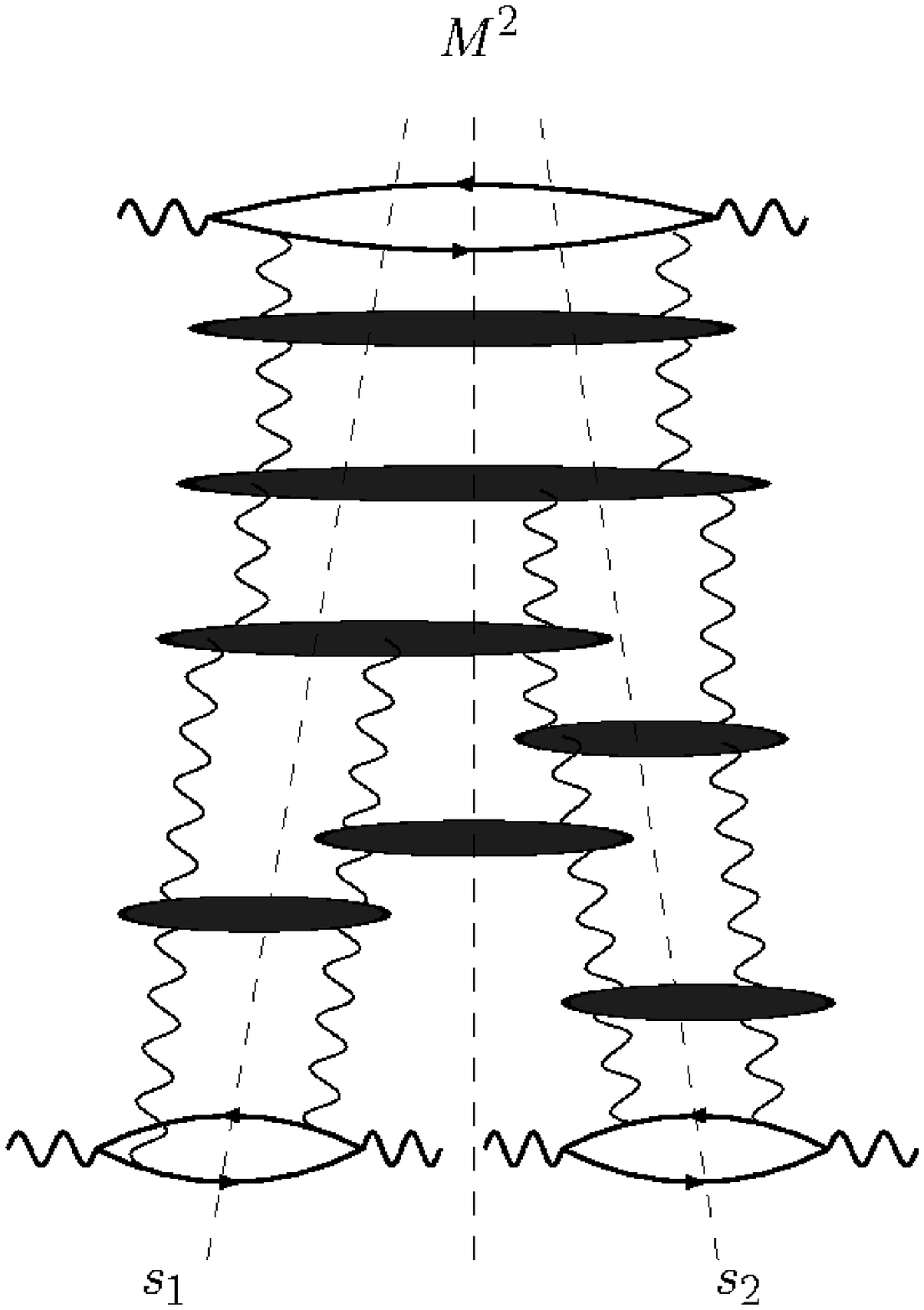}}
\\
\parbox{6cm}{\center (a)}\parbox{8cm}{\center (b)}
\caption{\small (a) The 'pair of pants' topology. (b) Typical contributions to the triple energy discontinuity for scattering of three virtual photons. For $\mathcal{N}=4$ SYM, the virtual photons are replace by $R$-currents.}
  \label{fig:trousexxr}
$\,$
\end{figure}

In the second part of this thesis we go then one step further and
consider color factors with the topology of a
pair-of-pants, Fig.\ref{fig:trousexxr}a.  As a suitable process for its
study we consider scattering of a highly virtual photon on two virtual
photons in the so-called triple Regge limit, Fig.\ref{fig:trousexxr}b.
This process has been considered before in \cite{Bartels:1994jj} for
$N_c = 3$ and has been shown to yield the transition from two to four
reggeized gluons.
 
In principal it is straight forward to determine the large $N_c$ limit of
the yet existing result \cite{Braun:1997nu,Braun:1997ax,Braun:1997sp},
but the connection with the expansion in terms of topologies is not
apparent: Instead of the 'pants' it rather seems to belong to three
disconnected cylinders. We therefore re-derive the result of the
large-$N_c$ limit of \cite{Bartels:1994jj} by summing only those
diagrams which belong to the topology of Fig.\ref{fig:trouserr} and
demonstrate that 'reggeizing' and 'irreducible' terms of
\cite{Bartels:1994jj} can be attributed to distinct classes of
diagrams on the surface of the pair-of-pants.  The study of these
color diagrams allows especially for a new view on the reggeization of
the gluon and on the triple-Pomeron-vertex: whereas reggeization can
be understood as a feature of planar QCD, the triple-Pomeron-vertex
requires a non-planar structure, which reminds of the non-planar
Mandelstam cross diagram \cite{Mandelstam:1963cw}.

Especially due to the advance of the AdS/CFT correspondence with
string coupling $g_s \sim 1/N_c$ on the gravity side, it seems natural
to use this topological approach for studying the Regge-limit of
$\mathcal{N}=4$ SYM.  A program to study the high-energy limit on both
sides of the correspondence has been initiated in
\cite{Bartels:2008zy}. It was found that it is convenient to use
$R$-currents which result from the global $SU_R(4)$ symmetry of
$\mathcal{N}=4 $ SYM to formulate correlators which are well-defined
both on the gauge theory side and on the string side of the
correspondence.  For the elastic scattering of two $R$-currents, both
sides of the correspondence have been investigated: On the string
side, the leading contribution in the zero slope limit is given by
Witten diagrams with graviton exchange. For the gauge theory side, the
impact factor of the $R$-current consists of the sum of a fermion and
a scalar loop in the adjoint representation of the gauge group, while
the interaction is mediated in the high energy limit by the
BFKL-Pomeron.

The study has then been extended on the gauge theory side to the
six-point function of $R$-currents in the triple-Regge-limit in
\cite{behm}. For finite $N_c$ one finds apart from the
triple-Pomeron-vertex and the reggeizing terms, a new contribution
which appears due to the adjoint representation of all particles in
$\mathcal{N}=4$ SYM and has no counterpart in QCD. We then consider
also this process with the topological expansion of color factors and
suggest a possible interpretation of the result on the gravity side of
the AdS/CFT-correspondence.

This thesis is organized as follows: Chapter 2 gives an introduction
to the effective action. We summarize the ideas that underlie its
derivation, discuss the building blocks of the effective and outline
the points that need clarification for a determination of virtual
corrections from the effective action. Chapter 3 is dedicated to the
discussion of the elastic quark-quark scattering amplitude. We match
effective theory amplitudes at low orders in the coupling with QCD
amplitudes and propose a regularization method for longitudinal
integrations. The proposed method is then applied to the derivation of
the reggeized gluon for the negative signatured part of the elastic
scattering amplitude.  Rules for longitudinal integrations of loops
with 2 reggeized gluons are derived and applied to a derivation of the
BFKL-equation. In Chapter 4 we extend our study to the $2 \to 3$ and
$2 \to 4$ production amplitude. We derive a certain part of the
one-loop corrections to the production vertex and show that they
provide the onset of corrections needed for a fulfillmentin of the
Steinmann-relations. Furthermore we derive the Reggeon-Particle-2
Reggeons vertex from the effective and apply it the the various parts
of the $2 \to 3$ and $2 \to 4$ production amplitude with positive and
mixed signature. In Chapter 5 we extend our results to exchanges with
$n> 2$ reggeized gluons. As a reference process we use again elastic
scattering of two quarks. We present a recipe for the derivation of
the pole-prescription of higher induced vertices and explicitly
determine the pole structure up to the third order in the
Yang-Mills-coupling. We formulate a scheme that allows to carry out
longitudinal loop integrals for loops containing $n>2$ reggeized
gluons and determine the quark-impact factors for coupling of three
and four reggeized gluons. We further derive vertices that describe in
the elastic amplitude transitions from one-to-three and two-to-four
reggeized gluons. We demonstrate that these vertices are for the state
of four reggeized gluons, together with pairwise interactions of
reggeized gluons, in accordance with the $2 \to 4$ Reggeon transition
vertex of \cite{Bartels:1994jj}. In Chapter 6 we discuss the $2 \to 4$
Reggeon transition vertex or triple-Pomeron-vertex from the point of
view of the large $N_c$ expansion.  After a brief review of the planar
limit (bootstrap and reggeization) and of the cylinder topology (BFKL)
we investigate the 6-point amplitude in the triple-Regge-limit, which
belongs to the pair of pants topology. We identify the
triple-Pomeron-vertex function and show that it belongs to a specific
set of graphs in color space which we identify as the analogue of the
Mandelstam diagram. In Chapter 7 we extend this study to the high
energy behavior of a six-point correlator of $R$-currents in
$\mathcal{N}=4$ SYM. We find three distinct classes, where one of them
yields the triple-Pomeron-vertex. The final Chapter 8 contains our
conclusions. In Appendix A we present an alternative regularization
scheme for the effective action. Appendix B contains details about the
evaluation of integrals.

 \chapter{The gauge invariant effective action of high energy QCD}
 \label{cha:calc}

 In this chapter we give a short but comprehensive introduction to
 the principles that underlie the effective action. In
 Sec.\ref{sec:heqcd} we recall general facts about QCD scattering
 amplitudes at high energies and we motivate the formulation of an
 effective action. In Sec.\ref{sec:prodqmrk} we consider
 tree-amplitudes in the so-called Quasi-Multi-Regge-Kinematics which
 lay the basis for the derivation of the effective action. In
 Sec.\ref{sec:effact} we present the effective action and its field
 content and give the (unregularized) Feynman-rules. In
 Sec.\ref{sec:virtual} we formulate the points which require
 clarification and which will be studied in the following chapters.

\section{High-energy-QCD amplitudes within the LLA}
\label{sec:heqcd}
The formulation of the effective action is based on the fact that scattering amplitudes reggeize in the high-energy-limit of QCD.
In particular, the reggeized gluon is to be understood as the effective degree-of-freedom in the $t$-channel of high-energy-QCD amplitudes.

\begin{figure}[thbp]
  \centering
  \parbox{6cm}{\center \includegraphics[height=4cm]{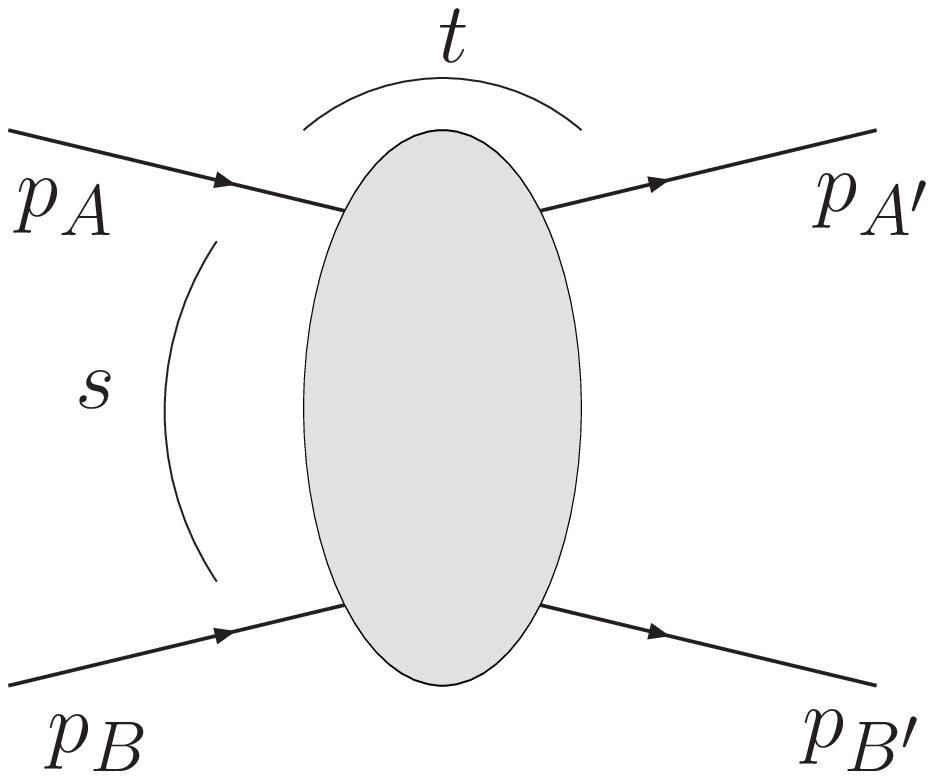}} 
 \parbox{6cm}{\center \includegraphics[height=6cm]{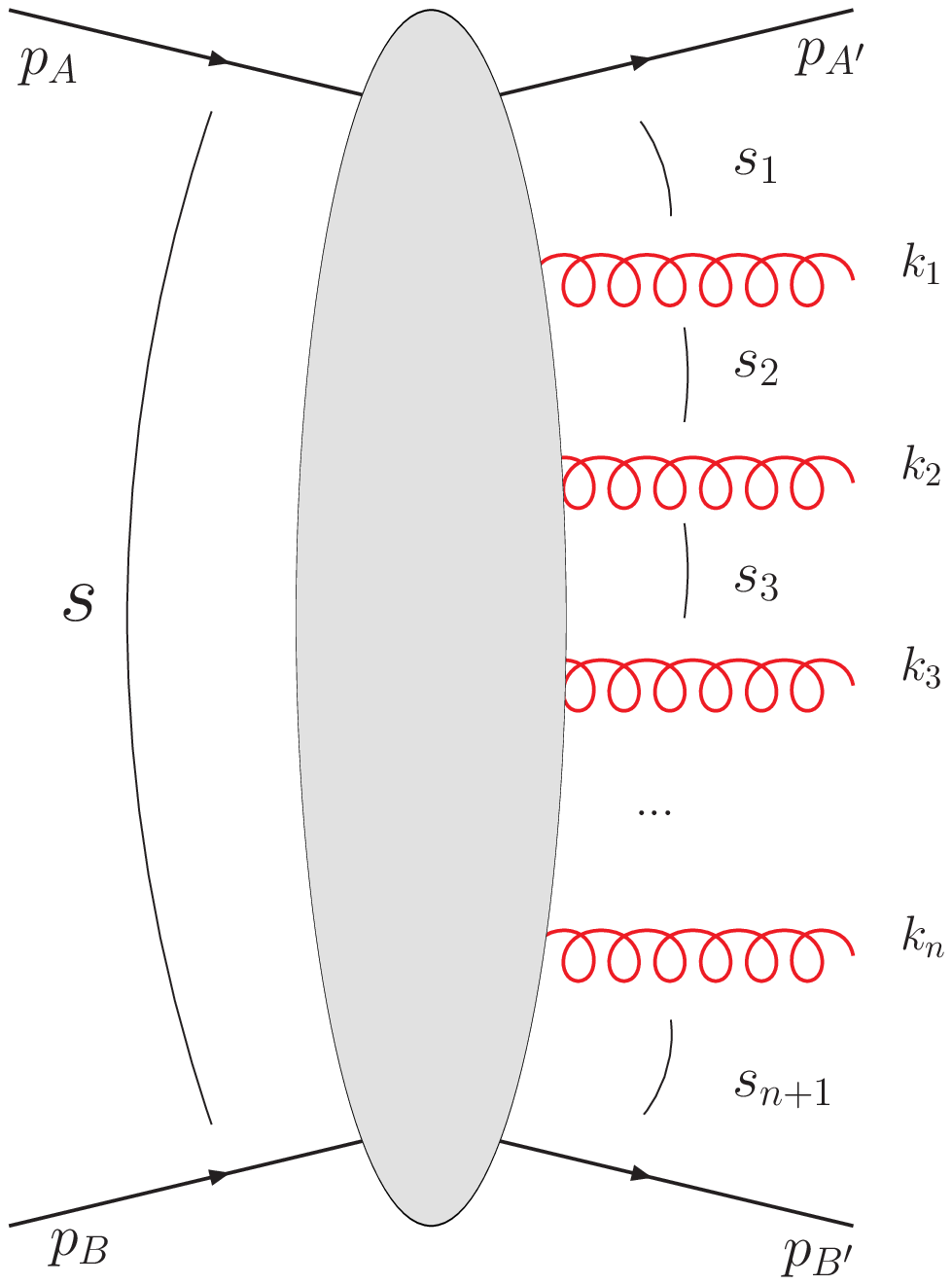}} \\
\parbox{6cm}{\center (a)} \parbox{6cm}{\center (b)} 
  \caption{\small The elastic (a) and the $n$-particle production amplitude (b) in the Regge-limit and in the Multi-Regge-Kinematics respectively.}
  \label{fig:elprod}
\end{figure}

This observation arises at first from the analysis of the elastic
scattering amplitude within the Leading Logarithmic Approximation
(LLA). There, perturbative terms of the order $\sim (\alpha_s \ln
s)^n$ are resummed to all orders, as the smallness of the strong
coupling constant $\alpha_s$ is compensated by the large logarithm
$\ln s$.  As a result of such an analysis it has been found
\cite{Lipatov:1976zz,Fadin:1975cb,Kuraev:1976ge, Kuraev:1977fs,
  Balitsky:1978ic} that the elastic scattering amplitude,
Fig.\ref{fig:elprod}a, factorizes within the LLA into the exchange of
a single reggeized gluon and its couplings to external particles.  The
elastic scattering amplitude of particles with color indices $A, B$
and helicities $\lambda_A, \lambda_B$, takes the following Regge-form,
\begin{align}
  \label{regge-form}
  \mathcal{M}_{2 \to 2} (s,t) 
=&  2 g \delta_{\lambda_A\lambda_{A'}} T^c_{AA'}  \frac{s^{1 + \beta(t)}}{t} g \delta_{\lambda_B\lambda_{B'}} T^c_{BB'},
\end{align}
with $j(t) = 1 + \beta(t)$ the gluon Regge trajectory.  As usually,
the Mandelstam variable $s = (p_A + p_B)^2$ yields the squared
center-of-mass energy of the elastic amplitude while $t = (p_A -
p_A')^2$ gives the squared momentum transfer. The $T^c_{AA'}$ are
generators of the $SU(N_c)$ gauge group in the representation
corresponding to the scattering particles and $g$ is the Yang-Mills
coupling constant.  Remarkably, the complete elastic scattering
amplitude, Eq.~(\ref{regge-form}), is within the LLA given by the
Born-amplitude times the Reggeon-factor $s^{\beta(t)}$, which gathers
all higher order corrections.
Moreover, reggeization of the gluon occurs not only for the elastic
scattering amplitude, but also for the high energy limit of a general
QCD-amplitude, where additionally $n$ gluons are produced,
Fig.\ref{fig:elprod}b.  Within the LLA, production of the $2 + n$ real
particles takes place in the Multi-Regge-Kinematics (MRK) where the
squared pair-energy of two neighboring produced gluons, $s_r = (k_r +
k_{r-1})^2$, is significantly larger than the momentum transfers in
the various $t$-channels, which are all of the same order of
magnitude:
\begin{align}
  \label{multiregge}
  s &= (p_A + p_B)^2 \gg s_r = (k_r + k_{r-1})^2 \gg {\bm{q}}_r^2 = -t_r ,
&
k_r = q_{r+1} - q_r, 
\end{align}
and we refer to Fig.\ref{fig:elprod}b for the nomenclature.
As the elastic scattering amplitude, Eq.(\ref{regge-form}), the  amplitude for the production
of $n$ gluons takes within the LLA  the following  factorizing Regge-form: 
\begin{align}
  \label{multiregge_form}
  \mathcal{M}_{2 \to 2 + n} =
&
2 s  g \delta_{\lambda_A\lambda_{A'}} T^{c_1}_{AA'}  
\frac{s_1^{ \beta({\bm q}_1^2)}}{{t}_1} 
\left[
 g T_{c_1c_2}^{d_1}  C^\mu(q_1, q_2)\epsilon_\mu^*(k_1) 
\right]
 \frac{s_2^{ \beta({\bm q}_2^2)}}{{t}_2}   \ldots 
\frac{s_{n+1}^{ \beta({\bm q}^2_{n+1})}}{{t}_{n+1}}
g \delta_{\lambda_B\lambda_{B'}} T^{c_{n+1}}_{BB'},
\end{align}
with $T^c_{ab} = -if_{abc}$ are $SU(N_c)$ generators in the adjoint
representation. Similar to the elastic amplitude, the $n$-particle
production amplitude has the remarkable feature that the all order
resummed expression is within the LLA given by the Born-amplitude,
times a Reggeon-factor $s_r^{\beta(-{\bm q}_r^2)}$, for every squared
sub-center-of-mass energy $s_r$.  One therefore observes that within
the LLA the complete interaction between $s$-channel produced
particles in the (Multi-) Regge-Kinematics can be reduced to the
exchange of a single reggeized gluon.  It is then this observation
which 
leads to the statement that 
 the reggeized gluon is  the relevant degree
of freedom in the $t$-channel of any QCD-amplitudes at high energies.
In Eq.(\ref{multiregge_form}), $ C_\mu(q_1, q_2) $ is the effective
Reggeon-Reggeon-Particle vertex, which describes the production of a
gluon by a reggeized gluon, and $\epsilon_\mu(k_r) $, the polarization
vector of the produced gluon.  This production vertex is an effective
vertex, which contains apart from the three-gluon-vertex also
so-called induced parts: 
\begin{align}
\label{eq:cmu}
  2C^\mu(q_1, q_2) = \gamma^{\mu-+} - 2\frac{q_1^2}{q_2^-}(n^-)^\mu  - 2\frac{q_2^2}{q_1^+}(n^+)^\mu 
\end{align}
where
\begin{align}
  \label{eq:three_gluon_project}
\gamma^{\mu-+} = q_1^+(n^-)^\mu +  q_2^-(n^+)^\mu- 2(q_1 + q_2)^\mu ,
\end{align}
is the light-cone projection of the three-gluon-vertex.  The above
light-cone basis is defined as
\begin{align}
  \label{eq:light-cone}
n^- &= \frac{p_A}{E}, & n^+ &= \frac{p_B}{E}, & E&=\sqrt{s}/2, & n^+\cdot n^- &= 2.
\end{align}
Light-cone vectors  are therefore  proportional to momenta  $p_A$ and $p_B$  of the scattering particles\footnote{In case of massive scattering particles, one should first construct a suitable Sudakov-basis of light-like momenta and then use those for the above definition.}. With 
\begin{align}
  \label{eq:light_cone_comp}
k^{\pm} = (n^{\pm})_\mu k^\mu,
\end{align}
 the  Sudakov-decomposition of an arbitrary four-momentum is given by
\begin{align}
  \label{eq:sudakov}
    k = k^+\frac{n^-}{2} +  k^-\frac{n^+}{2} + k_\perp.
\end{align}
For the description of high-energy-QCD amplitudes it is further useful to introduce rapidity, which is defined as
\begin{align}
   \label{eq:def_rap}
Y_k = \frac{1}{2} \ln \frac{k^+}{k^-}.
\end{align}
In terms of  rapidity, the 
 MRK, Eq.~(\ref{multiregge}), translates for instance into strong
ordering of the produced particles in rapidity i.e. produced particles
are separated significantly in their relative rapidity.  The
projections of the three-gluon-vertex in Eq.(\ref{eq:cmu}) on
light-cone directions results from the polarization tensor of
$t$-channel gluons in the high-energy limit: Connecting $s$-channel
particles widely separated in rapidity, its polarization tensor
reduces in a covariant gauge to its longitudinal part $ n^+_\mu
n^-_\nu/2$, while all other polarizations are suppressed by powers of
$s$.

To illustrate the  origin of the  induced
terms in Eq.~(\ref{eq:cmu}), we recall that scattering of
two quarks in the Regge-limit with production of one additional gluon
at central rapidities is obtained by the following set  of Feynman
diagrams:
\begin{align}
  \label{eq:prod_vertex}
 \parbox{1.5cm}{\includegraphics[width=1.5cm]{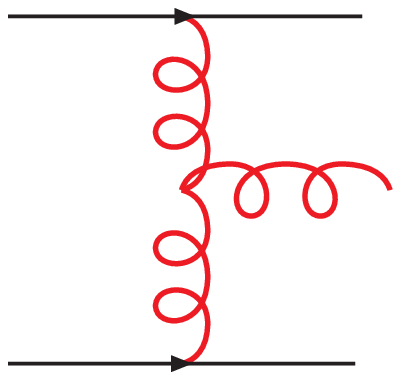}} 
         &+&
        \parbox{1.5cm}{\includegraphics[width=1.5cm]{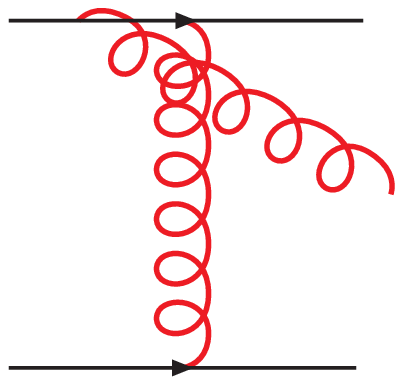}}
         &+ &
         \parbox{1.5cm}{\includegraphics[width=1.5cm]{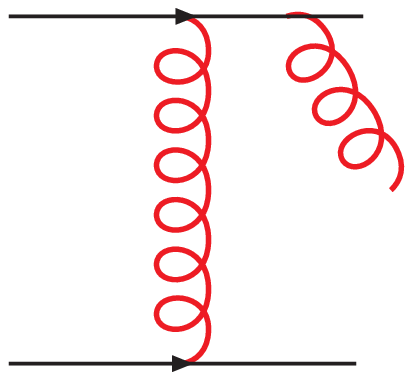}} 
         &+&
         \parbox{1.5cm}{\includegraphics[width=1.5cm]{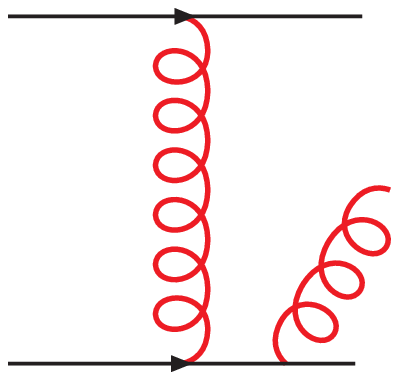}} 
         &+&
         \parbox{1.5cm}{\includegraphics[width=1.5cm]{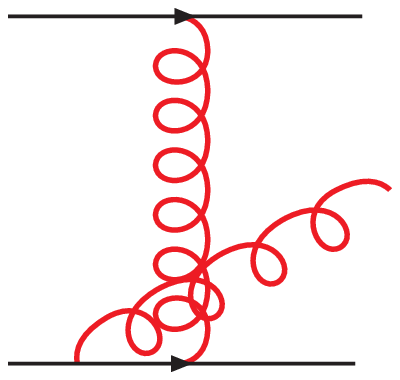}} 
=
\parbox{1.5cm}{\includegraphics[width=1.5cm]{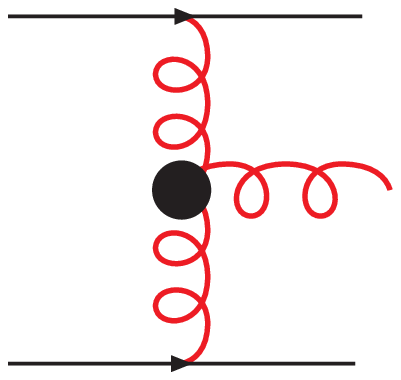}.}
        \end{align}
In the above sum, the vertex to the right is then to be associated
with the production vertex, Eq.~(\ref{eq:cmu}), which gathers all contributions to the production of a particle at central rapidities.
The induced terms of Eq.(\ref{eq:cmu}) arise from the diagrams 
in Eq.(\ref{eq:prod_vertex}), where the produced gluon
couples directly to the upper and lower quark respectively.
Supplementing the three-gluon vertex with these induced contributions,
the production vertex turns out to be gauge invariant,
\begin{align}
  \label{eq:C_mu_gauge_inv}
 C_\mu(q_1, q_2) k^\mu &= 0, & k = q_1 - q_2,  
\end{align}
which makes it possible to chose an arbitrary gauge for each of the
produced gluons. Similarly to the production vertex,  effective
vertices can be formulated for the coupling of 
scattering particles to  reggeized gluons. In the case where the
particle $A$ in Fig.\ref{fig:elprod} is given by a gluon, we have for
the coupling of the reggeized gluon in the $t$-channel to the
$s$-channel gluons
\begin{align}
  \label{eq:pprI}
  2 p_A^+ g\delta_{\lambda_A\lambda_{A'}}  =  \epsilon_{\lambda_A}(p_A)^\mu
\epsilon_{\lambda_A'}(p_A')^\nu \Gamma^{\mu\nu}_{GGR}(p_A, q),
\end{align}
where $\Gamma^{\mu\nu}_{GGR}$ is the effective Gluon-Gluon-Reggeon
(GGR) vertex. Similar to  the production vertex it is given as the sum of the
three-gluon-vertex and an induced contribution
\cite{Lipatov:1991nf,Kirschner:1994xi,Kirschner:1994gd}:
\begin{align}
  \label{eq:pprII}
\Gamma^{\mu\nu}_{GGR}(p_A, q) = \gamma^{\mu\nu+} -\frac{q^2}{p_A^+} (n^+)^\mu (n^+)^\nu,
\end{align}
with
\begin{align}
  \label{eq:tresgluon}
 \gamma^{\mu\nu+} = g^{\mu\nu} 2 p_A^+ - (n^+)^\mu (p_A + q)^\nu -  (p_A - 2q)^\mu (n^+)^\nu.
\end{align}
  To illustrate the different contributions to the effective vertices,
it proves useful to introduce a diagrammatic language which
describes the various interactions between the reggeized gluon
(denoted by wavy lines) and the standard QCD-gluon (curly lines).
The effective GGR-vertex  appears then as the following sum
\begin{align}
  \label{eq:ggr_diagram}
\parbox{2cm}{\includegraphics[width =2cm]{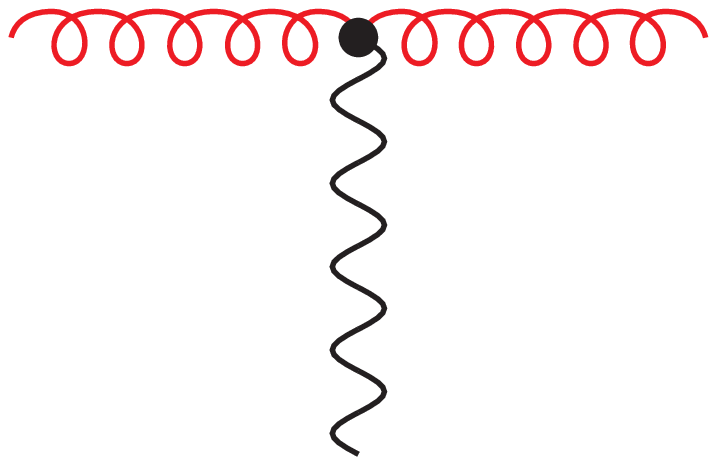}} = 
igT^c_{aa'} \Gamma^{\mu\nu}_{GGR}(p_A, q) = 
\parbox{2cm}{\includegraphics[width =2cm]{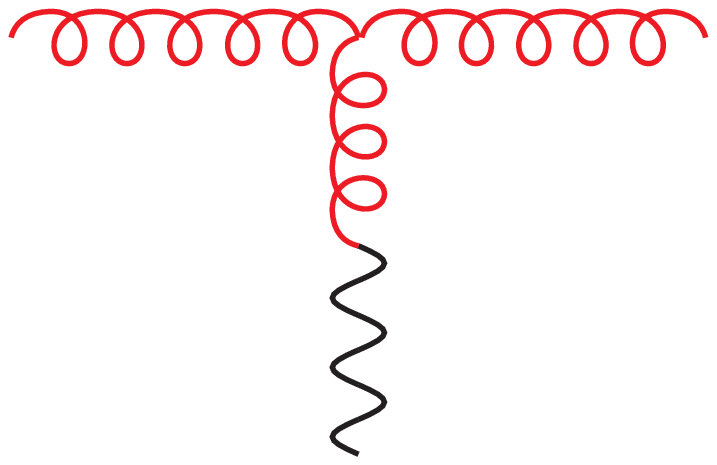}}+ \parbox{2cm}{\includegraphics[width =2cm]{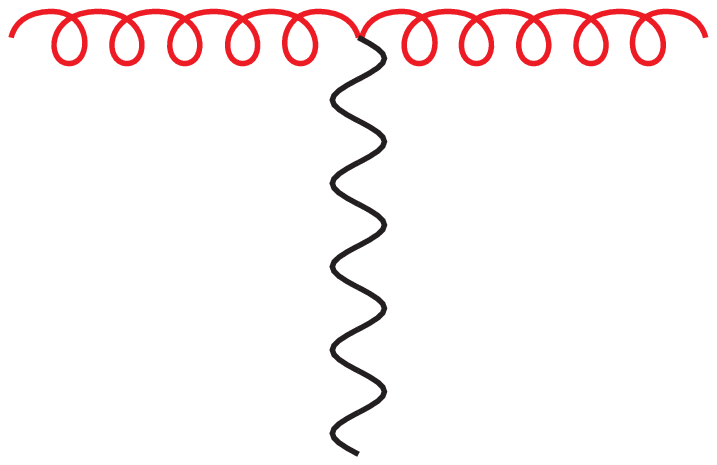}},
\end{align}
where the first diagram on the right hand side denotes coupling of the
reggeized gluon  to the three-gluon-vertex, and the second
diagram denotes the new induced contribution,
\begin{align}
  \label{inducedvertex11}
\parbox{1.5cm}{\includegraphics[width = 1.5cm]{ggr_d2.eps}} = \Delta^{\mu\nu+}_{aa'c} (p_A^+ )
  &= -ig {{q}}^2 \frac{T^c_{aa'} }{p_A^+}  (n^+)^{\mu} (n^+)^{\nu}.
\end{align}
The vertex  has the property
\begin{align}
  \label{eq:Ward_rpp}
\Gamma^{\mu\nu}_{RGG}(p_A, q)\cdot p_{A'}^\nu = p_A^+ p_A^\mu - p_A^2 (n^+)^\mu.
\end{align}
If the initial gluon $A$ is on the mass-shell ($p_A^2 = 0$) and its
polarization vector satisfies the Lorentz-condition
$\epsilon_{\lambda_A}(p_A)\cdot p_A = 0$, this vertex is gauge
invariant. It is therefore possible to use arbitrary gauges also for
this vertex.  The following picture arises: In the Regge-limit,
interaction between $s$-channel particles that are significantly
separated in rapidity is mediated by a reggeized gluon, which couples
to the $s$-channel particles by gauge invariant Reggeon-particle
vertices.  This implies that the reggeized gluon itself is invariant
under local gauge transformations, apart from the fact that it carries
color indices and transforms globally in the adjoint representation of
$SU(N_c)$.

\section[Gluon production within the (Quasi-) Multi-Regge-Kinematics]{Gluon production within the (Quasi-) Multi-Regge- \\ \mbox{Kinematics}}
\label{sec:prodqmrk}

The above results  suggest to reformulate QCD at high energies as
an effective field theory, that describes the interaction of QCD
particles  with the reggeized gluon. The idea to
formulate a calculus that describes the interaction of Reggeons and
usual particles goes back to the work of Gribov \cite{Gribov:1968fc}.
For QCD, a first effective action for the Multi-Regge-Kinematics has
been derived in
\cite{Lipatov:1991nf,Kirschner:1994xi,Kirschner:1994gd}.  In the
following we consider the more general effective action
\cite{Lipatov:1995pn,Lipatov:1996ts}, which allows to go beyond the
LLA and permits to derive loop corrections to the above presented
LLA-amplitudes. In following we  shortly sketch the derivation of
this action in presented in \cite{Lipatov:1995pn}: To this end one
uses that an action and its Lagrangian are classical objects and
therefore all information it contains about the interaction of its
fields is already present in tree-amplitudes.
\begin{figure}[t]
  \centering
  \parbox{5cm}{\includegraphics[height=4cm]{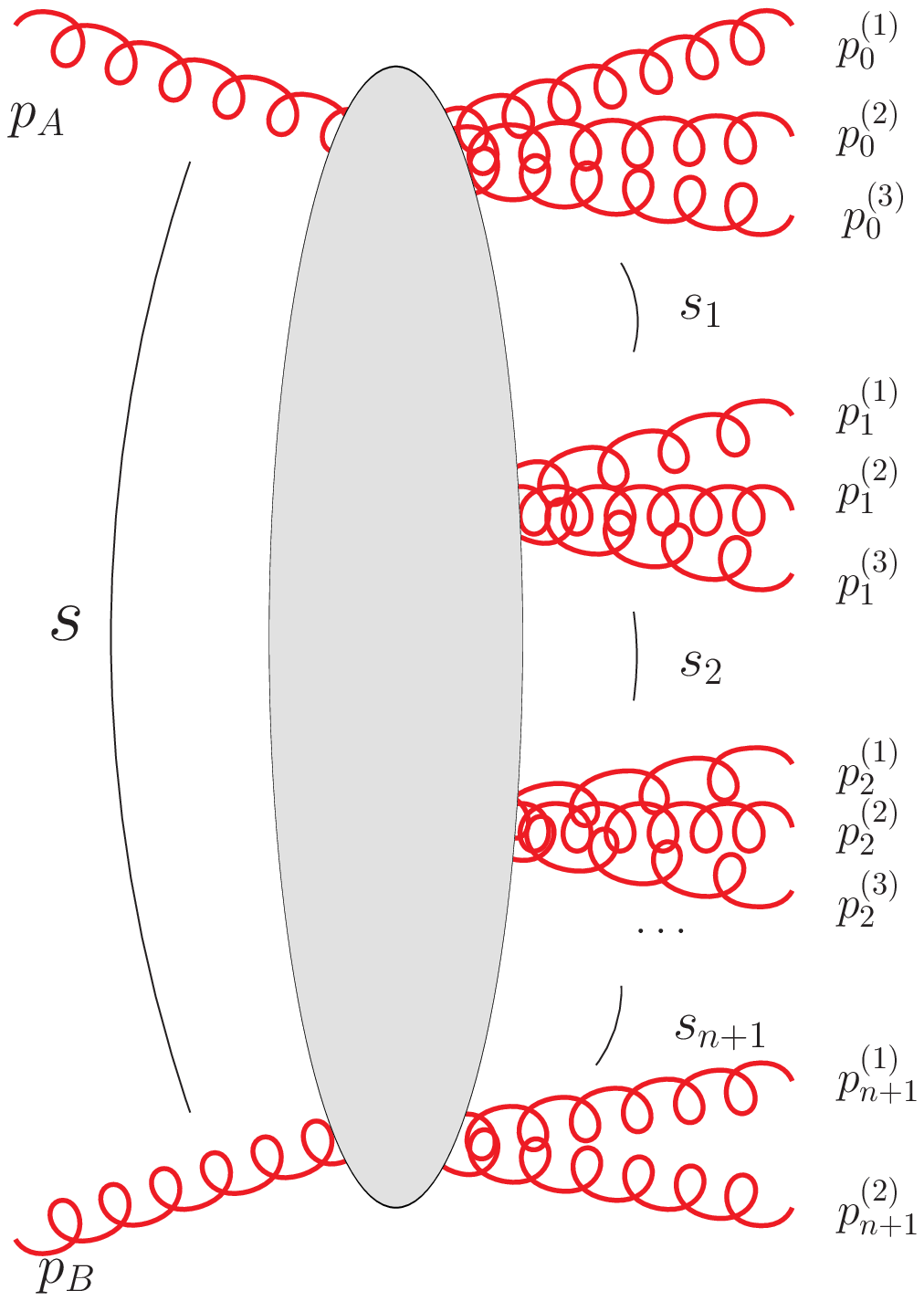}} 
 \parbox{5cm}{\includegraphics[height=3cm]{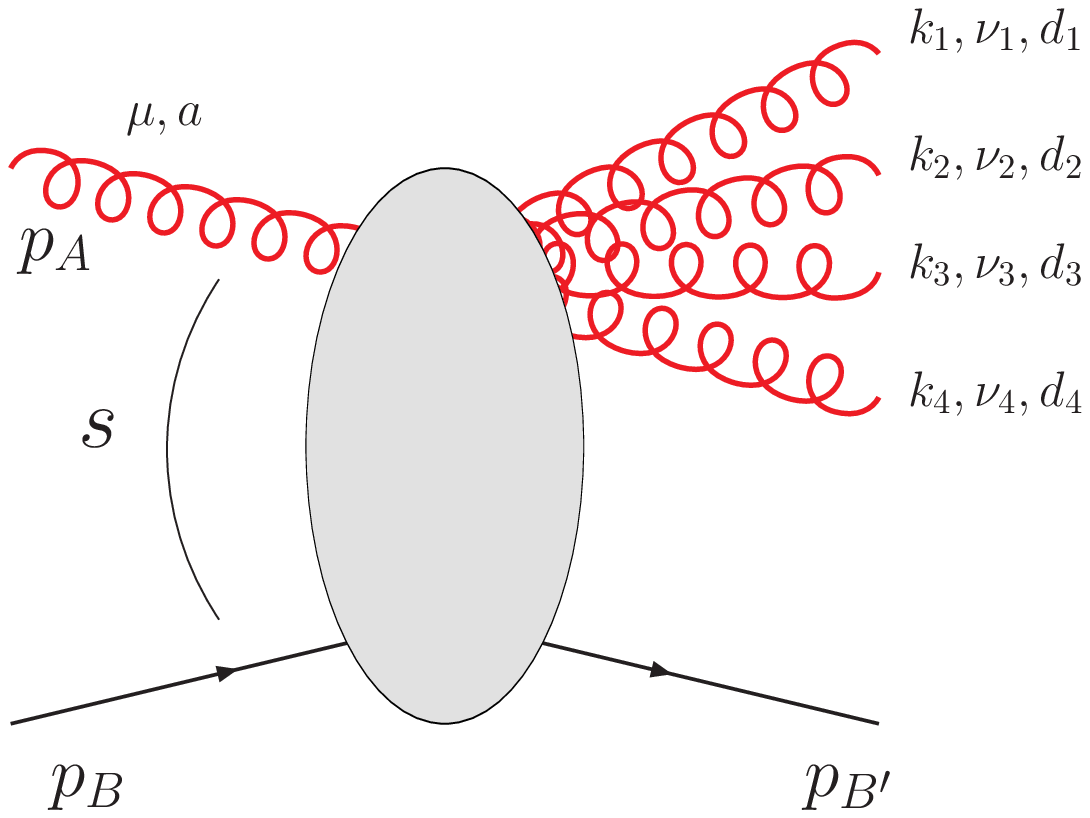}}  \\
\parbox{5cm}{\center (a)}\parbox{5cm}{\center (b)}
  \caption{\small a: The Quasi-Multi-Regge-Kinematics b: Quasi-elastic scattering of the quark B on the gluon A, where the gluon fragments into a cluster of gluons, nearby in rapidity.}
  \label{fig:qmrk}
\end{figure}
To obtain the relevant tree-amplitudes of high-energy QCD beyond the
LLA, it is needed to extend the previously discussed MRK to the
Quasi-Multi-Regge-Kinematics (QMRK) \cite{Fadin:1989kf}, where the
final state particles are separated into several groups of an
arbitrary number of particles with fixed invariant mass.  Each of
these groups of particles  is produced in the
Multi-Regge-Kinematics with respect to each other and
Eq.(\ref{multiregge}) holds now for groups rather than single
particles. In particular
 (see Fig.\ref{fig:qmrk}a for the nomenclature). 
\begin{align}
  \label{eq:qmrk_form}
k_r & =  \sum_{j} p_r^{(j)} & k_r^2& = \text{fixed}, & r &=0, \ldots n+1 ,
\end{align}
with
\begin{align}
  \label{multireggestar}
  s &= (p_A + p_B)^2 \gg s_r = (k_r + k_{r-1})^2 \gg {\bm{q}}_r^2 = -t_r ,
&
k_r = q_{r+1} - q_r.
\end{align}
 In terms of rapidity,  produced particles  are grouped into rapidity clusters where particles inside of each cluster
are close in rapidity with respect to each other, whereas each cluster
is separated significantly in rapidity from the other clusters.

Similar to the amplitudes of the LLA, we attempt to arrive in the
following at a description, where the interaction between the
different clusters is mediated by the exchange of a single reggeized
gluons, which couples to the produced particles by means of gauge
invariant effective vertices.

For simplicity, one starts with the quasi-elastic process where the
gluon $A$ fragments into a number of gluons, Fig.\ref{fig:qmrk}b.
Omitting polarization vectors of external gluons, we start with the
following ansatz for the quasi-elastic amplitude (following closely
\cite{Lipatov:1995pn}) with production of $n$ gluons in the
fragmentation region of the gluon $A$:
\begin{align}
  \label{eq:quasi_ansatz}
\mathcal{M}^{\mu\nu_1 \ldots \nu_n}_{ad_1 \ldots d_n} = \Gamma^{\mu\nu_1 \ldots \nu_n+}_{ad_1 \ldots d_nc} \frac{1}{q^2} p_B^- gT^c_{BB'} \delta_{\lambda_B\lambda_{B'}}.
\end{align}
In the above formula, $\nu_i$ and $d_i$, with $i = 1, \ldots , n$,
denote Lorentz and color indices of the produced gluons, whereas $\mu$
and $a$ belongs to the initial gluon $A$.  We begin with the case
where two gluons are produced in the fragmentation region and attempt
to construct the effective 3-Gluon-Reggeon vertex building all possible
combinations of the usual three- and four gluon vertices and the
induced vertex Eq.~(\ref{inducedvertex11}).  However it turns out that
the set of these diagrams is not sufficient to build a gauge invariant
vertex and it is necessary to introduce a further induced vertex,
\begin{align}
  \label{inducedvertex22}
\Delta^{\mu\nu_1\nu_2+}_{ad_1 d_2c} (p_A^+, k_1^+, k_2^+)=
 \parbox{3.2cm}{\includegraphics[width = 3cm]{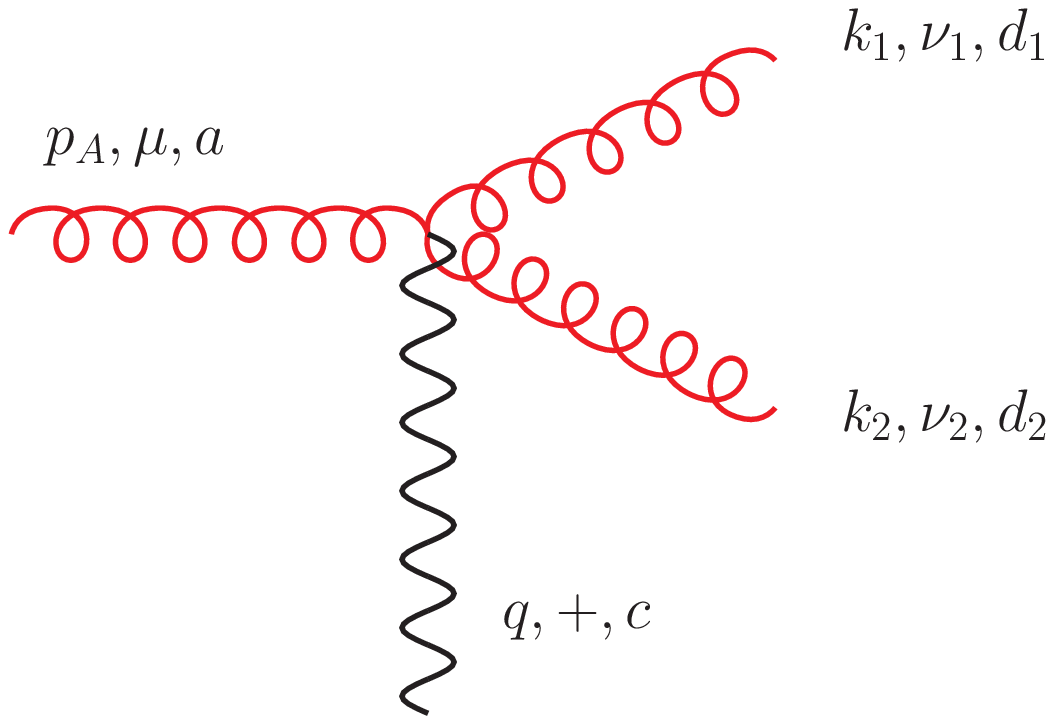}} &= ig^2 {{q}}^2 \left(\frac{T^d_{d_2d_1} T^c_{ad}}{k_2^+ p_A^+} + \frac{T^d_{d_2a}T^c_{d_1d} }{k_2^+ k_1^+}\right) (n^+)^{\mu} (n^+)^{\nu_1} (n^+)^{\nu_2},
\end{align}
which we call in the following the induced vertex of the second order,
in contrast to the induced vertex of the first order
Eq.~(\ref{inducedvertex11}).  The vertex Eq.~(\ref{inducedvertex22}) 
can  be shown to be Bose-symmetric making use of the
Jacobi-identity
\begin{align}
  \label{eq:jacobi}
T^d_{d_2a} T^c_{d_1d} - T^d_{d_2d_1} T^c_{ad} = T^d_{d_1a} T^c_{d_2d},
\end{align}
and  conservation of 'plus'- momenta in the quasi-elastic regime:
\begin{align}
  \label{eq:mom_consv}
p_A^+ + k_1^+ + k_2^+ = 0.
\end{align}
The complete 3-Gluon-Reggeon vertex arises  as the sum of the following contributions
\begin{align}
  \label{eq:gggr}
\Gamma^{\mu\nu_1\nu_2+}_{ad_1 d_2c} & =
  \parbox{2cm}{\includegraphics[width = 2cm]{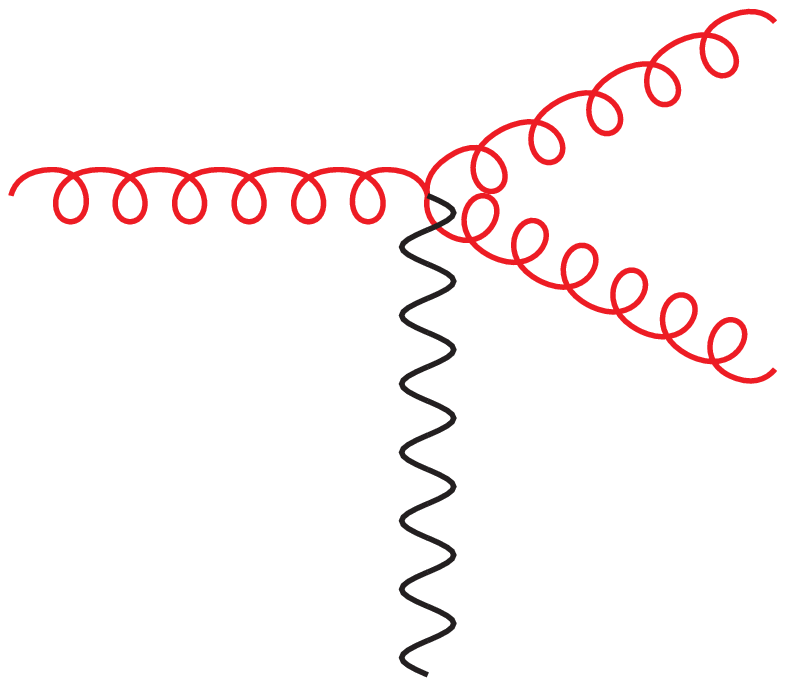}} 
  +
  \parbox{2cm}{\includegraphics[width = 2cm]{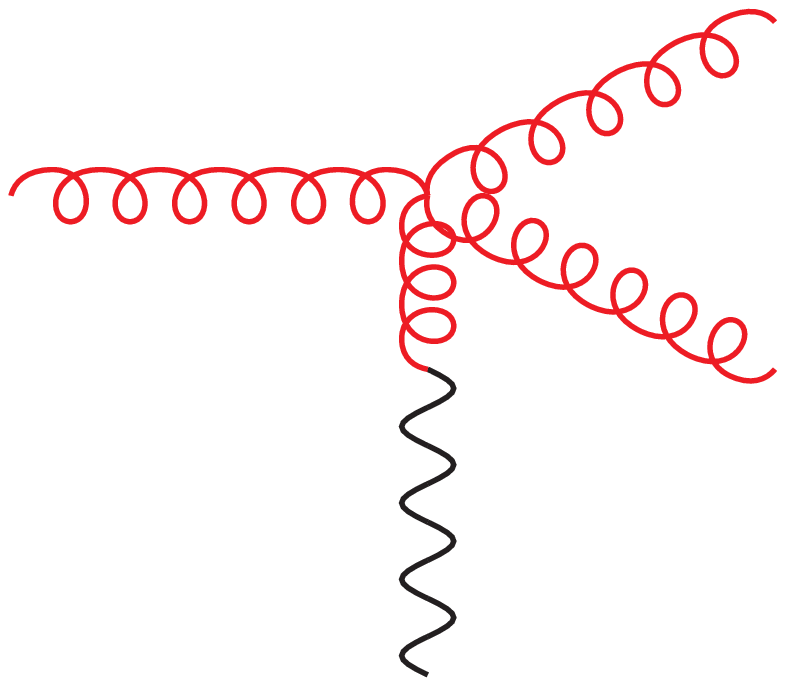}}
+
\parbox{2cm}{\includegraphics[width = 2cm]{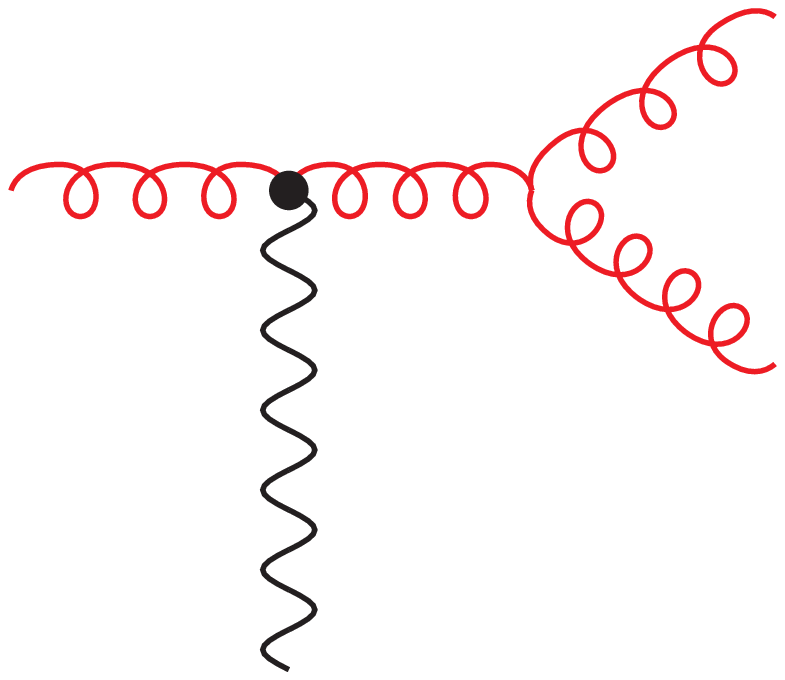}}
+
\parbox{2cm}{\includegraphics[width = 2cm]{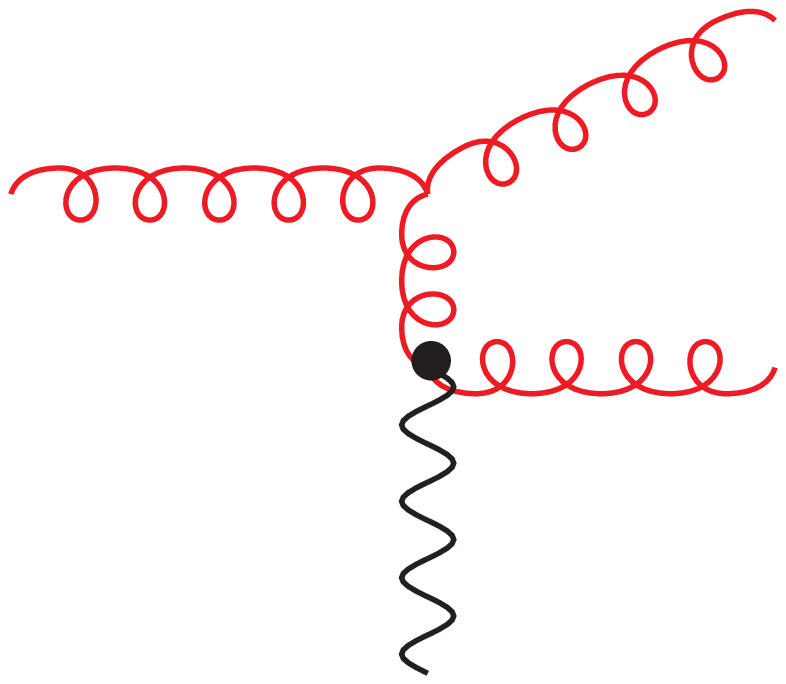}}
+
\parbox{2cm}{\includegraphics[width = 2cm]{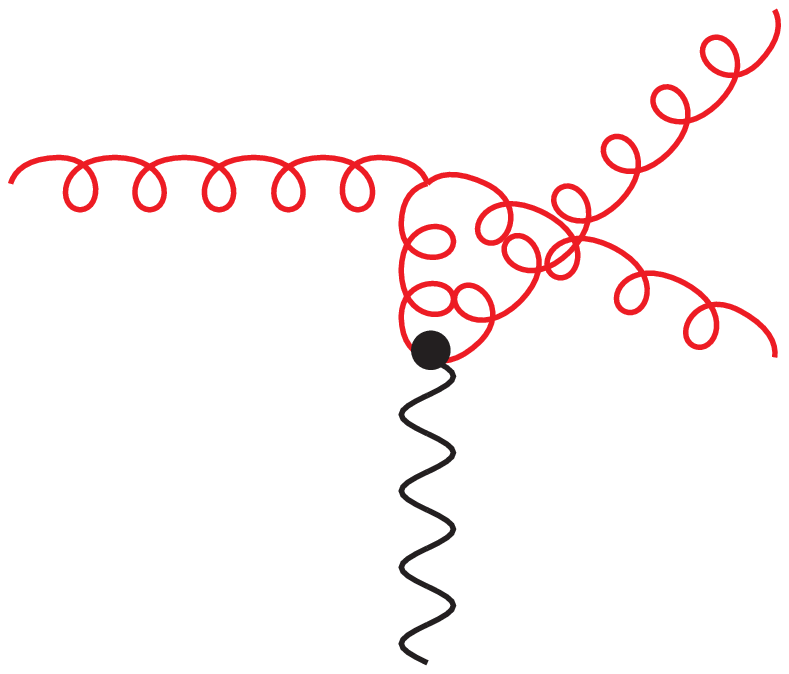}},
\end{align}
where we used the vertex Eq.~(\ref{eq:ggr_diagram}) to keep the
expression compact. This 3-Gluon-Reggeon vertex can be verified to be
gauge invariant \cite{Lipatov:1995pn,Antonov:2004hh}
\begin{align}
  \label{eq:gggr_gaugeInv}
(p_A)_{\mu} \Gamma^{\mu\nu_1\nu_2+}_{ad_1 d_2c}  = (k_i)_{\nu_i} \Gamma^{\mu\nu_1\nu_2+}_{ad_1 d_2c}  &= 0, &  i &= 1,2.
\end{align}
For the general case $n > 2$ it can be shown that gauge invariance
requires to introduce an infinite set of induced vertices which
satisfy the following recurrence relation
\begin{align}
  \label{eq:reccurrence}
&\Delta^{\nu_0\nu_1 \ldots \nu_r+}_{d_0d_1 \ldots d_n c} (k_0^+, k_1^+, \ldots, k_r^+) 
=
\notag \\ 
& \qquad =
\frac{(n^+)^{\nu_r}}{k_r^+} \sum_{i=0}^{r-1} T^{a}_{a_ra_i} \Delta^{\nu_0\nu_1 \ldots \nu_{r-1}+}_{d_0d_1 \ldots d_{i-1}d_{i+1} \ldots d_n c} (k_0^+,  \ldots, k^+_{i-1},k_i^+ + k^+_r,  k^+_{i+1} , \ldots, k_{r-1}^+) .
\end{align}
These vertices can be shown to be invariant under simultaneous
exchange of color, momenta and polarization of the gluons  making use
of the Jacobi-identity Eq.(\ref{eq:jacobi}) and the constraint
\begin{align}
  \label{eq:constrain}
\sum_{i=0}^r k_i^+ = 0.
\end{align}
We therefore arrive at a set of diagrammatic rules that allow to
construct production amplitudes in the quasi-elastic region,
corresponding to Fig.~\ref{fig:qmrk}b.  As far as the origin of the
newly introduced induced vertices is concerned, one should think of
them, similar to Eq.~(\ref{eq:prod_vertex}), as radiative corrections
that originate from particles close in rapidity to the particle B.

Furthermore, using symmetry arguments, fragmentation in the region of
the particle $B$, is described by a similar set of diagrams, with
'plus'-indices interchanged by the corresponding 'minus'-indices.  To
describe then production of gluons at central rapidities within the
Quasi-Multi-Regge-Kinematics, Fig.~\ref{fig:qmrk}, the usual
QCD-vertices need to be combined with induced vertices with both
'plus' and 'minus' indices.  That these combinations lead indeed to
gauge invariant production vertices has been verified explicitly for
up to three produced gluons in \cite{Lipatov:1995pn,Antonov:2004hh}.
We therefore arrived at a set of diagrammatic rules which allows to
build all tree-amplitudes within the QRMK.  In particular, these rules
yield gauge invariant vertices, which describe the production of
$s$-channel gluons, whereas the interaction between clusters of these
gluons is mediated by the (bare) reggeized gluon, which is invariant
under local gauge transformations.

\section{The effective action of high energy QCD}
\label{sec:effact}

It is then possible to derive an effective action  which reproduces the above set of rules. 
It has been  formulated in \cite{Lipatov:1995pn} and is given by
\begin{align}
  \label{effaction}
  S_{\text{eff}} &= \int \text{d}^4 x (\mathcal{L}_{\text{QCD}}(v_\mu, \psi) + \mathcal{L}_{\text{ind}} (v_\pm,  A_\pm)).
\end{align}
It contains apart from the fields $v_\mu(x)$ and
$\psi(x)$ corresponding to gluons and quarks respectively the reggeized gluon field
$A_\pm(x)$. As reggeized gluons connect
clusters with significantly different rapidities, these fields obey the
constraint
\begin{align}
  \label{eq:1kin_constraint}
\partial_\mp A_\pm = 0.
\end{align}
Furthermore they are  invariant, $\delta A_\pm = 0$,
under infinitesimal, local gauge transformations
\begin{align}
  \label{eq:gauge_gluon}
\delta v_\mu &= [D_\mu, \chi], & \delta \Psi &= -\chi \Psi,
\end{align}
where $D_\mu$ denotes the covariant derivative and $\chi$ the parameter of the gauge transformations which decreases at $x \to \infty$. On the other hand, the fields $A_{\pm}$  belong  globally to the  adjoint representation of $SU(N_c)$ and transform for constant $\chi$ as 
\begin{align}
  \label{eq:gauge_inv_rg2}
\delta A_\pm = g[A_\pm, \chi].
\end{align}
The Lagrangian of the
effective action consists  besides of the usual QCD -Lagrangian of
an additional induced term which contains the kinetic term for the
reggeized gluon field and its coupling to gluons. 
 \begin{align}
   \label{eq:1efflagrangian}
\mathcal{L}_{\text{ind}}  (v_\pm,  A_\pm)
=&\tr\left[\left(A_-(v) - A_- \right)\partial^2_\sigma A_+\right]
+\tr\left[\left(A_-(v) - A_- \right)\partial^2_\sigma A_+\right]
\end{align}
where the induced Reggeon-gluon vertices take the following form
\begin{align}
\label{eq2:efflagrangian}
A_\pm(v) =
&
-\frac{1}{g}\partial_\pm U(v_\pm)
 = 
-\tr\frac{1}{g} \partial_\pm \mathcal{P}\exp\bigg(-\frac{1}{2} g \int_{-\infty}^{x^\pm}dx'^\pm v_\pm(x')\bigg)   \notag \\
=&  v_\pm - g  v_\pm(1/\partial_\pm) v_\pm + g^2 v_\pm(1/\partial_\pm) v_\pm(1/\partial_\pm) v_\pm - \ldots.
 \end{align}
In the above expression, $v_\mu(x)$ and $A_\pm(x)$ are anti-Hermitian matrices in the fundamental representation of $SU(N_c)$-Lie algebra:
\begin{align}
  \label{eq:anitherm}
v_\mu(x) & = -it^av^a_\mu(x), &
A_\pm(x) & = -it^aA^a_\pm(x), &
[t^a,t^b] = if^{abc}t^c, \quad \tr(t^at^b)= \frac{1}{2}.
\end{align}
The above effective action  describes the interaction of reggeized
gluons with quarks and gluons, local in rapidity. In terms of the
amplitudes Fig.~\ref{fig:qmrk} locality in rapidity means that the
interactions between particles and reggeized gluons is always
restricted to a single rapidity cluster, whereas the interaction
between clusters is mediated by reggeized gluons alone. To make this
statement more quantitative, one further introduces a parameter $\eta$
which can be numerically big, but which is always significantly
smaller than the rapidity-interval over which the complete amplitude
extends $\Delta Y = \ln s \gg \eta$. Interactions between particles
and reggeized gluons are then restricted to a
rapidity interval $(Y_0 - \eta, Y_0 + \eta )$. Reggeized gluon fields
mediate the interaction over rapidity intervals
larger than $\eta$ and  never occur within
any rapidity cluster;  their propagator connects only parts of the
amplitude which differ in their relative rapidity more than the
parameter $\eta$. For the interaction of particles inside of a
cluster, $\eta$ takes therefore the role of an ultra-violet cut-off in
the relative longitudinal momenta, while it occurs as an infra-red
cut-off for the interaction between neighboring clusters. For physical
observables, the dependence on $\eta$ has  to cancel in
analogy to the dependence on a factorization scale in hard processes.

Due to the presence of a  term linear in $v_\pm$, the Euler-Lagrange
equations for the effective action have even in perturbation theory a
non-trivial classical solution $v_\pm = A_\pm + \mathcal{O}(g)$. The complete
classical solution is so far only known as a perturbative series in
$g$ \cite{Lipatov:1996ts}. In the following take the
perspective of \cite{Antonov:2004hh} and include into the
Feynman-rules a vertex that describes the direct transition between
reggeized gluons and gluons. We therefore do not subtract the gluon
field $v_\pm$ from the reggeized gluon $A_\pm$, but solve the
equations of motions perturbatively.

\subsubsection*{Feynman rules of the effective action}
The Feynman rules of the effective action were initially  given in
\cite{Antonov:2004hh}. However as there occurred a mixing of different
conventions, we find it useful to state the relevant expressions her
explicitly another time. Our Feynman-rules have been derived from the
Lagrangian according to the conventions of \cite{Peskin:1995ev} and
all momenta are taken to be in-going. The Feynman-rules of the
QCD-part of the effective Lagrangian can be obtained for instance from
\cite{Peskin:1995ev}. There are
an infinite number of induced vertices. In the following we state them
explicitly up to $\mathcal{O}(g^3)$, which covers all induced vertices
that will be needed in this thesis. With ${\bm q}^2 = -q^2_\perp$
denoting the Euclidean, two-dimensional squared momentum of the
reggeized gluon, the induced vertex of the zeroth order,
which describes the  direct transition of a QCD-gluon into a
reggeized gluon, is given by:
\begin{align}
  \label{inducedvertex0}
  \parbox{1.3cm}{\includegraphics[height = 2cm]{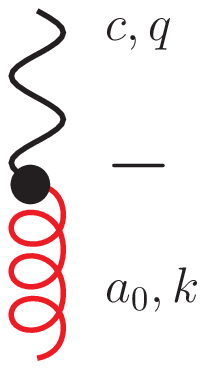}} &=\Delta^{\nu_0+}_{a_0 c} =  -i {\bm{q}}^2 \delta^{a_0 c} (n^-)^{\nu_0} ,
 &
 k_0^-  &= 0,
\end{align}
where $\nu_0$ is the polarization of the gluon. For the induced
  vertex of the first order we obtain:
\begin{align}
  \label{inducedvertex1}
  \parbox{3cm}{\includegraphics[height = 2cm]{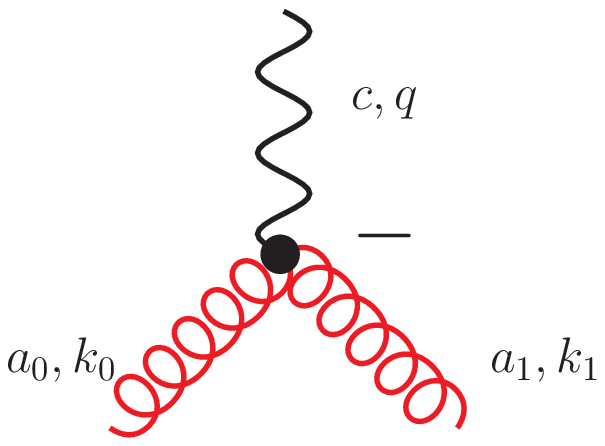}} &= 
\Delta^{\nu_0\nu_1+}_{a_0a_1 c} =
g{\bm{q}}^2 f^{a_0a_1 c}\frac{1}{k_0^-}(n^-)^{\nu_0}(n^-)^{\nu_1}
, &
 k_0^- + k_1^- &= 0,
\end{align}
and for the induced vertex of the second order,
\begin{align}
  \label{inducedvertex2}
  \parbox{2.4cm}{\includegraphics[height = 2cm]{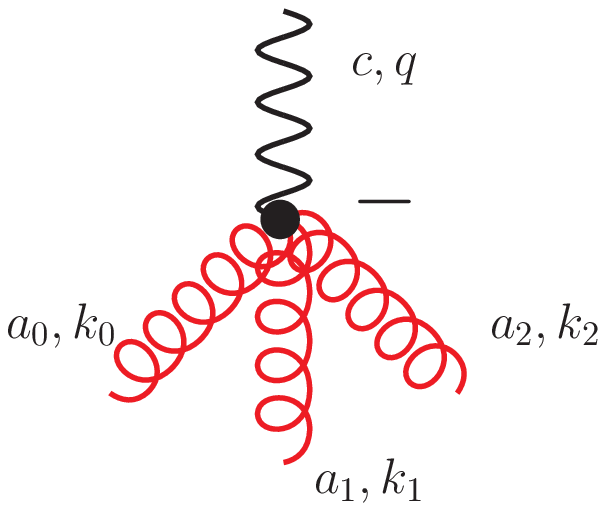}} &= 
\Delta^{\nu_0\nu_1\nu_2+}_{a_0a_1 a_2c} = 
ig^2 {\bm{q}}^2 \left(\frac{f^{a_2a_1 a} f^{a_0ac}}{k_2^- k_0^-} + \frac{f^{a_2a_0 a} f^{a_1ac}}{k_2^- k_1^-}\right) (n^-)^{\nu_0} (n^-)^{\nu_1} (n^-)^{\nu_2}
, \notag \\& \quad
 k_0^- + k_1^- + k_2^- = 0.
\end{align}
For the  induced vertex of the third order one finds
\begin{align}
  \label{eq:inducedvertex31}
\parbox{3cm}{\includegraphics[height = 2cm]{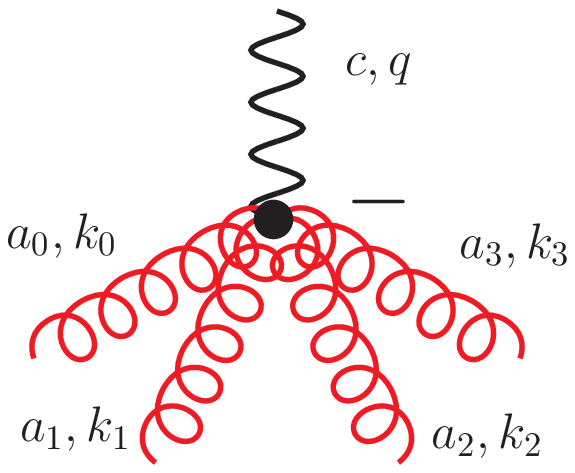}} =  \Delta^{\nu_0\nu_1\nu_2\nu_3+}_{a_0a_1 a_2a_3c},
\end{align} 
with
\begin{align}
  \label{inducedvertex32}
 \Delta^{\nu_0\nu_1\nu_2\nu_3+}_{a_0a_1 a_2a_3c}  = g^3 {\bm{q}}^2  
\bigg[
&
 \frac{f^{a_3a_2 c_2}}{k_3^-}
\bigg(
\frac{f^{c_2a_0c_1} f^{a_1c_1c}}{(k_0^- + k_1^-)k_1^-} +  \frac{f^{c_2a_1c_1} f^{a_0c_1c}}{(k_0^- + k_1^-)k_0^-}
\bigg) 
\notag \\
+& 
\frac{f^{a_3a_0 c_2}}{k_3^-}
\bigg(
\frac{f^{c_2a_1c_1} f^{a_2c_1c}}{(k_2^- + k_1^-)k_2^-} +  \frac{f^{c_2a_2c_1} f^{a_1c_1c}}{(k_2^- + k_1^-)k_1^-}
\bigg)
\notag \\
+ &
 \frac{f^{a_3a_1 c_2}}{k_3^-}
\bigg(
\frac{f^{c_2a_0c_1} f^{a_2c_1c}}{(k_2^- + k_0^-)k_2^-} +  \frac{f^{c_2a_1c_1} f^{a_0c_1c}}{(k_2^- + k_0^-)k_2^-}
\bigg)
\bigg] (n^-)^{\nu_0} (n^-)^{\nu_1}  (n^-)^{\nu_2}  (n^-)^{\nu_3} ,
\end{align}
and
\begin{align}
   k_0^- + k_1^- + k_2^-+  k_3^- = 0.
\end{align}
Note that in \cite{Antonov:2004hh} a general formula for the induced
vertex of the n-th order has been given. Above we stated only the
induced vertices for the Reggeon-field $A_-(x)$. There exists also the
same type of vertices for the field $A_+(x)$. They are obtained by
replacing all minus- by plus-labels. Besides the induced vertices, we
also obtain from the effective action as a new element the propagator
of the bare reggeized gluon
\begin{align}
  \label{reggeon-prop}
 \parbox{1cm}{\includegraphics[height=1.7cm]{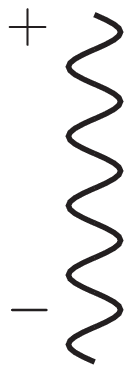}}  = \frac{i/2}{{\bm{q}}^2}
\end{align}
where the longitudinal part $q^+q^-$ vanishes due to the constraint
$\partial_\pm A_{\mp} = 0$.  As mentioned
before, the propagator of the reggeized gluon connects only parts of
the amplitude which are significantly separated in rapidity. With the
factorization parameter $\eta$ this constraint can be formulated by
supplementing the propagator of the reggeized gluon with  $\theta(Y_1 - Y_2 - \eta)$ where $Y_{1,2}$ are the rapidities
of the clusters which are connected by the reggeized gluon.

\section{Virtual corrections from the effective action}
\label{sec:virtual}

The above effective action  can
be used to determine tree amplitudes with Quasi-Multi-Regge-Kinematics
to arbitrary accuracy. In particular the production vertices needed
for the derivation of the BFKL-equation at NNLA have been derived in
\cite{Antonov:2004hh} and they have been used for the construction of
real corrections in \cite{Bartels:2008ce}.  However the potential
region of applicability of the effective action should be clearly
beyond the derivation of tree-amplitudes. Indeed it promises to be a
powerful tool that allows for the determination of the Regge-limit of
complete amplitudes, including virtual corrections.  In particular the
effective action should enable us to determine loop corrections to the
Regge-trajectory and to the vertices describing coupling of the
reggeized gluons to QCD-particles, which are required for the
determination of higher order corrections to the BFKL-equation.
Furthermore it should be possible to determine directly from the
effective action amplitudes where two and more reggeized gluons are
exchanged in the $t$-channel.  In particular, the effective action
seems to provide a self-consistent approach for unitarization of the
BFKL-Pomeron.

Attempting such calculations starting from the Feynman rules of the
effective action one soon encounters a number of complications which
allow not for a straight forward determination of the desired
quantities. All these complications are connected with the
determination of loop-corrections and therefore do not occur for
tree-amplitudes.  The
following points require clarification, in order to be able to use the
effective action for the determination of virtual corrections:
\begin{itemize}
\item First, one needs an adequate regularization that ensures
  locality in rapidity of the amplitudes of the effective action: For
  real particle production, the QMRK fixes the rapidity of the produced
  particles, which allows to organize produced particles into local
  rapidity clusters. Loop integrals extend on the other hand over the
  full range of rapidities, which can lead to a violation of the
  locality constraint.
  
\item Second, the operators that yield the induced vertices are
  singular and a suitable prescription for the occuring poles is
  needed: As far as the production of real particles in the QMRK is
  concerned, these poles are fixed to non-zero values by the QMRK,
  while integrating over longitudinal momenta, one enters naturally
  also the region, where the operators become singular.
  
\item Third, amplitudes in the Regge-limit are known to have a certain
  analytical structure.  This particularly concerns the occurrence of
  discontinuities in Mandelstam-variables. This point is closely
  connected to the ones mentioned above.  Starting from a too simple
  prescription it turns out that certain discontinuities are misses
  and one obtains incorrect results for more complicated quantities
  \cite{Hentschinski:2008rw}.
  
\item Fourth, attempting to determine integrations of loops that
  contain reggeized gluons, one finds integrals over light-cone
  momenta which are not convergent. It turns out that at one-loop a
  simple cut-off can be applied, corresponding to a principal values
  description \cite{Hentschinski:2008rw}, while for the exchange of
  more than two reggeized gluons this procedure fails.
  
\item Fifth, the determination of NLLA-corrections requires to
  consider renormalization of the effective theory. This implies to
  supplement the effective action by suitable counter-terms.

\end{itemize}
In the following we address all of these points besides the last one,
which is left for future studies.

\chapter{The elastic amplitude}
\label{cha:2to2}

\begin{figure}[htbp]
  \centering
  \parbox{4cm}{\includegraphics[width=4cm]{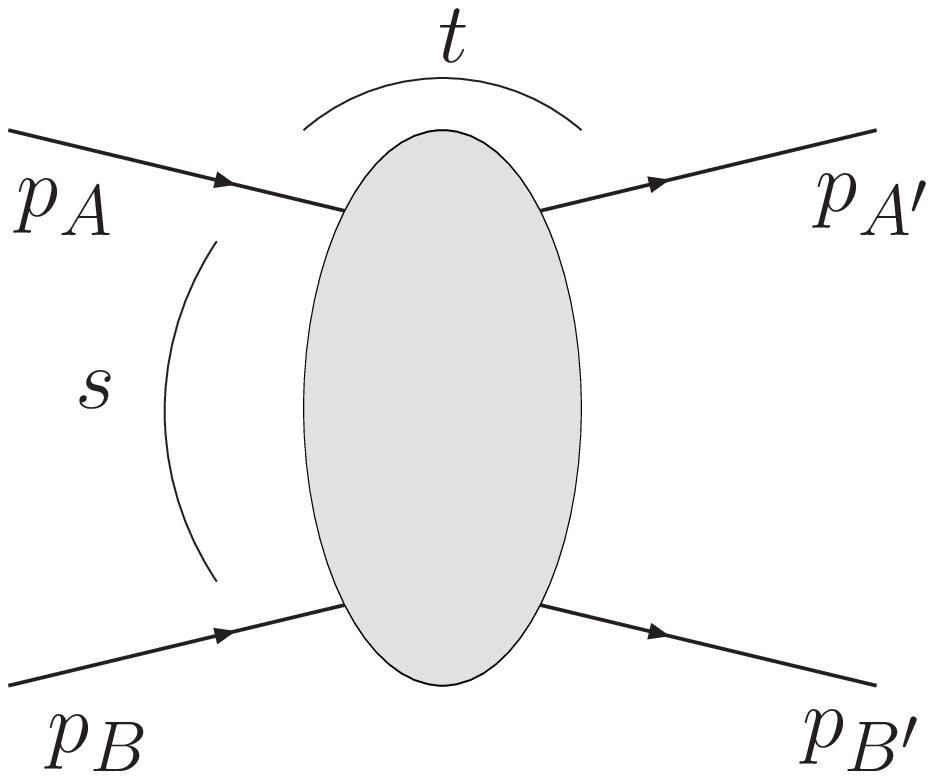}}
  \caption{\small The elastic  quark-quark scattering amplitude. }
  \label{fig:elastic}
\end{figure}

As a start of our studies of longitudinal integrations within the effective action, we  discuss the Regge limit $s \gg -t$ of the elastic
scattering  amplitude, Fig.\ref{fig:elastic}. 
In QCD, within the Leading and Next-to-Leading Logarithmic
Approximation (LLA) and (NLLA), the elastic amplitude or 4-point
amplitude is known to reggeize in the Regge-limit i.e. the interaction
between the scattering quarks $A$ and $B$ (and similarly for gluons)
is mediated by a single reggeized gluon.  For elastic-scattering
amplitudes with quantum numbers that due not allow for the exchange of
a single reggeized gluon,  the interactions is mediated by two or more
 reggeized gluons.  In all these processes, the reggeized
gluon is always associated with the 'Regge-limit' of a certain
sub-4-point-amplitude.  The elastic scattering amplitude is therefore
not only an interesting object to be studied within the effective
theory, but it constitutes also one of the basic building blocks of
the effective theory, as only the associated 4-point amplitude allows a proper
definition of the reggeized gluon.

The outline of this chapter is the following: In
Sec.\ref{sec:22negsig} we study reggeization of the gluon within the
effective action. We will derive a prescription for (longitudinal)
loop integrals of gluon-loops, to which at least one reggeized gluon
couples and demonstrate reggeization of the gluon in the effective theory by
resumming a certain class of diagrams. In
Sec.\ref{sec:tworeggeon_negsig} we  consider the  state of two
reggeized gluons as a first example for interactions between reggeized
gluons, which will lead us to the BFKL-equation.

The kinematics of the process that is studied here is the following:

As depicted in Fig.\ref{fig:elastic}, we consider scattering of two
highly energetic quarks, with almost light-like momenta $p_A$ and
$p_B$ which are close to the minus and the plus light-cone.  In 
particular $p_A^+  \gg p_A^- $
and $p_B^-  \gg p_B^+  $. The
two quarks have therefore mass $m_A^2 = p_A^+p_A^-$ and $m_B^2 =
p_B^+p_B^-$ which within the LLA can be taken to be of similar size of
the absolute value of the momentum transfer $\sqrt{|t|}$.  In
particular, to a good approximation we have $s = (p_A + p_B)^2 \simeq
p_A^+p_B^-$.  The error of this approximation is of the order $m^2/s$
and as the general accuracy of the effective action is not higher than
$1/s$, those corrections will be frequently neglected, whenever it
facilitates the calculations.
The difference in rapidity of quark A,  $Y_A = 1/2\ln(p_A^+/p_A^-)$, and
quark B, $Y_B = -1/2\ln(p_B^-/p_B^+)$, 
is given by 
\begin{align}
\label{eq:diff_rap}
  Y_{AB} = \frac{1}{2} \ln\left(\frac{p_A^+p_B^-}{p_A^-p_B^+}\right) =
\ln \left(\frac{s}{m_Am_B}\right),
\end{align}
where the last identity holds with power accuracy and is therefore
exact within the effective action. In the Regge-limit $s \gg -t,
m_A^2, m_B^2$ the two quarks are therefore separated by a large
difference in rapidity.  Furthermore we introduce a four-momentum $q =
p_A - p_A' = p_B' - p_B$ with $t = q^2$ which yields the momentum
transfer.  In the Regge-limit, $q^2 = q^2_\perp = - {\bm q}^2$, where
we indicate with bold letters two-dimensional, Euclidean momenta.  If
not indicated otherwise, we always use the Feynman-gauge in our
calculations.

 \section{ The reggeized gluon in the effective action }
 \label{sec:22negsig}

\begin{figure}[htbp]
   \centering
   \parbox{4cm}{\center \includegraphics[height=2cm]{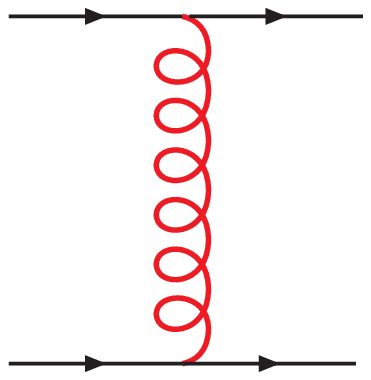}}
   \parbox{4cm}{\center \includegraphics[height=2cm]{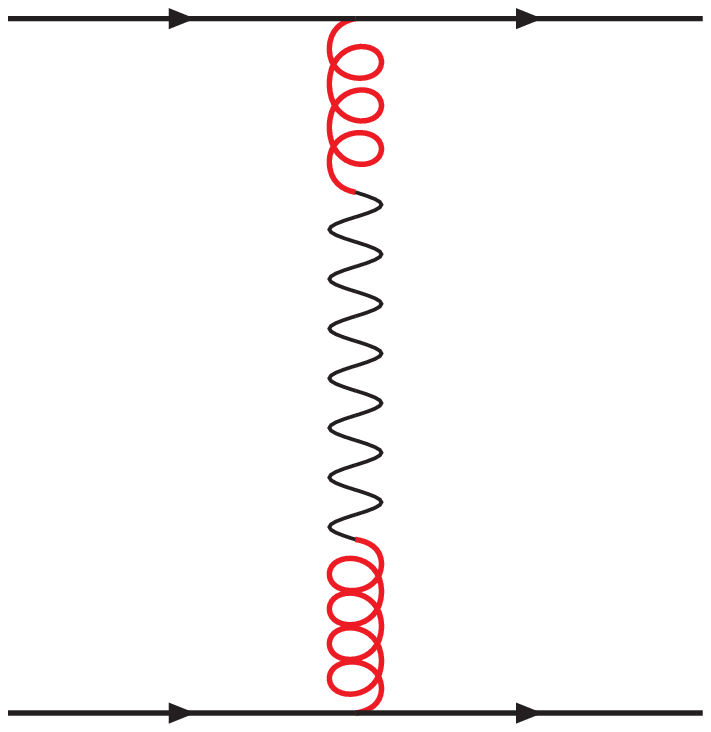}} \\
\parbox{4cm}{\center (a)} \parbox{4cm}{\center (b)} 
   \caption{\small Tree-level diagrams in QCD and in the effective theory. For the effective theory we will take from now on the convention that the reggeized gluon couples directly to the quark in an apparent way.}
   \label{fig:tree-level}
 \end{figure}

 In QCD, the scattering quarks interact at tree-level by exchange of a
 single gluon, Fig.\ref{fig:tree-level}a. As the Lagrangian of the
 effective action contains the complete QCD-Lagrangian, this graph
 exists also in the effective theory. However in the effective theory,
 this graph comes with the restriction that the scattering quarks are
 close in rapidity. More accurate, introducing a factorization
 parameter $\eta$ which separates interactions local in rapidity from
 non-local ones, this graph describes only the scattering process, if
 the difference of the scattering quarks' rapidities, $Y_{AB}$, is
 smaller than $\eta$.  In the Regge-limit, however, $Y_{AB} > \eta$
 (for any meaningful value of $\eta$): The quarks are non-local in
 rapidity and they interact within the effective action not by a
 QCD-gluon, but by a reggeized gluon, Fig.\ref{fig:tree-level}b.
 This case is of course highly trivial, but it illustrates in a simple
 way the underlying principle of the effective theory: Interactions
 local in rapidity are described by standard QCD-degree-of-freedoms,
 whereas non-local interactions by reggeized gluons alone.
 
For later reference, and as the coupling of one gluon to one reggeized gluon has no kinematical meaning, it is useful to define the effective quark-reggeized gluon-quark vertex:
\begin{align}
  \label{eq:quark_reggeon}
\parbox{2cm}{\includegraphics[width=2cm]{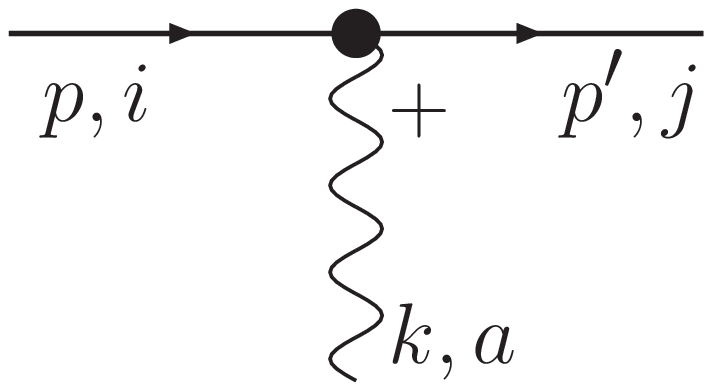}} &=
\parbox{2cm}{\includegraphics[width=2cm]{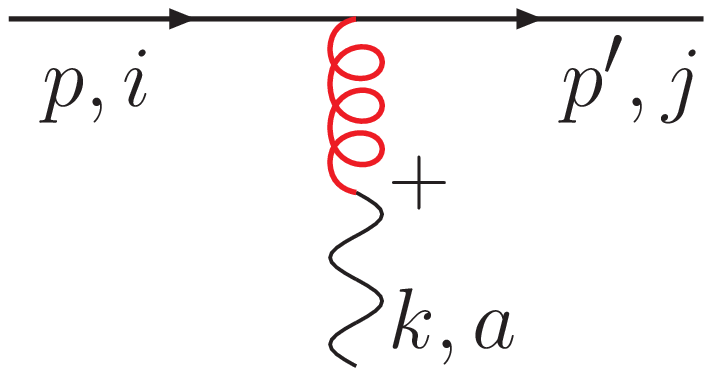}} = igt^a_{ij}\fdag{n^+}, && k^- = 0.
\end{align}
For on-shell quarks, the Quark-Quark-Reggeon vertex is in the Regge-limit given by
\begin{align}
  \label{eq:qark_reggeon_onshell}
\Gamma^{\pm,a \lambda\lambda'}_{QQR} & = 
 p_\pm \tilde{\Gamma}^{a \lambda\lambda'}_{QQR} =
\bar{u}_{\lambda'}(p')igt^a_{ij}\fdag{n^+} u_\lambda(p) \simeq   t_{ij}^a ig 2 p^\pm \delta_{\lambda\lambda'} .
\end{align}
which yields for the  tree-level amplitude in the Regge-limit
 \begin{align}
   \label{eq:22born}
  \mathcal{M}_{2 \to 2}^{ \text{tree}}(s, t) 
 &=
  -i\bar{u}_{A'} igt^a \gamma \cdot n^+ u_A \frac{-i/2}{q_\perp^2} \bar{u}_{B'} igt^a \gamma \cdot n^- u_B  \notag \\
&= 2 p_A^+p_B^- g t_{AA'}^c  \frac{1}{ -{\bm q}^2} g  t_{BB'}^c.
 \end{align}
In the following we shall discuss corrections to the
tree-level-diagram Eq.(\ref{eq:22born}). In doing so, we  focus on
corrections that yield large logarithms in the center of mass energy
$s$ and make use of simplifications due to the LLA.  In principle this  allows
to neglect any dependence on a transverse scale and similarly on the
factorization parameter  $\eta$. On the
other hand a consistent formulation of the effective theory requires
cancellation of any arbitrarily introduced scale and we will therefore
also demonstrated how such a cancellation can be achieved.

\subsection{Central rapidity diagrams and the gluon trajectory function}
\label{sec:traj}
\begin{figure}[htbp]
  \centering
   \parbox{3cm}{\center \includegraphics[height=3.5cm]{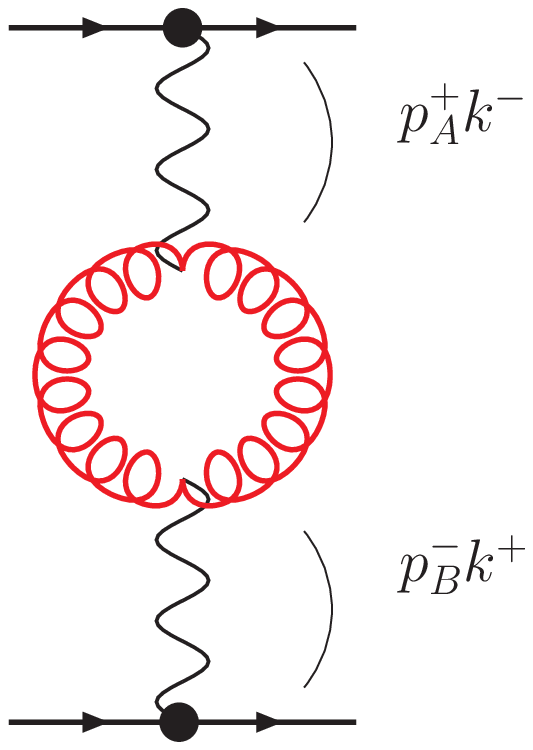}} 
   \parbox{2cm}{$\,$}
   \parbox{3cm}{\center \includegraphics[height=2.9cm]{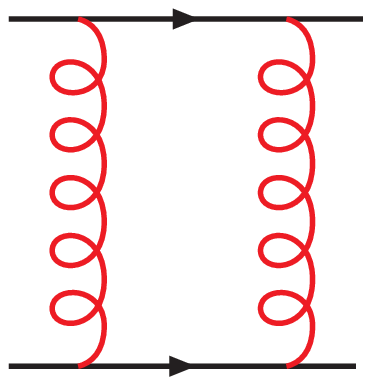}}
   \parbox{3cm}{\center \includegraphics[height=2.9cm]{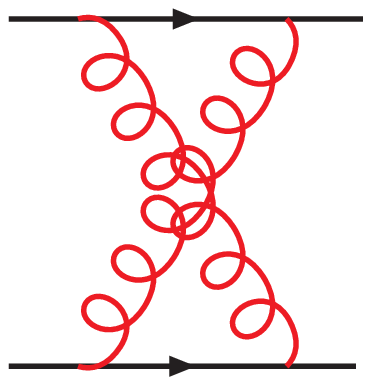}}
   \\
   \parbox{3cm}{\center CR} \parbox{2cm}{$\,$}  \parbox{3cm}{\center QCD1} \parbox{3cm}{\center QCD2}
  \caption{\small The central rapidity diagram of the effective action, where both reggeized gluons couple to gluon by an induced vertex. It arises for a particular kinematical limit out of the two QCD-diagrams to the right }
  \label{fig:one_reggeon}
\end{figure}

In this paragraph we   focus on the  central rapidity diagram CR in
 Fig.\ref{fig:one_reggeon}, where both  reggeized gluons couple to the gluon loop by an induced vertex. As we will see, this diagram yields  the gluon-trajectory-function, together with a large logarithm in $s$. 
That this type of diagram yields the gluon-trajectory and that 
their resummation leads to reggeization of the gluon, has been already
noted in \cite{Lipatov:1995pn}.  Apart from the diagram CR,
Fig.~\ref{fig:one_reggeon}, there are also central rapidity diagrams,
where the reggeized gluons couples to the gluon-loop by a three- or
four-gluon vertex, see Fig.\ref{fig:CRfurther}. Their loop integral
does not contribute within the LLA. They will be shortly addressed at
the end of this paragraph.  Using the unregulated Feynman rules  of the
effective action, see  Sec.~\ref{sec:effact}, the diagram CR of Fig.\ref{fig:one_reggeon} yields
the following expression:
 \begin{align}
   \label{eq:traj1}
 i \mathcal{M}^{\text{CR}}(s,t) =    s\tilde{\Gamma}_{QQR}^c    
   \!\!\int \!\!\frac{d k^+ dk^-}{2\pi} \!\! \int \!\!\frac{d^2 {\bm k}}{(2\pi)^3}     \frac{i/2}{{\bm{q}^2}}    \left(  \frac{g{\bm{q}}^2 f^{abc}}{-k^-}    \frac{ -i}{ k^2 + i \epsilon} \frac{-i}{(q -k)^2 + i \epsilon} \frac{g {\bm{q}}^2f^{abc'}}{k^+}
   \right) \frac{i/2}{{\bm{q}^2}}\tilde{\Gamma}_{QQR}^{c'}.
 \end{align}
 The term within the big bracket represents together with the integral
 the gluon loop, to which the reggeized gluons couple.  Note that the
 loop-integral in Eq.(\ref{eq:traj1}) contains implicitly an
 integration over the rapidity $Y_k= \ln (k^+/k^-)/2$ of the gluons
 inside the loop, which extends from minus to plus infinity.
 Furthermore, any rapidity-dependence inside of the integrand of
 Eq.(\ref{eq:traj1}) cancels and the expression is not restricted to
 central rapidities, but highly non-local in rapidity.  The derivation
 of the effective action on the other hand requires that interactions
 of usual QCD-degrees-of-freedom, quarks and gluons, are restricted to
 rapidity intervals $\Delta Y < \eta$, while any interactions, that
 extends over rapidity intervals $ \Delta Y > \eta$ is mediated by
 reggeized gluons alone.  For real particle production, from which the
 effective action has been originally derived, the rapidities of the
 produced particles are fixed and they are either close or not. For
 the loop integrals of virtual corrections on the other hand, we
 integrate at first over all rapidities which requires to implement a
 restriction to central rapidity values.  In principle this can be
 done by applying straight forward cut-offs to the
 rapidity/light-cone-momenta of the longitudinal loop integral.
 However this method has some drawbacks as it does not allow to
 include correctly phases: The imaginary part of the reggeized gluon
 is lost and therefore also its signature.

 To provide an appropriate prescription and regularization of the 
 integrals over longitudinal momenta $k^+$ and $k^-$, it is
 instructive to compare the above effective action diagram with the
 underlying QCD-amplitude, from which the effective action diagram
 arises.
 In the Regge-limit, at one loop, the relevant QCD-diagrams are given
 by diagrams QCD1 and QCD2,  Fig.~\ref{fig:one_reggeon},  with exchange of two $t$-channel gluons. Their sum is given by
\begin{align}
  \label{eq:QCD_raw}
i&\mathcal{M}_{\text{QCD}} =
\notag \\ &=
 \frac{1}{2}\int\!\!\! \frac{d^4 k}{(2\pi)^4}
\bar{u}_{A'}(p_A') 
\left(
            \frac{igt^{c_1}\gamma^{\mu_1}   i(\fdag{p_A} - \fdag{k})  igt^{c_2}\gamma^{\mu_2} }{(p_A - k)^2 + i\epsilon}    
            +   
            \frac{  igt^{c_2}\gamma^{\mu_2}  i(\fdag{p_A} - \fdag{q} + \fdag{k}) igt^{c_1}\gamma^{\mu_1}  }{(p_A -q + k)^2 + i\epsilon}  
\right)       u_A(p_A)
\notag \\ 
&  \qquad \qquad \qquad  \qquad \qquad \qquad\times
\frac{-ig_{\mu_1\nu_1}}{k^2 + i\epsilon}\frac{-ig_{\mu_2\nu_2}}{(q -k)^2 + i\epsilon} \times
\notag \\ & \qquad \qquad \qquad
\bar{u}_{B'}(p_B')  
\left(
            \frac{igt^{c_1}\gamma^{\nu_1}   i(\fdag{p_B} + \fdag{k})  igt^{c_2}\gamma^{\nu_2} }{(p_B + k)^2 + i\epsilon}    
            +   
            \frac{  igt^{c_2}\gamma^{\nu_2}  i(\fdag{p_B} )+ \fdag{q} - \fdag{k}) igt^{c_1}\gamma^{\nu_1}  }{(p_B +q - k)^2 + i\epsilon}  
\right)
         u_B(p_B).
\end{align}
Making use of simplifications that apply due to the Regge-limit  such as
neglecting small longitudinal components of external particles and
restricting to longitudinal polarization of the $t$-channel gluons,
which amounts to  replace the polarization tenser $g_{\mu\nu}$ by its
longitudinal components $ n^+_\mu n^-_\nu/2$, we arrive at
\begin{align}
  \label{eq:qcd1+qcd2}
i\mathcal{M}_{\text{QCD}} =  g^4{(p_A^+p_B^-)}
\int \frac{dk^+dk^-}{2\pi} \int \frac{d^2 {\bf{k}}}{(2\pi)^3}  
\left(
  \frac{(t^{c_1}t^{c_2})_{AA'}}{-k^- - \frac{{\bm k }^2+ m_A^2  + i\epsilon}{p_A^+ -k^+}}  + 
  \frac{(t^{c_2}t^{c_1})_{AA'}}{k^-  - \frac{({\bm q }- {\bm k})^2 + m_A^2  + i\epsilon}{p_A^+ +k^+}}  \right)
\notag \\
 \frac{1}{k^+k^- - {\bm k}^2 + i\epsilon}  \frac{1}{k^+k^- - ({\bm q} - {\bm k })^2 + i\epsilon}  
\left(
  \frac{(t^{c_1}t^{c_2})_{BB'}}{k^+ - \frac{({\bm q } - {\bm k})^2 +m_B^2 - i\epsilon}{p_B^- +k^-}}  + 
  \frac{(t^{c_2}t^{c_1})_{BB'}}{-k^+ - \frac{{\bm  k}^2  + m_B^2- i\epsilon}{p_B^- -k^-}}  \right).
\end{align}
To re-obtain the effective theory diagram CR, Eq.(\ref{eq:traj1}), we
further need to expand the terms inside the big brackets in
Eq.(\ref{eq:qcd1+qcd2}), in $p_A^+k^-$ and $p_B^-k^+$, which correspond to  the squared
center-of-mass-energies of the quark-gluon-sub-amplitudes. We obtain
\begin{align}
  \label{eq:expandee}
       \frac{t^{c_1}t^{c_2} }{-k^- - \frac{{\bm k }^2+ m_A^2  - i\epsilon}{p_A^+ -k^+}}  
       + 
       \frac{(t^{c_2}t^{c_1})}{k^-  - \frac{({\bm q }- {\bm k})^2 + m_A^2  - i\epsilon}{p_A^+ +k^+}}  
       =
      t^{c_1}t^{c_2}   \frac{p_A^+}{-p_A^+k^- + i\epsilon }  
       + 
      t^{c_2}t^{c_1} \frac{p_A^+}{p_A^+k^-   + i\epsilon}  + \mathcal{O}(\frac{1}{p_A^+k^-}) , \notag \\ 
      \frac{ t^{c_1}t^{c_2} }{k^+ - \frac{({\bm q } - {\bm k})^2 +m_B^2 + i\epsilon}{p_B^- +k^-}}  
      + 
      \frac{t^{c_2}t^{c_1}}{-k^+ - \frac{{\bm  k}^2  + m_B^2+ i\epsilon}{p_B^- -k^-}}  
      =
   t^{c_1}t^{c_2}  \frac{p_B^-}{p_B^-k^+  + i\epsilon }  + 
   t^{c_2} t^{c_1}  \frac{p_B^-}{-p_B^-k^+ + i\epsilon}   + \mathcal{O}(\frac{1}{p_B^-k^+}).
\end{align}
Building symmetric combinations of the $SU(N_c)$ generators,
$t^{c_1}t^{c_2} \to \{t^{c_1}, t^{c_2} \}/2$ the poles in $k^-$ and
$k^+$ cancel and we obtain no logarithm in $s$ from
Eq.(\ref{eq:qcd1+qcd2}).  In the effective action, this contribution
is contained in a different diagram (see
Sec.\ref{sec:tworeggeon_negsig}) and appears  as a  state of
two reggeized gluons.  For the antisymmetric color combination
$t^{c_1}t^{c_2} \to [t^{c_1}t^{c_2} ]/2 = if^{c_1c_2c} t^c$/2, the QCD
diagram coincides with Eq.~(\ref{eq:traj1}), as long as we take into
account only the most leading term of the expansion
Eq.(\ref{eq:expandee}).  This is however only meaningful, if the
squared center of mass energies of the quark-gluon-sub-amplitudes
$p_A^+k^-$ and $p_B^-k^+$, are larger than all other scales inside the
denominators on the left hand side of Eq.(\ref{eq:expandee}).  To give
meaning to Eq.(\ref{eq:traj1}), it is therefore necessary to implement a
certain lower cut-off on $p_A^+k^-$ and $p_B^-k^+$.  This can be achieved by making use of a property of the following Mellin-integral:
\begin{align}
    \label{eq:theta_Mellin}
     \lim_{\nu \to 0}    
     \int_{0 - i \infty}^{0+   i \infty} \frac{d \omega}{2\pi i} 
     \frac{1}{\omega + \nu} 
        \left(\pm \frac{p_A^+k^-}{\Lambda_a} \right)^{\omega} 
           = \left\{ 
             \begin{array}[h]{ll}
1 &  |p_A^+ k^-| > \Lambda_a \\ 
 & \\
0 & \text{otherwise} \qquad ,
             \end{array}
\right.
 \end{align}
 where the contour of integration runs along the imaginary axis, to
 the right of the pole at $-\nu$. The above method has compared to the
 the cut-off regularization the advantage, that it allows to give
 already the bare reggeized gluon the analytical properties of the
 reggeized gluon.  In particular, the parameter $-\nu$ acquires in the
 above the interpretation of an infinitesimal small Regge-trajectory
 and Eq.(\ref{eq:theta_Mellin}) allows to specify the phase of $
 p_A^+k^-$. This allows to introduce signature also for the bare
 reggeized gluon.  Depending on the choice of sign inside the bracket,
 Eq.(\ref{eq:theta_Mellin}), has a branch cut along the negative $(+)$
 or positive $(-)$ $p_A^+k^-$ axis.  These branch cuts should be
 associated with production thresholds of the underlying QCD-amplitude
 and they should therefore coincide with imaginary parts of the
 QCD-amplitude, Eq.(\ref{eq:expandee}).  Restricting to antisymmetric
 color exchange and neglecting terms of the order $k^+/p_A^+$, the
 combination of quark-propagators in the first line of
 Eq.(\ref{eq:expandee}) takes the following form:
 \begin{align}
   \label{eq:expandee_rl1}
         \frac{1}{2} \bigg[
\frac{-p_A^+ }{p_A^+k^- + {\bm k }^2 +  m_A^2  - i\epsilon }  
       &- 
       \frac{-p_A^+  }{- p_A^+k^-  + ({\bm q }- {\bm k})^2 + m_A^2  - i\epsilon}  \bigg].
  \end{align}
  The first term acquires an imaginary part or rather a discontinuity
  only for negative values of $p_A^+k^-$, while for the second term,
  the discontinuity is present for positive values of $p_A^+k^-$. In
  Eq.(\ref{eq:expandee_rl1}) only pole-discontinuities occur. However
  including higher order corrections to the quark-gluon
  sub-amplitudes, these poles turn into branch cuts. It is therefore
  natural to identify the imaginary parts of
  Eq.(\ref{eq:expandee_rl1}) with the branch-cuts of
  Eq.(\ref{eq:theta_Mellin}) which yields
\begin{align}
\label{eq:expandee_rl2}
     -\frac{p_A^+}{\Lambda_a}   \lim_{\nu \to 0} &      
     \int \frac{d \omega}{4\pi i} 
     \frac{1}{\omega + \nu} 
       \left[ \left(  \frac{p_A^+k^- -i\epsilon}{\Lambda_a} \right)^{\omega -1} 
       -  \left( \frac{-p_A^+k^- -i\epsilon}{\Lambda_a} \right)^{\omega -1} 
         \right] .
 \end{align}
 Note that the above prescription found for the branch cuts coincides
 with the Feynman rules for bypassing singularities
 \cite{Gribov:2003nw}: singularities on the real axis are circumvented
 by leading the contour below the negative real axis and above the
 positive real axis.  Together with the requirement that reggeized
 gluons carry negative signature, this rule allows  to derive the
 right $i\epsilon$-prescription without referring to the underlying
 QCD-Feynman diagram. We shall make use of this during later parts of
 this thesis.  From a diagrammatic point of view, the regularization
 Eq.(\ref{eq:theta_Mellin}) belongs to the propagator of the reggeized
 gluon as this is the only non-local part of the effective theory. In
 this sense, Eq.(\ref{eq:theta_Mellin}) gives a particular realization
 of the constraint that all interaction that is non-local in rapidity,
  is mediated by a reggeized gluon.
 
 To evaluate the loop integral in Eq.(\ref{eq:traj1}), we replace the
 pole $1/k^-$ by Eq.(\ref{eq:expandee_rl2}) and further perform a
 similar replacement for the pole $1/k^+$. As
 Eq.(\ref{eq:expandee_rl2}) yields a Reggeon with negative signature
 and trajectory $-\nu$ and as signature is generally conserved for the
 elastic amplitude \cite{Gribov:1968fc,Kirschner:1983di}, we expect
 that also the complete reggeized gluon has negative signature.  This
 is indeed the case. Performing the afore-mentioned replacements, we
 start with the following expression for the central-rapidity diagram
 CR:
\begin{align}
  \label{eq:CR_komplett}
 i\mathcal{M}^\text{CR}(s,t) =&    s \tilde{\Gamma}_{QQR}^c 
    \int \frac{dk^+ dk^-}{2\pi}  
\notag \\
& \frac{i/2}{{\bm q}^2}    
     \int  \frac{d \omega_1}{4\pi i} 
     \frac{1}{\omega_1 + \nu} 
      \left[  \frac{p_
A^+}{p_A^+k^- + i\epsilon}
       \left(   \frac{-p_A^+k^--i\epsilon}{ \Lambda_a} \right)^{\omega_1} 
     +
      \frac{p_A^+}{p_A^+k^- - i\epsilon} \left(\frac{p_A^+k^--i\epsilon}{\Lambda_a} \right)^{\omega_1}   \right] 
\notag \\
&
  \frac{(-g^2N_c) }{2}  \int \frac{d^2 {\bm k}}{(2\pi)^3}      \frac{-i}{ k^2 + i \epsilon} \frac{-i}{(q -k)^2 + i \epsilon} 2 {\bm{q}}^4 
    \notag \\
&
        \frac{i/2}{{\bm q}^2}   
     \int  \frac{d \omega_2}{4\pi i} 
     \frac{1}{\omega_2 + \nu} 
      \left[  \frac{p_B^-}{p_B^-k^+ + i\epsilon}
         \left(\frac{-p_B^+k^+-i\epsilon}{\Lambda_b} \right)^{\omega_2} 
     +
      \frac{p_B^-}{p_B^-k^+ - i\epsilon} \left(\frac{p_B^+k^+-i\epsilon}{\Lambda_b} \right)^{\omega_2}   \right] 
 \tilde{\Gamma}_{QQR}^c 
 \notag \\
=&  i\mathcal{M}_{2 \to 2}^{\text{tree}} (s,t)
 \int  \frac{d \omega_1}{2\pi i} \int  \frac{d \omega_2}{2\pi i} 
     \frac{1}{\omega_1 + \nu}  \left|\frac{p_A^+}{\Lambda_a} \right|^{\omega_1}  A (\omega_1,\omega_2)   
     \frac{1}{\omega_2 + \nu}  \left|\frac{p_B^-}{\Lambda_b} \right|^{\omega_2},
\end{align}
with
\begin{align}
  \label{eq:cr_bar}
A (\omega_1,& \omega_2 ) =  \frac{( ig^2N_c {\bm{q}}^2 ) }{2} \int \frac{dk^+ dk^-}{2\pi}  \int \frac{d^2 {\bm k}}{(2\pi)^3}      \frac{1}{ k^2 + i \epsilon} \frac{1}{(q -k)^2 + i \epsilon} 
\notag \\
\frac{1}{2}& \left[  (k^- -i\epsilon)^{\omega_1-1} -  (-k^- -i\epsilon)^{\omega_1-1} \right]
 \frac{1}{2}\left[(k^+ -i\epsilon)^{\omega_2-1}-(-k^+ -i\epsilon)^{\omega_2-1}   \right]. 
\end{align}
In our notation we suppressed  the dependence of $A(\omega_1, \omega_2)$ on $t = {\bm {q}}^2$.
\begin{figure}[htbp]
  \centering
  \parbox{6cm}{\includegraphics[width=5cm]{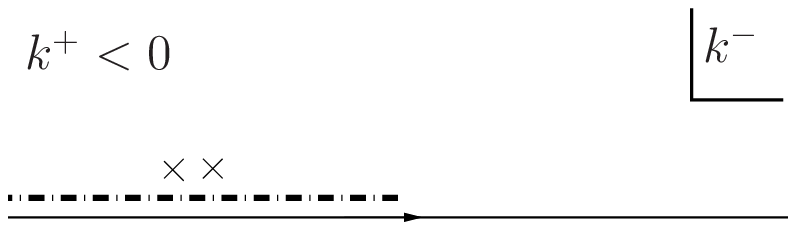}}
  \parbox{6cm}{\includegraphics[width=5cm]{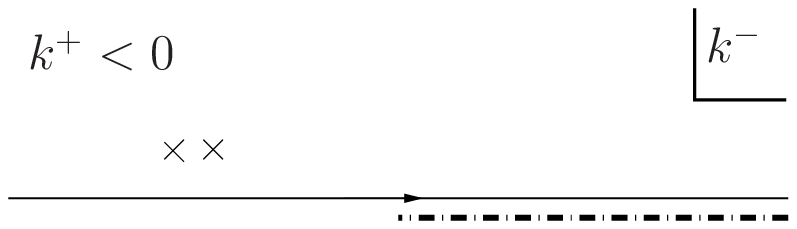}}
  \caption{\small The complex $k^-$-plane for $k^+ <0$   for the left and the right hand cut respectively. The crosses indicate the position of the poles from the gluon-propagators $1/(k^2 + i\epsilon)$ and $1/((q -k)^2 + i\epsilon)$.}
  \label{fig:cuts}  
\end{figure} 
To evaluate the longitudinal integrals we start with the integration
in $k^-$ and restrict ourselves first to the case $k^+ <0$. Then  only the cut $(-k^-)^{\omega_1}$ along the positive axis
contributes, while for the cut along the negative axis the contour of
integration can be enclosed without encircling any pole or branch-cut.
For $k^+ >0$, on the other hand, it is the branch cut along the
negative axis that contributes.  In both cases the $k^-$-integral can be 
evaluated by taking residues of the poles of the gluon-propagators and
we obtain
\begin{align}
  \label{eq:cr_bar_1int}
A (&\omega_1,\omega_2 ) = \frac{(-g^2N_c {\bm{q}}^2 ) }{2}  \int \frac{d^2 {\bm k}}{(2\pi)^3}  
\left[ \frac{( {\bm k}^2)^{\omega_1} }{ {\bm k}^2  [  ({\bm q} -{\bm k})^2 -{\bm k}^2 ]} 
-
\frac{ (({\bm q} -{\bm k})^2)^{\omega_1}}{ ({\bm q} -{\bm k})^2  [  ({\bm q} -{\bm k})^2 -{\bm k}^2] }\right]
  \notag \\
&  \frac{e^{-i\pi\omega_2} + 1}{4} 
  \left(\int_0^{\infty} \frac{dk^+ }{k^+}   (k^+ )^{\omega_2 -\omega_1} +  \int^0_{-\infty} \frac{dk^+ }{-k^+}   (-k^+ )^{\omega_2 -\omega_1}\right).
 \end{align}
 Within the LLA, we further replace\footnote{Note that this only applies  within the LLA, where all transverse scales are of the same order of magnitude. For higher accuracy,  a further non-zero contribution is obtained from the above calculation.} $({\bm k}^2)^{\omega_1} $ and $({\bm q}
-{\bm k})^2)^{\omega_1}$ by $ ({\bm q}^2)^{\omega_1}$, and we find:
\begin{align}
  \label{eq:cr_bar_2compact}
A (\omega_1,\omega_2) =  \frac{g^2N_c   }{2} & \int \frac{d^2 {\bm k}}{(2\pi)^3}  
 \frac{-{\bm{q}}^2}{ {\bm k}^2   ({\bm q} -{\bm k})^2  }  
 \frac{e^{-i\pi\omega_2} + 1}{2} 
({\bm q}^2 )^{\omega_2}  
  \int_0^{\infty} \frac{dk^+ }{k^+}  (k^+ )^{\omega_2-\omega_1}   .
\end{align}
To evaluate the integral over $k^+$, we first need to specify, which
of the two $\omega_i$-contours, $i=1,2$ is to the right of the other.
As there is no singularity to the left of the $\omega_1$ and
$\omega_2$ contours, we are free to chose $\Re$e$ \omega_2 > \Re$e$
\omega_1$. With such a choice, we obtain for the integral over $k^+$
\begin{align}
  \label{eq:cr_bar_3rho}
 \int_0^{\infty} \frac{dk^+ }{k^+}  (k^+ )^{\omega_2-\omega_1} = 
  &\frac{1}{\omega_2 -\omega_1}   
\lim_{\rho \to \infty} (\rho)^{\omega_2 -\omega_1}.
\end{align}
To give sense to this result, we need  to move the $\omega_2$
contour to left of the singularity at $\omega_2 = \omega_1$, i.e. we
pick up the residue at $\omega_2 = \omega_1$ while the remaining part
vanishes as now $\omega_2 < \omega_1$.  Of course we could have been
equally well started with $\Re$e$ \omega_2 < \Re$e$ \omega_1$, and one can verify that this leads to the identical result.
In short, the above $k^+$-integral yields nothing but a delta-function $2\pi i \delta(\omega_1 - \omega_2)$ and we obtain
\begin{align}
  \label{eq:a_bar_4rho}
A (\omega_1,\omega_2 ) =&    \beta({\bm q}^2)  
 \frac{\left[ ( - {\bm q}^2 -i\epsilon)^{\omega_2}  + (  {\bm q}^2 -i\epsilon)^{\omega_2}   \right] }{2}
2\pi i \delta(\omega_2 -\omega_1) ,
\end{align}
with 
\begin{align}
  \label{eq:beta_traj}
\beta({\bm q}^2) = \frac{g^2N_c  }{2}  \int \frac{d^2 {\bm k}}{(2\pi)^3}  
 \frac{- {\bm{q}}^2}{ {\bm k}^2   ({\bm q} -{\bm k})^2   },
\end{align}
the gluon trajectory function.
For the complete diagram we  obtain
 \begin{align}
   \label{eq:cr_omega_2}
 i\mathcal{M}^\text{CR}(s,t)
=& 
 i\mathcal{M}^{\text{tree}}_{2 \to 2} (s,t)
\beta({\bm q})
\int  \frac{d \omega_1}{2\pi i}      \frac{1}{(\omega_1 + \nu)^2}
 \frac{1}{2}\left[ 
    \left(-\frac{p_A^+p_B^- {\bm q}^2}{\Lambda_a\Lambda_b} \right)^{\omega_1} 
   +
   \left(\frac{p_A^+p_B^- {\bm q}^2}{\Lambda_a\Lambda_b} \right)^{\omega_1}
\right] 
\notag \\
=&  i\mathcal{M}^{\text{tree}}_{2 \to 2} (s,t) \beta({\bm q})
\frac{1}{2}\left[ 
  \ln \left(-\frac{p_A^+p_B^- {\bm q}^2}{\Lambda_a\Lambda_b} \right) 
   +
  \ln \left(\frac{p_A^+p_B^- {\bm q}^2}{\Lambda_a\Lambda_b} \right)
\right] .
 \end{align}
One can  define now parameters
\begin{align}
  \label{eq:1parameter}
\lambda_{a,b} = \frac{\Lambda_{a,b}}{m_{A,B} \sqrt{\bm{q}^2}},
\end{align}
which allow to write the above result as 
\begin{align}
  \label{eq:cr_rapi}
 \text{CR} = i\mathcal{M}^{\text{tree}}_{2 \to 2} (s,t) \beta({\bm q})
\frac{1}{2}\left[ 
  \ln \left(-\frac{p_A^+p_B^- }{m_Am_b} \right) 
   +
  \ln \left(\frac{p_A^+p_B^- }{m_A m_B} \right) - \ln \left(\lambda_a \lambda_b  \right)
\right]. 
\end{align}
Within the LLA, all transverse scale are  of the same order
of magnitude and we can set $\lambda_{a,b} = 1$. 
\begin{figure}[htbp]
  \centering
   \parbox{4cm}{\center \includegraphics[width=1.5cm]{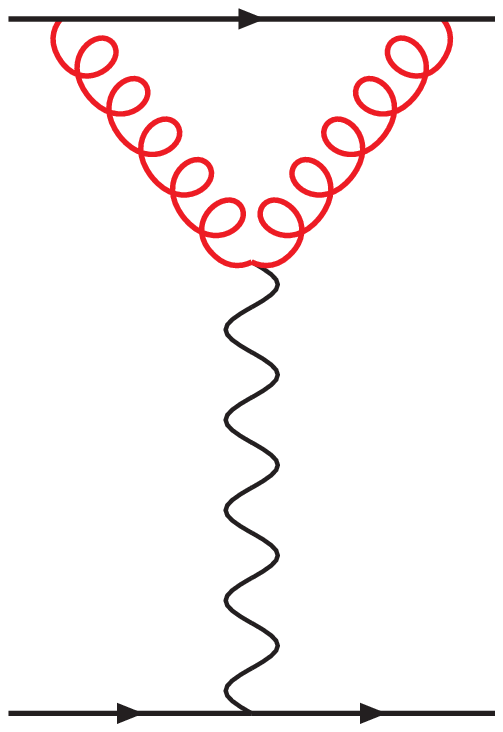}}
   \parbox{4cm}{\center \includegraphics[width=1.5cm]{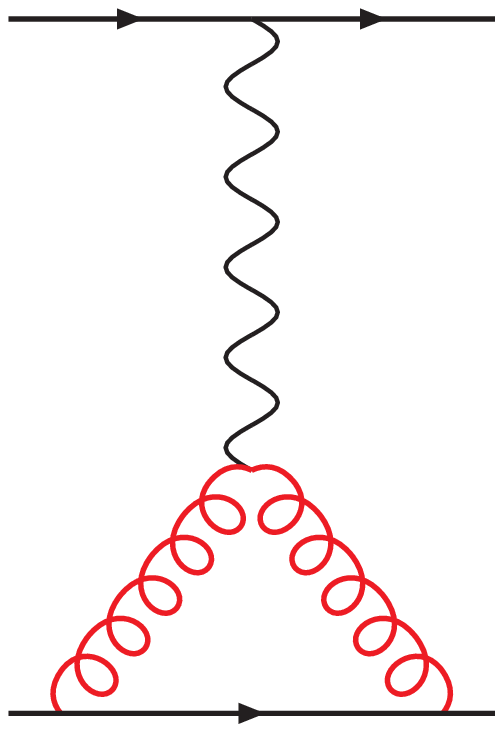}} \\
  \caption{\small Diagrams in the quasi-elastic regions of quark A and B resp., that cancel the cut-off dependence of the central rapidity diagram CR.}
  \label{fig:quasik}
\end{figure}
 In
general, the parameters $\lambda_{a,b} $ are supposed to cancel with
quasi-elastic diagrams, Fig.\ref{fig:quasik}. How this can occur in detail, will be addressed in Sec.\ref{sec:quasi-elastic}.

In the above calculation, a certain regularization scheme has been
used, that imposed certain lower cut-offs, $\Lambda_a$ and
$\Lambda_b$, on the squared center-of-mass energies of the
corresponding quark-gluon sub-amplitudes. This regularization scheme
is very plausible, as it exactly imposes the requirements that are
needed to obtain the effective theory diagram from the underlying
QCD-diagram. Furthermore, taking residues of the gluon propagators
during the above calculation, we explicitly set $k^+k^- = ({\bm q}
-{\bm k})^2 $ and $k^+k^-={\bm k}^2$. Together with the bounds
$p_A^+k^- > \Lambda_a$ and $p_B^-k^+ > \Lambda_b$ and with ${\bm k}^2
\simeq ({\bm q} - {\bm k})^2 \simeq {\bm q}^2$ in the LLA, the
rapidity of the gluon $Y_k = \ln(k^+/k^-)/2$ is then restricted to
central values , $Y_k \in [-\ln(\Lambda_a/\sqrt{{\bm q}^2 } p_A^+ ),
\ln(\Lambda_b/\sqrt{{\bm q}^2 }p_B^- ) ]$ as expected and required for
the central rapidity diagram.  Also the imaginary part and therefore
negative signature of the reggeized gluon are mapped obtained
correctly.  Nevertheless the method has some drawbacks: To obtain the
correct gluon trajectory function, it has been necessary to neglect a
dependence of the Reggeon-factors on transverse scales.  This
dependence is irrelevant within the LLA, but has to be included if one
goes beyond the LLA. At the present stage it is not clear, whether
such a contribution is meaningful or not within the NLLA.  In the
recent literature \cite{Balitsky:2008zza} it has been reported that
difficulties arise with certain LLA-regularization schemes if used for
NLLA-calculations.  The whole setup of \cite{Balitsky:2008zza} differs
from the present one. Particularly our method of regularization is at
first not related to the one to which \cite{Balitsky:2008zza} refers.
Nevertheless we want to note that it is also possible to introduce a
scheme which imposes the lower bound directly on the rapidity of the
gluon loop, rather than on (sub)-center-of-mass-energies. Some details
about this scheme are presented in App.\ref{cha:y_scheme} and we refer
there for details.  Within the LLA, both schemes yield identical
results and none of the schemes can be ruled out.

\begin{figure}[htbp]
  \centering
  \parbox{2.5cm}{\center \includegraphics[height=2.5cm]{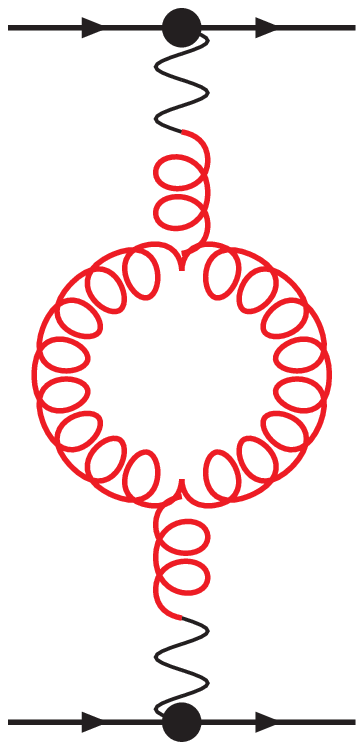}}
  \parbox{2.5cm}{\center \includegraphics[height=2.5cm]{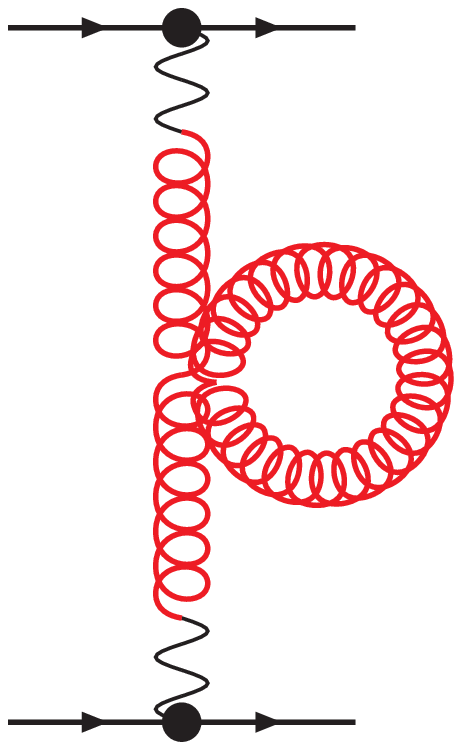}}
  \parbox{2.5cm}{\center \includegraphics[height=2.5cm]{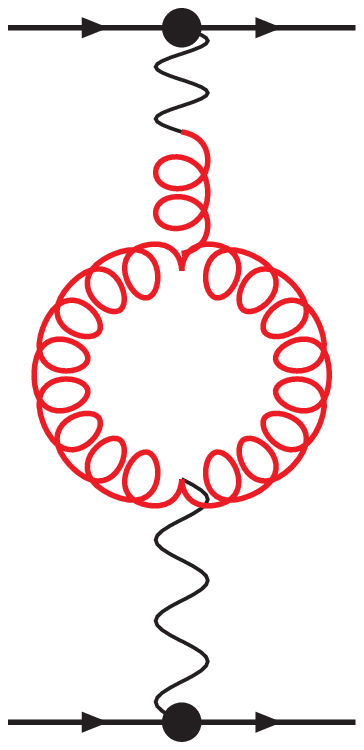}}
  \parbox{2.5cm}{\center \includegraphics[height=2.5cm]{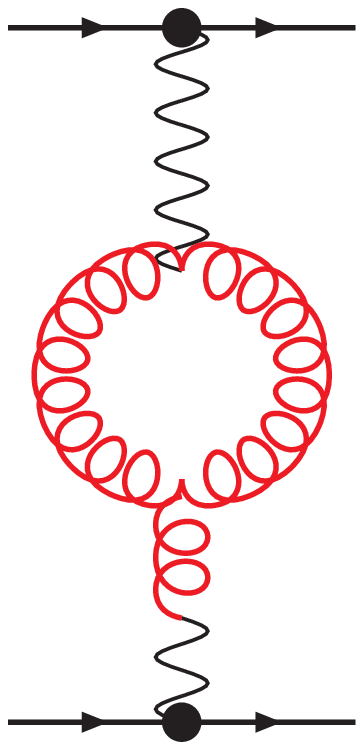}}
  \parbox{2.5cm}{\center \includegraphics[height=2.5cm]{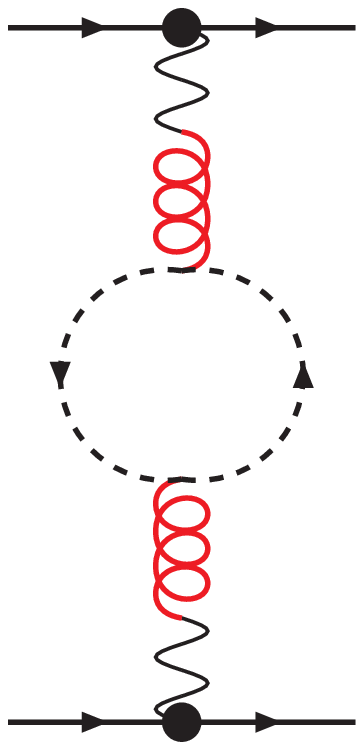}}
\parbox{2.5cm}{\center CR1}\parbox{2.5cm}{\center CR2}\parbox{2.5cm}{\center CR3}\parbox{2.5cm}{\center CR4}\parbox{2.5cm}{\center CR5}
  \caption{\small Further diagrams that occur  from the effective action for the central rapidities. Their sum  yields zero result.}
  \label{fig:CRfurther}
\end{figure}

Besides the diagram CR, which we discussed so far, there are also
diagrams, in which one or both reggeized gluons couple to the
gluon-loop by a three or a four-gluon vertex and there is furthermore a diagram that contains a ghost-loop.  These diagrams are shown in Fig.\ref{fig:CRfurther}. In their sum, these diagrams yield the  following
expression
 \begin{align}
   \label{eq:traj_QCD_diagrams}
\text{CR1} +\text{CR2} + \text{CR3} +\text{CR4} = s \tilde{\Gamma}^{c}_{QQR}   g^2 &  \frac{N_c}{2} \int \frac{dk^+ dk^-}{2\pi} \int \frac{d^2 {\bf k}}{(2\pi)^3}   \frac{i/2}{{\bm q}^2}
   \frac{1}{ k^2 + i \epsilon} \frac{1}{(q -k)^2 + i \epsilon}
    \notag \\
   & \qquad  \left( 2{\bm{k}}^2 + 2({\bm{q}} - {\bm{k}})^2 - 8 {\bm{q}}^2
   \right)  \frac{i/2}{{\bm q}^2}  \tilde{\Gamma}^{c}_{QQR}.
\end{align}
Unlike the diagram CR, Eq.~(\ref{eq:traj_QCD_diagrams}), contains no
pole in a light-cone momentum.  It is therefore possible 
to evaluate the integrals over $k^+$ and $k^-$ without introducing any regularization: The integrals are convergent, all singularities lie on the same side of the integration contour and the integral gives zero result.

On the other hand, 
the momentum structure of the above denominators is simplified due to
the coupling of the reggeized gluons. This is only justified, if the
the squared sub-center-of mass energies $p_A^+k^-$ and $p_B^-k^+$ are 
large.  Furthermore, introducing a regularization allows also to take
explicitly into account negative signature of the reggeized gluon. It
seems therefore plausible to introduce an analogous regularization by
a Mellin-integral also for these contributions.

Within the LLA, the question cannot be answered whether one should
include a regularization for these diagrams or not: In any case the
integral does not contribute within the LLA.  The answer to this
question lies at the NLLA: In that case, the integral still vanishes
if no regularization is included, while it is non-zero with
regularization. For the following discussion we take the convention
that the Mellin-integral should be also used for those diagram that do
not come with an explicit pole in $k^+$ and $k^-$.  The definite
answer however can be only given if corrections beyond the LLA are
considered.



\subsection{Reggeization of the gluon}
\label{sec:reggeizedgluon}

\begin{figure}[htbp]
  \centering
  \parbox{3cm}{\center \includegraphics[height=5cm]{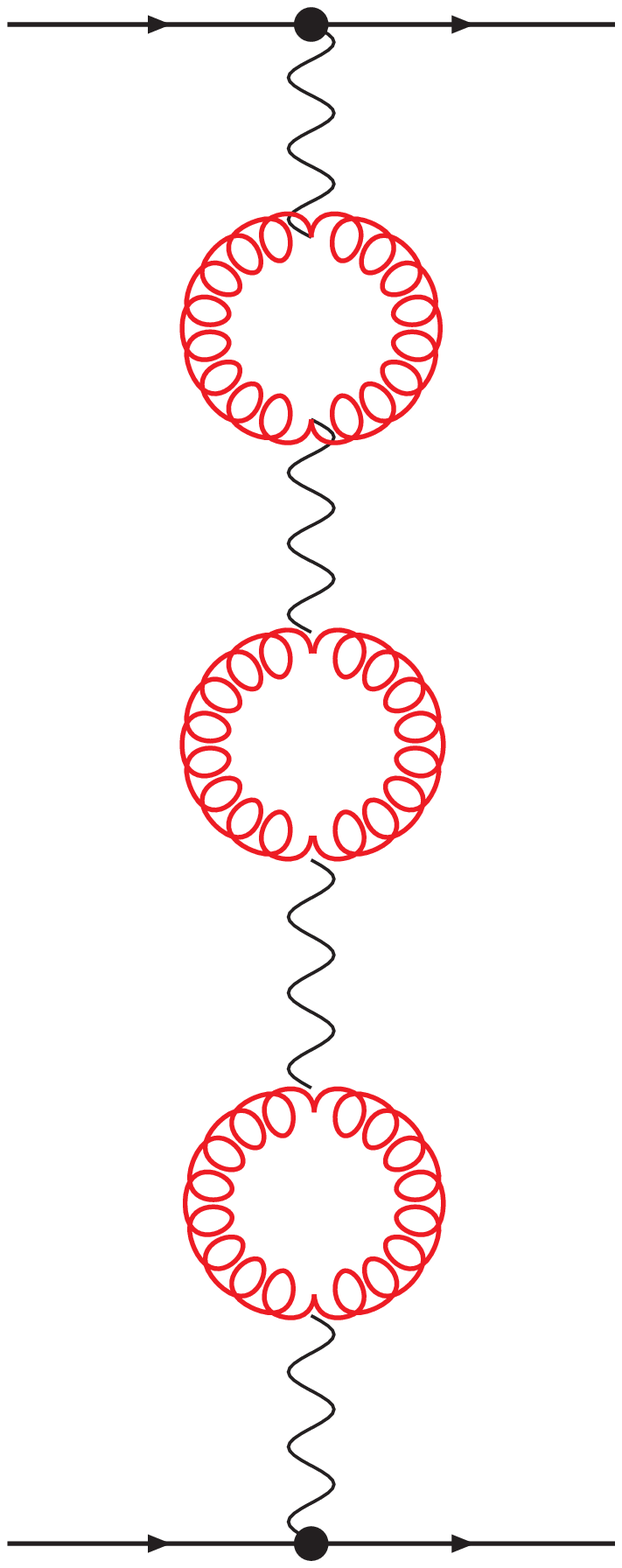}}  
  \parbox{3cm}{$\,$}   
  \parbox{3cm}{\center \includegraphics[height=5cm]{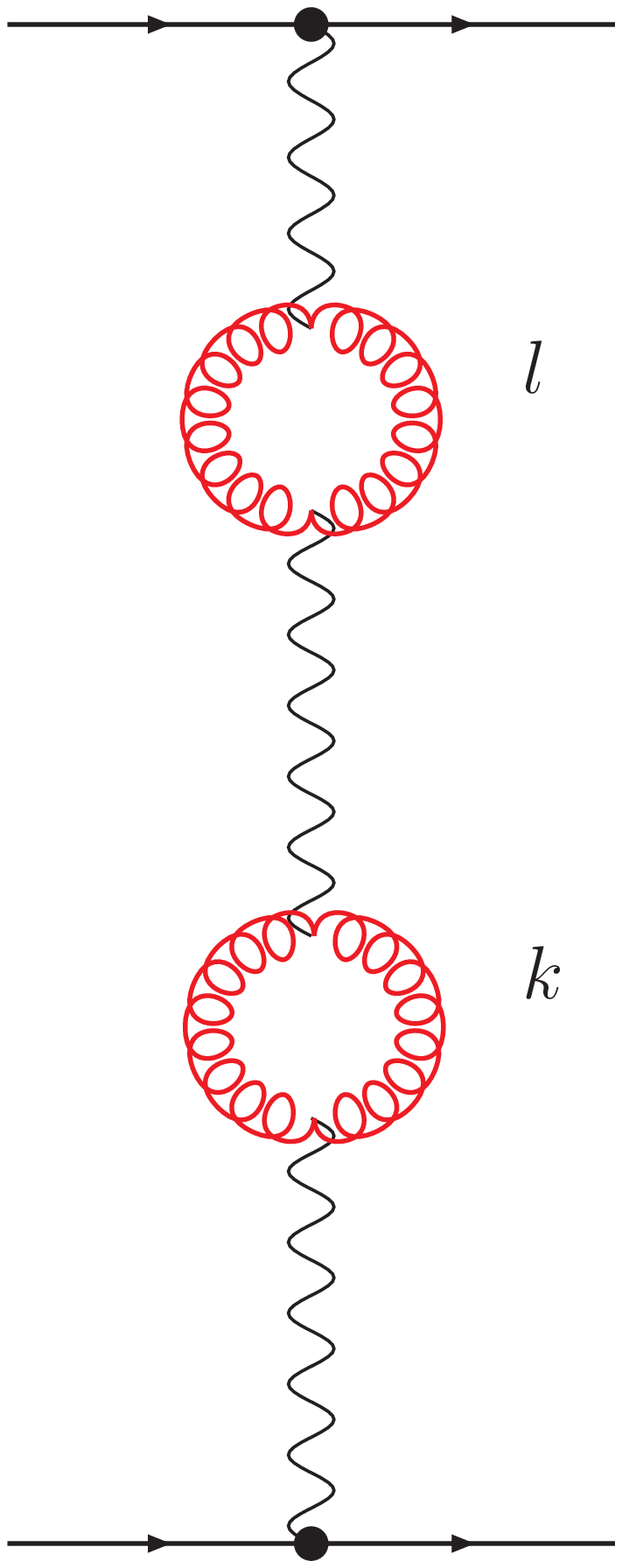}}\\
    \parbox{3cm}{\center (a)} \parbox{3cm}{$\,$}  \parbox{3cm}{\center (b)}
  \caption{ \small (a): Type of diagram that need to be resummed to obtain reggeization of the gluon. (b): The two-loop example }
  \label{fig:reggeization}
\end{figure}

\begin{figure}[htbp]
  \centering
  \parbox{3cm}{\center \includegraphics[height=3cm]{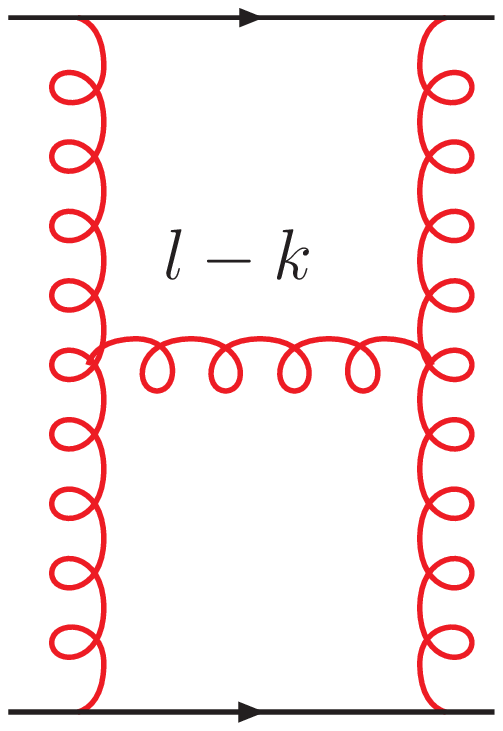}}
  \parbox{2cm}{$\,$}
  \parbox{3cm}{\center \includegraphics[height=5cm]{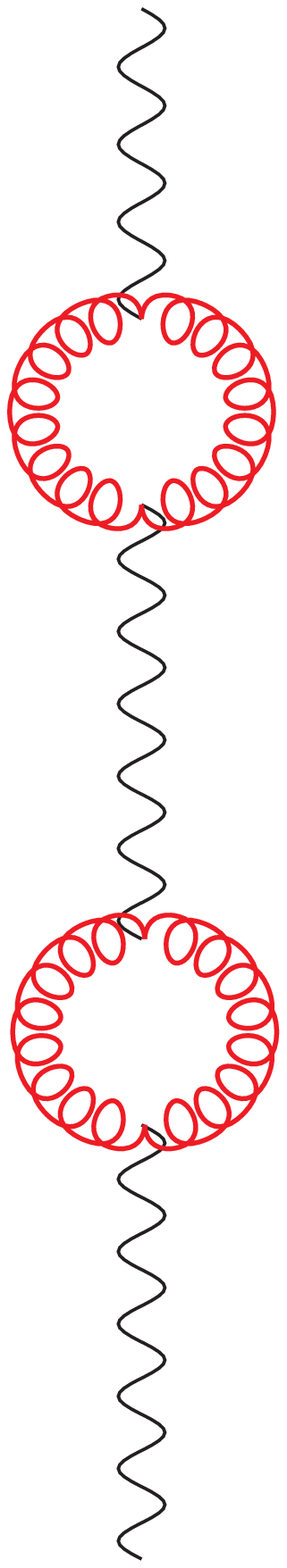}}
  \parbox{3cm}{\center \includegraphics[height=4cm]{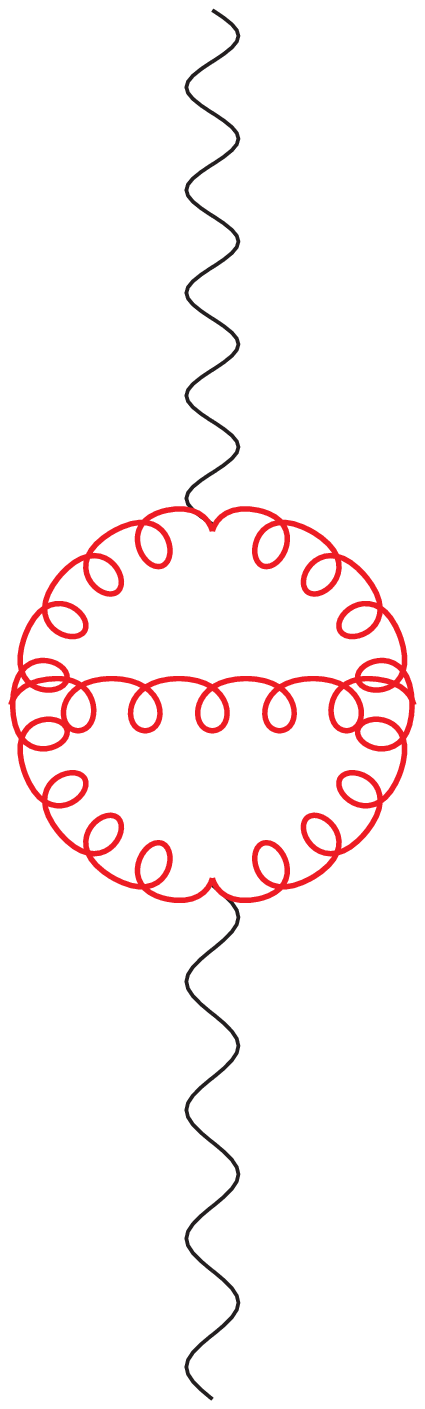}} \\
  \parbox{3cm}{\center (a)}  \parbox{2cm}{$\,$} \parbox{6cm}{\center (b)}
  \caption{\small (a):  QCD-ladder graph where the vertical gluon propagator leads  to the poles in $l^+$ and $k^-$. (b):Summing NLLA-diagrams, like the one  to the left one expects to obtain  a contribution that is proportional to the product of  two gluon trajectories functions. The cut-off $\Lambda_c$ separates this local (rapidity) contributions from the non-local contribution at the rhs. }
  \label{fig:two_loop_cutoff}
\end{figure}

With the leading non-trivial order of the reggeized gluon derived, we
demonstrate in the following that reggeization of the gluon arises
within the effective action by resumming diagrams like
Fig.~\ref{fig:reggeization}a, which contain an arbitrary number of
gluon loops.  We start with the diagram containing two gluon
loops, Fig.\ref{fig:reggeization}b: Every reggeized gluon couples to
the gluon loops by an induced vertex of the first order which yield
poles in the light cone momenta $l^+, l^-$ and $k^+, k^-$
respectively.  While at one-loop, both poles could be identified with
the most leading term of the expansion of the regarding
quark-propagators, in the two-loop case the poles in $l^+$ and $k^-$
arise due to a gluon propagator inside a QCD-ladder graph, as
illustrated in Fig.\ref{fig:two_loop_cutoff}a.  Without going into
details about the structure of the underlying QCD-graph, it is clear
that these poles arise  from an expansion of a QCD 4-point
amplitude, similar to the one-loop case.  This requires
to impose again a lower bound $\Lambda_c$ on the squared sub-center-of-mass energy $l^+k^-$. We therefore insert for the combination 'induced
vertex' times 'reggeized gluon propagator' times 'induced vertex' the
following expression into the diagram
\begin{align}
\label{eq:combiindpropind}
\parbox{3cm}{\includegraphics[width=2cm]{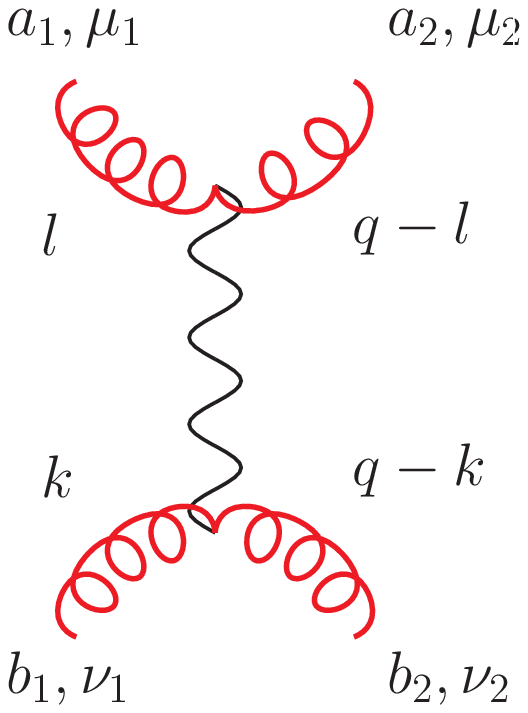}}  \left. \begin{array}{l}
{\displaystyle = f^{a_1a_2c}(n^+)^{\mu_1}(n^+)^{\mu_2} f^{b_1b_2c}(n^-)^{\nu_1}(n^-)^{\nu_2} \frac{i {\bm q}^2}{2} }  \\  \\
{\displaystyle   
\quad \times \int\frac{d \omega_3}{4\pi i} 
     \frac{\Lambda_c^{\omega_3}}{\omega_3 + \nu} 
      \left[ 
         \left( -l^+k^+-i\epsilon  \right)^{\omega_3-1} 
     -
      \left(l^+k^+-i\epsilon \right)^{\omega_3-1}   \right]. }
\end{array} \right. 
  \end{align}
  Again the cut-off $\Lambda_c$ needs to be canceled by a similar
  contribution in a different diagram.  While the cut-offs $\Lambda_a$
  and $\Lambda_b$ are canceled by diagrams in the quasi-elastic region
  of the scattering quarks (see Sec.\ref{sec:quasi-elastic}), we
  expect 'intermediate' cut-offs like $\Lambda_c$ to cancel with the
  corresponding part in higher order corrections to the trajectory
  function, see Fig.\ref{fig:two_loop_cutoff}b.
Making  repeatedly use of our result for the function $A(\omega_1, \omega_2)$, Eq.~(\ref{eq:a_bar_4rho}), we obtain at two-loop the following result
\begin{align}
   \label{eq:twoloop-mit2A}
\mathcal{M}^{\text{2loop}}_{2 \to 2}   =&    
 \mathcal{M}_{2 \to 2}^{\text{tree}}(s,t) 
 \times \int  \frac{d \omega}{2\pi i}  \frac{ \beta^2({\bm q}^2) }{(\omega + \nu)^{3}} 
\frac{1}{2} \left[ 
\left( -\frac{p_A^+p_B^- {\bm q}^4}{\Lambda_a\Lambda_b\Lambda_c}\right)^{\omega} + \left( \frac{p_A^+p_B^-{\bm q}^4}{\Lambda_a\Lambda_b\Lambda_c}\right)^{\omega}
\right].
 \end{align}
Within the LLA we  replace the cut-offs by a typical transverse scale of the process, $\Lambda_a\Lambda_b\Lambda_c = m_A m_B {\bm q}^2$, for instance, and 
with this choice, one obtains similarly for the diagram with $n$-loops
\begin{align}
  \label{eq:n_loop_traj_fertig}
i\mathcal{M}_{2 \to 2}^{\text{LLA,  n-loop}}(s,t) =& i\mathcal{M}_{2 \to 2}^{\text{tree}}(s,t) 
 \times \int  \frac{d \omega}{2\pi i}  \frac{ \beta^n({\bm q}^2) }{(\omega + \nu)^{n + 1}} 
\frac{1}{2} \left[ 
\left( -\frac{p_A^+p_B^-}{m_Am_B}\right)^{\omega} + \left( \frac{p_A^+p_B^-}{m_Am_B}\right)^{\omega},
\right].
\end{align}
Summing over the number $n$ of gluon  loops, we  obtain with
\begin{align}
  \label{eq:geometricprog}
\sum_{n=0}^{\infty}  \frac{ \beta^n({\bm q}^2) }{(\omega + \nu)^{n + 1}}  =  \frac{1}{\omega + \nu - \beta({\bm q}) } 
\end{align}
reggeization of the gluon:
\begin{align}
  \label{eq:allorder_traj_fertig}
\mathcal{M}_{2 \to 2}^{\text{LLA }}(s,t) =& \mathcal{M}_{2 \to 2}^{\text{tree}}(s,t) 
 \times \int  \frac{d \omega}{2\pi i}  \frac{1 }{\omega -  \beta({\bm q}^2) } 
\frac{1}{2} \left[ 
\left( -\frac{p_A^+p_B^-}{m_Am_B}\right)^{\omega} + \left( \frac{p_A^+p_B^-}{m_Am_B}\right)^{\omega}
\right].
\end{align}
Inserting Eq.~(\ref{eq:22born}) for the tree-level amplitude, the result can be written in a form that makes negative  signature of the reggeized gluon apparent 
\begin{align}
  \label{eq:sig_fac_extract}
\mathcal{M}_{2 \to 2}^{\text{LLA }}(s,t) &= 2g^2 \frac{s}{t} \int \frac{d \omega}{2\pi i}  \frac{1 }{\omega -  \beta({\bm q}^2) }  \left| \frac{s}{m_Am_B}\right|^{\omega}\frac{ \xi^{(-)}(\omega)}{2},&     \xi^{(-)}(\omega)=& e^{-i\pi\omega} + 1.
\end{align}
From the point of view of the effective action, the following representation is however more adequate:
 \begin{align}
  \label{eq:vertexsig_fac_extract}
i\mathcal{M}_{2 \to 2}^{\text{LLA }}(s,t) &=  \int \frac{d \omega}{4\pi i} \Gamma_{PPR}(p_A, q) \frac{i/2}{{\bm q}^2}  \frac{ \xi^{(-)}(\omega) }{\omega -  \beta({\bm q}^2) }  \left| \frac{p_A^+p_B^-}{s_R}\right|^{\omega}   \Gamma_{PPR}(p_B, q).
\end{align}
Here $ \Gamma_{PPR}(p_A, q)$ are the Particle-Particle-Reggeon
vertices which describe the coupling of a particle, a quark (P=Q) or a
gluon (P=G), to the reggeized gluon in the $t$-channel. Corrections
due to resummation of central rapidity-diagrams as illustrated above,
are then taken into account, by replacing every bare by the resummed
reggeized gluon. Due to the non-local nature of the reggeized gluon,
it is needed to consider a reggeized gluon always as an object
embedded into a minimal elastic amplitude which yields a well-defined
dependence of the reggeized gluon on the squared center-of-mass energy
of the sub-amplitude, $s= p_A^+p_B^-$, and its momentum transfer, $t =
-{\bm q}^2$. We note that for the reggeized gluons $s$ should be
always defined as the product of $p_A^+$ and $p_B^-$, also if
transverse components and other light-cone momenta of the particles
$A$ and $B$ are non-zero.  For the description of reggeized gluons it
is furthermore advisable to keep explicitly the Mellin-integral. Due
to Eq.~(\ref{eq:theta_Mellin}), it provides a natural implementation
of the non-locality constraint in rapidity of the reggeized gluon.

\subsection{Quasi-elastic corrections}
\label{sec:quasi-elastic}
\begin{figure}[htbp]
  \centering
   \parbox{4cm}{\center \includegraphics[height=2.5cm]{quasi_Ap.eps}}
   \parbox{4cm}{\center \includegraphics[height=2.5cm]{quasi_Bp.eps}} \parbox{6cm}{\center \includegraphics[width=1.5cm]{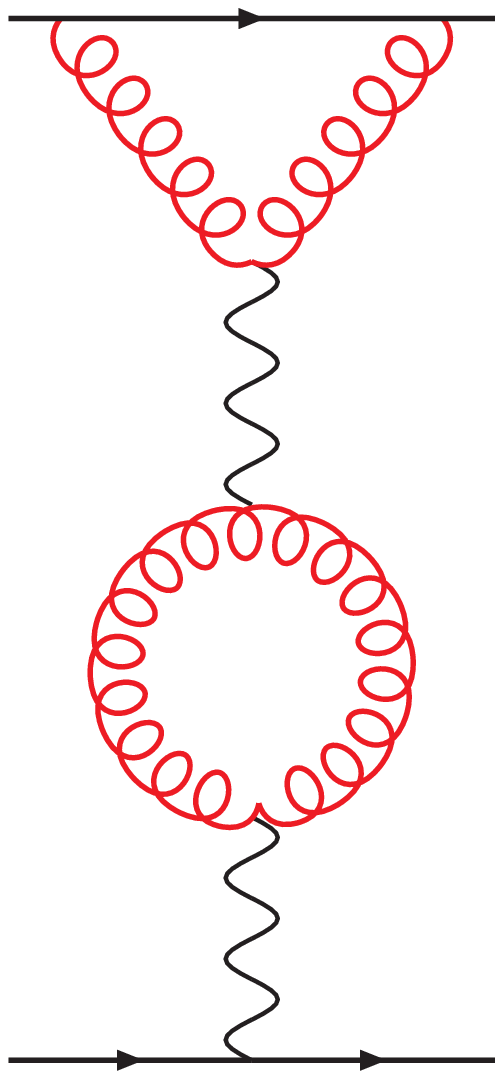}} \\
\parbox{4cm}{\center QEA} \parbox{4cm}{\center QEB} \parbox{6cm}{\center QEACR} 
  \caption{\small Diagrams in the quasi-elastic regions of quark A and B resp., where the reggeized gluon couples to the loop by an induced vertex. To the left a higher order example}
  \label{fig:quasi}
\end{figure}

Before finishing the discussion of the singe reggeized-gluon-exchange, we  demonstrate how a cancellation of the previously
introduced cut-offs can occur if one goes beyond the strict LLA.  The
quasi-elastic diagrams QEA and QEB, Fig.\ref{fig:quasi}, do not contribute within the LLA, but only
at NLLA. Also we do not attempt to determine the full
quasi-elastic-corrections to the quark-reggeized-gluon coupling, but are merely interested in the
logarithmic part that should allow a cancellation of the above
introduced parameter $\lambda_a$ and $\lambda_b$.
It is further natural to expect, that this contribution occurs inside the
diagrams QEA and QEB which contain an induced vertex. Like the central
rapidity diagram CR, these diagrams arise from the QCD-diagrams QCD1
and QCD2, Fig.\ref{fig:one_reggeon}, in a certain kinematical limit.
Diagrams where the reggeized gluon couple to the gluon-loop by a
three-gluon vertex, are expected to lead to usual QCD-vertex
corrections and are not of interest to us in the present context.
Also for the diagram QEA we attempt in the following to extract only
its leading logarithmic part, needed for the cancellation, while we
discard all other contributions.  For the diagram QEA in
Fig.\ref{fig:one_reggeon}, un-regularized Feynman-rules of the
effective action lead to the following expression
\begin{align}
  \label{quasi-elastic}
i\mathcal{M}^\text{QEA}(s,t) =& \int \frac{d^4 k}{(2\pi)^4}\bar{u}(p_A-q) igt^a\fdag{n}^+ \frac{i(\fdag{p}_A - \fdag{k} + m_A)}{(p_A -k)^2  - m_A^2+ i\epsilon} igt^b\fdag{n}^+ u(p_A) 
\notag \\
&
\frac{-i}{k^2 + i\epsilon} \frac{-i}{(q - k)^2 + i\epsilon} \frac{f^{abc}}{k^+} \frac{i/2}{{\bm q}^22} \bar{u}(p_B + q) ig t^c\fdag{n^+} u(p_B).
\end{align}
Due to on-shellness of the quark A and A', $q^- = ({\bm q}^2 +
m_A^2)/p_A^+ $ and $p_A^- =m_A^2/p_A^+ $ are small and neglected in
the following. 
For quasi-elastic corrections, the requirements imposed by locality in
rapidity are more involved than in the previously discussed case of
central-rapidity diagrams: Namely it is needed to combine two
constraints in that case: From the point of view of the loop build
from the quark and the gluons in Fig.\ref{fig:quasi}, diagram QEA,
rapidity of the gluon-loop $Y_k$ is restricted to the $[Y_A - \eta,
Y_A + \eta]$, with $Y_A$ the rapidity of the quark. While $Y_k$ is
bounded from above by the sheer structure of the integrand in
Eq.(\ref{quasi-elastic}), a regularization for the lower end is
required that ensures that this bound is fulfilled. Anticipating that
integral over longitudinal momenta will be performed by taking
residues of the gluon-propagator, the gluon can at least with LLA
accuracy be regarded as being close to the mass-shell also in the
present case. Imposing therefore a lower bound $p_B^-k^+ >
p_A^+p_B^-/\lambda_a$ will result in a constraint $Y_K > Y_A -
\ln\lambda_a $ which sets the rapidity of the gluon inside the loop
close to the one of the quark $A$. From the point of view of the
reggeized gluon, on the other hand, the requirement arises that the
rapidities of the gluon loop and the quark B respectively are
significantly separated. For the leading order diagram QEA,
Fig.\ref{fig:quasi}, this is fulfilled automatically as long as
scattering of the two quarks takes place in the Regge-limit. Including
however higher order corrections in the central-rapidity region as
diagram QEACR, Fig.\ref{fig:quasi}, it is needed to make this
constraint explicit.

 Making use of Eq.(\ref{eq:theta_Mellin}) and Eq.(\ref{eq:expandee_rl2})
we  propose to use the  following regulated expression:
\begin{align}
  \label{quasielastic-total}
i\mathcal{M}^\text{QEA}(s,t)
& = \int \frac{d\omega}{2\pi i}    \Gamma_{QQR}(\omega, \lambda_a ;p_A, q )  
\frac{i/2}{{\bm q}^2}  \frac{1}{\omega + \nu} \left|\frac{p_A^+p_B^-}{m_Am_B} \right|^\omega   
 \Gamma_{QQR}(p_B, q),
              \end{align}
with
\begin{align}
  \label{eq:gamma_bar}
 \Gamma_{QQR}(\omega, \lambda_a ;p_A, q ) = \Gamma_{QQR}(p_A, q ) \gamma_+ (\omega, \lambda_a),
\end{align}
and
\begin{align}
  \label{eq:qea_bar}
\gamma(\omega, \lambda_a)=    \frac{-g^2N_c{\bm q}^2}{2}  &
               \int \frac{dk^+ dk^-}{2\pi}  \int \frac{d^2 {\bm k}}{ (2\pi)^3}
 \frac{p_A^+ - k^+}{( p_A - k)^2  - m_A^2+ i\epsilon}                 \frac{1}{k^2+ i\epsilon} \frac{1}{(q - k)^2 + i\epsilon}  
\notag \\
 &   \int \frac{d\omega'}{4\pi i} \frac{p_B^-}{\omega' - \omega}  
               \left(\frac{\lambda_a}{p_A^+p_B^-}   \right)^{\omega'}       \left[\left({ p_B^- k^+  -i\epsilon}   \right)^{\omega'-1} 
      -
    \left({-p_B^- k^+  -i\epsilon} \right)^{\omega'-1}\right].
\end{align}
In the above expression we therefore use a combination of two
Mellin-integrals to impose the required bounds. Furthermore they allow
to write the complete expression in a factorized form, which allows to
include in a simple and straight forward way corrections from the
central rapidities or the quasi-elastic region of the quark B. The
Mellin-integral in $\omega'$ in Eq.(\ref{eq:qea_bar}) ensures then
that the gluon loop is close in rapidity to the quark A. Carrying out
this integral one picks up to pole at $\omega' = \omega $ and the
$\omega'$-factors in Eq.(\ref{eq:qea_bar}) combine with the
$\omega$-factors in Eq.(\ref{quasielastic-total}). Together the lead
to the constraint $p_B^-k^+ > \frac{m_Am_B}{\lambda_a}$ i.e. the
product $p_B^-k^+$ is required to be bigger than the typical
transverse scale, which seems to be a reasonable constraint, taking
into account the discussion of the previous paragraphs.

In Eq.~(\ref{eq:qea_bar}) the integral over $k^-$  is  only
non-zero if $p_A^+ > k^+ > 0$, and taking residues  of the
gluon propagators we obtain
\begin{align}
  \label{quasiel_cont1}
  \gamma_+(\omega, \lambda_a) =& \frac{-g^2N_c{\bm q}^2}{2} 
          \int \frac{d\omega'}{4\pi i}   \frac{1}{\omega' - \omega}   \int_0^{p_A^+} \frac{dk^+ }{k^+} [(- k^+)^{\omega'} + (k^+)^{\omega'}]  
               \left(\frac{  \lambda_a}{p_A^+}   \right)^{\omega'} 
    \notag \\
&
    \int \frac{d^2 {\bm k}}{ (2\pi)^3}
           \frac{1}{ ({\bm{q}} - {\bm{k}})^2 - \frac{k^+ }{k^+ - p_A^+}({\bm{k}}^2  + m_A^2)} 
      \frac{1}{ {\bm{k}}^2 - \frac{k^+}{k^+ - p_A^+} ({\bm{k}}^2 + m_A^2 )}       .
\end{align}
We are only interested in the leading logarithmic part of the
integral, which is proportional to $1/\omega$. It arises from the
region of integration with strong ordering of momenta, $p_A^+ \gg
k^+$. We obtain
\begin{align}
  \label{gamma__cont1}
  \gamma_+(\omega, \lambda_a) =& \int \frac{d\omega'}{4\pi i} \frac{1}{\omega' -
    \omega} [(-\lambda_a)^{\omega'} + (\lambda_a)^{\omega'}]
  \frac{\beta({\bm q})}{\omega'} = \frac{\beta({\bm q})}{\omega}
  \left[ \frac{(-\lambda_a)^{\omega} + (\lambda_a)^{\omega}}{2} -1
  \right],
\end{align}
where we set $\Re$e$\omega'>0$.  We  obtain for
Eq.(\ref{quasielastic-total})
\begin{align}
  \label{eq:quasielastic_inserted}
i\mathcal{M}^\text{QEA}(s,t)
& =  i\mathcal{M}^\text{tree}(s,t) \int \frac{d\omega}{2\pi i} 
\frac{1}{\omega + \nu} \left|\frac{p_A^+p_B^-}{m_Am_B} \right|^\omega   
  \frac{\beta({\bm q})}{\omega} \left[ \frac{(-\lambda_a)^{\omega} + (\lambda_a)^{\omega}}{2} -1   \right],
 \end{align}
and with  $\nu \to 0$ we find
\begin{align}
 \label{quasiel_fertiga}
\mathcal{M}^\text{QEA}=& \mathcal{M}_{2 \to 2}^{\text{tree}}(s,t) \beta({\bm q}) 
                    \frac{ \ln(- \lambda_a)  
+ \ln(\lambda_a)}{2}
\end{align}
and similarly for the diagram QEB 
\begin{align}
 \label{quasiel_fertigb}
\mathcal{M}^\text{QEB}=& \mathcal{M}_{2 \to 2}^{\text{tree}}(s,t) \beta({\bm q}) 
                    \frac{ \ln(-  \lambda_b)  + \ln( \lambda_b)}{2}.
\end{align}
As a consequence,  in  the sum of central rapidity and quasi-elastic diagrams, the dependence on the the parameters  $\lambda_a$ and $\lambda_b$ cancels as required 
\begin{align}
  \label{eq:central_quasi}
 \mathcal{M}^\text{QEA} + \mathcal{M}^\text{CR} + \mathcal{M}^\text{QEB} = \mathcal{M}^{\text{tree}}_{2 \to 2} (s,t) \beta({\bm q})
\frac{1}{2}\left[ 
  \ln \left(-\frac{p_A^+p_B^- }{m_Am_B} \right) 
   +
  \ln \left(\frac{p_A^+p_B^- }{m_A m_B} \right) 
\right]. 
\end{align}
It is also straight-forward, to verify with
Eq.(\ref{eq:quasielastic_inserted}) cancellation of the parameter
$\lambda_a$, if corrections that lead to reggeization of the gluon
like diagram QEACR in Fig.\ref{fig:quasi}, are additionally included. With central-rapidity diagrams resummed to all orders (and
treating those within the LLA) Eq.(\ref{eq:quasielastic_inserted})
turns into
\begin{align}
  \label{eq:quasielastic_inserted_resummed}
i\mathcal{M}^\text{QEA}(s,t)&= i\mathcal{M}^\text{tree}(s,t) \int \frac{d\omega}{2\pi i} 
\frac{1}{\omega -\beta({\bm q})} \left|\frac{p_A^+p_B^-}{m_Am_B} \right|^\omega   
  \frac{\beta({\bm q})}{\omega} \left[ \frac{(-\lambda_a)^{\omega} + (\lambda_a)^{\omega}}{2} -1   \right]
\notag \\
&=i\mathcal{M}^\text{tree}(s,t) \int \frac{d\omega}{2\pi i} 
\frac{1}{\omega -\beta({\bm q})} \left|\frac{p_A^+p_B^-}{m_Am_B} \right|^\omega   
 \xi^{(-)}(\omega) \frac{\beta({\bm q})}{\omega} \left[ |\lambda_a|^{\omega}  - \frac{1}{\xi^{(-)}(\omega)}   \right].
 \end{align}
In the second line of
 Eq.~(\ref{eq:quasielastic_inserted_resummed}), all corrections that
 formally belong to the Quark-Quark-Reggeon vertex at NLLA are
 contained in the squared bracket to the left. As all of 
 parameters inside the bracket are small in comparison with the difference in  rapidity $Y_{AB}$, we therefore expand this bracket  up to order $g^2$, which corresponds to the order of the additional correction of the additionally included  quasi-elastic diagram. We  expand up to order $\omega$ and arrive at
\begin{align}
  \label{eq:quasielastic_expand}
i\mathcal{M}^\text{QEA}(s,t)&=
 i\mathcal{M}^\text{tree}(s,t) \int \frac{d\omega}{2\pi i} 
\frac{1}{\omega -\beta({\bm q})} 
\left|\frac{p_A^+p_B^-}{m_Am_B} \right|^\omega  \xi^{(-)}(\omega) 
  \frac{\beta({\bm q})}{\omega} 
 \frac{\ln(-\lambda_a) + \ln(\lambda_a)}{2}  .
 \end{align}
 The above expression has now to cancel the corresponding factor
 $\lambda_a$ in the diagrams of Sec.\ref{sec:reggeizedgluon}, where
 this dependence is kept explicitly. From Eq.(\ref{eq:twoloop-mit2A})
 and Eq.(\ref{eq:allorder_traj_fertig}) we find, converting
 $\Lambda_A$ into $\lambda_a$ according to Eq.(\ref{eq:1parameter})
 and dropping all other cut-off factors, the following expression:
\begin{align}
  \label{eq:reggeized_lambda}
& i\mathcal{M}^\text{tree}(s,t) \int \frac{d\omega}{4\pi i} 
\frac{1}{\omega -\beta({\bm q})} \left[ \left(\frac{ -p_A^+p_B^-}{m_Am_B \lambda_a} \right)^\omega   +   \left(\frac{p_A^+p_B^-}{m_Am_B \lambda_a} \right)^\omega \right] \notag \\
 =& i\mathcal{M}^\text{tree}(s,t) \int \frac{d\omega}{4\pi i} 
 \left[ \left(\frac{ -p_A^+p_B^-}{m_Am_B}  \right)^\omega   +   \left(\frac{p_A^+p_B^-}{m_Am_B } \right)^\omega \right]
\left[ \frac{1}{\omega -\beta({\bm q})} 
 - \frac{\omega}{\omega -\beta({\bm q}) }\frac{\ln(-\lambda_a)  + \ln(\lambda_a)}{2}   \right].
 \end{align}
 In the second line corrections beyond the LLA have been expanded up
 to order $\omega$. Adding Eq.\eqref{eq:reggeized_lambda} and Eq.
 \eqref{eq:quasielastic_inserted_resummed} , the pole $1/(\omega-
 \beta({\bm q}))$ cancels for the part proportional to $\ln
 (\lambda_a) $, in accordance with the present accuracy of our
 calculations, and the integral over $\omega$ yields zero result for
 the $\lambda_a$-dependent part.
 
 We therefore conclude this section on the exchange of a single
 reggeized gluon with the observation that a cancellation of the
 factorization parameter can be achieved, if corrections beyond the
 LLA are included.


\section{ Exchange of two reggeized gluons}
\label{sec:tworeggeon_negsig}

In the following we consider the first example of an amplitude in the
effective theory, with interaction of reggeized gluons. From a
technical view-point this means, that we turn to loop-integrations
with at least two reggeized gluons inside the loop.  As a particular
process we consider the elastic amplitude with exchange of two
reggeized gluons.  In the Reggeon-calculus
\cite{Gribov:1968fc,Baker:1976cv}, this kind of diagrams are known as
\emph{Regge-cuts} as they lead to branch cuts in the complex angular
momentum.  In the present study they are of two-fold interest: At
first they allow to determine the Regge-limit of elastic amplitudes,
where external quantum numbers do not allow for the exchange of a
single reggeized gluon. This is for instance the case if we project
the external quark-states on the color singlet or if we explicitly
require the amplitude to have positive signature. Furthermore it is
principally possible that those diagrams provide sub-leading
corrections to the reggeized gluon. 

\subsection{The Born-diagrams}
\label{sec:born}

\begin{figure}[htbp]
  \centering
   \parbox{4cm}{\center \includegraphics[width=2.5cm]{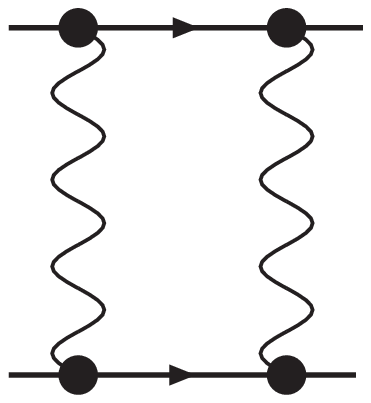}}
   \parbox{4cm}{\center \includegraphics[width=2.5cm]{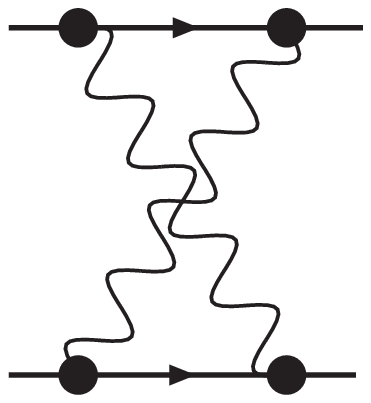}}\\
\parbox{4cm}{\center 2R1}\parbox{4cm}{\center 2R2}
  \caption{\small Diagrams with  exchange of two reggeized gluons. Coupling of the reggeized gluon to the quark is described by the effective vertex Eq.(\ref{eq:quark_reggeon}).}
  \label{two_reggeons}
\end{figure}

\begin{figure}[htbp]
  \centering
   \parbox{4cm}{\center \includegraphics[width=2.5cm]{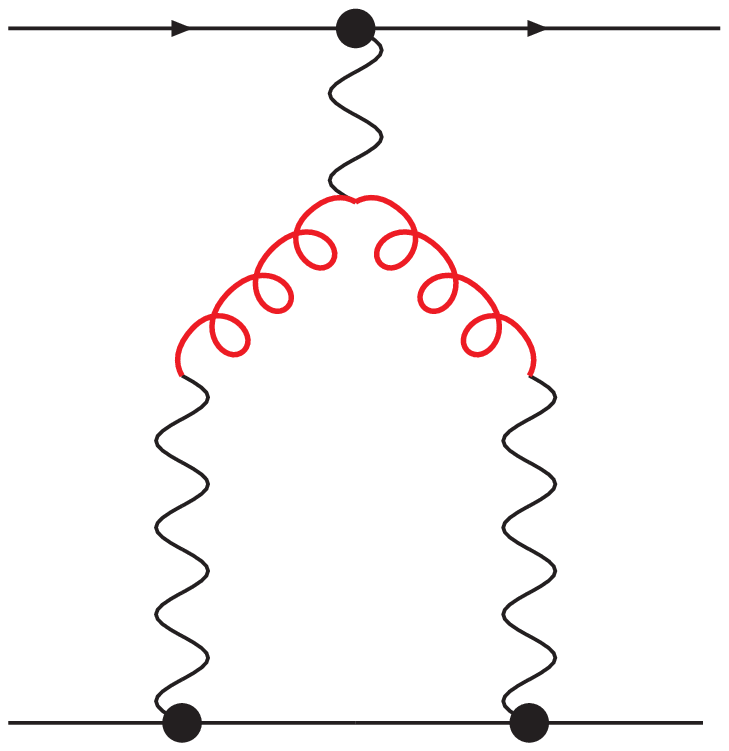}}
 \parbox{4cm}{\center \includegraphics[width=2.5cm]{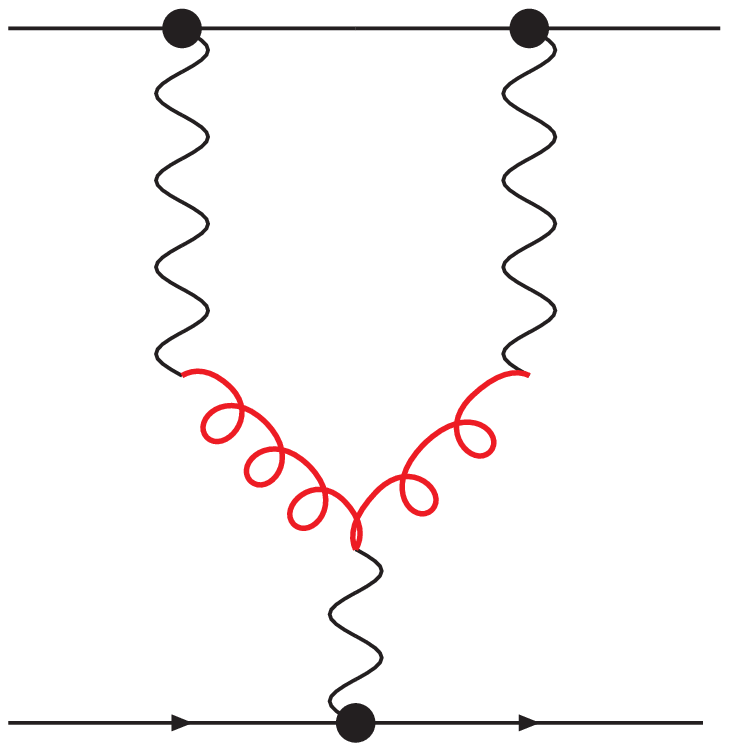}}
 \parbox{4cm}{\center \includegraphics[width=2.5cm]{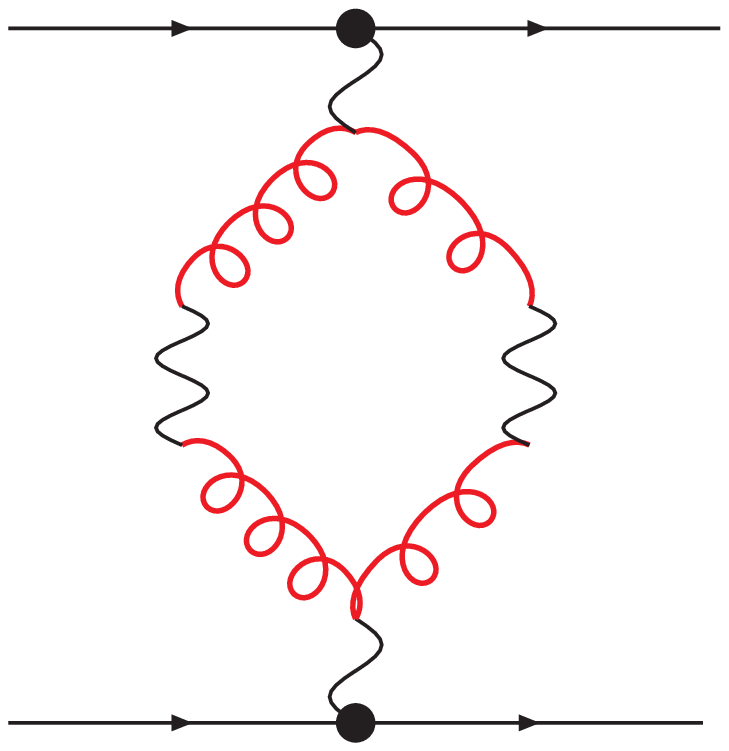}}\\
\parbox{4cm}{\center (a)}\parbox{4cm}{\center (b)}\parbox{4cm}{\center (c)}
  \caption{\small Feynman diagrams of the effective action provide besides diagrams 2R1 and 2R1 also the following set of diagrams, where one reggeized gluon splits up into two by an induced vertex. Diagrams where the induced vertex is replaced by a three-gluon-vertex exist as well, but there immediate zero due to the general properties of the reggeized gluon fields $A_\pm(x)$. }
  \label{fig:two_reggeon_subtract}
\end{figure}

To leading order in $g^2$,  the relevant diagrams are given by 2R1 and 2R2, as depicted in
Fig.\ref{two_reggeons}.
Another set of diagrams that occurs in principal as well is given by the graphs in Fig.\ref{fig:two_reggeon_subtract}. 
These diagrams take a special role in the discussion of the two-Reggeon exchange and will be addressed further down in this paragraph. For the sum of the diagrams 2R1 and 2R2 un-regularized 
Feynman rules of the effective action yield  the following expression:
\begin{align}
   \label{eq:22symmart_raw}  
  {\mathcal{M}}_{2 \to 2}^{\text{B}|\text{2R}}&(s,t) 
\notag   \\        
 =& 
            i|p_A^+||p_B^-| \int \frac{d^2 k_\perp}{(2 \pi)^2}  
        \bigg[(ig)^2 \int \frac{d \mu_A}{(-2 \pi i)} \left( 
 \frac{(t^{a_1} t^{a_2})_{AA'} }{\mu_A - {\bm k}^2 - m_A^2 + i \epsilon} 
 +
 \frac{ (t^{a_2} t^{a_1})_{AA'}}{-\mu_A - ({\bm q} -{\bm k})^2 -m_A^2 + i \epsilon} 
 \right)\bigg]
\notag \\
 \times &
 \frac{ 1 }{{\bm k}^2 (\bm{q} - \bm{k})^2} 
 \bigg[ (ig)^2\int \frac{d \mu_B}{2 \pi} \left( 
 \frac{(t^{a_1} t^{a_2})_{BB'} }{\mu_B - {\bm k}^2 - m_B^2 + i \epsilon} 
 +
 \frac{ (t^{a_2} t^{a_1})_{BB'}}{-\mu_B - ({\bm q} -{\bm k})^2 -m_B^2 + i \epsilon} 
 \right)\bigg],
 \end{align}
 where we defined $\mu_A = -p_A^+k^-$ and $\mu_B =p_B^-k^+ $.  The
 longitudinal loop integral factorizes and the integration can be
 carried out separately for each of the quark  impact factors,  which  are given by the big squared brackets in the first and second line of
 Eq.(\ref{eq:22symmart_raw}).  This kind of factorization is typical for loop-integrations of reggeized gluons. We therefore obtain
\begin{align}
   \label{eq:22symmart}
   \mathcal{M}^{\text{B}|\text{2R}}_{2 \to 2} 
            &= 
           2\pi i {|p_A^+||p_B^-|}\int \frac{d^2 k_\perp}{(2 \pi)^3} \frac{1}{{\bm k}^2 (\bm{q} - \bm{k})^2} 
        A^{a_1a_2}_{AA'} ( {\bm k}, {\bm q} - {\bm k}) 
        A^{a_1a_2}_{BB'} (  {\bm k}, {\bm q} - {\bm k}),
 \end{align}
with the  quark-impact factor for two reggeized gluons   given by
 \begin{align}
   \label{eq:22quarkimpaA}
    A^{a_1a_2}_{AA'}& ( {\bm k}, {\bm q} - {\bm k}) 
=
-g^2  \int \frac{d \mu_A}{( -2 \pi i)} \left( 
 \frac{(t^{a_1} t^{a_2})_{AA'} }{\mu_A - {\bm k}^2 - m_A^2 + i \epsilon} 
 +
 \frac{ (t^{a_2} t^{a_1})_{AA'}}{-\mu_A - ({\bm q} -{\bm k})^2 -m_A^2 + i \epsilon} 
 \right),
\\
 A^{a_1a_2}_{BB'}& ( {\bm k}, {\bm q} - {\bm k}) 
 = 
-g^2  \int \frac{d \mu_B}{( -2 \pi i)} \left( 
 \frac{(t^{a_1} t^{a_2})_{BB'} }{\mu_B - {\bm k}^2 - m_B^2 + i \epsilon} 
 +
 \frac{ (t^{a_2} t^{a_1})_{BB'}}{-\mu_B - ({\bm q} -{\bm k})^2 -m_B^2 + i \epsilon} 
 \right).
  \label{eq:22quarkimpaB}
 \end{align}
 As the impact factors do neither depend on $p_A^+$ nor on $p_B^-$, it
 is immediately clear, that the two reggeized gluon exchange
 amplitude, Eq.(\ref{eq:22symmart}), is symmetric under $s \to -s$ and
 therefore allows only for positive signature exchange in the
 $t$-channel.  From now on we restrict to the impact factor of the
 upper quark A, while the result for the quark B follows by symmetry.
 We then decompose color into a symmetric and an anti-symmetric part
 and obtain
 \begin{align}
   \label{eq:decompose_quarkif}
   A^{a_1a_2}_{AA'} ( {\bm k}, {\bm q} - {\bm k})  &=
  \frac{1}{2} \{t^{a_1}, t^{a_2}\}_{AA'}  A^{(+)}_{(2;0)}  ( {\bm k}, {\bm q} - {\bm k}) 
   +
   \frac{1}{2} [t^{a_1}, t^{a_2}]_{AA'}  A^{(-)}_{(2;0)} ( {\bm k}, {\bm q} - {\bm k}),
 \end{align}
with
\begin{align}
  \label{eq:if_sym}
 A^{(+)}_{(2;0)}  ( {\bm k}, {\bm q} - {\bm k})=&
-{g^2} \!\!  
\int \frac{d \mu_A}{(-2 \pi i)} \left( 
 \frac{1}{\mu_A - {\bm k}^2 - m_A^2 + i \epsilon} 
 +
 \frac{ 1}{-\mu_A - ({\bm q} -{\bm k})^2 -m_A^2 + i \epsilon} 
 \right),
\end{align}
and
\begin{align}
  \label{eq:if_asym}
 A^{(-)}_{(2;0)} ( {\bm k}, {\bm q} - {\bm k}) =&
-g^2  \!\!
\int \frac{d \mu_A}{( -2 \pi i)} \left( 
 \frac{1 }{\mu_A - {\bm k}^2 - m_A^2 + i \epsilon} 
 -
 \frac{1}{-\mu_A - ({\bm q} -{\bm k})^2 -m_A^2 + i \epsilon} 
 \right).
\end{align}
We start with the symmetric case: In the sum of the two denominators, the
singularity in $\mu_A$ cancels and the integral over $\mu_A$ is
 convergent.  Enclosing therefore for the \emph{sum} of the
two terms the contour of integration at infinity, we obtain
\begin{align}
  \label{eq:if_sym_fertig}
 A^{(+)}_{(2;0)}({\bm k}, {\bm q} - {\bm k}) =& -{g^2} ,
\end{align}
and
\begin{align}
  \label{eq:possig_cont}
 \mathcal{M}^{\text{B}|\text{2R}}_{2 \to 2} =&   \frac{i\pi}{2} \{t^{a_1}, t^{a_2}\}_{AA'}  \{t^{a_1}, t^{a_2}\}_{BB'}  g^4 |p_A^+ p_B^-|\int \frac{d^2 {\bm k}}{(2 \pi)^3} \frac{1}{{\bm k}^2 ({\bm q} - {\bm k})^2}.  
\end{align}
The discussion in the anti-symmetric case requires more care. The
logarithmic singularity in $\mu_A$ does not cancel and the integral
over the two terms appears to be ill-defined. To understand the origin
of this singularity, we return once again to the QCD-diagrams QCD1 and
QCD2 in Fig.\ref{fig:one_reggeon}. Taking into account simplifications
due to the Regge-limit, their sum is given by Eq.(\ref{eq:qcd1+qcd2}),
which we state here once again:
\begin{align*}
  \mathcal{M}_{\text{QCD}} =  g^4{(p_A^+p_B^-)}
\int \frac{dk^+dk^-}{2\pi i} \int \frac{d^2 {\bm{k}}}{(2\pi)^3}  
\left(
  \frac{(t^{c_1}t^{c_2})_{AA'}}{-k^- - \frac{{\bm k }^2+ m_A^2  + i\epsilon}{p_A^+ -k^+}}  + 
  \frac{(t^{c_2}t^{c_1})_{AA'}}{k^-  - \frac{({\bm q }- {\bm k})^2 + m_A^2  + i\epsilon}{p_A^+ +k^+}}  \right)
\notag \\
 \frac{1}{k^+k^- - {\bm k}^2 + i\epsilon}  \frac{1}{k^+k^- - ({\bm q} - {\bm k })^2 + i\epsilon}  
\left(
  \frac{(t^{c_1}t^{c_2})_{BB'}}{k^+ - \frac{({\bm q } - {\bm k})^2 +m_B^2 - i\epsilon}{p_B^- +k^-}}  + 
  \frac{(t^{c_2}t^{c_1})_{BB'}}{-k^+ - \frac{{\bm  k}^2  + m_B^2- i\epsilon}{p_B^- -k^-}}  \right)
\end{align*}
For the symmetric color combination, $k^-$ and $k^+$ are in the
integral fixed by the poles of the quark-propagators, inside the big
brackets. Consequently $k^+k^- \ll {\bm k}^2, ({\bm q } - {\bm k})^2$
and we find the expressions corresponding to the symmetric projection
of the diagrams 2R1 and 2R2.  For antisymmetric $t$-channel color on
the other hand, the leading pole in the expansion of the big brackets
in $\mu_A =- p_A^+k^-$ and $\mu_B=-p_B^- k^+$ does not cancel.  It was
shown in Sec.\ref{sec:traj} that it rather leads to the induced vertex
which enters for $\mu_A, \mu_B > \Lambda_a, \Lambda_b$ the central
rapidity and quasi-elastic diagrams.  In particular, integrating in
Eq.(\ref{eq:if_asym}) this pole-part over the full range of $\mu_A$
and $\mu_B$, we would count this contribution twice, due to its
presence in quasi-elastic and central-rapidity diagrams.  To avoid
this unjustified double-counting, this pole needs to be subtracted
from Eq.(\ref{eq:if_asym}) for large values of $\mu_A$ and $\mu_B$
respectively and doing so, integrals over $\mu_A$ and $\mu_B$ turn out
to be convergent.

\begin{figure}[htbp]
  \centering
  \parbox{2cm}{\includegraphics[height=2cm]{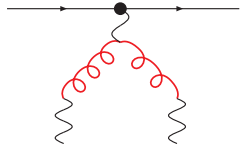}}
  \caption{\small Coupling of a reggeized gluon to two reggeized gluons by an induced vertices. These diagrams do not contribute to physical amplitudes, however they provide the diagrammatic form of the subtraction terms in the antisymmetric color sector.}
  \label{fig:threeR}
\end{figure}

The particular form of the term we should subtract is given by the
graphs in Fig.\ref{fig:two_reggeon_subtract} or for the impact alone
by Fig.\ref{fig:threeR}, which describes the splitting of a single
reggeized gluon into two reggeized gluons by an induced vertex.
Applying the same factorization as for the impact factor, the diagram Fig.\ref{fig:threeR} yields, including a lower bound on $\mu_A$ due to the presence of the reggeized gluon,
\begin{align}
  \label{eq:subtraction_diagram}
 A^{S a_1a_2}_{AA'} = \frac{-g^2}{2} [t^{a_1},t^{a_2}]_{AA'}  \int \frac{d \mu_A}{(-2\pi i)}  \int_{0 -i\infty}^{0 + i\infty} \frac{d \omega}{2\pi i}
 \frac{\Lambda_a^{-\omega}}{\omega + \nu} \bigg[ \left({\mu_A -i\epsilon } \right)^{\omega -1} -  \left({-\mu_A -i\epsilon }\right)^{\omega-1} \bigg].
\end{align}
Making use of 
\begin{align}
  \label{eq:theta_trick}
 \int_{0 -i\infty}^{0 + i\infty}  \frac{d \omega}{2\pi i}
 \frac{1}{\omega + \nu}  \bigg(\frac{\mu_A }{\Lambda_a} \bigg)^{\omega}  
= 
1
-
  \int_{0 -i\infty}^{0 + i\infty}  \frac{d \omega}{2\pi i} \frac{-1}{\omega - \nu}  \bigg(\frac{\mu_A }{\Lambda_a} \bigg)^{\omega} 
\end{align}
where on the right-hand-side, the pole in $\omega$ is  now to the left of the contour, we find, subtracting Eq.(\ref{eq:subtraction_diagram}) from Eq.(\ref{eq:if_asym})
\begin{align}
  \label{eq:if_asym_regularized2}
 A^{a_1a_2,(-)}_{AA'} &({\bm k}, {\bm q} - {\bm k})  - A^{S a_1a_2}_{AA'}({\bm k}, {\bm q} - {\bm k}) =
\notag \\
=& \frac{ig^2}{2} [t^{a_1}, t^{a_2}]_{AA'}
\int \frac{d \mu_A}{2 \pi}
 \bigg\{
 \int_{0 -i\infty}^{0 + i\infty} \frac{d \omega}{2\pi i} \frac{-\Lambda_a^\omega }{\omega - \nu} \bigg[ \left( {\mu_A -i\epsilon } \right)^{\omega-1} -
 \left({-\mu_A -i\epsilon } \right)^{\omega-1} \bigg]
\notag \\
   +&
   \bigg(\frac{1 }{\mu_A - {\bm k}^2 - m_A^2 + i \epsilon} 
   -
   \frac{1}{\mu_A + i\epsilon}     
\bigg)
-
\bigg(   \frac{1}{-\mu_A - ({\bm q} -{\bm k})^2 -m_A^2 + i \epsilon} 
-  \frac{1}{\mu_A + i\epsilon} 
 \bigg)\bigg\}.
\end{align}
In the last line all integrals are now convergent. Taking
residues, the result of the two brackets cancels. In the
first line, there is no singularity in the the
$\omega$-plane to the left of the $\omega$-contour which allows to
move the $\omega$-contour to the left until $\omega$ acquires a
negative real part and the integral over $\mu_A$ gets  convergent.
As there is no further singularity in the $\mu_A$-plane, we  close the $\mu_A$-contour without encircling any
singularity at infinity and   Eq.(\ref{eq:if_asym_regularized2}) vanishes.

This is the expected  result as at high center of mass energies
$s$, the QCD amplitude with anti-symmetric color exchange reggeizes and is 
described by the exchange of a single reggeized gluon.  In particular
the amplitude with antisymmetric color in the $t$-channel is given by
the sum of quasi-elastic and central-rapidity diagrams alone.

The method developed here for longitudinal integrations in the context
of loops containing two or more reggeized gluons, turns out to be a
rather general one: The part of the amplitudes that can be constructed
from direct coupling the reggeized gluons by induced vertices is
already contained in the corresponding central-rapidity diagrams.

A comment is in order concerning diagrams like
Fig.\ref{fig:two_reggeon_subtract}: With our subtraction-mechanism,
those diagrams are subtracted by themselves and do not contribute.
This subtraction mechanism can be further automatized by adding a
subtraction term to the Lagrangian, that subtracts these contributions
automatically. Details about the subtraction term will be presented in
Sec.\ref{sec:impa4_bkp}.

Another comment is in order concerning the decoupling of the
anti-symmetric two Reggeon state of the quark: To study reggeization
of the gluon, one usually makes use of $s$-channel dispersion
relations (see for instance \cite{Forshaw:1997dc} for a pedagogical
introduction). This requires to determine the $s$-discontinuity of the
elastic amplitude (i.e. its imaginary part).  Applying  a bootstrap,
the discontinuity is expressed as a bound state of  two reggeized
gluons.  However, after resumming corrections within the LLA by the
BFKL-equation to all orders, the two Reggeon state drops out and one
is left with the exchange of a single reggeized gluon, for details see
\cite{Bartels:1978fc,Bartels:1991bh,Forshaw:1997dc}. As a consequence,
also in the dispersion-relation based approach, the anti-symmetric two
Reggeon state decouples  from the quark.  In the effective
action the reggeized gluon comes explicitly with negative signature
and therefore contains also the $s$-discontinuity of the elastic
amplitude.

\subsection{Higher order corrections: the BFKL-equation}
\label{sec:bfkl}
\begin{figure}[htbp]
  \centering
  \parbox{3.5cm}{\center \includegraphics[height=2cm]{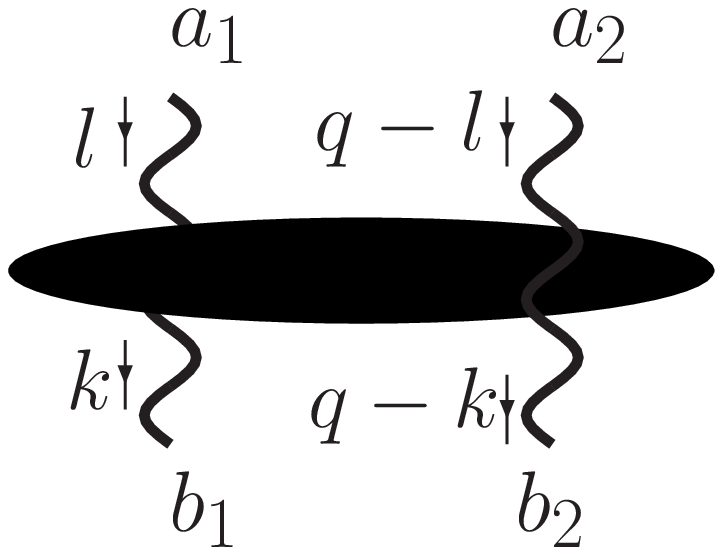}} =
 \parbox{2.5cm}{\center \includegraphics[height=1.5cm]{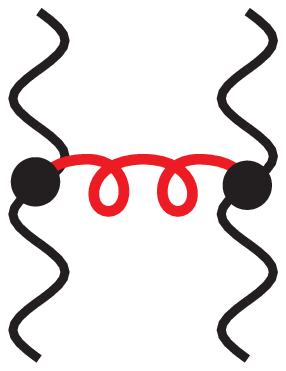}} +
 \parbox{2.5cm}{\center \includegraphics[height=1.5cm]{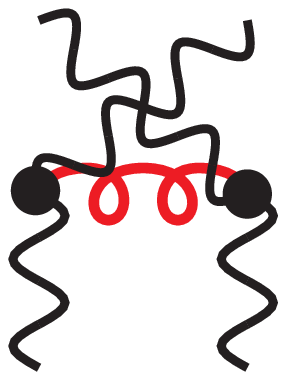}} +
 \parbox{2.5cm}{\center \includegraphics[height=1.5cm]{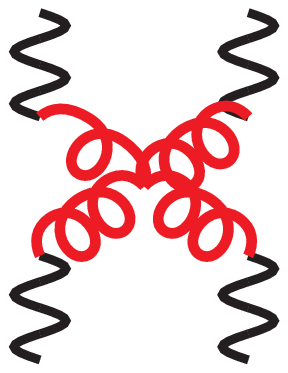}} \\
 \parbox{3.9cm}{\center $\,$ }\parbox{2.5cm}{\center K22A }\parbox{3.5cm}{\center K22B}\parbox{2.5cm}{\center K22C }
  \caption{\small Interaction of two reggeized gluons which yields to 2-to-2 reggeized gluons transition kernel.}
  \label{fig:22kernel}
\end{figure}

\begin{figure}[htbp]
  \centering \parbox{2cm}{\center
    \includegraphics[height=2cm]{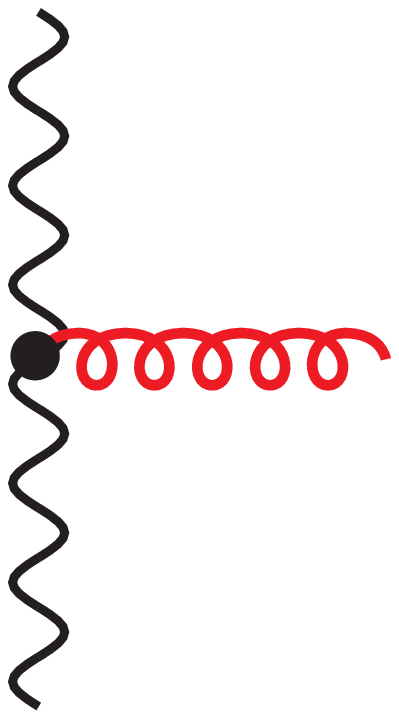}} = \parbox{2cm}{\center
    \includegraphics[height=2cm]{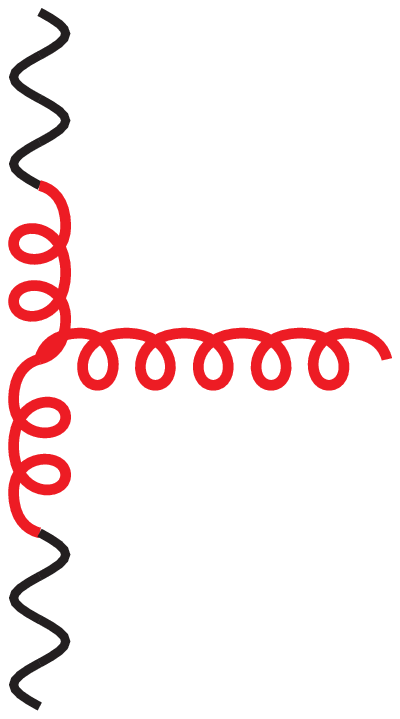}} + \parbox{2cm}{\center
    \includegraphics[height=2cm]{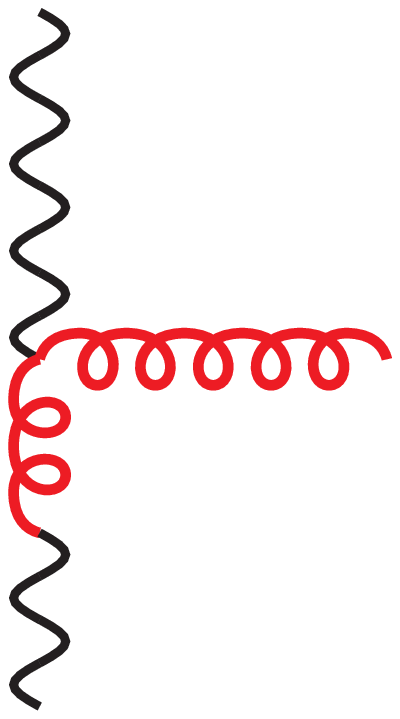}} + \parbox{2cm}{\center
    \includegraphics[height=2cm]{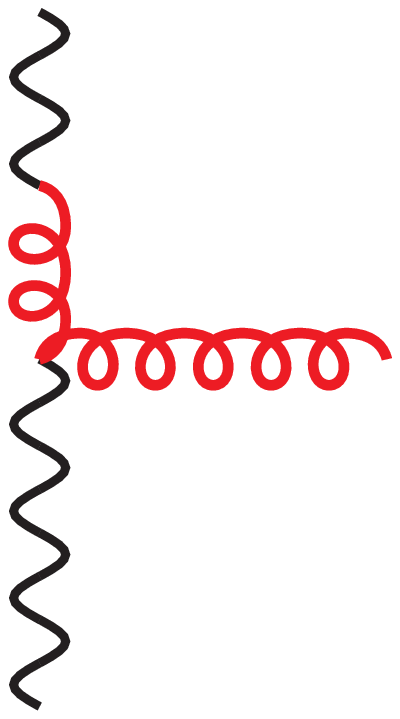}}
  \caption{\small The effective particle production or Lipatov vertex in the effective action formalism.}
  \label{fig:lipprod_vertex}
\end{figure}

Higher order corrections to the diagrams in Fig.\ref{two_reggeons} are
two-fold: On the one hand they involve corrections due to reggeization
of the gluon, which are taken into account by replacing bare reggeized
gluons by their resummed counterparts
Eq.(\ref{eq:allorder_traj_fertig}). On the other hand there are
corrections due to interactions between reggeized gluons. In
particular, every interaction between two reggeized gluons yields a
logarithm in $s$ and must be taken into account with the LLA.
Interactions between the two reggeized gluons are then resummed by the
BFKL-equation, which we will derive in the following. Furthermore,
there exist also corrections due to exchange of more than two
reggeized gluons. For the elastic amplitude, these corrections are
formally sub-leading within in the LLA. They are addressed in
Cha.\ref{cha:v24}.  Within the LLA interactions between two reggeized
gluons is given by the diagrams in Fig.\ref{fig:22kernel}.
There we use  the gauge invariant production or Lipatov vertex
 \begin{align}
  \label{eq:1lipatov}
  \parbox{1cm}{\includegraphics[height = 1cm]{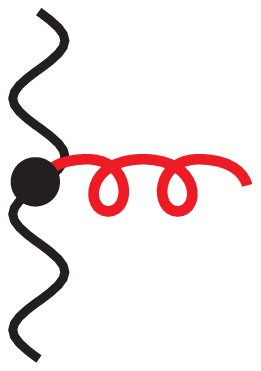}} =    2 gf^{acb}& C_\mu(l, k), 
\end{align}
which is an effective vertex. Within the effective action it arises as
a combination of the diagrams in Fig.\ref{fig:lipprod_vertex}
which yields
\begin{align}
\label{eq:2lipatov}
 C_\mu(l, k) &= \left( \frac{l^+}{2} + 
     \frac{{\bm{l}}_1^2}{k^- }   \right) (n^-)^\mu + \left( \frac{k^-}{2} + \frac{{\bm{k}}_2^2}{ l^+ } \right) (n^-)^\mu - (l+ k)^\mu_\perp.
\end{align}
The production vertex is gauge invariant; especially the relation
\begin{align}
  \label{eq:3lipatov}
 C_\mu(l, k) \cdot (l - k)^\mu = 0,
\end{align}
 holds also if
the gluon is off-shell.  For the individual diagrams in
Fig.\ref{fig:22kernel} one obtains the following expressions
\begin{align}
  \label{eq:k22A}
\text{K22A} &= -i8g^2T^{b_1}_{a_1c}T^{b_2}_{ca_2} \left[\frac{1}{2} + 
    \left(
      -{\bm q}^2 
      +
      \frac{({\bm l} - {\bm q})^2{\bm k}^2 }{-l^+k^-} 
      +
      \frac{({\bm k} - {\bm q})^2{\bm l}^2 }{-l^+k^-}
    \right) \frac{1}{-l^+k^- - ({\bm{l } - {\bm k}})^2 + i\epsilon}
 \right], \\
  \label{eq:k22B}
\text{K22B} &= -i8g^2T^{b_2}_{a_1c}T^{b_1}_{ca_2} \left[\frac{1}{2} + 
    \left(
      -{\bm q}^2 
      +
      \frac{{\bm l}^2{\bm k}^2 }{l^+k^-} 
      +
      \frac{({\bm k}\! -\! {\bm q})^2( {\bm l}\! -\! {\bm q})^2 }{l^+k^-}
    \right) \frac{1}{l^+k^- \! - ({\bm{l } -{\bm q} \!+ \!{\bm k}})^2 + i\epsilon}
 \right], \\
  \label{eq:k22C}
\text{K22C} &= i4g^2 \left( T^{b_1}_{a_1c}T^{b_2}_{ca_2}  + T^{b_2}_{a_1c}T^{b_1}_{ca_2} \right),
\end{align}
where $T^b_{ac} = if^{abc}$ are generators in the adjoint
representation of $SU(N_c)$. In the sum of the
three diagrams, K22C cancels the constant term in K22A and K22B and we
obtain
\begin{align}
  \label{eq:K22}
\text{K22A} + \text{K22B} + \text{K22C} =& -i8g^2 \bigg[
       -{\bm q}^2   \bigg(
         \frac{ T^{b_1}_{a_1c}T^{b_2}_{ca_2}}{-l^+k^- - ({\bm{l } - {\bm k}})^2 + i\epsilon} 
         +
         \frac{T^{b_2}_{a_1c}T^{b_1}_{ca_2}}{l^+k^- \! - ({\bm{l } -{\bm q} \!+ \!{\bm k}})^2 + i\epsilon} 
         \bigg)
\notag \\
& +
T^{b_1}_{a_1c}T^{b_2}_{ca_2}  \left(
      \frac{({\bm l} - {\bm q})^2{\bm k}^2 }{-l^+k^-} 
      +
      \frac{({\bm k} - {\bm q})^2{\bm l}^2 }{-l^+k^-}
    \right) \frac{1}{-l^+k^- - ({\bm{l } - {\bm k}})^2 + i\epsilon}
\notag \\
&+
T^{b_2}_{a_1c}T^{b_1}_{ca_2}\left(
  \frac{{\bm l}^2{\bm k}^2 }{l^+k^-} 
      +
      \frac{({\bm k}\! -\! {\bm q})^2( {\bm l}\! -\! {\bm q})^2 }{l^+k^-}
    \right) \frac{1}{l^+k^- \! - ({\bm{l } -{\bm q} \!+ \!{\bm k}})^2 + i\epsilon}
\bigg].
\end{align}
\begin{figure}[htbp]
  \centering
  \parbox{3cm}{\center \includegraphics[height=2cm]{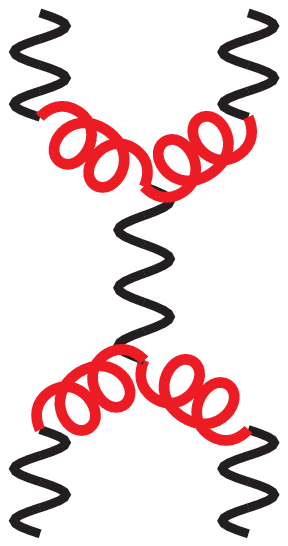}} \\
 \parbox{3cm}{\center K22S}
  \caption{\small Subtraction term for the interaction of two-reggeized gluons. It takes into account the contributions of the two-Reggeon-interaction with anti-symmetric $t$-channel color that are already contained in the reggeized gluon.}
  \label{fig:bfklsubtraction}
\end{figure}
Similar to the two-gluon quark-impact-factor in
Sec.\ref{sec:tworeggeon_negsig}, if we project on the antisymmetric
color channel, we find in the first line, once we integrate over
light-cone momenta, a logarithmic singularity in $l^+k^-$.  For values
of $l^+k^-$ larger than $\Lambda_c$, this pole is already included in
the 2-loop reggeized gluon, Sec.\ref{sec:reggeizedgluon} and
particularly Eq.(\ref{eq:combiindpropind}). The corresponding
contribution needs therefore to be subtracted, and  the precise form of
the subtraction term is obtained from the graph K22S in
Fig.\ref{fig:bfklsubtraction},   which yields the following expression:
\begin{align}
  \label{eq:bfklsubtraction}
\text{K22S} &= -i8g^2 {{\bm q}^2} [T^{b_1}, T^{b_2}]_{a_1a_2} 
\int \frac{d\omega}{4\pi i} 
   \frac{1}{\omega+ \nu} \left[  \left(\frac{l^+k^--i\epsilon}{\Lambda_c} \right)^{\omega -1} -  \left(\frac{-l^+k^--i\epsilon}{\Lambda_c} \right)^{\omega -1}  \right] 
.
\end{align}
Making use of Eq.(\ref{eq:theta_trick}), we obtain altogether for the interaction kernel
\begin{align}
  \label{eq:K22manip}
&\text{K22A} + \text{K22B} + \text{K22C} - \text{K22S} =-i8g^2 \bigg[
 T^{b_1}_{a_1c}T^{b_2}_{ca_2}   
      \frac{-{\bm q}^2  ({\bm l } - {\bm k})^2
      +
      ({\bm l} - {\bm q})^2{\bm k}^2  
      +
     ({\bm k} - {\bm q})^2{\bm l}^2 }
     {(-l^+k^-)(-l^+k^- - ({\bm{l } - {\bm k}})^2 + i\epsilon)}
 \notag \\
&
+ T^{b_2}_{a_1c}T^{b_1}_{ca_2}   
         \frac{ -{\bm q}^2 ({\bm l} - {\bm q} + {\bm k}) 
      +
      {\bm l}^2{\bm k}^2 
      +
      ({\bm k}\! -\! {\bm q})^2( {\bm l}\! -\! {\bm q})^2 } {(l^+k^-)(l^+k^- \! - ({\bm{l } -{\bm q} \!+ \!{\bm k}})^2 + i\epsilon)}
\notag \\
& \qquad \qquad \qquad  +{\bm q}^2 [T^{b_1}, T^{b_2} ]_{a_1a_2} \int \frac{d\omega}{4\pi i} 
   \frac{\Lambda_c}{\omega- \nu} \left[  \left(l^+k^- \!-\!i\epsilon\right)^{\omega -1} \!\!\!\! - \! \left(-l^+k^-\!-\!i\epsilon \right)^{\omega -1}  \right]  \bigg]
\end{align}
where in the last line the $\omega$-contour is to the left of the pole at $\omega=\nu$ and the $i \epsilon$-prescription of the poles is given by
\begin{align}
  \label{eq:polelk}
\frac{1}{l^+k^-} = \left(\frac{1/2}{l^+k^-\! +\! i\epsilon} \!+\! \frac{1/2}{l^+k^-\! -\! i\epsilon}   \right) 
= \frac{1}{4} \left( \frac{p_B^-}{p_B^-l^+\! +\! i\epsilon} \!+\!
                    \frac{p_B^-}{p_B^-l^+ \!-\!i\epsilon}\right) 
 \left( \frac{p_A^+}{p_A^+k^-\! +\! i\epsilon} 
                 \! +\! \frac{p_A^+}{p_A^+k^-\! -\!i\epsilon}\right).
\end{align}
To obtain the one-loop correction to the quark-quark-scattering
amplitude with exchange of two reggeized gluons within the LLA, we
insert the reggeized gluon interaction vertex Fig.\ref{fig:22kernel}
into the Born diagram Fig.\ref{fig:qq_1k22}a, which leads to the graph
of Fig.\ref{fig:qq_1k22}b.
\begin{figure}[htbp]
  \centering
\parbox{4cm}{\center \includegraphics[height=2.5cm]{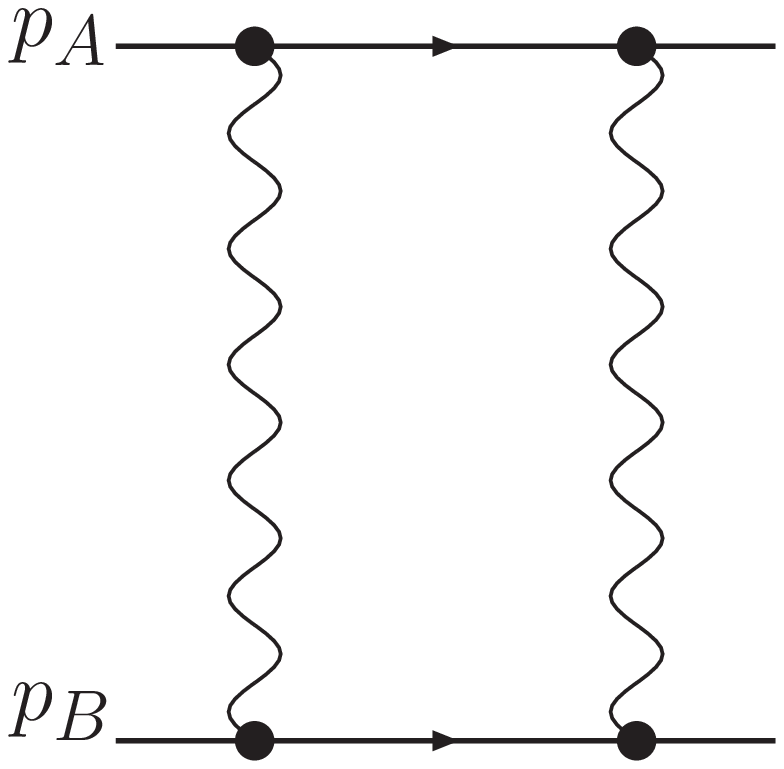}}  
\parbox{4cm}{\center \includegraphics[height=2.5cm]{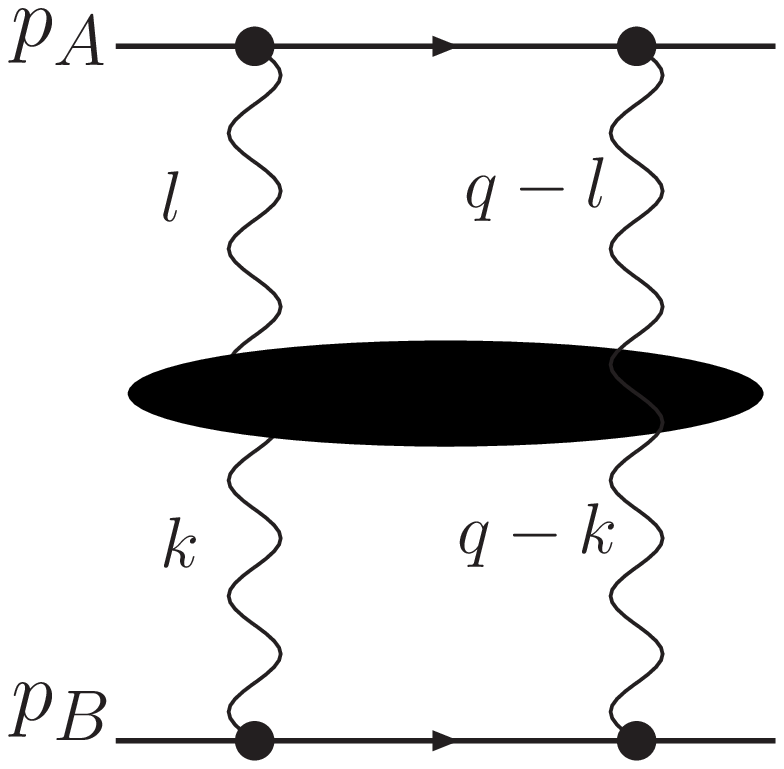}} \\
\parbox{4cm}{\center (a)}\parbox{4cm}{\center (b)}
  \caption{\small Quark-quark-scattering at Born-level and with insertion of one interaction kernel between the reggeized gluons. }
  \label{fig:qq_1k22}
\end{figure}
To evaluate the integration over longitudinal loop momenta $l$ and
$k$, the following peculiarity of a theory of reggeized gluons has to
be taken into account: Every single reggeized gluon belongs to a
certain 4-point sub-amplitude inside the complete amplitude and
presence of the reggeized gluons requires, that the center of mass
energy of the sub-amplitude is significantly larger than all other
scales that occur for this particular sub-amplitude.
\begin{figure}[htbp]
  \centering
  \parbox{6cm}{\center \includegraphics[height=3cm]{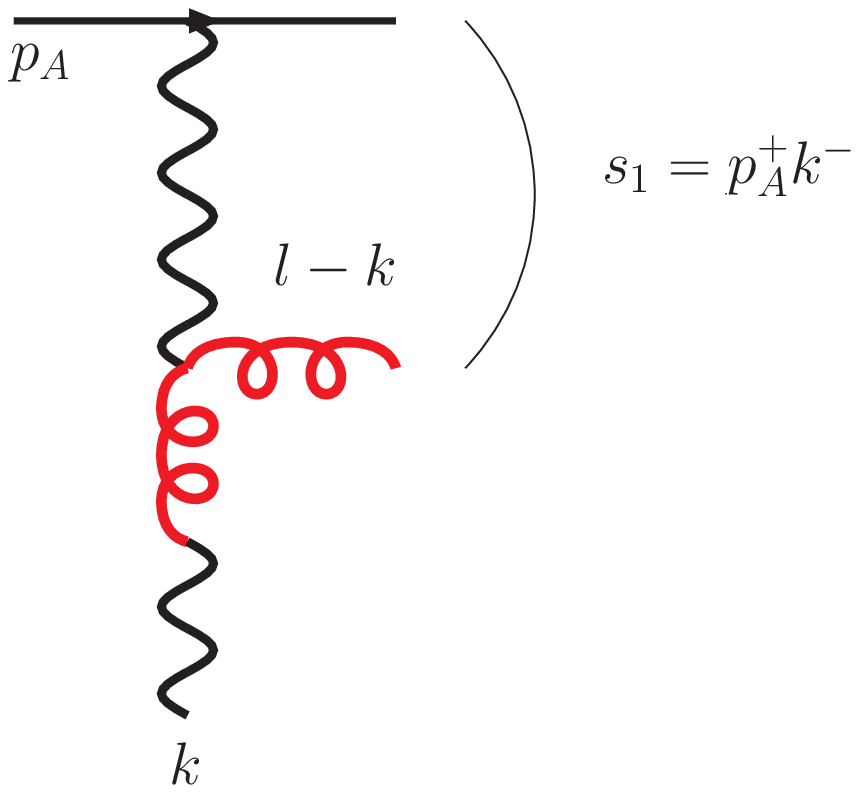}}
 \parbox{6cm}{\center \includegraphics[height=3cm]{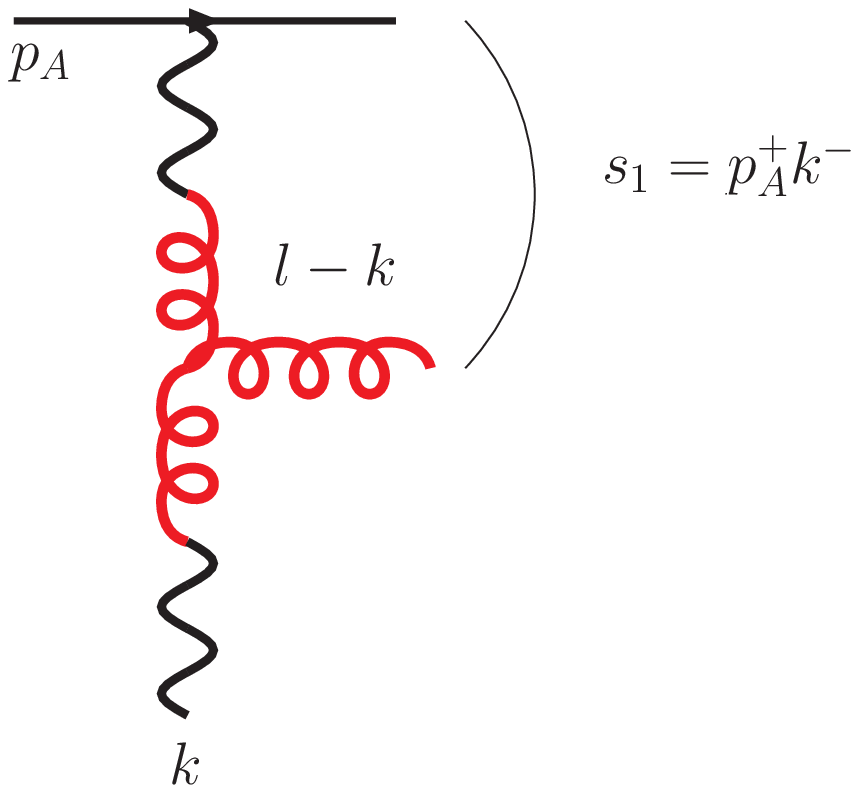}} 
\\
 \parbox{6cm}{\center (a)}  \parbox{6cm}{\center (b)}
  \caption{\small 4-point sub-amplitudes with a Regge-pole i.e. the interaction is mediated by the reggeized gluon, involving an induced vertex of the first order (a) and the three-gluon vertex (b).}
  \label{fig:indus_prod}
\end{figure}
Example of such sub-amplitudes are shown in Fig.\ref{fig:indus_prod}.
Imposing these lower bounds is not only demanded due to the presence
of the reggeized gluons, but they also appear naturally if one
attempts to derive the graphs of the effective theory from the
underlying QCD-graphs, similar to Sec.\ref{sec:22negsig}.  For graphs
that contain combinations of induced vertices, as
Fig.\ref{fig:qcd_eff1}a, this occurs in the same way as for the
reggeized gluon in Sec.\ref{sec:traj}. There, the induced vertices
could be shown to arise as the leading term of the expansion of the
corresponding quark-propagators, Eq.(\ref{eq:expandee_rl2}).  To give
meaning to the expansion, we need to introduce lower bounds on the
center-of-mass energies of the sub-amplitudes.

Apart from diagrams with induced vertices, the interaction kernel
Fig.\ref{fig:22kernel} contains also contributions due to the
three-gluon-vertices, of which the QCD counter-part is shown in
Fig.\ref{fig:qcd_eff1}b.  Also in that case, the effective action
graphs imply simplification that only apply if the the sub-energies
$s_1 = p_A^+k^-$ and $s_2 = p_B^-l^+$ are large. In the effective
theory diagram at the right hand side of Fig.\ref{fig:qcd_eff1}b, the
propagator of the gluon with momentum $l -k$ is considerably
simplified compared to the corresponding gluon propagator in full QCD
to the left: Within the effective theory, 'small' longitudinal momenta
$l^-$ and $k^+$ are neglected against the corresponding 'large'
momenta $k^-$ and $l^+$ respectively. From Sec.\ref{sec:born} we know,
that products $p_A^+l^-$ and $p_B^-k^+$ are of the order of the
transverse scale and consequently neglecting $l^-$ and $k^+$ in the
propagator is only justified if $s_1 =p_A^+k^-$ and $s_2 =p_B^-l^+$
are considerably larger than the transverse scale.

\begin{figure}[htbp]
  \centering
  \parbox{2.5cm}{\center \includegraphics[height=2.5cm]{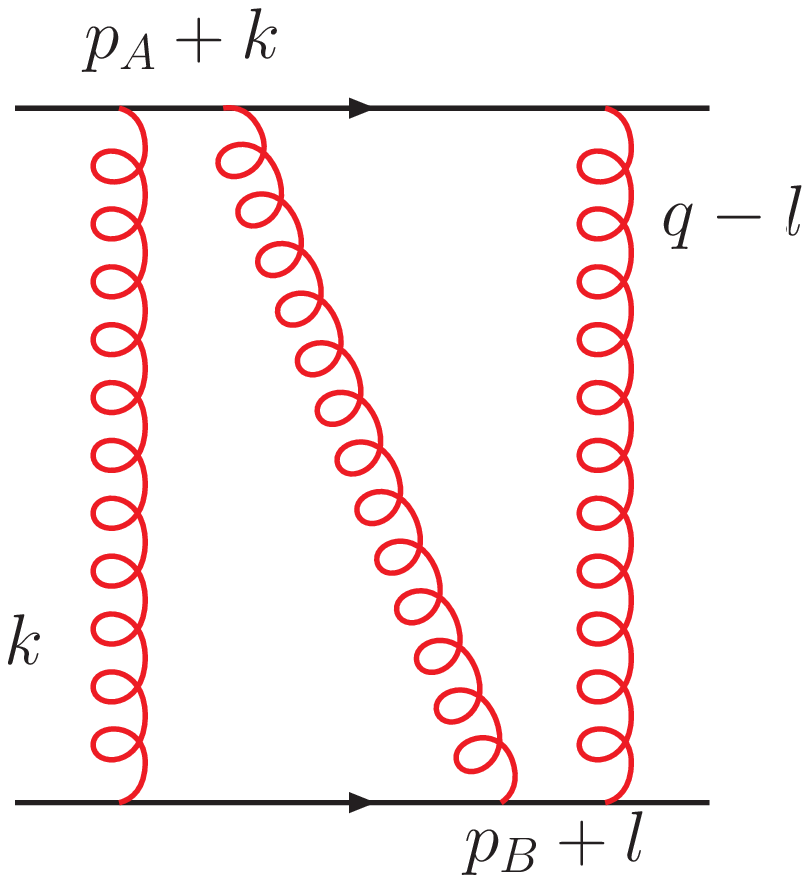}} 
  $\to$
  \parbox{2.5cm}{\center \includegraphics[height=2.5cm]{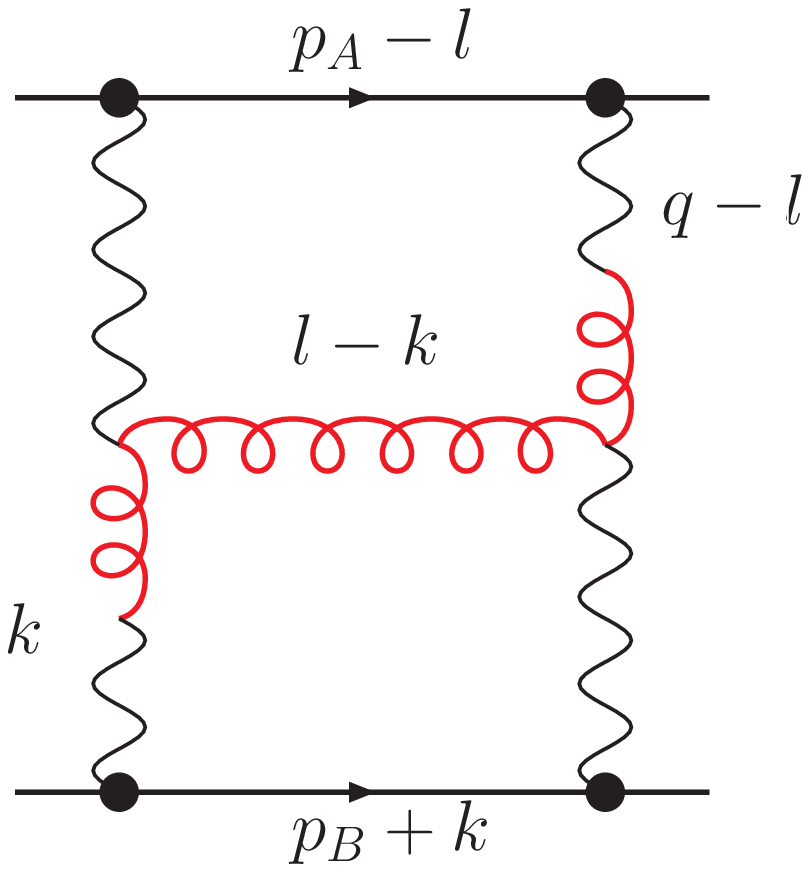}}
  \parbox{1.5cm}{$\,$}
  \parbox{2.5cm}{\center \includegraphics[height=2.5cm]{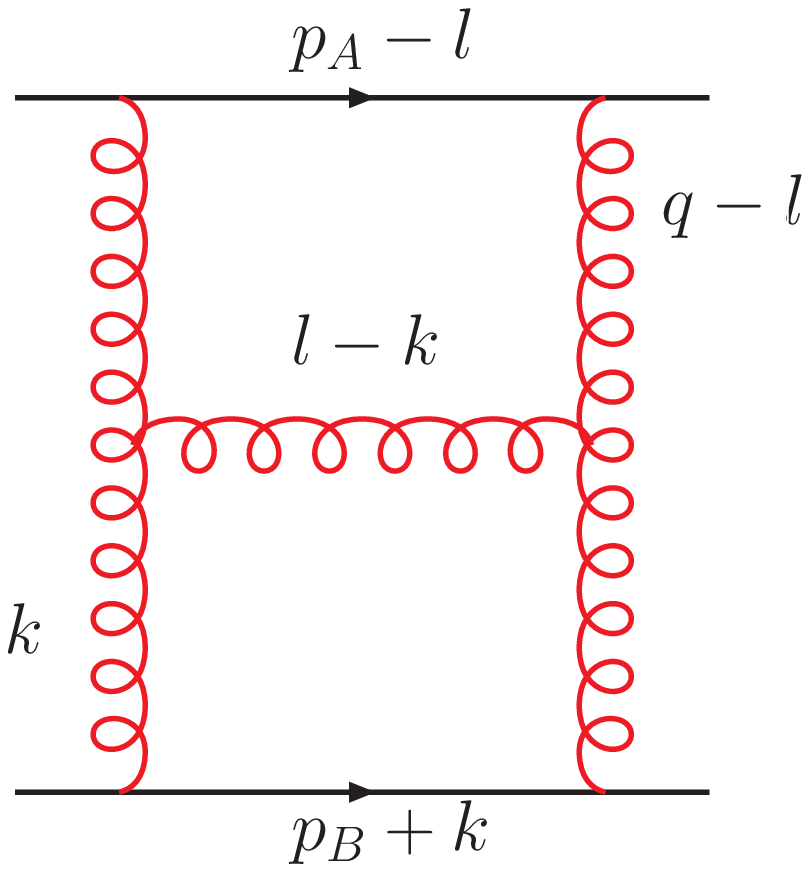}} 
  $\to$
  \parbox{2.5cm}{\center \includegraphics[height = 2.5cm]{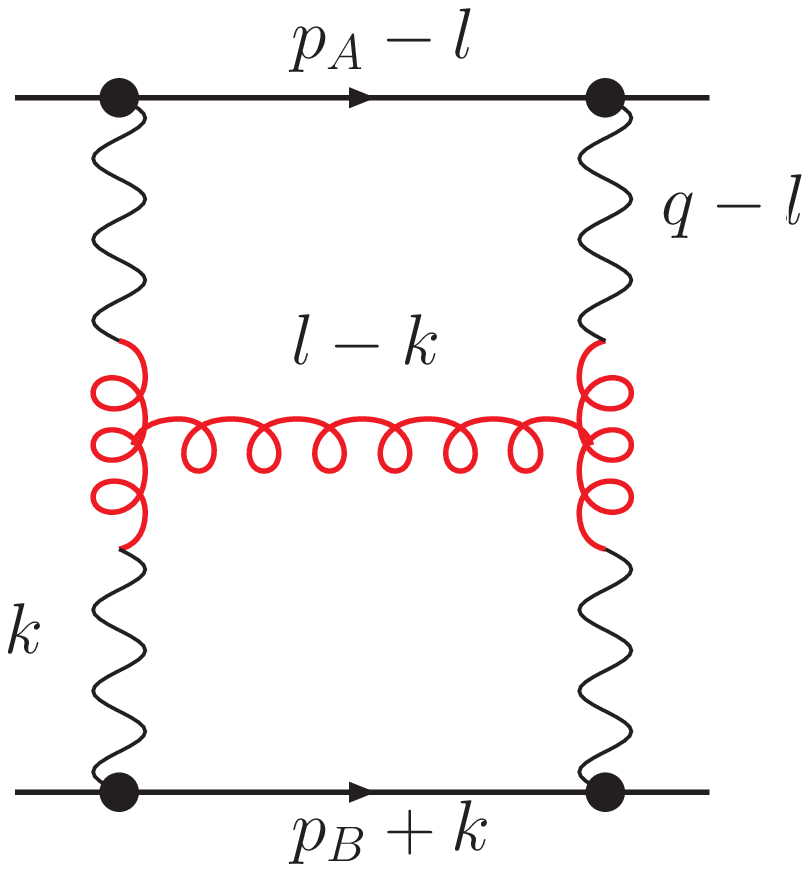}} \\
\parbox{5.5cm}{\center (a)} \parbox{1.5cm}{$\,$} \parbox{5.5cm}{\center (b)} 
  \caption{\small Emergence of effective action graphs out of usual QCD-diagrams. (a): Combination of two induced vertices, which arise out of the quark-propagators $(\fdag{p_A} -\fdag{l})^{-1}$ and $(\fdag{p_B} + \fdag{k})^{-1}$.  (b): Combination of two three-gluon vertices. In both cases the gluon  propagator in the effective theory is simplified due to the particular kinematics, compared to the full QCD diagram. }
  \label{fig:qcd_eff1}
\end{figure}

\begin{figure}[htbp]
  \centering
\parbox{4cm}{\center \includegraphics[height=2.5cm]{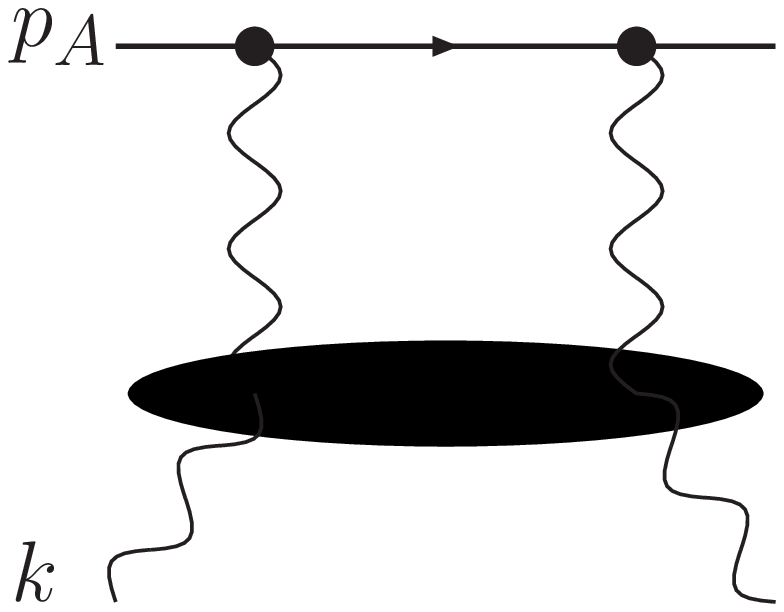}}
\parbox{4cm}{\center \includegraphics[height=2.5cm]{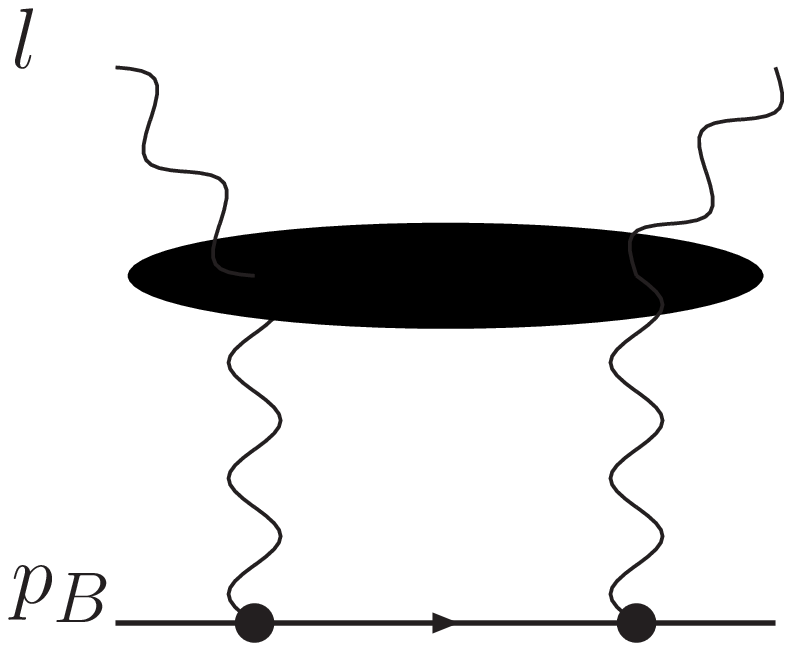}} 
  \caption{\small  Sub-amplitudes of the second diagram of Fig.\ref{fig:qq_1k22}: For each of the two sub-amplitudes, presence of the two reggeized gluons requires that the corresponding squared sub-center-of-mass energies, $s_1 = (p_A - k)^2$ and $s_2 = (p_B + l)^2$ are significantly larger than a lower cut-off $\Lambda$.}
  \label{fig:qq_2k22_subs}
\end{figure}

In the case of the 2-2 reggeized gluon transition we are now in the
comfortable situation that the constraint due to both reggeized gluons
can be gathered into one single constraint for the sub-amplitudes
depicted in Fig.\ref{fig:qq_2k22_subs}.  As in Sec.\ref{sec:traj}, the
(inverse) Mellin-integrals, Eq.(\ref{eq:theta_Mellin}), can be used to
impose the lower cut-offs.  As explained in Sec.\ref{sec:traj} this
factors lead to branch cuts in the complex plane along the real axis,
and both their phase and a prescription how to lead the contour of
integration around these branch-cuts has to be given.
For the diagrams containing induced vertices, the result of
Sec.\ref{sec:traj} applies immediately. They contain a pole with a
certain $i\epsilon$-prescription, and poles slightly below the real
axis are associated with a branch-cut along the positive semi-axis,
while poles slightly above the real axis come with a branch cut along
the negative semi-axis.  At first the  interaction vertex that come
at first with no pole. However, such a pole appears after the subtraction of the
diagram K22S in
Eq.(\ref{eq:K22manip}). Furthermore, this pole comes with the same pole-prescription as the
poles of the induced vertices.  We will treat these poles in the following
like poles arising from  induced vertices  and particularly require that  their $i\epsilon$
-prescription coincides with the $i\epsilon$-prescription  of the branch-cuts of the Mellin integral, Eq.(\ref{eq:theta_Mellin}). 
Poles coming with a $+i\epsilon$  will be associated with a branch cut
along the positive semi-axis. Poles with a $-i\epsilon$ with a
branch cut along the negative semi-axis,  in accordance with the
Feynman-prescription for bypassing singularities on the real-axis.
For every sub-amplitude of Fig.\ref{fig:qq_2k22_subs}, the poles in $l^+$ and $k^-$ are therefore replaced by the following expressions
\begin{align}
  \label{eq:phases1}
\frac{1}{l^+} \qquad &\to \qquad p_B^- \int \frac{d\omega_2 }{4\pi i} \frac{  \Lambda_b^{\omega_2} }{\omega_2 +\nu } 
\left[    \left(p_B^-l^+ -i\epsilon \right)^{\omega_2-1 }   - 
 \left(-p_B^-l^+ -i\epsilon  \right)^{\omega_2-1}  \right],
\notag \\
\frac{1}{k^-} \qquad & \to \qquad  p_A^+\int \frac{d\omega_1 }{4\pi i} \frac{ \Lambda_a^{\omega_1}   }{\omega_1 +\nu } 
\left[ \left( p_A^+k^- -i\epsilon \right)^{\omega_1-1} - \left( -p_A^+k^-
-i\epsilon \right)^{\omega_1-1}  
\right].
\end{align}
This result can be also obtained by simply inserting the resummed reggeized gluon from Eq.(\ref{eq:vertexsig_fac_extract}) into our expression. From Eq.(\ref{eq:vertexsig_fac_extract}), every  reggeized gluon
carries a phase factor $\sim (e^{-i\pi \omega} + 1)$, with $\omega$
the Mellin variable/the trajectory function of the regarding reggeized
gluon. The overall phase is therefore given by the product of the two
phase factors which can be rewritten (following closely
\cite{Gribov:1968fc,Baker:1976cv}) as
\begin{align}
  \label{eq:sig_manipus}
\frac{e^{-i\pi\omega} + 1}{2 } \frac{e^{-i\pi\omega'} + 1}{2 }=  \frac{e^{-i\pi( \omega + \omega')} + 1}{2}  \gamma_{\omega;\omega' }^{(-,-)},
\end{align}
with
\begin{align}
  \label{eq:gamma}
\gamma_{\omega; \omega'}^{(-,-)}  & =  \frac{\cos(\pi \omega/2 )\cos(\pi \omega'/2 )}{ \cos(\pi(\omega + \omega') /2) } .
\end{align}
To low orders in $g^2$, and within the LLA, $\gamma_{\omega_1;
  \omega_2}^{(-,-)} =1$ and our above choice for the phases,
Eq.~(\ref{eq:phases1}) is indeed in accordance with the phase
structure of the reggeized gluons.  Only starting from the NNLLA,
corrections due the real factor $\gamma_{\omega_1; \omega_2}^{(-,-)}$
need to be included.
In the following, it will be convenient for us to introduce the abbreviations 
\begin{align}
  \label{eq:abrev_mom}
{\bm l }_1 &= {\bm l},   & {\bm l}_2 &= ( {\bm q} - {\bm l }), \notag \\
{\bm k}_1 &= {\bm k},   & {\bm k}_2 &= ( {\bm q} - {\bm k }). 
\end{align}
The one-loop correction to the elastic quark-quark-scattering
amplitude with exchange of two reggeized gluons is then given by:
\begin{align}
  \label{eq:bfkl1mal}
 \mathcal{M}^{\text{1L}(+)}_{2 \to 2}   = 2 \pi i |p_A^+||p_B^-|
\int \frac{d \omega_1}{2 \pi i} \int \frac{d \omega_2}{2 \pi i}
    A^{a_1a_2}_{(2;0)} 
       & \otimes_{{\bm l}}  \frac{1}{\omega_1 - \beta({\bm l}_1)  - \beta({\bm l}_2) } \left| \frac{p_A^+}{\Lambda_a} \right|^{\omega_1}
  \notag \\
  B^{a_1a_2;b_1b_2 }(\omega_1,\omega_2)     &
  \otimes_{{\bm k}}\frac{1}{\omega_2 - \beta({\bm k}_1)  - \beta({\bm k}_2) }
\left| \frac{p_B^-}{\Lambda_b} \right|^{\omega_2}
  A^{b_1b_2}_{(2;0)} .
\end{align}
In analogy to the function $A(\omega_1, \omega_2)$ in the case of the reggeized gluon, Eq.~(\ref{eq:cr_bar}), we define a function $B(\omega_1, \omega_2)$. As for the reggeized gluon, this function can be defined within  different schemes.  Generally it can be written as the sum of the following three terms
\begin{align}
  \label{eq:defBeins}
 B^{a_1a_2;b_1b_2 }(\omega_1,\omega_2)  =& T^{b_1}_{a_1c}T^{b_2}_{ca_2} B^{(12) }(\omega_1,\omega_2) 
 + T^{b_2}_{a_1c}T^{b_1}_{ca_2} B^{(21) }(\omega_1,\omega_2)
\notag \\
&[T^{b_1}, T^{b_2}]_{a_1a_2}  B^{(\text{R}) }(\omega_1,\omega_2),
\end{align}
where we suppressed in our notation the dependence on momenta and cut-offs.  A  convolution symbol which includes the propagators of the reggeized gluons is defined as follows:
\begin{align}
  \label{eq:convolution}
\otimes_{\bm{k}} = \int \frac{d^2 {\bm k}}{(2 \pi)^3} \frac{1}{{\bm k}_1^2 {\bm k}_2^2},
\end{align}
with  ${\bm k}_1 + {\bm k}_2 = {\bm q}$. We  obtain
\begin{align}
  \label{eq:defBij} 
B^{(ij)}(\omega_1,\omega_2) =&   \frac{i}{4}\mathcal{K}_{2 \to 2} ({\bm l}_1,  {\bm l}_2,{\bm k}_i,  {\bm k}_j  )  \int \frac{dl^+ k^-}{2\pi} \frac{({\bm l }_1 - {\bm k}_i)^2 }
          {-l^+k^- - ({\bm{l }_1 - {\bm k}_i})^2 + i\epsilon} 
\notag \\
 & 
\left[ (k^- -i\epsilon)^{\omega_1\!-\!1} 
 - 
(-k^- -i\epsilon)^{\omega_1\!-\!1}
\right]
 \left[ (l^+ -i\epsilon)^{\omega_2\!-\!1}
   - (- l^+ -i\epsilon)^{\omega_2\!-\!1}    \right],
\end{align}
with
\begin{align}
  \label{eq:kernel22_1}
   \mathcal{K}_{2 \to 2} ({\bm l}_1,  {\bm l}_2,{\bm k}_1,  {\bm k}_2  ) &=
 \frac{g^2}{2} \bigg( {\bm q}^2  - 
      \frac{{\bm l}_2^2 {\bm k}_1^2 }{({\bm l}_1 - {\bm k}_1)^2} 
      +
      \frac{{\bm k}_2^2{\bm l}_1^2 }{({\bm l}_1 - {\bm k}_1)^2}
     \bigg). 
\end{align}
$B^{(\text{R})}$ contains the contribution due to the  last line  of  Eq.(\ref{eq:K22manip}) and can be shown to vanish. 
The longitudinal integrals of Eq.(\ref{eq:defBij}) are evaluated similar to Sec.\ref{sec:traj} and yield:
\begin{align}
  \label{eq:evalfB} 
B^{(ij)}(\omega_1,\omega_2 ) &= -2\pi i \delta(\omega_1 - \omega_2) 
\frac{ \left[- ({\bm l}_1 - {\bm k}_i)^2   \right]^{\omega_1 }    + \left[ ({\bm l}_1 - {\bm k}_i)^2 \right]^{\omega_1}}{2}
 \mathcal{K}_{2 \to 2} ({\bm l}_1,  {\bm l}_2,{\bm k}_i,  {\bm k}_j  ).
\end{align}
Making use of simplifications due to the LLA, we set
$\Lambda_a\Lambda_b/({\bm l}_1 - {\bm k}_i)^2 = m_A m_B $.   Introducing further
\begin{align}
  \label{eq:bpm}
B^{(\pm)}(\omega_1,\omega_2, p_A^+, p_B^-) =   B^{(12)}(\omega_1,\omega_2, p_A^+, p_B^-)  \pm   B^{(21)}(\omega_1,\omega_2, p_A^+, p_B^-),
\end{align}
we find
\begin{align}
  \label{eq:defBeins}
 B^{a_1a_2;b_1b_2 }(\omega_1,\omega_2, p_A^+, p_B^-)  &= \{T^{b_1},T^{b_2}\}_{a_1a_2} B^{(+) }(\omega_1,\omega_2, p_A^+, p_B^-) 
\notag \\ &
 + [T^{b_2},T^{b_2}]_{a_1a_2} B^{(-) }(\omega_1,\omega_2, p_A^+, p_B^-).
\end{align}
From the previous paragraph we know, that the quark impact factor for  two reggeized gluons couples only to the symmetric color configuration in the
$t$-channel. However there exists also the case of a non-zero
coupling of a  two reggeized gluon-state with  antisymmetric color (we will
encounter an example in Sec.\ref{sec:--}).
Generalizing our results in  analogy to the reggeized gluon Sec.~\ref{sec:reggeizedgluon} to the elastic quark-quark-scattering amplitude with exchange of two reggeized gluons, with $n$ interactions we obtain for the every  inserted  interaction between two reggeized gluons, a factor
\begin{align}
  \label{eq:bfkl_factor}
 \otimes_{\bm k} \mathcal{K}_{2 \to 2}^{\{a \} \to \{b \} }
\frac{1}{\omega - \beta({\bm k}_{1})- \beta({\bm k}_{2}) } ,
\end{align}
with
\begin{align}
  \label{eq:Ksym}
\mathcal{K}_{2 \to 2}^{\{a \} \to \{b \} } =
 f^{a_1b_1c}f^{cb_2a_2} 
     \mathcal{K}_{2 \to 2}( {\bm l}_1, {\bm l}_2 ;  {\bm k}_1, {\bm k}_2  ) 
+
 f^{a_1b_2c}f^{cb_1a_2} 
     \mathcal{K}_{2 \to 2}( {\bm l}_1, {\bm l}_2 ;  {\bm k}_2, {\bm k}_1  ).
\end{align}
 The all order amplitude is therefore within the LLA given by
\begin{align}
  \label{eq:bfkl1mal_done}
 \mathcal{M}^{\text{LLA}|\text{2R}}_{2 \to 2}  & =  i\pi  |p_A^+||p_B^-|
\int \frac{d \omega}{2 \pi i}  \left[  \left(\frac{ -p_A^+ p_B^-}{m_A^+m_B^-}   \right)^{\omega}  +  \left(\frac{ p_A^+ p_B^-}{m_A^+m_B^-}   \right)^{\omega}           \right] \phi(\omega, t),
\end{align}
where the partial wave $\phi(\omega, t) $ is a convolution of the
upper and the lower quark-impact factors and the BFKL-Green's function:
\begin{align}
\phi(\omega, t) =
  &  A^{a_1a_2}_{(2;0)} ({\bm l}_1, {\bm l}_2)  
       \otimes_{\bm l}  
G^{a_1a_2;b_1b_2}_{\text{BFKL}} ({\bm l}_1, {\bm l}_2 ;{\bm k}_1, {\bm k}_2)
    \otimes_{\bm k} 
  A^{a_1a_2 }_{(2;0)} ({\bm k}_1, {\bm k}_2) ,
\end{align}
where  the  BFKL-Green's function $ G^{a_1a_2;b_1b_2}_{\text{BFKL}} $ is given as the solution of the BFKL-equation
\begin{align}
  \label{eq:bfkl}
 (\omega - \beta ( {\bm l}^2_1) - \beta ( {\bm l}^2_2) ) G^{a_1a_2;b_1b_2}_{\text{BFKL}} ({\bm l}_1, {\bm l}_2 ;{\bm k}_1, {\bm k}_2)
  & =   \delta^{a_1b_1} \delta^{a_2b_2}  
(2\pi)^3 \delta^{(2)}({\bm l} - {\bm k})
\notag \\ 
&+ \left( \mathcal{K}_{2 \to 2}^{\{c \} \to \{b \} } \otimes  G^{c_1c_2;b_1b_2}_{\text{BFKL}} \right)({\bm l}_1, {\bm l}_2 ;{\bm k}_1, {\bm k}_2).
\end{align}
As can be easily verified, $\mathcal{M}^{\text{LLA}|\text{2R}}_{2 \to 2} $ is invariant under
substitutions $p_A, p_B \to - p_A, -p_B$ and  has therefore
positive signature, as expected.  
Furthermore, with
\begin{align}
  \label{eq:gamma_id}
e^{-i\pi \omega} + 1 = 2 i \frac{e^{-i\pi \omega} - 1}{\sin \pi\omega} \cos^2(\omega\pi/2),
\end{align}
and using that $\cos^2(\omega\pi/2) \to 1$ for the LLA the amplitude
can brought into the form, usually stated for the Sommerfeld-Watson
representation, of an amplitude with positive signature:
  \begin{align}
  \label{eq:possig}
 \mathcal{M}^{\text{LLA}(+)}_{2 \to 2}  & =  -2\pi
\int \frac{d \omega}{2 \pi i} |s|^{1 + \omega}  \frac{\xi^{(+)}(\omega)}{\sin\pi\omega} \phi(\omega, t) &
\xi^{(+)}(\omega)&  ={e^{-i\pi\omega}-1}
\end{align}
with $\xi^{(+)} $ the signature factor for positive signature
exchange. The negative signature part of the quark-quark scattering
amplitude is completely contained in the reggeized gluon: Decomposing
the quark-impact factor $A^{a_1a_2}_{(2;0)}$ into a color singlet part
$A^{(\bf 1)}_{(2;0)}$ and a color octet part $A^{(\bf 8)}_{(2;0)}$ and
convoluting them with the BFKL-Green's function, one obtains in the
octet-case from the BFKL-equation a pole-solution, which inserted into
Eq.~(\ref{eq:possig}) exhibits the reggeized gluon. Within the
effective action the antisymmetric octet ${\bf 8_A}$ state of two
reggeized gluons decouples from the quark, while the antisymmetric
resummed reggeized gluon arises due to resummation of corrections to
the bare reggeized gluon, as illustrated in Sec.\ref{sec:22negsig}.
It is in the effective theory a elementary degree of freedom. In
particular its derivation does not make use of the bootstrap relations
\cite{Kuraev:1977fs,Kuraev:1976ge,Lipatov:1976zz,Fadin:1975cb},\cite{Bartels:1978fc,Bartels:1991bh}.
The state of two reggeized gluons in the symmetric octet ${\bf 8_S}$
on the other hand couples to the quark and correspondingly there
exists a pole solution to the BFKL-equation, the so-called
'd'-Reggeon. Unlike the antisymmetric 'f'-Reggeon, it is not a
fundamental degree of freedom in the effective theory, but arises as a
state of two reggeized gluons.

\section{Conclusion }
\label{sec:mod_feyn}

In the present chapter, the elastic scattering amplitude has been
studied from the point of view of the effective action and rules for
the occurring longitudinal integrations have been given.  In short,
interaction between two particles is only described by a reggeized
gluon if the center-of-mass energy of the minimal elastic
(sub-)amplitude that can be associated with its exchange is larger
than a certain lower bound; otherwise the interaction is mediated by a
usual QCD-gluon. The propagator of the reggeized gluon depends
therefore not only on the momenta of the reggeized gluon (as usual),
but furthermore also on the momenta of the particle it couples to.  In
order to incorporate that lower bound, it turned out that the
Mellin-integral Eq.~(\ref{eq:theta_Mellin}) provides a suitable
regularization method. It has the nice properties that it both allows
to include negative signature of the reggeized gluon and higher order
loop correction in an easy and straight-forward way. In particular,
imposing this bound turned out to be sufficient to keep the
interactions of QCD-particles in the effective action local in
rapidity.  Using this rules, we showed within the LLA that the leading
part of the elastic amplitude is within the effective action
completely summarized in the exchange of a single (resummed) reggeized
gluon with negative signature.  Due to the introduced cut-off, the
reggeized gluon carries a dependence on an arbitrary mass-scale.
Within the LLA, this mass-scale can be chosen to be of the size of the
typical transverse scale and this dependence is not of interest.
Beyond the LLA however, it is essential that this scale cancels for
physical amplitudes. It was demonstrated that such a cancellation can
be achieved for the elastic amplitude by taking into account
quasi-elastic corrections to the quark-reggeized gluon couplings,
which are  formally beyond the LLA.

For the part of the elastic amplitude with positive signatured, on
the other hand, interaction between the scattering quarks is mediated
by the state of two reggeized gluon and interaction between the
two reggeized gluons is resummed by the famous BFKL-equation.
For the longitudinal integrals occurring due to loops containing
reggeized gluons, two points are worth to be noted:

At first, integrals over longitudinal components show generally a
logarithmic divergence which at first does not allow for an evaluation
of the integrals. The occurrence of this divergence can be traced back
to a part in these integrals which is already contained in the induced
vertices. This part therefore needs to be subtracted in order to avoid
to count this contribution twice. This subtraction term can be
obtained from graphs which contain induced vertices to which
\emph{only} reggeized gluons couple.

For loop-corrections to the exchange of reggeized gluons one further
has to take into account the following: Defining for every single
reggeized gluon a 4-point-sub-amplitude inside the complete amplitude,
presence of the reggeized gluon requires that the center-of-mass
energy of the sub-amplitude is significantly larger than all other
scales.  To implement this lower bound we used again the
Mellin-integral. It is allows to include signature of the reggeized
gluon and to include corrections due to reggeization of the gluon.  In
particular the position of the resulting branch-cut is determined by
the signature factor of the reggeized gluon. For loop integrations as
in the above it is convenient to combine the two signature factors
into a single one, making use of Eq.(\ref{eq:sig_manipus}). If
additional poles in light-cone momenta occur, either from induced
vertices or due to a subtraction by a subtraction-diagram, it turns
out to be mandatory that the $i\epsilon$ prescription of the
branch-cut coincides with the one of the poles.

 \chapter{Loop corrections to production processes from the effective action }
\label{cha:2to3}

In the foregoing chapter, from the study of the elastic amplitude
within the LLA, a set of rules has been derived, that can be used to
determine the interaction of reggeized gluons within the effective
action.  A non-trivial testing-ground for these rules is given by
loop-corrections to production amplitudes in the
Multi-Regge-Kinematics. As for the elastic amplitudes, loop
corrections lead within the LLA to logarithms in the various
center-of-mass-energies of the initial and final state particles.
These logarithms yield in the physical region discontinuities, which
can be associated with production thresholds of produced particles.
There exists a non-trivial constraint on the occurrence of these
discontinuities, the Steinmann-relations \cite{Steinmann:1960}. The
proposed regularization by the Mellin-integral on the other hand
introduces automatically discontinuities and it is a priori not clear,
to which extend this is in accordance with the Steinmann-relations or
not.

The outline of this chapter is the following: In
Sec.~\ref{sec:anal_rep} we explain the constraints due to the
Steinmann-relations and introduce the analytic representations of the
$2 \to 3$ and $2 \to 4$ amplitude.  In Sec.~\ref{sec:born+possig} we
discuss corrections to the production amplitude with negative
signature in all $t$-channels, while in the Sec.~\ref{sec:mixed} the
various production amplitudes with positive and mixed signature are
presented.

\section{Analytical representation of production amplitudes}
\label{sec:anal_rep}
\begin{figure}[htbp]
  \centering
  \parbox{5cm}{\includegraphics[width=4.5cm]{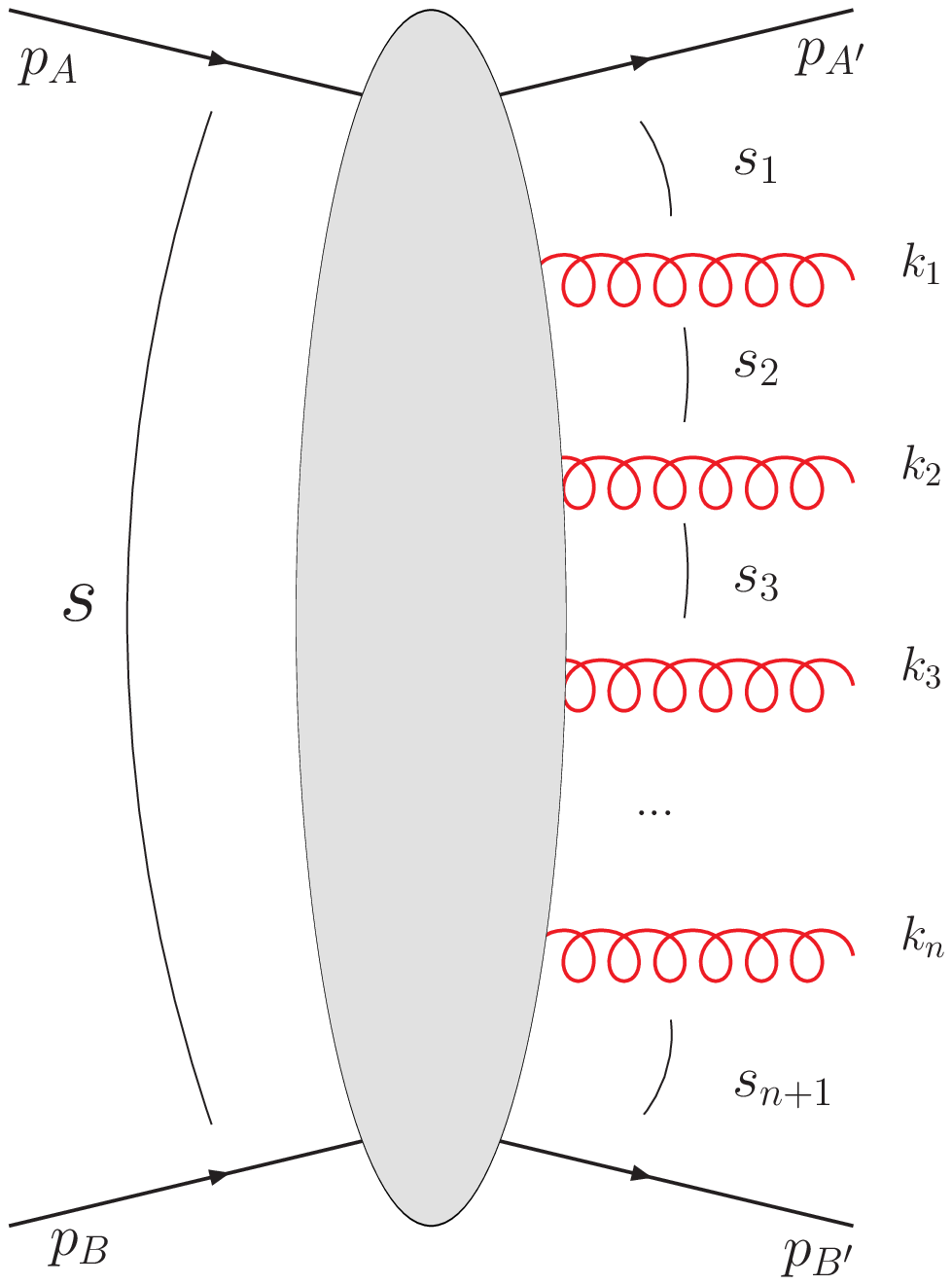}}
  \caption{\small The Multi-Regge-Kinematics of the $n+2$-particle production amplitude.}
  \label{fig:mrk}
\end{figure}

In the following section we give a summary of the analytical
representations of the $2\to 3$ and the $2 \to 4$ production amplitude
in the high energy limit, which generalizes the representation
 of the elastic amplitude. In general, we
consider in the following the $n$-particle production amplitude in the
Multi-Regge-Kinematics (see also Fig.\ref{fig:mrk}) which is defined as
\begin{align}
  \label{eq:mrk_def2}
 s, s_r & \to \infty, &  s_r/s &\to 0, \\
t_r, \kappa_r & = k_r^+k_r^- = {\bm k}_r^2   = \text{fixed},
\end{align}
with
\begin{align}
  \label{eq:mrk_def1}
 s & = (p_A + p_B)^2,   & s_r &= (k_r + k_{r -1})^2, \notag \\
k_r &= q_r - q_{r -1}.  & t_r &= q_{r}^2.
\end{align}
 A key element in the study of production amplitudes  are the
Steinmann-relations \cite{Steinmann:1960} which forbid the existence of
simultaneous energy discontinuities in overlapping channels.
\begin{figure}[htbp]
  \centering
  \parbox{7cm}{\includegraphics[width=7cm]{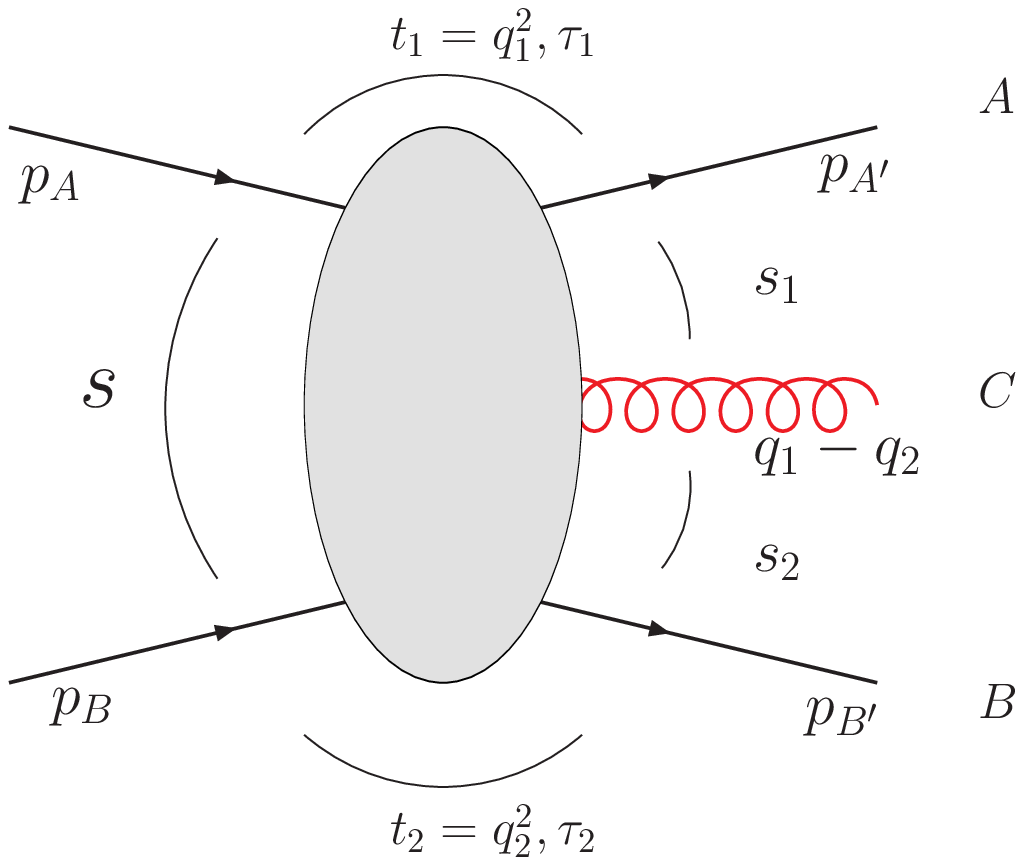}}
  \caption{\small The $2\to3$production amplitude}
  \label{fig:prod_raw23}
\end{figure}
 As an
example, we consider the $2\to 3$ production amplitude in
Fig.~\ref{fig:prod_raw23}:  The produced particle $C$  can either build a resonance or bound state with the
particle $A'$ \emph{or} the particle $B'$, but never with both of
them. In more technical terms this means, that in the physical region, the
production amplitude can either have a discontinuity in $s_1$ -
corresponding to a bound state of particles $A'$ and $C$ - or in
$s_2$, which corresponds to a bound state of particles $C$ and $B'$,
but never in both of them at a time. On the other hand  the particles $A'$, $B'$ and $C$ can form a single resonance
state, which corresponds to the  discontinuity in $s$. 
The discontinuity structure of
the $2 \to 3$ amplitude in the high energy limit is further
illustrated in Fig.~\ref{fig:disc23}.
\begin{figure}[htbp]
  \centering
  \parbox{4.5cm}{\center \includegraphics[width=4cm]{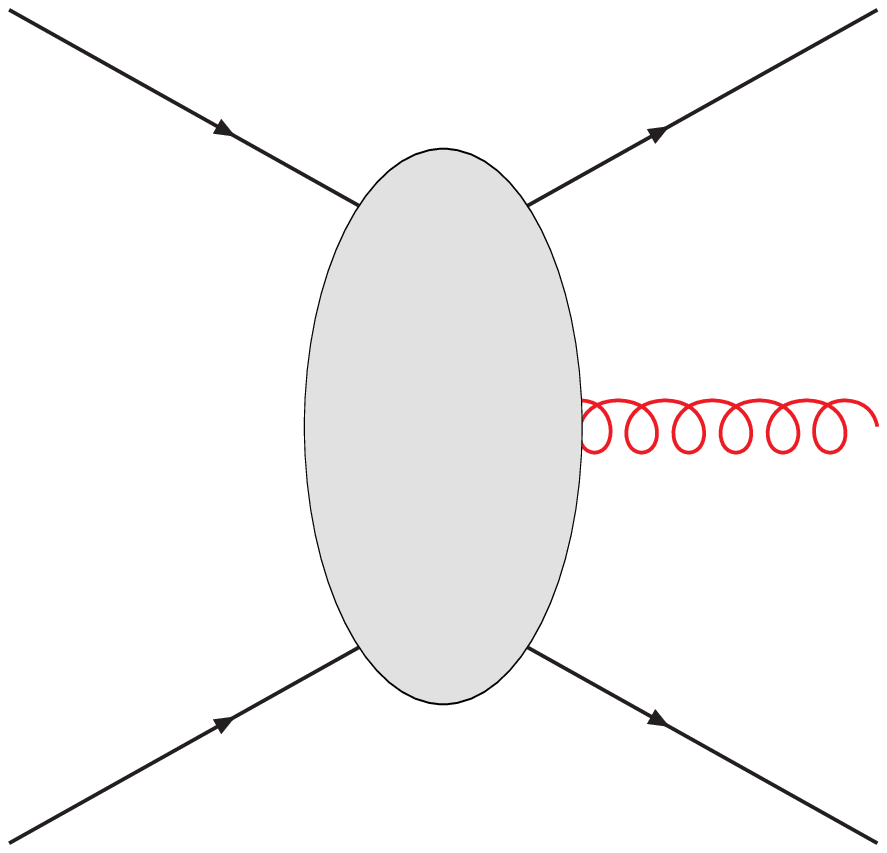}} 
=
  \parbox{4.5cm}{ \center \includegraphics[width=4cm]{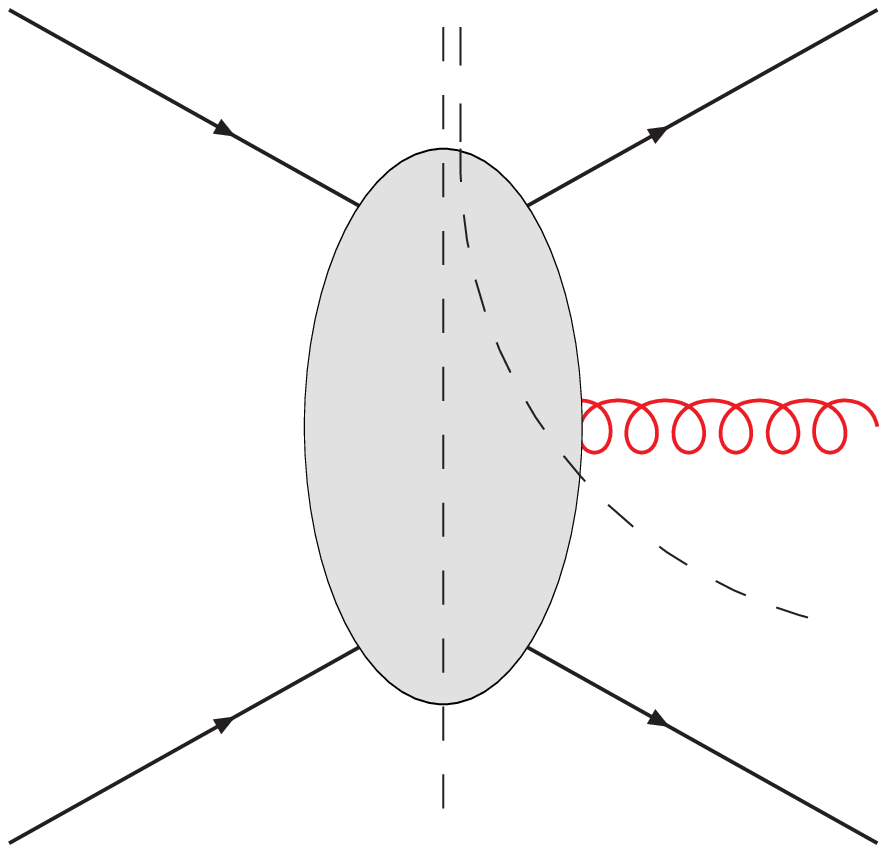}}
  + 
  \parbox{4.5cm}{\center \includegraphics[width=4cm]{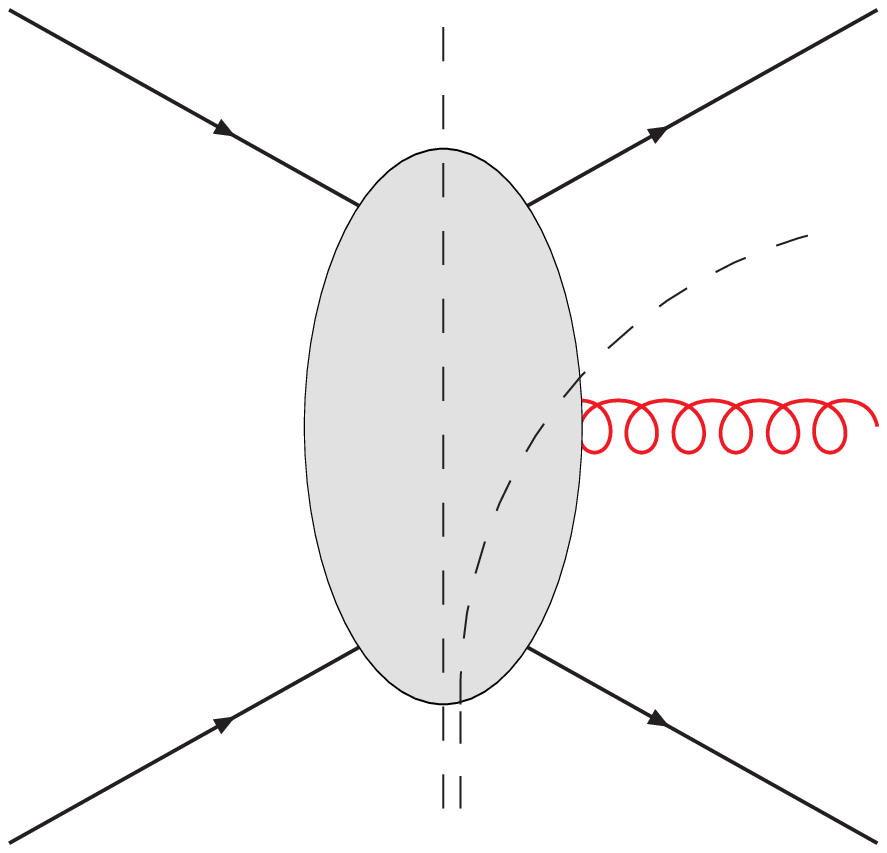}}
  \caption{\small Analytic structure of the $2 \to 3$ production amplitude. It is given as a sum of two terms which have either allow for cuts in $s$ and $s_1$ or in $s$ and $s_2$}
  \label{fig:disc23}
\end{figure}
 In the
double-Regge-kinematics
an analytic representation  
can be formulated, which respects the Steinmann-rules.
In the case of the elastic scattering amplitude, such a representation is given by 
\begin{align}
\label{eq:anal_22}
\mathcal{M}^{(\tau)}_{2 \to 2} =   s \int \frac{d\omega}{2\pi i}   |s|^{\omega} \xi^{(\tau)}(\omega) \frac{F(\omega; t)}{\sin (\pi \omega)},
 \end{align}
with 
\begin{align}
  \label{eq:2sigi}
\xi^{(\tau)}(\omega) &=  {e^{-i\pi \omega} -\tau},
&\tau &= \pm 1,
\end{align}
with $\tau$ the signature of the amplitude. As there is just one
direct channel, the Steinmann relations provide no further constraint
on the discontinuities in $s$. For $2 \to 3$ production amplitude, an
analogous analytic representation is given as the sum of two terms,
which allow either for a discontinuity in $s$ and $s_1$ or in $s$ and
$s_2$,
\begin{align}
  \label{eq:anal_23}
\mathcal{M}_{2 \to 3}  
= 
s \int \frac{d\omega_1}{2\pi i}\int \frac{d\omega_2}{2\pi i} & \bigg(
|s|^{\omega_2} |s_1|^{\omega_1 - \omega_2} 
\xi^{(\tau_2)}_{(\omega_2)} \xi^{(\tau_1,\tau_2)}_{(\omega_1,\omega_2)} 
\frac{V_1 (\omega_1, \omega_2 ; t_1, t_2, \kappa_1)}{\sin \pi\omega_{12}}
\notag \\
&+
|s|^{\omega_1} |s_2|^{\omega_2 - \omega_1}
\xi^{(\tau_1)}_{(\omega_1)} \xi^{(\tau_2,\tau_1)}_{(\omega_2,\omega_1)} 
\frac{ V_2 (\omega_1, \omega_2 ; t_1, t_2, \kappa_1)}{\sin \pi \omega_{21}} \bigg),
\end{align}
where we defined 
\begin{align}
  \label{eq:omega_ij}
\omega_{ij} = \omega_i -\omega_j,
\end{align}
and $\tau_{1,2} = \pm 1$ is the signature
of the $t_1$ and $t_2$-channel respectively. Furthermore a generalized signature factor is defined as 
\begin{align}
  \label{eq:doule_sig}
 \xi^{(\tau_1,\tau_2)}_{(\omega_1,\omega_2)}  = e^{-i\pi (\omega_1 - \omega_2)} + \tau_1\tau_2
\end{align}
while the partial waves $V_1$ and $V_2$ are real. 
 Representations like
Eq.~(\ref{eq:anal_23}) have been first derived from  models which
contain only Regge-poles like massive $\phi^3$-theory
\cite{Drummond:1969ft,Weis:1972ir},  the dual Veneziano 6-point amplitude
\cite{Brower:1974yv} and from generalized Froissart-Gribov partial
wave representations \cite{White:1974vy}. 
As for the elastic amplitude, the above representation can be used to
determine the $2 \to 3$ production amplitude within the LLA by taking
 discontinuities in $s$, $s_1$ and/or $s_2$. This allows for the determination
of $V_1$ and $V_2$ from on-shell, tree-level  production amplitudes,
making use of unitarity in  all sub-channels. Such a program has been
 carried out in \cite{Bartels:1978fc} and
\cite{Bartels:1980pe} and also recently  in
\cite{Bartels:2008ce},  where  the analytical properties of
Bern-Dixon-Smirnov (BDS)-amplitudes \cite{Bern:2005iz} in the high-energy limit have been examined.

In the case of the $2\to 4$ amplitude, the  analytic representation in the triple-Regge-limit requires already five terms. It is given by
\begin{align}
  \label{eq:anal_24}
\mathcal{M}_{2 \to 4}  
=& 
s \int \frac{d\omega_1}{2\pi i} \int \frac{d\omega_2}{2\pi i} \int \frac{d\omega_3}{2\pi i}
\notag \\
& \bigg(
|s_1|^{\omega_1 -\omega_2} |s_{012}|^{\omega_2 - \omega_3} |s|^{\omega_3}
 \xi^{(\tau_1,\tau_2)}_{(\omega_1,\omega_2)}  \xi^{(\tau_2,\tau_3)}_{(\omega_2,\omega_3)} \xi^{(\tau_3)}_{(\omega_3)}
\frac{W_1 (\omega_1, \omega_2 , \omega_3; t_1, t_2, t_3, \kappa_1, \kappa_2)}{\sin \pi\omega_{12} \sin\pi\omega_{23}}
\notag \\
&+|s_3|^{\omega_3 -\omega_2} |s_{123}|^{\omega_2 - \omega_1} |s|^{\omega_1}
 \xi^{(\tau_3,\tau_2)}_{(\omega_3,\omega_2)}  \xi^{(\tau_2,\tau_1)}_{(\omega_2,\omega_1)} \xi^{(\tau_1)}_{(\omega_1)}
\frac{W_2 (\omega_1, \omega_2 , \omega_3; t_1, t_2, t_3, \kappa_1, \kappa_2)}{\sin \pi\omega_{32} \sin\pi\omega_{21}}
\notag \\
&+|s_2|^{\omega_2 -\omega_1} |s_{012}|^{\omega_1 - \omega_3} |s|^{\omega_3}
 \xi^{(\tau_2,\tau_1)}_{(\omega_2,\omega_1)}  \xi^{(\tau_1,\tau_3)}_{(\omega_1,\omega_3)} \xi^{(\tau_3)}_{(\omega_3)}
\frac{W_3 (\omega_1, \omega_2 , \omega_3; t_1, t_2, t_3, \kappa_1, \kappa_2, \kappa_{123})}{\sin \pi\omega_{21} \sin\pi\omega_{13}}
\notag \\
&+|s_2|^{\omega_2 -\omega_3} |s_{123}|^{\omega_3 - \omega_1} |s|^{\omega_1}
 \xi^{(\tau_2,\tau_3)}_{(\omega_2,\omega_3)}  \xi^{(\tau_3,\tau_1)}_{(\omega_3,\omega_1)} \xi^{(\tau_1)}_{(\omega_1)}
\frac{W_4 (\omega_1, \omega_2 , \omega_3; t_1, t_2, t_3, \kappa_1, \kappa_2, \kappa_{123})}{\sin \pi\omega_{23} \sin\pi\omega_{31}}
\notag \\
&+|s_3|^{\omega_3 -\omega_2} |s_{1}|^{\omega_1 - \omega_2} |s|^{\omega_2}
 \xi^{(\tau_3,\tau_2)}_{(\omega_3,\omega_2)}  \xi^{(\tau_1,\tau_2)}_{(\omega_1,\omega_2)} \xi^{(\tau_2)}_{(\omega_2)}
\frac{W_5 (\omega_1, \omega_2 , \omega_3; t_1, t_2, t_3, \kappa_1, \kappa_2)}{\sin \pi\omega_{32} \sin\pi\omega_{12}}
 \bigg).
\end{align}
where we defined   
\begin{align}
  \label{eq:s123}
s_{012} & = (p_A + k_2)^2 =- p_A^+q_3^- 
&
s_{123}  & = (p_B + k_1)^2 = p_B^-q_1^+  \notag \\
\kappa_{123} &= ({\bm k}_1 + {\bm k_2})^2 = ({\bm q}_1 - {\bm q}_3)^2
\end{align}
Taking discontinuities in $s_1$, $s_2$, $s_3$ $s_{012}$ and $s_{123}$,
the partial waves $W_i$ can be determined from unitarity as outlined
in \cite{Bartels:1978fc, Bartels:1980pe, Bartels:2008ce}. In the
following we attempt a derivation of the LLA $2 \to 3$ and $2 \to 4$
production amplitude from the effective action. Particular focus will
be given on the analytic structure of the results and too which extend
the Steinmann-relations can be regarded to be fulfilled, if logarithms
are resummed within the effective action, employing the LLA.

\section{The production amplitude with negative signature in all $t$-channels}
\label{sec:born+possig}
We start with the case where the signature in all $t$-channels is
negative. From the discussion in the previous chapter and from general
reasoning we know, that this includes the part that is leading in the 
Leading Logarithmic Approximation.

\subsection{The production amplitude at Leading Logarithmic Accuracy}
\label{sec:mrklla}

With the produced particles in the Multi-Regge-Kinematics separated by
a large difference in rapidity, it is immediately clear that the
interaction between $s$-channel produced particles is mediated by
reggeized gluons alone, while the leading contribution is given by the
exchange of a single reggeized gluon with negative signature.
Production of the real particles is on the other hand described by the
production vertex, which has been introduced already in the context of
the BFKL-equation in Eq.(\ref{eq:1lipatov}):
\begin{align}
  \label{eq:1lipatov_repeat}
  \parbox{1cm}{\includegraphics[height = 1cm]{bfklreel.eps}} =    2 igT^{c}_{ba}& C_\mu(q_1, q_2) ,
\end{align}
\begin{figure}[t]
  \centering
  \parbox{4cm}{\includegraphics[width=4cm]{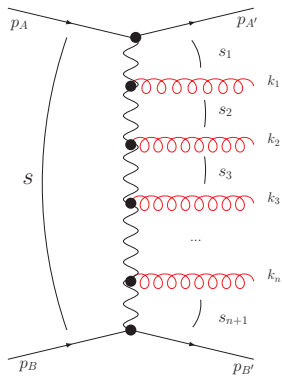}}
  \caption{\small Production amplitude in the MRK to leading order. All $t$-channel exchanges are reggeized and contain the maximal number of possible logarithms}
  \label{fig:mrk_reggeon}
\end{figure}
with 
\begin{align}
\label{eq:2lipatov_repeat}
 C_\mu(q_1, q_2) &= \left( \frac{q_1^+}{2} + 
     \frac{{\bm{q}}_1^2}{q_2^- }   \right) (n^-)^\mu + \left( \frac{q_2^-}{2} + \frac{{\bm{q}}_2^2}{ q_1^+ } \right) (n^-)^\mu - (q_1 + q_2)^\mu_\perp,
\end{align}
where we take $q_1$ incoming and $q_2$ outgoing from the vertex.  We
note here, that if contracted with a physical polarization vector, the
production amplitude can be shown to depend only on transverse
momenta, and the dependence on $q_1^+$ and $q_2^-$ can be
eliminated.  The production amplitude in the Multi-Regge-Kinematics,
Fig.\ref{fig:mrk} is therefore at Leading-Logarithmic-Accuracy given
by diagrams like Fig.\ref{fig:mrk_reggeon}. Using our result from
Sec.\ref{sec:22negsig}, we would at first  insert in
Fig.\ref{fig:mrk_reggeon} for every reggeized gluon that connects two
$s$-channel gluons with momentum $k_{r-1}$ and momentum $k_r$, the following expression:
\begin{align}
 \label{eq:prop_regg_mrk}
\frac{i/2}{{\bm q}_r^2}\int \frac{d\omega_r}{2\pi i} \frac{1}{\omega_r
- \beta({\bm q}^2_r)} \frac{1}{2} \left [ \left( \frac{- s_r}{(
\kappa_{r_1}\kappa_r)^{1/2}} \right)^{\omega_r} + \left( \frac{s_r}{(
\kappa_{r_1}\kappa_r)^{1/2}} \right)^{\omega_r} \right]
\end{align}
The above expression however has a non-zero discontinuity in $s_r$.
Correspondingly, Fig. \ref{fig:mrk_reggeon} has for every squared
sub-energy $s_r$, $r = 1, \ldots , n$ a non-zero discontinuity in
contradiction to the Steinmann-relations which forbid simultaneous
singularities in over-lapping channels.  At first it seems that the
effective action is in conflict with the Steinmann-relations. This
problem can only be solved if the complete all order production vertex
has an internal phase structure, which balances the phase structure of
the reggeized gluon. Within the LLA only the leading part of the
complete production vertex is taken into account, which is completely
real.  We therefore conclude that as long we take into account only
the leading part of the production vertex, we should simultaneously
only include the most leading, real part of
Eq.(\ref{eq:prop_regg_mrk}). This requires to suppress
for the production amplitude at LLA the imaginary parts of the
reggeized gluons, which is of the same order as higher order
correction to the production vertex.  As a consequence, within the
LLA, we should in Fig.\ref{fig:mrk_reggeon} insert instead of
Eq.(\ref{eq:prop_regg_mrk})
\begin{align}
 \label{eq:prop_regg_mrk_a}
\frac{i/2}{{\bm q}_r^2}\int \frac{d\omega_r}{2\pi i} 
\frac{1}{\omega_r- \beta({\bm q}_r)}   \left| \frac{ s_r}{(
\kappa_{r_1}\kappa_r)^{1/2}} \right|^{\omega_r} ,
\end{align}
which means to take into account for every single reggeized gluon only
the leading real part.

\subsection{The Reggeon-Reggeon-particle vertex at 1-loop }
\label{sec:--}

In this paragraph we attempt to determine the one-loop corrections to
the production vertex and verify, whether it shows the on-set of
corrections needed for the complete production amplitude to satisfy
the Steinmann-relations.

From a technical point of view, the 1-loop-correction to the
production vertex shares many features with the 1-loop correction to
the quark-reggeized gluon coupling of Sec.\ref{sec:quasi-elastic}. As
there, the 1-loop correction to the production-vertex belongs formally
to the NLLA, while all large logarithms in $s_1$ and $s_2$ are already
resummed by  central-rapidity diagrams of the adjacent reggeized
gluons. Furthermore, in analogy to Sec.\ref{sec:quasi-elastic}, we are
in the following only interested in a certain part of the complete
1-loop corrections. Whereas in Sec.\ref{sec:quasi-elastic} we mainly
focused on the cancellation of the factorization parameter that separates the
central-rapidity diagrams from the quasi-elastic diagrams, we are in
the following especially  interested in the part that carries the
internal phase structure of the vertex.

It turns out that this phase structure is particularly connected with
a certain set of diagrams, to which we will restrict in the following.
In particular, presence of the non-trivial phase structure is
immediately connected to the presence of discontinuities of the
complete production amplitude: In the physical region of the different
$s$-channels (for the $2\to 3$ amplitude these are $s$, $s_1$ and
$s_2$), these discontinuities occur only in diagrams which can be
'cut' in the corresponding $s$-channel energy variables, which
requires in the underlying QCD-diagram the presence of at least two
$t$-channel gluons.  In the effective theory those states are for
negative signature gathered in induced vertices\footnote{This is
  certainly true at this level of accuracy. We however note that also
  the exchange of three, five, etc reggeized gluons turns out to carry
  negative signature, which however constitutes in this case a higher
  order effect.} and we therefore restrict in the following to
diagrams that involve at least one induced vertex of the first or
higher order.

\begin{figure}[htbp]
  \centering
   \parbox{3cm}{\includegraphics[height=3cm]{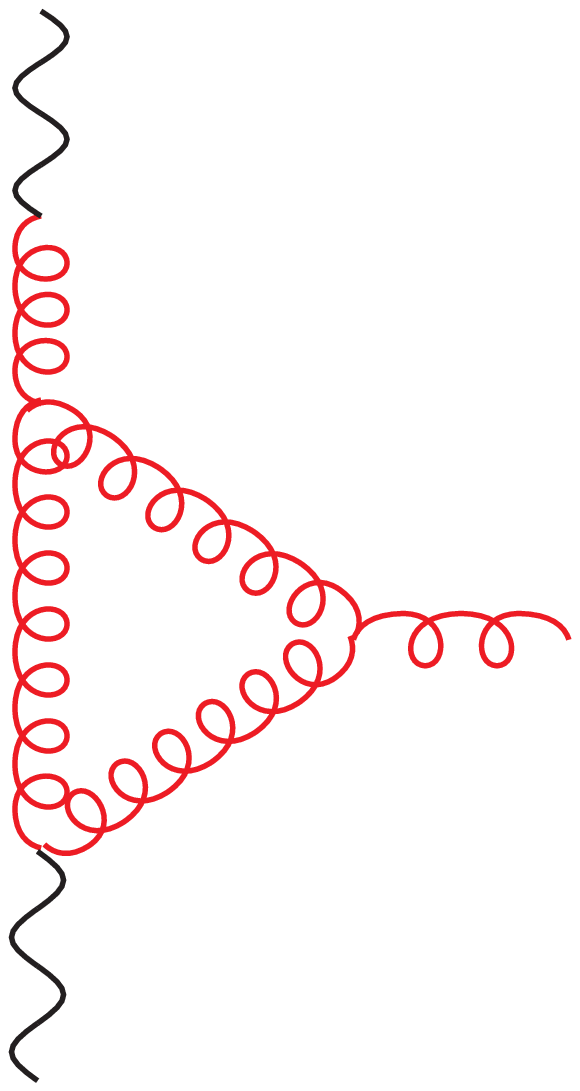}\\ {diagram (B1)}}
\parbox{1cm}{}
 \parbox{3cm}{\includegraphics[height=3cm]{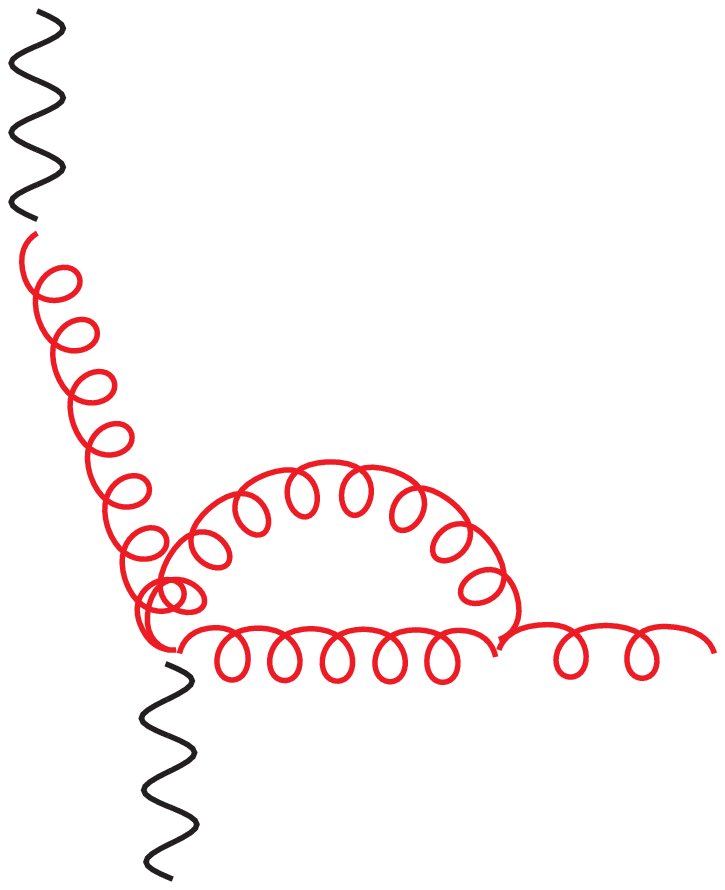}\\ {diagram (B2)}}
\parbox{1cm}{}
\parbox{3cm}{\includegraphics[height=3cm]{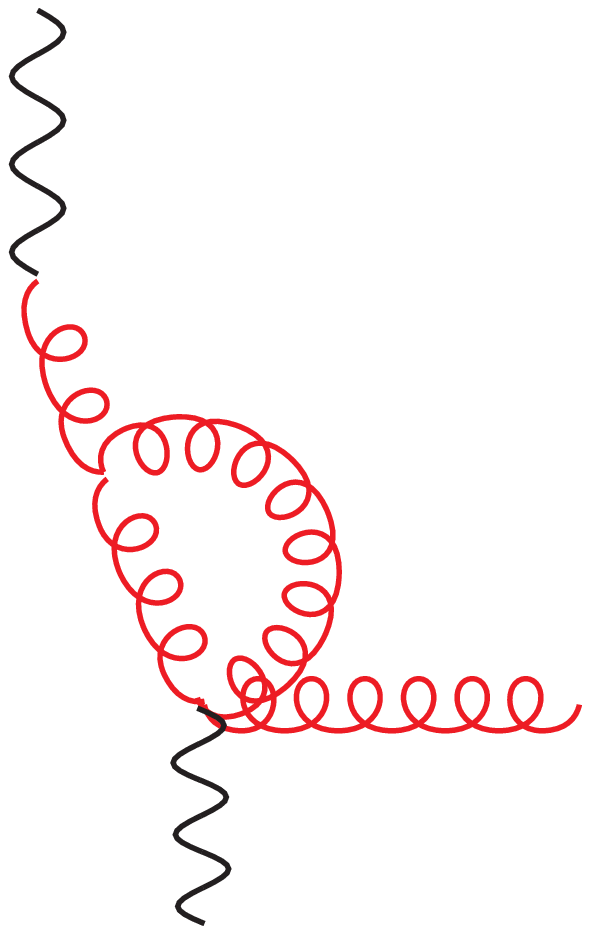}\\ {diagram (B3)}}
\parbox{1cm}{}
\parbox{3cm}{\includegraphics[height=3cm]{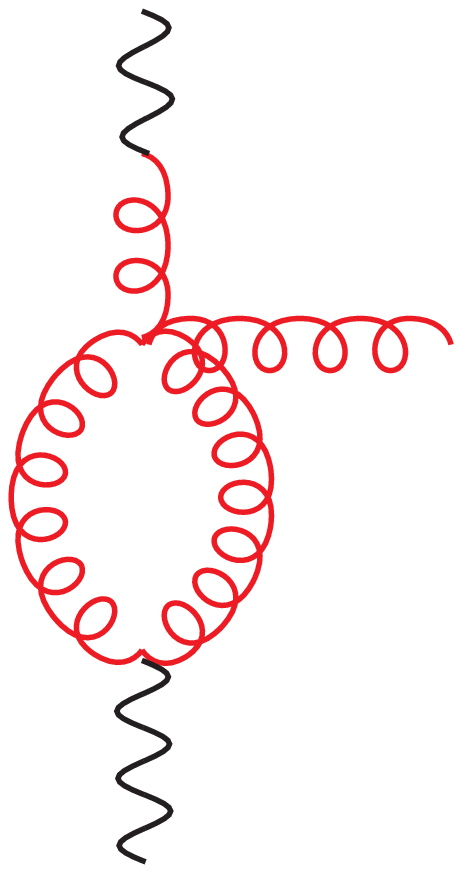}\\ {diagram (B4)}}
  \caption{ \small Feynman-diagrams contributing to the 1-loop corrections to the Reggeon-Reggeon-Particle vertex,  which contain  one induced vertex,  occurring in the coupling to the lower reggeized gluon. The corresponding  diagrams where the upper, minus reggeon couples by a induced vertex to the gluon loop, are not shown explicitly and can be obtained from the former ones by corresponding replacements}
  \label{fig:feyndiag}
\end{figure}
\begin{figure}[htbp]
  \centering
   \parbox{3cm}{\includegraphics[height=3cm]{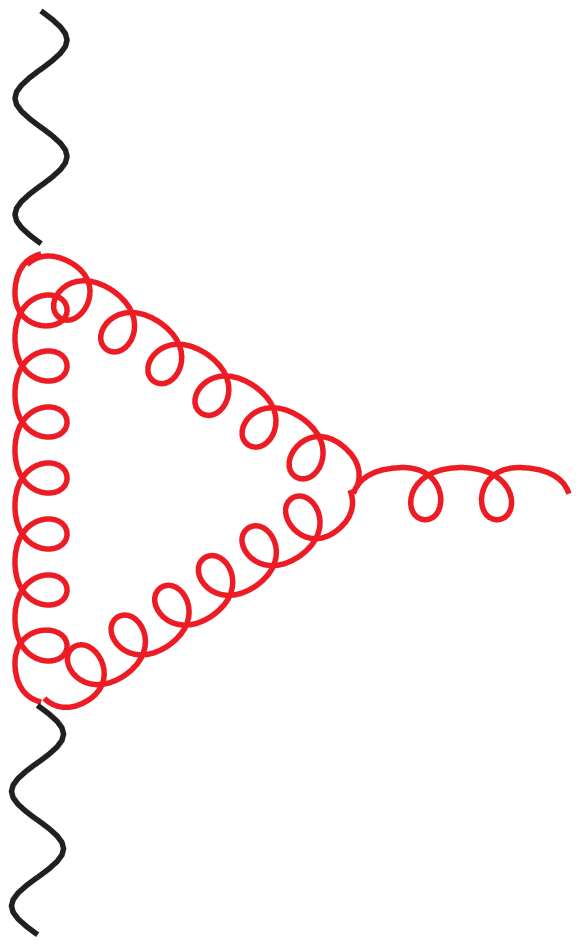}\\ {diagram (AB1)}}
\parbox{1cm}{}
 \parbox{3cm}{\includegraphics[height=3cm]{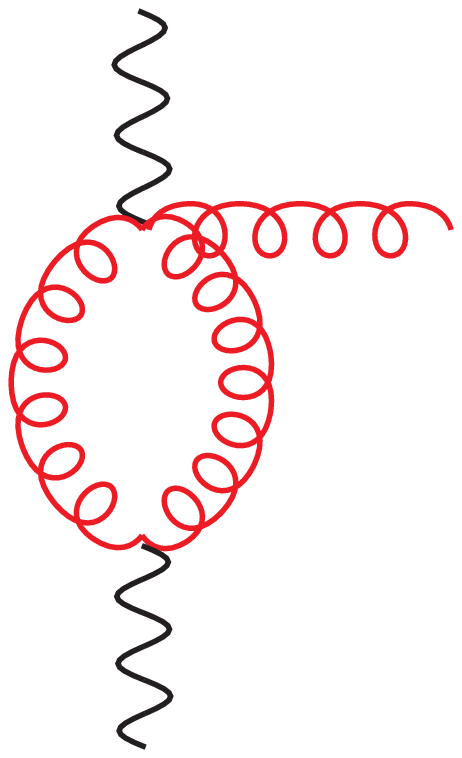}\\ {diagram (AB2)}}
\parbox{1cm}{}
\parbox{3cm}{\includegraphics[height=3cm]{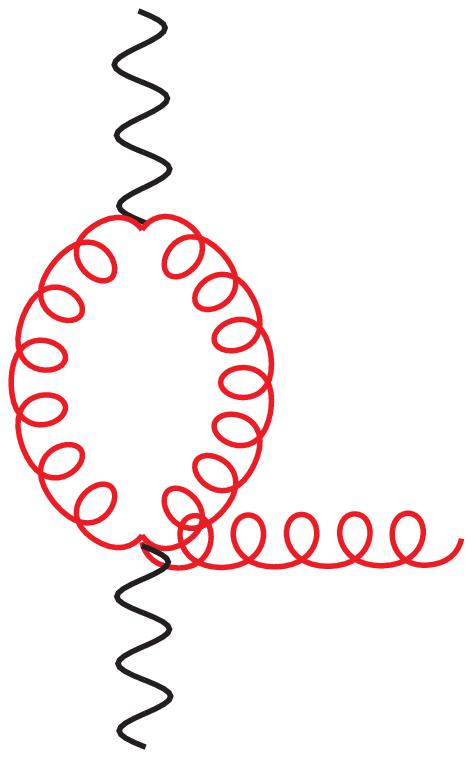}\\ {diagram (AB3)}}
\parbox{1cm}{}
  \caption{\small Feynman-diagrams  contributing to the 1-loop corrections to the Reggeon-Reggeon-Particle vertex where the reggeized gluons  couple from below and from above by induced vertices to the gluon loop.}
  \label{fig:feyndiag2}
\end{figure}

Relevant examples of diagrams are listed in Fig.\ref{fig:feyndiag} and
Fig.\ref{fig:feyndiag2}.  In all these diagrams, a reggeized gluon
couples from above and below to a gluon-loop.  In the following we
concentrate on the $2\to 3$ production amplitude.  The corrections
obtained for the production vertex  can similarly
be used for amplitudes with more produced particles, like the $2\to 4$
production amplitude, and we will return to this particular case at
the end of this paragraph.

As in  Sec.~\ref{sec:quasi-elastic}, loop-corrections to
Particle-Reggeon vertices require to impose certain restrictions on
the center-of-mass energy of the quark-gluon sub-amplitude. 
In order to do so, we factorize the production amplitude, similar to the treatment of the quasi-elastic corrections in Sec.~\ref{sec:quasi-elastic}, in the following way:
\begin{align}
  \label{eq:fac_23}
i&\mathcal{M}_{2 \to 3} = s \int \frac{d\omega_1}{2\pi i } \int \frac{d\omega_2}{2\pi i } 
\tilde{\Gamma}^a_{QQR}(\omega_1;  q_1; \lambda_{11}) 
 \times  \frac{i/2}{{\bm q}_1^2} \frac{1}{\omega_1 -\beta({\bm q}^2_1)} 
\left|\frac{s_1}{s_{R_1}\lambda_{11}\lambda_{12}}  \right|^{\omega_1} 
 \notag \\
&
 \times \Gamma_{RRG}^{acb,\mu}(\omega_1, \omega_2;q_1, q_2; \lambda_{12}, \lambda_{21} )
 \times \frac{i/2}{{\bm q}_2^2} \frac{1}{\omega_2 -\beta({\bm q}^2_2)} 
\left|\frac{s_2}{s_{R_2} \lambda_{21}\lambda_{22} }  \right|^{\omega_2}\!\! \times  \tilde{\Gamma}^b_{QQR}(\omega_2;  q_2;\lambda_{22} ).
\end{align}
$s_{R_1}$ and $s_{R_2}$ are the regarding transverse scales of the
reggeized gluons, which can be set to be $s_{R_1} = m_A\kappa_1^{1/2}$
and $s_{R_2} = m_B\kappa_1^{1/2}$ in accordance with our previous
conventions. The precise choice does indeed not matter here and we
only require a certain fixed scale, of the order of the transverse
scales of the process.  $\lambda_{ij}$ $i,j = 1,2$ are factorization parameters  of the
reggeized gluon and of the particle-reggeized gluon vertices
respectively.

\begin{figure}[t]
  \centering
  \parbox{3.5cm}{\center \includegraphics[width=3cm]{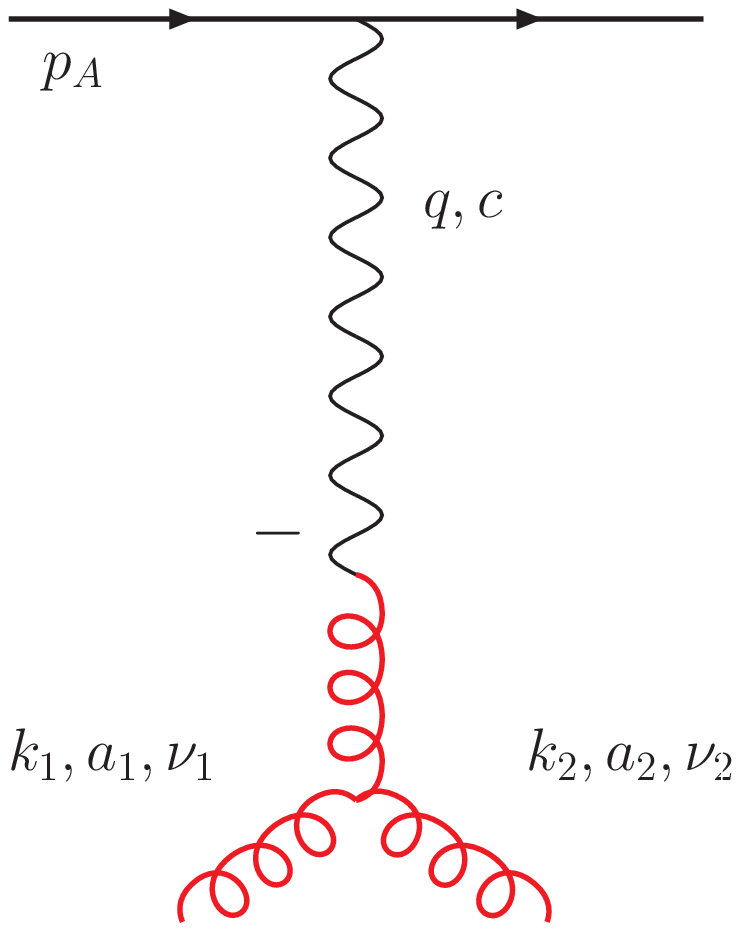}}
  \parbox{3.5cm}{\center \includegraphics[width=3cm]{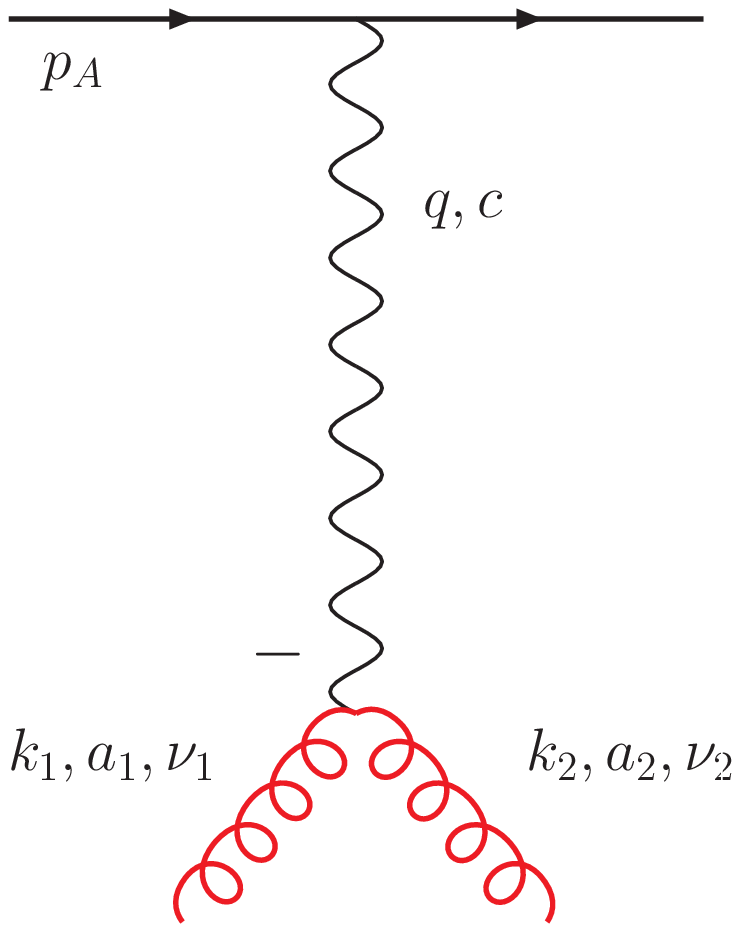}} 
  \parbox{3.5cm}{\center \includegraphics[width=3cm]{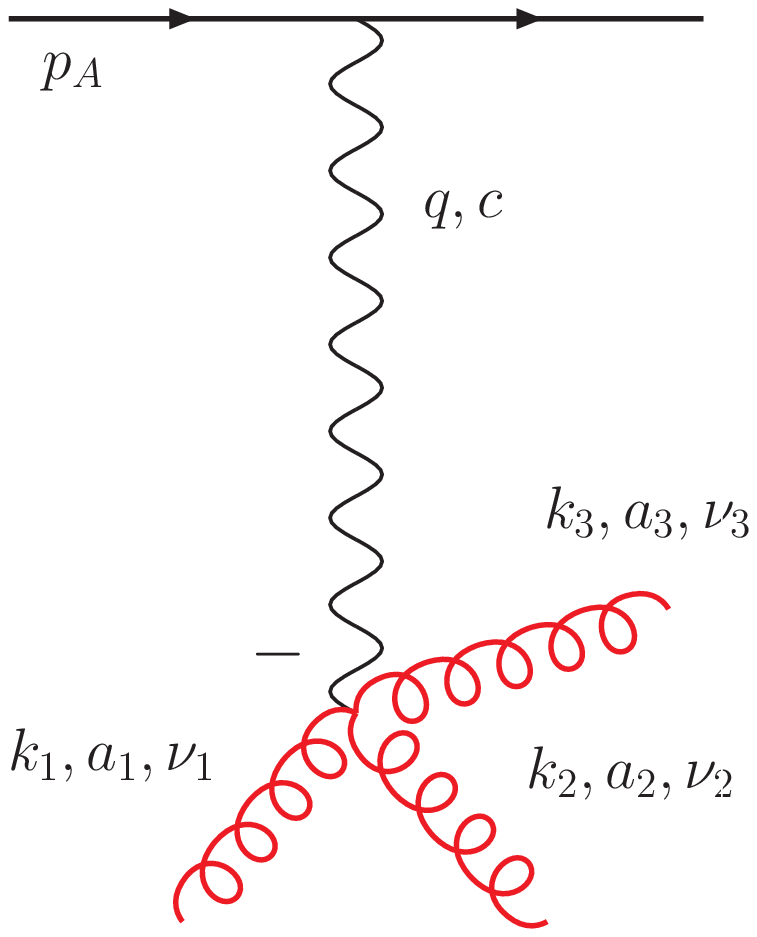}}
  \parbox{3.5cm}{\center \includegraphics[width=3cm]{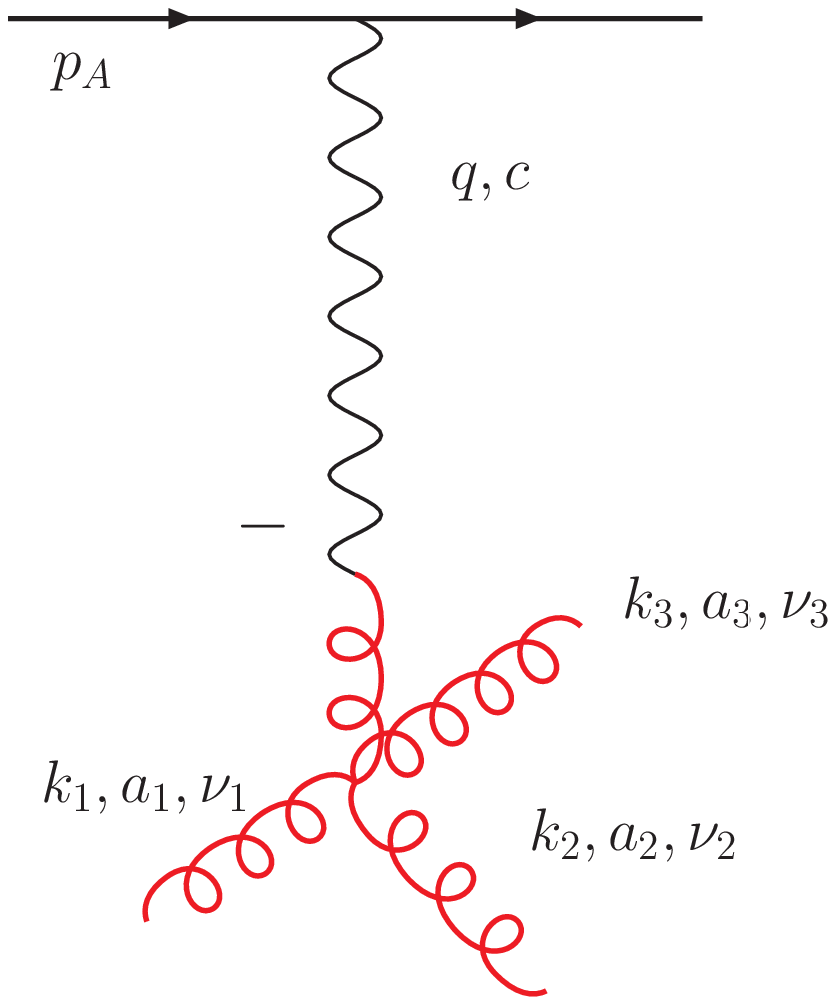}} \\
  \parbox{3.5cm}{\center (a)}
  \parbox{3.5cm}{\center (b)}
  \parbox{3.5cm}{\center (c)}
  \parbox{3.5cm}{\center (d)}
  \caption{\small Quark-gluon sub-amplitudes of the various couplings of reggeized gluons to the gluon loop in Fig.\ref{fig:feyndiag} and Fig.\ref{fig:feyndiag2}.}
  \label{fig:couplings}
\end{figure}

From Sec.~\ref{sec:quasi-elastic} we have the following expression for
the sum of Born and one-loop contributions of the quark-reggeized gluon
vertex (while as far one-loop corrections are concerned, only the
leading logarithmic part of diagrams like Fig.~\ref{fig:quasi} is
included)
\begin{align}
  \label{eq:qqr_born+1loop}
\Gamma^a_{QQR}(\omega_1;  q_1; \lambda_{11})  = \Gamma^a_{QQR}(  q_1 )
 \left( \frac{\omega_1 - \beta({\bm q}^2_1)}{\omega_1}  +\frac{\beta({\bm q}^2_1)}{\omega_1}  \bigg[   (-\lambda_{11})^{\omega_1} + (\lambda_{11})^{\omega_1}     \bigg] \right),
\end{align}
where the first term leads to a vanishing expression, once inserted
into the final result.  For the diagrams like Fig.~\ref{fig:feyndiag}
and Fig.~\ref{fig:feyndiag2} that yield the one-loop correction to the
production vertex we make an ansatz in analogy to
Eq.~(\ref{eq:gamma_bar}) of Sec.~\ref{sec:quasi-elastic}: The
situation is most straight forward for diagrams which involve only
couplings of a reggeized gluon to two QCD-gluons, like diagram B1 in
Fig.~\ref{fig:feyndiag}.  In that case we start with the following
expression
\begin{align}
  \label{eq:Gamma_B1}
 \Gamma^{acb,\mu}_{\text{B1}}& (\omega_1, \omega_2;q_1, q_2; \lambda_{12}, \lambda_{21} )=
\int \frac{d^4 k}{(2\pi)^4} 
\bar{\gamma}_{\text{3G},\nu_1-\nu_2}^{e_1ae_2}(\omega_1; -k, q_1, k-q_1)
\frac{-i}{(q_1 - k)^2} 
\notag \\
&  {\gamma}_{\text{3G},\nu_1\mu\nu_2}^{e_2be_3}( q_1-k, q_2 - q_1, k-q_2)
 \frac{-i}{(q_2 - k)^2}  \bar{\Delta}_{\text{+},\nu_1\nu_2}^{e_1ae_2}(\omega_2; q_2-k, -q_2, k)\frac{-i}{k^2},
\end{align}
where ${\gamma}_{\text{3G}}$ is the usual three-gluon-vertex, while
the three-gluon-vertex $\bar{\gamma}_{\text{3G}}(\omega_1)$ and the
induced vertex $\bar{\Delta}_{\text{+} }(\omega_2)$ are modified due
to their coupling to the reggeized gluons. They require negative
signature and a lower bound in their corresponding squared
center-of-mass energy.  For the modified three-gluon-vertex, the
sub-amplitude that belongs to the reggeized gluon is depicted in
Fig.~\ref{fig:couplings}a, which has a center-of-mass energy
$p_A^+k^-$.  As for the quasi-elastic region in
Sec.~\ref{sec:quasi-elastic} we impose for vertex corrections the
lower bound $p_A^+k^- > s_1/\lambda_{12}$, which we implement as usual
by the Mellin-integral:
\begin{align}
  \label{eq:threegluon_modified}
\bar{\gamma}_{\text{3G},\nu_1-\nu_2}^{e_1ae_2}(\omega_1; -k, q_1, k-q_1) 
=\int \frac{d\omega_1'}{4\pi i} 
\frac{1}{\omega_1' -\omega_1}
&\left[
\left( \frac{-p_A^+k^- -i\epsilon}{s_1/\lambda_{12}}  \right)^{\omega_1'}
+
\left( \frac{p_A^+k^- -i\epsilon}{s_1/\lambda_{12}}  \right)^{\omega_1'}
   \right] 
\notag \\
&{\gamma}_{\text{3G},\nu_1-\nu_2}^{e_1ae_2}( -k, q_1, k-q_1)
\end{align}
Similarly one has the following expression for the induced vertex of the first order,  Fig.~\ref{fig:couplings}b 
\begin{align}
  \label{eq:threegluon_modified}
\bar{\Delta}_{\text{+},\nu_1\nu_2}^{e_1ae_2}(\omega_1; -k, q_1, k-q_1) 
=\int \frac{d\omega_1'}{4\pi i} 
\frac{1}{\omega_1' -\omega_1}
&\left[
\left( \frac{-p_A^+k^- -i\epsilon}{s_1/\lambda_{12}}  \right)^{\omega_1'}
+
\left( \frac{p_A^+k^- -i\epsilon}{s_1/\lambda_{12}}  \right)^{\omega_1'}
   \right] 
\notag \\
& g{\bm q}_1^2 f^{e_1ae_2}\frac{1}{-k^-} n^-_{\nu_1}n^-_{\nu_2},
\end{align}
where the $i\epsilon$-prescription of the pole in $k^-$ has to
coincide as usually with the $i\epsilon$-prescription of the
branch-cuts due to the Mellin-integral. For the sub-amplitudes
Fig.~\ref{fig:couplings}c and Fig.~\ref{fig:couplings}d which contain
more than three gluons on their lower end, the definition of the
center-of-mass energy on which the lower bound is to be imposed is
less apparent: Choosing at random one of the three external gluon
momenta to define the center-of-mass energy of the
reggeized gluon, breaks Bose-symmetry of both  the induced vertex of
the second order and  the four-gluon-vertex. While the momentum of
the produced gluon is special, as it is fixed already by the
Multi-/Double-Regge-Kinematics and as it only indirectly contributes
to the loop integral, keeping such a symmetry is particularly desirable for the two gluons that enter the gluon-loop. 
 For the
induced vertex of the second order, the right choice can  be
deduced from the pole-structure, which can be always stated in a form, that it contains only one pole, with the loop-momentum $k^-$: With the unregularized vertex given
by
\begin{align}
  \label{eq:indu2_unalles}
\Delta^{ce_1e_2a}_{\mu\nu_1\nu_2} = -i{\bm q}_1^2 \left[
   \frac{f^{ce_1e}f^{e_2ea}}{q_2^- (-k^-)} 
   +
   \frac{f^{ce_1e}f^{e_2ea}}{q_2^- (k^- - q_2^-)}
 \right] n^-_{\mu} n^-_{\nu_1} n^-_{\nu_2},
\end{align}
we shall use in the following
\begin{align}
  \label{eq:indu2_reggular}
\bar{\Delta}_{\text{-},\nu_1\nu_2\mu}^{e_1ae_2}(\omega_1) 
=&\frac{-i{\bm q}_1^2}{q_2^-} n^-_{\mu}
\int \frac{d\omega_1'}{4\pi i} 
\frac{1}{\omega_1' -\omega_1}
  \left[\frac{f^{ce_1e}f^{e_2ea}}{ (-k^-)}
\left(
\left( \frac{-p_A^+k^- -i\epsilon}{s_1/\lambda_{12}}  \right)^{\omega_1'}
+
\left( \frac{p_A^+k^- -i\epsilon}{s_1/\lambda_{12}}  \right)^{\omega_1'}
   \right) \right.
\notag \\
+& \left.   \frac{f^{ce_1e}f^{e_2ea}}{ (k^- - q_2^-)} \left(
\left( \frac{-p_A^+(k^- -q_2^-) -i\epsilon}{s_1/\lambda_{12}}  \right)^{\omega_1'}
+
\left( \frac{p_A^+(k^- - q_2^-) -i\epsilon}{s_1/\lambda_{12}}  \right)^{\omega_1'}
   \right) 
\right] n^-_{\nu_1} n^-_{\nu_2},
\end{align}
for our calculations. We note here, that in general it is non-trivial
that the $i\epsilon$ prescription of the pole due to  the second induced
vertex  coincides with the $i\epsilon$ prescription of the
pole of the first induced vertex.  We will address the $i\epsilon$-prescription of the higher induced vertices in Sec.~\ref{sec:indu_presc} and anticipating the result obtained there, it can be confirmed that this is indeed true, if one of the three gluons that couple to the induced vertex is real.

For the reggeized gluon coupling to the four-gluon-vertex,
Fig.~\ref{fig:couplings}d, we  propose a similar modification.  As
for the induced vertex of the second order, we propose for every
occurring color structure a different regularization, such that the
complete vertex has the desired symmetry. With the (unmodified)
four-gluon vertex in Fig.~\ref{fig:couplings}d given by
\begin{align}
  \label{eq:4gl_unmodi}
{\gamma}_{\text{4G},-\mu\nu_1\nu_2}^{ace_1e_2} = -ig^2 \big[ &
 f^{ace}f^{e_1e_2e} 
\left(n^-_{\nu_1} g_{\mu\nu_2} - n^-_{\nu_2} g_{\mu\nu_1} \right)
+
 f^{ae_1e}f^{ce_2e} 
\left(n^-_{\nu_2} g_{\mu\nu_1} - n^-_{\mu} g_{\nu_1\nu_2} \right)
\notag \\
&+
f^{ae_2e}f^{ce_1e} 
\left(   n^-_{\nu_1} g_{\mu\nu_2} -  n^-_{\mu} g_{\nu_1\nu_2} \right) \big],
\end{align}
we use the following modified version in our  calculations
\begin{align}
  \label{eq:4gl_modi}
\tilde{\gamma}_{\text{4G},-\mu\nu_1\nu_2}^{ace_1e_2}&(\omega_1) = -ig^2
\int \frac{d\omega_1'}{2\pi i} \frac{1}{\omega_1' - \omega_1} 
\notag \\ 
& \left[ 
 f^{ace}f^{e_1e_2e} 
\left(n^-_{\nu_1} g_{\mu\nu_2} - n^-_{\nu_2} g_{\mu\nu_1} \right)
\left(
\left( \frac{-p_A^+q_2^- -i\epsilon}{s_1/\lambda_{12}}  \right)^{\omega_1'}
+
\left( \frac{p_A^+q_2^- -i\epsilon}{s_1/\lambda_{12}}  \right)^{\omega_1'}
   \right) \right.
\notag \\
&+ 
 f^{ae_1e}f^{ce_2e} 
\left(n^-_{\nu_2} g_{\mu\nu_1} - n^-_{\mu} g_{\nu_1\nu_2} \right)\left(
\left( \frac{-p_A^+k_1^- -i\epsilon}{s_1/\lambda_{12}}  \right)^{\omega_1'}
+
\left( \frac{p_A^+k_1^- -i\epsilon}{s_1/\lambda_{12}}  \right)^{\omega_1'}
   \right)
\notag \\
&+
f^{ae_2e}f^{ce_1e}  \left.
\left(  n^-_{\nu_1} g_{\mu\nu_2}   - n^-_{\mu} g_{\nu_1\nu_2}  \right) 
\left(\left( \frac{-p_A^+k_2^- -i\epsilon}{s_1/\lambda_{12}}  \right)^{\omega_1'}
+
\left( \frac{p_A^+k_2^- -i\epsilon}{s_1/\lambda_{12}}  \right)^{\omega_1'}
   \right)
 \right],
\end{align}
with  $k_1 = k$ and $k_2 = q_2 - k$. The above expression is
 symmetric under simultaneous exchange of momenta, polarization
and color of the three gluons as needed. 
In the evaluation of the longitudinal integrations we then restrict to
the leading logarithmic part of the integrals, similar to 
 Sec.~\ref{sec:quasi-elastic}.  We note  that  in principle  also the sub-leading part can show 
some sort of divergence, which however only occurs if the different
transverse momenta are not of the same order of magnitude. 
Those kind of divergences is thus related to transverse logarithms which are not taken into account within the LLA.

 For the evaluation of the
leading part, knowledge of two integrals $I_1$ and
$I_2$ is sufficient.  Integral $I_1$ can be found in slightly different
form already in the analysis of massive $\phi^3$-theory of
\cite{Drummond:1969ft}.  It is given by
\begin{align}
  \label{eq:1doublecut}
I_1 =& \frac{1}{2} \int \frac{d k^+k^-}{2\pi} \int \frac{d^2 {\bm k}}{(2\pi)^3}\frac{-i}{k^2 + i\epsilon} \frac{-i}{(q_1 - k)^2 + i\epsilon} \frac{-i}{(q_2 - k)^2 + i\epsilon} \frac{1}{k^+k^-}
\notag \\
& \quad \left[\left( \frac{-p_A^+k^- -i\epsilon}{s_1/\lambda_{12}}\right)^{\omega_1'} 
+
\left( \frac{p_A^+k^- -i\epsilon}{s_1/\lambda_{12}}\right)^{\omega_1'} 
\right]
\left[\left( \frac{-p_B^- k^+-i\epsilon}{s_2/\lambda_{21}}\right)^{\omega_2'}
 +
\left( \frac{p_B^- k^+-i\epsilon}{s_2/\lambda_{21}}\right)^{\omega_2}
 \right] 
\notag \\
=&   \lambda_{12}^{\omega_1'}\lambda_{21}^{\omega_2'} 
       \left[ \kappa_1^{-\omega_1}
    \xi_{\omega_1'}^{(-)} \xi_{\omega_2' \omega_1'}^{(-,-)}  \frac{F_1(\omega_1', \omega_2', t_1, t_2, \kappa_1)}{\sin \pi(\omega_{2}' - \omega_1')}   
+ 
\kappa_1^{-\omega_2'}
 \xi_{\omega_2'}^{(-)} \xi_{\omega_1' \omega_2'}^{(-,-)}  \frac{F_1(\omega_2', \omega_1', t_1, t_2, \kappa_1)}{\sin \pi(\omega_1' - \omega_2')} 
 \right],
\end{align}
with signature factors defined as in Eqs.~(\ref{eq:2sigi}) and (\ref{eq:doule_sig}) and
\begin{align}
  \label{eq:F1}
   F_1(\omega_1', \omega_2', t_1, t_2, \kappa) =&
 \frac{-i \pi}{2} \int \frac{d^2 {\bm k}}{(2\pi)^3}    \int_0^\infty d\lambda \lambda^{2 -\omega_1'} e^{-i\pi \omega_1'/2} \int \prod_{i=1}^3 
 dx_i \,\delta(1 - \sum_{i=1}^3 x_i) x_2^{\omega_2'- \omega_1'}
\notag \\ &
 e^{-i\lambda(x_1 {\bm{k}}^2 + x_2 ({\bm{q}}_1 -{\bm{k}}^2) + x_3 ({\bm{q}}_2 - \bm{k})^2 )} \frac{M(\omega_2'| \omega_2' - \omega_1' + 1| i\kappa_1 \lambda x_2x_3)}{\Gamma(\omega_2' - \omega_1' + 1) \Gamma(1 - \omega_2')},
\end{align}
where $M(a|b|z) = {}_1F_1(a|b|z)$ is a confluent hypergeometric
function (Kummer's function \cite{Abramowitz::1949ft}). The LLA
corresponds to the limit $\omega_1', \omega_2' \to 0$ which can be
taken safely as $M(a|b|z)/\Gamma(b)$ is an entire function of all its
variables. We find
\begin{align}
  \label{eq:F1_LLA}
 F_1(\omega_1', \omega_2', t_1, t_2, \kappa) = \frac{-\pi}{2} \int \frac{d^2 {\bm k}}{(2\pi)^3 }\frac{({\bm k}^2)^{\omega_1'}}{ {\bm{k}}^2  ({\bm{q}}_1 -{\bm{k}}^2)  ({\bm{q}}_2 - \bm{k})^2},
\end{align}
where $({\bm k}^2)^{\omega_1} $ represents a typical transverse scale
of the production amplitude such as ${\bm q}^2_1$, ${\bm q}^2_2$ or
$\kappa_1$ which is not specified in the LLA.  The second integral is
very similar to Eq.~(\ref{eq:1doublecut}), but has a slightly simpler
structure
\begin{align}
  \label{eq:2doublecut}
I_2 =& \frac{-i}{2} \int \frac{d k^+k^-}{2\pi} \int \frac{d^2 {\bm k}}{(2\pi)^3}\frac{-i}{k^2 + i\epsilon} \frac{-i}{(q_2 - k)^2 + i\epsilon} \frac{1}{k^+k^-}
\notag \\
& \quad \left[\left( \frac{-p_A^+k^- -i\epsilon}{s_1/\lambda_{12}}\right)^{\omega_1'} 
+
\left( \frac{p_A^+k^- -i\epsilon}{s_1/\lambda_{12}}\right)^{\omega_1'} 
\right]
\left[\left( \frac{-p_B^- k^+-i\epsilon}{s_2/\lambda_{21}}\right)^{\omega_2'}
 +
\left( \frac{p_B^- k^+-i\epsilon}{s_2/\lambda_{21}}\right)^{\omega_2'}
 \right] 
\notag \\
=&  \lambda_{12}^{\omega_1'}
\lambda_{21}^{\omega_2'} 
      \kappa_1^{-\omega_2'}
 \xi_{\omega_2'}^{(-)} \xi_{\omega_1' \omega_2'}^{(-,-)}  \frac{F_2(\omega_2', \omega_1',  t_2, \kappa_1)}{\sin \pi(\omega_{1}' - \omega_2')}\bigg|_{\Re\text{e}\omega_2' > \Re\text{e}\omega_1'}.
 \end{align}
Note that the evaluation of  $I_2$ requires the $\omega_2'$-contour to be to the right of the $\omega_1'$ contour.  $F_2$ is given by
\begin{align}
  \label{eq:F1}
   F_2(\omega_1', \omega_2',  t_2, \kappa) =&
 \frac{- \pi}{2} \int \frac{d^2 {\bm k}}{(2\pi)^3}    \int_0^\infty d\lambda \lambda^{1 -\omega_1'} e^{-i\pi \omega_2'/2} \int \prod_{i=1}^2 
 dx_i \,\delta(1 - \sum_{i=1}^2 x_i) x_2^{\omega_1'- \omega_2'}
\notag \\ &
 e^{-i\lambda(x_1 {\bm{k}}^2 + x_2 ({\bm{q}}_2 -{\bm{k}}^2) } \frac{1}{\Gamma(\omega_2' - \omega_1' + 1) \Gamma(1 - \omega_2')},
\end{align}
which reduces in the limit $\omega_1', \omega_2' \to 0$ to
\begin{align}
  \label{eq:F2_LLA}
 F_2(\omega_1', \omega_2', t_1, t_2, \kappa) =
 \frac{\pi}{2} \int \frac{d^2 {\bm k}}{(2\pi)^3 } 
\frac{({\bm k}^2)^{\omega_1'}}{ {\bm{k}}^2    ({\bm{q}}_2 - \bm{k})^2}.
\end{align}
We then write  the  production vertex as the sum of its Born and 1-loop corrections $\delta\Gamma^\mu_{RRG}$
\begin{align}
  \label{eq:production_born+loop}
 \Gamma_{RRG}^{acb,\mu}(\omega_1, \omega_2;q_1, q_2; \lambda_{12}, \lambda_{21} ) = & 2gf^{acb} C^{\mu} (q_1, q_2) \frac{\xi^{-}(\omega_1)}{2} \frac{\xi^{-}(\omega_2)}{2} 
\notag \\ &
+  \delta\Gamma_{RRG}^{acb,\mu}(\omega_1, \omega_2;q_1, q_2; \lambda_{12}, \lambda_{21} ),
\end{align}
where signature factors of the adjacent reggeized gluons are absorbed
in the production vertex, in accordance with our decomposition
Eq.~(\ref{eq:fac_23}). For the 1-loop corrections we find within our accuracy
\begin{align}
  \label{eq:prod_1loop}
 \delta&\Gamma_{RRG}^{acb,\mu}(\omega_1, \omega_2;q_1, q_2; \lambda_{12}, \lambda_{21} )
 =
\int \frac{d\omega_1'}{4\pi i }\int \frac{d\omega_1'}{4\pi i } 
            \frac{1}{\omega'_1 - \omega_1}  \frac{1}{\omega'_2 - \omega_2}  \lambda_{12}^{\omega_1'} \lambda_{21}^{\omega_2'} 
\notag \\
 &\left[  
   \kappa_1^{-\omega_2'}  \xi_{\omega_2'}^{(-)} \xi_{\omega_1' \omega_2'}^{(-,-)}  \frac{ V_a(q_1, q_2)  }{\omega_1' - \omega_2'} \bigg|_{\Re\text{e}\omega_1' > \Re\text{e}\omega_2'}
+
\kappa_1^{-\omega_1'}  \xi_{\omega_1'}^{(-)} 
 \xi_{\omega_2' \omega_1'}^{(-,-)}
 \frac{ V_a(q_2, q_1)     }{\omega_2' - \omega_1'}
 \bigg|_{\Re\text{e}\omega_2' > \Re\text{e}\omega_1'}    
\right. 
\notag \\
 &\left. \left(
     \kappa_1^{-\omega_1'}  \xi_{\omega_1'}^{(-)} 
 \xi_{\omega_2' \omega_1'}^{(-,-)}
 \frac{1 }{\omega_2' - \omega_1'}  
+
\kappa_1^{-\omega_2'}  \xi_{\omega_2'}^{(-)} \xi_{\omega_1' \omega_2'}^{(-,-)}  \frac{1}{\omega_1' - \omega_2'}
    \right) V_b(q_1, q_2) \right],
\end{align}
where
\begin{align}
  \label{eq:Va}
V_a(q_1, q_2) =&  \quad \beta({\bm{q}}_1)\,  2gf^{acb} C^\mu(q_1, q_2) 
\\ \label{eq:Vb}
V_b(q_1, q_2) =&  \frac{N_c g^2}{2}  \int \frac{d^2 \bm{k}}{(2\pi)^3}  \frac{ {{\bm{q}}_2^2 {\bm{q}}_1^2} }{ {\bm{k}}^2  ({\bm{q}}_1 -{\bm{k}}^2)  ({\bm{q}}_2 - \bm{k})^2}  \,2g f^{acb} C^\mu(q_1 - k, q_2 - k)
\end{align}
and we approximated 
$$\frac{\pi}{\sin\pi(\omega_1' - \omega_2')} \simeq \frac{1}{\omega_1'
  - \omega_2'},$$
in accordance with the LLA. Note that
Eq.~(\ref{eq:prod_1loop}) resembles closely the analytic
representation of the $2\to3$ production amplitude
Eq.~(\ref{eq:anal_23_fac}). However, whereas
Eq.~(\ref{eq:anal_23_fac}) represents the analytic structure of the
$2\to 3$ production \emph{amplitude}, Eq.(\ref{eq:prod_1loop})
constitutes merely the 1-loop correction to the production
\emph{vertex} (with the Born-term not yet included) and is only exact
up to $\mathcal{O}(g^3)$. Eq.~(\ref{eq:anal_23_fac}) can only be
obtained from the effective action, if corrections concerning the
phase structure of the production vertex are resummed to all orders in
perturbation theory.  Nevertheless the strong coincidence in the
structure of Eq.~(\ref{eq:anal_23_fac}) and Eq.(\ref{eq:prod_1loop})
is encouraging as it suggests that the effective action indeed yields
the correct analytic structure of the production amplitude. Evaluating
the integrals over $\omega_1'$, $\omega_2'$ in
Eq.(\ref{eq:prod_1loop}) and expanding in 'small' logarithms $\ln
\lambda_{1,2}$ and $\ln \kappa_1$, we obtain
\begin{align}
  \label{eq:verte_corr}
 \delta\Gamma_{RRG}^{acb,\mu}&(\omega_1, \omega_2;q_1, q_2; \lambda_{12}, \lambda_{21} )
 = \frac{\xi^{-}(\omega_1)}{2} \frac{\xi^{-}(\omega_2)}{2}
\bigg[  \frac{\ln -\lambda_{12} + \ln\lambda_{12}}{2}  V_a(q_1, q_2) \notag \\
&+  \frac{\ln -\lambda_{21} + \ln\lambda_{21}}{2} V_a(q_2, q_1) +  \frac{\ln-\kappa_1 + \ln\kappa_1}{2} V_b(q_2, q_1),
\bigg],
\end{align}
where we extracted the complete signature factors of the reggeized
gluons that couple to the production vertex, whereas all other factors
have been expanded up to $\mathcal{O}(\omega_i)$. Together with the Born term we obtain
\begin{align}
  \label{eq:prod_compl}
\Gamma_{RRG}^{acb,\mu}&(\omega_1, \omega_2 ; q_1, q_2; \lambda_{12}, \lambda_{21} )
 = 2gf^{acb} \times
\notag \\
 &\frac{\xi^{-}(\omega_1)}{2} \frac{\xi^{-}(\omega_2)}{2}
\bigg[
 \left( 1 +\frac{\ln -\lambda_{12} + \ln\lambda_{12}}{2}\beta({\bm q}_1) 
+ 
 \frac{\ln -\lambda_{21} + \ln\lambda_{21}}{2}\beta({\bm q}_2) \right)C^{\mu}(q_1, q_2)
\notag \\
&+
 \frac{\ln-\kappa_1 + \ln\kappa_1}{2} \frac{N_c g^2}{2}  \int \frac{d^2 \bm{k}}{(2\pi)^3}  \frac{ {{\bm{q}}_2^2 {\bm{q}}_1^2} }{ {\bm{k}}^2  ({\bm{q}}_1 -{\bm{k}}^2)  ({\bm{q}}_2 - \bm{k})^2}  \, C^\mu(q_1 - k, q_2 - k)
   \bigg].
 \end{align}
 Inserting the result into Eq.~(\ref{eq:fac_23}) and expanding there
 the Regge-factors and the quark-reggeized gluon vertex, Eq.~(\ref{eq:qqr_born+1loop}),  in $\lambda_{11}$, $\lambda_{12}$,
 $\lambda_{21}$ and $\lambda_{22}$ we find that to the desired
 accuracy the dependence on these parameters vanishes for the
 complete amplitude:
 \begin{align}
  \label{eq:insert_23}
\mathcal{M}_{2 \to 3} =& 2s\int \frac{d\omega_1}{2\pi i } \int \frac{d\omega_2}{2\pi i } 
gt^a 
 \times  \frac{1}{t_1} \frac{1}{\omega_1 -\beta({\bm q}^2_1)} 
\left|\frac{s_1}{s_{R_1}} \right|^{\omega_1} \frac{\xi^{-}(\omega_1)}{2}
 \notag \\
&
 \times gT_{ba}^c \tilde{\Gamma}^\mu_{{RRG}}(q_1, q_2)
  \times \frac{1}{t_2} \frac{1}{\omega_2 -\beta({\bm q}^2_2)} 
\left|\frac{s_2}{s_{R_2}  }  \right|^{\omega_2}\frac{\xi^{-}(\omega_2)}{2}\!\! \times  gt^b,
\end{align}
with
\begin{align}
  \label{eq:gamma_tilde}
\tilde{\Gamma}^\mu_{{RRG}}(q_1, q_2) =   C^{\mu}(q_1, q_2) +  
 \frac{\ln-\kappa_1 + \ln\kappa_1}{2} \frac{N_c g^2}{2}  \int \frac{d^2 \bm{k}}{(2\pi)^3}  \frac{ {{\bm{q}}_2^2 {\bm{q}}_1^2}  C^\mu(q_1 - k, q_2 - k) }{ {\bm{k}}^2  ({\bm{q}}_1 -{\bm{k}}^2)  ({\bm{q}}_2 - \bm{k})^2}  \, ,
 \end{align}
and $t_{1,2} = -{\bm q}_{1,2}^2$.
We shall confront this result at first with the analytic representation of the $2 \to 3$ amplitude Eq.~(\ref{eq:anal_23}), which we rewrite as
\begin{align}
  \label{eq:anal_23_fac}
\mathcal{M}_{2 \to 3}  
= &
s \int \frac{d\omega_1}{2\pi i}  \int \frac{d\omega_2}{2\pi i} 
 |s_1|^{\omega_1 } |s_2|^{\omega_2} \notag \\
&\left(\kappa_1^{-\omega_2} \xi^{(\tau_2)}_{(\omega_2)} \xi^{(\tau_1,\tau_2)}_{(\omega_1,\omega_2)} 
\frac{V_1 (\omega_1, \omega_2 ; t_1, t_2, \kappa_1) }{\sin\pi\omega_{12}}
+
\kappa_1^{-\omega_1}
\xi^{(\tau_1)}_{(\omega_1)} \xi^{(\tau_2,\tau_1)}_{(\omega_2,\omega_1)} 
 \frac{V_2 (\omega_1, \omega_2 ; t_1, t_2, \kappa_1)}{\sin \pi\omega_{21}} \right).
\end{align}
Following closely \cite{Bartels:1974tj}, we introduce for Eq.~(\ref{eq:anal_23_fac}) factors 
\begin{align}
  \label{eq:2phifactor_xis}
  \phi^{\omega_1}_{\omega_1\omega_2} &= \frac{1}{ \xi_{\omega_1}^{(-)}
    \xi_{\omega_2}^{(-)}}  \xi_{\omega_1}^{(-)} \xi_{\omega_2 \omega_1}^{(-,-)}
&\text{and} &&
 \phi^{\omega_2}_{\omega_1\omega_2} &= \frac{1}{ \xi_{\omega_1}^{(-)}
    \xi_{\omega_2}^{(-)}} \xi_{\omega_2}^{(-)} \xi_{\omega_1 \omega_2}^{(-,-)}
\end{align}
and rewrite them as
\begin{align}
  \label{eq:1phifactor_xis}
\phi^{\omega_i}_{\omega_1\omega_2} 
&=
e^{i\pi\omega_i} - \frac{1}{ \xi_{\omega_1}^{(-)}  \xi_{\omega_2}^{(-)}} (e^{-i\pi\omega_i} - e^{i\pi\omega_i}).
\end{align}
Eq.~(\ref{eq:anal_23_fac}) takes then the form
\begin{align}
  \label{eq:anal_23_fac_manip}
\mathcal{M}_{2 \to 3}  
= &
s \int \frac{d\omega_1}{2\pi i}  \int \frac{d\omega_2}{2\pi i} 
 |s_1|^{\omega_1 } \frac{\xi^{(-)}_{\omega_1}}{2}  |s_2|^{\omega_2} \frac{\xi^{(-)}_{\omega_2}}{2}  \notag \\
& \frac{\kappa_1^{-\omega_2} \phi^{\omega_2}_{\omega_1\omega_2}    V_1 (\omega_1, \omega_2 ; t_1, t_2, \kappa_1)  - 
\kappa_1^{-\omega_1}\phi^{\omega_1}_{\omega_1\omega_2} 
V_2 (\omega_1, \omega_2 ; t_1, t_2, \kappa_1) }{\sin\pi\omega_{12}}.
\end{align}
Assuming from now on to have to this accuracy for negative signature
only exchange of Regge-poles in both $t$-channels, we define functions
$ \tilde{V}_i(t_1, t_2, \kappa_1) $ by
\begin{align}
  \label{eq:V12}
 V_i (\omega_1, \omega_2 ; t_1, t_2, \kappa_1) =& gt^a \frac{1}{\omega_1 - \beta({\bm q}_1^2)} \frac{1}{{t}_1} \times  T^{c}_{ba} \pi \tilde{V}_i ( t_1, t_2, \kappa_1)
 \times
\frac{1}{\omega_2 - \beta({\bm q}_2^2))} \frac{1}{t_2} gt^b.
\end{align}
The functions $\tilde{V}_i$ can  be determined within the LLA
\cite{Bartels:1978fc,Bartels:2008ce} from taking discontinuities in
Eq.~(\ref{eq:anal_23}) and making use of unitarity\footnote{The
  following choice coincides  with the one of
  \cite{Bartels:2008ce} for the functions $V_1$ and $V_2$ up to a
  factor $\pi$ which has been extracted in the present case }:
\begin{align}
  \label{eq:vtilde1}
\tilde{V}_1(t_1, t_2, \kappa_1)  &= 
 \beta({\bm q}_1^2)C^{\mu}(q_1, q_2) +  
 \frac{N_c g^2}{2}  \int \frac{d^2 \bm{k}}{(2\pi)^3}  \frac{ {{\bm{q}}_2^2 {\bm{q}}_1^2} }{ {\bm{k}}^2  ({\bm{q}}_1 -{\bm{k}}^2)  ({\bm{q}}_2 - \bm{k})^2}  \, C^\mu(q_1 - k, q_2 - k) \\
 \label{eq:vtilde1}
\tilde{V}_2(t_1, t_2, \kappa_1)  &= 
 \beta({\bm q}_2^2)C^{\mu}(q_1, q_2) +  
 \frac{N_c g^2}{2}  \int \frac{d^2 \bm{k}}{(2\pi)^3}  \frac{ {{\bm{q}}_2^2 {\bm{q}}_1^2} }{ {\bm{k}}^2  ({\bm{q}}_1 -{\bm{k}}^2)  ({\bm{q}}_2 - \bm{k})^2}  \, C^\mu(q_1 - k, q_2 - k).
\end{align}
Taking into account that $\kappa_1 =$fixed in the Double-Regge-Kinematics, we  expand 
\begin{align}
  \label{eq:kappa_expand}
\kappa_1^{\omega_i}\phi^{\omega_i}_{\omega_1\omega_2} = 1 + \omega_i\frac{\ln -\kappa_1 + \ln \kappa_1}{2} + \ldots
\end{align}
where the dots indicate terms of higher order in $\omega_1$, $\omega_2$
which is equivalent to terms of higher order in $g^2$. With
$\sin\pi\omega_{12} \simeq \pi \omega_{12}$ due to the LLA, one can
then verify that Eq.~(\ref{eq:anal_23_fac_manip}) and
Eq.~(\ref{eq:insert_23}) agree with each other at the desired
accuracy.  Eq.~(\ref{eq:insert_23}) also satisfies the
Steinmann-relations  in this sense. Whereas
Eq.~(\ref{eq:anal_23_fac_manip})/Eq.~(\ref{eq:anal_23}) satisfy the
Steinmann-relations by construction, Eq.~(\ref{eq:insert_23}) does so
only within the achieved accuracy. The complete inner analytic
structure of the production vertex that is needed in a field theory of
Reggeons in order to satisfy the Steinmann-relations,  is 
only obtained to the level of accuracy the phase structure of the
production vertex has been determined. This should  not be understood
as a failure of the effective theory or its prescription of
longitudinal integrations, but  is natural for an analysis bases on
the LLA, as the logarithms inside the production vertex which yield
the analytic structure are small and not included at LLA. On the other
hand this tells us that  care is needed when dealing with the
phase structure of reggeized gluons within the effective action: It
can be only trusted up to the level of accuracy one goes beyond the
LLA, in particular when reggeized gluon - particle vertices have a
inner phase structure, as it is the case for the production amplitude.




Similar observations hold for the $2 \to 4$ production amplitude. This
case is generally more involved than the $2 \to 3$ production
amplitude: Apart from the Regge-pole exchange in all three
$t$-channels, the further exchange of a  state of two or more
reggeized gluons in the $t_2$-channel is required by unitarity.  For
negative signature, on the other hand, this contribution can be shown
to be of higher order in $g^2$ from the analysis of the analytic
representation Eq.~(\ref{eq:anal_24}), which is confirmed by the
results from the effective action.  For the present level of accuracy,
the interaction is for negative signature mediated by
Regge-pole-exchange alone. Recent interest in the analytical structure
of LLA-production amplitudes in the planar limit is triggered by the
study of BDS-amplitudes.  In that case the Regge-cut contribution is
already relevant at the present level of accuracy
\cite{Bartels:2008ce}.

For the effective action, we start with the following expression,
which generalizes Eq.~(\ref{eq:fac_23}):
\begin{align}
  \label{eq:fac_24}
i&\mathcal{M}_{2 \to 4} = s \int \frac{d\omega_1}{2\pi i } \int \frac{d\omega_2}{2\pi i } \int \frac{d\omega_3}{2\pi i } 
\tilde{\Gamma}^a_{QQR}(\omega_1;  q_1; \lambda_{11}) 
 \times  \frac{i/2}{{\bm q}_1^2} \frac{1}{\omega_1 -\beta({\bm q}_1)} 
\left|\frac{s_1}{s_{R_1}\lambda_{11}\lambda_{12}}  \right|^{\omega_1} 
 \notag \\
&
 \times \Gamma_{RRG}^{ac_1e,\mu_1}(\omega_1, \omega_2;q_1, q_2; \lambda_{12}, \lambda_{21} )
 \times \frac{i/2}{{\bm q}_2^2} \frac{1}{\omega_2 -\beta({\bm q}_2)} 
\left|\frac{s_2}{s_{R_2} \lambda_{21}\lambda_{23} }  \right|^{\omega_2}\!\! \times \left( \xi^{(-)}_{\omega_2}\right)^{-1}
\notag \\
&
\times \Gamma_{RRG}^{ec_2b,\mu_2}(\omega_2, \omega_3;q_2, q_3; \lambda_{23}, \lambda_{32} )
 \times \frac{i/2}{{\bm q}_3^2} \frac{1}{\omega_3 -\beta({\bm q}_3)} 
\left|\frac{s_3}{s_{R_3} \lambda_{32}\lambda_{33} }  \right|^{\omega_2}\!\! \times
 \tilde{\Gamma}^b_{QQR}(\omega_2;  q_2;\lambda_{33} ).
\end{align}
According to our conventions, the signature factors of the reggeized
gluons are absorbed into the production vertices which requires
introduction of the factor$(\xi^{(-)}_{\omega_2})^{-1}$. We further
note, that within our achieved accuracy the above obtained
loop-correction to the production amplitude can  only be inserted for
one of the production amplitudes, while for the other one should
simple use the Born result. Making use of the results
Eq.~(\ref{eq:qqr_born+1loop}) and Eq.~(\ref{eq:prod_compl}) we 
obtain
 \begin{align}
  \label{eq:insert_24}
&\mathcal{M}_{2 \to 4} = 2 s\int \frac{d\omega_1}{2\pi i } \int \frac{d\omega_2}{2\pi i }  \int \frac{d\omega_3}{2\pi i }
gt^a 
 \times  \frac{1}{t_1} \frac{1}{\omega_1 -\beta({\bm q}^2_1)} 
\left|\frac{s_1}{s_{R_1}} \right|^{\omega_1} \frac{\xi^{-}(\omega_1)}{2}
 \notag \\
& \bigg[ 
 \times  gT^{c_1}_{eq} C^{\mu_1}(q_1, q_2)
  \times \frac{1}{{t}_2} \frac{1}{\omega_2 -\beta({\bm q}^2_2)} 
\left|\frac{s_2}{s_{R_2}  }  \right|^{\omega_2}\frac{\xi^{-}(\omega_2)}{2}\!\! \times  
gT^{c_2}_{be} C^{\mu_2}(q_2, q_3)
\notag \\
&+
\times gT^{c_1}_{ea} \delta\tilde{\Gamma}^{\mu_1}_{{RRG}}(q_1, q_2)
  \times \frac{1}{{t}_2} \frac{1}{\omega_2 -\beta({\bm q}^2_2)} 
\left|\frac{s_2}{s_{R_2}  }  \right|^{\omega_2}\frac{\xi^{-}(\omega_2)}{2}\!\! \times  
gT^{c_2}_{be} C^{\mu_2}(q_2, q_3)
\notag \\
&+
\times gT^{c_1}_{ea} C^{\mu_1}(q_1, q_2)
  \times \frac{1}{{t}_2} \frac{1}{\omega_2 -\beta({\bm q}^2_2)} 
\left|\frac{s_2}{s_{R_2}  }  \right|^{\omega_2}\frac{\xi^{-}(\omega_2)}{2}\!\! \times  
gT^{c_2}_{be} \delta\tilde{\Gamma}^{\mu_2}_{{RRG}}(q_2, q_3)
\bigg]
\notag \\
& \times \frac{1}{{t}_3} \frac{1}{\omega_3 -\beta({\bm q}^2_3)} 
\left|\frac{s_3}{s_{R_3}  }  \right|^{\omega_3}\frac{\xi^{-}(\omega_3)}{2}\!\! \times  
gt^b
\end{align}
where 
\begin{align}
  \label{eq:delta_gamma}
\delta \tilde{\Gamma}^\mu_{{RRG}}(q_1, q_2) =  
 \frac{\ln-\kappa_1 + \ln\kappa_1}{2} \frac{N_c g^2}{2}  \int \frac{d^2 \bm{k}}{(2\pi)^3}  \frac{ {{\bm{q}}_2^2 {\bm{q}}_1^2} }{ {\bm{k}}^2  ({\bm{q}}_1 -{\bm{k}}^2)  ({\bm{q}}_2 - \bm{k})^2}  \, C^\mu(q_1 - k, q_2 - k)
\end{align}
is the one-loop contribution of Eq.~(\ref{eq:gamma_tilde}). This
results needs then to be confronted with the analytical representation
of the $2 \to 4$ production amplitude, Eq.~(\ref{eq:anal_24}). As for
the $2\to3$ amplitude, the partial waves $W_i$, $i = 1, \ldots, 5$ can
be determined at LLA by taking discontinuities of
Eq.~(\ref{eq:anal_24}) and making use of unitarity in the
(sub-)channels (see \cite{Bartels:1978fc,Bartels:2008ce}). Extracting
furthermore in Eq.~(\ref{eq:anal_24}) signature factors for each
$t$-channel and expanding the remainder up to the above level of
accuracy, we find  agreement of both expressions. In particular, Regge-cut contribution cancel for negative signature at this level of accuracy and Eq.~(\ref{eq:insert_24}) yields the correct result.

\section{Positive and mixed signatured production amplitudes}
\label{sec:mixed}

So far we concentrated our analysis on production amplitudes in which
all $t$-channels carry negative signature. From a technical point of
view, the analysis of production amplitudes with positive and mixed
signature is slightly simpler: We don't need to extract the analytic
structure from NLLA-calculations. It is sufficient to consider the
interaction of reggeized gluons instead, which is similar  to the
derivation of the BFKL-equation in Sec.\ref{sec:tworeggeon_negsig}.

\begin{figure}[htbp]
  \centering
  \parbox{4cm}{\center \includegraphics[width=3cm]{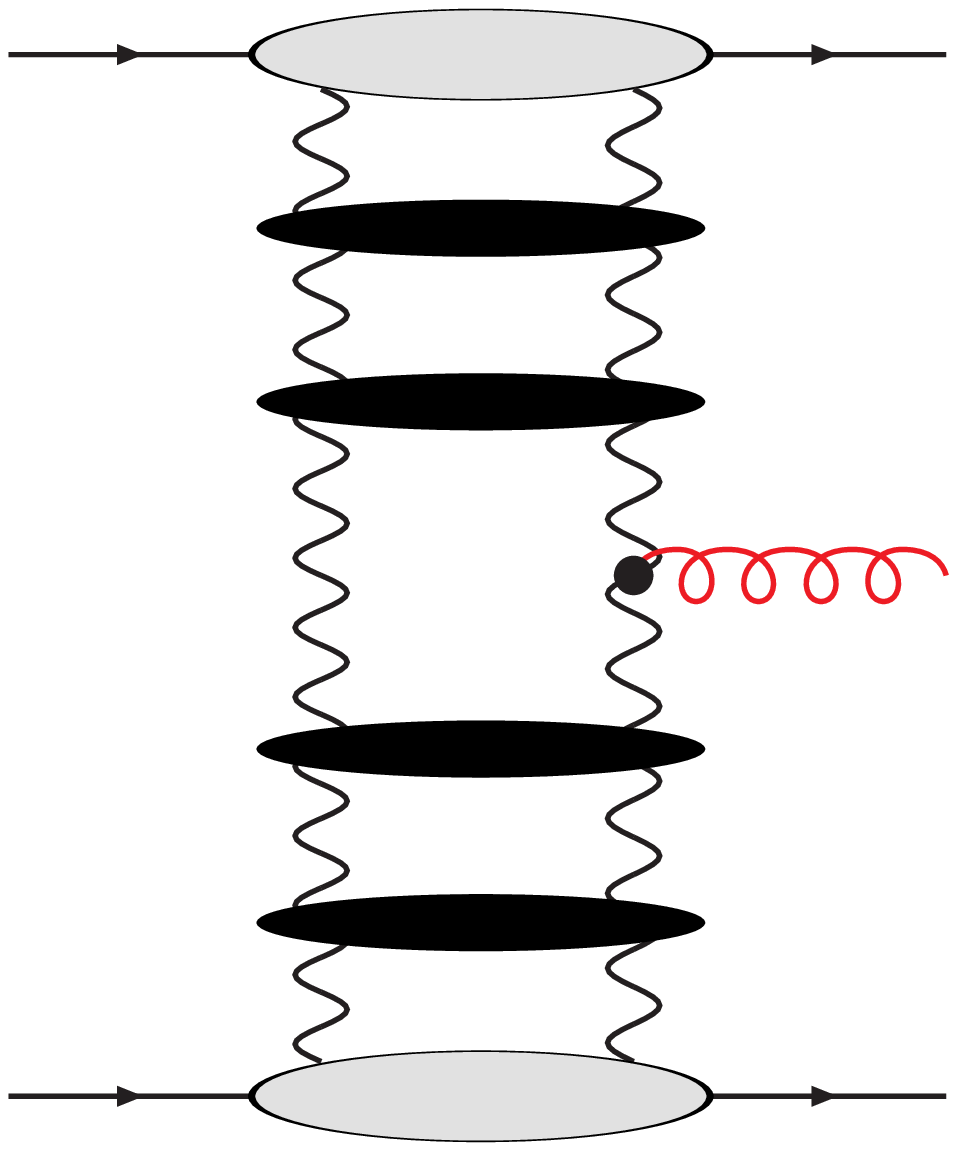}}
\parbox{4cm}{\center \includegraphics[width=3cm]{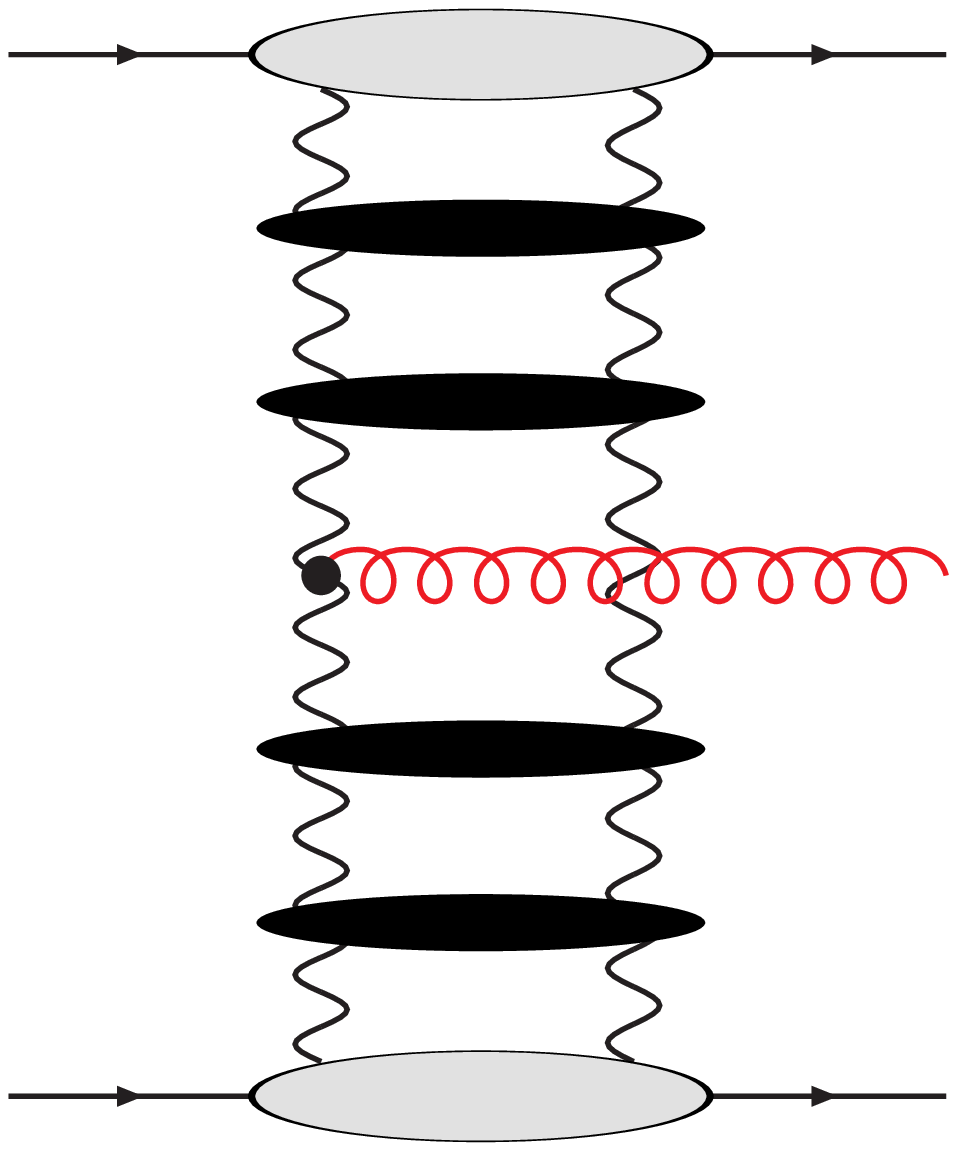}}
\parbox{4cm}{\center \includegraphics[width=3cm]{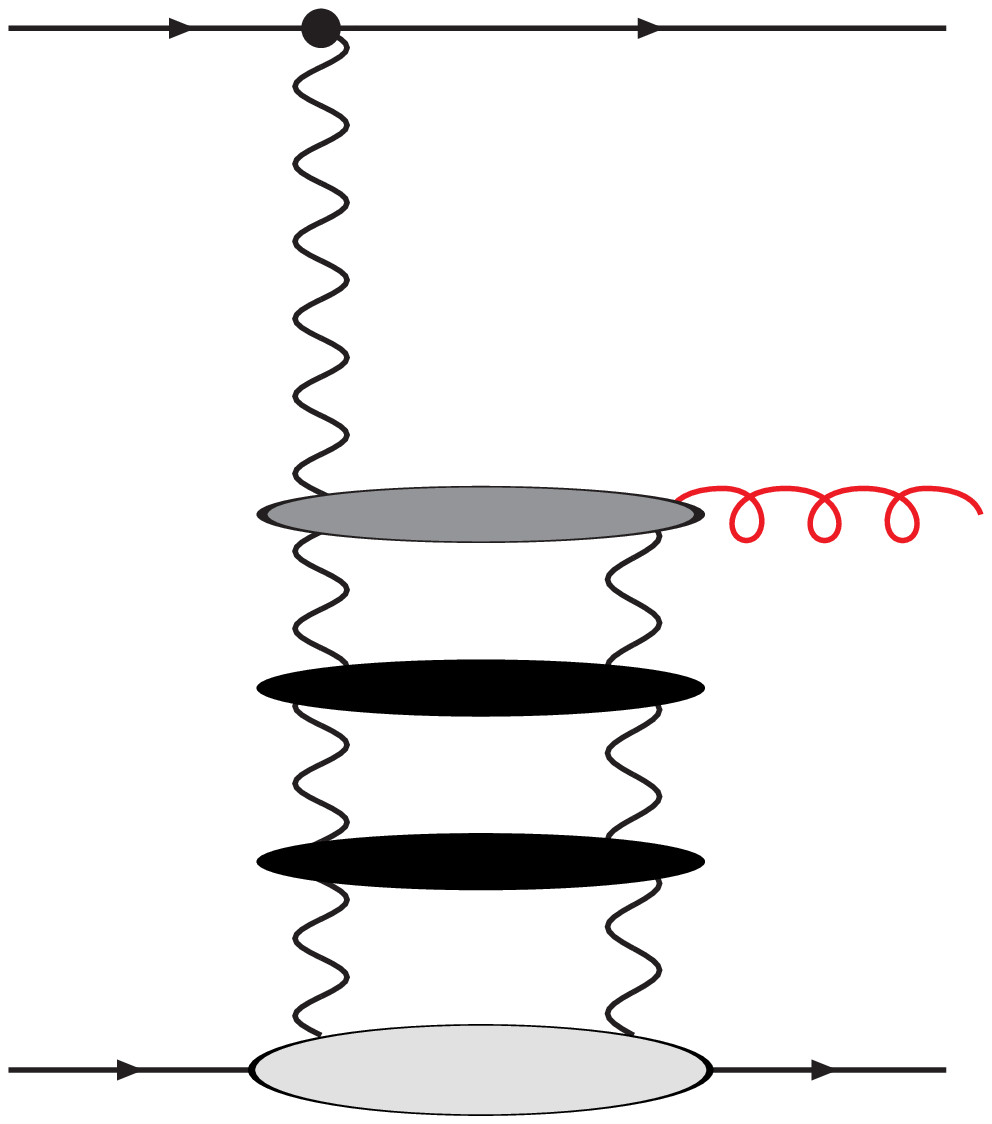}} \\
\parbox{4cm}{\center (a)}\parbox{4cm}{\center (b)}\parbox{4cm}{\center (c)}
  \caption{\small $2\to3$ production amplitude with positive ($\tau_1 = \tau_2 = + 1$) and mixed ($\tau_1 = -1$, $\tau_2 = +1$). The grey blobs on on the top on the bottom of the diagram denote the quark-impact factors, whereas interaction between reggeized gluons is described by the BFKL-Kernel. In (c) a new vertex appears: The Reggeon-Particle-Reggeon-Reggeon (RP2R) vertex.  }
  \label{fig:mixed23}
\end{figure}

The different contributions that can occur for the $2 \to 3$
production amplitude are summarized in Fig.\ref{fig:mixed23}. In the
case where both $t$-channels carry positive signature, ($\tau_1 =
\tau_2 = + 1$ ), Fig.\ref{fig:mixed23}a and Fig.\ref{fig:mixed23}b,
interaction in both $t$-channels is mediated by a  state of two
reggeized gluons, which is described by the BFKL-equation. The
production of the gluon at central rapidities is described on the other
hand by the leading order production vertex
Eq.(\ref{eq:1lipatov_repeat}).
The third graph in Fig.\ref{fig:mixed23}c corresponds to mixed
signature, where the $t_1$ channel carries negative signature ($\tau_1
= -1$) and the $t_2$-channel positive signature ($\tau_2 = +1$). The
case where $\tau_1 = +1$ and $\tau_2 = -1 $ exists as well and follows
from the above case by symmetry. For mixed symmetry, a new element occurs, the Reggeon-Particle-2 Reggeon vertex ((RP2R)-vertex), which we will derive from the effective action in Sec.\ref{sec:RPRR}. 
\begin{figure}[htbp]
  \centering
  \parbox{3.5cm}{\center \includegraphics[width=3cm]{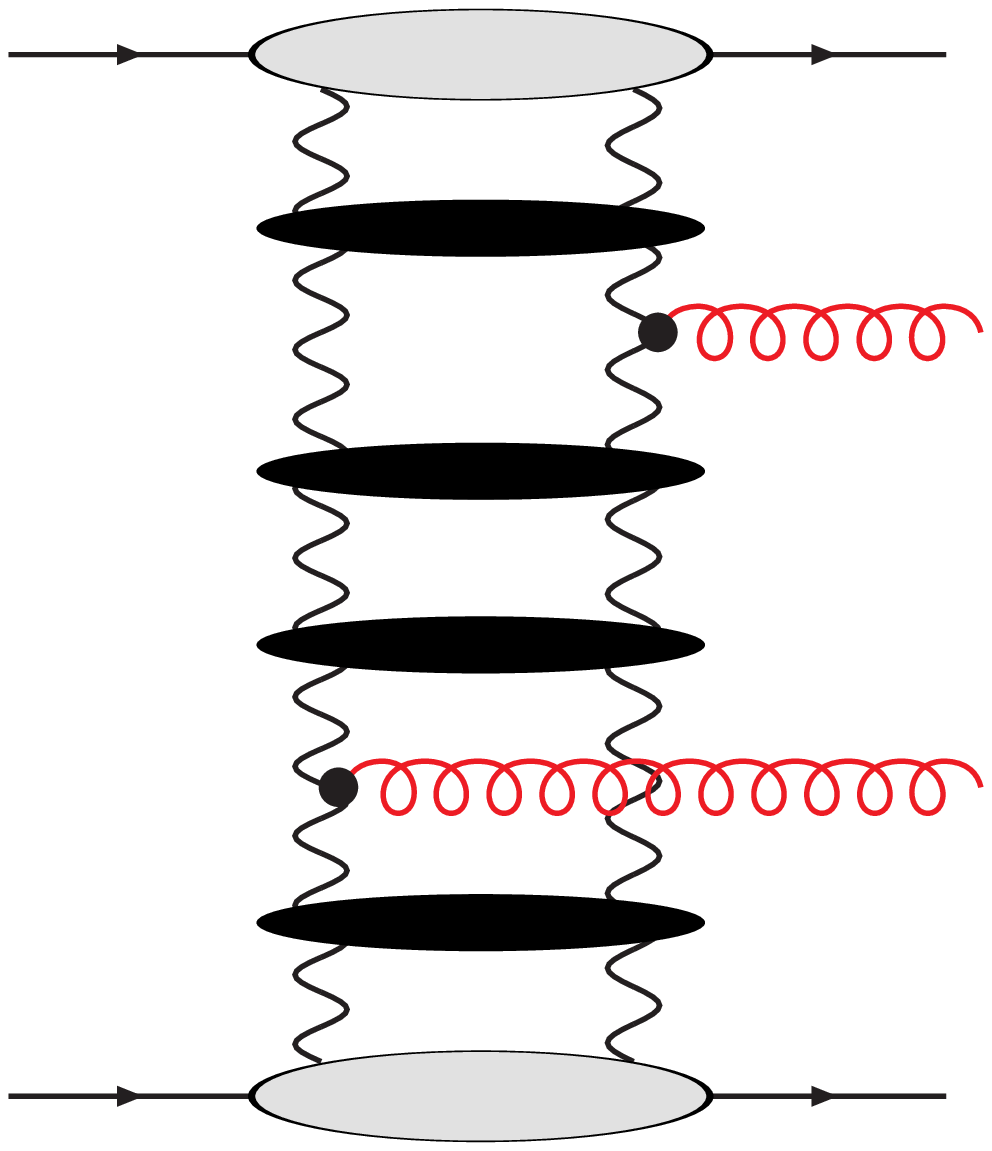}}
  \parbox{3.5cm}{\center \includegraphics[width=3cm]{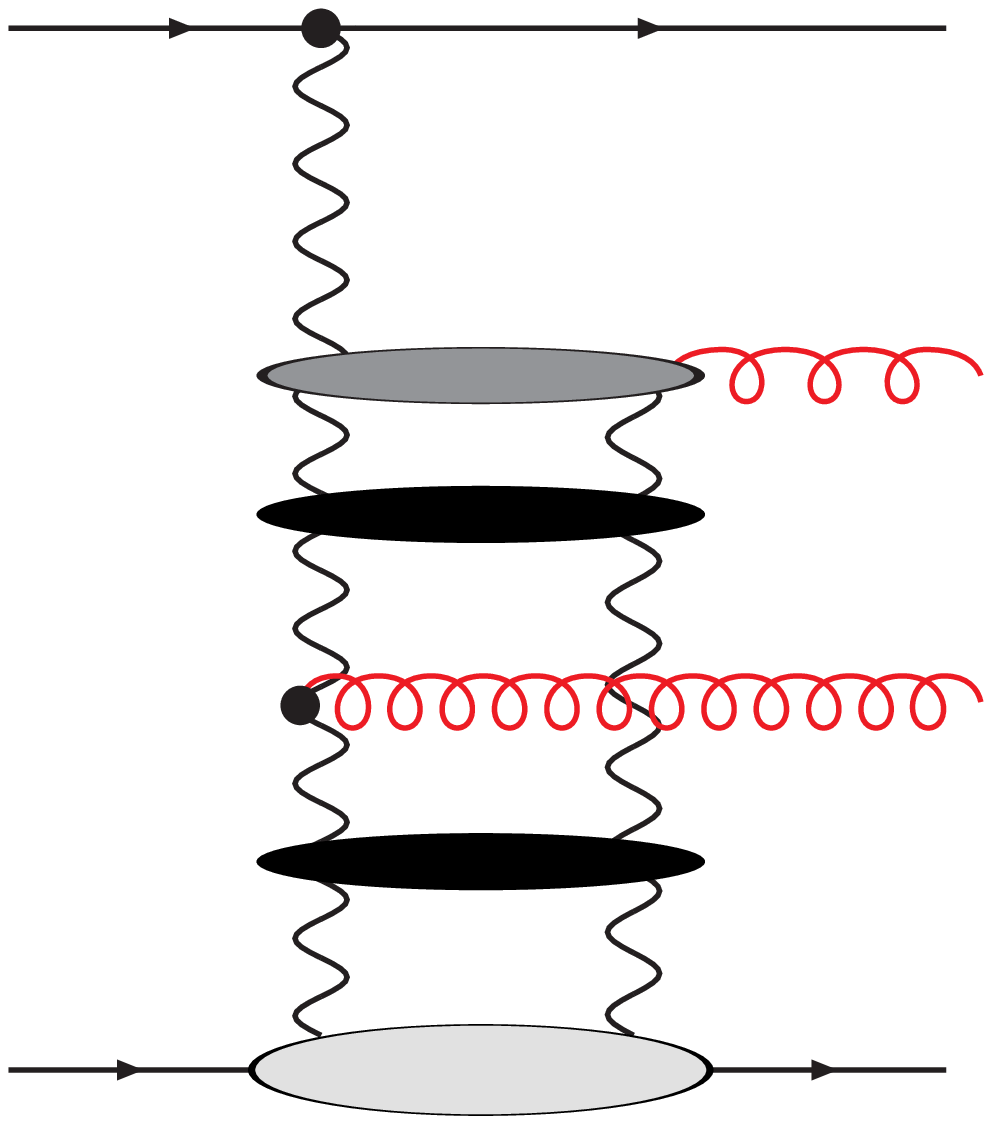}}
  \parbox{3.5cm}{\center \includegraphics[width=3cm]{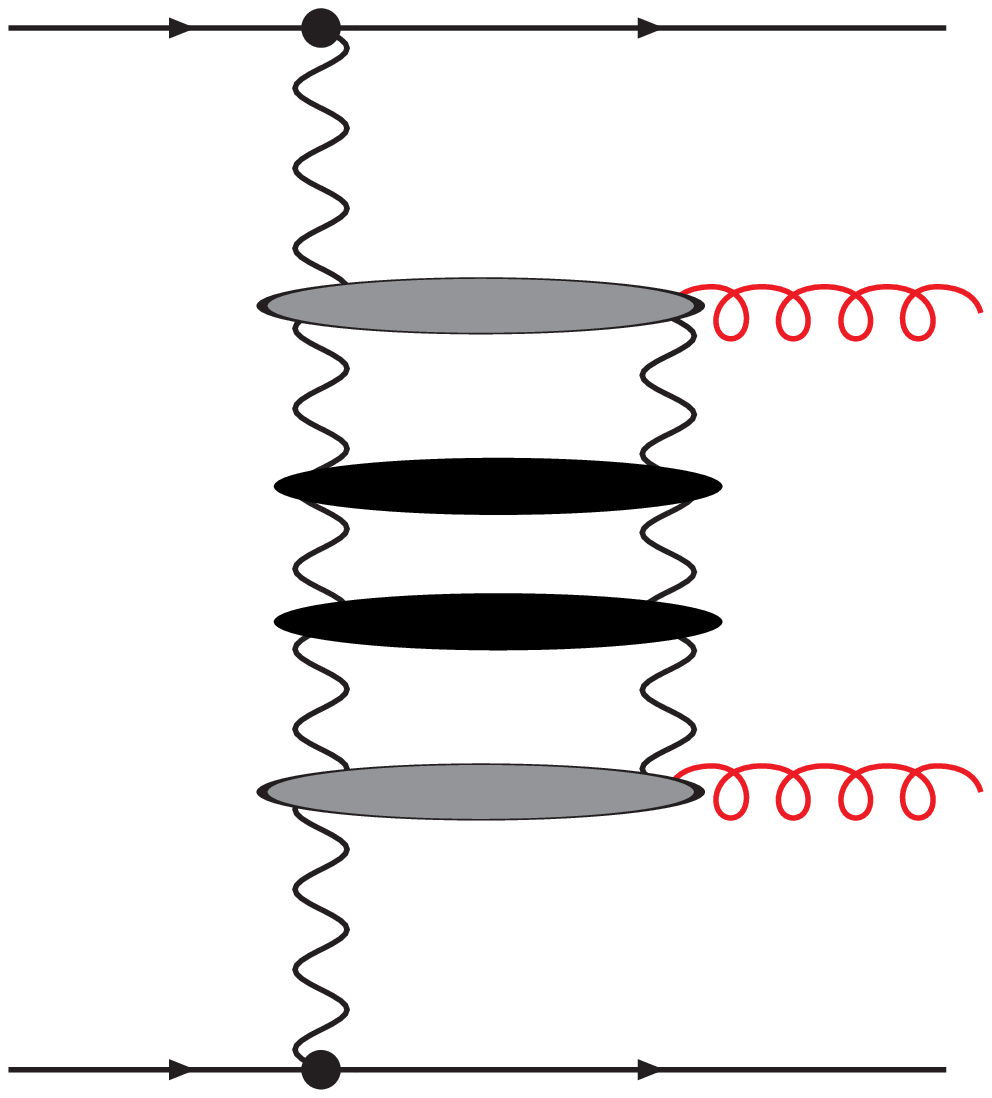}}
 \\
\parbox{3.5cm}{\center (a)}\parbox{3.5cm}{\center (b)}
\parbox{3.5cm}{\center (c)}
  \caption{\small Contributions to the $2\to 4$ production amplitude with positive (a) and mixed (b,c,d) signature. Explicitly we have the following signature configurations a: $(+,+,+)$, b: $(-, +, +)$, c: $(-, +, -)$ 
}
  \label{fig:mixed24}
\end{figure}
Knowledge of this vertex is then also sufficient to build all positive
and mixed signatured $2 \to 4$ production amplitudes,
Fig.\ref{fig:mixed24}, apart from the combination $(+,-,+)$, which
explicitly requires the exchange of three reggeized gluons, which has
not been addressed so far. Apart from that it is sub-leading compared
to the other signature configurations and we will not include it in
our present analysis.

\subsection{The Reggeon-Particle-2Reggeon (RP2R)-vertex}
\label{sec:RPRR}

In the following we will discuss the derivation of the 
Reggeon-Particle-2Reggeon (RP2R)-vertex from the effective
action. To do so, we will apply the rules for the interactions of  reggeized gluons of  Sec.\ref{sec:tworeggeon_negsig}.

We further   that the derivation of the RP2R-vertex from the effective
action was also addressed in \cite{Braun:2006sk}. During their
derivation, the authors of \cite{Braun:2006sk} find a certain
ambiguity concerning the evaluation of longitudinal integrals. 
In the following we demonstrate that such an ambiguity does not occur, if one applies the rules derived in Sec.\ref{sec:tworeggeon_negsig}.
In contrast to our general analysis, which is based on covariant
Feynman gauge, the authors of \cite{Braun:2006sk} use in their
analysis the  light-cone gauge, $V_+ = 0$, which makes it difficult to
compare directly between the two calculations. Nevertheless we believe
that the problem of \cite{Braun:2006sk} arises due to a
misinterpretation of the Feynman-rules of the effective action and due
to the absence of subtraction terms, as introduced in
Sec.\ref{sec:tworeggeon_negsig} and not due to  a peculiarity of the
light-cone gauge.

As for the correction to the production amplitude in Sec.\ref{sec:--},
we will use in the following the $2 \to 3$ production amplitude,
Fig.\ref{fig:mixed23}c, as a reference process.  There all corrections
above the RP2R-vertex are resummed within the LLA by the upper
reggeized gluon, whereas below such resummation is performed by the
BFKL-Green's function.  We factorize this amplitude in the following
way
\begin{align}
  \label{prodamp_mixedsig0}
  i\mathcal{M}^{(-,+)}_{2 \to 3} =   p_A^+ |p_B^-|  \int \frac{d\omega_1}{2\pi i} \int \frac{d\omega_2}{2\pi i}
\frac{1}{2}\left[ (-s_2)^{\omega_2}   + s_2^{\omega_2}\right]   
& \,\Gamma^a_{QQR}( q_1) \times      \frac{1}{\omega_1 - \beta({\bm q}^2_1)}  \frac{i/2}{{\bm{q}}_1^2} 
  \notag \\
 \times  \int \frac{d^2 {\bm k}}{(2\pi)^2} \Gamma^{acb_1b_2, \mu}_{RP2R} (\omega_1; p_A^+, q_1^+;  {\bm k}, {\bm q}_2 \!- \!{\bm k})
& 
\times \frac{1 }{{\bm k}^2 ({\bm q}_2 - {\bm k})^2}\times  A^{b_1b_2} (\omega_2;  {\bm k}, {\bm q}_2 - {\bm k} ),
  \end{align}
with
\begin{align}
  \label{eq:AGbfkl}
A^{b_1b_2}_2 (\omega_2;  {\bm k}, {\bm q}_2 - {\bm k} ) = G_{\text{BFKL}}^{b_1b_2; b_{1}'b_2'}  ( \omega_2;  {\bm k}, {\bm q}_2 - {\bm k} ;  {\bm l}, {\bm q}_2 - {\bm l} )   ) \otimes_{\bm l}
  A_{(2,0)}^{b_1'b_2'}( {\bm l}, {\bm q}_2 - {\bm k} )
\end{align}
where $G_{\text{BFKL}}$ is the BFKL-Green's function Eq.~(\ref{eq:bfkl}).

\begin{figure}[htbp]
  \centering
  \parbox{2.5cm}{\center \includegraphics[width=2cm]{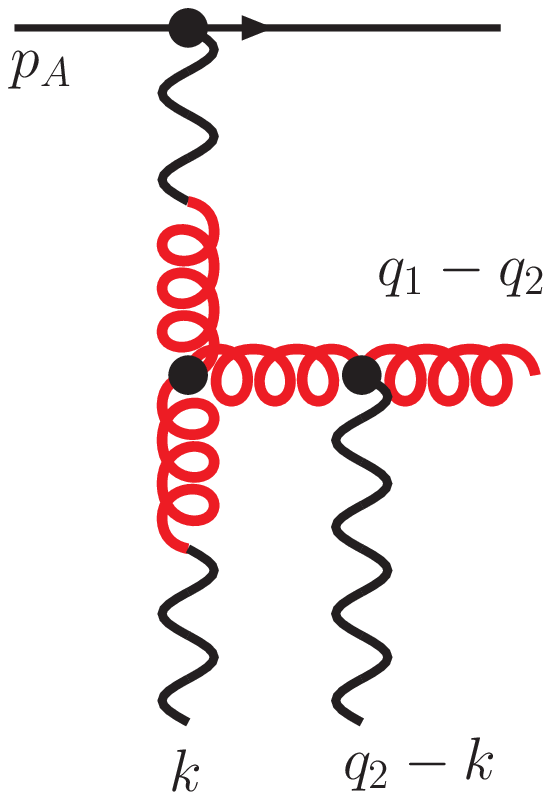}}
  \parbox{2.5cm}{\center \includegraphics[width=2cm]{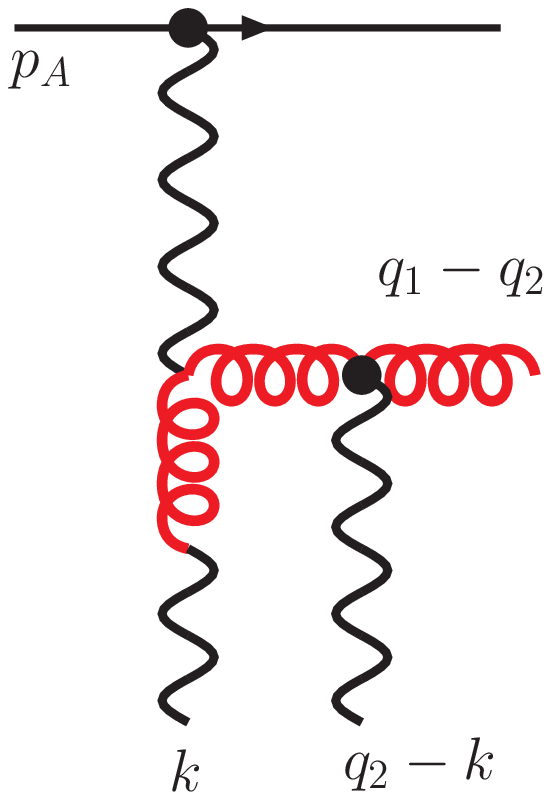}}
  \parbox{2.5cm}{\center \includegraphics[width=2cm]{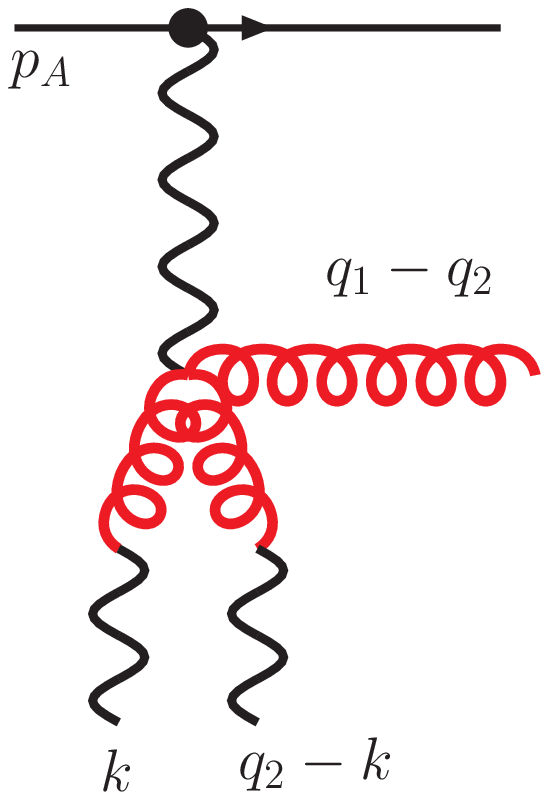}}
  \parbox{2.5cm}{\center \includegraphics[width=2cm]{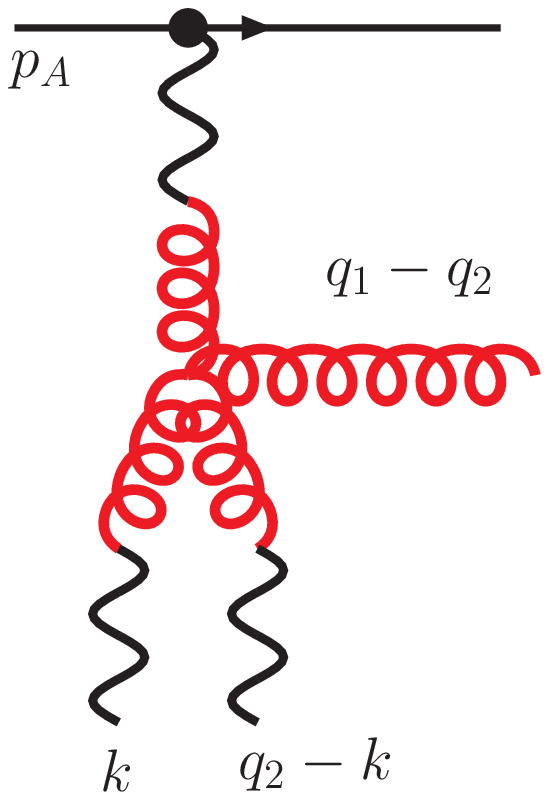}}
  \parbox{2.5cm}{\center \includegraphics[width=2cm]{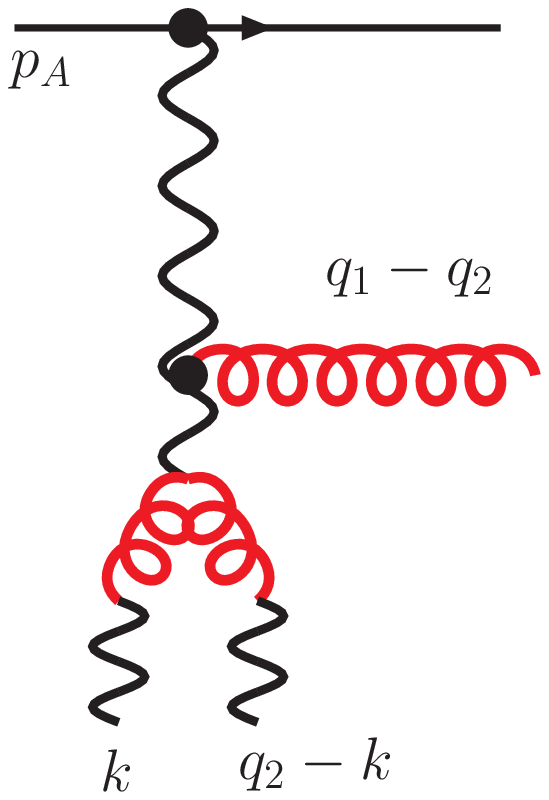}} \\
  \parbox{2.5cm}{\center (a)} 
  \parbox{2.5cm}{\center (b)} 
  \parbox{2.5cm}{\center (c)} 
  \parbox{2.5cm}{\center (d)}  
  \parbox{2.5cm}{\center (e)} 
  \caption{\small Different kinds of effective Feynman-graphs, that contribute to the RP2R-vertex. The last graph is a subtraction graph, that subtracts the logarithmic singularity of the first graph.}
  \label{fig:rprr}
\end{figure}

As usually for loops containing two or more reggeized gluons, the
longitudinal integration factorizes and it is sufficient to consider
the diagrams of Fig.\ref{fig:rprr}.
Apart from the diagrams that are depicted in Fig.\ref{fig:rprr}, there
occurs also the combination where in Fig.\ref{fig:rprr}a and
Fig.\ref{fig:rprr}b the lower two reggeized gluons are interchanged.
Fig.\ref{fig:rprr}e is a subtraction graph and comes with a relative
minus sign. The necessity of including such a subtraction graph is
motivated similar to Sec.\ref{sec:tworeggeon_negsig}: There, for both
the quark-impact factor and the BFKL-interaction-kernel a subtraction
graph has been introduced to resolve a counting problem with the
reggeized gluon and to make the integral over light-cone momenta
finite.  For exactly the same reason, one also requires a subtraction
graph for the RP2R-vertex.

To evaluate the integration over longitudinal momenta it turns out to
be advisable not to consider individual diagrams, but to group the sum
of all diagrams first according to their color- and then according to
their transverse momentum structure. In this way, already some of the
singularities cancel by themselves. Furthermore the transverse
momentum structure can be classified according to their dependence on
transverse reggeized gluon momenta. Generally one finds that terms
that depend only on the sum of the two momenta are less convergent.
This is to be expected, as it is this class of contributions that
initially gave rise to the induced vertices. As a consequence, we
expect these divergences to be removed by a subtraction term, like
Fig.~\ref{fig:rprr}e.

In the present case, we also drop all terms that vanish for production
of a real particle i.e.  that are either proportional to $(q_1 -
q_2)^2 = 0$ or to $(q_1 - q_2)^\mu$ and therefore orthogonal to
physical polarizations. The transverse momentum structure reduces for
the unintegrated RP2R-vertex to two classes of terms: For the first
class the numerator depends only on the sum of the transverse momenta
of both reggeized gluons,
\begin{align}
   \label{rprr_raw1}
 (T^{b_1}T^{b_2})_{ac}  {G}_{(A)}^{\mu}  &=  ig^2    
  (T^{b_1}T^{b_2})_{ac}  
  \frac{q_1^+  }{-q_1^+ k^- - ({\bm{q}}_1 - {\bm{k}})^2 + i\epsilon}  
   C_\mu(q_1, q_2).
  \end{align}
For the second class, the numerator explicitly depends on the  momenta of the individual  reggeized gluons, ${\bm k}$ and ${\bm q}_2 - {\bm k}$,
\begin{align}
  \label{rprr_raw2}
(T^{b_1}T^{b_2})_{ac}  {G}_{(B)}^{\mu}  &=  ig^2    
 (T^{b_1}T^{b_2})_{ac}  
 \frac{ {\bm q}_1^2}{-k^-}  \frac{C_\mu(q_1 - k, q_2 - k)  }{-q_1^+ k^- - ({\bm{q}}_1 - {\bm{k}})^2 + i\epsilon} ,
 \end{align} 
with
\begin{align}
  \label{eq:Cmu}
C_\mu(q_1 - k, q_2 - k) = 
(\frac{q_1^+}{2} + \frac{({\bm{q}}_1 - {\bm{k}})^2}{q_2^-}) (n^-)^\mu
+
(\frac{q_2^-}{2} + \frac{({\bm{q}}_2 - {\bm{k}})^2}{q_1^+})(n^+)^\mu 
- (q_1 + q_2 - 2k)_\perp^\mu,
\end{align}
not dependent on  $k^-$.  For each of the above
terms there exists also a corresponding 'crossed' term which is
obtained by simultaneously interchanging momenta and color labels of
the two lower reggeized gluons. 

Attempting to integrate both classes over $k^-$, we find that the
second class, Eq.~(\ref{rprr_raw2}), yields a convergent integrand,
where the first class does not.  Concentrating at first on the second
class, one realizes that it comes with an additional pole in $k^-$,
compared to the first class. Even though by grouping sums of diagrams,
the relation to individual diagrams is less pronounced, it is
nevertheless clear, that this pole is due to an induced vertex and
needs to be treated with the corresponding prescription.

To evaluate the $k^-$-integral over Eq.~(\ref{rprr_raw2}) , we further
add the Reggeon factor  of the upper reggeized gluon and obtain
the following integral
\begin{align}
  \label{rprr_raw22}
 (T^{b_1}T^{b_2})_{ac}  {\Gamma}_{(B)}^{\mu}   =   ig^2    
 (T^{b_1}T^{b_2})_{ac}   \int \frac{d k^-}{2\pi} & \bigg[ 
 \frac{1}{2}\left[ \left( \frac{ p_A^+k^-\!\! -\!i\epsilon}{\Lambda_a} \right)^{\omega_1}
        + \left(\frac{ -p_A^+k^-\!\! -\!i\epsilon}{\Lambda_a} \right)^{\omega_1} \right]
\notag \\
&  \frac{ {\bm q}_1^2}{-k^-}  \frac{C_\mu(q_1 - k, q_2 - k)  }{-q_1^+ k^- - ({\bm{q}}_1 - {\bm{k}})^2 + i\epsilon} ,
 \end{align} 
where the $i\epsilon$-prescription of the pole in $k^-$ is meant to coincide with the $i\epsilon$-prescription of the branch-cut it is multiplied with. 
The integral is easily evaluated and we find
\begin{align}
  \label{eq:rprr22_evalf}
 (T^{b_1}T^{b_2})_{ac}  {\Gamma}_{(B)}^{\mu}  =   \frac{g^2}{2}(T^{b_1} T^{b_2})_{ac} \frac{q_1^+}{|q_1^+|}\left| \frac{p_A^+}{q_1^+} \right|^{\omega_1}  
           \left( \frac{({\bm q}_1 - {\bm k})^2}{\Lambda_a}  \right)^{\omega} 
           \frac{{\bm{q}}_1^2}{(\bm{q}_1\! -\! {\bm{k}})^2}  C_\mu(q_1\! -\! k, q_2\! -\! k) ,
\end{align}
with $\Lambda_a \simeq ({\bm q}_1 - {\bm k})^2$ due to the LLA.  Next
we turn to the symmetric term, Eq.(\ref{rprr_raw1}). Unlike Eq.(\ref{rprr_raw2}) it provides at first not  a convergent integrand. On the other hand,
this term has the same numerator as the subtraction term and we can
expect that after subtraction of Fig.~\ref{fig:rprr}e we will obtain a
convergent integrand. To this end we add and subtract a term with the
same pole structure as the subtraction graph and obtain for
Eq.~(\ref{rprr_raw1})
\begin{align}
   \label{rprr_raw1_+-}
 (T^{b_1}T^{b_2})_{ac}  {G}_{(A)}^{\mu}  &=  ig^2    
  (T^{b_1}T^{b_2})_{ac}  \bigg[
   \frac{ C_\mu(q_1, q_2)}{2}\left(  \frac{q_1^+ }{-q_1^+k^- + i\epsilon} -  \frac{q_1^+ }{q_1^+k^- + i\epsilon}  \right)
\notag \\
& +
\frac{ ({\bm{q}}_1 - {\bm{k}})^2   C_\mu(q_1, q_2) }{-q_1^+ k^- - ({\bm{q}}_1 - {\bm{k}})^2 + i\epsilon}    
\frac{1}{2} \left(  \frac{q_1^+ }{-q_1^+k^- + i\epsilon} -  \frac{q_1^+ }{q_1^+k^- + i\epsilon}  \right) 
  \bigg] .
  \end{align}
  Apart from the above contribution, we also have the contribution,
  where the two lower reggeized gluons are interchanged 
  and which takes the following
  form:
\begin{align}
   \label{rprr_raw1_+-X}
& (T^{b_2}T^{b_1})_{ac}  {G}_{(AX)}^{\mu}  =  ig^2    
  (T^{b_2}T^{b_1})_{ac}  \bigg[
  \frac{ C_\mu(q_1, q_2)}{2}\left(  \frac{q_1^+ }{q_1^+(k^--q_2^-) + i\epsilon} - \frac{q_1^+ }{q_1^+(q_2^- - k^-) + i\epsilon}   \right)
\notag \\
 & + 
 \frac{ ({\bm{q}}_1 - {\bm{k}})^2   C_\mu(q_1, q_2) }{q_1^+( k^- - q_2^-) - ({\bm{q}}_1 - {\bm q})2 +{\bm{k}})^2 + i\epsilon}    
\frac{1}{2} \left(  \frac{q_1^+ }{q_1^+(k^--q_2^-) + i\epsilon} - \frac{q_1^+ }{q_1^+(q_2^- - k^-) + i\epsilon}  \right) 
  \bigg] .
  \end{align}
  The two terms in the first line of Eq.~(\ref{rprr_raw1_+-}) and
  Eq.~(\ref{rprr_raw1_+-X}) agree  with each other in their
  momentum part, up to the factor $q_2^-$ in the denominator. For the integral over $k^-$
  this factor can be removed by a corresponding shift in the
  integration variable or, alternatively, by adding and  subtracting
  a corresponding term which does \emph{not} contain this factor
  $q_2^-$. One can then  verify that the difference between the terms
  with and without $q_2^-$ is convergent and vanishes, once it is
  integrated over $k^-$. After those manipulations, the momentum
  factors in the first lines of Eq.~(\ref{rprr_raw1_+-}) and
  Eq.~(\ref{rprr_raw1_+-X}) agree exactly, while they come with a
  relative minus sign and different ordering of color factors. For their
  sum, the color factor reduces  to the commutator and the sum 
  agrees in both  color- and in momentum- structure with the
  subtraction graph, Fig.~\ref{fig:rprr}e. As a consequence, the sum
  of the first lines of the rhs of Eq.~(\ref{rprr_raw1_+-}) and
  Eq.~(\ref{rprr_raw1_+-X}) is canceled by the subtraction graph and
  we are left with the contributions in the second line of
  Eq.~(\ref{rprr_raw1_+-}) and Eq.~(\ref{rprr_raw1_+-X}) respectively,
  
  For their integral over $k^-$, we then restrict to
  Eq.~(\ref{rprr_raw1_+-}) alone. Together with the Reggeon-factor of the upper reggeized gluon we find:
\begin{align}
  \label{rprr_raw11}
 (T^{b_1}T^{b_2})_{ac}  {\Gamma}_{(A)}^{\mu}   =   ig^2    
 (T^{b_1}T^{b_2})_{ac}   \int \frac{d k^-}{2\pi} & 
 \frac{1}{2}\left[ \left( \frac{ p_A^+k^-\!\! -\!i\epsilon}{\Lambda_a} \right)^{\omega_1}
        + \left(\frac{ -p_A^+ k^-\!\! -\!i\epsilon}{ \Lambda_a} \right)^{\omega_1} \right]
\notag \\
&     \frac{q_1^+ }{-q_1^+k^- }  \frac{ ({\bm{q}}_1 - {\bm{k}})^2   C_\mu(q_1, q_2) }{-q_1^+ k^- - ({\bm{q}}_1 - {\bm{k}})^2 + i\epsilon}.    
 \end{align} 
 Note that the $i\epsilon$-prescription of the above
 '$\omega$'-factors and of the poles in Eq.~(\ref{rprr_raw1_+-})
 coincides, as for the sum of the poles the dependence on $q_1^+$ is
 absent.  The integral is easily evaluated and yields 
\begin{align}
  \label{eq:rprr22_evalf}
 (T^{b_1}T^{b_2})_{ac}  {\Gamma}_{(A)}^{\mu}  =  - \frac{g^2}{2}(T^{b_1}, T^{b_2})_{ac} \frac{q_1^+}{|q_1^+|}\left| \frac{p_A^+}{q_1^+} \right|^{\omega_1}  
                       C_\mu(q_1 , q_2 ) ,
\end{align}
where we already approximated $\Lambda_a \simeq ({\bm q}_1 - {\bm
  k})^2$ in accordance with the LLA.  To obtain the complete
RP2R-vertex, we  add up our results Eq.~(\ref{eq:rprr22_evalf})
and Eq.~(\ref{eq:rprr22_evalf}) and complete them by supplementing
their counterparts, obtained from exchanging color and momenta of the
two lower reggeized gluons. Altogether we obtain
\begin{align}
   \label{eq:Gmma_gesamt}
 \Gamma^{acb_1b_2, \mu}_{RP2R}& (\omega_1; p_A^+, q_1^+;  {\bm k}, {\bm q}_2 \!- \!{\bm k}) = 
   -\frac{q_1^+}{|q_1^+|} \left| \frac{p_A^+}{q_1^+} \right|^{\omega_1}       
 \tilde{\Gamma}^{ac_1b_1b_2, \mu}_{RP2R}(q_1, q_2,  {\bm k}).
 \end{align}
From the point of view of Reggeon-field theory,
$\tilde{\Gamma}_{RP2R}$ constitutes the proper RP2R-vertex, as it does
not contain any dependence on $s$-channel energy-variables.
$\tilde{\Gamma}_{RP2R}$ can then be written as the sum of two terms,  symmetric ($\Gamma^S$) and antisymmetric  ($\Gamma^A$) under the exchange of the lower reggeized gluons,
\begin{align}
  \label{eq:gamma_tilde_farbe}
\tilde{\Gamma}^{ac_1b_1b_2, \mu}_{RP2R}(q_1, q_2,  {\bm k}) =   \left[ \{ T^{b_1}, T^{b_2}\}_{ac} \tilde{\Gamma}^{S, \mu}(q_1, q_2,  {\bm k}) +  [ T^{b_1}, T^{b_2}]_{ac_1} \tilde{\Gamma}^{A, \mu}(q_1, q_2,  {\bm k})
\right],
\end{align}
where 
\begin{align}
  \label{eq:gamma_S}
\tilde{\Gamma}_{RP2R}^{S;\mu} ( q_1, q_2, {\bm k} ) = {g^2} \bigg[  C_\mu(q_1 , q_2 )  
-& 
     \frac{1}{2}\frac{{\bm{q}}_1^2}{(\bm{q}_1\! -\! {\bm{k}})^2}  C_\mu(q_1\! -\! k, q_2\! -\! k)  
\notag \\
 &- 
 \frac{1}{2}\frac{{\bm{q}}_1^2}{(\bm{q}_1\! -\!{\bm q_2}\!+\! {\bm{k}})^2}  C_\mu(q_1\! -\! q_2\! +\! k, \! k)
\bigg]
\end{align}
and
\begin{align}
  \label{eq:gamma_A}
\tilde{\Gamma}_{RP2R}^{A;\mu} ( q_1, q_2, {\bm k} ) = -{g^2} \bigg[  
     \frac{1}{2}\frac{{\bm{q}}_1^2}{(\bm{q}_1\! -\! {\bm{k}})^2}  C_\mu(q_1\! -\! k, q_2\! -\! k)  
 - \frac{1}{2}\frac{{\bm{q}}_1^2}{(\bm{q}_1\! -\!{\bm q_2}\!+\! {\bm{k}})^2}  C_\mu(q_1\! -\! q_2\! +\! k, \! k)
\bigg].
\end{align}
Note that the latter term gives a first example of a  coupling of a QCD-particle to 
a  state of two reggeized gluons, which is antisymmetric under
the combined exchange of color and momenta.


\subsection{Mixed and positive signature for the $2\to3$ amplitude}
\label{sec:mixed23}
\begin{figure}[htbp]
  \centering
  \parbox{4cm}{\center \includegraphics[width=3cm]{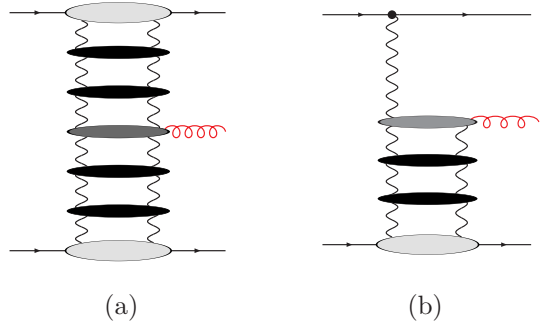}}
\parbox{4cm}{\center \includegraphics[width=3cm]{mixed23.eps}} \\
\parbox{4cm}{\center (a)}\parbox{4cm}{\center (b)}
  \caption{\small Diagrams corresponding to the  $2\to3$ production amplitude with mixed ($\tau_1 = -1$, $\tau_2 = +1$) and  positive ($\tau_1 = \tau_2 = + 1$)  signature. Making use of the diagrammatic rules  described in the text, these  diagrams allow for the derivation of the corresponding partial waves. }
  \label{fig:mixed23_m}
\end{figure}

With the results of the previous paragraph, we have now all
ingredients  to construct the various $2\to 3$ production
amplitudes with mixed and positive signature of Fig.\ref{fig:mixed23}.
We start with the mixed signature case $(\tau_1 = -, \tau_2 = +)$
whereas the case $(\tau_1 = +, \tau_2 = -)$ follows  by
symmetry. Inserting Eq.~(\ref{eq:Gmma_gesamt}) into Eq.~(\ref{prodamp_mixedsig0}) we obtain for the $2\to 3$ production amplitude the following expression:
\begin{align}
  \label{prodamp_mixedsig1}
 i \mathcal{M}^{(-,+)}_{2 \to 3} =  2\pi s \frac{s_2}{|s_2|} \int \frac{d\omega_1}{2\pi i} \int \frac{d\omega_2}{2\pi i} \left|{s}\right|^{\omega_1}|s_2|^{\omega_2 -\omega_1} \frac{1}{2} \left(e^{-i\pi\omega_2} + 1 \right)
V_{(-,+)}(\omega_1, \omega_2; {\bm q}_1, {\bm q}_2)
  \end{align}
with
\begin{align}
  \label{eq:V_2}
V_{(-,+)} (\omega_1, \omega_2; {\bm q}_1, {\bm q}_2) &
=   
gt^a \times      \frac{1}{\omega_1 - \beta({\bm q}^2_1)}  \frac{1}{t_1}  \times   \tilde{\Gamma}^{ac_1b_1b_2, \mu}_{RP2R}(q_1, q_2,  {\bm l}) \epsilon_\mu(q_1 - q_2)
  \notag \\
 &   \otimes_{\bm l}   G_{\text{BFKL}}^{b_1b_2; b_{1}'b_2'}  ( \omega_2;  {\bm l}, {\bm q}_2 - {\bm l} ;  {\bm k}, {\bm q}_2 - {\bm k} )    \otimes_{\bm k}
  A_{(2,0)}^{b_1'b_2'}( {\bm k}, {\bm q}_2 - {\bm k} )
%
  \end{align}
  For the partial wave\footnote{Note that the partial waves appearing in the following all carry a signature dependence unlike the partial waves in the analytical representations. It is due to the fact that within the effective action sometimes two or more partial waves of the analytical representations are combined, which makes the final result different for every signature configuration}  $V_{(-,+)}$ we note that it can be
  obtained directly from Fig.\ref{fig:mixed23_m}b, by  applying a simple set of
  diagrammatic rules,  which we present in the following. It should be  noted that these are \emph{not} Feynman-rules of the
  effective action, but serve only to construct partial waves given the factorization Eq.(\ref{prodamp_mixedsig1}) in energy-factors and partial wave. 
In the diagrammatic language, we have for the coupling of  the external quarks to the  single and the  state of reggeized gluons respectively:
  \begin{align}
    \label{eq:qqr}
\parbox{2cm}{\includegraphics[height=1cm]{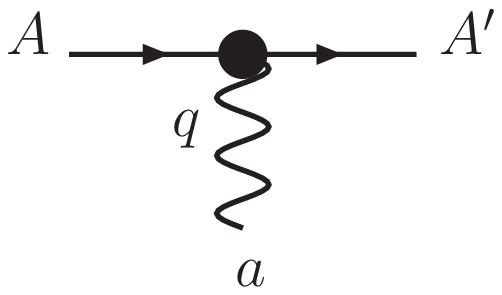}} &= gt^a_{AA'}
&
\parbox{2.5cm}{\includegraphics[height=1cm]{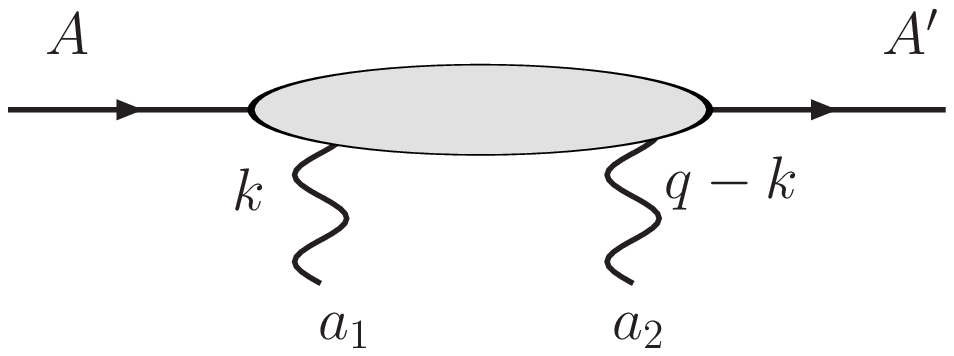}} &=  A_{(2,0)}^{a_1a_2}( {\bm k}, {\bm q} - {\bm k} )
  \end{align}
Propagation of the  single reggeized gluon and  the  state of two reggeized gluon corresponds to
\begin{align}
    \label{eq:regg}
\parbox{1cm}{\includegraphics[height=2cm]{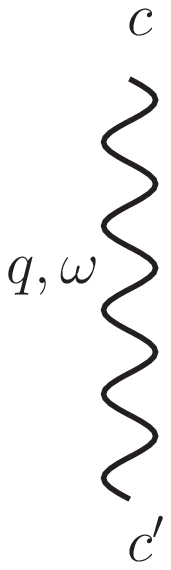}} &=   \frac{1}{\omega - \beta({\bm q}^2)}  \frac{1}{-{\bm{q}}^2} 
&
\parbox{2cm}{\includegraphics[height=2cm]{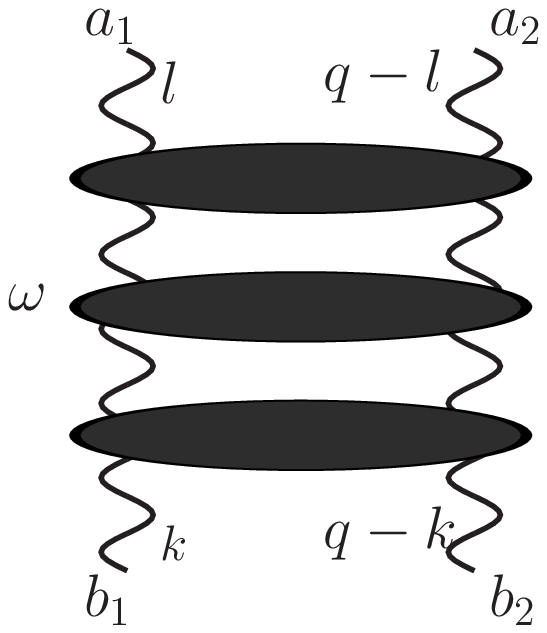}} &=   G_{\text{BFKL}}^{a_1a_2; b_{1}b_2}( \omega;  {\bm l}, {\bm q} - {\bm l} ;  {\bm k}, {\bm q} - {\bm k} )  .
  \end{align}
Further the RP2R-vertex is given by
\begin{align}
    \label{eq:rprr}
\parbox{3cm}{\includegraphics[width=3cm]{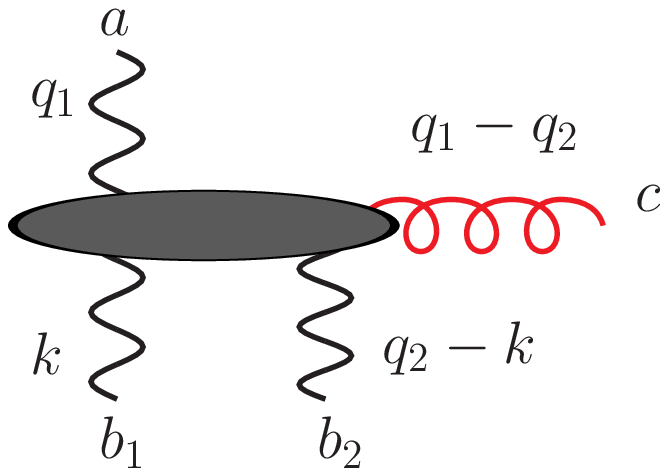}} =  \tilde{\Gamma}^{ac_1b_1b_2, \mu}_{RP2R}(q_1, q_2,  {\bm l}) \epsilon_\mu(q_1 - q_2)
  \end{align}
and  every loop of two reggeized gluons requires the following  convolution symbol (already implicitly contained inside the BFKL-Green's function)
\begin{align}
  \label{eq:conv}
\otimes_{\bm k} = \int \frac{d^2 {\bm k}_1}{(2\pi)^3} \frac{1 }{{\bm k}_1^2 {\bm k}_2^2}
\end{align}
which contains both transverse loop integration and transverse reggeized gluon
propagators. As usually, transverse momenta obey the constraint ${\bm
  q}_i = {\bm k}_1 + {\bm k}_2$ with $t_i = {\bm q}_i^2 $ the momentum
transfer of the corresponding $t$-channel.  Using these rules,
construction of the partial wave Eq.(\ref{eq:V_2}) from diagram
Fig.\ref{fig:mixed23_m}b is straight-forward.

For $V_{(-,+)}$ we then only note that, due to the symmetry of the quark-impact factor $A^{b_1b_2}_2(\omega_2)$ under 
  exchange of transverse momenta and color of the reggeized gluons,  the RP2R-vertex is projected on its  symmetric part  Eq.(\ref{eq:gamma_S}), whereas the anti-symmetric part Eq.(\ref{eq:gamma_A}) decouples in that case.
  Taking a closer look at the energy-dependence of the
  production amplitude Eq.~(\ref{prodamp_mixedsig1}), we
  observe at first the absence of singularities in overlapping production channels
  and conclude that the Steinmann-relations are fulfilled for
  Eq.~(\ref{prodamp_mixedsig1}). In a next step we  confront
  Eq.(\ref{prodamp_mixedsig1}) with the analytic representation of the
  $2\to3$ production amplitude, Eq.~(\ref{eq:anal_23}). 
  
  At first we note that  Eq.~(\ref{eq:anal_23}) does
  not contain the partial wave $V_1$: As outlined in
  \cite{Bartels:1980pe}, for the signature configuration ($\tau_1 = -,
  \tau_2 = +$), the energy-factors in front of $V_1$ are sub-leading
  in comparison to those appearing in front of $V_2$ and consequently  the partial wave $V_1$ is absent in our LLA-analysis.
  
  For the second term we first observe that the behavior of
  Eq.~(\ref{prodamp_mixedsig1}) under substitutions $s \to -s$ and
  $s_2 \to -s_2$ is in full agreement with the signature structure of
  Eq.~(\ref{eq:anal_23}).
 As far the explicit form of the signature factors is concerned, we  apply Eq.~(\ref{eq:gamma_id}) to rewrite Eq.~(\ref{prodamp_mixedsig1}) into the following form
\begin{align}
  \label{prodamp_mixedsig2}
  \mathcal{M}^{(-,+)}_{2 \to 3} =  -2\pi s \int \frac{d\omega_1}{2\pi i} \int \frac{d\omega_2}{2\pi i} \left|{s}\right|^{\omega_1}|s_2|^{\omega_2 -\omega_1}
 \frac{ e^{-i\pi\omega_2} - 1 }{\sin (\pi\omega_2)} 
V_{(-,+)}(\omega_1, \omega_2; {\bm q}_1, {\bm q}_2).
  \end{align}
  Nevertheless, the signature factors of Eq.~(\ref{prodamp_mixedsig2})
  coincides with those in Eq.~(\ref{eq:anal_23}) only in the limit
  $\omega_1 \to 0$.
  \begin{figure}[htbp]
  \centering
  \parbox{5cm}{\center \includegraphics[height=4cm]{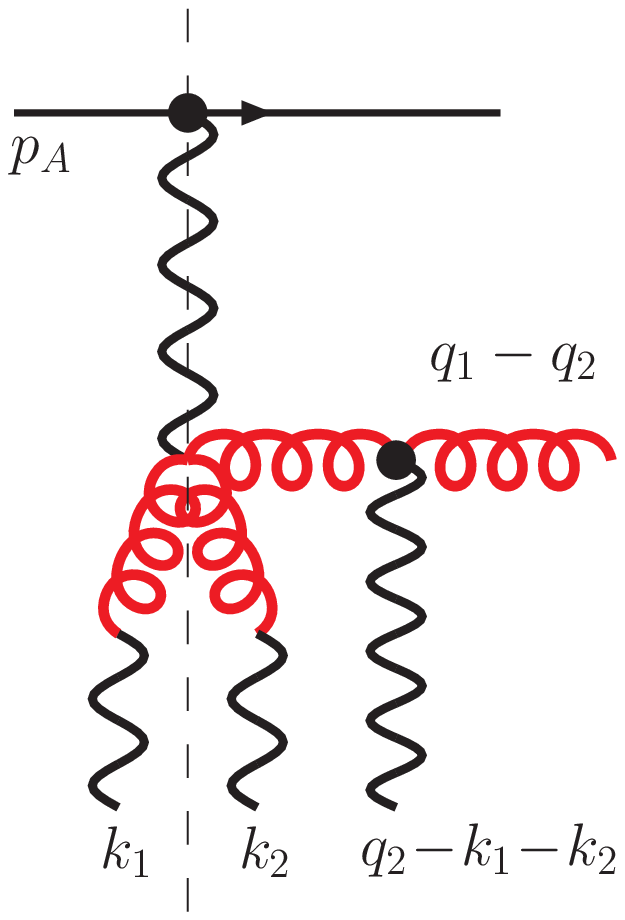}} 
\parbox{5cm}{\center \includegraphics[height=4cm]{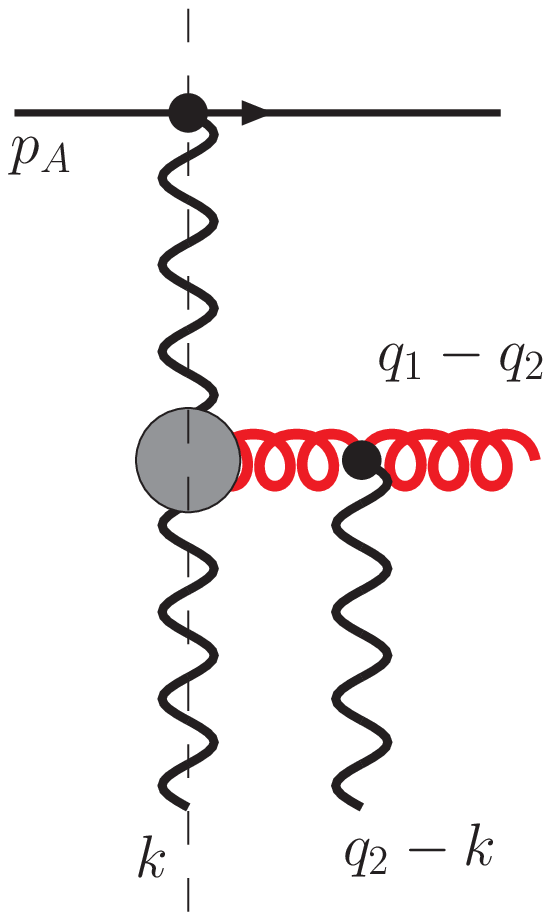}}\\
\parbox{5cm}{\center (a)}\parbox{5cm}{\center (b)}
  \caption{\small Higher contributions which yield a non-zero $s$-discontinuity for symmetric (a) and anti-symmetric (b) color in the $t_2$-channel. While diagram (a) involves a coupling of the two lower reggeized gluons to the upper reggeized gluons, diagram (b) contains the production vertex at next-to-leading order, which allows for a $s$-channel cut. }
  \label{fig:s_cut_minusplus}
\end{figure}
This means that within our accuracy, the upper reggeized gluon is
completely real and contains no imaginary part.  We note that also
this observation coincides with the analysis of the mixed signature
production amplitude by taking discontinuities: As outlined in
\cite{Bartels:1980pe}, the signature configuration ($\tau_1 = -,
\tau_2 = +$) is determined within the LLA by the
$s_{2}$-discontinuity, whereas the discontinuity in $s$ is of higher
order and does not occur at leading order and the LLA.  We expect
first corrections to the signature factor to occur from diagrams like
Fig.\ref{fig:s_cut_minusplus} which contain elements that allow for
'cutting' of the upper reggeized gluon.

The $2 \to 3$ production amplitude with positive signature,
Fig.~\ref{fig:mixed23_m}b is obtained within the LLA, by inserting the
RPR-production vertex Eq.~(\ref{eq:1lipatov_repeat}) either on the
first Fig.~\ref{fig:mixed23}a or on the second Fig.~\ref{fig:mixed23}b
reggeized gluon.  Construction of the phase structure requires some
care and we shall shortly outline the basic steps of the underlying
analysis. To this end we start with the Born diagram, where only
reggeized gluons are exchanged and no interaction kernel between the
reggeized gluons is included. Leaving for simplicity all details about
the transverse momentum structure of the diagram aside, we obtain,
restricting only to the part dependent on  Mellin-variables the following
expression
\begin{align}
  \label{eq:Mellin_prop}
\int \frac{d\tilde{\omega}}{2\pi i} 
\frac{s^{\tilde{\omega}} \xi^{(-)} (\tilde{\omega})}{\tilde{\omega} -\beta } 
\int \frac{d\tilde{\omega}_1}{2\pi i} 
\frac{s_1^{\tilde{\omega}_1} \xi^{(-)} (\tilde{\omega}_1)}{\tilde{\omega}_1 - \beta_1}
\int \frac{d\tilde{\omega}_2}{2\pi i} 
\frac{s_2^{\tilde{\omega}_2} \xi^{(-)} (\tilde{\omega}_2)}{\tilde{\omega}_2 - \beta_2},
\end{align}
where we further abbreviated the trajectory functions as
\begin{align}
  \label{eq:traj_abrev}
\beta & \equiv \beta({\bm k}), & \beta_1 & \equiv \beta({\bm q}_1 - {\bm k}), & \beta_2  & \equiv \beta({\bm q}_2 - {\bm k}).
\end{align}
Using
\begin{align}
  \label{eq:kappa_gym}
s_2 = \frac{s\kappa}{s_1}
\end{align}
and $\kappa^{-\tilde{\omega}_2} \simeq 1$ within the LLA and further
introducing new Mellin-variables $\tilde{\omega} \to \omega_2 =
\tilde{\omega} + \tilde{\omega_2}$, we arrive at
\begin{align}
  \label{eq:Mellin_prop2}
\int \frac{d \omega_2}{2\pi i} \int \frac{d\tilde{\omega}_1}{2\pi i} 
\int \frac{d\tilde{\omega}_2}{2\pi i}
\frac{s^{{\omega}_2} \xi^{(-)} (\omega_2 - \tilde{\omega}_2)    \xi^{(-)} (\tilde{\omega}_2)   }{(\omega_2 - \tilde{\omega}_2 -\beta)   ( \tilde{\omega}_2 - \beta_2)} 
\frac{s_1^{\tilde{\omega}_1 - \tilde{\omega}_2} \xi^{(-)} (\tilde{\omega}_1)}{\tilde{\omega}_1 - \beta_1}.
 \end{align}
The Double-Regge-Kinematics allows now for the $\tilde{\omega}_2$ contour to be closed only to the right-hand-side, which corresponds to the  Residue at  $\tilde{\omega}_2 = \omega_1 - \beta$. Substituting furthermore $\tilde{\omega}_1 \to \omega_1 + \beta$ we finally arrive at
\begin{align}
  \label{eq:Mellin_prop3}
\int \frac{d{\omega}_1}{2\pi i}
\int \frac{d \omega_2}{2\pi i}
\frac{s^{{\omega}_2} \xi^{(-)} (\beta)    \xi^{(-)} ( \omega_2 -\beta )   }
{(\omega_2 - \beta_2 -\beta) } 
\frac{s_1^{{\omega}_1 - {\omega}_2} \xi^{(-)} ({\omega}_1 - \beta)}{(\omega_1 - \beta_1 - \beta)}.
 \end{align}
 With Eq.~(\ref{eq:sig_manipus}) and Eq.~(\ref{eq:gamma_id}) the
 product $\xi^{(-)}(\beta) \xi^{(-)}(\omega_2 - \beta)$ converts
 within the LLA into $\sim \xi^{(+)} (\omega_2)/\sin \pi\omega_2$. In
 particular allowing furthermore for interactions  between the
 reggeized gluons, this  allows  then for a derivation of the
 BFKL-Green's function along the lines of Sec.~\ref{sec:bfkl} for the
 $t_2$-channel. As far as the $t_1$-channel is concerned we  note
 that within the LLA $s^{\omega_2}s^{\omega_1 -\omega_2} =
 s^{\omega_1}s^{\omega_2 -\omega_1}$, making use of
 Eq.~(\ref{eq:kappa_gym}). With
 Eq.~(\ref{eq:sig_manipus}) and Eq.~(\ref{eq:gamma_id})  the
 product $\xi^{(-)}(\beta) \xi^{(-)}(\omega_1 - \beta)$ can be converted
 into $\sim \xi^{(+)} (\omega_1)/\sin \pi\omega_1$, which then allows
 for the derivation of the BFKL-Green's function in the $t_1$-channel.

For the formulation of the corresponding partial wave, we  further require a new ingredient in our diagrammatic language, namely the RRPRR-vertex, which at this level of accuracy is built up of two disconnected contributions, mainly given by insertions of the production vertex Eq.~(\ref{eq:1lipatov_repeat}). We have
\begin{align}
    \label{eq:rrprr}
\parbox{3cm}{\includegraphics[width=3cm]{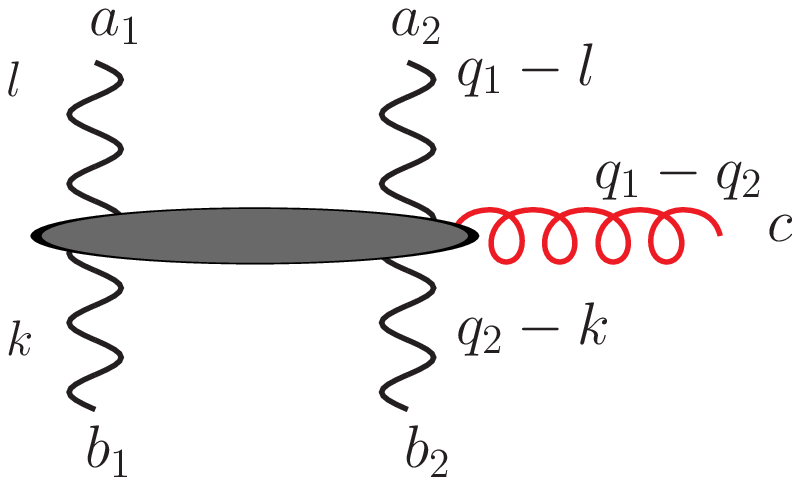}} =& 
 {\bm l}^2 (2\pi)^2\delta^{(2)}({\bm l} - {\bm k}) \quad  
\delta^{a_1b_1}T_{b_2a_2}^c g C^\mu(q_1 - l, q_2 - l)\epsilon_\mu(q_1 - q_2)
\notag \\
 +
 ( {\bm q}_1 - {\bm l})^2(2\pi)^2&  \delta^{(2)}({\bm l} - {\bm k}  - {\bm q}_1 + {\bm q}_2) \quad 
T_{b_1a_1}^c \delta^{a_2b_2}
g  C^\mu( l,l - q_1 + q_2 )\epsilon_\mu(q_1 - q_2)
  \end{align}
  Note that the production vertices do not depend on the longitudinal
  momenta, in particular not on $l^+$ and $l^-$.
The $2 \to 3$ production amplitude with positive signature in both
  $t$-channels is within the LLA  given by
  \begin{align}
    \label{eq:23++}
\mathcal{M}_{2 \to 3}^{(+,+)} =& -2\pi s
\int \frac{d\omega_1}{2\pi i}\int \frac{d\omega_2}{2\pi i}
|s|^{\omega_1} |s_2|^{\omega_2 - \omega_1} \frac{e^{-i\pi\omega_1} - 1}{\sin\pi\omega_1} V_{(+,+)} (\omega_1, \omega_2; q_1, q_2) 
\notag \\
\simeq &
-2\pi s
\int \frac{d\omega_1}{2\pi i}\int \frac{d\omega_2}{2\pi i}
|s|^{\omega_2} |s_1|^{\omega_1 - \omega_2} \frac{e^{-i\pi\omega_2} - 1}{\sin\pi\omega_2} V_{(+,+)} (\omega_1, \omega_2; q_1, q_2) ,
  \end{align}
  where the identity holds with LLA accuracy and the partial wave $
  V_{(+,+)} $ can be obtained making use of our diagrammatic rules.
  Further we note that the above expression is in accordance with the
  Steinmann-relations and within the considered accuracy also with the
  analytic representation of the $2\to 3$ production amplitude,
  Eq.~(\ref{eq:anal_23}). As explained in detail in
  \cite{Bartels:1980pe}, the $2 \to 3$ production amplitude with
  positive signature in both $t$-channels is proportional to the
  $s$-discontinuity, whereas the discontinuity in $s_1$ and $s_2$
  contributes only at higher orders, in accordance with the above
  findings. As a consequence the above partial wave cannot be
  associated with any of the partial waves in Eq.~(\ref{eq:anal_23}),
  but has to be understood as their combined sum.


\subsection{Mixed and positive signature for the $2\to4$ amplitude}
\label{sec:mixed24}
\begin{figure}[htbp]
  \centering
  \parbox{3.5cm}{\center \includegraphics[width=3cm]{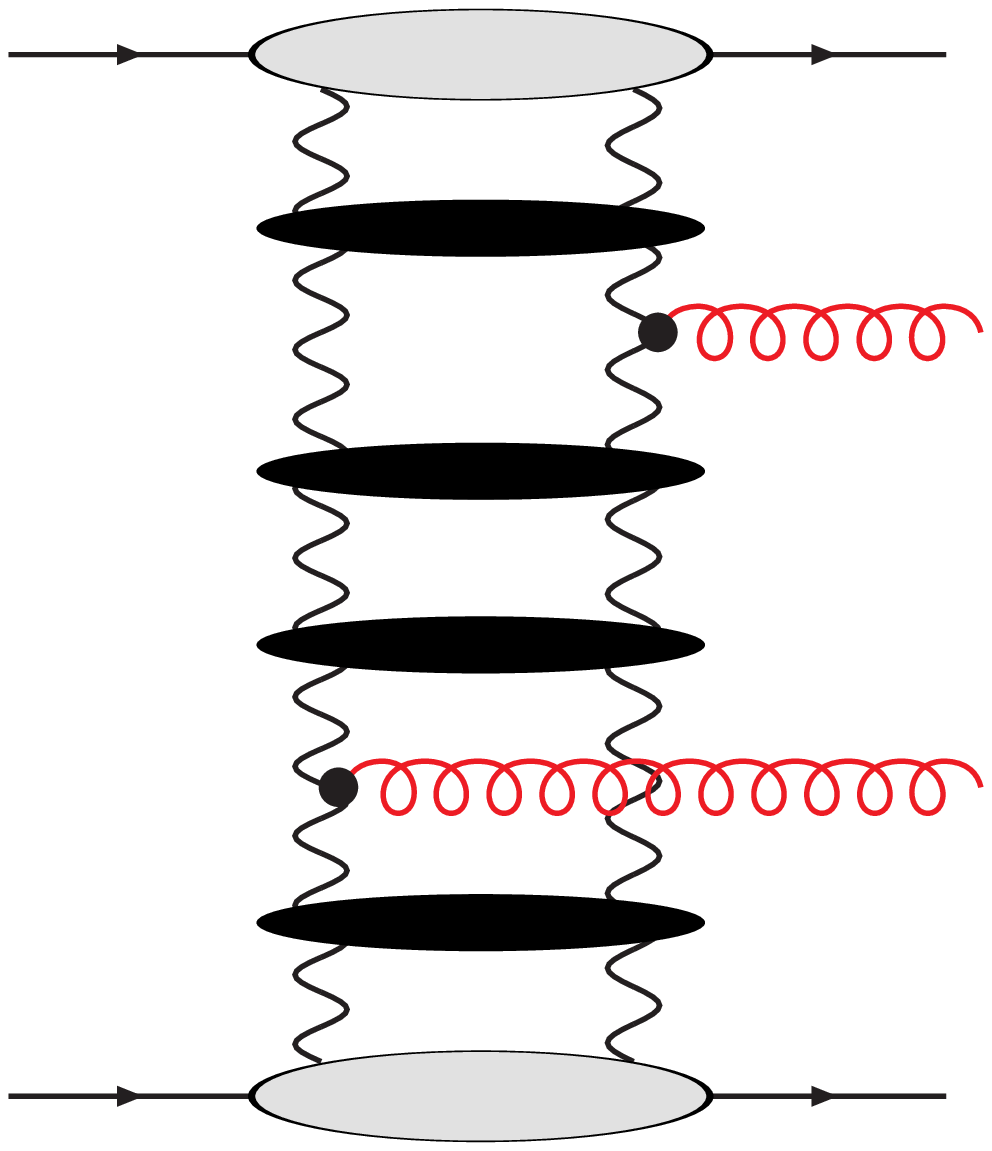}}
  \parbox{3.5cm}{\center \includegraphics[width=3cm]{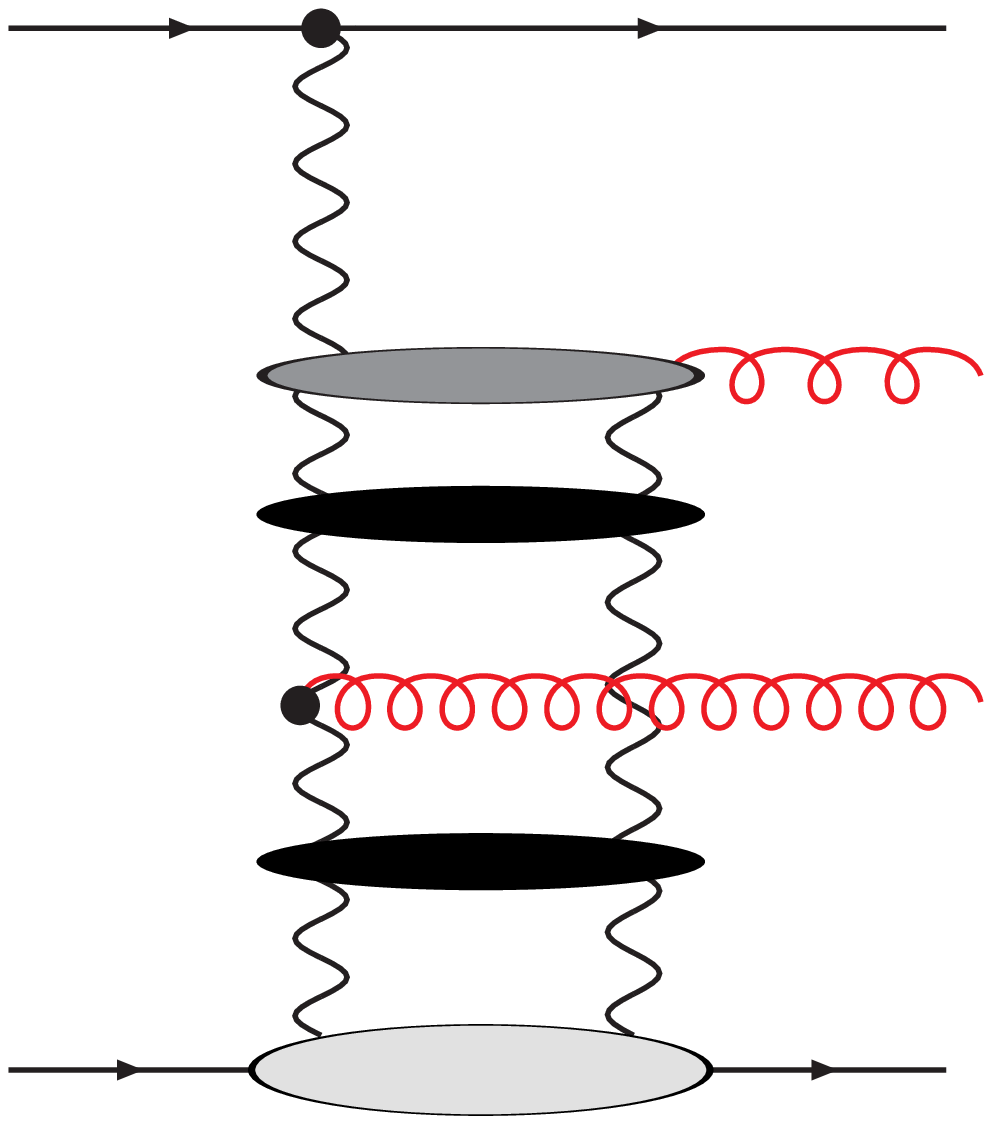}}
  \parbox{3.5cm}{\center \includegraphics[width=3cm]{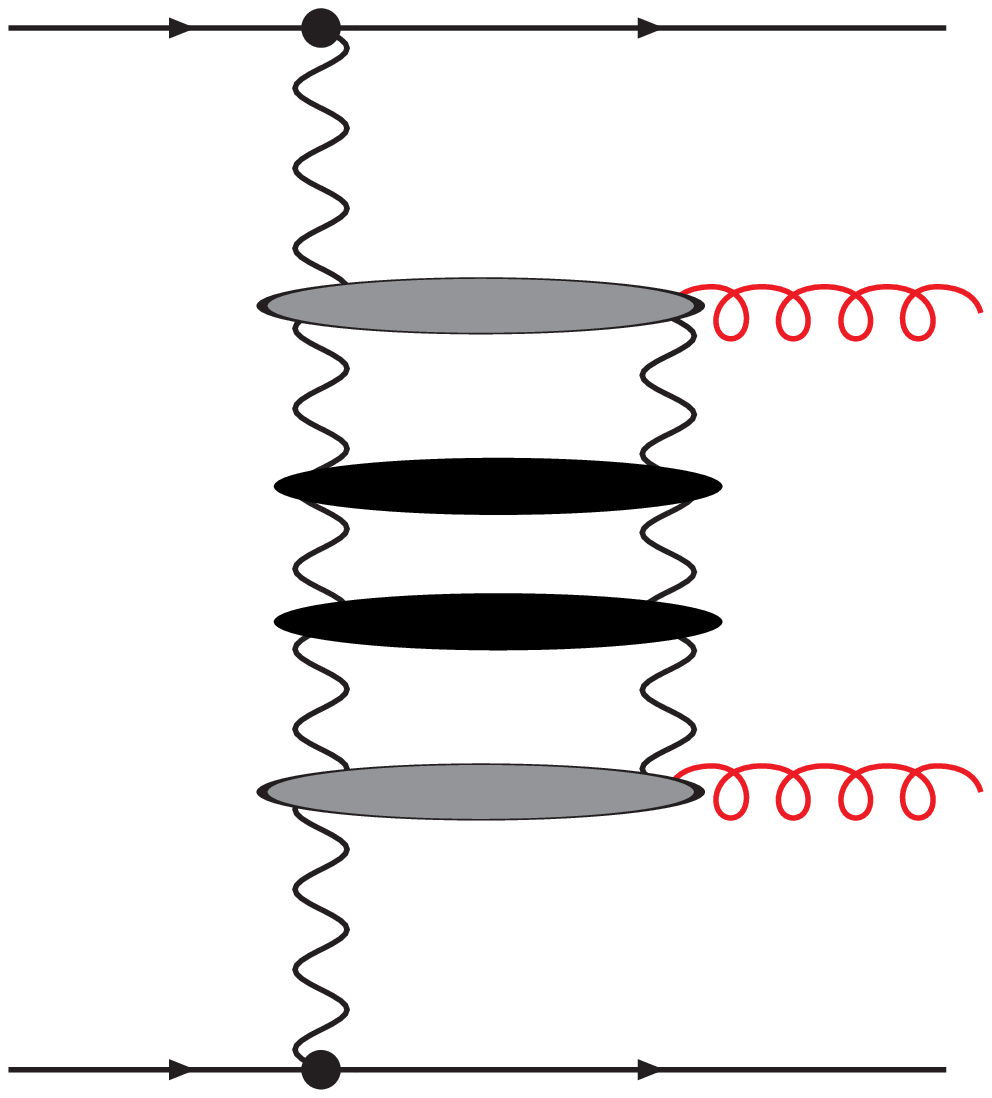}}
 \\
\parbox{3.5cm}{\center (a)}\parbox{3.5cm}{\center (b)}
\parbox{3.5cm}{\center (c)}
  \caption{\small Contributions to the $2\to 4$ production amplitude with positive (a) and mixed (b,c,d) signature. Explicitly we have the following signature configurations a: $(+,+,+)$, b: $(-, +, +)$, c: $(-, +, -)$, d: $(+, -, +)$. }
  \label{fig:mixed24_m}
\end{figure}

Next we turn to the various $2\to4$ production amplitudes with
positive and mixed signature.  As before, the individual partial waves
can be obtained applying  the diagrammatic rules given above to
the graphs in Fig.~\ref{fig:mixed24_m}.
We start with the case, $(+,+,+)$ where signature in all three
$t$-channels is positive, Fig.~\ref{fig:mixed24_m}a. Within the LLA,
this amplitude is found \cite{Bartels:1980pe} to be proportional to
the discontinuity in $s$. For the analytic representation this means
that the amplitude is within the LLA proportional to the sum of all
five partial waves in Eq.~(\ref{eq:anal_24}). From the effective
action we find within the LLA the following result
\begin{align}
  \label{eq:24+++}
\mathcal{M}^{(+,+,+)}_{2 \to 4} = & -2\pi s \int \frac{d\omega_1}{2\pi i} \int \frac{d\omega_2}{2\pi i}\int \frac{d\omega_3}{2\pi i}
|s|^{\omega_3}|s_{012}|^{\omega_2 - \omega_3} |s_1|^{\omega_1 - \omega_2} \frac{e^{-i\pi\omega_3} -1}{\sin\pi\omega_3} \notag \\
& \times 
W_{(+,+,+)}(\omega_1, \omega_2, \omega_3; q_1,q_2,q_3)
\end{align}
where to leading accuracy the following identity can be shown to hold 
\begin{align}
  \label{eq:lla_id}
|s|^{\omega_3}|s_{012}|^{\omega_2 - \omega_3} |s_1|^{\omega_1 - \omega_2} \frac{e^{-i\pi\omega_3} -1}{\sin\pi\omega_3} 
&\simeq
|s|^{\omega_2}|s_{1}|^{\omega_1 - \omega_2} |s_2|^{\omega_3 - \omega_2} \frac{e^{-i\pi\omega_2} -1}{\sin\pi\omega_2}
\notag \\
& \simeq
 |s|^{\omega_1}|s_{123}|^{\omega_2 - \omega_1} |s_3|^{\omega_3 - \omega_2} \frac{e^{-i\pi\omega_1} -1}{\sin\pi\omega_1},
\end{align}
and using further identities like 
\begin{align}
  \label{eq:lla_id2}
|s_{123}|^{\omega_2 - \omega_1} |s_3|^{\omega_3 - \omega_2} \simeq |s_{123}|^{\omega_3 - \omega_1} |s_2|^{\omega_2 - \omega_3},
\end{align}
that equally hold with the LLA, it is easily shown that the
energy dependence  of Eq.~(\ref{eq:24+++}) is in
agreement with the leading terms of Eq.~(\ref{eq:anal_24}). The
partial wave $W_{(+,+,+)} $ on the other hand can be obtained from
Fig.~\ref{fig:mixed24_m}, making use of the diagrammatic rules.

Next we turn to the signature configuration $(-, +, +)$,
Fig.~\ref{fig:mixed24_m}b. Within the LLA this amplitude is
proportional to its discontinuity in $s_{123}$. For the analytic
representation Eq.~(\ref{eq:anal_24}), it is hence to leading accuracy
proportional to the sum of partial waves $W_2$ and $W_4$. From the
effective action we obtain
\begin{align}
  \label{eq:24-++}
\mathcal{M}^{(-,+,+)}_{2 \to 4} = & -2\pi s \int \frac{d\omega_1}{2\pi i} \int \frac{d\omega_2}{2\pi i}\int \frac{d\omega_3}{2\pi i}
|s|^{\omega_1}|s_{123}|^{\omega_2 - \omega_1} |s_3|^{\omega_3 - \omega_1} \frac{e^{-i\pi\omega_2} -1}{\sin\pi\omega_2} \notag \\
& \times 
W_{(-,+,+)}(\omega_1, \omega_2, \omega_3; q_1,q_2,q_3),
\end{align}
where within the LLA
\begin{align}
  \label{eq:lla_id}
  |s_{123}|^{\omega_2} |s_3|^{\omega_3 - \omega_1}
  \frac{e^{-i\pi\omega_2} -1}{\sin\pi\omega_2} \simeq |s_{123}|^{\omega_3}
  |s_2|^{\omega_2 - \omega_3} \frac{e^{-i\pi\omega_3}
    -1}{\sin\pi\omega_3}.
\end{align}
To the given level of accuracy, the energy dependence of the effective
action is therefore found to be in accordance with the one of the
analytic representation Eq.~(\ref{eq:anal_24}).

The signature configuration $(-,+,-)$, Fig.~\ref{fig:mixed24_m}c, is
proportional to its discontinuity in $s_2$ and therefore proportional
to the sum of the partial waves $W_3$ and $W_4$. From the effective
action analysis we obtain
\begin{align}
  \label{eq:24-+-}
\mathcal{M}^{(-,+,-)}_{2 \to 4} = & -2\pi s \int \frac{d\omega_1}{2\pi i} \int \frac{d\omega_2}{2\pi i}\int \frac{d\omega_3}{2\pi i}
|s|^{\omega_1}|s_{123}|^{\omega_3 - \omega_1} |s_2|^{\omega_2 - \omega_3} \frac{e^{-i\pi\omega_2} -1}{\sin\pi\omega_2} \notag \\
& \times
W_{(-,+,-)}(\omega_1, \omega_2, \omega_3; q_1, q_2, q_3),
\end{align}
while the  LLA allows for the following manipulation 
\begin{align}
  \label{eq:lla_id4}
  |s|^{\omega_1}|s_{123}|^{\omega_3 - \omega_1} |s_2|^{\omega_2 - \omega_3}
  \frac{e^{-i\pi\omega_2} -1}{\sin\pi\omega_2}
  \simeq 
  |s|^{\omega_3}|s_{012}|^{\omega_1 - \omega_3} |s_2|^{\omega_2 - \omega_1} 
  \frac{e^{-i\pi\omega_2} -1}{\sin\pi\omega_2}.
\end{align}
The partial wave $W_{(-,+,-)}$ is obtained with the help of the
diagrammatic rules.

\begin{figure}[htbp]
  \centering
  \parbox{5.5cm}{\center \includegraphics[height=4cm]{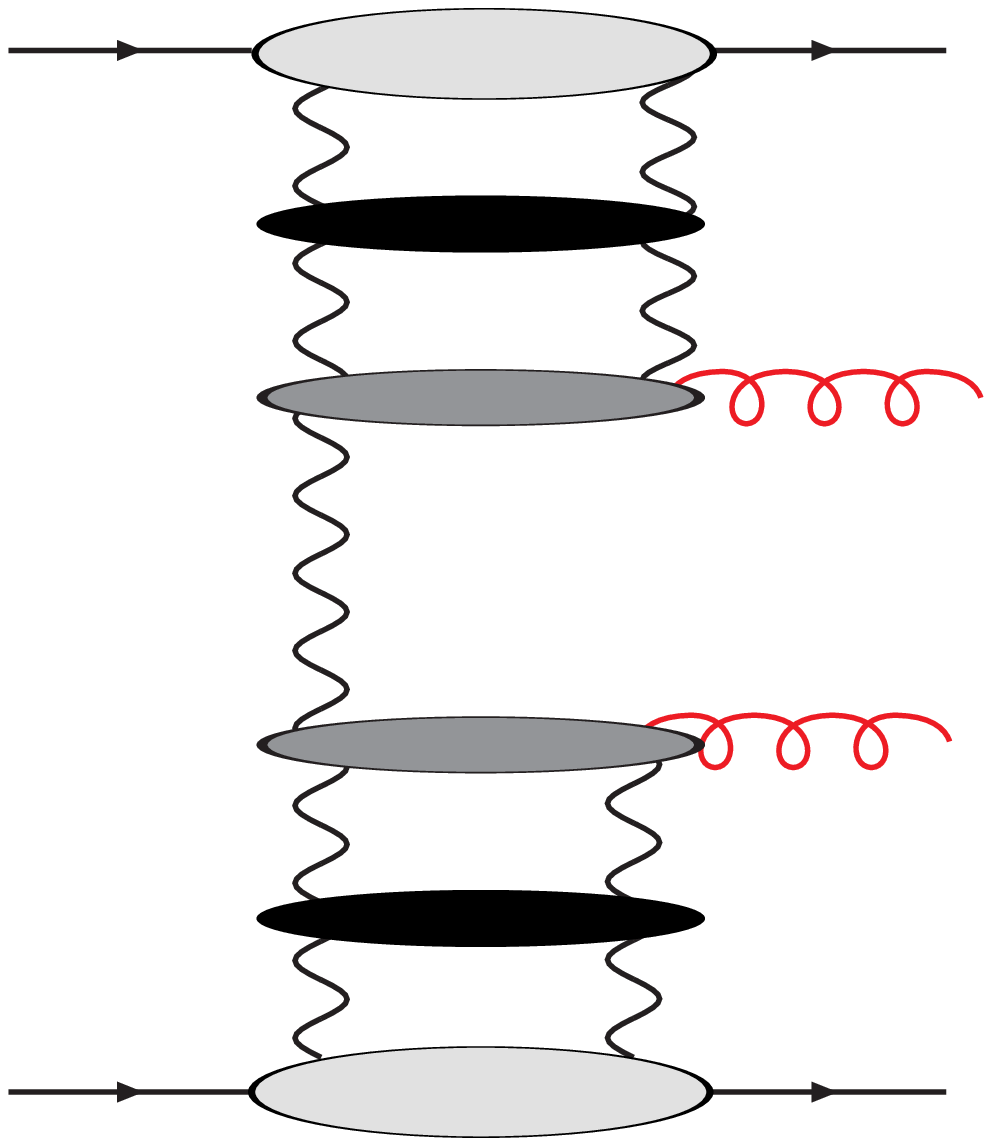}}
 \parbox{7.5cm}{\center \includegraphics[height=4cm]{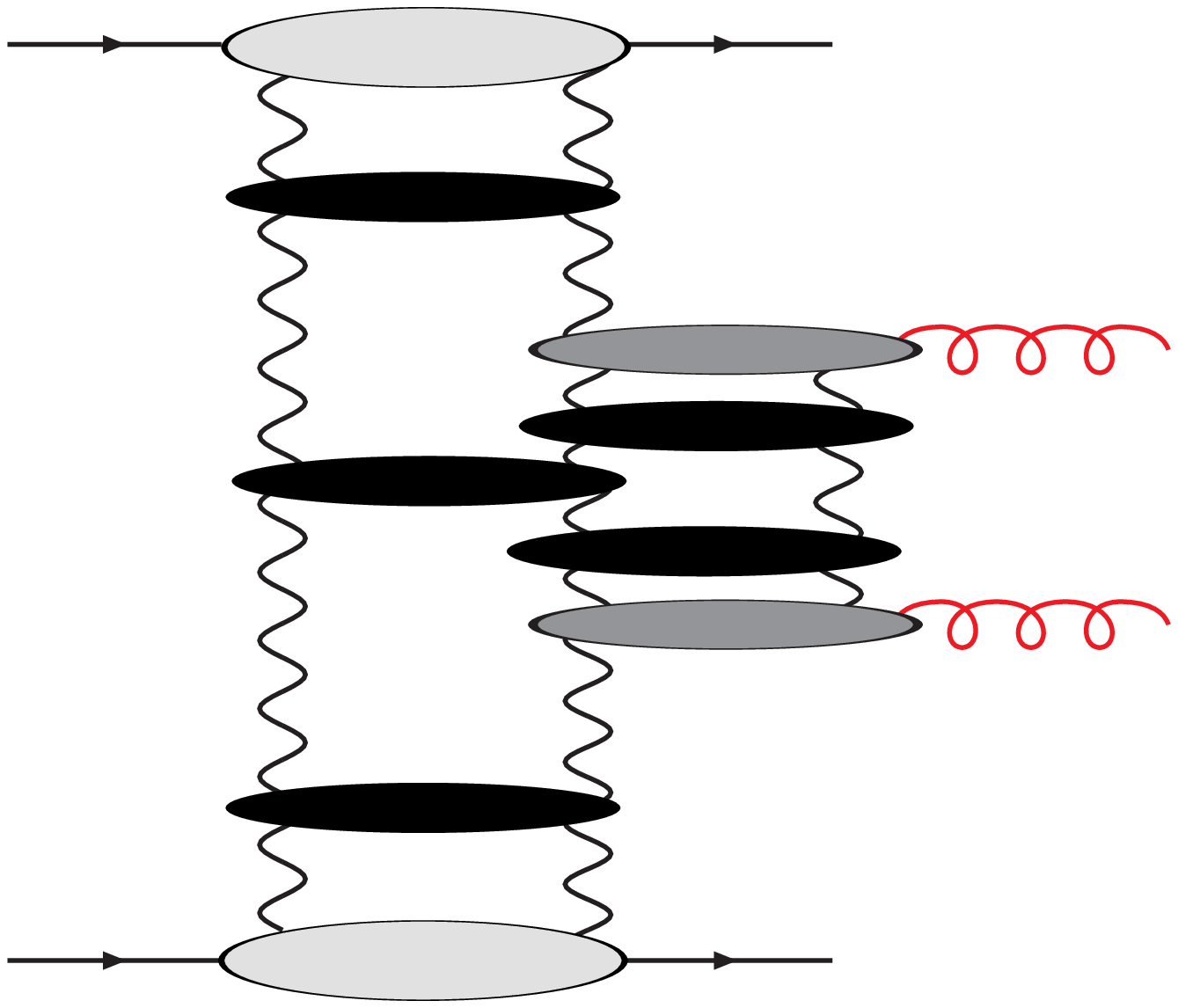}}
\\
\parbox{5.5cm}{\center (a)}   \parbox{7.5cm}{\center (b)}
  \caption{\small Graphs that contribute to the production amplitude with the signature configuration (+,-,+) }
  \label{fig:+-+}
\end{figure}

The production amplitude with the signature configuration $(+,-,+)$ on
the other hand is of higher order.  This appears from the analysis of
both the effective action and the analytic representation.  From
latter we find that to leading accuracy the partial waves $W_3$, $W_4$
and $W_5$ contribute, which can be obtained by taking the
double-discontinuity in $s_1$ and $s_3$ ($W_5$) and in $s$ and $s_2$
($W_3$, $W_4$ ). For the effective action, these contributions
correspond to the diagrams in Fig.~\ref{fig:+-+}a and
Fig.~\ref{fig:+-+}b respectively.  Whereas Fig.~\ref{fig:+-+}a
contains only  states of two reggeized gluons and can in
principle be constructed, Fig.~\ref{fig:+-+}b requires the exchange of
a state of three reggeized gluons, which has not been addressed
so far from the effective action. The description of these states
within the effective will be addressed in the following chapter, where
we however refer to the elastic amplitude as a reference process.

The same applies to the Regge-cut contributions for the signature
configuration $(-,-,-)$, found in the analysis of \cite{Bartels:1978fc},
which has not been discovered in Sec.~\ref{sec:born+possig}, as it is
of higher order: Also this contribution requires within the effective
action the  state of three reggeized gluons.

\section{Conclusion}
\label{sec:concoprod}

In the present chapter the results of Cha.\ref{cha:2to2} have been
applied to the study of production amplitudes. The analysis of
production amplitudes with negative signature revealed that the phase
structure of the reggeized gluon, Eq.(\ref{eq:vertexsig_fac_extract}),
must be taken with some care: The Reggeon-Particle-Reggeon vertex,
which describes the production of a gluon, carries itself a inner
phase structure. The phase structure of the complete amplitude is
therefore only obtained correctly if simultaneously with the phase
structure of the reggeized gluons, the inner phase structure of the
production vertex is included.  Within the LLA, only the leading part
of the production vertex is included, which is real.  Including then
nevertheless the complete signature factor of the reggeized gluon, one
finds simultaneous discontinuities in over-lapping channels which
violate Steinmann-relations.  For the final result, phases of the
reggeized gluons should be therefore only trusted with the same
amount of accuracy as the corresponding inner phases of production
vertices are taken into account.

We then demonstrated by an explicit calculation that the 1-loop
corrections to the production vertex yield the onset of the required
corrections. In particular they could be shown to be in accordance
with the analytic representation of the production amplitudes, which
satisfy the Steinmann-relations by construction.

For the construction of production amplitudes with positive and mixed
signature, the Reggeon-Particle-2Reggeon (RP2R) vertex has been
determined.  Together with the production vertex,
Eq.(\ref{eq:2lipatov_repeat}), the RP2R-vertex allowed for the
construction of all signature configurations of the $2 \to 3$ and $2
\to4$ production amplitude, apart form the configuration $(+,-,+)$
which requires information about the  state of three-reggeized
gluons. The basis for constructing amplitudes with states of more than
two reggeized gluons will be laid in following chapter. There we will
consider states of three and four reggeized gluons within the elastic
scattering amplitude. The results can then be applied also to the
construction of production amplitudes.

\chapter{Exchanges of $n > 2$ reggeized gluons in the elastic amplitude}
\label{cha:v24}
In this chapter we aim to arrive on a description of  states of $n >2$ within the effective action.   At the end of the previous
chapter it turned out  that in order to describe all signature
configuration of $n$-particle production amplitudes within the
effective action, knowledge about the states with more than two
reggeized gluons is inevitable. Apart from the$ (+,-,+)$ configuration of
the $2 \to 4$ production amplitude, which requires the state of three
reggeized gluons already at leading order, unitarity requires also the
presence of the three-Reggeon-cut for the $(-,-,-)$ configuration.
States with many reggeized gluons are generally connected with
unitarity corrections and it is generally argued that inclusion of the
exchange of an arbitrary high number of reggeized gluons yields
unitarization of the BFKL-Pomeron.

The main goal of this chapter is to provide the necessary tools for
the study of exchanges of more than two reggeized gluons within the
effective action. As in the previous chapters, we will be mainly
concerned with the issue of longitudinal integrations.  In particular
the subtraction mechanism for reggeized gluon loops, introduced in
Cha.~\ref{cha:2to2} is needed to be generalized and we determine the pole prescription for the higher induced vertices.

As a reference process we consider as in Cha.~\ref{cha:2to2}, elastic
scattering of two quarks at high center of mass energies with exchange
of three and four reggeized gluons.  We then determine the coupling of
three an four reggeized gluons to the quark. Apart from the
interaction between the reggeized gluons themselves, the complete
description of the state of three and four reggeized gluons requires
vertices that describe transitions from one-to-three and two-to-four
reggeized gluons respectively are included. Combining them with
pairwise interactions between reggeized gluons, they can be shown to
yield for the four reggeized gluon state in the overall color singlet
the 2-4 reggeized gluon transition vertex of \cite{Bartels:1994jj}.

The outline of this chapter is the following: In
Sec.~\ref{sec:indu_presc} we determine the pole structure of induced
vertices up to the third order. In Sec.~\ref{sec:impa4_bkp} we discuss
the quark-impact factor and propose a all order generalization of the
subtraction scheme for the two reggeized gluon exchange in
Sec.~\ref{sec:tworeggeon_negsig}. In Sec.~\ref{sec:v23} we discuss the
properties of transitions that change the number of reggeized gluons
while in Sec.\ref{sec:v24bartelswuest} we compare our findings with
the analysis of \cite{Bartels:1994jj}.

\section{The pole prescription of induced vertices}
\label{sec:indu_presc}

In the following section we investigate more closely  the pole structure of
the induced vertices. In particular we will explicitly determine  the
pole structure for induced vertices up to the third order, while the
formalism that is presented allows in principle also for the
determination of the pole structure of induced vertices of arbitrary
high order.  For  the exchange of up to four
reggeized gluons, this pole structure is of two-fold need: It is
needed for the proper definition of the longitudinal integrations and
for the  higher induced vertices that occur as a building block of the
reggeized gluon transition vertices.

\subsection{The induced vertex of the first order}
\label{sec:color_basis}

As preparation for the
discussion of the pole structure of higher induced vertices,   we briefly review the
derivation of the pole structure of the induced vertex of the first
order in Sec.\ref{sec:22negsig}. There, the following pole
prescription has been found by comparing the effective theory
diagram with the underlying QCD-diagram and  expanding the quark-propagators of
the QCD-amplitude in the center of mass energy of the quark-2 gluon
sub-amplitude $p_A^+k^-$:
 \begin{align}
   \label{eq:reprod_expandee}
 \frac{(t^{c_1}t^{c_2})_{AA'}}{-k^- + i\epsilon/p_A^+}  
       + 
       \frac{(t^{c_2}t^{c_1})_{AA'}}{k^-   + i\epsilon/p_A^+} .
 \end{align}
 To arrive at the color structure of the induced vertex,
 Eq.~(\ref{inducedvertex1}), we further decomposed the products of
 $SU(N_c)$ generators, $t^{c_1}t^{c_2} $ and $t^{c_2}t^{c_1}$ into two
 terms, symmetric and anti-symmetric under exchange of the gluon color
 labels $c_1$ and $c_2$, and  identified the antisymmetric part with the
 induced vertex of the first order, Eq.~(\ref{inducedvertex1}). Note,
 that equally well we could have decided to decompose the color
 structure in a different way, as long as one of the elements would
 yield the commutator. It would have been  this term that would
 have had determined the pole-prescription of the induced vertex
 instead. An alternative possible choice would have been for instance,
 to keep the combination $t^{c_1}t^{c_2}$, while  the second term
 would have been rewritten as the first term plus the commutator. Such
 a choice has been used for instance in the analysis of QCD-amplitudes
 of \cite{Treleani:1994at, Cheng:1979px}. 
Since the pole structure of the subtraction terms changes simultaneously with the pole structure of the induced vertices, 
 the overall result for the sum of all different
 contributions remains  unchanged, while individual results
 alter.
Taking a different choice for the decomposition of the
 color structure in Eq.~(\ref{eq:reprod_expandee}) than the originally
 chosen symmetric/anti-symmetric decomposition, leads to a reggeized
 gluon without signature. The symmetric/anti-symmetric decomposition
 corresponds therefore at least at this level to the decomposition of
 the quark-2 gluon amplitude into positive and negative signature, where the part with  negative signature is associated with the
 reggeized gluon.
 
 For the effective action a different choice than the
 symmetric/anti-symmetric decomposition, has the further 
 disadvantage that it leads to a dependence of the pole structure on
 the value of\footnote{ The dependence of the individual terms
   in Eq.~(\ref{eq:reprod_expandee}) corresponds to a special kind of
   a Mandelstam-Leibbrandt prescription
   \cite{Mandelstam:1982cb,Leibbrandt:1983pj,Bassetto:1991ue}, where
   the conjugated momentum is not as usually the plus-momentum of the
   gluon, $k^+$, , but the plus-momentum of the quark, $p_A^+$.
   Restricting to the planar part of scattering amplitudes, only the
   first term in Eq.~(\ref{eq:reprod_expandee}) is present and a
   prescription of this kind is needed.} $p_A^+$, whereas with the
 symmetric/antisymmetric choice
\begin{align}
  \label{eq:ident}
f^{c_1c_2c} \frac{1}{-k^-} \equiv
\frac{1}{2}\frac{f^{c_1c_2c} }{-k^- + i\epsilon/p_A^+}  
       -
     \frac{1}{2}  \frac{f^{c_1c_2c} }{k^-   + i\epsilon/p_A^+}
=
f^{c_1c_2c} \frac{1}{2} \left(  \frac{1}{ k^- - i\epsilon} +  \frac{1}{ k^- + i\epsilon} \right),
\end{align}
the dependence on $p_A^+$ cancels. Such a dependence is unsatisfactory
from a point of view of Regge-factorization. From a technical point of
view, it would at least complicate the regularization proposed for the
two reggeized gluon state in Sec.~\ref{sec:tworeggeon_negsig}.
Furthermore, with the $i\epsilon$ prescription of the pole independent
of the particle to which the associated reggeized gluon couples, it
is possible to determine the pole-prescription directly from the
effective action.

To this end we recall that, in the effective action, induced vertices
arise from the term
\begin{align}
  \label{eq:ind_vertex_L}
\mathcal{L}^{\text{GR}}_{\text{ind}} = A_\pm(v)\partial^2_{\sigma} A_\mp && \textrm{with} && A\pm(v) = v_\pm - gv_\pm \frac{1}{\partial_\pm} v_\pm + g^2v_\pm \frac{1}{\partial_\pm} v_\pm \frac{1}{\partial_\pm} v_\pm + \ldots
\end{align}
where $v_\pm(x) = -it^av^a_\pm(x)$ and $A_\pm(x) = -it^aA^a_\pm(x)$
are matrices in color space.  To give meaning to the operators
$1/\partial_\pm$ at zero, it is necessary to introduce a prescription for these poles  in the
effective action.  A possible choice is
given by\footnote{Such choices can be also found in the literature, for
  instance in the context of soft-effective theories, see
  \cite{Beneke:2002ph} } $\partial_\pm \to\partial_\pm -\epsilon$,
which agrees especially  with Eq.~(\ref{eq:reprod_expandee})
for $p_A^+ > 0$. In general, path-ordered exponents with such a
pole-prescription are used to describe to leading accuracy the
propagation of a highly energetic particle interacting in a soft-gluon
field \cite{ Balitsky:1996ub}.

With the proposed  prescription, we obtain from the effective action  the following term for the first induced vertex
\begin{align}
 \label{eq:indu_1L_eps}
 i{\bm q}^2 2 \left( \frac{\tr(t^{a_1} t^{a_2} t^c)}{k_2^- + i\epsilon} + \frac{\tr(t^{a_2} t^{a_1} t^c)}{k^-_1 + i\epsilon}\right)n^-_{\nu_1} n^-_{\nu_2} ,
\end{align}
with $k_2^- = -k_1^-$. In the above and also in the following we
restrict  for simplicity to $A_-(v)$. All results apply
equally to $A_+(v)$ with the corresponding substitutions. As for
Eq.(\ref{eq:reprod_expandee}), the color structure does not reduce to
the commutator as long $\epsilon \neq 0$. It is therefore necessary
to rewrite the color structure in a symmetric/anti-symmetric color
basis and we find
\begin{align}
 \label{eq:indu_1L_eps}
  {\bm q}^2 \left[ \frac{f^{a_1a_2c}}{2} \left(  \frac{1}{k_1^- + i\epsilon} + \frac{1}{k_1^- - i\epsilon} \right) + \frac{d^{a_1a_2c}}{2} \text{sgn}(\epsilon) 2\pi\delta(k^-_1)   
\right]
  n^-_{\nu_1} n^-_{\nu_2} .
\end{align}
Due to the regularization of the pole, we therefore obtain a second
color structure proportional to the symmetric structure constant
$d^{a_1a_2c}$ for the induced vertex of the first order, which is not
present in the original formulation, Eq.~(\ref{inducedvertex1}).  At
first it might seem plausible to include also the second, symmetric
term into the definition of the induced vertices. It  would
describe a  reggeized gluon with positive signature in addition to the
  reggeized gluon with negative signature. Furthermore,  this term does not
contribute for real-production amplitudes, as it is 
 proportional to the delta-function. It therefore leads  not to  a contradiction with the derivation of the
effective action. We shall nevertheless reject the possibility to include this term for the following four reasons:
\begin{itemize}
\item[(i)] Inclusion of such terms is physically not motivated:
  Starting from the quark-gluon amplitude as given for instance by
  Eq.~(\ref{eq:reprod_expandee}), the physical idea behind induced
  vertices is to gather the most divergent part of the sub-amplitude
  that does not vanish for large values of the product $p_A^+k^-$.
  This does not include contributions proportional to a
  delta-function.

\item[(ii)] While the term proportional to the commutator is
  independent (up to normalization) of the precise representation of
  the gluon-fields, the symmetric part differs for different
  representations of the gluon-field. For the adjoint representation for instance, the
  symmetric term in Eq.~(\ref{eq:indu_1L_eps}) vanishes. 
Unlike the  commutator it is therefore not always possible to extract such a term from an underlying QCD-amplitude\footnote{That there exists indeed a dependence on the  representation of the scattering particles has  also been observed in \cite{behm}. There an additional contribution for scattering particles in the adjoint representation has been found, which is not present for the  fundamental representation. }.
  
\item[(iii)] The symmetric term depends on the sign of $\epsilon$ or,
  in terms of  the underlying QCD amplitude Eq.~(\ref{eq:reprod_expandee}), on
  the sign of the light-cone-momentum $p_A^+$.
  
\item[(iv)] Including  symmetric terms, the effective action is no longer
  hermitian \cite{private_Lipatov}, while an
  hermitian action is required for a unitary $S$-matrix.
\end{itemize}

We therefore conclude that it is preferable to keep only terms with
antisymmetric color structure while terms with symmetric color
structure should be dropped for the effective action.  With the
substitution $\partial_\pm \to \partial_\pm -\epsilon$, the correct
$i\epsilon$-prescription arises therefore not immediately from the
effective action, but it is necessary to add some further, external
information.  In particular, to obtain the correct pole prescription
for the induced vertices, one should first use the pole prescription
$\partial_\pm \to\partial_\pm -\epsilon$ for the poles in
Eq.~(\ref{eq:ind_vertex_L}), then derive the induced vertices in
momentum space, decompose the traces of generators in a suitable
symmetric/anti-symmetric basis, drop all symmetric terms and identify
the terms containing only commutators with the induced vertex.

It would be desirable to have a better prescription of the operators
Eq.~(\ref{eq:ind_vertex_L}), that allows to obtain the correct
pole-prescription directly from the effective action, without the need
to drop any terms by hand.  This can be obtained at least
perturbatively by using the above recipe and then inserting 
 in the action the term  that yields the obtained result.  As
far as the  first induced vertex is concerned, it is
straight-forward to give such a pole-prescription, 
\begin{align}
 \label{eq:indu_1L_epsalt}
   - gv_- \frac{1}{2} \left( \frac{1}{\partial_- -\epsilon} + \frac{1}{\partial_- + \epsilon} \right) v_- \partial^2_{\sigma} A_\mp ,
\end{align}
which yields immediately the correct induced vertex, without the need
to drop any term.  For future reference we then further introduce a
function
\begin{align}
\label{eq:indu1_pole_g}
           g_1(k^-_1, k_2^-)&=  \frac{1}{2} \bigg(\frac{1}{k^-_1 +i\epsilon} +\frac{1}{k^-_1 -i\epsilon}  \bigg) = \frac{1}{[k_1^-]} &&\text{with} & k_1^- + k_2^- &= 0,
\end{align}
where, 
\begin{align}
  \label{eq:def_cauchypv}
\frac{1}{[k^-]} = 
           \frac{1}{2} \bigg(\frac{1}{ k^- -i\epsilon } +
             \frac{1}{ k^- +i\epsilon } \bigg)  ,
\end{align}
is the Cauchy principal value,
such that the induced vertex of the first order with pole-prescription is given by
\begin{align}
\label{eq:indu1_pole_g?indu}
\Delta^{a_1a_2c}_{\nu_1\nu_2-} = 
{\bm q}^2 gf^{a_1a_2c}  g_1(k^-_1)          
n^-_{\nu_1}n^-_{\nu_2}.
\end{align}

\subsection{The induced vertex of the second order}
\label{sec:indu2}
From the effective action without pole prescription the second induced vertex is given by
\begin{align}
  \label{inducedvertex2_reloaded}
  \parbox{2.4cm}{\includegraphics[height = 2cm]{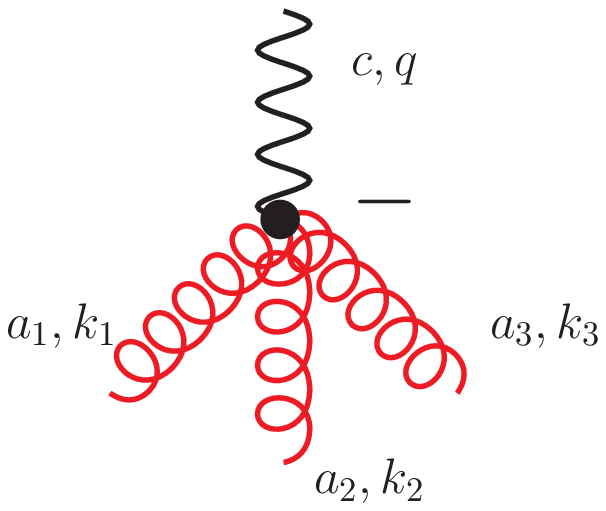}} &= 
\Delta^{\nu_1\nu_2\nu_3+}_{a_1a_2 a_3c} = 
ig^2 {\bm{q}}^2 \left(\frac{f^{a_3a_2 a} f^{a_1ac}}{k_3^- k_1^-} + \frac{f^{a_3a_1 a} f^{a_2ac}}{k_3^- k_2^-}\right) (n^-)^{\nu_1} (n^-)^{\nu_2} (n^-)^{\nu_3}
 \notag \\& \quad
 k_1^- + k_2^- + k_3^- = 0.
\end{align}
As far the prescription of the poles is concerned, a first guess might be, 
 that it is suitable to describe every single pole of the second induced vertex by a Cauchy principal value. However, such a choice needs to be rejected as it violates Bose-symmetry of the  vertex. This is due to the fact 
 the  eikonal identity  that applies for Cauchy principal values holds not in a purely algebraic sense, but yields an extra term proportional to two delta-functions  (see also \cite{Bassetto:1991ue})
\begin{align}
  \label{eq:eik_id_cpv}
\frac{1}{[k_1^-][k_1^- + k_2^-]}  + \frac{1}{[k_1^-][k_1^- + k_2^-]} = \frac{1}{[k_1^-][ k_2^-]} + \pi^2 \delta(k_1^-)\delta(k_2^-).
\end{align}
It is therefore not possible to simple 'guess' the correct pole
structure of the second induced vertex, but it needs to be derived
explicitly, along the lines stated above.  In particular, to derive
the pole-prescription, we need to find an appropriate symmetry basis
that generalizes the decomposition in commutator and anti-commutator
of the previous paragraph.  This basis needs to contain elements with
a double-commutator, which then yield  the color structure of
the induced vertex Eq.~(\ref{inducedvertex2_reloaded}). Furthermore to obtain a  reggeized
gluon with negative signature,
it is desirable to find a prescription that is invariant under
substitutions $\epsilon \to -\epsilon$.  In the following we use  a short-cut notation, which
makes the underlying structure of the above expressions more apparent.
\begin{align}
  \label{eq:com}
[1,2] = [t^{a_1},t^{a_2} ]
\end{align}
represents the commutator of two generators with color indices $a_1$ and $a_2$ and
\begin{align}
 \label{eq:sym}
S_n (1 \ldots n) = \frac{1}{n!} \sum_{i_1, \ldots i_n}t^{a_{i_1}} \ldots  t^{a_{i_n}}
\end{align}
with the sum over all permutations of the numbers $1, \ldots, n$,
yields symmetrization of $n$ generators. In this notation,  a  suitable 'basis' is 
given by
  \begin{align}
  \label{eq:change3_symbolic}
[[3,1],2], \quad [[3,2],1],  
\quad S_2\left([1,2]3 \right), \quad S_2\left([1,3]2 \right), \quad S_2\left([2,3]1 \right),
 \quad S_3\left(123 \right).
\end{align}
Here  terms that do not contain
commutators are maximally symmetrized.
 Note that Eq.~(\ref{eq:com})
yields again a generator and therefore expressions like $ S_2([1,3]2 )
$ which contain symmetrization of commutators are well defined. In
particular, in such a expressions, a commutator should be always dealt
with like a single generator,
for instance
\begin{align}
  \label{eq:s132}
S_2\left([1,3]2 \right) = \frac{1}{2} \left( t^{a_1}t^{a_3}t^{a_2}  - t^{a_3}t^{a_1}t^{a_2} 
+ t^{a_2}t^{a_1}t^{a_3} - t^{a_2}t^{a_3}t^{a_1} \right).
\end{align}
The set Eq.~(\ref{eq:change3_symbolic}) contains then two
double-anti-symmetric, three mixed-symmetric and one totally symmetric
element. There exists also a third double-antisymmetric element
$[[1,2],3]$, which can be expressed in terms of the other two by means
of the Jacobi-identity. The above decomposition shares some properties
with the usual Young-tableau decomposition, while the definition of
anti-symmetrization differs in the present case.

To derive the pole prescription of the induced vertex of the second order, we start from the effective action with the following expression
\begin{align}
 \label{eq:indu_2L_epsa}
   g^2v_-& \frac{1}{\partial_- -\epsilon} v_- \frac{1}{\partial_- -\epsilon} v_- \partial^2_{\sigma} A_+
\end{align}
which yields
\begin{align}
  \label{eq:indu_2L_eps}
 -ig^2 2{\bm q}^2& n^-_{\nu_1} n^-_{\nu_2}  n^-_{\nu_3} 
 \notag \\
  \bigg(&
 \frac{\tr(t^{a_1} t^{a_2} t^{a_3} t^c)}{(k_2^- + k_3^-+ i\epsilon)(k_3^- + i\epsilon)} 
 +
\frac{\tr(t^{a_2} t^{a_1} t^{a_3} t^c)}{(k_1^- + k_3^-+ i\epsilon)(k_3^- + i\epsilon)}
+
\frac{\tr(t^{a_1} t^{a_3} t^{a_2} t^c)}{(k_2^- + k_3^-+ i\epsilon)(k_2^- + i\epsilon)} \notag \\
& +
\frac{\tr(t^{a_3} t^{a_1} t^{a_2} t^c)}{(k_1^- + k_2^-+ i\epsilon)(k_2^- + i\epsilon)}
 +
\frac{\tr(t^{a_3} t^{a_2} t^{a_1} t^c)}{(k_1^- + k_2^-+ i\epsilon)(k_1^- + i\epsilon)}
 +
\frac{\tr(t^{a_2} t^{a_3} t^{a_1} t^c)}{(k_1^- + k_3^-+ i\epsilon)(k_1^- + i\epsilon)}
\bigg)
\end{align}
with $k_1^- + k_2^- + k_3^- = 0$. Leaving aside for the moment the factor
$-ig^2 2{\bm q}^2 n^-_{\nu_1} n^-_{\nu_2} n^-_{\nu_3} $ and also the
projection on the color octet, $\tr (\,\cdot \, t^c)$, where the dot represents
 any product of $SU(N_c)$ generators,  we obtain for the basis Eq.(\ref{eq:change3_symbolic}) from Eq.(\ref{eq:indu_2L_eps})
\begin{align*}
    -\frac{1}{6} \left[[3,1],2\right] \bigg(&
            \frac{2}{( k^-_3 -i\epsilon) ( k^-_1 +i\epsilon) } 
            +
            \frac{2}{( k^-_3 +i\epsilon) ( k^-_1 -i\epsilon) } 
              +
            \frac{1}{( k^-_2 +i\epsilon) ( k^-_3 -i\epsilon) } 
\notag \\            
  &+
            \frac{1}{( k^-_2 -i\epsilon) ( k^-_3 +i\epsilon) } 
            +
            \frac{1}{( k^-_1 +i\epsilon) ( k^-_2 -i\epsilon) } 
            +  
            \frac{1}{( k^-_1 -i\epsilon) ( k^-_2 +i\epsilon) } \bigg) \notag \\
 -\frac{1}{6} \left[[3,2],1\right]\bigg(&
            \frac{2}{( k^-_3 -i\epsilon) ( k^-_2 +i\epsilon) } 
            +
            \frac{2}{( k^-_3 +i\epsilon) ( k^-_2 -i\epsilon) } 
              +
            \frac{1}{( k^-_1 +i\epsilon) ( k^-_3 -i\epsilon) } \bigg)
\notag \\            
  &+
            \frac{1}{( k^-_1 -i\epsilon) ( k^-_3 +i\epsilon) } 
            +
            \frac{1}{( k^-_2 +i\epsilon) ( k^-_1 -i\epsilon) } 
            +  
            \frac{1}{( k^-_2 -i\epsilon) ( k^-_1 +i\epsilon) } \bigg) \notag \\
 -\frac{1}{2} S_2\left([1,2]3 \right)\bigg(&
            \frac{1}{( k^-_3 -i\epsilon) ( k^-_1 +i\epsilon) } 
            -
            \frac{1}{( k^-_3 +i\epsilon) ( k^-_1 -i\epsilon) } 
            +
            \frac{1}{( k^-_2 -i\epsilon) ( k^-_3 +i\epsilon) } 
\notag \\            
  &-
            \frac{1}{( k^-_2 +i\epsilon) ( k^-_3 -i\epsilon) } 
            +
            \frac{1}{( k^-_1 +i\epsilon) ( k^-_2 -i\epsilon) } 
            -  
            \frac{1}{( k^-_1 -i\epsilon) ( k^-_2 +i\epsilon) } \bigg) \notag \\
- \frac{1}{2} S_2\left([1,3]2 \right)\bigg(&
            \frac{1}{( k^-_2 -i\epsilon) ( k^-_1 +i\epsilon) } 
            -
            \frac{1}{( k^-_2 +i\epsilon) ( k^-_1 -i\epsilon) } 
            +
            \frac{1}{( k^-_3 -i\epsilon) ( k^-_2 +i\epsilon) } 
\notag \\            
  &-
            \frac{1}{( k^-_3 +i\epsilon) ( k^-_2 -i\epsilon) } 
            +
            \frac{1}{( k^-_1 +i\epsilon) ( k^-_3 -i\epsilon) } 
            -  
            \frac{1}{( k^-_1 -i\epsilon) ( k^-_3 +i\epsilon) } \bigg)
\end{align*}
\begin{align}
 \label{eq:eikonal_inbasis}
 -\frac{1}{2}S_2\left([2,3]1 \right)  \bigg(&
            \frac{1}{( k^-_3 -i\epsilon) ( k^-_1 +i\epsilon) } 
            -
            \frac{1}{( k^-_3 +i\epsilon) ( k^-_1 -i\epsilon) } 
            +
            \frac{1}{( k^-_2 +i\epsilon) ( k^-_3 -i\epsilon) } 
\notag \\            
  &-
            \frac{1}{( k^-_2 -i\epsilon) ( k^-_3 +i\epsilon) } 
            +
            \frac{1}{( k^-_1 -i\epsilon) ( k^-_2 +i\epsilon) } 
            -  
            \frac{1}{( k^-_1 +i\epsilon) ( k^-_2 -i\epsilon) } \bigg)
 \notag \\
 -S_3\left(123 \right)  \bigg(&
            \frac{1}{( k^-_3 -i\epsilon) ( k^-_1 +i\epsilon) } 
            +
            \frac{1}{( k^-_3 +i\epsilon) ( k^-_1 -i\epsilon) } 
            +
            \frac{1}{( k^-_2 -i\epsilon) ( k^-_3 +i\epsilon) } 
\notag \\            
  &+
            \frac{1}{( k^-_2 +i\epsilon) ( k^-_3 -i\epsilon) } 
            +
            \frac{1}{( k^-_1 +i\epsilon) ( k^-_2 -i\epsilon) } 
            +  
            \frac{1}{( k^-_1 -i\epsilon) ( k^-_2 +i\epsilon) } \bigg) \notag \\
\end{align}
To obtain the induced vertices with pole-prescription, we drop all
symmetric terms that come with an $S_2$ or an $S_3$ symbol and keep
only the terms coming with  double anti-symmetric structure, which yield the color structure of Eq.(\ref{inducedvertex2_reloaded}). Making
use of the Jacobi-Identity, Bose-symmetry of the induced vertex is
easily verified.  The pole structure can  further be simplified by
means of the eikonal identity
\begin{align}
  \label{eq:eikonal_id}
\frac{1}{ k_1^- + i\epsilon} \frac{1}{ k_1^- + k_2^- + i\epsilon}  + \frac{1}{ k_2^- + i\epsilon} \frac{1}{ k_1^- + k_2^- + i\epsilon} = \frac{1}{ k_1^- + i\epsilon} \frac{1}{ k_2^- + i\epsilon},
\end{align}
which holds not only in a purely algebraic sense, but is with the
above pole prescription also true in the sense of the theory of
distributions \cite{Bassetto:1991ue}, unlike the eikonal identity for
poles with principal value prescription, Eq.(\ref{eq:eik_id_cpv}).
Evaluating  commutators and adding the
color-octet projection together with the common factor of 
Eq.(\ref{eq:indu_2L_eps}), which altogether  corresponds to the following substitution
\begin{align}
  \label{eq:substution_symb_f}
 \left[[3,1],2\right]  & \to -ig^2{\bm q}^2 n^-_{\nu_1} n^-_{\nu_2} n^-_{\nu_3}f^{a_3a_2a}f^{a_1ac}, \notag \\
 \left[[3,2],1\right]& \to
-ig^2{\bm q}^2 n^-_{\nu_1} n^-_{\nu_2} n^-_{\nu_3}f^{a_3a_1a}f^{a_2ac} , 
\end{align}
we obtain for the second induced vertex
from Eq.~(\ref{eq:eikonal_inbasis})
\begin{align}
  \label{eq:double_comm_simpli}
\Delta^{\nu_1\nu_2\nu_3+}_{a_1a_2 a_3c} 
  =  -ig^2{\bm q}^2 n^-_{\nu_1} n^-_{\nu_2} n^-_{\nu_3}  \left[ f^{a_3a_2a}f^{a_1ac}
   g_2(k_3^-,k_2^-, k_1^-)
          + f^{a_3a_1a}f^{a_2ac}
       g_2(k_3^-,k_1^-, k_2^-)\right],
\end{align}
where
\begin{align}
  \label{eq:g2}
g_2(k_3^-,k_1^-, k_2^-)=   -\frac{1}{6}\bigg[&
            \frac{1}{ k^-_3 -i\epsilon } 
            \bigg( \frac{2}{  k^-_2 +i\epsilon} 
                                       + \frac{1}{  k^-_2 -i\epsilon} \bigg) 
            +
             \frac{1}{ k^-_3 +i\epsilon } 
            \bigg( \frac{2}{  k^-_2 -i\epsilon} 
                                       + \frac{1}{  k^-_2 +i\epsilon} \bigg) 
                             \bigg].
\end{align}
It is then easily verified that Eq.(\ref{eq:double_comm_simpli})
reduces to Eq.~(\ref{inducedvertex2}) in the limit $\epsilon = 0$.
Eq.(\ref{eq:double_comm_simpli}) can be further rewritten in a more
symmetric way, making use of the identity,
\begin{align}
  \label{eq:delta_func}
\frac{1}{k^- + i\epsilon} - \frac{1}{k^- -i\epsilon} = -2\pi i \delta(k^-),
\end{align}
which leads to
\begin{align}
  \label{eq:cpv_rep}
g_2(k_3^-,k_1^-, k_2^-) = 
    \bigg[&  \frac{-1}{[k_3^-][k_2^-]} -\frac{\pi^2}{3}\delta(k_2^-)\delta(k_3^-) \bigg].
\end{align}
With Eq.~(\ref{eq:eik_id_cpv}), 
Bose-symmetry of  Eq.(\ref{eq:cpv_rep}) is easily proven. 
 In particular, taking into account $k_1^- + k_2^- + k_3^- = 0$, we find
\begin{align}
  \label{eq:identity_poles}
g_2(k_3^-,k_2^-, k_1^-) = - g_2(k_1^-,k_3^-, k_2^-) - g_2(k_3^-,k_1^-, k_2^-).
\end{align}
It is therefore possible to replace Eq.(\ref{eq:indu_2L_epsa}) in the effective action by
\begin{align}
  \label{eq:alt_pole:indu2}
 \frac{1}{6}\bigg[  gv_-& \bigg[
\frac{1}{\partial_- -\epsilon } v_- \left( \frac{2}{\partial_- -\epsilon} + \frac{1}{\partial_- + \epsilon}   \right) \notag \\
& + 
\frac{1}{\partial_- +\epsilon } v_- \left( \frac{2}{\partial_- +\epsilon} + \frac{1}{\partial_- - \epsilon}   \right)
 \bigg]
  v_- 
\partial^2_{\sigma} A_+.
\end{align}
Every term proportional to a trace of four $SU(N_c)$ generators (as on
the right hand side of Eq.(\ref{eq:indu_2L_eps})), will have the pole
prescription of Eq.(\ref{eq:double_comm_simpli}). This prescription on
the other hand is  equivalent to Eq.(\ref{eq:cpv_rep}) and
satisfies Eq.(\ref{eq:identity_poles}). Therefore, any symmetric term
drops out, similar to the  case, where the
$i\epsilon$-prescription is absent and we obtain
Eq.(\ref{eq:double_comm_simpli}) directly from the effective action.

\subsection{The induced vertex of the third order}
\label{sec:indu3}

The derivation of the pole-prescription of the induced vertex of the
third order follows the same lines: Again we will chose a symmetry
basis built up of multiple commutators and symmetrization in the
remaining color indices. However the number of terms increases
rapidly with the number of gluons involved, and therefore the resulting
expression become large. We therefore present in the following only the
main results. In
particular, the color basis now  consists   24 terms and is given by
 \begin{align}
  \label{eq:change4_symbolic}
  \begin{array}[h]{c|c|c|c|c}
\left[\left[[4,1 ],2  \right],3   \right] &
S_2\left([1,2],[3,4]\right) &
S_2\left([[1,2] ,3]4\right) &
S_3\left([1,2]34\right) &
S_4\left(1234\right) \\
\left[\left[[4,1 ],3  \right],2   \right] &
S_2\left([1,3],[2,4]\right) &
S_2\left([[3,2] ,1]4\right) &
S_3\left([1,3]24\right)  &
 \\
\left[\left[[4,2 ],1  \right],3   \right] &
S_2\left([1,4],[2,3]\right)  &
S_2\left([[1,2] ,4]3\right)
 &
S_3\left([1,4]23\right) &
 \\
\left[\left[[4,2 ],3  \right],1   \right] &
& S_2\left([[4,2] ,1]3\right)
 &
S_3\left([2,3]14\right) &  \\
\left[\left[[4,3 ],1  \right],2   \right] &
& S_2\left([[1,3] ,4]2\right)
 &
S_3\left([2,4]13\right) &
 \\
\left[\left[[4,3 ],2  \right],1   \right] &
& S_2\left([[4,3] ,1]2\right)
 &
S_3\left([3,4]12 \right) &
\\
 && S_2\left([[2,3] ,4]1\right)
 &
&
\\ && S_2\left([[4,3] ,2]1\right)
 && \\
  \end{array}
\notag\\ &.
\end{align}
As for the first and second induced vertex, we start from the effective action with the following expression
\begin{align}
  \label{eq:indu3_starteff}
 -g^3v_-& \frac{1}{\partial_- -\epsilon} v_- \frac{1}{\partial_- -\epsilon} v_- \frac{1}{\partial_- -\epsilon} v_-\partial^2_{\sigma} A_+.
\end{align}
We obtain from this
expression 24 terms that each come  with four $SU(N_c)$ generators
projected on the color octet by the projector $\tr(.t^c)$ where  every
term corresponds to a specific ordering of the four generators. These
products of generators can then be expressed in terms of  the color basis
Eq.(\ref{eq:change4_symbolic}), which similarly to
Eq.(\ref{eq:eikonal_inbasis}), leads to certain combination of poles
for every basis-element. In general, each of the 24 basis elements
comes with 24 pole-terms.  For the induced vertex of the third order, we then drop all
symmetric terms, and keep only those proportional to a triple
commutator, of the first column of Eq.(\ref{eq:change4_symbolic}).  Making use of the eikonal identity,  Eq.(\ref{eq:eikonal_id}), we arrive at the
following pole-prescription
\begin{align}
  \label{eq:pole_indu3}
\Delta^{\nu_1\nu_2\nu_3+}_{a_1a_2 a_3c}& = 
-g^3 {\bm q}^2
n^-_{\nu_1} n^-_{\nu_2} n^-_{\nu_3} n^-_{\nu_4} \times \notag \\
\bigg[
& f^{a_4a_1d_2}f^{d_2a_3d_1}f^{d_1a_2c} 
 g_3(k^-_4, k_1^-,  k^-_3, k^-_2)
+
f^{a_4a_1d_2}f^{d_2a_2d_1}f^{d_1a_3c}
 g_3(k^-_4, k^-_1,  k^-_2, k^-_3)
\notag \\
+&
f^{a_4a_2d_2}f^{d_2a_1d_1}f^{d_1a_3c}
 g_3(k^-_4, k^-_2,  k^-_1 , k^-_3)
+
f^{a_4a_2d_2}f^{d_2a_3d_1}f^{d_1a_1c}
 g_3(k^-_4,+ k^-_2, k^-_3 , k^-_1)
\notag \\
+&
f^{a_4a_3d_2}f^{d_2a_1d_1}f^{d_1a_2c}
 g_3(k^-_4,  k^-_3, k^-_1 , k^-_2)
+
f^{a_4a_3d_2}f^{d_2a_2d_1}f^{d_1a_1c}
 g_3(k^-_4,  k^-_3, k^-_2, k^-_1) 
\bigg]
\end{align}
where the function $g_3(k_4^-,k_1,  k_3^-, k_2^-)$ is defined as
\begin{align}
\label{eq:functionf}
g_3(k_4^-, k_1^-,  k_3^-, &k_2^-) =
\notag \\
=-\frac{1}{12}\bigg\{&\frac{1}{k_4\!+\!i\epsilon} \bigg[
               \frac{1}{k_2^-\!\! +\!\! i\epsilon} \left( \frac{1}{k_2^-\!\! +\!\! k_3^- \!\!+\!\! i\epsilon }  +   \frac{1}{k_2^-\!\! +\!\! k_3^-\!\! - \!\!i\epsilon }\right) 
               +  
\frac{1}{k_2^-\!\! -\!\! i\epsilon} \left(   
               \frac{3}{k_2^-\!\! +\!\! k_3^-\!\! -\!\! i\epsilon }
               +
               \frac{1}{k_2^-\!\! + \!\!k_3^-\!\! +\!\! i\epsilon }
\right)
                                           \bigg]
\notag \\
& \frac{1}{k_4\!-\!i\epsilon} \bigg[
               \frac{1}{k_2^-\!\! -\!\! i\epsilon} \left( \frac{1}{k_2^-\!\! +\!\! k_3^- \!\!-\!\! i\epsilon }  +   \frac{1}{k_2^-\!\! +\!\! k_3^-\!\! + \!\!i\epsilon }\right) 
               +  
\frac{1}{k_2^-\!\! +\!\! i\epsilon} \left(   
               \frac{3}{k_2^-\!\! +\!\! k_3^-\!\! +\!\! i\epsilon }
               +
               \frac{1}{k_2^-\!\! + \!\!k_3^-\!\! -\!\! i\epsilon }
\right)
                                           \bigg]
\bigg\}
\end{align}
Similarly to Eq.(\ref{eq:eik_id_cpv}), the function  $g_3 (k_2^-, k_2^- + k_3^-, k_2^-)$  can be written as combination of Cauchy-principal values and delta-functions:
\begin{align}
  \label{eq:f_cpv}
g_3(k_4^-, k_1^-,  k_3^-, k_2^-)
 =\bigg(&
          \frac{-1}{[k_4^-][k_2^- + k_3^-][k_2^-]}
            -
            \frac{\pi^2}{3} \delta(k_2^-) \delta(k_3^-) \frac{-1}{[k_4^-]} 
             \notag \\
             &-
            \frac{\pi^2}{3} \delta(k_2^-) \delta(k_4^-) \frac{1}{[ k_3^-]} 
         - 
          \frac{\pi^2}{3} \delta(k_2^- + k_3^-) \delta(k_4^-) \frac{1}{[k_2^-]} \bigg).
\end{align}
With Eq.(\ref{eq:eik_id_cpv}) it can then be shown that identities like
\begin{align}
  \label{eq:ids_indu4}
g_3(k_4^-, k_1^-,  k_3^-, k_2^-) = g_3(k_2^-, k_3^-,  k_1^-, k_4^-) +g_3(k_2^-, k_3^-,  k_4^-, k_1^-)+g_3(k_2^-, k_1^-,  k_3^-, k_4^-)
\end{align}
hold, which allow to prove Bose-symmetry of the induced vertex with
the above pole-prescription. Furthermore, due to those identities, we can
state the pole-prescription for the induced vertex of the third order
already at the action level, similar to the case of first and the
second induced vertex; Eq.(\ref{eq:indu3_starteff}) is to replace by: 
\begin{align}
  \label{eq:alt_pole:indu3}
 \frac{1}{12}\bigg[  gv_-\bigg(
\frac{1}{\partial_- -\epsilon } v_-& \bigg[ \frac{1}{\partial_- -\epsilon } v_-\left( \frac{1}{\partial_- -\epsilon} + \frac{1}{\partial_- + \epsilon}   \right) \notag \\
&
 + 
\frac{1}{\partial_- +\epsilon } v_- \left( \frac{3}{\partial_- +\epsilon} + \frac{1}{\partial_- - \epsilon}   \right)
 \bigg]
\notag \\ 
\frac{1}{\partial_- +\epsilon } v_-& \bigg[ \frac{1}{\partial_- +\epsilon } v_-\left( \frac{1}{\partial_- +\epsilon} + \frac{1}{\partial_- - \epsilon}   \right) \notag \\
&
 + 
\frac{1}{\partial_- -\epsilon } v_- \left( \frac{3}{\partial_- -\epsilon} + \frac{1}{\partial_- + \epsilon}   \right)
 \bigg]
 v_- 
\partial^2_{\sigma} A_+.
\end{align}
Obviously Eq.(\ref{eq:functionf}) and Eq.(\ref{eq:f_cpv}) follow a
certain symmetry pattern and taking further into account Eq.
Eq.(\ref{eq:double_comm_simpli}) and Eq.(\ref{eq:def_cauchypv}) it
seems plausible that there exists a general formula that fixes the
poles of all induced vertices. This suggests that the poles of the induced vertex of the
$n$th order with $n + 1$ gluons are described by  
the following  function $g$
\begin{align}
  \label{eq:speculation}
g(&k_{n+1}^-, k_n^-,  \ldots, k_2^-) =
\frac{1}{c_n}\bigg\{ 
\left( \frac{1}{k_{n+1}\!+\!i\epsilon} +  \frac{1}{k_{n+1}\!-\!i\epsilon}  \right)  
\left( \frac{1}{k_2^-\!\! +\!\ldots\! + \!\! k_{n-1}^-  \!\!+\!\! i\epsilon }  +
 \frac{1}{k_2^-\!\! +\!\ldots\! + \!\! k_{n-1}^-  \!\!-\!\! i\epsilon }  
\right) 
\notag \\
&\times \cdots \times
           \left(  \frac{1}{k_2^-\!\! +\!\! i\epsilon} - \frac{1}{k_2^-\!\! +\!\! i\epsilon}  \right)
               +  
\frac{n}{( k_{n+1}\!-\!i\epsilon) (k_2^-\!\! +\! \dots \! +\!\! k_{n-1}^- \!\!+\!\! i\epsilon) (k_2^-\!\! +\! \dots \! +\!\! k_{n-2}^- \!\!+\!\! i\epsilon) \cdots  (k_2^-\!\! +\!\! i\epsilon)}
\notag \\
& \qquad \qquad \qquad \qquad \frac{n}{( k_{n+1}\!+\!i\epsilon) (k_2^-\!\! +\! \dots \! +\!\! k_{n-1}^- \!\!-\!\! i\epsilon) (k_2^-\!\! +\! \dots \! +\!\! k_{n-2}^- \!\!-\!\! i\epsilon) \cdots  (k_2^-\!\! +\!\! i\epsilon)}
\bigg\},
\end{align}
where 
\begin{align}
  \label{eq:defk}
 c_n &= 2^n +2(n-1)& \text{with}&& c_1 = 2, c_2 = 6, c_3 = 12 .
\end{align}
The above formula is in accordance with the induced vertices of the
first, second and third order.  Whether this function really yields
the correct pole structure for the induced vertex of the fifth order,
for instance, and whether it yields for a general induced vertex
Bose-symmetry remains however to be verified.

\section{Quark-impact factors and subtraction terms}
\label{sec:impa4_bkp}

In the present section we begin our study of the elastic quark-quark
scattering amplitude where the interaction of the quarks involves the
exchange of three and four reggeized gluons in the $t$-channel.

\subsection{The quark-impact factor with three gluons}
\label{sec:quark_impa3}
\begin{figure}[htbp]
  \centering 
\parbox{6cm}{\center \includegraphics[height=2.7cm]{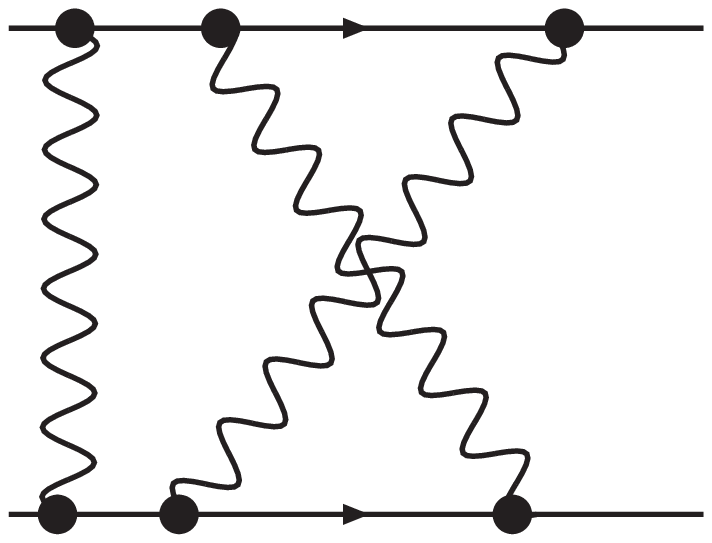}} 
\parbox{6cm}{\center \includegraphics[height=2.7cm]{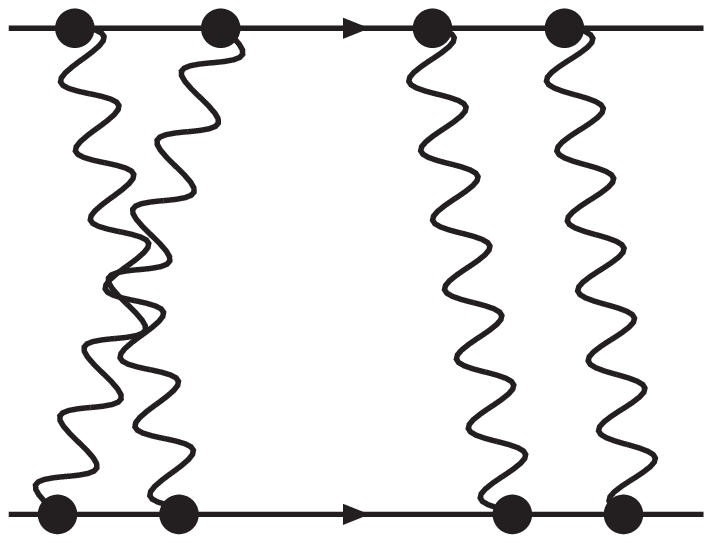}}  \\
  \parbox{6cm}{\center (a)}  \parbox{6cm}{\center (b)}
  \caption{\small Typical diagrams for the exchange of three (a) and four (b) reggeized gluons. In both cases, longitudinal integrations factorize, in  analogy to the exchange of two reggeized gluons}
  \label{fig:3regg}
\end{figure}
To leading order in $g$, the state of three reggeized gluons appears
for the first time with diagrams like Fig.~\ref{fig:3regg}a.  Similar
to the exchange of two reggeized gluons, the longitudinal integrals can
be factorized between the two scattering quarks and it is therefore
sufficient to consider for the longitudinal integral the impact factor
of the quark alone. The Born-amplitudes with exchange of three
reggeized gluons is therefore given by the following convolution
\begin{align}
  \label{eq:born_3gluon}
      \mathcal{M}_{2 \to 2}^{\text{B}|\text{3R}} = 
\frac{2(2\pi)^2}{3!} p_A^+p_B^-   A^{a_1a_2a_3}_{(3;0)} ({\bm k}_1,{\bm k}_2,{\bm k}_3 ) A^{a_1a_2a_3}_{(3;0)} ({\bm k}_1,{\bm k}_2,{\bm k}_3 ).
\end{align}
The convolution symbol is defined to contain both transverse integrals and the transverse propagators of the reggeized gluons,
\begin{align}
  \label{eq:convolf3}
\otimes_{\bm{k}_{\{123\}}} = \int \frac{d^2 {\bm k}_1 }{(2\pi)^3}  \int \frac{d^2 {\bm k}_2 }{(2\pi)^3}  \frac{1}{{\bm k}_1^2 {\bm k}_2^2{\bm k}_3^2},
\end{align}
with the constraint ${\bm k}_1 +{\bm k}_2 + {\bm k}_3 = {\bm q} $
implied.  The factor $1/3!$ is a symmetry factor, while the dependence
of the impact factor on the external color labels of the quarks has
been suppressed in our notation.  As the impact factors can be shown
not to depend on $p_A^+$ and $p_B^-$ and are therefore independent of
$s$, we note that the amplitude Eq.~(\ref{eq:born_3gluon}) is real and
carries negative signature. This is expected, as we consider the
exchange of a  state of three reggeized gluons, which has
 negative signature itself. Signature of a Regge-cut is given by the product of signature of contributing Regge-poles.   The 3-gluon quark impact factors
is then obtained from the following integral:
\begin{align}
  \label{eq:quark_3gluon}
 i A^{a_1a_2a_3}_{AA',(3;0)}&  ({\bm k}_1,{\bm k}_2,{\bm k}_3 )  = 
          (ig)^3\int\frac{d \mu_1}{(-2\pi i)}\int\frac{d \mu_2}{(-2\pi i)}
    \bigg[
         (t^{a_1}t^{a_2}t^{a_3})_{AA'} F_3(p_A, p_A' ; k_1, k_2, k_3) \notag \\ 
         &+
         (t^{a_2}t^{a_1}t^{a_3})_{AA'} F_3(p_A, p_A' ;k_2, k_1, k_3)
         + 
         (t^{a_1}t^{a_3}t^{a_2})_{AA'} F_3(p_A, p_A' ;k_1, k_3, k_2)
\notag \\         
         &         
 +
         (t^{a_3}t^{a_1}t^{a_2})_{AA'} F_3(p_A, p_A' ;k_3, k_1, k_2)
         +
         (t^{a_2}t^{a_3}t^{a_1})_{AA'} F_3(p_A, p_A' ;k_2, k_3, k_1)
\notag \\         
         &         
+
         (t^{a_3}t^{a_2}t^{a_1})_{AA'} F_3(p_A, p_A' ;k_3, k_2, k_1)
          \bigg],
\end{align}
with
\begin{align}
  \label{eq:grossF_def}
 F_3& (p_A, p_A' ;k_1, k_2, k_3) =
\notag \\
 &\frac{1}{(\mu_1 + p_A^+p_A^- -  ( {\bm p}_A - {\bm k}_1)^2  -m_A^2+ i\epsilon)(\mu_1 + \mu_2 + p_A^+p_A^-  - ( {\bm p}_A - {\bm k}_1 - {\bm k}_2)^2 -m_A^2 + i\epsilon)} .
\end{align}
In Eq.~(\ref{eq:grossF_def}) we defined, similar to
Sec.\ref{sec:tworeggeon_negsig}. $\mu_i = -p_A^+k_i^-$ with $i =
1,2,3$ with the constraint $\mu_1 + \mu_2 + \mu_3 = 0$. We want to
note here that the statement concerning signature of the 
state of three reggeized gluons, can be immediately connected to the
variable change $\mu_i = -p_A^+k_i^-$. Whereas for the state
of two reggeized gluons, with only one integral over $k^-$, the
variable change yields a Jacobian factor $1/|p_A^+|$, for three
reggeized gluons, the Jacobian is $1/(p_A^+)^2$ instead. While
corresponding factors in the numerator lead always to an amplitude
proportional to $ s$, the precise structure of the Jacobian factors in
the denominator determines the behavior of the amplitude under the
substitution $s \to -s$ and hence its signature. We further note 
that the above
function $F_3$ is more general than needed for the determination of
the quark-impact factor. For instance for the quark-impact factor the transverse quark-momentum is zero,
${\bm p}_A = 0$, and $p_A^+p_A^- = m_A^2$ due to on-shellness of the quarks and cancels with the corresponding term. 
Later we shall encounter examples which require the general form of $F_3$, while the integral over the $\mu_i$ can be treated in complete analogy to the present case.   Actually, as far as the
evaluation of the integral over $\mu_i$ is concerned, we are mainly
interested in the $\mu_i$-dependence and it is therefore convenient to
define quantities
\begin{align}
  \label{eq:mij}
m_{i}^2 &= - p_A^+p_A^- +  ( {\bm p}_A - {\bm k}_i)^2  -m_A^2
\notag \\
m_{ij}^2 &=  -p_A^+p_A^-  + ( {\bm p}_A - {\bm k}_i - {\bm k}_j)^2-m_A^2
\notag \\
m_{ijk}^2 &=  -p_A^+p_A^-  + ( {\bm p}_A - {\bm k}_i - {\bm k}_j -{\bm k}_k)^2-m_A^2
 \qquad \text{etc.}
\end{align}
 that gather the (for the moment)  redundant information such that
\begin{align}
  \label{eq:grossF_defe}
 F_3(p_A, p_A' ;k_1, k_2, k_3) = &\frac{1}{(\mu_1 - m_{1}^2 +  i\epsilon)(\mu_1 + \mu_2 - m_{12}^2 + i\epsilon)} 
\end{align}
We further note that $F_3$ contains only contributions due to the
direct coupling of the reggeized gluons to the quark and no contributions
due to subtraction terms are included at this stage.  Attempting 
a straight forward integration of $F_3$ over $\mu_1$ and $\mu_2$ one
 realizes soon that this integrals are not well-defined: Starting with
the integration over $\mu_1$, the integral is convergent and yields
zero result, as all singularities lie on same side of the integration
contour. Starting with  $\mu_2$ the integral turns out to be
divergent and cannot be evaluated.  Changing furthermore integrations
variables from $\mu_2$ to $\mu_3 = -\mu_1 - \mu_2$, also the first
integral turns out to be divergent and can no longer be calculated. It
is however well known that integrals like Eq.~(\ref{eq:quark_3gluon})
can give nevertheless a meaningful result from the QED-limit, where all
$SU(N_c)$-generators are replaced by a unit matrix. This allows for
cancellation of all singularities  between  individual
terms. In QED, this is known as the 'eikonal formula'(see for instance
\cite{Cheng:1987ga} and references therein) and it can be demonstrated
that the integral over the sum of all terms leads to a finite result.

In QCD, on the other hand, such a cancellation of divergences between
individual terms is prohibited by the color structure of the
individual terms. It is therefore necessary to find a different solution to
the problem. An analogous problem was  already present for the
state of two reggeized gluons in Sec.\ref{sec:tworeggeon_negsig}. In that case, the expression
corresponding to the above $F_3$ is given by
\begin{align}
  \label{eq:F2}
 F_2(k_1, k_2) =  \frac{1 }{\mu_1 - m_1^2   + i \epsilon}.
\end{align}
with $m_1^2$ defined by Eq.~(\ref{eq:mij}). It is then possible to decompose this function further, 
\begin{align}
  \label{eq:F2_decompose}
 F_2(k_1, k_2) = F_{2}^R(k_1, k_2) + g_1(\mu_1,\mu_2),
\end{align}
where
\begin{align}
  \label{eq:F2R}
F_{2}^R(k_1,k_2) = \frac{1}{2} \left( \frac{1}{\mu_1 + i\epsilon} +  \frac{1}{\mu_1 - i\epsilon} \right)
\frac{ m_1^2 }{\mu_1 - m_1^2  + i \epsilon} ,
\end{align}
yields a convergent integrand in $\mu_1$, while the divergent part is
contained in the function $g_1(\mu_1)$, which is the pole part of the
first induced vertex, Eq.(\ref{eq:indu1_pole_g}).  Making
 use of a subtraction graph, the function $g_1(\mu_1, \mu_2)$ in
Eq.~(\ref{eq:F2_decompose}) is canceled and one is left with the
convergent integral over $F_2^R$.

So far the subtraction terms have been obtained from the diagram,
where the state of two reggeized gluons arises from a direct decay of
a single reggeized gluon into two gluons, by means of an induced
vertex (compare Fig.~\ref{fig:threeR} and Fig.~\ref{fig:rprr}e). This term
was then subtracted from the original expression. As we will see in
short, also the for $F_3$ the divergent parts of the integrand can be
isolated and removed by  means of corresponding subtraction terms.
However in that case it becomes rather uncomfortable to first
determine all possible direct couplings of reggeized gluons to each
other and then to subtract them in the appropriate way. It seems
therefore advisable to supplement the effective Lagrangian by a
subtraction term. It generates these subtraction diagrams
automatically and the  correct choice of sign is implemented
automatically. Furthermore, this term also allows to remove the
direct couplings of the reggeized gluons to each other by induced
vertices, as they are now meaningless\footnote{It has been thought
  that those terms might be needed for more complicated processes,
  like the six-point-amplitude for instance. So far however we could
  not encounter any need for those transitions in any amplitude
  studied by us so far and we therefore suggest to remove these terms
  from the Lagrangian. Indeed we believe that those terms would
  instead lead to an over-counting which could be only cured by
  introducing another cut-off, which to our understanding is not necessary. }. The term to be added to the
Lagrangian is  given by
\begin{align}
  \label{eq:lagrangian_subtract} 
\mathcal{L}_{\text{subtract}}(A_+, A_-) = -2 \tr (A_+(A_+) - A_+ )\partial_\sigma^2 A_- -2 \tr (A_-(A_-) - A_- )\partial_\sigma^2 A_+ ,
\end{align} 
with
\begin{align}
\label{eq2:efflagrangianrr}
A_\pm(A_\pm) =
&
-\frac{1}{g}\partial_\pm U(A_\pm)
 = 
-\tr\frac{1}{g} \partial_\pm \mathcal{P}\exp\bigg(-\frac{1}{2} g \int_{-\infty}^{x^\pm}dx'^\pm A_\pm(x')\bigg)  .
 \end{align}
The complete effective action then reads
\begin{align}
  \label{effaction_modi}
  S_{\text{eff}} &= \int \text{d}^4 x [\mathcal{L}_{\text{QCD}}(v_\mu, \psi) + \mathcal{L}_{\text{ind}} (v_\pm, A_\pm) + \mathcal{L}_{\text{subtract}}(A_+, A_-) ].
\end{align}
As this term only depends on the reggeized gluon fields, it does not
contribute to the tree-diagrams of the production amplitudes in the
Quasi-Multi-Regge-Kinematics and it is therefore not in conflict with the original formulation of the effective action. 
\begin{figure}[htbp]
  \centering  
\parbox{3cm}{\center \includegraphics[height=1.3cm]{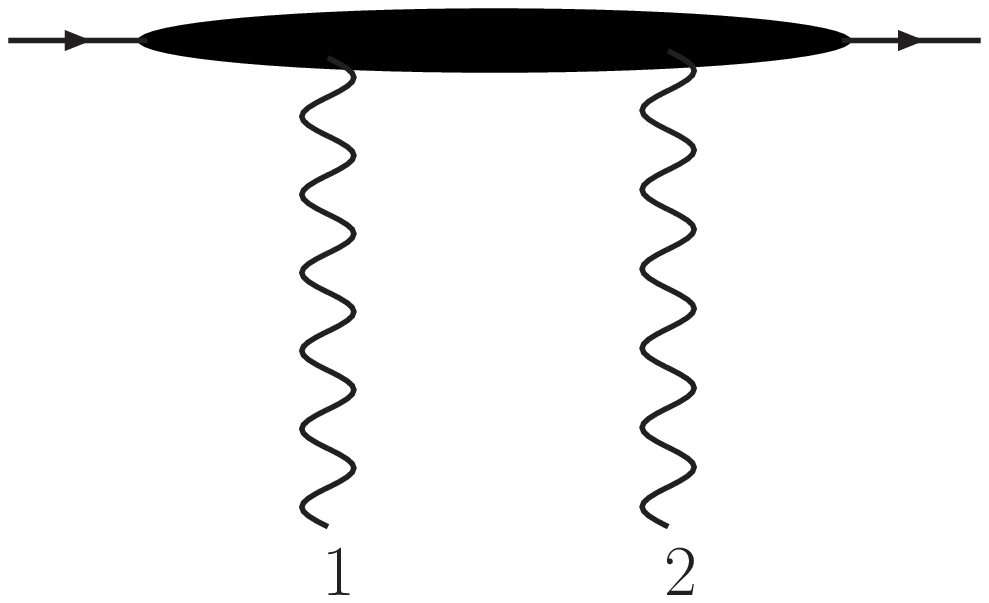}} {\large = }
\parbox{3cm}{\center \includegraphics[height=1.3cm]{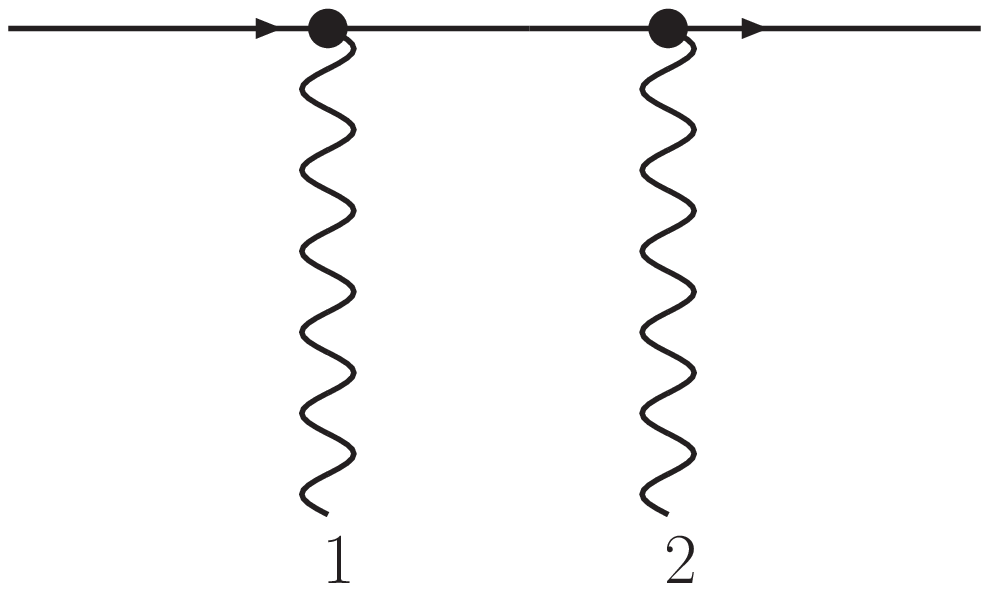}} {\large + } 
\parbox{3cm}{\center \includegraphics[height=1.3cm]{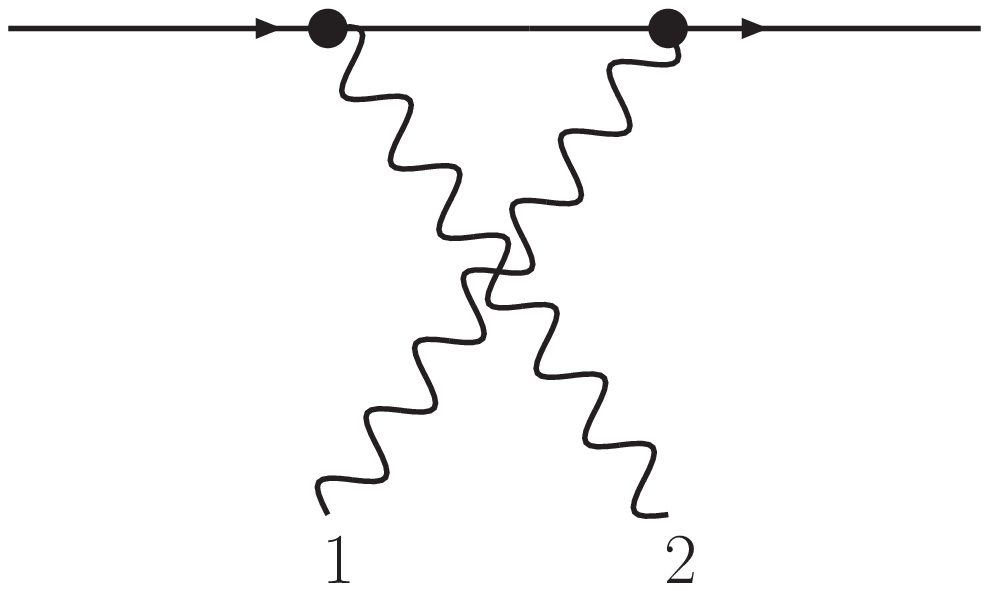}}  {\large + }
\parbox{3cm}{\center \includegraphics[height=1.3cm]{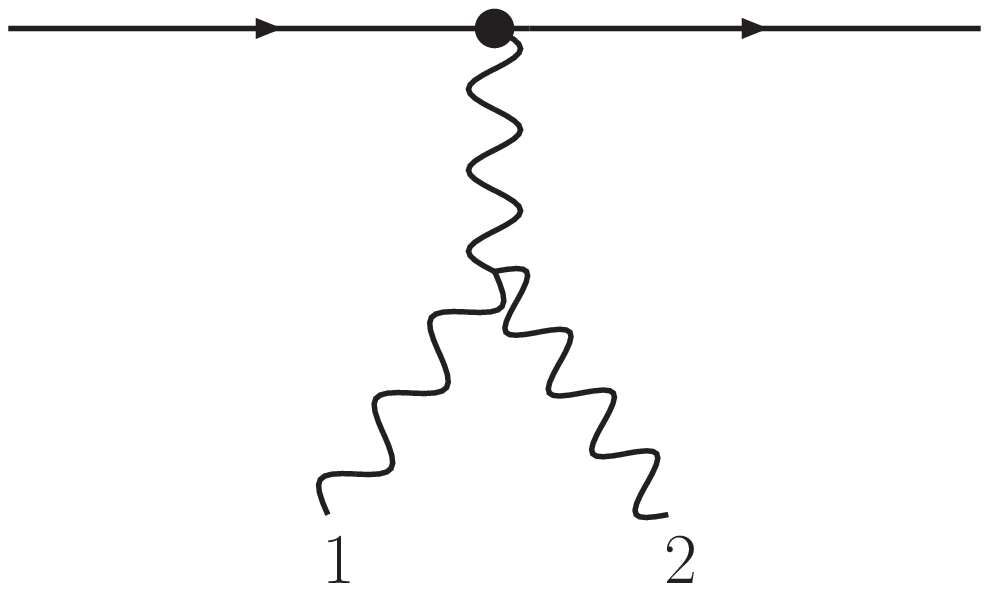}} 
   \caption{\small The two reggeized gluon impact factor is given as the sum of the direct couplings of the two reggeized gluons to the quark, minus the subtraction graph, integrated over $\mu_1 = -p_A^+k_1^-$.  }
  \label{fig:ghost_graph}
\end{figure}
The resulting subtraction diagrams then remove the direct couplings of
reggeized gluons by induced vertices  and parts of the
particle-$n$-reggeized gluons amplitudes, that are already contained
in the induced vertices and which would lead otherwise to a
double-counting of parts of the underlying QCD-diagrams.  We therefore remove with this term two different contribution which explains the overall factor of two in  Eq.(\ref{eq:lagrangian_subtract}).

For the two-reggeized gluon impact factor, all contributing diagrams
are then given in Fig.\ref{fig:ghost_graph}.  Note that
Eq.~(\ref{eq:lagrangian_subtract}) already contains the minus sign
needed for the subtraction and therefore the subtraction graphs appear
in Fig.\ref{fig:ghost_graph} with a relative plus. The diagram with
direct coupling of three reggeized gluons denotes the subtraction
term. Direct coupling of a reggeized gluon by an induced vertex on the
other hand are from now understood to be removed from the effective
action by the subtraction terms and appear no longer.

\begin{figure}[htbp]
\parbox{2.3cm}{\center \includegraphics[height=1cm]{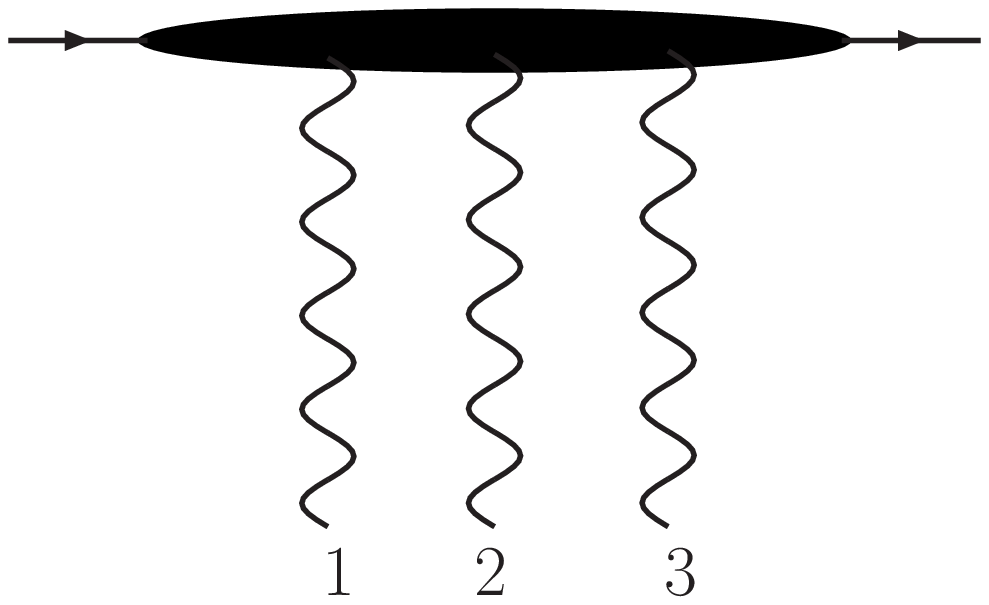}} {\large = }
\parbox{2.3cm}{\center \includegraphics[height=1cm]{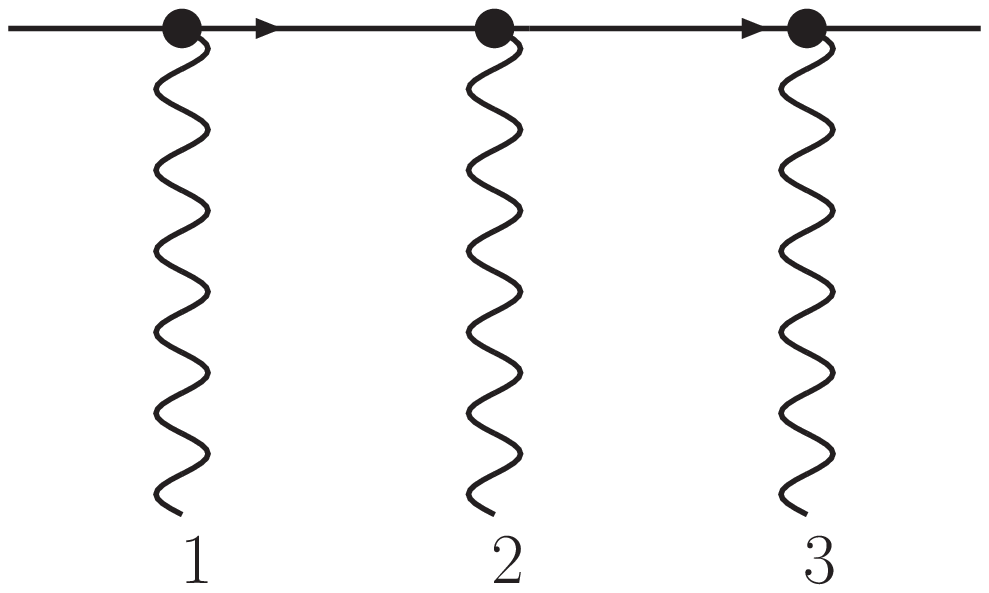}}  {\large +}
\parbox{2.3cm}{\center \includegraphics[height=1cm]{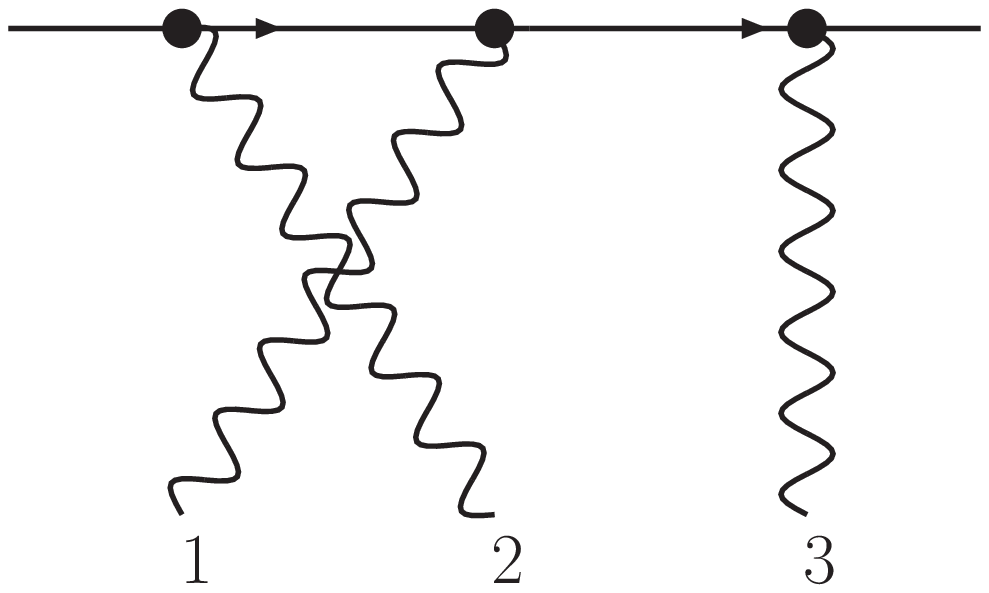}}  {\large +}
\parbox{2.3cm}{\center \includegraphics[height=1cm]{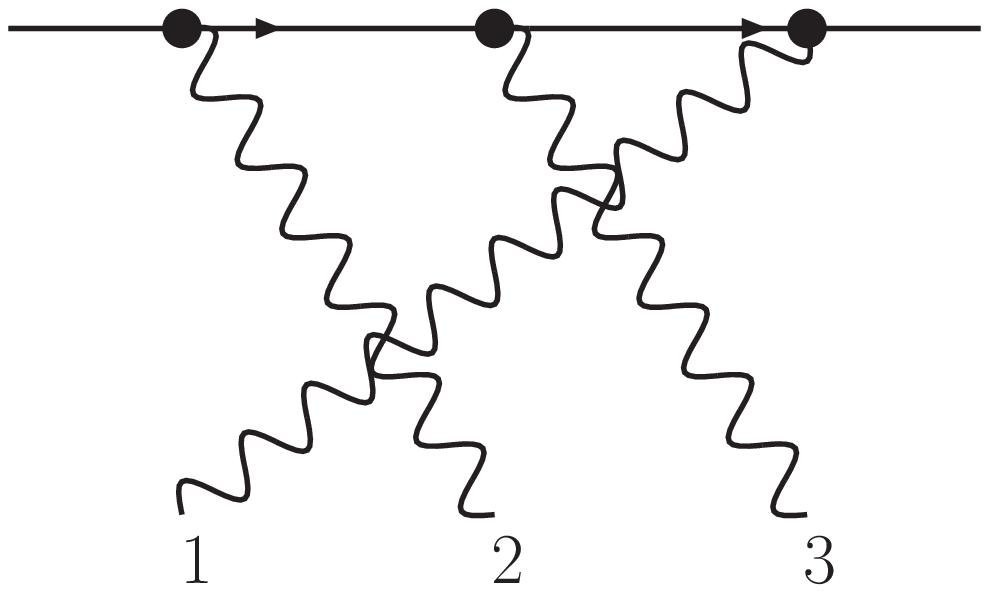}}  {\large +}
\parbox{2.3cm}{\center \includegraphics[height=1cm]{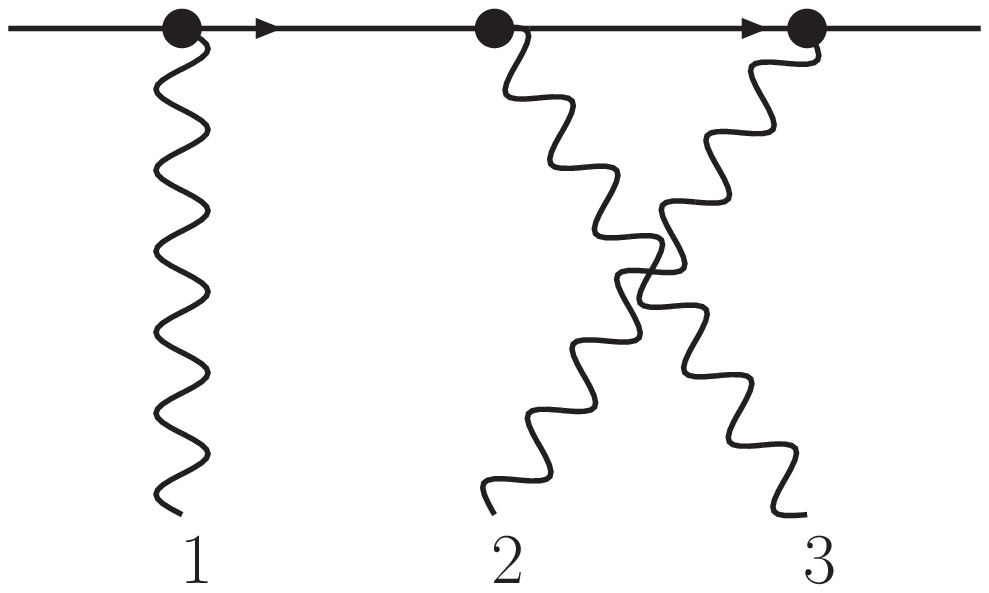}}  {\large +} \\
\parbox{2.9cm}{\center $\,$}
\parbox{2.3cm}{\center \includegraphics[height=1cm]{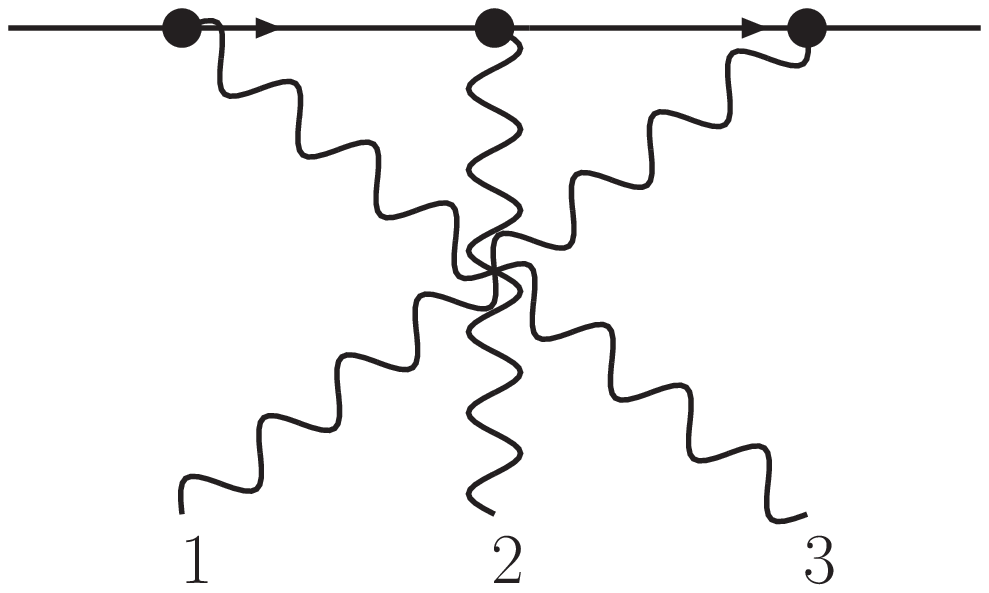}}  {\large +}
\parbox{2.3cm}{\center \includegraphics[height=1cm]{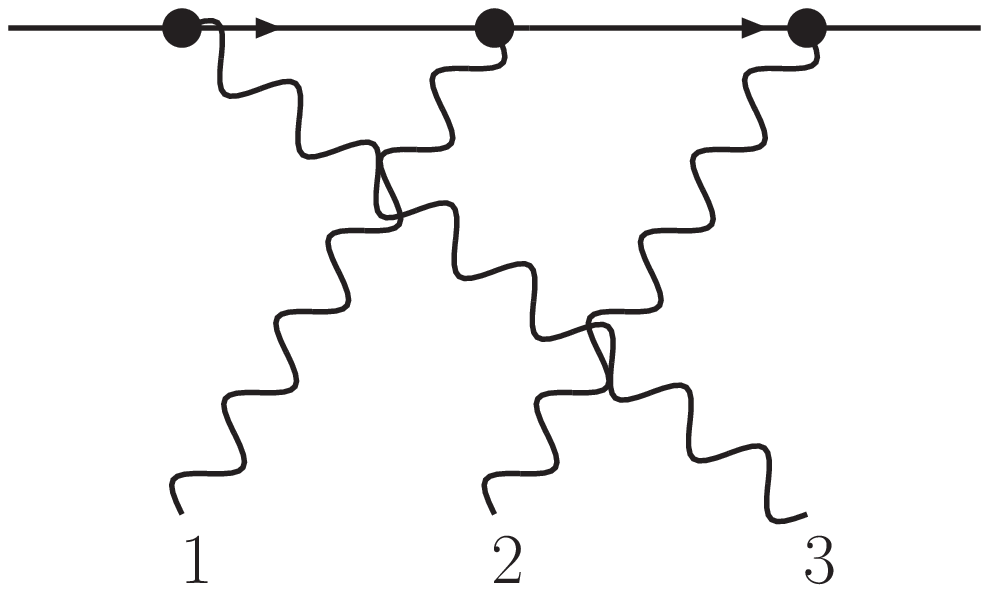}}  {\large +} 
\parbox{2.3cm}{\center \includegraphics[height=1cm]{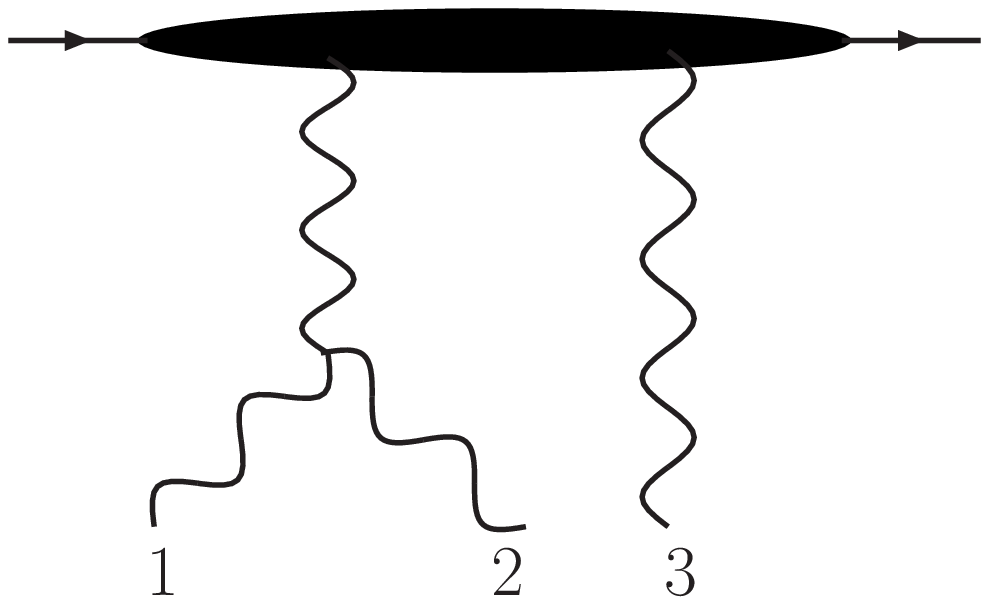}} {\large +}
\parbox{2.3cm}{\center \includegraphics[height=1cm]{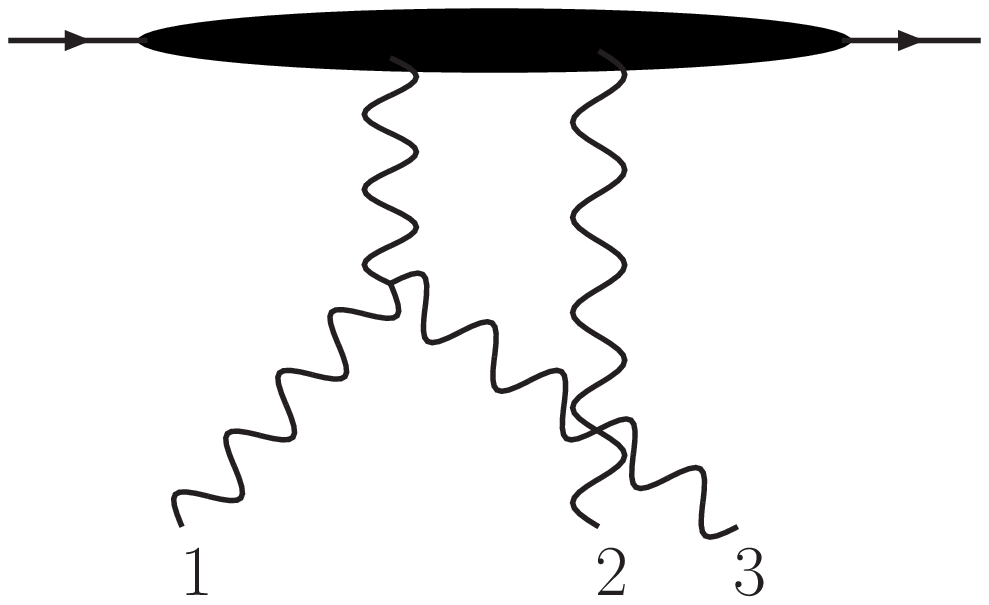}} {\large +}\\
\parbox{2.9cm}{\center $\,$}
\parbox{2.3cm}{\center \includegraphics[height=1cm]{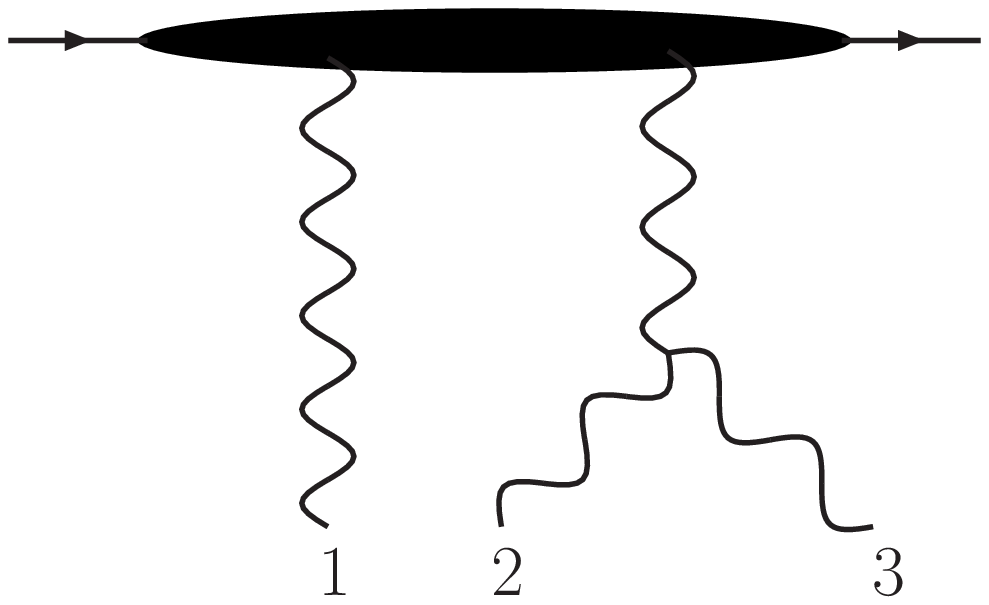}} {\large +}
\parbox{2.3cm}{\center \includegraphics[height=1cm]{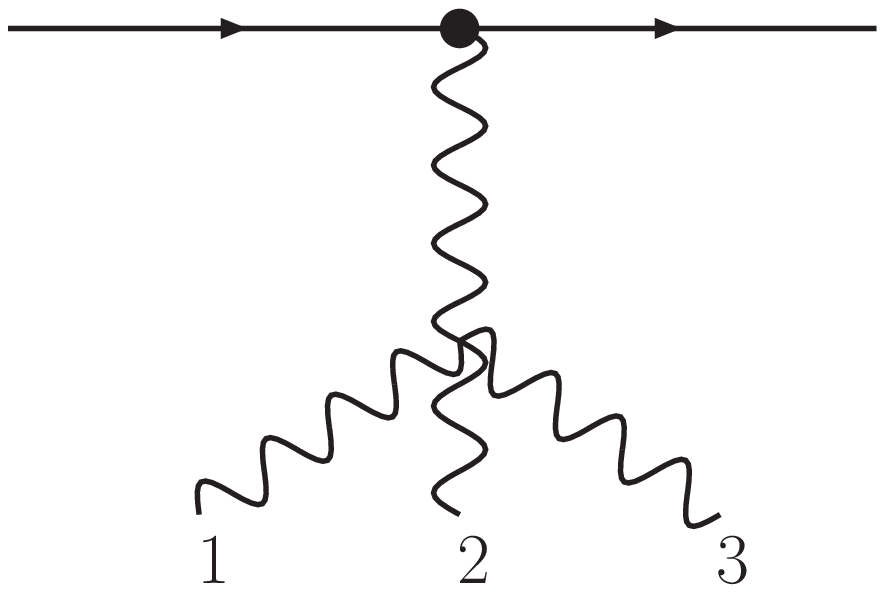}}
  \caption{\small Diagrams that contribute to the three reggeized gluon quark-impact-factor. For the sum of the subtraction graphs, we used the two-reggeized gluon to compactify the presentation. }
  \label{fig:impa3}
\end{figure}

For the three-reggeized-gluon impact-factor, the relevant graphs,
including subtraction diagrams are depicted in Fig.~\ref{fig:impa3},
where we used the two-reggeized gluon impact factor,
Fig.~\ref{fig:ghost_graph}, to compactify the representation.

To evaluate the integrations in Eq.~(\ref{eq:quark_3gluon}) one can use  for the function $F_3$  the
following decomposition
\begin{align}
  \label{eq:grossF_decomp}
 F_3 (k_1, k_2, k_3) &=  F^R_3 (k_1, k_2, k_3) + g_2(\mu_1, \mu_2,  \mu_3)
\notag \\
 & F_2^R(k_1, k_2 + k_3)  g_1(\mu_1 + \mu_2, \mu_3) +  F_2^R(k_1 +k_2,  k_3)  g_1(\mu_1 , \mu_2 + \mu_3) ,
\end{align}
with the regularized $F_3$-function
\begin{align}
  \label{eq:grossFR}
 F^R_3 &(k_1, k_2, k_3) =   \frac{{m}_1^2{m}_{12}^2}{(\mu_1 - {m}_1^2 + i\epsilon)(\mu_1 +\mu_2 - {m}_{12}^2 + i\epsilon)}
\notag \\
   & \times
\frac{1}{6} \bigg[
\frac{1}{\mu_1\! +\!i\epsilon} 
 \left(\frac{2}{\mu_1\! +\! \mu_2 \!-\!i\epsilon} + \frac{1}{\mu_1 \!+ \!\mu_2\! +\!i\epsilon} \right)  +  \frac{1}{\mu_1 \!-\!i\epsilon} \left(\frac{2}{\mu_1\! +\! \mu_2 \!+\!i\epsilon} + \frac{1}{\mu_1\! +\! \mu_2\! -\!i\epsilon} \right)\bigg]
 ,
\end{align}
where we suppressed in our notation the dependence of $F_3$ on $p_A,
p_A'$. For all terms in Eq.~(\ref{eq:grossF_decomp}) that come with  a
function $g_2$ or $g_3$, it can be demonstrated that the integral over
$\mu_1$ and $\mu_2$ yields zero result if these terms are combined
with the  subtraction diagrams. The integral over
$F_3^R$ on the other hand is convergent,
\begin{align}
  \label{eq:int_F3}
\int\frac{d \mu_1}{(-2\pi i)}\int\frac{d \mu_2}{(-2\pi i)}  F^R_3 (k_1, k_2, k_3) = \frac{1}{6}.
\end{align}
We obtain hence for the quark-impact-factor
\begin{align}
  \label{eq:quark_3gluon_fertig}
  iA^{a_1a_2a_3}_{AA',(3;0)} & ({\bm k}_1,{\bm k}_2,{\bm k}_3 ) = 
  (ig)^3 \frac{1}{3!}\sum_{i_1 \ldots i_3} t^{a_{i_1}}t^{a_{i_2}}t^{a_{i_3}},
\end{align}
where the sum is over all permutations of numbers $1, \ldots, 3$.  An
important check on the correctness of this result is given by the
QED-limit, where all $SU(N_c)$-generators are replaced by 1, which can  easily be verified to be fulfilled.

\subsection{The quark-impact factor with four gluons}
\label{sec:quark_impa4}

The determination of the quark-impact factor follows a similar scheme: Starting from 
\begin{align}
  \label{eq:born_3gluon}
      \mathcal{M}_{2 \to 2}^{\text{B}|\text{4R}} = -i\frac{2(2\pi)^3}{4!}
| p_A^+||p_B^-| 
 & A^{a_1a_2a_3a_4}_{(4;0)} ({\bm k}_1,{\bm k}_2,{\bm k}_3 , {\bm k}_4) 
\otimes_{{\bm k}_{1234}}
 A^{a_1a_2a_3a_4}_{(4;0)} ({\bm k}_1,{\bm k}_2,{\bm k}_3,{\bm k}_4 ),
\end{align}
where the convolution symbol is defined as
\begin{align}
  \label{eq:convopl4}
\otimes_{{\bm k}_{1234}} =
\int \frac{d^2 {\bm k}_1 }{(2\pi)^3}  \int \frac{d^2 {\bm k}_2 }{(2\pi)^3}\int \frac{d^2 {\bm k}_3 }{(2\pi)^3}   \frac{1}{{\bm k}_1^2 {\bm k}_2^2{\bm k}_3^2{\bm k}_4^2},
\end{align}
with the constraint ${\bm k}_1 + \ldots + {\bm k}_4 = {\bm q}$.
As appropriate for the  state of four reggeized gluons we observe that the above amplitude carries positive signature.  The four gluon impact factor results from the following integral, 
\begin{align}
  \label{eq:quark_4gluon}
  A^{a_1a_2a_3}_{AA',(3;0)} & ({\bm k}_1,{\bm k}_2,{\bm k}_3 )  = 
          (ig)^4\int\frac{d \mu_1}{(-2\pi i)} \int\frac{d \mu_2}{(-2\pi i)}\int\frac{d \mu_3}{(-2\pi i)}
\notag \\   
&  \sum_{i_1, \ldots i_4}
         (t^{a_{i_1}}t^{a_{i_2}}t^{a_{i_3}}t^{a_{i_4}} )_{AA'} 
\left[ F_4(k_{i_1}, k_{i_2}, k_{i_3},k_{i_4} ) +  F^S_4(k_{i_1}, k_{i_2}, k_{i_3},k_{i_4} )
\right],
\end{align}
where the sum is over all permutations of the numbers $1, \ldots, 4$. The function $F_4$ is defined as 
\begin{align}
  \label{eq:def_F4}
F_4&( p_A, p_A'; k_1, k_2, k_3, k_4) = 
\frac{1}{\mu_1  - m_1^2 \!+\! i\epsilon}
\frac{1}{\mu_1 + \mu_2  - m_{12}^2 \!+\! i\epsilon}
\frac{1}{\mu_1 + \mu_2 + \mu_3 -   m_{123}^2\! + \!\!i\epsilon } ,
\end{align}
and contains all contributions due to the direct coupling of the
reggeized gluon to the quark. $F_4^S$ on the other hand contains all
 contributions that arise due to the presence of the subtraction diagrams and we defined as previously $\mu_i =-p_A^+k_i^- $ with $i=1, \ldots 4$ and $\sum_{i=1}^4\mu_i= 0$.

Similarly to $F_3$, the function $F_4$ can be decomposed into a convergent part $F_4^R$ and parts in which the divergent contributions are gathered:
\begin{align}
  \label{eq:decomp_F4}
F_4& (k_1, k_2, k_3, k_4) = F^R_4(1, 2, 3, 3)  + g_4(1,2,3,4)
\notag \\
& + 
F_3^R(1, 2, 34) g_1(123 , 4) 
+
F_3^R(1, 23, 4)  g_1( 12,   34)
+
F_3^R(12,  3, 4)  g_1(1,  234)
\notag \\
& +
F_2^R(123, 4) g_2(1,2,3) + F_2^R(1, 234) g_2(2,3,4) + F_2^R(12, 34) g_1(1,234) g_1(123,4).
\end{align}
where we introduced the following short-cut notation
\begin{align}
  \label{eq:short-cut}
F_3^R(12, 3, 4) &= F_3^R(k_1 + k_2,  k_3, k_4) 
&
g(1, 234) =
& g_1(\mu_1,  \mu_2 +   \mu_3 + \mu_4)
\end{align}
where a string of numbers corresponds to the sum of the momenta with the corresponding indices.
The convergent contribution is given by
\begin{align}
  \label{eq:F4R}
F_4^R& (k_1, k_2, k_3, k_4) = \frac{1}{24} 
\notag \\
&\bigg[
\bigg(
\frac{4}{\mu_1 - {m}_1^2 + i\epsilon} - \frac{1}{\mu_1 -i\epsilon} - \frac{3}{\mu_1 + i\epsilon}  
 \bigg)
\bigg( \frac{1}{\mu_1 + \mu_2 - m_{12}^2 + i\epsilon} - \frac{1}{\mu_1 + \mu_2 -i\epsilon}
\bigg)
\notag \\
& \bigg(
\frac{1}{-\mu_4 - m_{123}^2 +i\epsilon} + \frac{1}{\mu_4 + i\epsilon}
\bigg)
\notag \\
&+
\bigg(
\frac{4}{\mu_1 - m_1^2 + i\epsilon} - \frac{1}{\mu_1 + i\epsilon} - \frac{3}{\mu_1 - i\epsilon}  
 \bigg)
\bigg( \frac{1}{\mu_1 + \mu_2 - m^2_{12} + i\epsilon} - \frac{1}{\mu_1 + \mu_2 +i\epsilon}
\bigg)
\notag \\
& \bigg(
\frac{1}{-\mu_4 - m_{123}^2 +i\epsilon} + \frac{1}{\mu_4 - i\epsilon}
\bigg)
\notag \\
&+
\bigg(
\frac{8}{\mu_1 - {m}_1^2 + i\epsilon} - \frac{5}{\mu_1 - i\epsilon} - \frac{3}{\mu_1 + i\epsilon}  
 \bigg)
\bigg( \frac{1}{\mu_1 + \mu_2 - m_{12}^2 + i\epsilon} - \frac{1}{\mu_1 + \mu_2 +i\epsilon}
\bigg)
\notag \\
& \bigg(
\frac{1}{-\mu_4 - m_{123}^2 +i\epsilon} + \frac{1}{\mu_4 + i\epsilon}
\bigg)
\notag \\
&+
\bigg(
\frac{8}{\mu_1 - {m}_1^2 + i\epsilon} - \frac{5}{\mu_1 + i\epsilon} - \frac{3}{\mu_1 - i\epsilon}  
 \bigg)
\bigg( \frac{1}{\mu_1 + \mu_2 - m_{12}^2 + i\epsilon} - \frac{1}{\mu_1 + \mu_2 -i\epsilon}
\bigg)
\notag \\
& \bigg(
\frac{1}{-\mu_4 - m_{123}^2 +i\epsilon} + \frac{1}{\mu_4 - i\epsilon}
\bigg)
\bigg].
\end{align}
The integral over divergent parts of Eq.(\ref{eq:decomp_F4}),  which is contained in the functions  $g_i$, $i = 1,2,3$,  vanishes, when combined with the
contributions from subtraction diagrams $F_4^S$.  The
integral over $F^R_4$ yields on the other hand the following result
\begin{align}
  \label{eq:int_F4R}
\int\frac{d \mu_1}{(-2\pi i)} &\int\frac{d \mu_2}{(-2\pi i)}\int\frac{d \mu_3}{(-2\pi i)} F_4^R(k_1, k_2, k_3, k_4) = \frac{1}{24},
\end{align}
and the four gluon quark-impact factor is given by
\begin{align}
  \label{eq:quark_4gluonfertig}
  A^{a_1a_2a_3}_{AA',(4;0)}  ({\bm k}_1,{\bm k}_2,{\bm k}_3 )  = 
       g^4 \frac{1}{4!}\sum_{i_1 \ldots i_4} t^{a_{i_1}}t^{a_{i_2}}t^{a_{i_3}}t^{a_{i_4}},
\end{align}
where the sum is over all permutations of numbers $1, \ldots, 4$. Again, taking he QED-limit,  $t^{a_i} \to 1$, we find coincidence of the above result with the QED-'eikonal-formula'.

\section{Number changing Reggeon transition kernels}
\label{sec:v23}

Compared to the exchange of a single reggeized gluon, with negative
signature, and the  state of two-reggeized gluons, with positive
signature,  states of three and four reggeized gluons are
suppressed by a relative factor of $g^2$ compared to the corresponding
leading part.  Resumming therefore loop corrections to the exchange of
three and four reggeized gluons that are maximally enhanced by
logarithms in the center of mass energy $\sqrt{s}$, we are therefore
strictly speaking working within Generalized LLA (GLLA) as defined in
\cite{Bartels:1980pe,Bartels:1999aw}, where all loop-corrections are
resummed that are enhanced by a maximal power of logarithms in $s$.
One part of the corrections is given by the pairwise interaction of
the reggeized gluons by a two-to-two kernel, as derived in
Sec.\ref{sec:tworeggeon_negsig}, which are known as BKP-corrections
\cite{Bartels:1980pe,Kwiecinski:1980wb}.  Apart from them, a complete
description of the state of $n>2$ reggeized gluons within the GLLA
requires to include transition vertices. In the present case these are
transitions from one-to-three and two-to-four reggeized gluons.
Transitions from one-to-two and two-to-three are on the other hand not
allowed by singature conservation \cite{Gribov:1968fc}.  In the
following we will see that signature conservation is for the elastic
amplitude satisfied automatically and is not needed to be imposed as
an external constraint.

\subsection{The transition of 1-2  and 1-3 reggeized gluons}
\label{sec:v12}
\begin{figure}[htbp]
  \centering
  \parbox{4cm}{\center \includegraphics[height = 3.5cm]{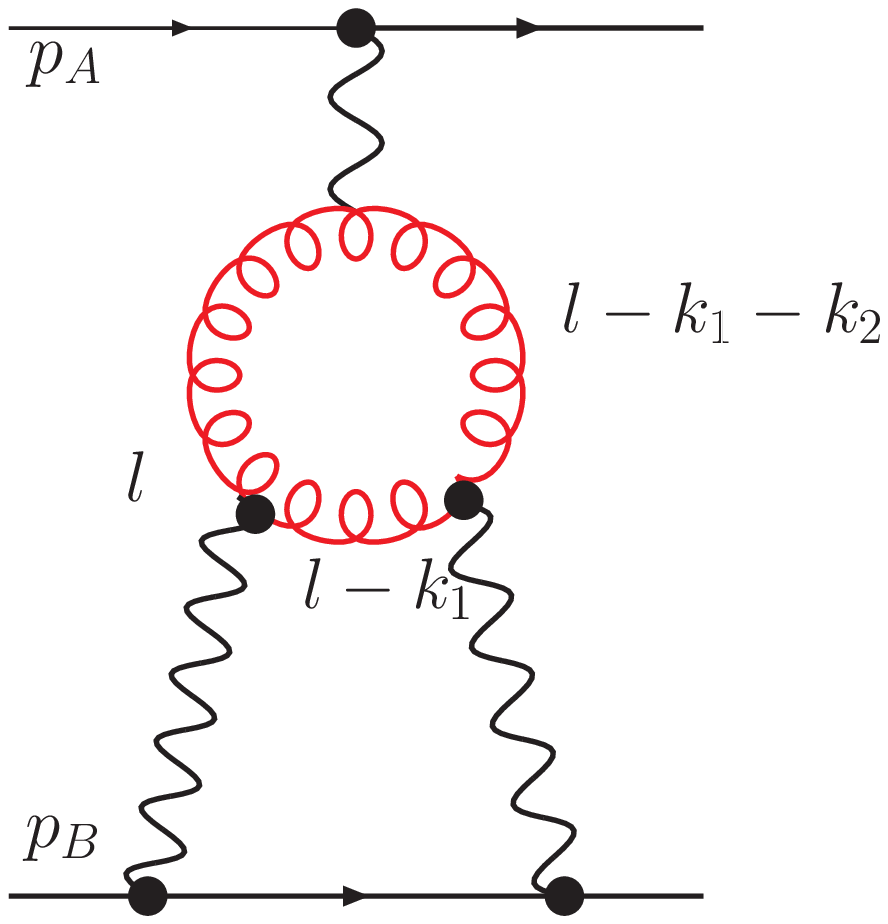}}
  \parbox{4cm}{\center \includegraphics[height = 3.5cm]{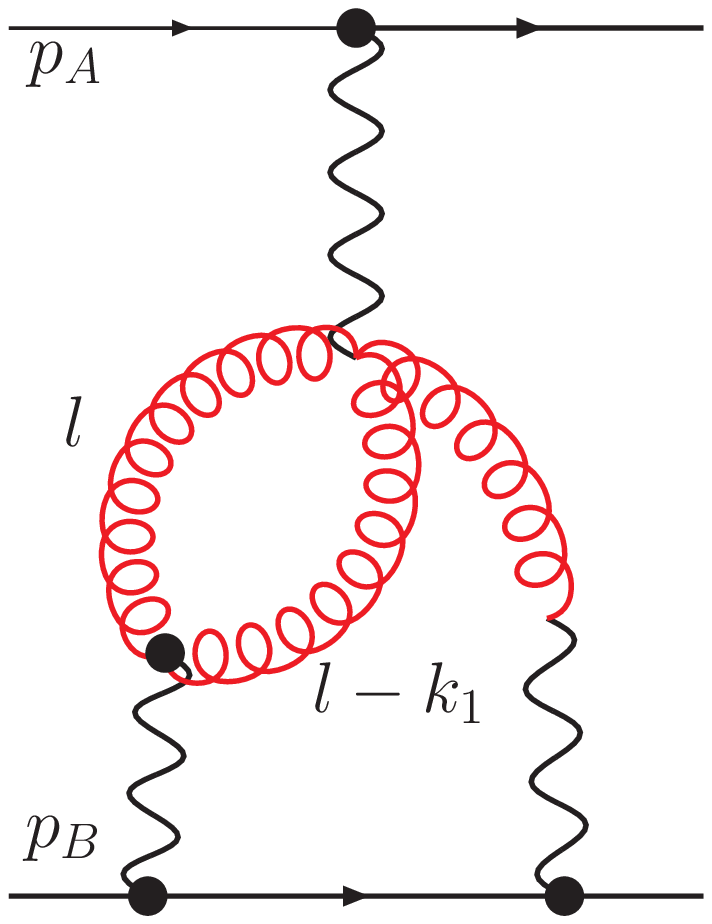}}
  \parbox{4cm}{\center \includegraphics[height = 3.5cm]{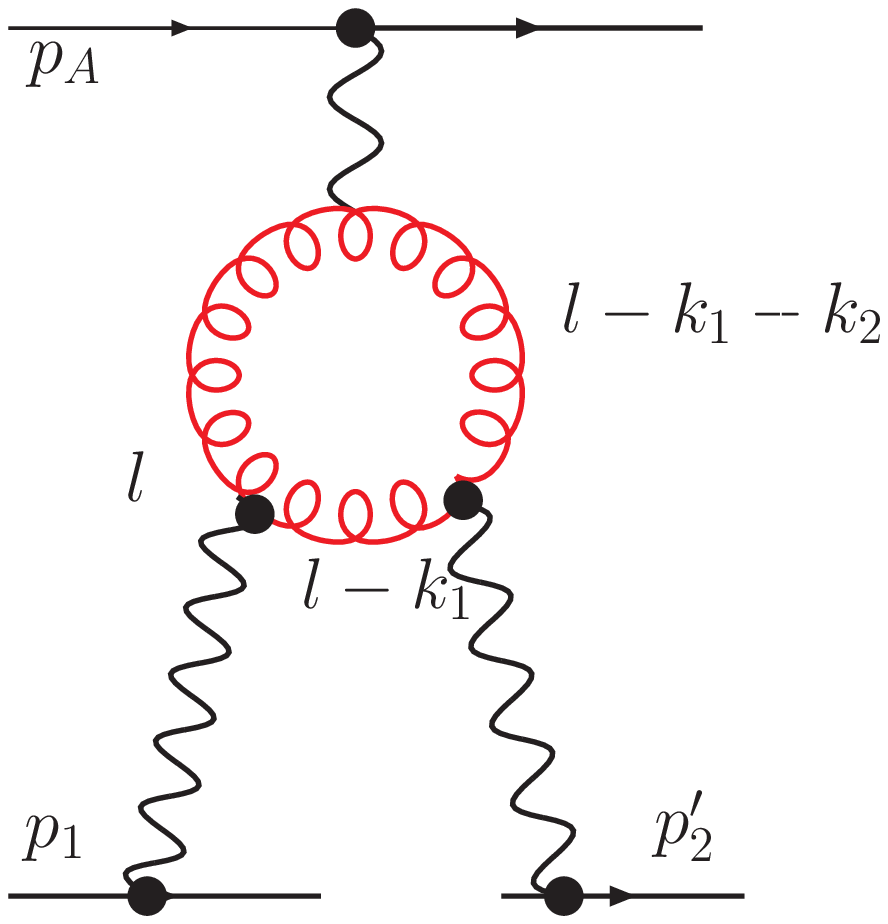}} \\
\parbox{4cm}{\center (a)}\parbox{4cm}{\center (b)}\parbox{4cm}{\center (c)}
  \caption{\small Diagrams with an induced vertex of the first (a) and the second (b) order that contribute to the 1-2 transition. While the 1-2 transition has to vanish within the elastic amplitude, it might contribute to the triple-Regge-limit of the 6-point amplitude (c).}
  \label{fig:onetwo}
\end{figure}

\begin{figure}[htbp]
  \centering
  \parbox{4cm}{\center \includegraphics[height = 3.5cm]{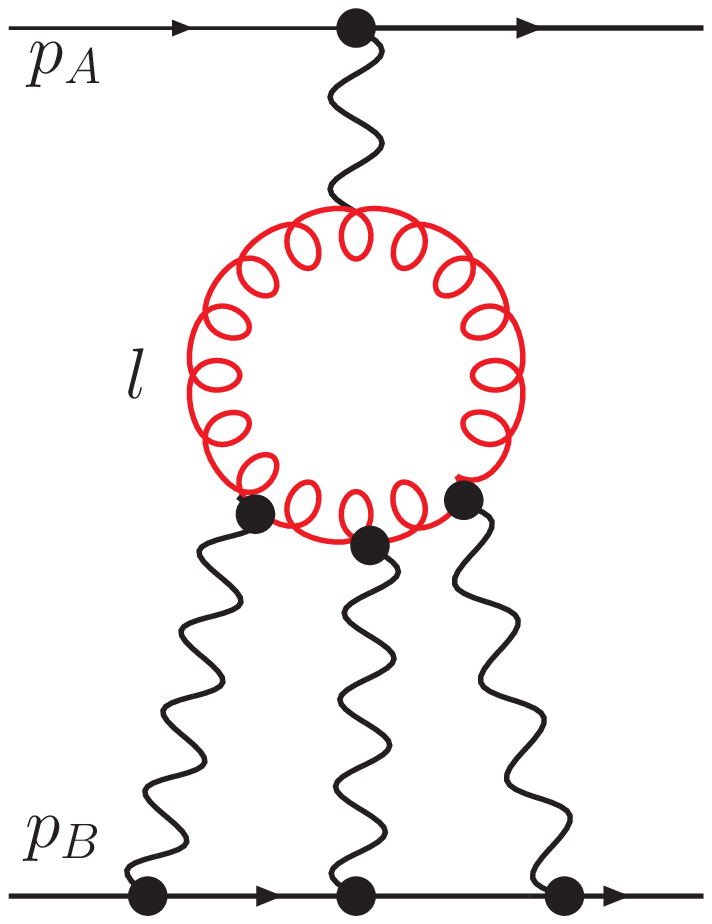}}
  \parbox{4cm}{\center \includegraphics[height = 3.5cm]{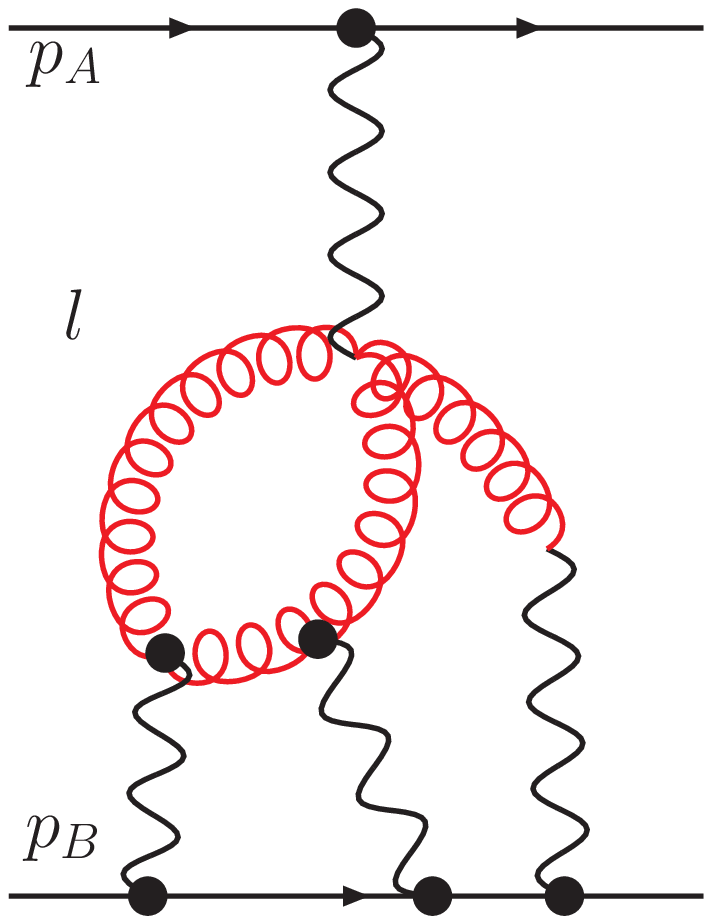}}
  \parbox{4cm}{\center \includegraphics[height = 3.5cm]{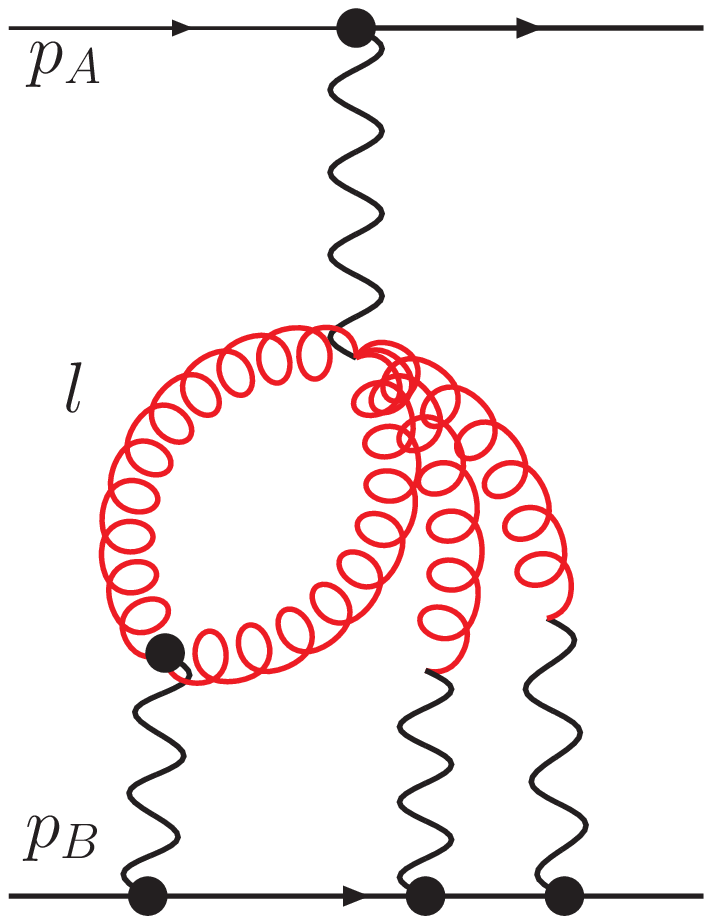}} \\
\parbox{4cm}{\center (a)}\parbox{4cm}{\center (b)}\parbox{4cm}{\center (c)}
  \caption{\small Diagrams with an induced vertex of the first (a), the second (b) and the third  order (c) that contribute to the 1-3 transition. }
  \label{fig:onethree}
\end{figure}

As far as the description of the relevant longitudinal integrations is
concerned, transitions from one to two and one to three reggeized
gluons share many features.  In the following we restrict then to the
case where a single reggeized gluons, coming form the quark A, couples
to two and three reggeized gluons respectively, which connected to the
quark B, see also Fig.\ref{fig:onetwo} and Fig.\ref{fig:onethree}. All
other cases can then be easily obtained from this expression.

Both for the one-to-two and the one-to-three transitions, the
reggeized gluons couple to each other by a gluon loop, 
similar to the central rapidity diagram of Sec.~\ref{sec:22negsig}
that yields the gluon trajectory function.  Apart from the
coupling of the 'upper' to the 'lower' reggeized gluons by a
gluon-loop, there occur also various subtraction diagrams.
As far as integrations over longitudinal momenta are concerned, we
have as usual the constraint, that every 4-point sub-amplitude, where
the interaction is mediated by a reggeized gluon, requires a large
center of mass energy. This constraint is imposed by the usual Mellin
integrals. They further allow to include due to reggeization of the
gluon and corrections to the state of two and three reggeized gluon
respectively in an easy and straight-forward way LLA-corrections.  As
far as the upper reggeized gluon is concerned, the description is
identical to the case of the gluon trajectory function in
Sec.~\ref{sec:22negsig}. The case, where the reggeized
gluon couples to the gluon loop by an induced vertex of higher order,
as in Fig.~\ref{fig:onetwo}b and Fig.~\ref{fig:onethree}b-c, requires however special care.  There,
the reggeized gluon is seemingly part of a 5- or a 6-point
sub-amplitude and not of a 4-point amplitude.
We follow also in the following the solution proposed in
Sec.\ref{sec:born+possig}, where a similar problem occurred.  The
natural choice to define the relevant 4-point-amplitude is given by
imposing the constraint on the gluon momenta that enter the
gluon-loop: In the essential region of integration of the gluon-loop,
loop momenta are close to the mass-shell and in this sense, the
constraint on the center-of-mass energy can be related to a
corresponding constraint on the rapidity-variable. The gluons that do
not enter the gluon loop, on the other hand, couple directly to
reggeized gluons and are therefore generally off-shell and do not
impose any constraint on rapidity.  As far as the correct choice of
the momenta inside the loop is concerned, we recall from
Sec.\ref{sec:born+possig} that it is always possible to state higher
induced vertices in a form, that there is always only one pole that
contains the loop momenta and it is then this pole-momentum that
should be used to construct the corresponding center-of-mass energy.

For the lower reggeized gluons, one has to distinguish between the
case where only one reggeized gluon couples to the gluon-loop from
below, as in Fig.~\ref{fig:onetwo}b and Fig.~\ref{fig:onethree}c, and
the case where a state of two or three reggeized gluons couples to the
gluon-loop as in Fig.~\ref{fig:onetwo}a and
Fig.~\ref{fig:onethree}a-b. While in the former case we proceed in the
same way as for the gluon trajectory, Sec.~\ref{sec:22negsig}, in the
latter case we combine the phases of the individual reggeized gluons,
making repeatedly use of Eq.~(\ref{eq:sig_manipus}).

For the general evaluation of the diagrams we note that there occurs a
large amount of cancellations between individual diagrams, which we
partly trace back to identities between the induced vertices like
Eq.~(\ref{eq:reccurrence}). In particular it seems to be possible to find a
general argument related to gauge invariance, that allows to exclude
certain contributions from the very beginning which would facilitate
the calculations considerably. Unfortunately, such a general rule has
not been found yet. Nevertheless it seems advisable to
achieve as many cancellations as possible between the individual
diagrams, which in particular cancels certain divergent parts and to
start  only after-wards the evaluation of the integrals.

As a next step we evaluate the longitudinal part of the
reggeized-gluon-loop-integral, which can be attributed to the
transition vertex: With the loop momenta of the gluon loop denoted by
$l$ and the loop momenta of the reggeized gluon loop by $k_i$ where
$i=1$ for the one-to-two transition and $i=1,2$ for the one-to-two
transition (see Fig.~\ref{fig:onetwo} and Fig.~\ref{fig:onethree}), we
substitute the minus momenta of the reggeized gluon loops, $k_i^-$, by
$k_i^- \to \mu_i = -l^+k_i^-$. This yields Jacobian factors $1/|l^+|$
and $1/(l^+)^2$ for the one-to-two and the one-to-three transition
respectively.  As we will see in short, these Jacobian factors connect
again directly to the signature of the  states of reggeized
gluons.  The integrals over the $\mu_i$ are then easily evaluated,
taking into account contributions due to subtraction diagrams and
following closely the procedure described in Sec.~\ref{sec:impa4_bkp}
for the evaluation of the impact factors.  For the the one-to-two
transition we then arrive typically at integrals of the following form
\begin{align}
  \label{eq:12_raw1}
 \int \frac{d^4 l}{(2\pi)^4}  &
\left[ 
       \left(   \frac{-p_A^+l^--i\epsilon}{ \Lambda_a} \right)^{\omega_1}  
     +
      \left(\frac{p_A^+l^--i\epsilon}{\Lambda_a} \right)^{\omega_1}   \right]
 \left[  
         \left(\frac{-p_B^+l^+-i\epsilon}{\Lambda_b} \right)^{\omega_2} 
     +
       \left(\frac{p_B^+l^+-i\epsilon}{\Lambda_b} \right)^{\omega_2}   \right]
\notag \\
& 
\frac{1}{|l^+| } \frac{1}{l^-} \quad  \frac{1 }{l^+l^-  -  {\bm l}^2  +  i\epsilon} \quad 
\frac{1}{l^+l^- - ({\bm l}  - {\bm k}_1  + {\bm k}_2 )^2  +i\epsilon } 
\end{align}
where the precise structure of transverse momenta differs for
different diagrams. Furthermore there exists a contribution that comes
with an additional factor $l^+l^-$ in the numerator. In any case the
integrals over $l^+$ and $l^-$ are convergent and substituting
simultaneously $l^+ \to - l^+$ and $l^- \to -l^-$, the above integral
changes sign and therefore yields zero and as a consequence, the
transition from one to two reggeized gluons inside the elastic
amplitude vanishes.  We however note, that the reason for vanishing of
this integral is given by the Jacobian factor $|l^+|$ that occurs as a
consequence of the longitudinal loop integral of the reggeized gluon
loop. Note that the above result allows for the one-to-two transition
to occur inside a 6-point amplitude, Fig.~\ref{fig:onetwo}c, for
instance, where a integral over the relative momenta of the reggeized
gluons is not present.

 For the transition of one to three reggeized gluons
we arrive after evaluating integrals over $\mu_1$ and $\mu_2$ at integrals of the following form
\begin{align}
  \label{eq:13_raw1}
 \int \frac{d^4 l}{(2\pi)^4}  &
\left[ 
       \left(   \frac{-p_A^+l^--i\epsilon}{ \Lambda_a} \right)^{\omega_1}  
     +
      \left(\frac{p_A^+l^--i\epsilon}{\Lambda_a} \right)^{\omega_1}   \right]
 \left[  
         \left(\frac{-p_B^+l^+-i\epsilon}{\Lambda_b} \right)^{\omega_2} 
     +
       \left(\frac{p_B^+l^+-i\epsilon}{\Lambda_b} \right)^{\omega_2}   \right]
\notag \\
& 
\frac{1}{l^+ } \frac{1}{l^-} \quad  \frac{1 }{l^+l^-  -  {\bm l}^2  +  i\epsilon} \quad 
\frac{1}{l^+l^- - ({\bm l}  - {\bm k}_1  + {\bm k}_2 )^2  +i\epsilon } 
\end{align}
Due to the different Jacobian factors, the integral does not vanish
but can be evaluated in analogy to the central rapidity diagram of
Sec.~\ref{sec:22negsig}.  Taking into account simplifications due to
the LLA, we arrive at the following result for elastic quark-quark
scattering amplitude with a transition from one to three reggeized
gluons:
\begin{align}
\label{eq:M_U13}
   \mathcal{M}_{2 \to 2} =   \frac{ (2\pi)^2}{3!}  p_A^+p_B^-
\int \frac{d \omega}{2 \pi i} 
 \left[  \left(\frac{ -p_A^+ p_B^-}{m_A^+m_B^-}   \right)^{\omega}  +  \left(\frac{ p_A^+ p_B^-}{m_A^+m_B^-}   \right)^{\omega}           \right]
 \phi^{\text{NC}}_{1 \to 3 }(\omega, t)
\end{align}
where
\begin{align}
  \label{eq:phi_13}
\phi^{\text{NC}}_{1 \to 3 }(\omega, t) =  g t^a \frac{1}{{\bm q}^2}\frac{1}{\omega - \beta({\bm q}^2)} \int \frac{d^2 {\bm k}_1 }{(2\pi)^3}  \int \frac{d^2 {\bm k}_2 }{(2\pi)^3}& \,
U_{1 \to 3}^{a;b_1b_2b_3}({\bm q} ; {\bm k}_1, {\bm k}_2,  {\bm k}_3) 
\notag \\
&\frac{1}{{\bm k}_1^2 {\bm k}_2^2{\bm k}_3^2}
A^{b_1b_2b_3}_{(3;0)} (\omega| {\bm k}_1,{\bm k}_2,{\bm k}_3 )  
\end{align}
with $U_{1 \to 3}$ to 1-to-3 reggeized gluon transition vertex.  To
display the transverse momentum structure of the vertex, we make use
of so-called Reggeon-momentum diagrams (for an detailed introduction
to these diagrams see for instance \cite{Bartels:1999aw}), which allow
to illustrate the transverse momentum structure in a diagrammatic way.
In every diagram there is a lowest vertex, which determines the
resulting expression. While the sum of momenta that leaves the vertex
below determines the numerator, the two lines that enter this vertex
from above determine the two denominators. Further, closed loop imply
an integration over the transverse loop momentum.  For the 1-to-3
reggeized gluon transition vertex, the following diagrams occur, which
correspond to the following expression in momentum space:
\begin{align}
  \label{eq:deft123}
\parbox{1.5cm}{ \center\includegraphics[height=1.1cm]{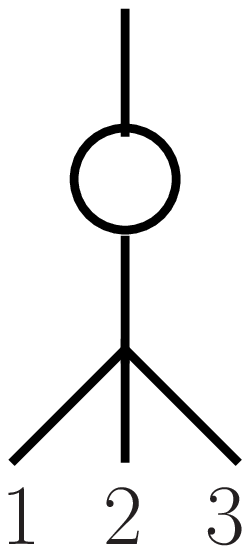}}& = \int \frac{d^2 {\bm l}}{(2\pi)^3} \frac{( {\bm k}_1 + {\bm k}_2 + {\bm k}_3)^2 }{{\bm l}^2 ( {\bm k}_1 + {\bm k}_2 + {\bm k}_3 - {\bm l})^2 }, \\
\parbox{1.5cm}{\center \includegraphics[height=1.1cm]{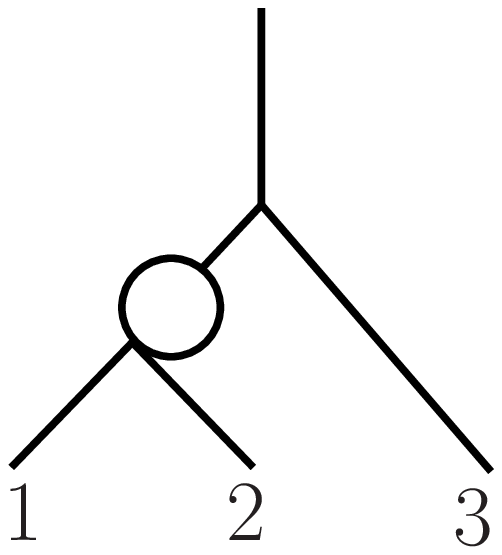}} &= 
\int \frac{d^2 {\bm l}}{(2\pi)^3} \frac{( {\bm k}_1 + {\bm k}_2 )^2 }{{\bm l}^2 ( {\bm k}_1 + {\bm k}_2  - {\bm l})^2 },
\\ 
\parbox{1.5cm}{\center \includegraphics[height=1.1cm]{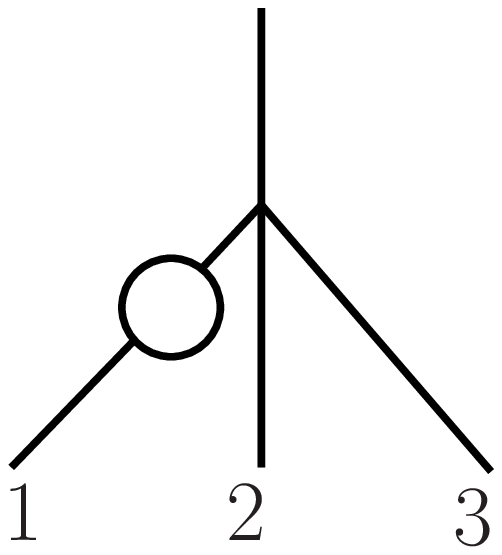}} &= 
\int \frac{d^2 {\bm l}}{(2\pi)^3} \frac{ {\bm k}_1^2 }{{\bm l}^2 ( {\bm k}_1   - {\bm l})^2 }.
\end{align}
The complete transition vertex is then given by
\begin{align}
  \label{eq:U13}
U_{1 \to 3}^{a;b_1b_2b_3}& ({\bm q}; {\bm k}_1, {\bm k}_2,  {\bm k}_3) =
\notag \\
& \tr\left(T^a T^{b_{1}}T^{b_{2}} T^{b_{3}}  \right) 
 U ({\bm q}; {\bm k}_{1},  {\bm k}_{2} ,  {\bm k}_{3}) 
+ \tr\left(T^a T^{b_{2}}T^{b_{1}} T^{b_{3}}  \right) 
 U ({\bm q}; {\bm k}_{2},  {\bm k}_{1} ,  {\bm k}_{3}) 
\notag \\
& + \tr\left(T^a T^{b_{1}}T^{b_{3}} T^{b_{2}}  \right) 
 U ({\bm q}; {\bm k}_{1},  {\bm k}_{3} ,  {\bm k}_{2}) ,
\end{align}
where the sum is over all permutations of numbers $1, \ldots, 3$ with the trace and generators in the adjoint representation of $SU(N_c)$ and
\begin{align}
  \label{eq:U}
  U ({\bm q}; {\bm k}_{1},  {\bm k}_{2} ,  {\bm k}_{3})  = g^4 \frac{{\bm q}^2}{6}\left[ \,\parbox{.9cm}{\includegraphics[height=1.1cm]{t123.ps}} 
-
\parbox{1.5cm}{ \includegraphics[height=1.1cm]{s12m3.ps}}
+
\parbox{1.5cm}{ \includegraphics[height=1.1cm]{s1m23.ps}} \right].
\end{align}

\subsection{The transition of 2-3  and 2-4 reggeized gluons }
\label{sec:v23connect}

\begin{figure}[htbp]
  \centering
  \parbox{4cm}{\center\includegraphics[height=3cm]{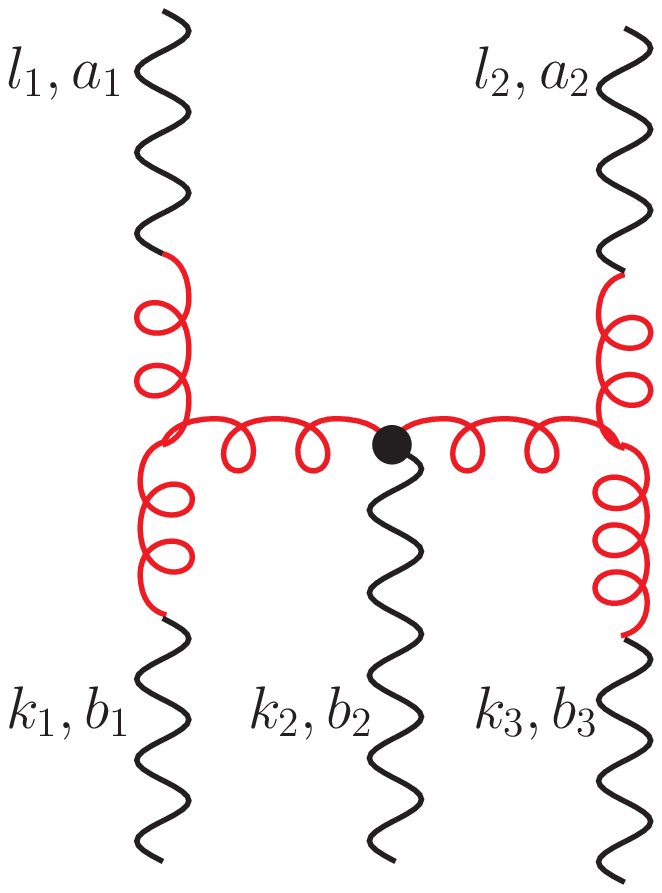}}
 \parbox{4cm}{\center \includegraphics[height=3cm]{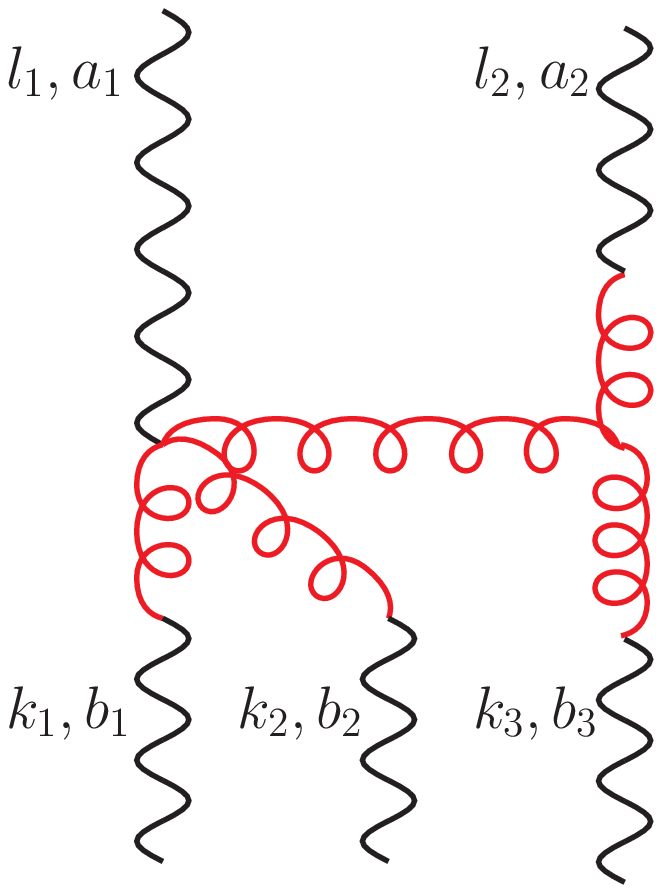}}
 \parbox{4cm}{\center \includegraphics[height=3cm]{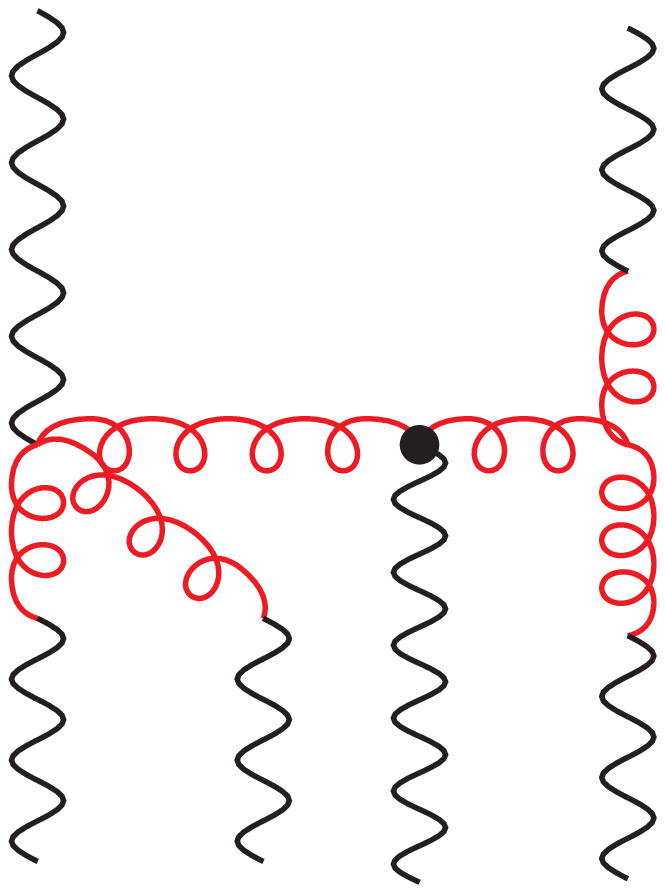}} \\
\parbox{4cm}{\center (a)}\parbox{4cm}{\center (b)}\parbox{4cm}{\center (c)}
  \caption{\small Diagrams that contribute to the transition form two-to-three (a,b) and two-to-four reggeized gluons.}
  \label{fig:trans234}
\end{figure}

In accordance with the discussion of the two-to-two transition in
Sec.\ref{sec:tworeggeon_negsig} and the Reggeon-Particle-2-Reggeon
vertex in Sec.\ref{sec:mixed}, it is also in the present case
advisable to consider for the discussion of longitudinal integrations
not individual diagrams, but to group the sum of all diagrams first
according to their color- and then according to their transverse
momentum structure. In this way cancellation of certain
divergent parts between individual diagrams is achieved, which is otherwise difficult to be obtained. 
Diagrams that contribute to the two-to-three reggeized gluons transitions are shown in  Fig.\ref{fig:trans234}a-b, while a typical contribution to the  two-to-four reggeized gluon transition can be found in  Fig.\ref{fig:trans234}c.

In the case of the $2-3$ transition, the obtained unintegrated transition vertex is  for \emph{one} ordering of color and momenta indices given by
\begin{align}
  \label{eq:unint23}
il^+  &16 (T^{b_1}T^{b_2}T^{b_3})_{a_1a_2} \bigg[ \bigg(  {\bm q}^2 
-
 {\bm l}_1^2 ({\bm q} - {\bm k}_1)^2  g_1(\mu_1, -\mu_2-\mu_3) 
- 
 {\bm l}_2^2 ({\bm q} - {\bm k}_3)^2 g_1(\mu_1 + \mu_2, -\mu_3) 
\notag \\
&
+
 {\bm l}_1^2  {\bm k}_2^2 {\bm l}_2^2   g_1(\mu_1, -\mu_2-\mu_3)  g_1(\mu_1 + \mu_2, -\mu_3) \bigg) F_3(l_1, l_2 ; k_1, k_2, k_3) 
\notag \\
&
 + {\bm l}_1^2 {\bm k}_3^2  g_2(\mu_3, \mu_2, \mu_1)  F_2(l_1,l_2; k_1 + k_2, k_3)
+
 {\bm l}_2^2 {\bm k}_1^2  g_2(\mu_1, \mu_2, \mu_3) F_2(l_1, l_2 ;k_1, k_2 + k_3)\bigg]
\end{align}
where $\mu_i = -l^+k^-_i$, $i = 1,2,3$ with $\mu_1 + \mu_2 + \mu_3 =
0$. Further ${\bm k}_1 +{\bm k}_2 +{\bm k}_3 = {\bm q} = {\bm l}_1 +
{\bm l_2} $, while $l_1^+ = -l_2^+ \equiv l^+ $ and $l_1^- = 0 =
l_2^-$.    The appearance of the functions $g_i$ $i=1,2$ in
Eq.(\ref{eq:unint23}) is directly associated with the occurrence of
induced vertices in the underlying Feynman-diagrams. For instance the
first term in the first line of Eq.(\ref{eq:unint23}) comes from a
combination with no induced vertex, while the term in the second line
of Eq.(\ref{eq:unint23}) comes due to a combination of two induced
vertices of the first order.  Terms in third line on the other hand
come from a combination of an induced vertex of the second order with
a reggeized gluon that couples without an induced vertex to the gluon.
In the case of the two-to-four transition a very similar expression
can be derived, which contains as a new element also the function
$g_3$, corresponding to an induced vertex of the third order.

\begin{figure}[htbp]
  \centering
  \parbox{3cm}{\center \includegraphics[width=2cm]{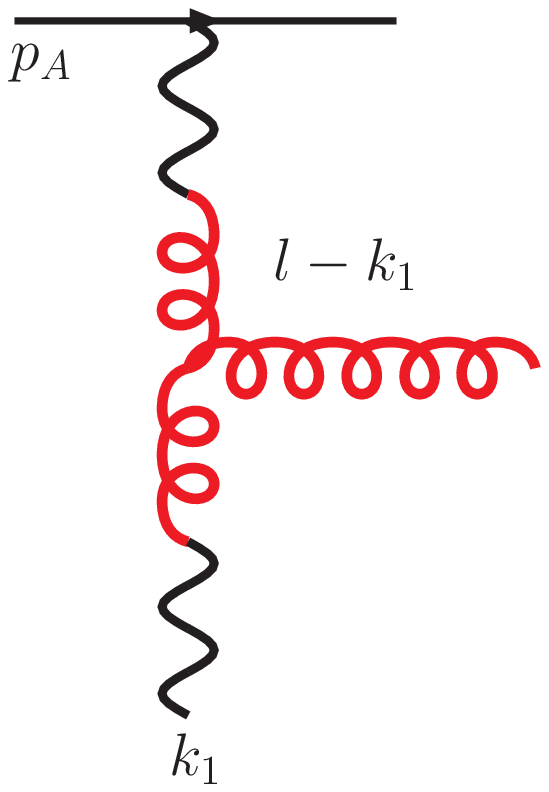}}
  \parbox{3cm}{\center \includegraphics[width=2cm]{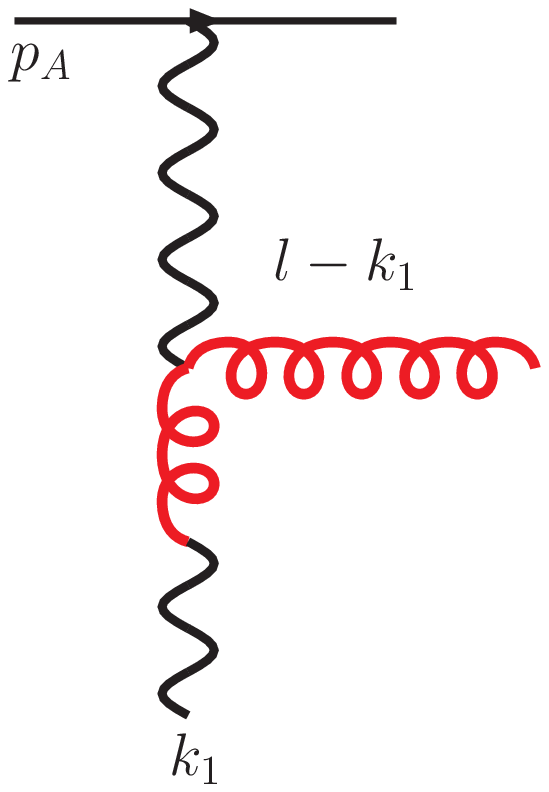}}
  \parbox{3cm}{\center \includegraphics[width=2cm]{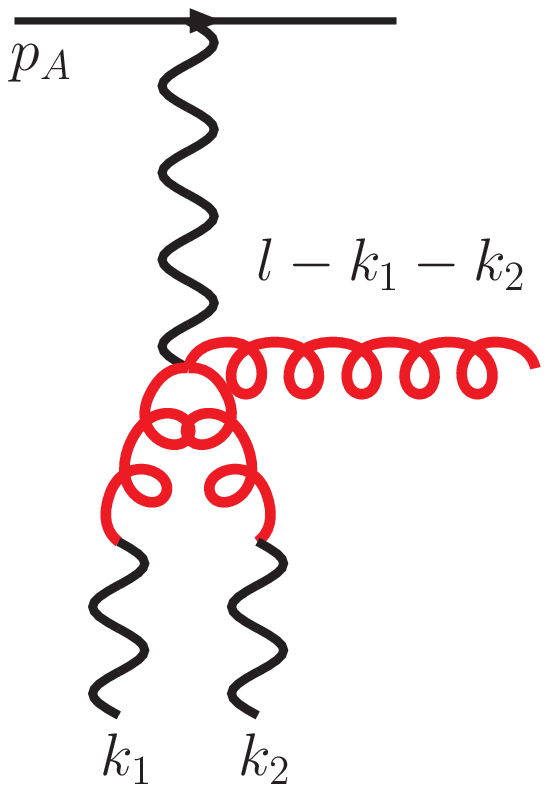}}
  \parbox{3cm}{\center \includegraphics[width=2cm]{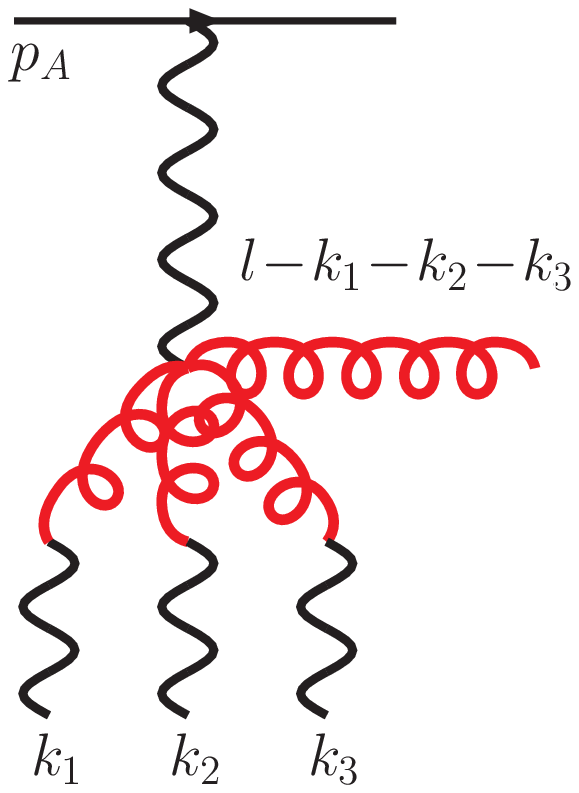}} \\
  \parbox{3cm}{\center (a)} 
  \parbox{3cm}{\center (b)} 
  \parbox{3cm}{\center (c)} 
  \parbox{3cm}{\center (d)} 
  \caption{\small Different kind of center of mass energies that need to be large, in order for the interaction to be correctly described by a  reggeized gluon. }
  \label{fig:regularize}
\end{figure}

Also in the present case, every 4-point amplitude where the
interaction is mediated by a reggeized gluon, requires a large center
of mass energy, and a corresponding lower bound is imposed making use
of the Mellin-integral.  For the graphs Fig.\ref{fig:regularize}a-b, where the reggeized gluon couples to the gluon by a three-gluon-vertex or an induced vertex of the first order, 
the situation is similar to the 2-2 transition and the squared
center-of-mass energy is given by $p_A^+k^-_1$. However, unlike the
case of the 2-2 transition, it is generally no longer possible to
combine the two Mellin-integrals that belong to the two upper
reggeized gluons into a single one for the whole sub-amplitude.
Instead, every reggeized gluon is treated separately and are only
combined into a single Mellin-integral, after the longitudinal
integrations have been carried out. For graphs like
Fig.\ref{fig:regularize}c-d , which contain  induced vertices of
higher order, a problem similar to the one in Sec.~\ref{sec:v12}
arises: The reggeized gluon is seemingly part of a 5- or a 6-point
sub-amplitude and not a 4-point amplitude.  In analogy with the
procedure in Sec.~\ref{sec:v12} we will use in the following the
momentum of the 'produced' QCD gluon to construct the center-of-mass
energy, on which the lower bound is imposed and which enters the
Mellin integral.  In particular for Fig.~\ref{fig:regularize}c and
Fig.~\ref{fig:regularize}d, this results into the center-of-mass
energies $s^{(c)}_A = p_A^+ (k_1^- + k_2^-)$ and $s^{(d)}_A = p_A^+
(k_1^- + k_2^- + k_3^-)$ respectively.  Anticipating that all
longitudinal integrals are convergent, once subtraction graphs are
included, this then implies that the transition is fixed at central
rapidities, as required.

Taking into account further subtraction diagrams, longitudinal
integrals turn out to be convergent and can be evaluated similarly to
the two-to-two transition in Sec.~\ref{sec:tworeggeon_negsig}, with
the methods described in Sec.~\ref{sec:impa4_bkp}. Especially, using
the methods of Sec.~\ref{sec:impa4_bkp}, we arrive in the course of
the analysis on the 'regularized' integrands $F_2^R$, $F_3^R$ and
$F_4^R$ which all come with certain poles in light-cone momenta, and
for a correct evaluation of the integral it is mandatory that the
$i\epsilon$ prescription of these poles coincides with the one of the
branch-cuts appearing due to the Mellin-integral.  In the case of the
2-3 transition, after including subtraction diagrams, the first term
in the first line of Eq.(\ref{eq:unint23}) will be proportional to the
function $F_3^R$ of Eq.~(\ref{eq:grossFR}), plus terms with vanishing
integral. The integral is then given by
\begin{align}
  \label{eq:23_f3r}
\int \frac{d l^+}{l^+} & \left[  (-p_B^-l^+ -i\epsilon)^{\omega}  
   +(p_B^-l^+ -i\epsilon)^{\omega_2}    \right] \int \frac{d\mu_1}{(-2\pi i)} \int \frac{d\mu_2}{(-2\pi i)}
\frac{{m}_1^2{m}_{12}^2}{(\mu_1 - {m}_1^2 + i\epsilon)(\mu_1 +\mu_2 - {m}_{12}^2 + i\epsilon)}
\notag \\
    \times
\frac{1}{6} &\bigg[ \left(\frac{p_A^+\mu_1}{l^+} -i\epsilon\right)^{\omega_1-1} 
 \left[\left(\frac{p_A^+(\mu_1 + \mu_2)}{l^+} -i\epsilon\right)^{\omega_2-1}  - 2 \left(\frac{-p_A^+(\mu_1 + \mu_2)}{l^+} -i\epsilon\right)^{\omega_2-1} 
 \right]  
\notag \\
 -&
  \left(\frac{-p_A^+\mu_1}{l^+} -i\epsilon\right)^{\omega_1-1} 
 \left[  2 \left(\frac{p_A^+(\mu_1 + \mu_2)}{l^+} -i\epsilon\right)^{\omega_2-1} -
\left(\frac{-p_A^+(\mu_1 + \mu_2)}{l^+} -i\epsilon\right)^{\omega_2-1} 
 \right] \bigg],
\end{align}
where we note that the factor $1/l^+$ arises due to a Jacobian factor
from the transition $k_i^- \to \mu_i = -l^+k_i^-$, $i = 1,2$ and
$m_1^2 = ({\bm l} - {\bm k}_1)^2$ and $m_{12}^2 = ({\bm l} - {\bm k}_1
- {\bm k}_2)^2$. With the substitutions $l^+ \to - l^+$ the integral
changes sign and correspondingly the integral vanishes. A similar
observation can be made for all integrals connected with the
transition from two to three reggeized gluons. The two-to-three
transition vanishes therefore, if inserted into the elastic amplitude.
On the other hand, if it would occur inside in a 6-point amplitude for
instance, it is generally possible to obtain a non-zero result for
this transition.

Integrals occurring for the two-to-four transition are very 
similar to Eq.~(\ref{eq:23_f3r}), the main difference being that they
contain three integrations over $\mu_i$, $i = 1, \ldots 4$ and that
the corresponding Jacobian yields therefore a factor of $1/|l^+|$, instead. As
a consequence, the integrals do not vanish in that case, in accordance
with signature conservation. 

For the elastic scattering amplitude with one transition from two to four reggeized gluons
 we then find  the following expression:
\begin{align}
  \label{eq:4R1}
 \mathcal{M}_{2 \to 2}   =\frac{ (2\pi)^3}{4!}   s
 \int \frac{d \omega}{2 \pi i}    \left(\frac{ s}{s_R}   \right)^{\omega}  \frac{\xi^{(+)}}{\sin \pi\omega}  \phi_{4}^{\text{NC}}(\omega, t),
\end{align}
with 
\begin{align}
  \label{eq:phi24}
\phi_4^{\text{NC}}(\omega, t) =  &
 A^{a_1a_2}_{(2;0)} ( {\bm l}_1,{\bm l}_2 ) \frac{1}{\omega - \sum_i^2 \beta({\bm l}_i^2)} \otimes_{{\bm l}_{12}}  U_{2\to 4}^{a_1a_2; b_1b_2b_3b_4} ( {\bm l}_1,{\bm l}_2; {\bm k}_1,{\bm k}_2,{\bm k}_3,{\bm k}_4 )
\notag \\ & \qquad \qquad  \frac{1}{\omega - \sum_i^4 \beta({\bm k}_i^2)} \otimes_{{\bm k}_{\{1234 \}}}
 A^{a_1a_2a_3a_4}_{(4;0 )} ({\bm k}_1,{\bm k}_2,{\bm k}_3,{\bm k}_4 ).
\end{align}
  The transition
 is given as the sum of a connected and a
disconnected part
\begin{align}
  \label{eq:trans24}
  U_{2\to 4}^{a_1a_2; b_1b_2b_3b_4} ( {\bm l}_1,{\bm l}_2; {\bm
    k}_1,{\bm k}_2,{\bm k}_3,{\bm k}_4 ) = &
  U_{\text{connect}}^{a_1a_2; b_1b_2b_3b_4} ( {\bm l}_1,{\bm l}_2;
  {\bm k}_1,{\bm k}_2,{\bm k}_3,{\bm k}_4 ) \notag \\&+
  U_{\text{disconnect}}^{a_1a_2; b_1b_2b_3b_4} ( {\bm l}_1,{\bm l}_2;
  {\bm k}_1,{\bm k}_2,{\bm k}_3,{\bm k}_4 )
\end{align}
where the latter arises as all possible combinations of the 1-3 transition Eq.(\ref{eq:U13}) of the previous section:
\begin{align} 
  \label{eq:disconnected}
 U_{\text{disconnect}}^{a_1a_2; b_1b_2b_3b_4} &( {\bm l}_1,{\bm l}_2; {\bm k}_1,{\bm k}_2,{\bm k}_3,{\bm k}_4 ) =
\notag \\
 = \frac{1}{2}\sum_{i_1,i_2} \bigg[ &
\delta^{a_{i_1}b_{4}}
(2\pi)^3 \delta^{(2)}( \bm{l }_{i_2} - \bm{k }_{4} )
U_{1 \to 3}^{a_{i_2};b_{1}1b_{2}b_{3}}({\bm l}_{i_1}; {\bm k}_{1}, {\bm k}_{2},  {\bm k}_{3})
\notag \\ &
+
\delta^{a_{i_1}b_{3}}
(2\pi)^3 \delta^{(2)}( \bm{l }_{i_2} - \bm{k }_{3} )
U_{1 \to 3}^{a_{i_2};b_{1}1b_{2}b_{4}}({\bm l}_{i_1}; {\bm k}_{1}, {\bm k}_{2},  {\bm k}_{4})
\notag \\ &
+
\delta^{a_{i_1}b_{2}}
(2\pi)^3 \delta^{(2)}( \bm{l }_{i_2} - \bm{k }_{2} )
U_{1 \to 3}^{a_{i_2};b_{1}1b_{3}b_{4}}({\bm l}_{i_1}; {\bm k}_{1}, {\bm k}_{3},  {\bm k}_{4})
\notag \\ &
+
\delta^{a_{i_1}b_{1}}
(2\pi)^3 \delta^{(2)}( \bm{l }_{i_2} - \bm{k }_{1} )
U_{1 \to 3}^{a_{i_2};b_{2}1b_{3}b_{4}}({\bm l}_{i_1}; {\bm k}_{2}, {\bm k}_{3},  {\bm k}_{4}) \bigg]
\end{align}
where the sum is over permutations of the numbers $1,2$.  To state the
connected part, we again make  use of a diagrammatic notation in
terms of Reggeon momentum diagrams. For the connected part, the
following types of expressions are needed:
\begin{align}
  \label{eq:def_rmfd}
\parbox{1.5cm}{\includegraphics[height=1.3cm]{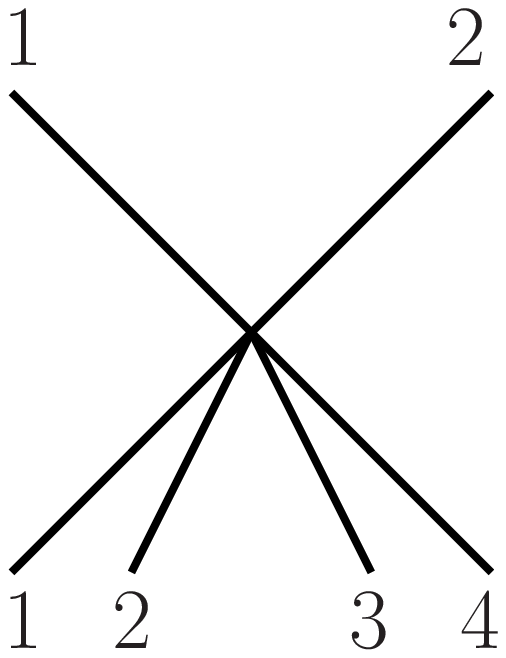}} &= 
\frac{( {\bm k}_1+ {\bm k}_2+ {\bm k}_3+ {\bm k}_4)^2}{{\bm l}_1^2{\bm l}_2^2} 
& 
\parbox{1.5cm}{\includegraphics[height=1.3cm]{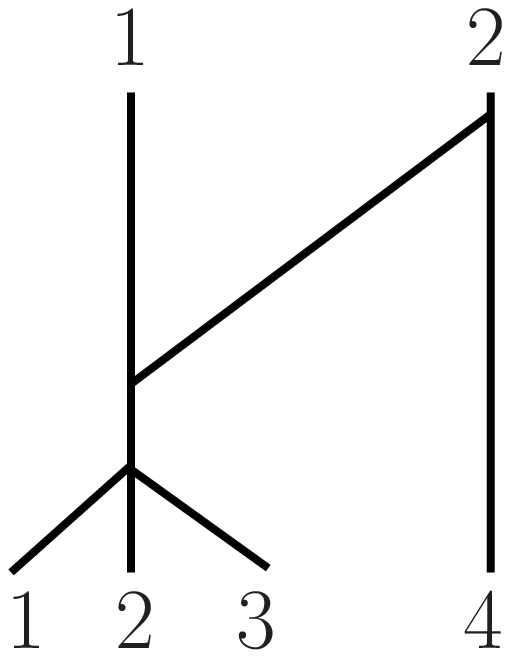}} &= 
\frac{( {\bm k}_1+ {\bm k}_2+ {\bm k}_3)^2 }{{\bm l}_1^2 ({\bm l}_2 - {\bm k}_4)^2 } 
 \notag \\
\parbox{1.5cm}{\includegraphics[height=1.3cm]{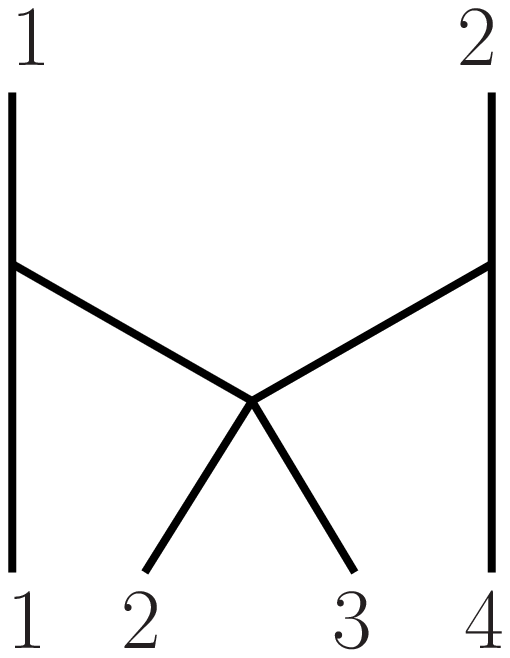}} &= 
\frac{( {\bm k}_2+ {\bm k}_3)^2 }{({\bm l}_1 - {\bm k}_1)^2( {\bm l}_2 - {\bm k}_4)^2} 
& 
\parbox{1.5cm}{\includegraphics[height=1.3cm]{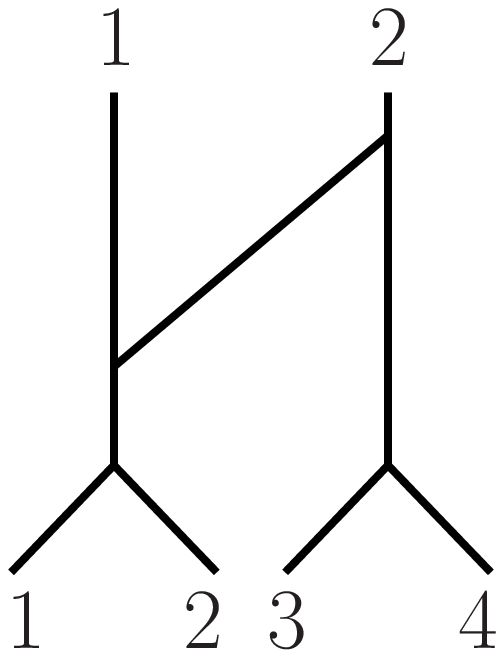}} &= 
\frac{ ( {\bm k}_1+ {\bm k}_2)^2  }{{\bm l}_1^2({\bm l}_2  - {\bm k}_3- {\bm k}_4)^2 }  \notag \\
\parbox{1.5cm}{\includegraphics[height=1.3cm]{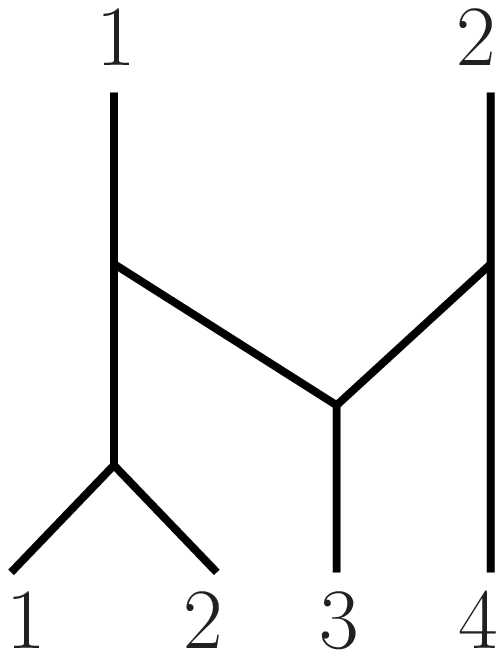}} &= \frac{{\bm k}_3^2 }{( {\bm l}_1 - {\bm k}_1 - {\bm k}_2)^2({\bm l}_2  - {\bm k}_4)^2} 
\end{align}
The transition vertex is then given by
\begin{align} 
  \label{eq:connected}
U_{\text{connect}}^{a_1a_2; b_1b_2b_3b_4}& ( {\bm l}_1,{\bm l}_2; {\bm k}_1,{\bm k}_2,{\bm k}_3,{\bm k}_4 ) 
 = \sum_{j_1, \ldots, j_4}
\left( T^{b_{j_1}} T^{b_{j_2}}T^{b_{j_3}} T^{b_{j_4}}  \right)_{a_1a_2} 
 U_C ( {\bm l}_{i_1},{\bm l}_{i_2}; {\bm k}_{j_1},  {\bm k}_{j_2} ,  {\bm k}_{j_3},{\bm k}_{j_4}) 
\end{align}
where the sum is  over all permutations of numbers  $1, \ldots, 4$ and  generators are in the adjoint representation of $SU(N_c)$. The function $U_C$ is given by
\begin{align}
  \label{eq:UC}
 U_C ( {\bm l}_{1},{\bm l}_{2}; {\bm k}_{1},  {\bm k}_{2} ,  {\bm k}_{3},{\bm k}_{4}) &=
\notag \\
g^4{\bm l}_1^2{\bm l}_2^2 & \left[  \frac{1}{24}\parbox{1.5cm}{\includegraphics[height=1.3cm]{arf.ps}} 
- \frac{1}{12} \parbox{1.5cm}{\includegraphics[height=1.3cm]{brf123.ps}}
- \frac{1}{12} \parbox{1.5cm}{\includegraphics[height=1.3cm]{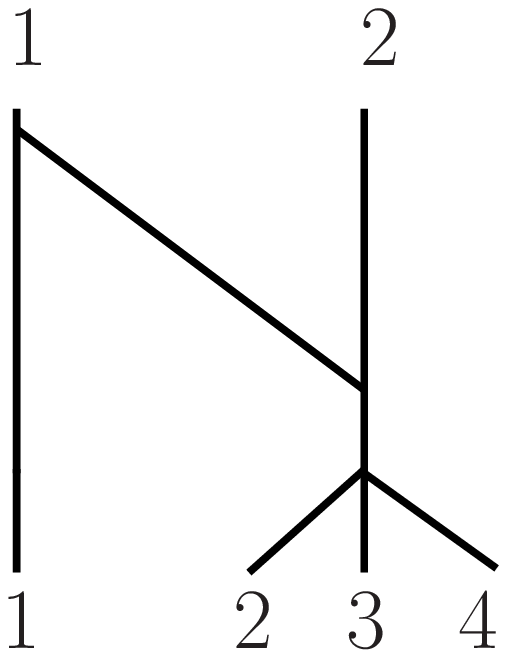}}
+
\frac{1}{8} \parbox{1.5cm}{\includegraphics[height=1.3cm]{crf1234.ps}}\right.
\notag \\
&
+ \left.
\frac{1}{24}  \parbox{1.5cm}{\includegraphics[height=1.3cm]{drf1234.ps}}
-
\frac{1}{24} \parbox{1.5cm}{\includegraphics[height=1.3cm]{erf1234.ps}}
+
\frac{1}{24}  \parbox{1.5cm}{\includegraphics[height=1.3cm]{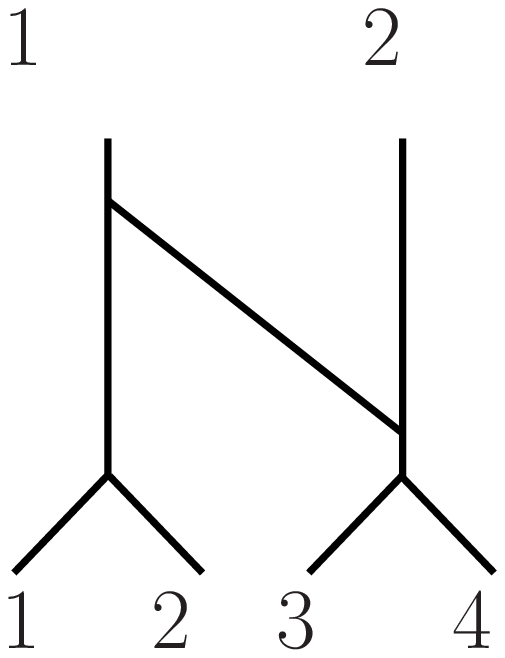}}
-
\frac{1}{24} \parbox{1.5cm}{\includegraphics[height=1.3cm]{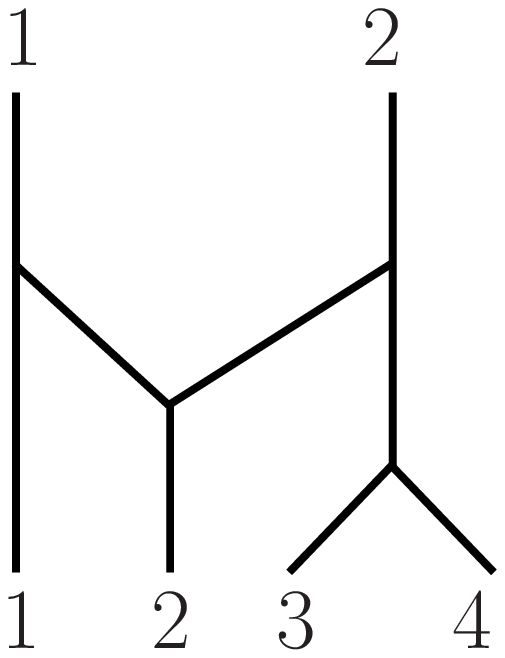}}
\right]
\end{align}

\section{The state of three and four reggeized gluons}
\label{sec:v24bartelswuest}
For a description of the complete state of three and four reggeized
gluons, the above derived transition vertices need to be combined with
the pairwise interactions by two-to-two transition kernels between the
individual reggeized gluons.

\subsection{Integral equations }
\label{sec:bkp}

\begin{figure}[htbp]
  \centering
  \parbox{3.1cm}{\center \includegraphics[height = 2.5cm]{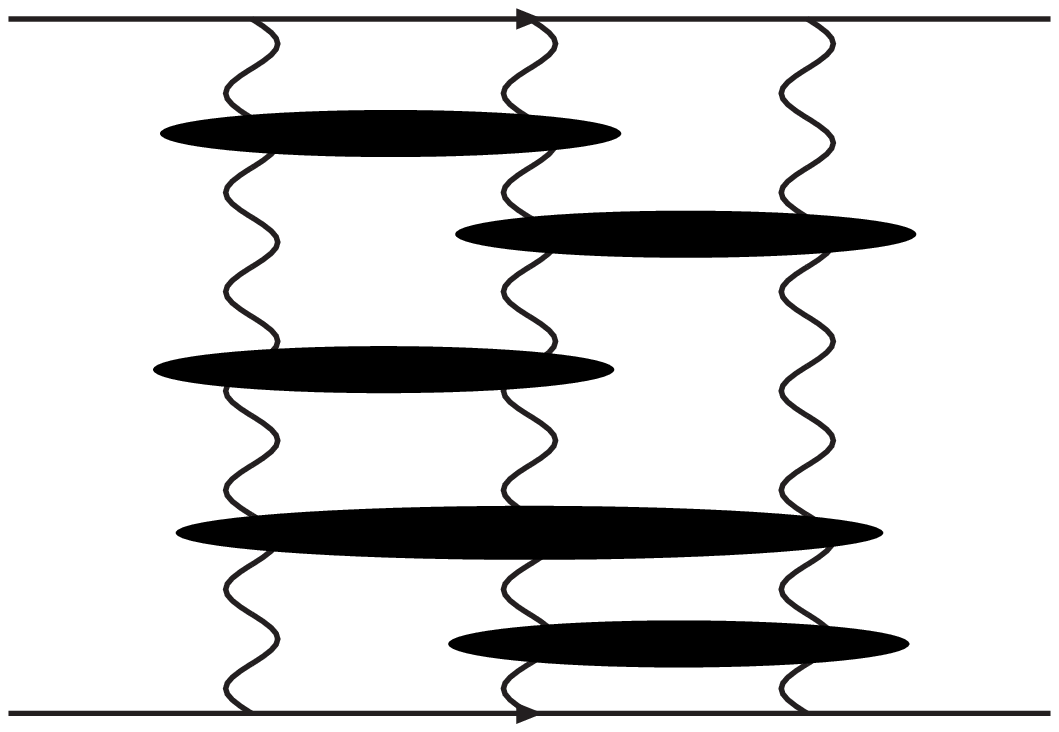}}
  \parbox{3.1cm}{\center \includegraphics[height = 2.5cm]{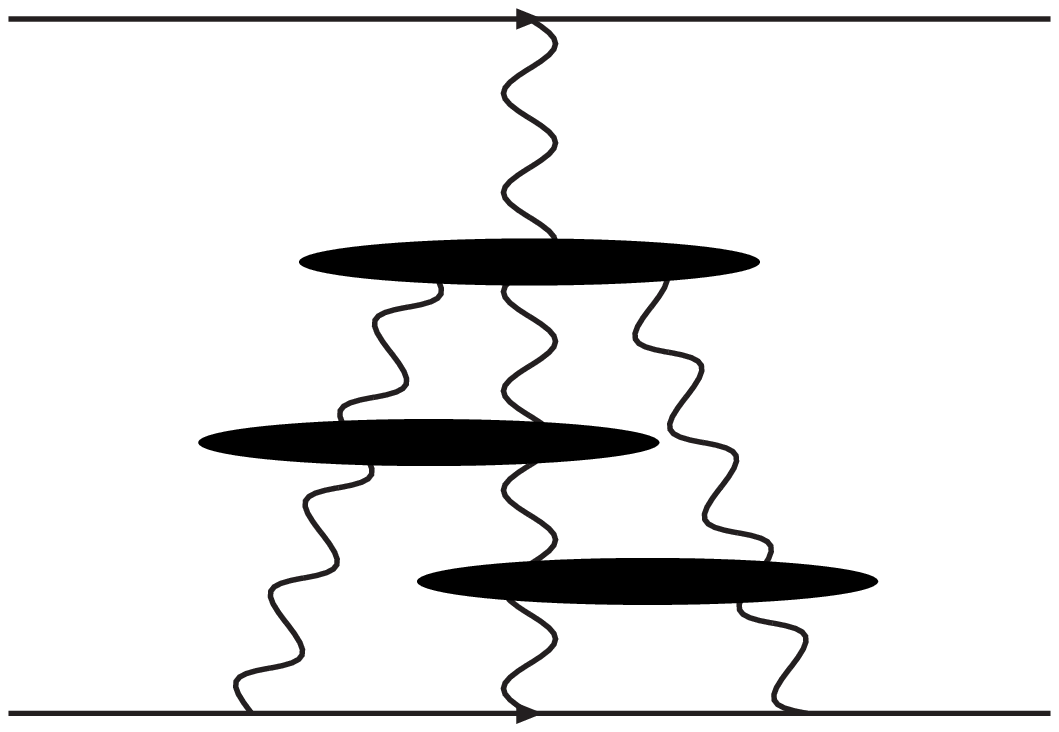}}
  \parbox{3.1cm}{\center \includegraphics[height = 2.5cm]{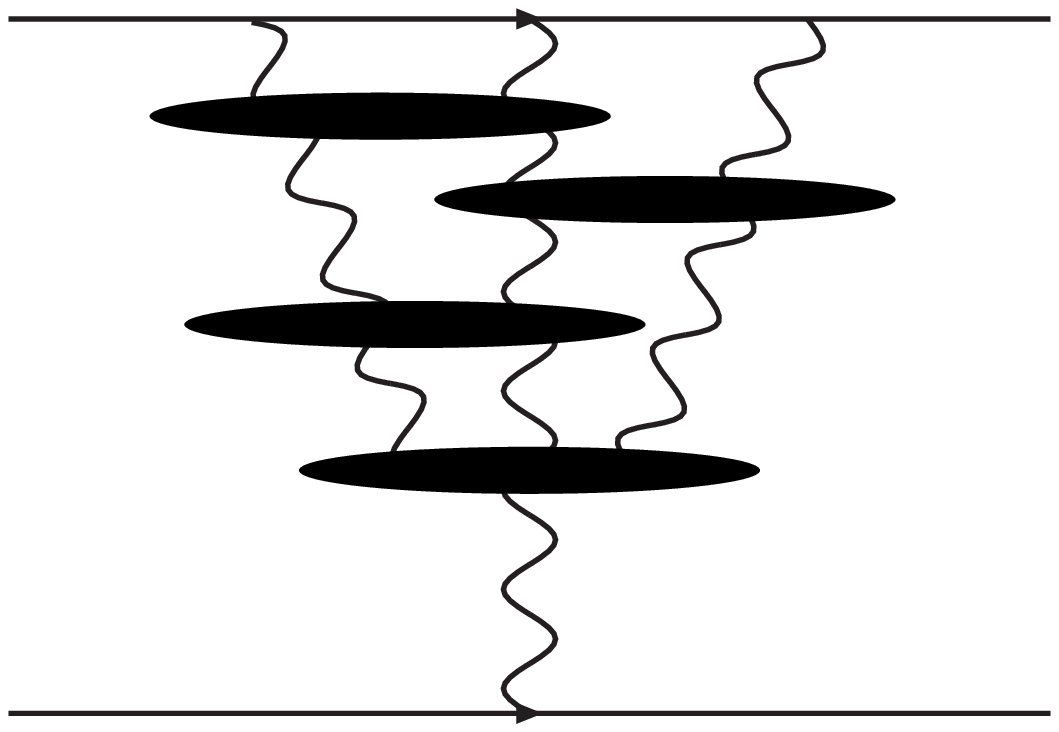}} 
  \parbox{3.1cm}{\center \includegraphics[height = 2.5cm]{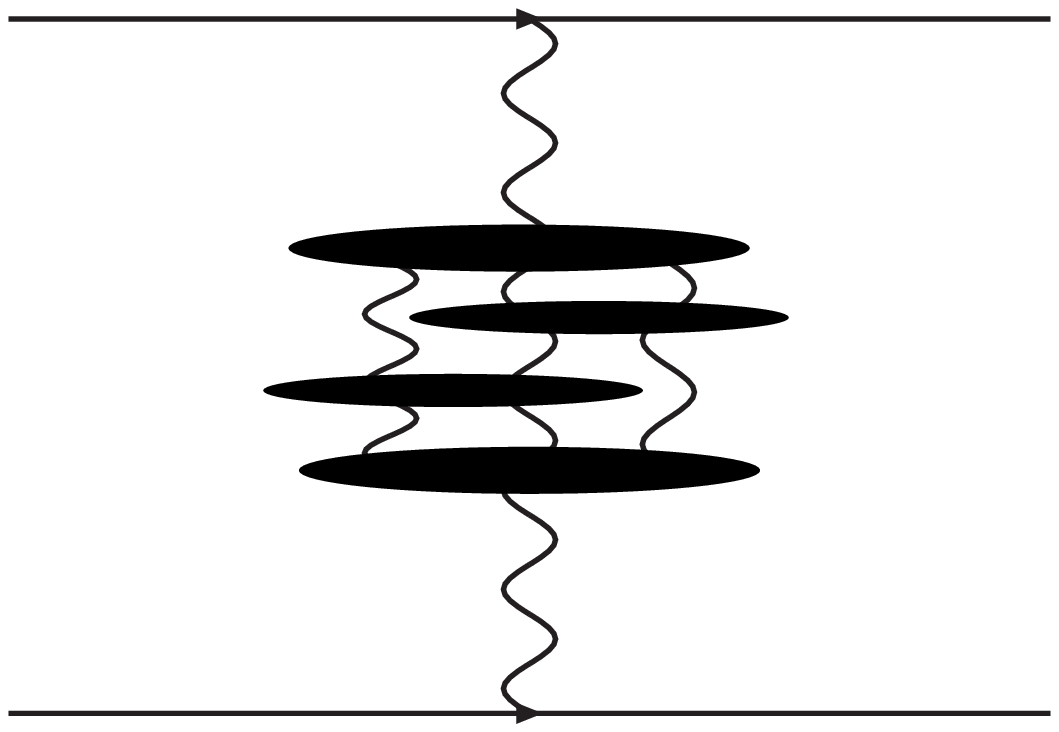}} \\
\parbox{3.1cm}{ \center (a)}\parbox{3.1cm}{ \center (b)}\parbox{3.1cm}{ \center (c)}\parbox{3.1cm}{ \center (d)}
  \caption{\small Different types of loop corrections to the state of three reggeized gluons  (a) three reggeized gluons coupling to both quarks and  diagrams containing  one   (b)(c) and two (d) transition(s) between the one and three reggeized gluon state.}
  \label{fig:loop_34regg}
\end{figure}

For three reggeized gluons, the relevant contributions within the GLLA are depicted in Fig.\ref{fig:loop_34regg}, with all other corrections suppressed at least by an additional factor of $g^2$.

For every diagram it is possible to identify at a certain stage a state of three reggeized gluons. We then factorize the amplitude at this point into two quark-three reggeized gluon amplitudes. Expressing the complete amplitude within the GLLA as the following Mellin-transform
\begin{align}
  \label{eq:bkpmal_done}
 \mathcal{M}^{\text{LLA}|\text{3R}}_{2 \to 2}  & =  \frac{ (2\pi)^2}{3!}  s
\int \frac{d \omega}{2 \pi i}   \left(\frac{ s}{s_R}   \right)^{\omega} 
 \xi^{(-)}(\omega)  \phi_3(\omega, t), 
\end{align}
we write the partial wave $ \phi_3(\omega, t)$ the following convolution:
\begin{align}
  \label{eq:phi_3bkp}
\phi_3(\omega, t) =   A^{a_1a_2a_3}_{3} (\omega| {\bm k}_1,{\bm k}_2,{\bm k}_3 ) 
(\omega - \sum_{i}^3\beta({\bm k}_i^2)) \otimes_{{\bm k}_{123}}
 A^{a_1a_2a_3}_{3} ({\bm k}_1,{\bm k}_2,{\bm k}_3 )
\end{align}
where $ A^{a_1a_2a_3}_{3}(\omega) $ denote the corresponding
quark-3reggeized gluon amplitudes which is defined to contain a factor $1/(\omega -  \sum_{i}^3\beta({\bm k}_i^2))$ for the external reggeized gluons. The state of three
reggeized gluons can both arise from a direct coupling to the quark
and by a transition from a $1 -3$ transition. It is therefore within
the GLLA resummed by the following integral equation
\begin{align}
  \label{eq:bkp}
(\omega - \sum_{i = 1}^3 \beta({\bm k}_i^2))  A^{b_1b_2b_3}_{3}(\omega| {\bm k}_1,{\bm k}_2,{\bm k}_3 ) 
= A^{a_1a_2a_3}_{(3;0)} + 
U_{1 \to 3}^{a;b_1b_2b_3}   A_1^a(\omega) +
\sum \mathcal{K}_{2 \to 2}^{\{b \} \to \{a \} }\otimes A^{a_1a_2a_3}_{3 },
\end{align}
where the sum on the right-hand-side contains all possible insertions
of the 2-2 kernel and
\begin{align}
  \label{eq:A1}
A^a_1(\omega |{\bm q}) = gt^a \frac{1}{\omega - \beta({\bm q}^2)} \frac{1}{{\bm q}^2}
\end{align}
is the amplitude describing coupling of a single reggeized gluon to the quark.

\begin{figure}[htbp]
  \centering
  \parbox{5cm}{\center \includegraphics[height=3cm]{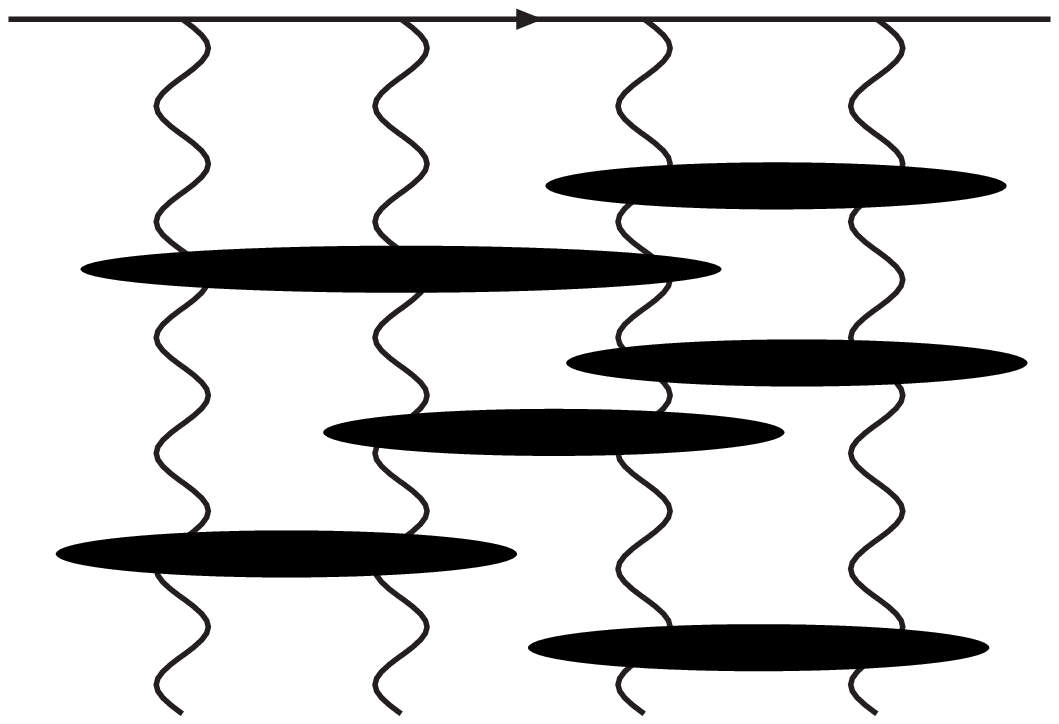}}
  \parbox{5cm}{\center \includegraphics[height=3cm]{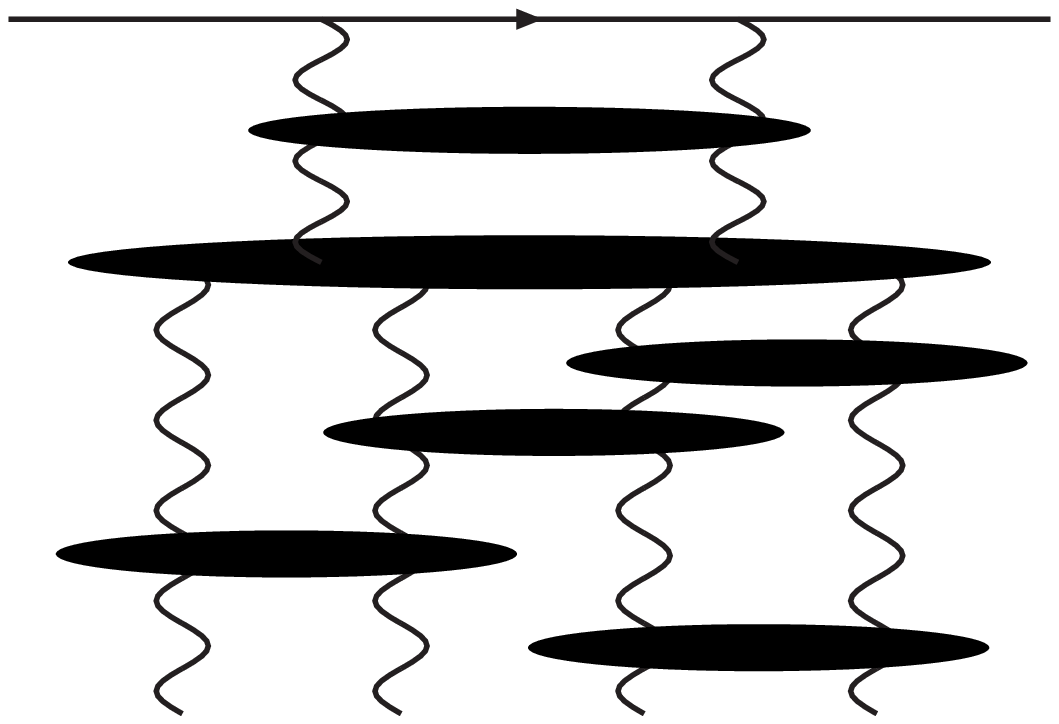}} \\
\parbox{5cm}{\center (a)}\parbox{5cm}{\center (b)}
  \caption{\small Different contributions to the state of four reggeized gluons:  (a) : direct coupling of the four reggeized gluons to the quark (b): transition from the two to the four reggeized gluon state}
  \label{fig:4regg}
\end{figure}
The partial wave that describes the exchange of four reggeized gluons
can be factorized in an analogous way. The different contributions are
depicted in Fig.\ref{fig:4regg}: The state of four reggeized gluons
may either arise from a direct coupling of the four reggeized gluon to
the upper quark (in the case of Fig.\ref{fig:4regg}) or by a
transition between the state of two and four reggeized gluons. The
elastic scattering amplitude with exchange of four reggeized gluons
within the GLLA given by
\begin{align}
  \label{eq:4R}
 \mathcal{M}^{\text{LLA}|\text{4R} }_{2 \to 2}   =\frac{ (2\pi)^3}{4!}   s
 \int \frac{d \omega}{2 \pi i}    \left(\frac{ s}{s_R}   \right)^{\omega}  \frac{\xi^{(+)}}{\sin \pi\omega}  \phi_4(\omega, t).
\end{align}
The partial wave $\phi_4(\omega, t)$  is then factorized into two quark-four-reggeized gluons amplitudes in the following way:
\begin{align}
  \label{eq:phi_4bkp}
\phi_4(\omega, t) =  
 A^{a_1a_2a_3a_4}_{4} (\omega| {\bm k}_1,{\bm k}_2,{\bm k}_3,{\bm k}_4 ) 
(\omega - \sum_{i = 1}^4 \beta({\bm k}_i^2))\otimes_{{\bm k}_{1234}}
 A^{a_1a_2a_3a_4}_{4} (\omega |{\bm k}_1,{\bm k}_2,{\bm k}_3,{\bm k}_4 ).
\end{align}
As for the state of three reggeized gluons, also the quark-four-reggeized gluons amplitude is non-amputated and  contains a factor $1/(\omega - \sum_{i = 1}^4 \beta({\bm k}_i^2) )$ for its external reggeized gluons. It  satisfies the following integral equation
\begin{align}
  \label{eq:4bkp}
(\omega - \sum_{i = 1}^4 \beta({\bm k}_i^2))  A^{b_1b_2b_3b_4}_{4 }(\omega| {\bm k}_1,{\bm k}_2,{\bm k}_3,{\bm k}_4 ) 
=&
 A^{b_1b_2b_3b_4}_{(4;0)} + 
 U^{a_1a_2;b_1b_2b_3b_4}_{2\to4} \otimes  A^{a_1a_2}_{2}(\omega)
\notag \\
& \qquad  + 
\sum \mathcal{K}_{2 \to 2}^{\{a \} \to \{b \} }\otimes A^{a_1a_2a_3a_4}_{4}(\omega).
\end{align}
Compared to existing results in the literature (to which we compare in
the following paragraph) the above integral equations have the special
feature that they are not restricted to the overall color singlet. In
particular they allow for a search of further Regge-poles in the
color-octet channel. In the following we shall however not do so, but
compare instead our results with the yet existing results for the
color singlet channel.

\subsection{The four reggeized gluon state in the color singlet}
\label{sec:singlet}

In the following section, we compare the above obtained result with
existing results in the literature.  In particular the question arises in which  sense the integral equations describing the state of four reggeized gluons are
related to the analysis of
\cite{Bartels:1992ym,Bartels:1993ih,Bartels:1994jj}. While the  analysis presented there  was carried out for  $N_c = 3$, the result was reproduced for arbitrary, finite $N_c$ in \cite{Bartels:1999aw} and we will refer if we compare the exact form of results.

To compare with \cite{Bartels:1994jj} we consider quark-reggeized
gluons amplitudes, as shown in Fig.\ref{fig:4regg}.  In
\cite{Bartels:1994jj} the analysis has been carried out for state of
four reggeized gluons coupling to a virtual photon, and not to a
quark. In particular this restricts the four reggeized gluon state to
the overall color singlet. In turn, we shall therefore only take the
color-singlet part of the scattering quark A into account.

\subsection{The Born-result}
\label{sec:bornyellow}

To lowest order in the coupling constant, the contribution due to the
transition of two-to-four reggeized gluons is absent and we encounter
only the contribution due to the quark impact factor with four
reggeized gluons, Eq.~(\ref{eq:quark_4gluonfertig}). To 
compare with the presentation in \cite{Bartels:1994jj, Bartels:1999aw}
we project the scattering quark onto the color singlet and write the
four gluon impact-factor Eq.(\ref{eq:quark_4gluonfertig}) as a
superposition of the two-reggeized gluon quark-impact factor
Eq.~(\ref{eq:decompose_quarkif}), in analogy with the analysis
of \cite{Bartels:1994jj, Bartels:1999aw}.  Projected onto the
color-singlet, the two-gluon impact factor  reads
\begin{align}
  \label{eq:twoimpa}
   B^{a_1a_2}_{(2;0)}({\bm k}_1, {\bm k}_2) = \delta^{a_1a_2}  B_{(2;0)}({\bm k}_1, {\bm k}_2) = \frac{-g^2}{2N_c}\delta^{a_1a_2}.
\end{align}
where we denote here impact factors and partial waves by a letter 'B' in order stress that in the following case the quarks are explicitly projected on their color singlet.
The transverse momenta dependence of the above function is of course
 trivial. We however attempt to stay as close as possible to
the analysis of \cite{Bartels:1994jj} and that is why it is convenient
for us to introduce formally such a function. Furthermore, let us define color tensors,
\begin{align}
  \label{eq:d1234}
d^{a_1a_2a_3a_4} =&  \tr( t^{a_1} t^{a_2} t^{a_3} t^{a_4})  +
\tr( t^{a_4} t^{a_3} t^{a_2} t^{a_1})  ,
\end{align}
 following  \cite{Bartels:1994jj, Bartels:1999aw}. In the present case, it is then furthermore useful to introduce analogous color tensors, one which is  symmetric under  the exchange  of the first two and the last two pairs of indices 
 \begin{align}
  \label{eq:d2}
d^{a_1a_2a_3a_4}_2 =& \frac{1}{2} \big[d^{a_1a_2a_3a_4} + d^{a_2a_1a_3a_4} \big],
\end{align}
and one which is  symmetric under exchange of any pair of color indices
\begin{align}
  \label{eq:ds tensor}
d^{a_1a_2a_3a_4}_S =& \frac{1}{3}\big[ d^{a_1a_2a_3a_4} + d^{a_2a_1a_3a_4} +
 d^{a_1a_3a_2a_4} \big].
\end{align}
The impact factor for four reggeized gluons coupling to the quark, as derived from the effective action, can then be  expressed,  for the overall color singlet,  in terms of the two reggeized gluon impact factor in the following way:
\begin{align}
  \label{eq:fourimap} 
B^{a_1a_2a_3a_4}_{(4;0)}&({\bm k}_1, {\bm k}_2, {\bm k}_3, {\bm k}_4) =
\notag \\
-g^2& d^{a_1a_2a_3a_4}_S \left[ B_{(2;0)}(123,4) + B_{(2;0)}(124,3)  +
B_{(2;0)}(134,2)  + B_{(2;0)}(234,1) \right] 
\notag \\
+
g^2 &d^{a_1a_2a_3a_4}_2 B_{(2;0)}(12,34) 
+
 g^2 d^{a_1a_3a_2a_4}_2 B_{(2;0)}(13,24) 
+
 g^2 d^{a_1a_4a_2a_3}_2 B_{(2;0)}(14,23),
\end{align}
where we introduced the following  short-hand notation
\begin{align}
  \label{eq:shorthand}
B_{(2;0)}(12,34)  = B_{(2;0)} ({\bm k}_1 + {\bm k}_2, {\bm k}_3 +{\bm k}_4),
\end{align}
where a string of numbers represent a sum of transverse momenta with the
corresponding indices. In \cite{Bartels:1994jj}, the corresponding
result was obtained from taking the triple discontinuity of the
six-point amplitude, as outlined in the first part of this thesis.
With $D_{(2;0)}$ and $D_{(4;0)}$ denoting the virtual photon impact
factor with two and four reggeized gluons, the result for the four
gluon impact factor is given by
\begin{align}
  \label{eq:4impabartels}
D^{a_1a_2a_3a_4}_{(4;0)}&({\bm k}_1, {\bm k}_2, {\bm k}_3, {\bm k}_4) =
\notag \\
-g^2& d^{a_1a_2a_3a_4}  \left[ D_{(2;0)}(123,4) + D_{(2;0)}(234,1)  +
 - D_{(2;0)}(14,23)  \right] 
\notag \\
-g^2& d^{a_2a_1a_3a_4}  \left[ D_{(2;0)}(134,2) + D_{(2;0)}(124,3)  -
D_{(2;0)}(12,34)  + D_{(2;0)}(13,24) \right] .
\end{align}
Comparing the two results, one observes, apart from the different
impact factors, the following difference compared to
Eq.(\ref{eq:4impabartels}): Eq.(\ref{eq:fourimap}) lacks a certain antisymmetric color structure.
While, for example, the term $D_{(2;0)}(12,34)$ comes with a color
tensor
\begin{align}
  \label{eq:colortensor12@34}
d^{a_2a_1a_3a_4} = d^{a_1a_2a_3a_4}_2 +\frac{1}{4} f^{a_1a_2k}f^{a_3a_4k},
\end{align}
the analogous term $B_{(2;0)}(12,34)$ in Eq.(\ref{eq:fourimap}) lacks
the second color antisymmetric structure on the left hand side of
Eq.(\ref{eq:colortensor12@34}).
\begin{figure}[htbp]
  \centering
   \parbox{6cm}{\center \includegraphics[height=3cm]{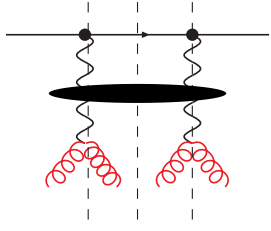}}
  \caption{\small  A contribution of the effective action that can yield a contribution to the triple discontinuity with antisymmetric color.}
  \label{fig:disc1234}
\end{figure}
The reason lies in the different definitions of the two quantities:
Whereas Eq.(\ref{eq:4impabartels}) arises from the
triple-discontinuity of a six-point amplitude, which naturally
includes both symmetric and anti-symmetric color structure,
Eq.(\ref{eq:fourimap}) constitutes an impact factor of the effective
theory. In Sec.\ref{sec:tworeggeon_negsig} and Sec.\ref{sec:impa4_bkp}
it has been shown that for the effective action the configurations
with antisymmetric color are completely contained in the reggeized
gluons, due, in particular, to the presence of the induced vertices.
The state of two and four reggeized gluons coupling to the quark is on
the other hand restricted to completely symmetric color
configurations.  In particular, reggeized gluons in the effective
action carry phase factors and therefore have a non-zero discontinuity
themselves. The triple energy discontinuity for the antisymmetric
color configuration should therefore arise in the effective action
form graphs like Fig.\ref{fig:4regg}c.  Indeed, obtaining in
Eq.(\ref{eq:fourimap}) also the antisymmetric color states, would have
been a strong hint for a possible over-counting in the effective
action.

\subsection{The  four reggeized gluons and the triple Pomeron vertex}
\label{sec:oneloopblack}

To resum the state of four reggeized gluons, Fig.\ref{fig:4regg}, we then use the integral equation Eq.(\ref{eq:4bkp}) 
\begin{align}
  \label{eq:inteq4}
(\omega - \sum_{i = 1}^4 \beta({\bm k}_i^2))  B^{b_1b_2b_3b_4}_{4 }(\omega| {\bm k}_1,{\bm k}_2,{\bm k}_3,{\bm k}_4 ) 
=&
 B^{b_1b_2b_3b_4}_{(4;0)} + 
 U^{a_1a_2;b_1b_2b_3b_4}_{2\to4} \otimes  B^{a_1a_2}_{2}(\omega)
\notag \\
& \qquad  + 
\sum \mathcal{K}_{2 \to 2}^{\{a \} \to \{b \} }\otimes B^{a_1a_2a_3a_4}_{4}(\omega).
\end{align}
In the following analysis we then follow closely the approach taken in \cite{Bartels:1994jj} and split up the four-reggeized-gluon amplitude into a reggeizing part $B_4^R$ and into an irreducible part $B_4^I$,
\begin{align}
  \label{eq:a4ri}
B_4 = B_4^R + B_4^I.
\end{align}
The reggeizing part is then given as the superposition of two-reggeized gluon amplitudes $B_2(\omega)$ 
\begin{align}
  \label{eq:A4R}
B^{R;a_1a_2a_3a_4}_{4}&(\omega| {\bm k}_1, {\bm k}_2, {\bm k}_3, {\bm k}_4) =
\notag \\
-g^2& d^{a_1a_2a_3a_4}_S \left[ B_{2}(\omega| 123,4) + B_{2}(\omega|124,3)  +
B_{2}(\omega|134,2)  + B_{2}(\omega|234,1) \right] 
\notag \\
+
g^2 &d^{a_1a_2a_3a_4}_2 B_{2}(\omega|12,34) 
+
 g^2 d^{a_1a_3a_2a_4}_2 B_{2}(\omega|13,24) 
+
 g^2 d^{a_1a_4a_2a_3}_2 B_{2}(\omega|14,23)
\end{align}
where the particular form is obtained by replacing all two-reggeized
gluon impact-factors $B_{(2;0)}$ in the four-reggeized gluon impact
factor Eq.(\ref{eq:fourimap}) by their corresponding quark-2
reggeized gluon amplitudes, following \cite{Bartels:1994jj}. We then insert the ansatz
Eq.(\ref{eq:a4ri}) into the integral equation Eq.(\ref{eq:inteq4}) and
obtain as such a new integral equation for $B_4^I$. Using  the
BFKL-equation for $B_2(\omega)$,
\begin{align}
  \label{eq:bfkl_a}
(\omega - \sum_{i = 1}^2 \beta({\bm k}_i^2))  B_{2 }(\omega| {\bm k}_1,{\bm k}_2 ) 
=&
 B^{b_1b_2}_{(2;0)} 
 + (-N_c)
\mathcal{K}_{2 \to 2}^{\{a \} \to \{b \} }\otimes B^{a_1a_2}_{2}(\omega)
\end{align}
and employing various color identities (see for instance
\cite{Bartels:1999aw} for a suitable collection), we finally arrive at the 
following integral equation for $B_4^I$,
\begin{align}
  \label{eq:a4iinteq4}
(\omega - \sum_{i = 1}^4 \beta({\bm k}_i^2))  B^{I; b_1b_2b_3b_4}_{4}(\omega| {\bm k}_1,{\bm k}_2,{\bm k}_3,{\bm k}_4 ) 
=&
 V^{a_1a_2; b_1b_2b_3b_4}_{2\to 4} \otimes  B^{a_1a_2}_{2}(\omega)
\notag \\
& \qquad  + 
\sum \mathcal{K}_{2 \to 2}^{\{a \} \to \{b \} }\otimes B^{I;a_1a_2a_3a_4}_{4}(\omega)
\end{align}
where the kernel $ V_{2\to 4}$ agrees with the arbitrary $N_c$ version
\cite{Bartels:1999aw} of the two-to-four reggeized gluon vertex of
\cite{Bartels:1994jj}.

\section{Conclusion}
\label{sec:conci}
In the present chapter we extended our study of the effective action to
the exchange of states of $n>2$ reggeized gluons in the $t$-channel of
the elastic scattering amplitude. As a particular example, the state
of three and four reggeized gluons has been considered. 
 While the former has
negative signature and provides a correction to the exchange of a
single reggeized gluon, the latter can be shown to have positive
signature and gives a correction to the state of two reggeized gluons.

A first step to arrive at these results requires to determine the pole
prescription of higher induced vertices. They can be obtained by
 the following recipe: Starting from a simple pole
prescription $1/\partial \to 1/\partial -\epsilon$ in the effective
action, one decomposes the color structure into a basis of color
tensors, which is constructed from  multiple entangled
commutators of generators and symmetrization in the remaining parts.
The most antisymmetric terms, given in terms of multiple entangled
commutators alone, coincide in their color structure with induced
vertices without pole-prescription.  We then identified the momentum
structure proportional to these maximal antisymmetric color structure
as the induced vertex with pole-prescription, while we drop all other
terms.  The resulting vertices turns out to be
Bose-symmetric.

For the longitudinal loop integrals of loops containing more than two
reggeized gluons we found that, also in that case bad convergence
properties of the occurring integrals ask for suitable subtraction
diagrams.  We proposed to add a subtraction term to the effective
Lagrangian that generates the subtraction diagrams.  Including these
diagrams, a derivation of quark impact-factors with three and four
reggeized gluons and of vertices which describe the transition from
one to three and two to four reggeized gluons becomes possible.
Transition vertices from one-to-two and two-to-three reggeized gluons
on the other hand, are not allowed due to signature conservation:
Their vanishing is obtained as a result from the effective action and
is not needed to be imposed as an external constraint.

In a next step we presented integral equations for the state of three
and four reggeized gluons, which resum corrections within the
generalized LLA. Apart from the derived non-zero transition vertices,
they resum pairwise interactions between reggeized gluons by
two-to-two transition kernels.  We then investigate more closely the
integral equation for the quark-four-reggeized-gluons amplitude with
the quark projected onto the color singlet. Following a similar
analysis by Bartels and W\"usthoff in \cite{Bartels:1994jj}, the
four-reggeized gluon amplitude was split up into a reggeizing and an
irreducible part. While the former takes the form of a state of two
reggeized gluons that finally decays with various symmetric color
tensors, the latter arises from the two-to-four Reggeon transition
vertex derived by Bartels and W\"usthoff in \cite{Bartels:1994jj}. The
symmetric color tensors of the reggeizing contributions reflect the
choice of the color basis of the induced vertices in
Sec.\ref{sec:indu_presc}.  Contributions with antisymmetric color
tensors are contained inside higher order corrections to the reggeized
gluon and occur therefore not as a part of the state of four reggeized
gluons.

With the above discussion of the state of three and four reggeized
gluons we finish our discussion of longitudinal integrations within
the effective action. Let us therefore shortly summarize our central
results: The effective action \cite{Lipatov:1995pn,Lipatov:1996ts}
factorizes high-energy-QCD amplitudes into sub-amplitudes which are
local in rapidity. Non-local interaction between these sub-amplitudes
is mediated by the exchange of reggeized gluons. This statement implies
a non-trivial constraint on the longitudinal momenta of QCD-particles
inside the local sub-amplitudes.  In practice, this can be achieved by
imposing a lower bound on the 'center-of-mass' energy of the reggeized
gluons that connects the local sub-amplitudes. To give a proper
definition of the center-of-mass energy of a reggeized gluon, it is
necessary to consider the latter always as an object embedded into a
minimal sub-amplitude that can be associated with its exchange. The
prime example is the case where the reggeized gluon couples on both
ends to a Particle-Particle-Reggeon vertex, which allows for a
straight-forward definition of the center-of-mass energy of the
reggeized gluon. The lower bound on the center of mass energy can be
then imposed through an (inverse) Mellin-integral. It has the
advantage that it allows to include easily both higher order
corrections due to reggeization of the gluon and signature of the
reggeized gluon.  In the following we state the  example where the
reggeized gluon couples on both ends to a gluon by a
Gluon-Gluon-Reggeon vertices $ \Gamma_{GGR}$:
\begin{align}
  \label{eq:vertexsig_fac_extract_re}
\parbox{2cm}{\includegraphics[width=2cm]{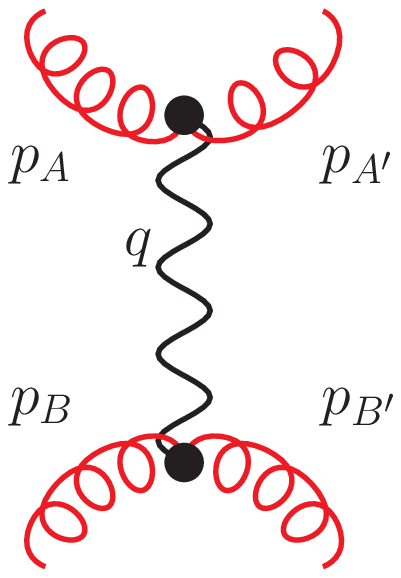}} 
 &=  \int \frac{d \omega}{4\pi i} \Gamma^+_{GGR}(p_A, q) \frac{i/2}{{\bm q}^2}  \frac{1 }{\omega -  \beta({\bm q}^2) } 
 \left[ \left( \frac{-s -i\epsilon}{s_R}\right)^{\omega}  +\left( \frac{s-i\epsilon}{s_R}\right)^{\omega} \right]  \Gamma^-_{GGR}(p_B, q).
\end{align}
The squared center-of-mass energy $s$ is always defined as the product
of the large light-cone-momenta of the scattering particles, in the
present case $s = p_A^+p_B^-$, also if transverse or other light-cone
momenta of the scattering particles are non-zero. This  is
particularly suitable as it preserves Bose-symmetry of the external
gluons: Due to the general property of the reggeized gluon fields
\begin{align}
  \label{eq:prop_reggfield}
\partial_-A_+ = 0 = \partial_+A_-,
\end{align}
we have $p_A^+ = -p_{A'}^+$ and $p_B^- = -p_{B'}^-$. $s_R$ is the
lower bound that is imposed on $s$: With the contour of integration
of the $\omega$-integral parallel to the imaginary axis, to the right
of the singularity at $\omega = \beta({\bm q}^2)$, the integral yields
zero, unless $s > s_R$.  $ \beta({\bm q}^2)$ is the gluon trajectory
function, which for the bare reggeized gluon needs to be replaced by a
parameter $-\nu$, $\nu > 0$, which is taken in the limit $\nu \to 0$.
Other couplings of the reggeized gluon to particles and further reggeized gluons involve the four-gluon-vertex and higher induced vertices. In these cases $s$ is no longer uniquely determined. Implementations for these cases can be found in Sec.~\ref{sec:born+possig} and Sec.~\ref{sec:v23}.



For states of two or more reggeized gluons the above regularization
applies equally. Sometimes it is however useful to combine phases of
individual reggeized gluons to a single phase, see
Sec.~\ref{sec:tworeggeon_negsig} for instance.  For the evaluation of
longitudinal integrals of loops containing reggeized gluons we propose
to introduce so-called subtraction-diagrams. They remove an
over-counting inside the diagrams of the effective action and provide
well-defined convergent integrals for states of $n \geq 2$ reggeized
gluons.  They can be derived directly from the Lagrangian of the
effective theory, by supplementing it by a subtraction term,
\begin{align}
  \label{eq:lagrangian_subtractrr} 
\mathcal{L}_{\text{subtract}}(A_+, A_-) = -2 \tr (A_+(A_+) - A_+ )\partial_\sigma^2 A_- -2 \tr (A_-(A_-) - A_- )\partial_\sigma^2 A_+ ,
\end{align} 
with
\begin{align}
\label{eq2:efflagrangianrrr}
A_\pm(A_\pm) =
&
-\frac{1}{g}\partial_\pm U(A_\pm)
 = 
-\tr\frac{1}{g} \partial_\pm \mathcal{P} \exp\bigg(-\frac{1}{2} g \int_{-\infty}^{x^\pm}dx'^\pm A_\pm(x')\bigg)  .
 \end{align}
The complete effective action then  reads
\begin{align}
  \label{effaction_modirr}
  S_{\text{eff}} &= \int \text{d}^4 x [\mathcal{L}_{\text{QCD}}(v_\mu, \psi) + \mathcal{L}_{\text{ind}} (v_\pm, A_\pm) + \mathcal{L}_{\text{subtract}}(A_+, A_-) ].
\end{align}
A last point concerns the pole-prescription of the induced vertices of
$\mathcal{L}_{\text{ind}} $, which also enters the subtraction terms:
A suitable pole-prescription has been determined in
Sec.~\ref{sec:indu_presc} and explicit expressions can be found there
, see Eqs.~(\ref{eq:indu1_pole_g?indu}), (\ref{eq:double_comm_simpli})
and (\ref{eq:pole_indu3}).  The presented choice of the pole
prescription yields induced vertices that coincide in their color
structure with the originally proposed induced vertices without
pole-prescription. Apart from that, they can be shown to obey
Bose-symmetry and to agree with negative signature of the reggeized
gluon.  As a final rule we want to emphasis that, whenever a
pole-momenta coincides with a momenta inside the Reggeon-factor in
Eq.~(\ref{eq:vertexsig_fac_extract_re}), it is mandatory that the
$i\epsilon$-prescription of the pole and the branch-cut coincide with
each other.

With this short summary of results concerning longitudinal integrals,
we conclude the discussion of the effective action. In the following
chapter we turn to the study high-energy amplitudes in the context of
the large $N_c$ expansion in terms of topologies. In particular we
will consider color factors with the topology of the pair-of-pants for
 the six-point-amplitude in the triple-Regge-limit.

\chapter{The triple Pomeron vertex  and the pair-of-pants topology}
\label{sec:pants}
\section{Introduction}
\label{sec:intro}

Starting from the classical paper by 't Hooft \cite{'tHooft:1973jz}
the large $N_c$ limit of gauge theories has remained in the center of
attention for more than 25 years. In high energy QCD-phenomenology for
instance, the large-$N_c$ expansion proves its usefulness as a
simplifying tool, if one attempts to resum higher order contributions,
enhanced by large logarithms. With the general structure of the color
factors being often too complex, the large $N_c$ limit allows for
resummation with an acceptable reduction of accuracy.

A new attraction of the large $N_c$ expansion results nowadays from
the fact that it organizes the color structure of Feynman diagrams in terms
of topologies of two-dimensional surfaces which resemble the loop
expansion of a closed string theory. With the advance of the AdS/CFT
correspondence \cite{Maldacena:1997re, Witten:1998qj,Gubser:1998bc } which, in the limit of large $N_c$,
connects $N=4$ Super-Yang-Mills theory with a closed string theory in
Anti-de-Sitter space and with string coupling proportional to
$N_c^{-1}$ this idea is more prevailing than ever.

The leading term of the large $N_c$ expansion is given by color
factors that have the topology of the sphere or equivalently the
plane, and one usually refers to these leading contributions as
'planar' diagrams.  Planar diagrams contribute, for example, to
gluon-gluon scattering amplitudes or to multi-gluon production
amplitudes.  On the other hand, there exist also processes where the
$N_c$ leading color factor has not the topology of the plane.  This
is, for instance, the case for the scattering of two electromagnetic
currents or virtual photons at high energies where the center of mass
energy squared $s$ is far bigger than the momentum transfer squared
and the virtualities of the photons.  In this case, the interaction
between the scattering currents is mediated by the exchange of gluons
between two quark loops, and, with quarks in QCD in the fundamental
representation of the gauge group $SU(N_c)$, the leading color factor
has the topology of a sphere with two boundaries, a cylinder.  Within
perturbative QCD such processes, at leading
\cite{Lipatov:1976zz,Kuraev:1976ge,Fadin:1975cb,
  Kuraev:1977fs,Balitsky:1978ic} and next-to-leading order accuracy
\cite{Fadin:1998py,Ciafaloni:1998gs }, are described by the
BFKL-Pomeron, which is a bound state of two reggeized gluons. Higher
order unitarity corrections are expected to involve bound states of
more than 2 reggeized gluons, so called BKP-states
\cite{Bartels:1980pe, Kwiecinski:1980wb}, which for the cylinder
topology have been found to be integrable
\cite{Lipatov:1993yb,Lipatov:1994xy,Faddeev:1994zg}.
 
In the present study we go one step further and consider the class of
process whose leading color factor has the topology of a sphere with
three boundaries.
\begin{figure}[htbp]
  \centering
  \begin{minipage}[h]{5cm}
    \includegraphics[width=4.5cm]{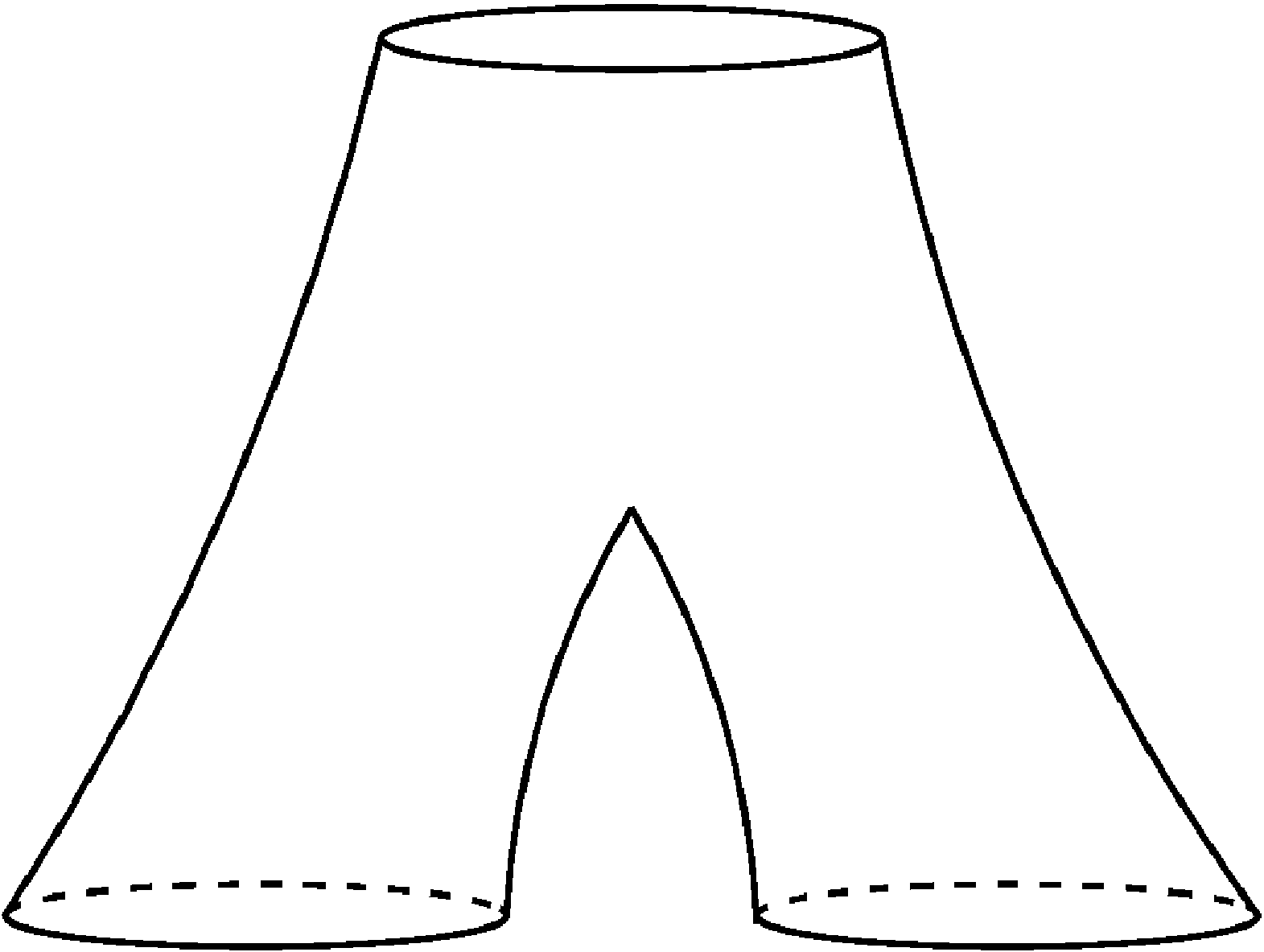}
  \end{minipage}
  \begin{minipage}[h]{.5cm}
$\,$
  \end{minipage}
\begin{minipage}[h]{8.5cm}
\parbox{4cm}{\includegraphics[height = 4.5cm]{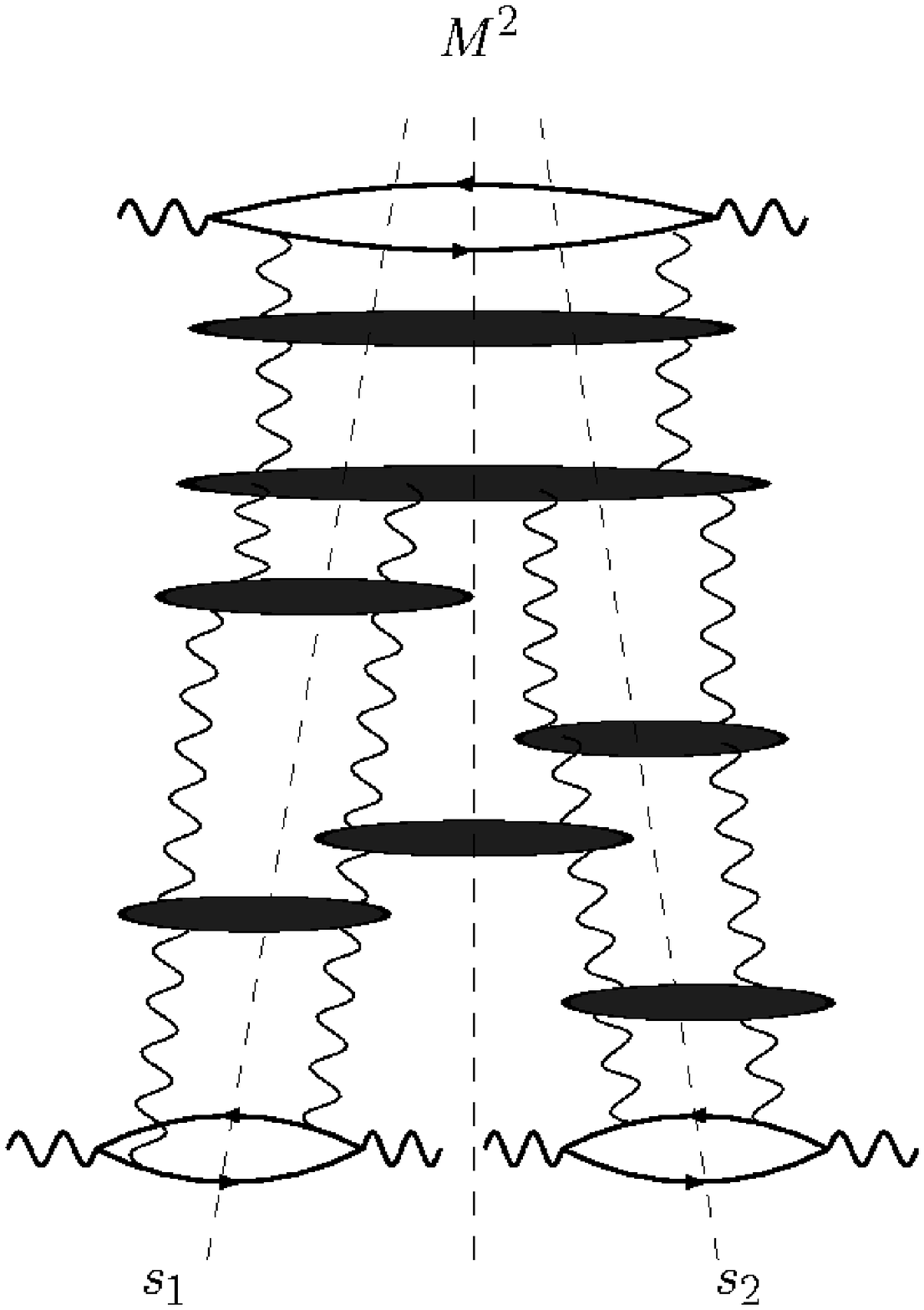}}
\parbox{4cm}{\includegraphics[height = 4.5cm]{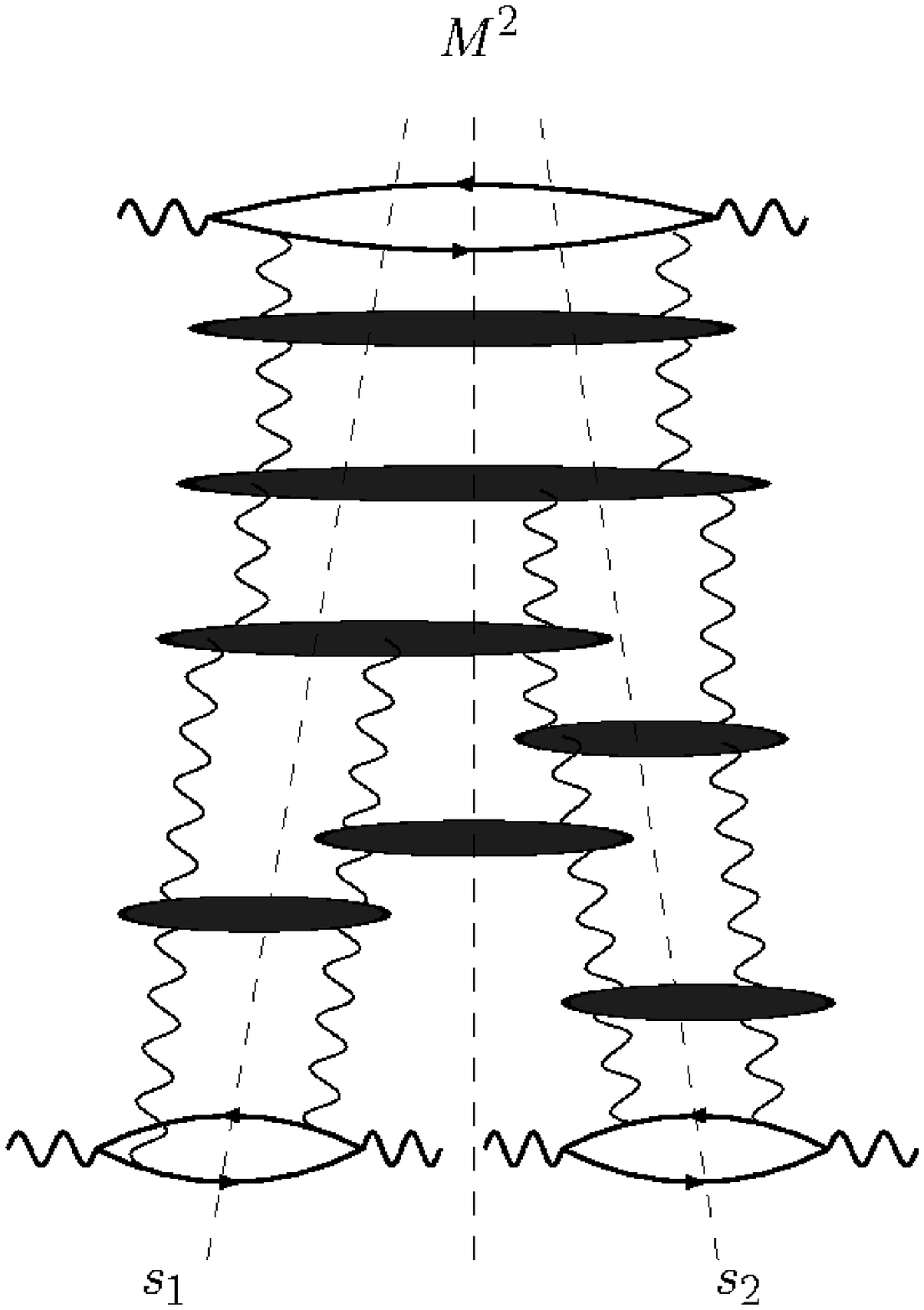}}
  \end{minipage}
\\
\begin{minipage}[h]{5cm}
\caption{\small The 'pair of pants' topology}
  \label{fig:trouserr}
\end{minipage}
 \begin{minipage}[h]{.5cm}
$\,$
  \end{minipage}
\begin{minipage}[h]{8.5cm}
\caption{\small Typical contributions to the triple energy discontinuity for scattering of three virtual photons }
\label{fig:triple_discon}
\end{minipage}
\end{figure}
Such a surface is depicted in Fig.\ref{fig:trouserr} and it is usually
referred to as the topology of a 'pair of pants'. As a suitable
example we consider the three-to-three process which describes the
scattering of a highly virtual photon on two virtual photons in the
triple Regge limit. Similar to the BFKL-Pomeron on the cylinder, we
resum all contributions which are maximally enhanced by logarithms of
energies. The class of diagrams selected in this way is illustrated in
Fig.\ref{fig:triple_discon}: the three photon impact factors introduce
three boundaries, and thus belong to the topology depicted in
Fig.\ref{fig:trouserr}.  Compared to the simple BFKL cylinder, the new
feature is the splitting of one cylinder into two cylinders which is
related to the 'triple Pomeron vertex' \cite{Bartels:1994jj}.  Within
the AdS/CFT correspondence, the electromagnetic current corresponds to
the R-current, and the high energy behavior of the 6-point amplitude,
on the string side, is expected to exhibit the triple graviton vertex.

In QCD this $3 \to 3$ process has been investigated before for $N_c=3$
in \cite{Bartels:1994jj}, the triple Pomeron vertex has been derived,
and it has been shown to be invariant under two-dimensional M\"obius
transformations \cite{Bartels:1995kf}.  It is straightforward to
repeat the analysis for arbitrary $N_c$ and to take the large-$N_c$
limit of these calculations.  The system of four reggeized gluons and
the triple Pomeron vertex have been further investigated particularly
for the limit $N_c \to \infty$ in \cite{Braun:1995hh,Braun:1997gm,Braun:1997nu}.
However the connection of these results with the expansion in terms of
topologies of two-dimensional surfaces is not apparent.  Instead of
the pair-of-pants, the color factor rather seems to correspond to
three disconnected cylinders.  In the present chapter, we therefore
demonstrate that by summing only diagrams with the topology of the
pair-of-pants, Fig.\ref{fig:trouserr}, one obtains the result of the
large-$N_c$ limit of \cite{Bartels:1994jj}.  In particular we find
that the 'reggeizing' and 'irreducible' terms of \cite{Bartels:1994jj}
can be attributed to distinct classes of diagrams on the surface of
the pair-of-pants.

This chapter is organized as follows. In section \ref{sec:elastic} we
briefly review the planar and cylinder topologies. In the high energy
limit, the planar diagrams satisfy the bootstrap condition of the
reggeizing gluon, whereas the cylinder diagrams lead to the famous
BFKL amplitude.  In section 3 we turn to the analytic form of the $3
\to 3$ scattering amplitude and give a general description of the
diagrams which need to be summed. In Section 4 we study the color
lines on the surface of the pants, arriving at the definition of two
distinct classes of diagrams in color space (named 'planar' and
'non-planar'). We also review the different momentum space kernels
which describe the interactions of reggeized gluons. In section 5 we
sum the 'planar' diagrams and arrive at the 'reggeized' amplitudes
introduced in \cite{Bartels:1994jj}, whereas in the section 6 we
investigate the 'non-planar' diagrams and re-derive the triple Pomeron
vertex of \cite{Bartels:1994jj}. In section 7 we briefly summarize our
results, and the final section 8 contains a few conclusions.

\section{Elastic amplitudes}
\label{sec:elastic}
In this section we begin by recalling the definition of the large
$N_c$ expansion, following the classical paper by 't Hooft
\cite{'tHooft:1973jz}. As a preparation for the study of the
pair-of-pants topology, we then reconsider the simplest examples of
diagrams, whose color factor have the topology of the plane and of the
cylinder. In the Regge-limit they yield the reggeized gluon and the
BFKL-Pomeron, respectively.

\subsection{The large $N_c$ expansion}
\label{sec:NCexp}

To study the large $N_c$ limit as  an expansion of
topologies of the color factor, one is asked to translate the
color factors of a scattering amplitude into the so-called double line notation. To
this end we use the $SU(N_c)$ generators in the fundamental representation, 
$(g^a)^i_j$, in the normalization\footnote{
  Note that this deviates by a factor $2$ from  
  the standard normalization $\tr(t^a t^b) = \delta^{ab}/2$.} 
$\tr(g^ag^b) = \delta^{ab}$, and we make use of the identity
\begin{align}
  \label{eq:structure_const}
f^{abc}  
&=
 \frac{1}{i\sqrt{2}}\big[\tr(g^ag^bg^c) - \tr(g^cg^bg^a) \big]  .
\end{align}
For all inner gluon lines in the adjoint representation, the indices
can then be expressed in terms of (anti-)fundamental indices by means
of the Fierz identity
\begin{align}
  \label{eq:fierz}
    (g^a)^i_j(g^a)^k_l  =  \delta^i_l \delta^k_j - \frac{1}{N_c} \delta^i_j\delta^k_l.
\end{align}
Making use of a diagrammatic notation, where a Kronecker-delta is
represented by a single line with an arrow, indicating the flow from
the upper to the lower index,
\begin{align}
  \label{eq:kronecker_delta}
\delta^i_j = \parbox{2cm}{\includegraphics[width=2cm]{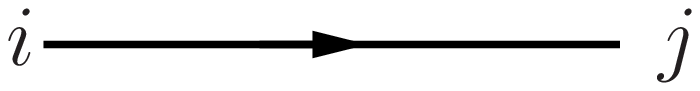}},
\end{align}
the structure constants $f^{abc}$ are  expressed as
\begin{align}
  \label{eq:structure_const1}
    f^{a_1a_2a_3} (g^{a_1})^{i_1}_{j_1} (g^{a_2})_{j_2}^{i_2} (g^{a_3})_{j_3}^{i_3}
 =
\frac{1}{i\sqrt{2}}   \left(\,\, \parbox{1cm}{\includegraphics[height=2cm]{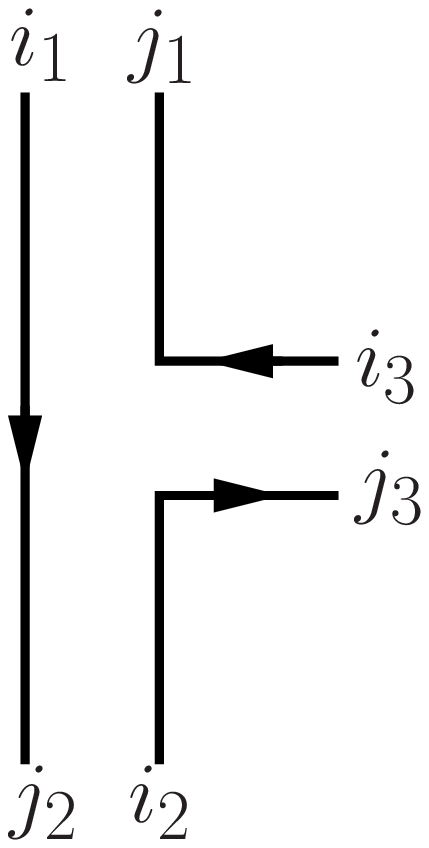}}
    -
    \parbox{1.3cm}{\includegraphics[height=2cm]{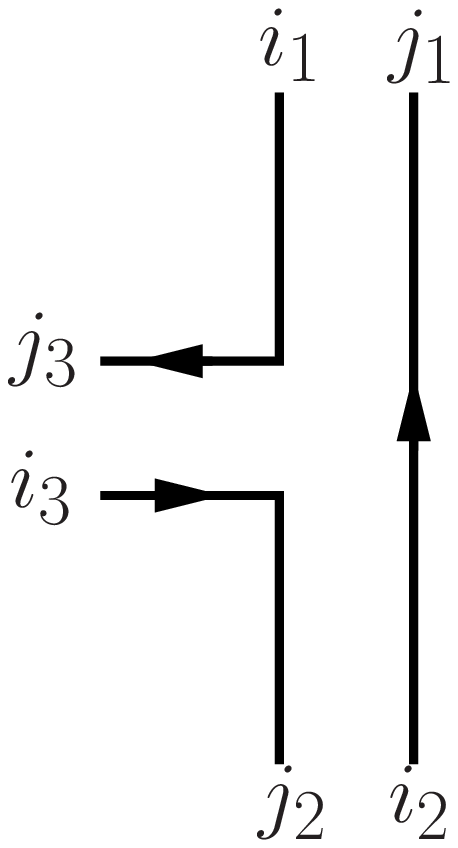}}
\right).
\end{align}
In order to arrive at the large $N_c$ expansion, one drops, in the
Fierz-identity Eq.(\ref{eq:fierz}), the second term. As a consequence,
the gluon is represented by a double line
\begin{align}
  \label{eq:double_line}
\delta^i_l \delta^k_j= \parbox{2cm}{\includegraphics[width=2cm]{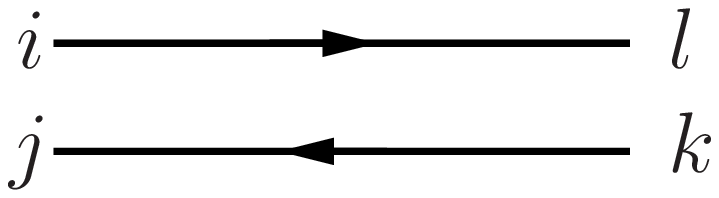}}.
\end{align}
The neglect of second term of the Fierz-identity Eq.(\ref{eq:fierz})
which serves to subtract the trace of the $SU(N_c)$ gluon implies that
we consider an $U(N_c)$ gluon rather than an $SU(N_c)$ gluon. While
this term is suppressed for large $N_c$, dropping this term is only
correct as long as we stay within the leading term of the expansion.
Tracelessness of the $SU(N_c)$ gluon can be taken into account by
introducing an additional $U(1)$ ghost field
\cite{Colemann:1985ich,Manohar:1998xv,Matevosyan:2008zz}, which subtracts the trace of
the $U(N_c)$-gluon. Since the $U(1)$ ghost-field commutes with the
$U(N_c)$ gluon-field, it couples only to quarks, but not to gluons.
Within the Leading Logarithmic Approximation, which we will use in the
following, the interaction between scattering objects is mediated only
by gluons, while quarks occur in the coupling of the gluons to the
scattering objects.  Corrections due to the $U(1)$ ghost in this
approximation therefore appear only in low orders of the strong
coupling and are taken into account easily.

Using the double line representation for gluons,
Eq.(\ref{eq:double_line}), and representing quarks in the fundamental
representation by single lines, the color factor of a vacuum Feynman
diagram turns into a network of double and single lines, which can be
drawn on a two-dimensional surface with Euler number $\chi = 2 -2h -b
$, where $h$ is the number of handles of the surface, and $b$ the
number of boundaries or holes.  A closed color-loop always delivers a
factor $N_c$. With quarks being represented by single lines, a closed
quark-loop, compared to a corresponding gluon-loop, is $1/N_c$
suppressed and leads always to a boundary.  For an arbitrary vacuum
graph $T$ one arrives at the following expansion in $N_c$
\begin{align}
  \label{eq:vacuumgraph}
T = \sum_{h,b}^\infty N_c^{2 - 2h -b} T_{h,b}(\lambda).
\end{align}
where
\begin{align}
  \label{eq:thooft}
\lambda &= g^2N_c 
\end{align}
is the 't Hooft-coupling which is held fixed, while $N_c$ is taken to
infinity.  The expansion Eq.(\ref{eq:vacuumgraph}) matches the loop
expansion of a closed string theory with the string coupling $1/N_c$.
For $N_c \to \infty$, the leading diagrams are those that have the
topology of a sphere: zero handles and zero boundaries, $h=b=0$. If
quarks are included, the leading diagrams have the topology of a disk,
i.e. the surface with zero handle and one boundary, $h=0, b=1$. The
disk fits on the plane, with the boundary as the outermost edge.
Diagrams with two boundaries and zero handles can be drawn on the
surface of a cylinder, those with three boundaries on the surface of a
pair-of-pants. Boundaries are also be obtained by removing, from the
sphere, one or more points: removing one point, one obtains the disk,
which can be drawn on the plane, and by identifying the removed point
with infinity, the graphs can be drawn on the (infinite) plane.
Removing two points we obtain the cylinder and so on. By definition,
the expansion Eq.(\ref{eq:vacuumgraph}) is defined for vacuum graphs.
However, from the earliest days on \cite{'tHooft:1974hx}, the large
$N_c$-limit has been also applied to the scattering of colored
objects.  In order to consider the topological expansion of an
amplitude with colored external legs, one needs to embed it into a
vacuum graph which then defines the topological expansion of an
amplitude with colored external legs.

In the following it will be convenient to define modified couplings 
\begin{align}
  \label{eq:modified_thooft}
\bar{g} = \frac{g}{\sqrt{2}},\hspace{3cm}  \bar{ \lambda} = 
\bar{g}^2N_c = \frac{g^2N_c}{2},
\end{align}
which absorb the factor $1/\sqrt{2}$. Due to our normalization of $SU(N_c)$
generators which differs from the more standard one, $\tr(t^at^b) =
\delta^{ab}/2$,
such a factor $1/\sqrt{2}$ arises for each quark-gluon coupling
and, because of Eq.(\ref{eq:structure_const1}), also for each   
gluon-gluon coupling.

\subsection{Planar amplitudes: Reggeization of the gluon}
\label{sec:reggeization}

In the present paragraph, we shall discuss the Regge-limit of planar
amplitudes (first addressed in \cite{Braun:1997ax}) which satisfy
the bootstrap condition of the reggeized gluon, the basic ingredient
for the further studies.  In order to study reggeization of the gluon,
one considers the scattering of colored objects.  To be definite, we
consider the scattering of a quark on an anti-quark. The disk-topology
of the color factors becomes apparent, if we connect color lines of
the incoming quark and antiquark with each other and of the outgoing
quark and antiquark (note that, if instead we would connect the color
lines of the ingoing and outgoing quark with each other and of the
incoming and outgoing antiquark, we would discover the cylinder
topology).  As far as the momentum part of the diagrams is concerned,
our method is the following: We consider the Regge limit $s \gg -t$ of
the elastic amplitude and make use of the Leading Logarithmic
Approximation (LLA), where we select all diagrams that are maximally
enhanced by a logarithm in $s$, i.e. that are proportional to
$\frac{\lambda}{N_c}(\lambda \ln s)^n$, and sum them up to all orders
in $\lambda$. Furthermore, in this approximation, all $t$-channel
particles are gluons (quark exchanges are power suppressed).

We start by considering tree- and one-loop-level diagrams within the
LLA, and convince ourselves that these leading order result, for
planar color graphs, satisfy the bootstrap condition and are thus in 
agreement with the reggeization of the gluon.  At
tree level, the (anti-)quarks interact by exchange of a single gluon.
In the standard adjoint notation, the color factor of the diagram is given
by $(g^{a})^i_j (g^{a})^l_k$, which in the double-line notation turns into
$\delta^i_k\delta^l_j$, while the $U(1)$-ghost can be disregarded for
the plane. We therefore have
\begin{align}
  \label{eq:tree_quarks}
 \parbox{1.5cm}{\includegraphics[width=1.5cm]{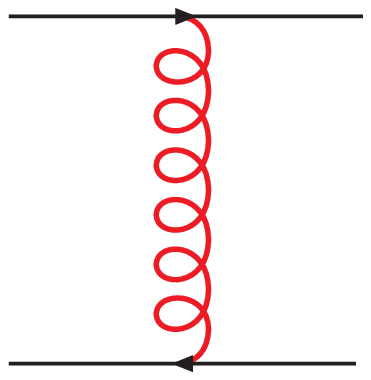}} & = \parbox{1.5cm}{\includegraphics[width=1.5cm]{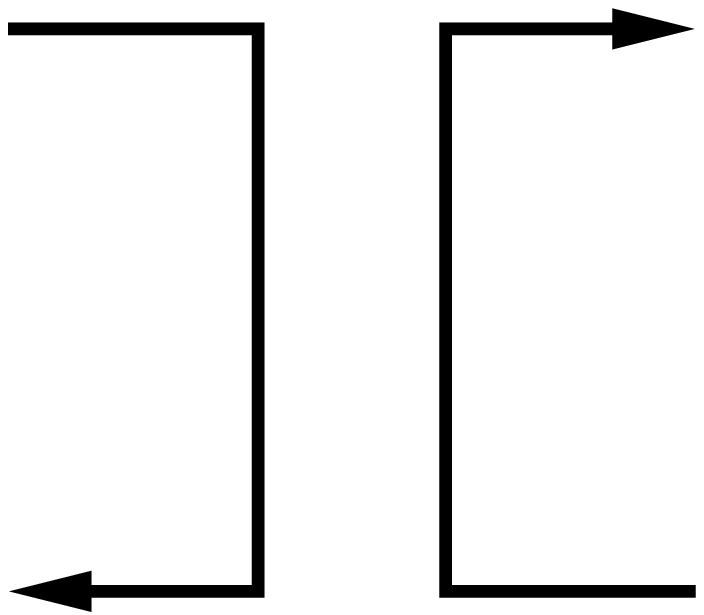}} \times T^{\text{tree}}_{q\bar{q}}(s,t), &
 T^{\text{tree}}_{q\bar{q}}(s,t) &= 2 \bar{g} \frac{-s}{t} \bar{g}.
\end{align}
From now on we adopt the following notation: in a Feynman 
diagram gluons are usually denoted by curly lines, and in this notation the diagram 
is understood to represent both the color factors and the momentum part.
Alternatively, when drawing a diagram with the double line notation for 
each gluon, this diagram represents only the color part, and the momentum 
part has to be written separately. 

\begin{figure}[htbp]
  \centering
  \begin{minipage}{.5\textwidth}
\parbox{1.7cm}{\includegraphics[width=1.7cm]{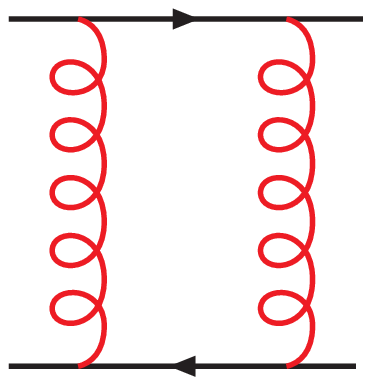}}  : $\quad$
     \parbox{2cm}{\includegraphics[height=1.2cm]{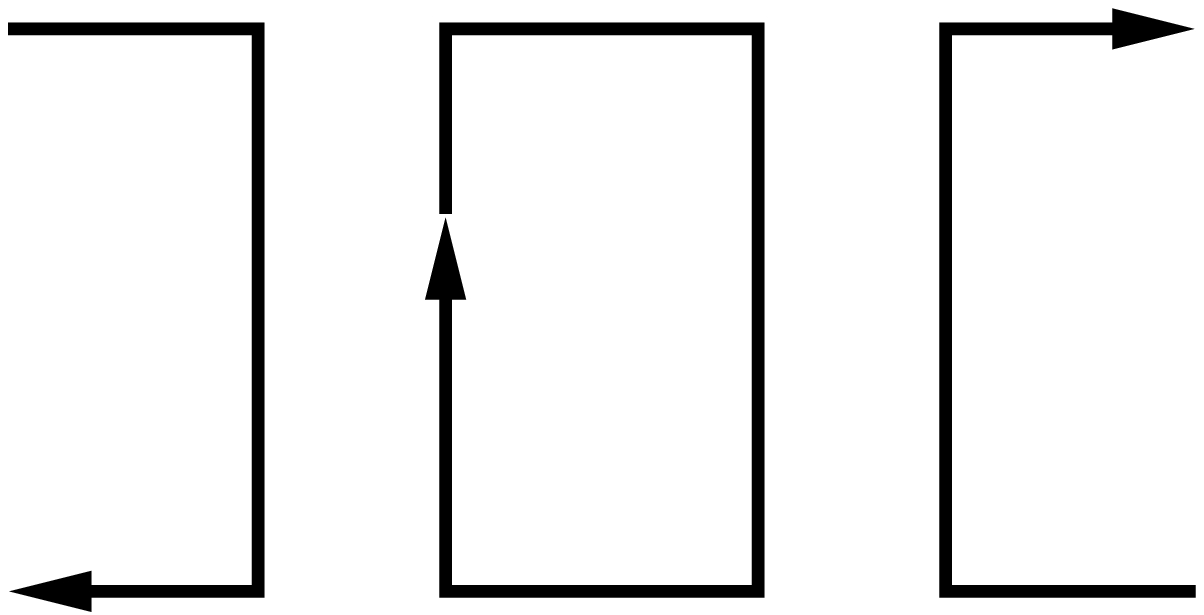}}
= {\large $N_c$}  \parbox{1cm}{\includegraphics[height=1.2cm]{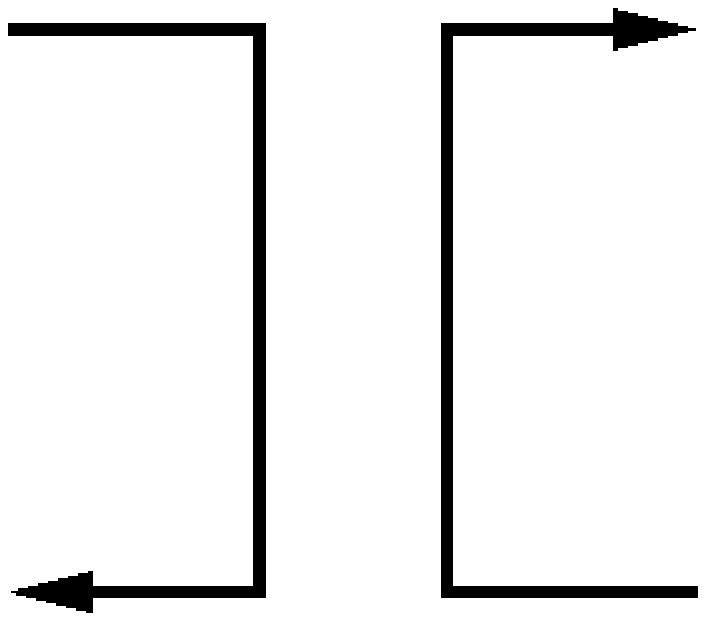}}
  \end{minipage}
  \begin{minipage}{.1\textwidth}
$\,$
  \end{minipage}
  \begin{minipage}{.3\textwidth}
\parbox{1.7cm}{\includegraphics[width=1.7cm]{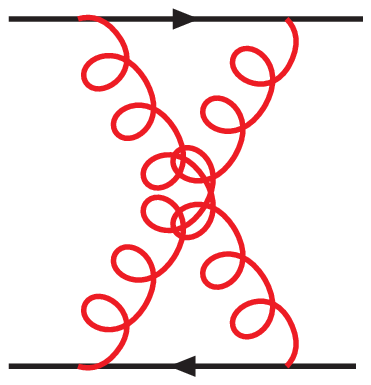}} : $\quad$
     \parbox{1.8cm}{\includegraphics[height=1.2cm]{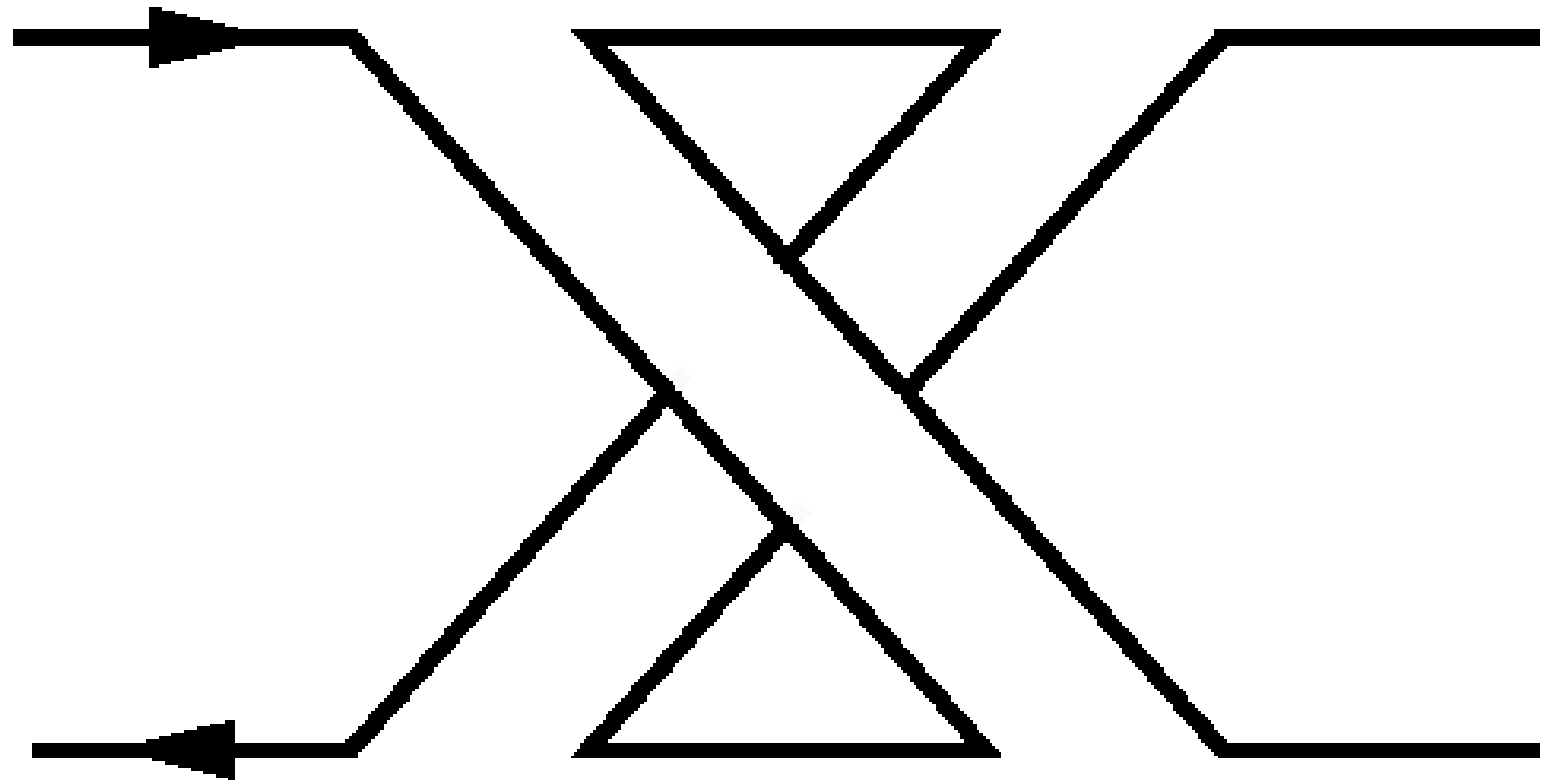}}
  \end{minipage}
  \caption{\small One-loop corrections to  quark-antiquark scattering. 
On the left hand side, 
the color factor of the planar Feynman diagram is $N_c$ times the Born color 
factor. On the right hand side,
the non-planar diagram does not fit onto the plane and is $N_c$ suppressed.}
  \label{fig:col_corr_1gl}
\end{figure}

Turning to higher order graphs, we have the
restriction that, with every insertion of an additional internal loop 
coming with a factor $g^2$, we have to 
produce a closed color loop, yielding a factor $N_c$ which then
combines to the 't Hooft coupling $\lambda = g^2 N_c$. Only such 
higher order term will be included into the planar approximation discussed in this 
section. 
This restriction holds also for other topologies in the expansion
Eq.(\ref{eq:vacuumgraph}): every insertion of an additional gluon
into the Born diagram provides automatically a factor $g^2$ and must
be compensated by a closed color loop in order to stay within the
considered coefficient $T_{h,b}$ of the expansion.  
For our planar loop, for each higher order graph the tensorial structure of the
color factor will be proportional to the one of the Born term.   
As far as the momentum part is concerned, the  1-loop diagrams for
quark-antiquark scattering that contribute within the LLA are shown  in Fig.\ref{fig:col_corr_1gl}.
For both diagrams, the momentum part is of the same order of magnitude,
i.e. it is proportional to $g^2 s \ln(-s)$ and $g^2 u \ln (-u)$,
respectively, with $u \simeq -s$ in the Regge limit. However, 
when counting closed color loops, one sees that only the
first diagram is leading in $N_c$. Also, when closing the color lines of the 
incoming quark and antiquark and of the outgoing particles, 
we observe that for the first diagram the 'closed' color factor, indeed, 
fits on the
disk with $h=0, b=1$, while the second 'closed' color factor has an
additional handle $h=1, b=1$.  We therefore find for the quark-antiquark
scattering amplitude at 1-loop:
\begin{align}
  \label{eq:beta_func}
T_{q\bar{q}}^{\text{1-loop}} (s, t) &= \ln (-s)\beta(t)T_{q\bar{q}}^
{\text{tree}} (s, t)
&& \mbox{with}&
\beta(- {\bf q}^2) &= \bar{\lambda} \int \frac{d^2 {\bf k}}{(2\pi)^3}\frac{- {\bf q}^2}{{\bf k}^2({\bf q} - {\bf k})^2},
\end{align}
where the color factor $N_c$ has already been included into 
the gluon trajectory function, and bold letters denote Euclidean momenta 
perpendicular  to the light-like momenta $p_1$ and $p_2$ of the scattering 
quark and antiquark; in particular  $t = -{\bf q}^2$. Eq.(\ref{eq:beta_func}) can be
understood as the order $\mathcal{O}(g^4)$ term of the expansion of the exchange of
a planar reggeized gluon:
\begin{align}
  \label{eq:qqbar_reggeized}
T_{q\bar{q}} (s, t) &= 2 \bar{g} (-s)^{1 +\beta(t)} \frac{1}{t} \bar{g}. 
\end{align}
For our further analysis we use the following analytic representation of the planar elastic amplitude in the
Regge limit:
\begin{align}
  \label{eq:elastic_rep}
 T_{2\to2}(s,t) =  s\int_{\sigma -i\infty}^{\sigma + i\infty} \frac{d \omega}{2\pi i} s^\omega \xi(\omega) \phi(\omega,t).
\end{align}
The partial wave amplitude $\phi(\omega, t)$ is a real-valued function, 
and phases are contained in the signature factor $\xi(\omega)$: 
\begin{align}
  \label{eq:sigfac_reggeon}
\xi(\omega) = -\pi\frac{e^{-i\pi\omega}}{\sin(\pi\omega)}.
\end{align}
Note the difference from the 'usual' signature factor $\xi \sim {e^{-i\pi\omega}} \pm 1$ 
which contains both right and left hand energy cuts:  
at one loop we have shown that only the Feynman diagram with the $s$-channel 
discontinuity belongs to the planar approximation whereas the $u$ discontinuity is absent. 
For quark-antiquark-scattering, this also holds for higher order terms.
We then take  the discontinuity in $s$ 
\begin{align}
  \label{eq:disc_elastic}
 \Im\text{m}_s  T_{2\to2}(s,t) = \mathrm{disc}_s  T_{2\to2}(s,t) =  \pi s\int \frac{d \omega}{2\pi i} s^\omega  \phi(\omega,t),
\end{align}
which, by unitarity 
\begin{align}
2 \Im\text{m}_s  T_{2\to2}(s,t) = \sum \int T_{2\to n}  T_{2\to n}^*
\label{unitarity}
\end{align}
relates the Mellin transform of the partial wave amplitude
$\phi(\omega, t)$  to the sum of production processes. To
leading order in $g$, the partial wave has the form
\begin{align}
  \label{eq:reggeon_PWlo}
\phi^{(0)}(\omega, t) = \frac{2}{\omega} \phi_{(2;0)} \otimes \phi_{(2;0)}
= \frac{2\bar{g}^2}{\omega t} \beta(t)
\end{align}
where $\phi_{(2;0)} = \bar{g}^2 N_c^{1/2}$ denotes the non-singlet two gluon impact 
factor for quark and antiquark, and the convolution symbol is defined as 
\begin{align}
\otimes = \int\frac{d^2 {\bf k}}{(2\pi)^3 {\bf
    k}_1^2 {\bf k}_2^2}
\end{align}
with ${\bf q} = {\bf k}_1+{\bf k}_2$ and $t= - {\bf q}^2$.
Higher order corrections involve diagrams where 
additionally real gluons are produced. To leading logarithmic
accuracy, real particle production takes place within the
Multi-Regge-Kinematics (MRK), where the produced particles are widely
separated in rapidity.  The leading order diagram, with one additional
$s$-channel gluon, is illustrated in Fig.\ref{fig:sdisc_real3}: 
\begin{figure}[htbp]
   \centering 
   \parbox{4cm}{\includegraphics[height=2cm]{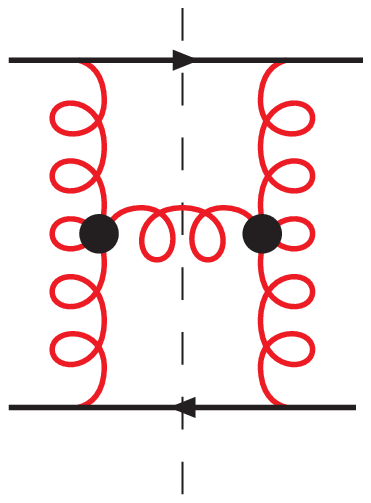}}
   \parbox{2.5cm} { \includegraphics[width=2.5cm]{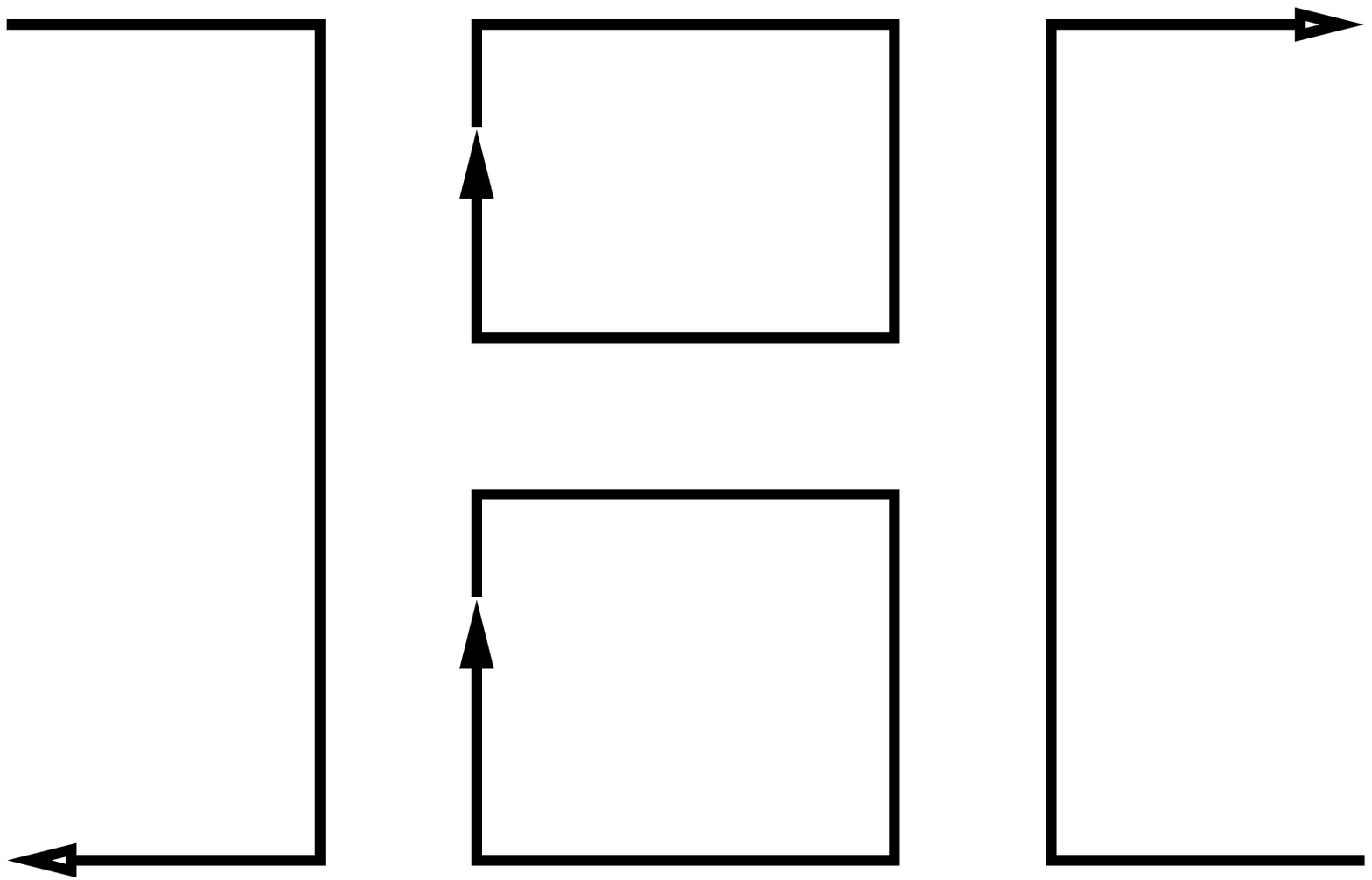}}
   \caption{\small $s$-channel discontinuity with three real particles. To the right  the corresponding planar color factor.}
   \label{fig:sdisc_real3}
\end{figure}
Here the particle production vertex (depicted by a dot) is an effective vertex for the production 
of one real gluon.
This production vertex is build by the following Feynman diagrams:
\begin{align}
  \label{eq:prod_vertex}
\parbox{1.5cm}{\includegraphics[width=1.5cm]{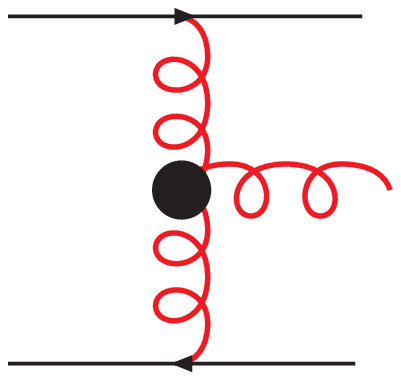}}
        & =         \parbox{1.5cm}{\includegraphics[width=1.5cm]{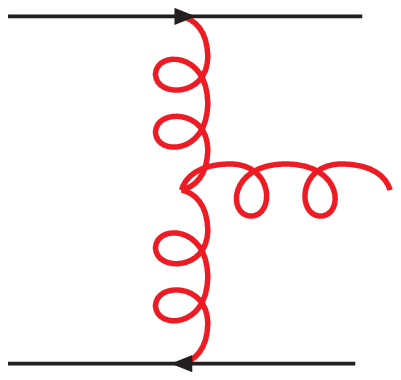}} 
         &+&
         \parbox{1.5cm}{\includegraphics[width=1.5cm]{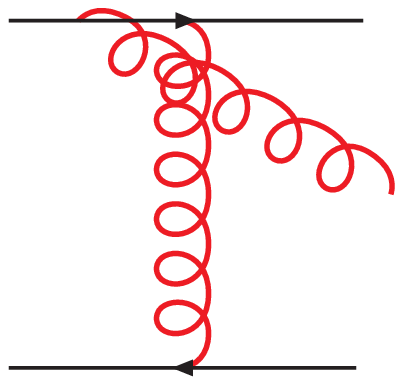}}
         &+ &
         \parbox{1.5cm}{\includegraphics[width=1.5cm]{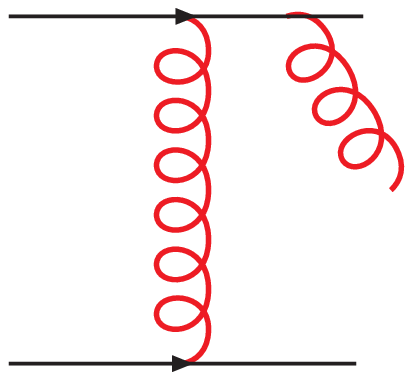}} 
         &+&
         \parbox{1.5cm}{\includegraphics[width=1.5cm]{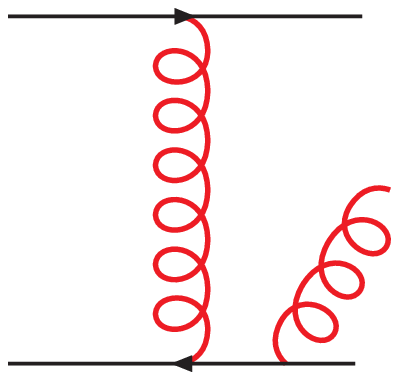}} 
         &+&
         \parbox{1.5cm}{\includegraphics[width=1.5cm]{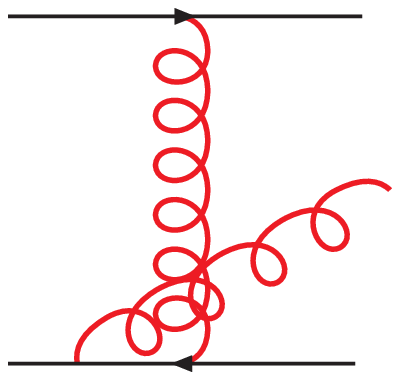}} 
\notag \\
&\frac{1}{i}\bigg(
         \parbox{1cm}{ \includegraphics[width=1cm]{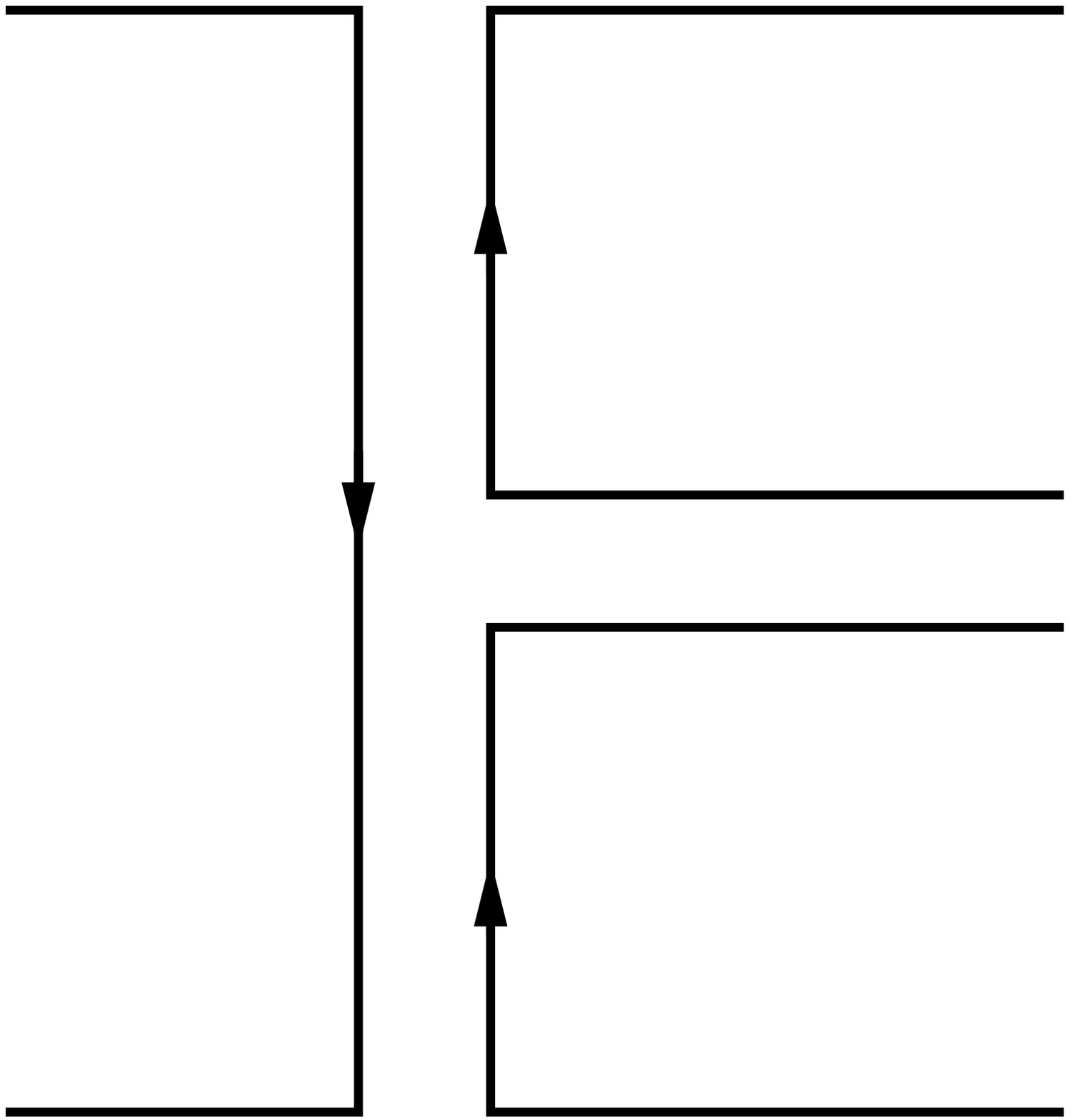}}
         -
         \parbox{1cm}{ \includegraphics[width=1.3cm]{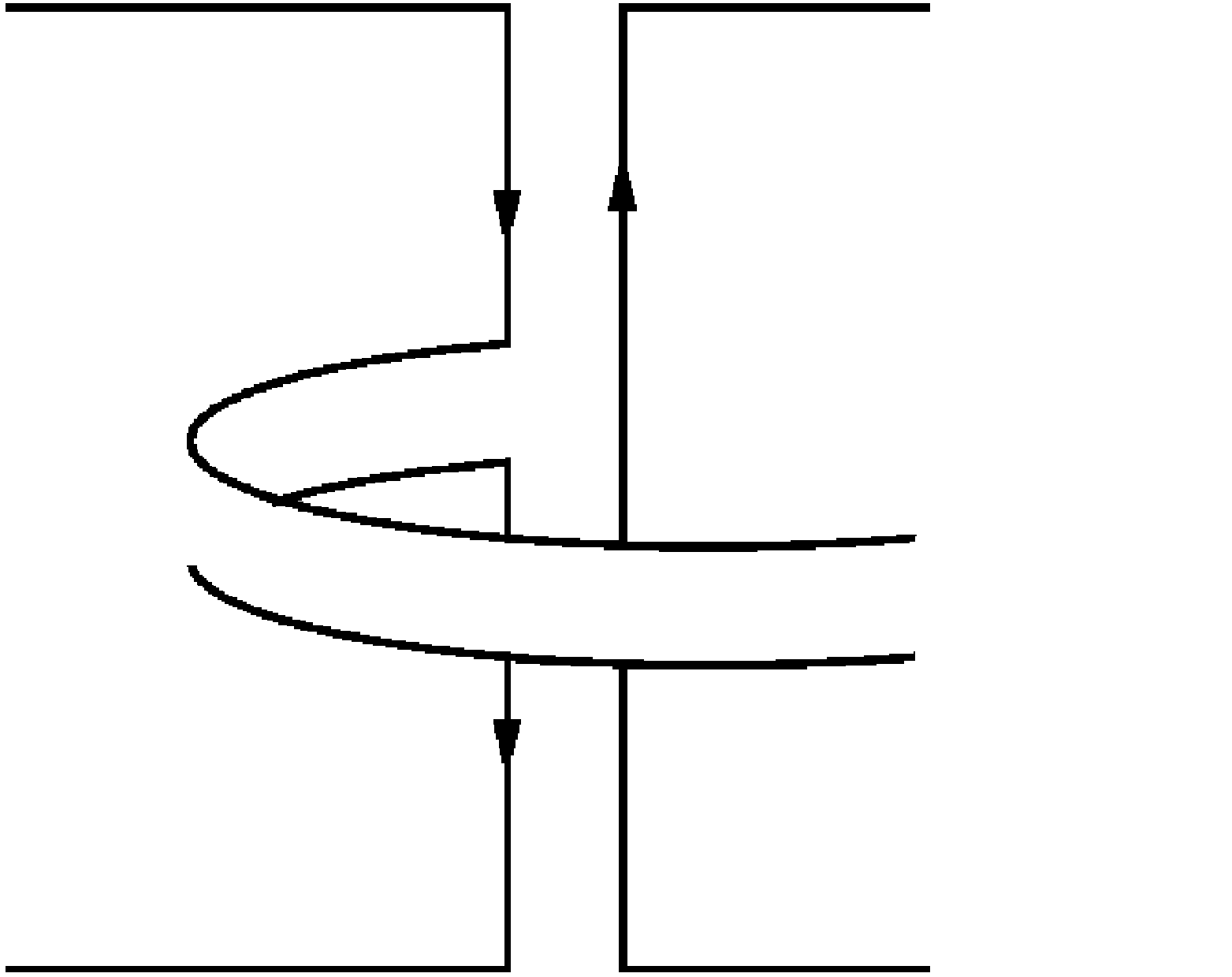}} 
         \bigg) 
&&
        \frac{-1}{i}   \parbox{1.2cm}{  \includegraphics[width=1cm]{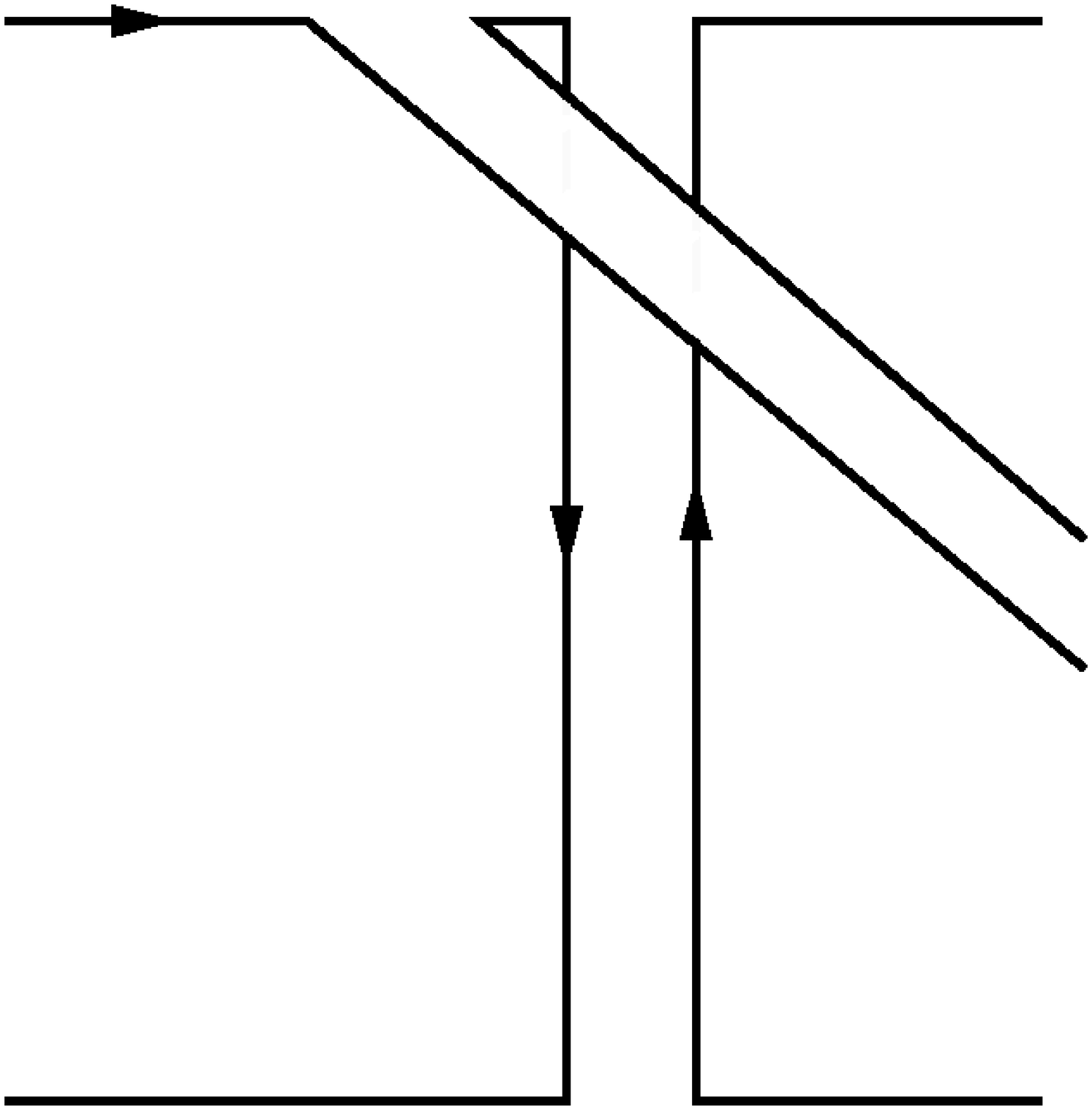}} 
        && 
\frac{1}{i}
        \parbox{1cm}{\includegraphics[width=1cm]{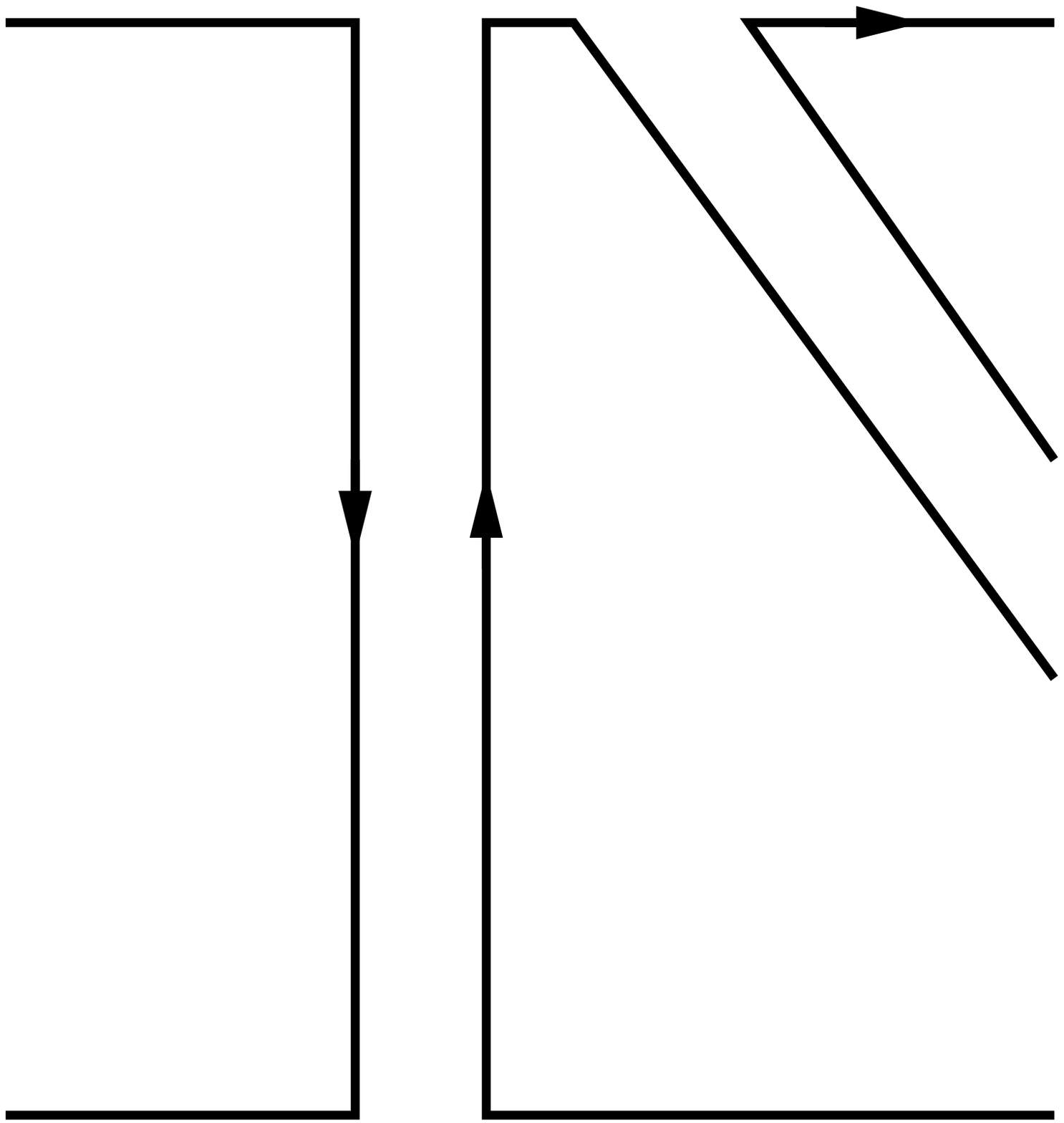}} 
        &&
 \frac{1}{i}
        \parbox{1cm}{\includegraphics[width=1cm]{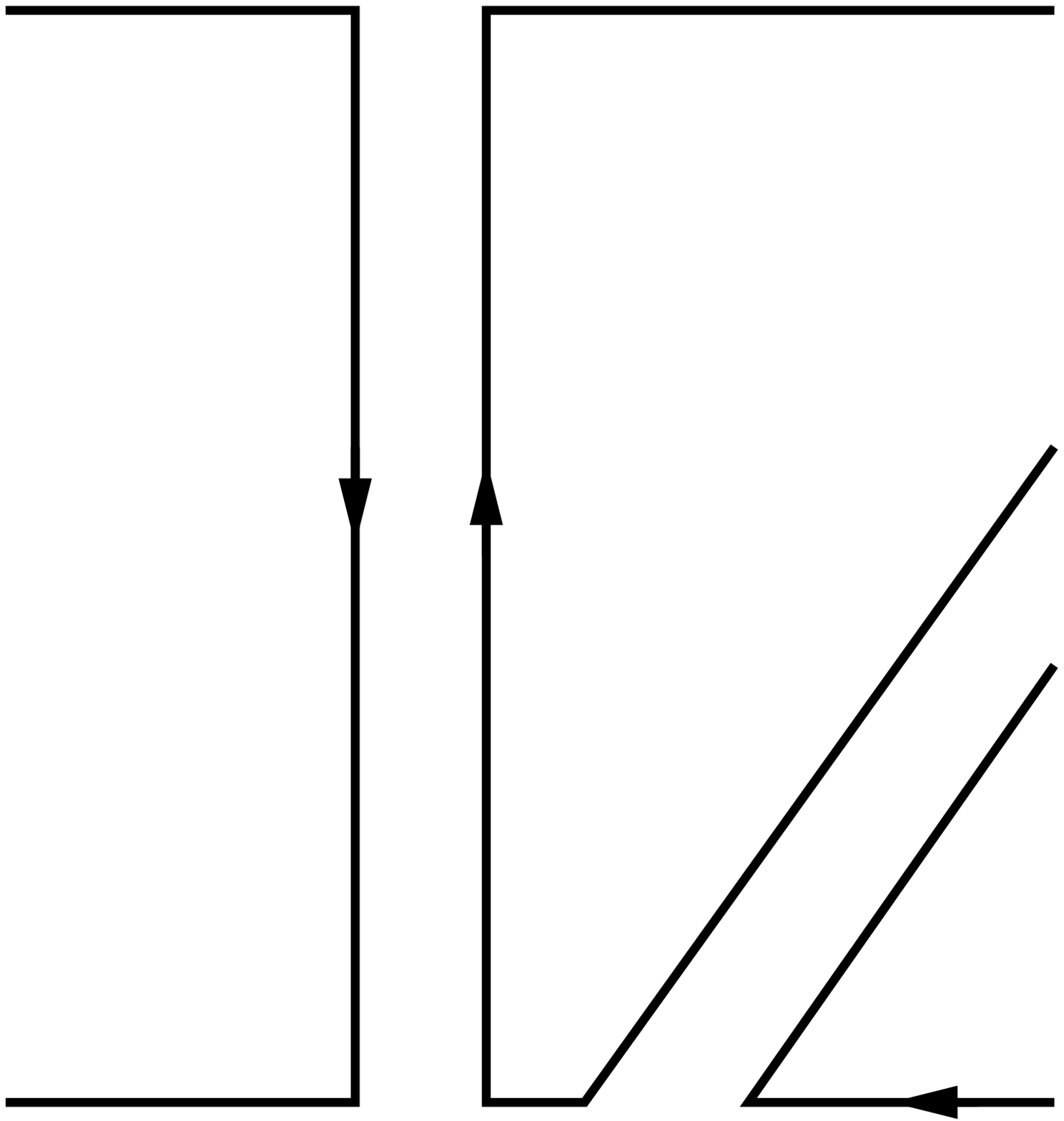}}
&&  \frac{-1}{i}
        \parbox{1cm}{\includegraphics[width=1cm]{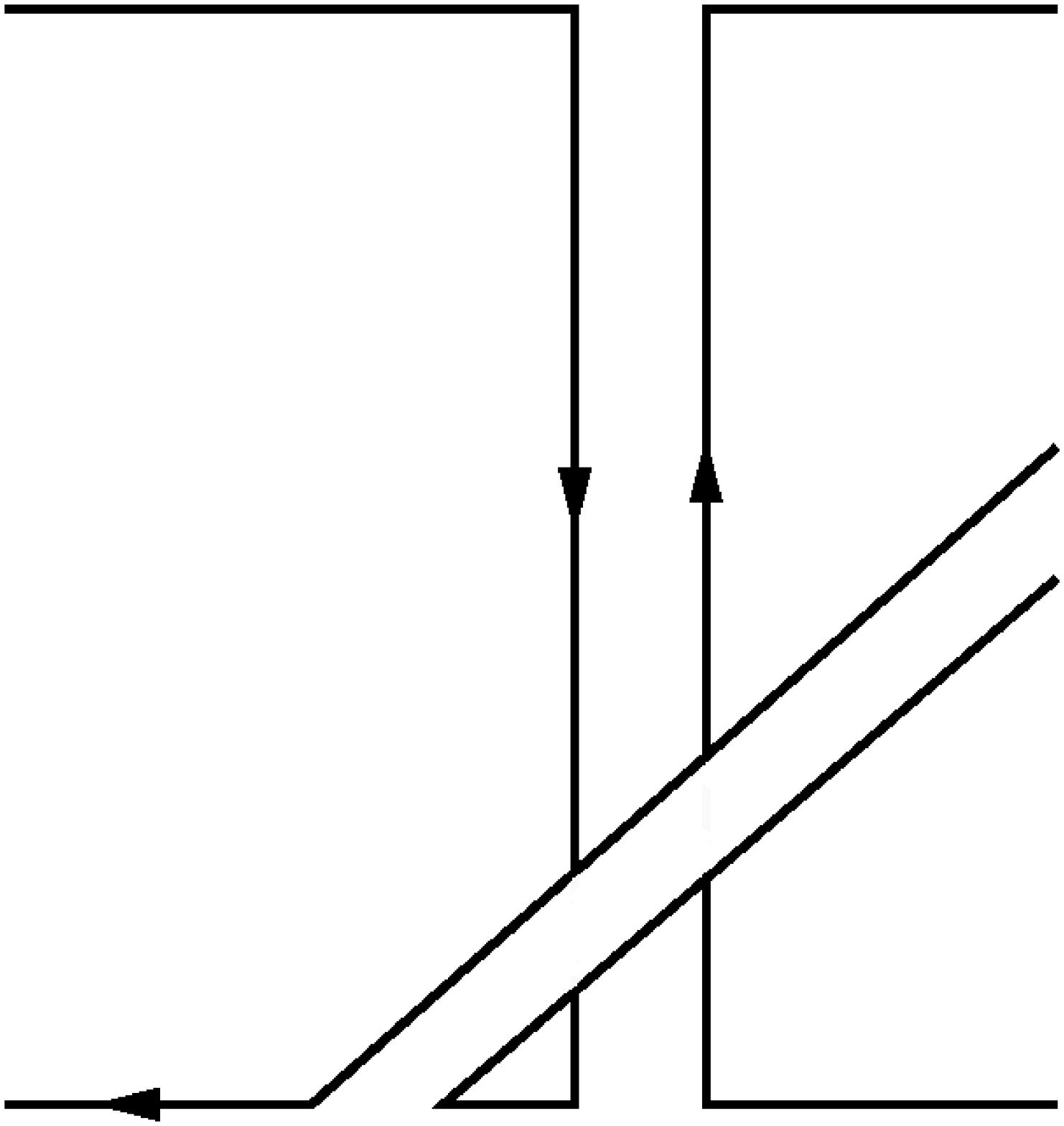}.}
\end{align}
In the second line we show, for each of the five different
Feynman-diagrams, the corresponding color factor in the double-line
notation. For the last four diagrams on the right hand side, where the
real gluon is emitted from the quark and the antiquark respectively,
we have shifted factors $i$ and $(-i)$ from the momentum part to the
color factor. With such a shift, the (remaining) momentum parts of the
second and the third diagram and the momentum parts of the fourth and
the fifth diagram coincide in the considered kinematical regime with
each other. Color factors and momentum parts of the production vertex
can therefore be written in a factorized form, and we obtain
\begin{align}
  \label{eq:lipatov_factor}
\parbox{1.5cm}{\includegraphics[width=1.2cm]{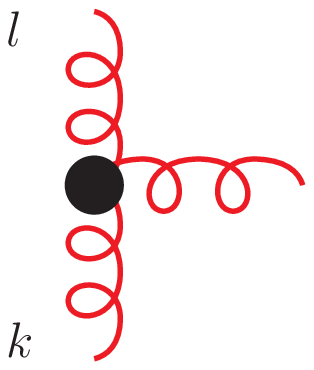}}
        & =
 \frac{{\bar g}}{i} 
     \left(\quad
      \parbox{1cm}{\includegraphics[height=1.2cm]{prod_gross1.eps}} \quad
      - \quad
      \parbox{1cm}{ \includegraphics[height=1.2cm]{prod_gross1b_bow1.eps}} 
     \quad  \right) \quad \times C(l, k)\cdot \epsilon_{(\lambda)},
\end{align}
where $\lambda$ denotes the helicity of the produced gluon. 
With the light-like momenta of the quark and the antiquark given by $p_1$ and $p_2$ respectively we have with $s \simeq 2p_1\cdot p_2$,
\begin{align}
  \label{eq:production_vertex}
        C^\mu(l, k) =    \left(\alpha_l  + 2\frac{{\bf   l}^2}{ \beta_k s}      \right)p_1^\mu  + \left(\beta_k  + 2 \frac{{\bf  k}^2}{ \alpha_l s}      \right) p_2^\mu  - (l_\perp^\mu + k_\perp^\mu), 
\end{align}
where  $ \alpha_l s = 2 p_2\cdot l $
and $\beta_k s = 2 p_1\cdot k $.  When squaring the production
amplitude, we find 
for the color part that only one of the four
possible combinations fits on the plane, namely the one shown on the
right hand side of Fig.\ref{fig:sdisc_real3}. 
From the point of view of Feynman-diagrams, the planar approximation 
includes only a sub-set of the diagrams in Eq.(\ref{eq:prod_vertex}), namely 
those with the color structure
\begin{align}
  \label{eq:prod_vertex_NC}
  \parbox{2cm}{\center \includegraphics[width=1.2cm]{prod_gross1.eps}} 
         +
         \parbox{2cm}{\center\includegraphics[width=1.2cm]{prod_gross2.eps}}
         + 
         \parbox{2cm}{\center \includegraphics[width=1.2cm]{prod_gross3.eps} }
\end{align}
However due to the factorization of momentum and color parts in the
MRK, Eq.(\ref{eq:lipatov_factor}), we still recover the complete
production vertex $C_\mu$ and therefore the BFKL kernel.  Making use
of our re-defined couplings $\bar{g}$ and $\bar{\lambda}$, and
including the factors $1/i$ of Eq.(\ref{eq:lipatov_factor}), we find,
for the plane, the $2\to2$ Reggeon transition kernel $K_{2 \to2}$ as:
\begin{align}
  \label{eq:22kernel_momis}
\parbox{2cm}{\includegraphics[width=1.8cm]{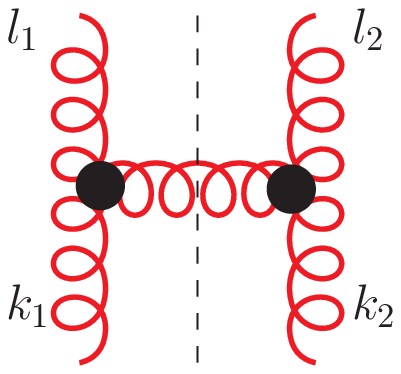}}  = \bar{\lambda} K_{2 \to 2}({\bf l}_1, {\bf l }_2;{\bf k}_1, {\bf k}_2) 
= - \bar{\lambda} \big[
 ({\bf k}_1 + {\bf k}_2)^2 - \frac{{\bf l}_1^2 {\bf k}_2^2}{({\bf k}_1 - {\bf l}_1)^2}  - 
\frac{{\bf l}_2^2 {\bf k}_1^2}{({\bf k}_1 - {\bf l}_1)^2} \big]
\end{align}
with the constraint ${\bf l}_1 + {\bf l}_2 ={\bf k}_1 + {\bf k}_2 ={\bf q}$.

\begin{figure}[htbp]
  \centering
  \parbox{4cm}{\center \includegraphics[height = 3.5cm]{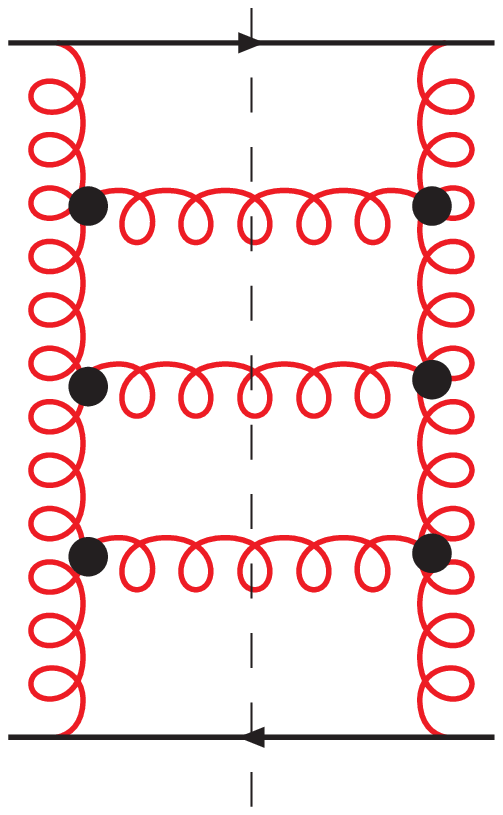}}
  \parbox{1cm}{$\,$}
  \parbox{4cm} { \center \includegraphics[height = 3cm]{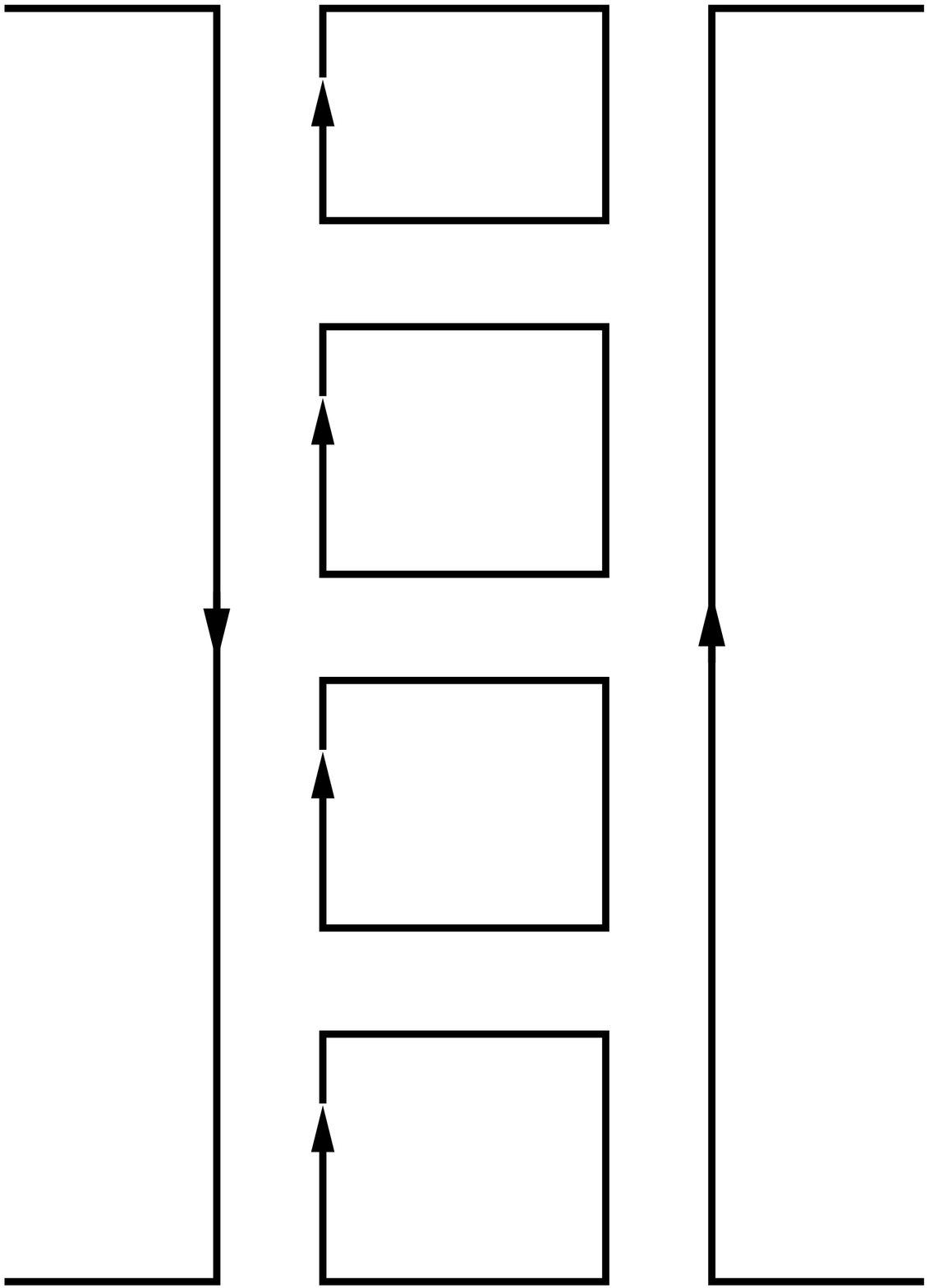}} 
  \caption{\small The s-channel discontinuity of the elastic quark-antiquark scattering amplitude for three gluon production.
On the right, the color structure of the planar diagrams.}
  \label{fig:elastic_disc}
\end{figure}

Higher order terms with the production of $n$-gluons within the MRK are taken into
account in a similar way. Making use of Regge-factorization of the production amplitudes, 
it can be shown that each emission of an additional real gluon is
taken into account by insertion of the effective production vertex 
Eq.(\ref{eq:lipatov_factor}), as illustrated in
Fig.\ref{fig:elastic_disc} for the squared production amplitude.  Similar 
to the production of one-gluon, for each of the inserted production vertices
only one of the four possible combinations of color factors fits onto the plane, 
and the resulting color factor can be found on the right-hand-side of Fig.\ref{fig:elastic_disc}.

In order to obtain the complete expression of order $\lambda^n$ of the partial wave $\phi$,  
we need to include the reggeization of the gluon. This is done in the same way as 
in the finite-$N_c$ case, except that at each step we have to restrict ourselves to the 
planar approximation. In lowest order we return, on the rhs of the unitarity equation (\ref{unitarity}), to the 
elastic intermediate state and include, in one of the two factors $T_{2\to2}$, the one loop result 
(\ref{eq:beta_func}). Together with the real one-gluon production discussed above, this provides, in 
(\ref{eq:qqbar_reggeized}), the complete imaginary part to order $g^6$ (or $\lambda^3$). Compared to the 
tree diagram expression (\ref{eq:tree_quarks}), (\ref{eq:qqbar_reggeized}) replaces the exchange of 
an elementary $t$-channel gluon by a reggeized gluon. In higher order $g$, the analogous replacement has to be done, 
on the rhs of (\ref{unitarity}), for all $t$-channel exchanges in the production amplitudes $T_{2 \to n}$.        

In order to prove the self-consistency of this procedure ,
one has to show that the BFKL bootstrap condition is satisfied in the planar approximation. 
For the partial wave in Mellin-space, reggeization of the $t$-channel gluons within the LLA requires to
introduce, for every $t$-channel state of two gluons, a Reggeon-propagator
$1/(\omega - \beta({\bf k}) - \beta({\bf q} - {\bf k}))$. In order to perform the resummation of all 
production processes in Fig.\ref{fig:elastic_disc}, it is most convenient to formulate the BFKL integral-equation in the planar approximation.  
To this end we first factorize off, at the lower end of the ladder, the antiquark-impact factor $\phi_{(2;0)}$.
We then define, for the scattering of a quark on a reggeized gluon,
the amplitude $ \phi_2(\omega|{\bf k}_1, {\bf k}_2)$, which contains the Reggeon propagator
$1/(\omega - \beta({\bf{k}}_1) - \beta({\bf{k}}_2))$ of the two
reggeized gluons at the lower end. The BFKL-equation for this amplitude reads:
\begin{align}
  \label{eq:bfkl-eq}
\big(\omega - \sum_i^2\beta({\bf{k}}_i)\big) \phi_2(\omega | {\bf k}_1, {\bf k}_2) &= 
 \phi_{(2;0)}( {\bf k}_1 + {\bf k}_2)
+
 \bar{\lambda}  K_{2 \to 2}
\otimes \phi_2(\omega| {\bf k}_1, {\bf k}_2),  
\end{align}
where the kernel coincides with the finite $N_c$ case.
With our impact factor $\phi_{(2;0) }$ which depends only on the sum of
the transverse momenta of the $t$-channel gluons we find for Eq. (\ref{eq:bfkl-eq}) 
the familiar pole solution
\begin{align}
  \label{eq:pole_sol}
\phi_2(\omega|{\bf k}_1 +{ \bf k}_2) =& \frac{\phi_{(2;0)} }{\omega - 
\beta( -({\bf k}_1 +{ \bf k}_2)^2 )}.
\end{align}
The partial wave for quark-antiquark scattering becomes 
\begin{align}
  \label{eq:pole_phi}
\phi (\omega, t) = 2 \int \frac{d^2{\bf l}}{(2 \pi)^3} \frac{1}{{\bf l}_1^2{\bf l}_2^2} \frac{\phi_{(2;0)} \phi_{(2;0)}  }{\omega - \beta(t )} = \frac{1}{t} \frac{2 \bar{g}^2  \beta(t)}{\omega - \beta(t ) }.
\end{align}
Inserting this into Eq.(\ref{eq:elastic_rep}) and using that in LLA $\sin
\pi\omega \simeq \pi\omega$, we find the reggeized
gluon, as proposed in Eq.(\ref{eq:qqbar_reggeized}).
This proves that, in the planar approximation, the bootstrap property is satisfied.

\subsection{Cylinder topology: The BFKL-Pomeron }
\label{sec:cylinder}
Next we turn to the cylinder topology which,
in the Regge-limit, leads to the BFKL-Pomeron.  In the expansion
Eq.(\ref{eq:vacuumgraph}), this corresponds to the term with two
boundaries and zero handles, $b=2, h= 0$.  Even though it would be
possible to carry out this analysis for quark-antiquark scattering, it
is more natural to do so for scattering of two highly virtual photons,
which provides an IR-finite and gauge-invariant amplitude.
Reggeization of the gluon will be an important ingredient in this
analysis, as we will see in short.
\begin{figure}[htbp]
  \centering
  \parbox{4.5cm}{\center \includegraphics[width=4cm]{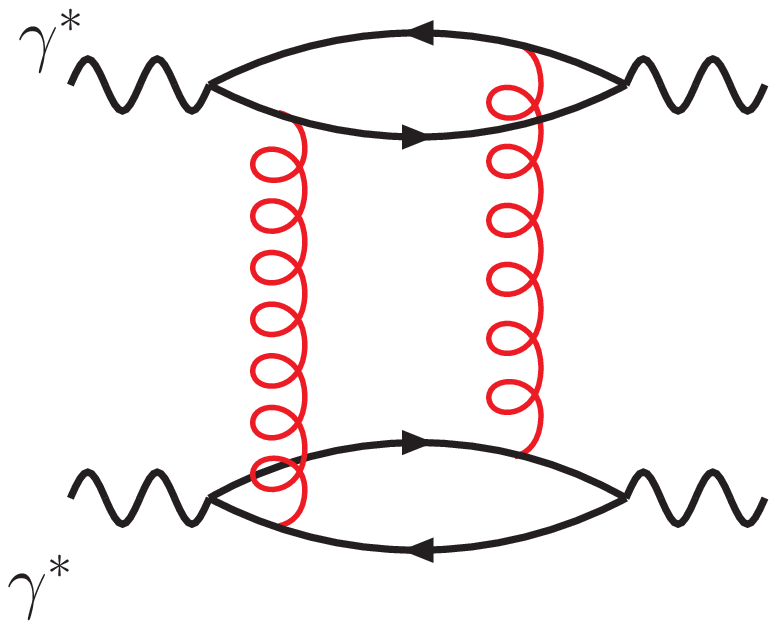}} 
  \parbox{4cm}{\center\includegraphics[width=3.5cm]{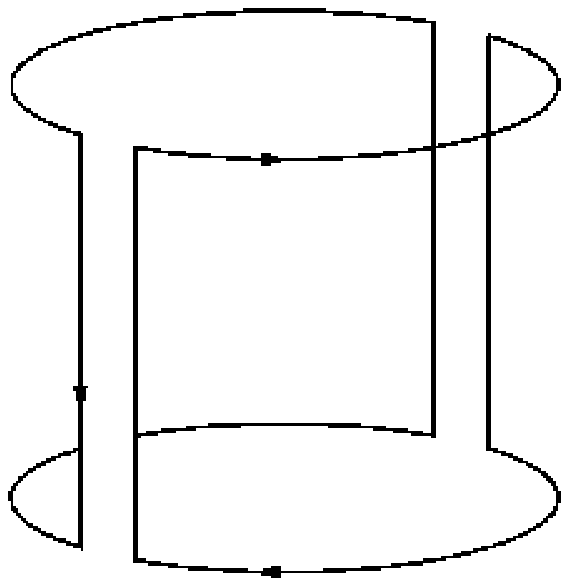}}
  \parbox{3.5cm}{\center \includegraphics[width=2.5cm]{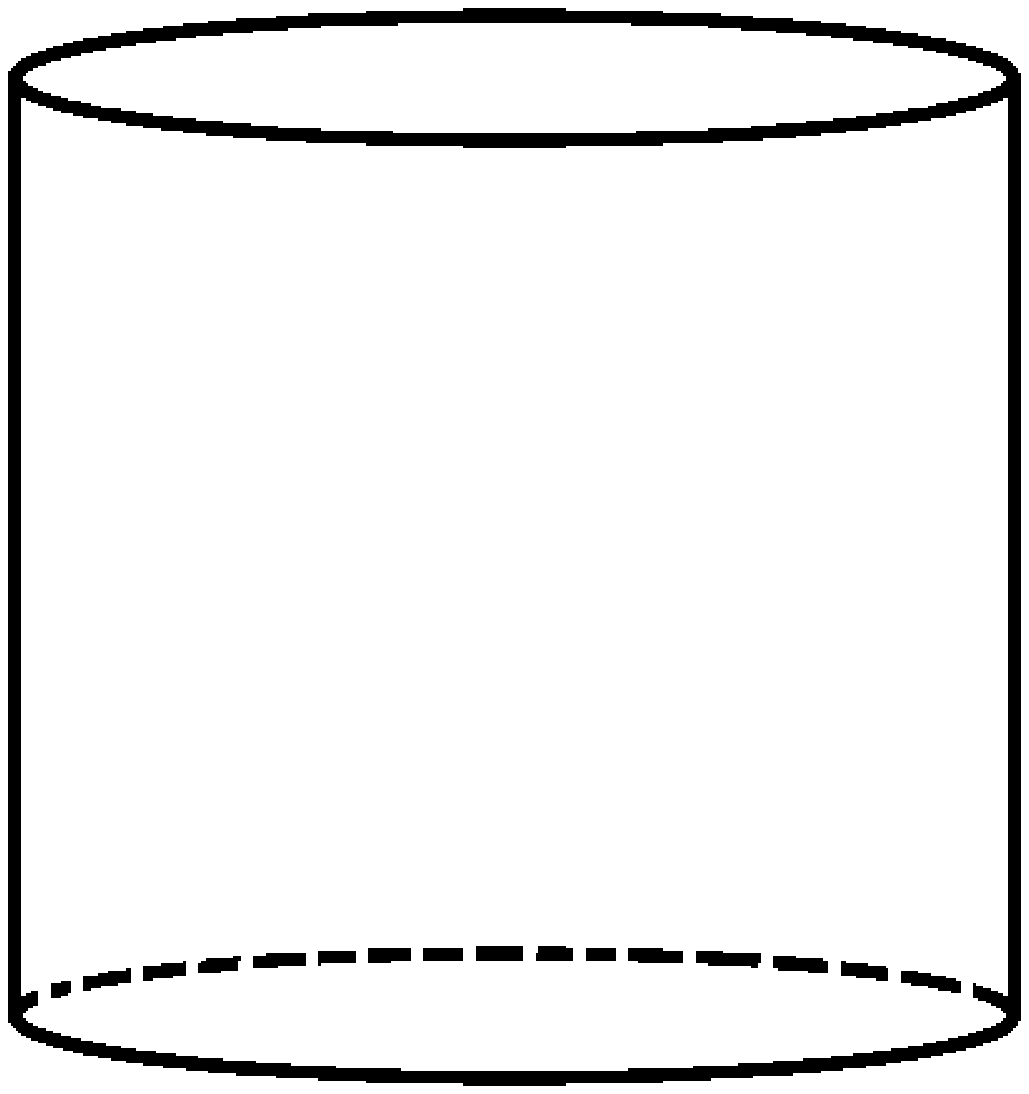}  } 
  \caption{\small To the left, a typical leading order diagram with its double-line color line factor in the middle.  The color factor fits on the cylinder surface, depicted to the right.}
  \label{fig:cyl}
\end{figure}
As a physical process with such a topology we consider the scattering
of two highly virtual photons in the Regge limit $s \gg |t|$, and
again we use the LLA to sum to all orders those diagrams which have
one power of $\ln s$ for each loop. In particular, the interaction
between the two photons is mediated by gluons in the $t$-channel, and
to lowest order in the electromagnetic coupling, each photon couples
to a quark-loop.  The quark-loops provide the two boundaries of the
cylinder. To leading order, the two photons interact by the exchange
of two $t$-channel gluons. A typical leading order diagram is shown in
Fig.\ref{fig:cyl}, together with its color factor. Unlike the planar
case, color factors of diagrams with both discontinuities in the $s$-
and in the $u$-channel fit on the cylinder. The
$t$-channel-interaction has therefore positive signature, similar to
the $N_c$ finite case,

To resum higher order terms we use again the analytic representation Eq.(\ref{eq:elastic_rep})
of the elastic amplitude in the Regge limit. Now the signature factor includes both right and left hand cuts: 
\begin{align}
  \label{eq:sigfac_possig}
\xi(\omega) = -\pi\frac{e^{-i\pi\omega} - 1}{\sin (\pi\omega)}
\end{align}
and belongs to positive signature in the $t$-channel. As in
the previous paragraph, we take the discontinuity in $s$ which allows
to determine the partial wave amplitude by considering real particle
production processes. Coupling of the two gluon state to the virtual
photon is described by the two gluon impact factor of the virtual
photon which is given by the sum of the following four Feynman diagrams:
\begin{align}
  \label{eq:gamma_impact}
{D}_{(2;0)} ({\bf k}_1, {\bf k}_2) = 
        \parbox{2.5cm}{\includegraphics[width=2.5cm]{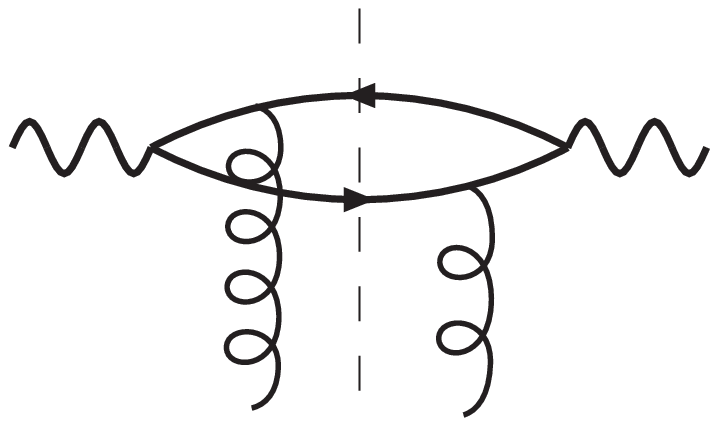}} 
        +
        \parbox{2.5cm}{\includegraphics[width=2.5cm]{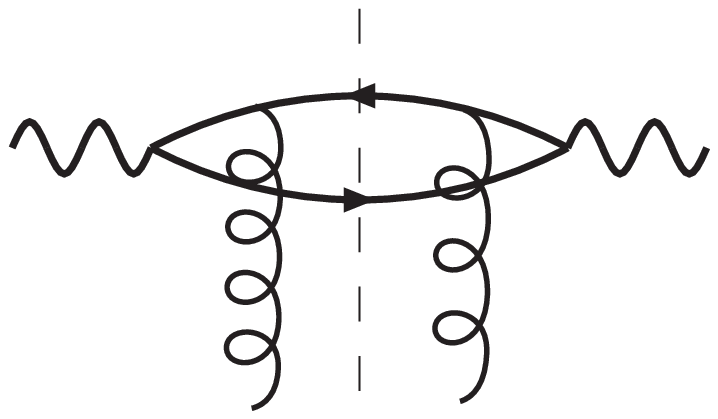}} 
        +
        \parbox{2.5cm}{\includegraphics[width=2.5cm]{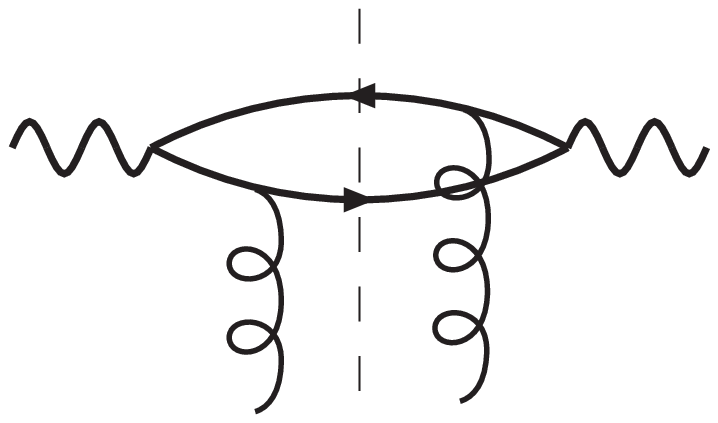}} 
        +
        \parbox{2.5cm}{\includegraphics[width=2.5cm]{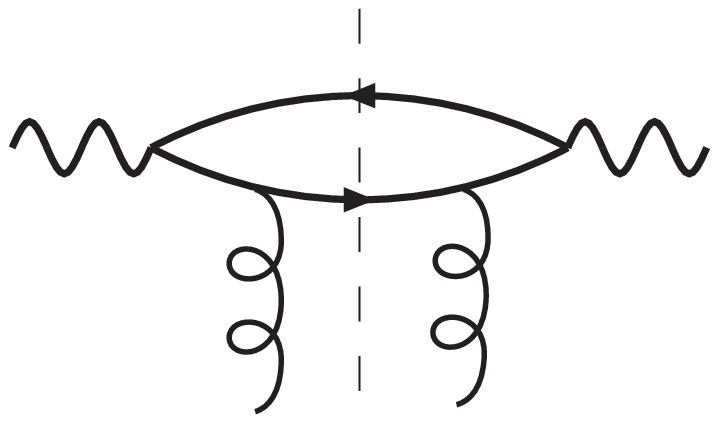}}
\end{align}
with the constraint ${\bf k}_1 + {\bf k}_2 = {\bf q}$.  It has  the important
properties to be symmetric under exchange of its transverse momenta
arguments, and to vanish whenever one of the transverse momenta
goes to zero\footnote{An analytic form of $D_{(2;0)}$ for the forward case can be found in \cite{Bartels:1994jj}. In particular, the  above $D_{(2;0)}$ is  $1/\sqrt{8}$ times  Eq.(2.4) of \cite{Bartels:1994jj}, where the factor $\sqrt{8}$ originates from $\sqrt{N_c^2 -1}$ with $N_c=3$.  }.  For the large $N_c$ treatment it is convenient to absorb a
factor $N_c$ into the impact factor by defining $\mathcal{D}_{(2;0)}
({\bf k}_1, {\bf k}_2) :=N_c{D}_{(2;0)} ({\bf k}_1, {\bf k}_2)$ which
is proportional to the 't Hooft coupling $\lambda$.  To leading
order in $\lambda$, the partial wave is  given by
\begin{align}
  \label{eq:cylind-topol-bfkl}
  \phi^{(0)} (\omega, t) = \frac{2}{\omega} \mathcal{D}_{(2;0)} \otimes \mathcal{D}_{(2;0)}.
\end{align}
Higher order corrections to the $s$-discontinuity are taken into account
by considering real gluon production in the Multi-Regge-kinematics.
Production of real gluons is described, in the same way as in the previous section, by
the particle production vertex in Eq.(\ref{eq:lipatov_factor}).  To
study the color factor on the cylinder, we start with
the Born term, Fig.\ref{fig:cyl} and insert one additional
$s$-channel gluon
\begin{figure}[htbp]
  \centering
  \parbox{4cm}{\includegraphics[height=3cm]{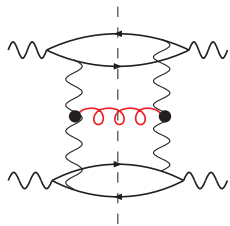}}
  \caption{\small $s$-discontinuity for the scattering of two virtual photons with one additional real    $s$-channel gluon.  Wavy lines for $t$-channel gluons denote reggeized gluons.}
  \label{fig:discgamma2}
\end{figure}
which leads us to the diagram Fig.\ref{fig:discgamma2}. For the color factors on the cylinder we find, 
compared to the plane an important difference: 
in addition to the combinations that fits  on the plane,
\begin{align}
  \label{eq:bfkl_sing1}
   \parbox{2cm}{\includegraphics[height=2cm]{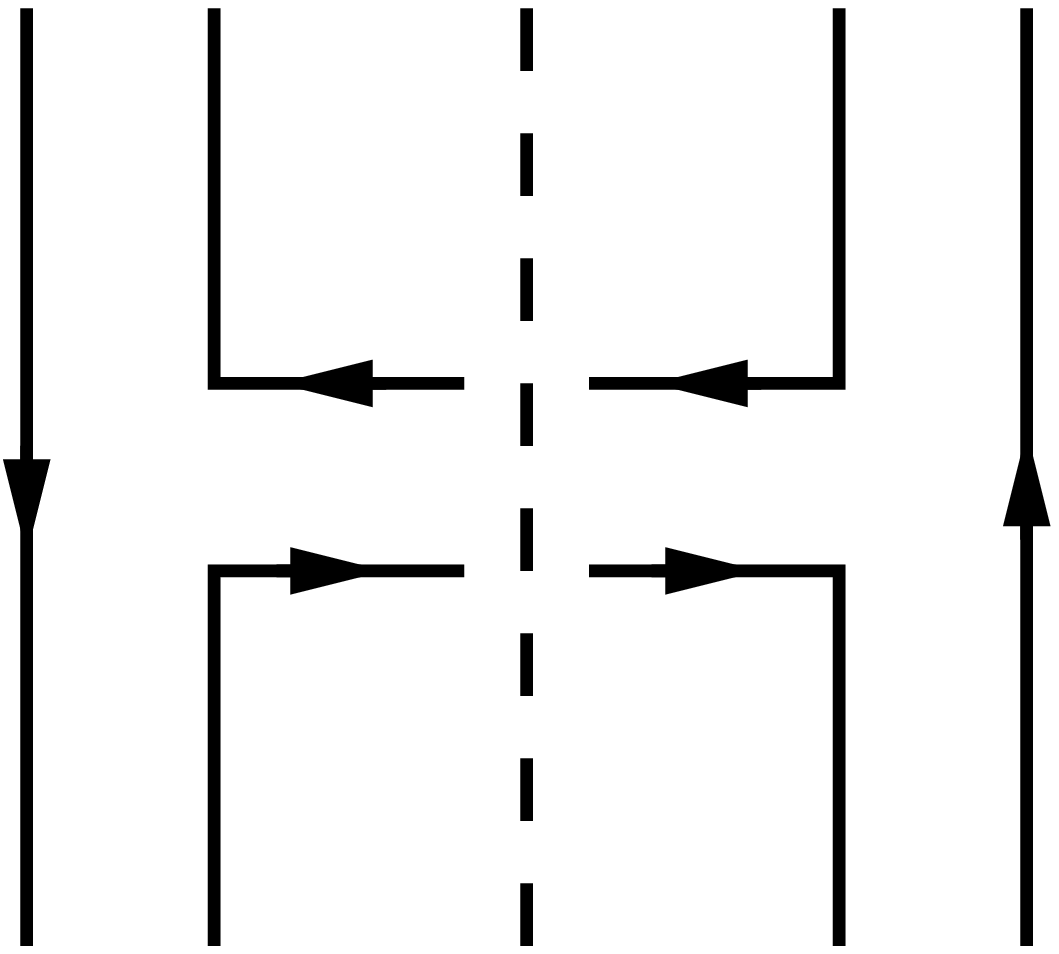}} \qquad \to \qquad  \parbox{3cm}{ \includegraphics[width=2.5cm]{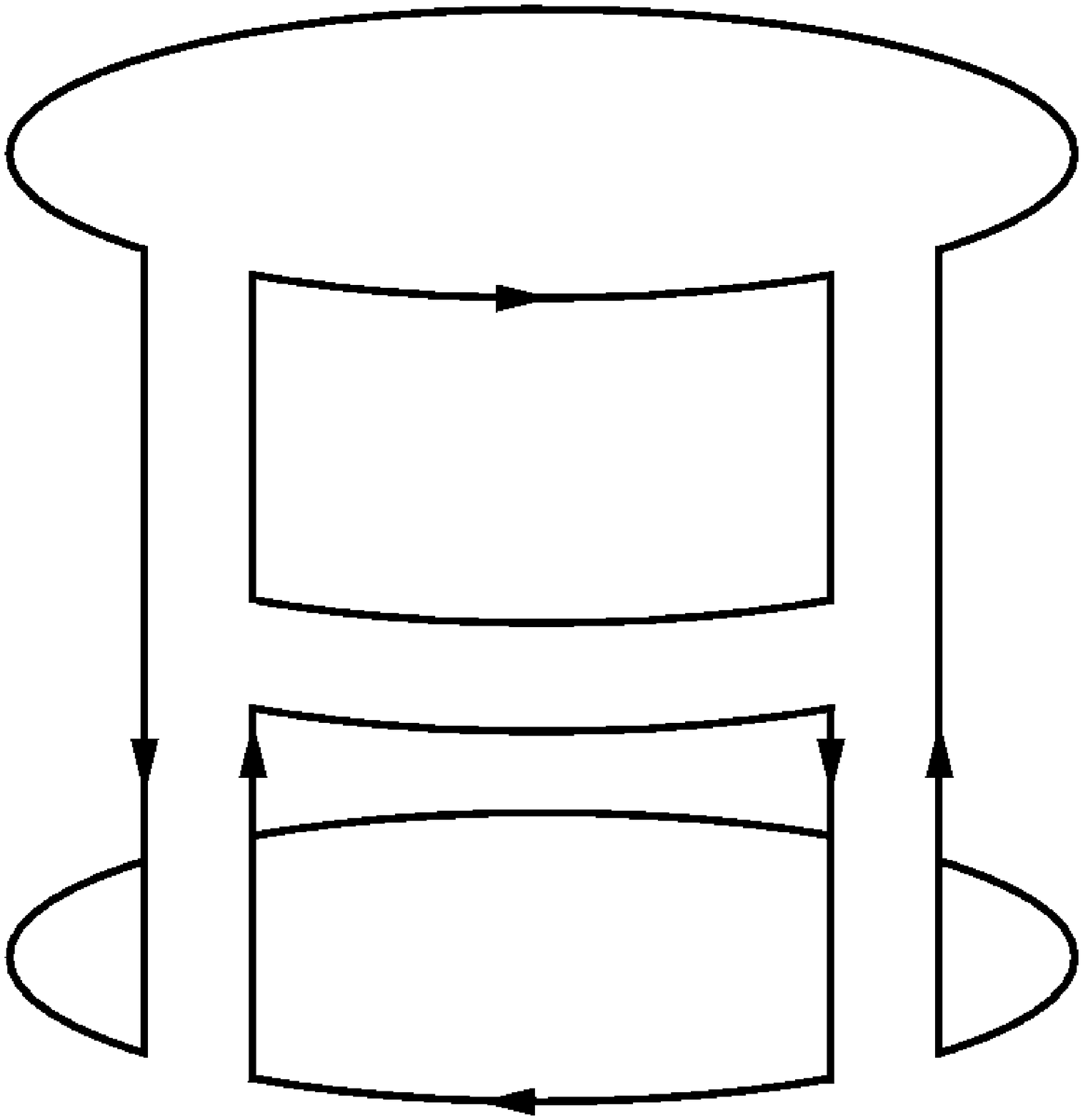}\quad ,}
\end{align}
on the cylinder also the following combination contributes:
\begin{align}
  \label{eq:bfkl_sing2}
 \parbox{3cm}{\includegraphics[height=2cm]{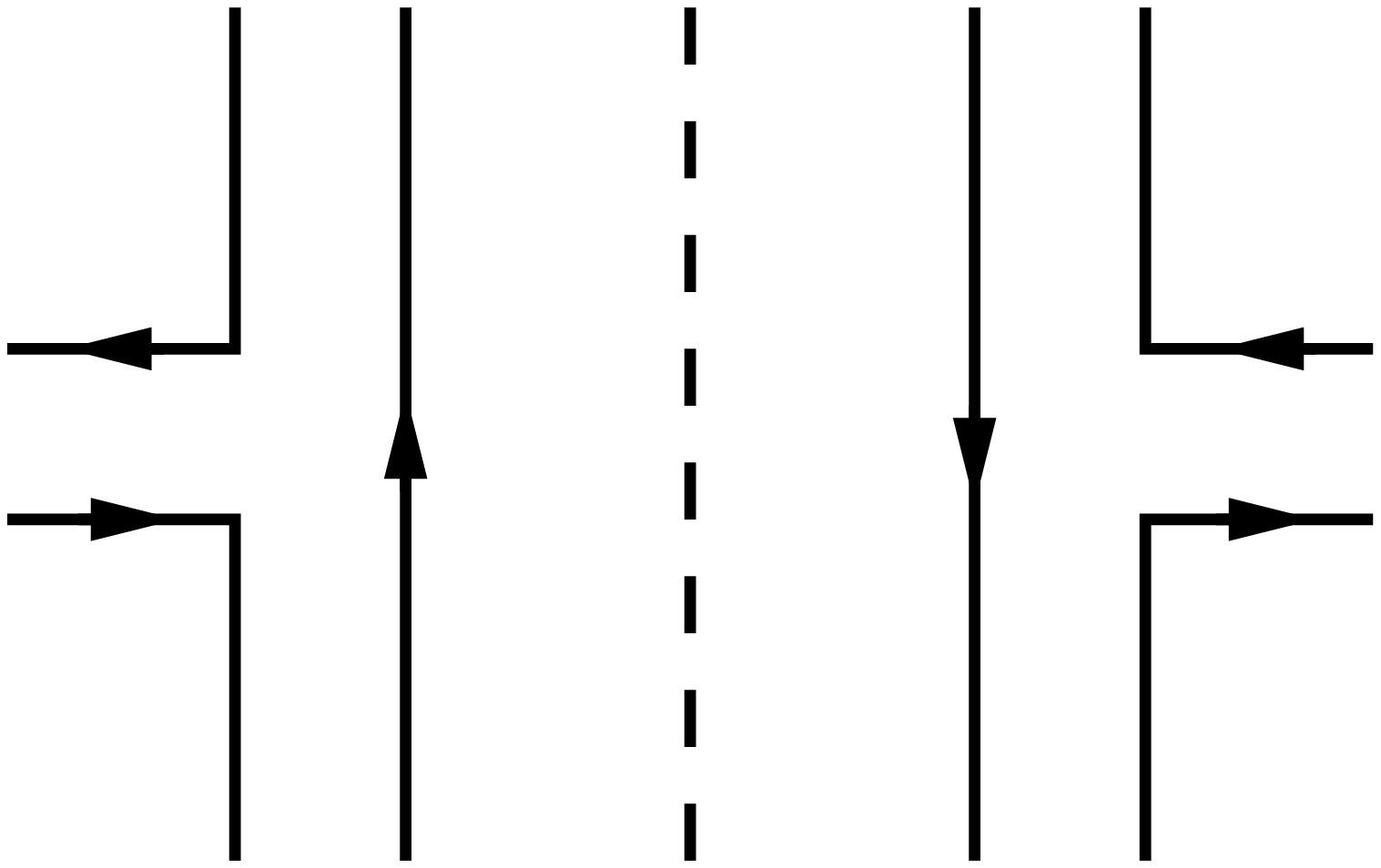}}\qquad \to\qquad \parbox{3.5cm}{ \includegraphics[width=2.5cm]{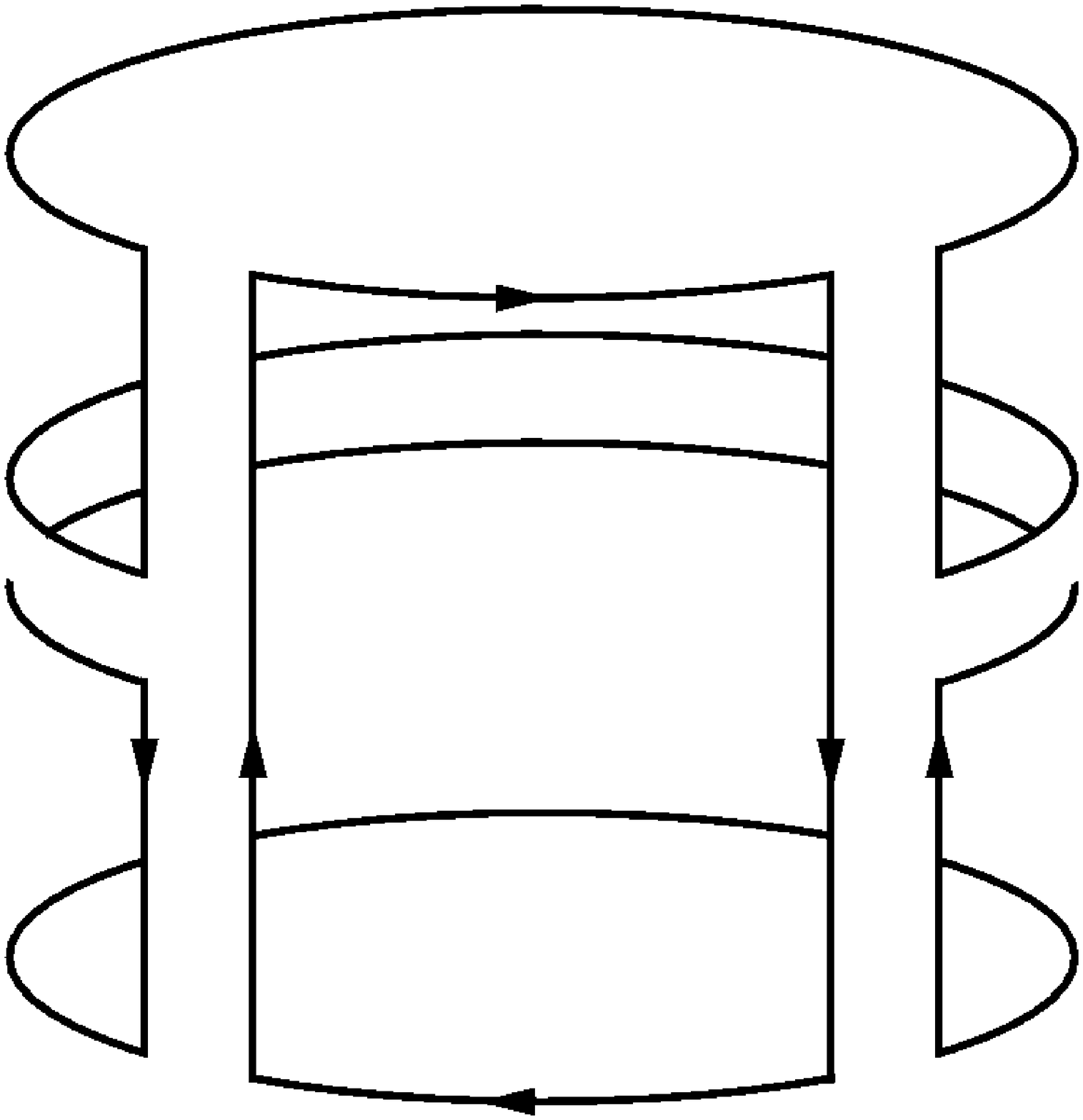}\quad.}
\end{align}
For the two-to-two Reggeon transition on the cylinder we therefore
obtain a factor two compared to the plane.  This counting generalizes
to diagrams with more than one real gluon: For each produced real
gluon the two combinations of color factors Eq.(\ref{eq:bfkl_sing1})
and Eq.(\ref{eq:bfkl_sing2}) need to be added.  An example with three
produced gluons is shown in Fig.\ref{fig:fore_back}:
\begin{figure}[htbp]
  \centering 
  \parbox{4.5cm}{\includegraphics[width=4cm]{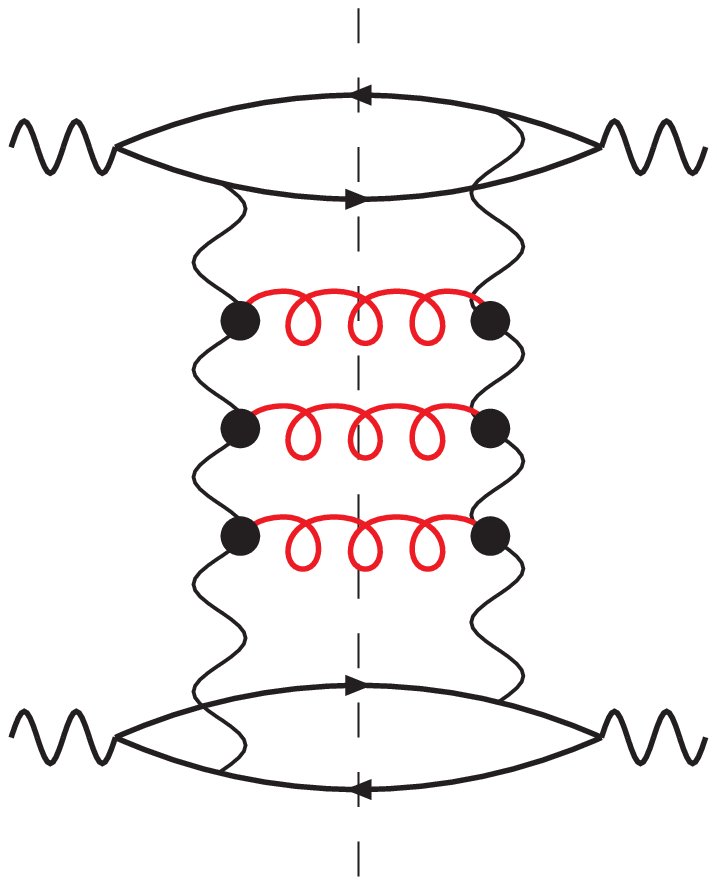}}
  \parbox{4.5cm}{\center \includegraphics[width=2.5cm]{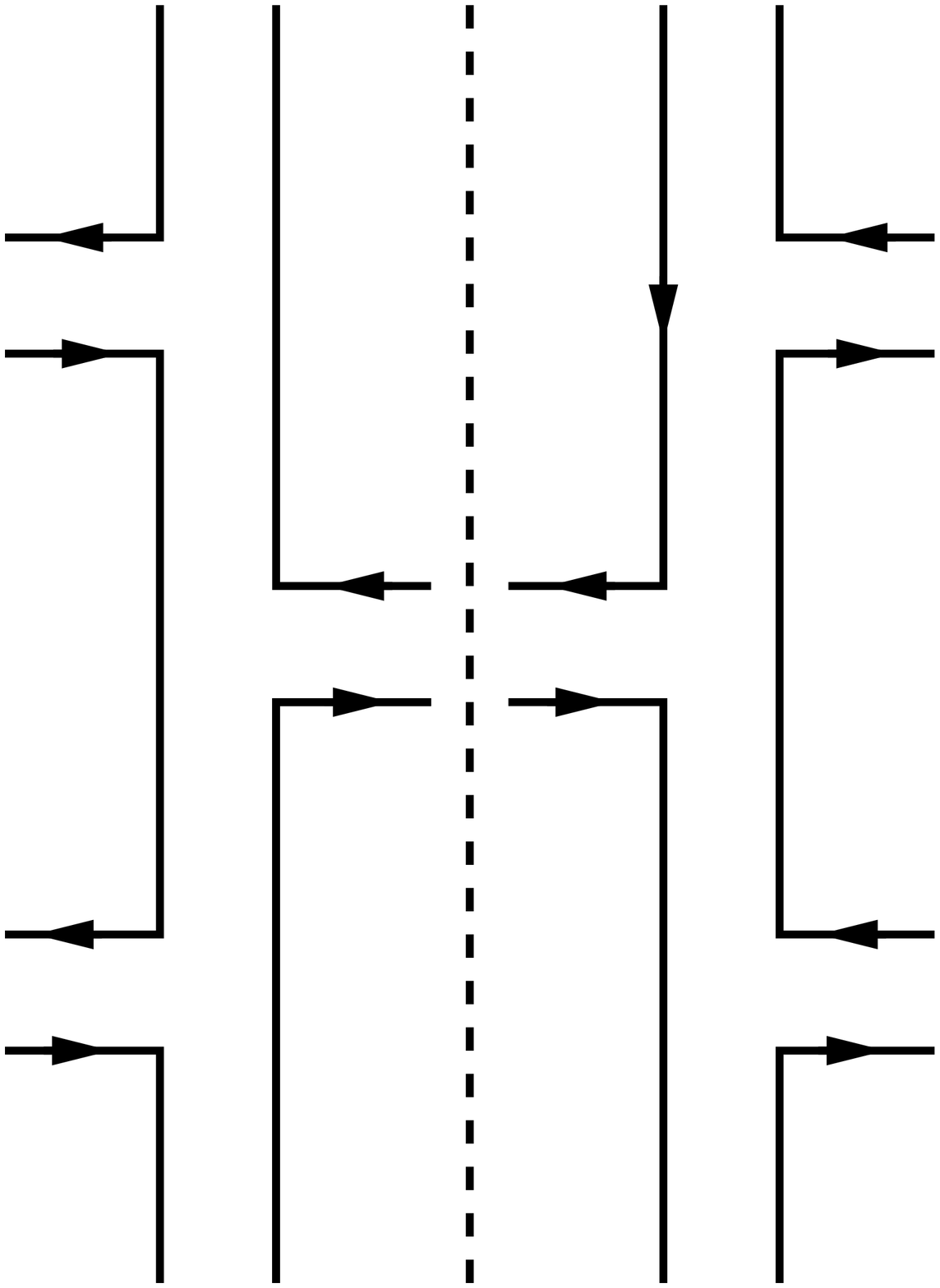}}
  \parbox{4.5cm}{\center \includegraphics[width=2.5cm]{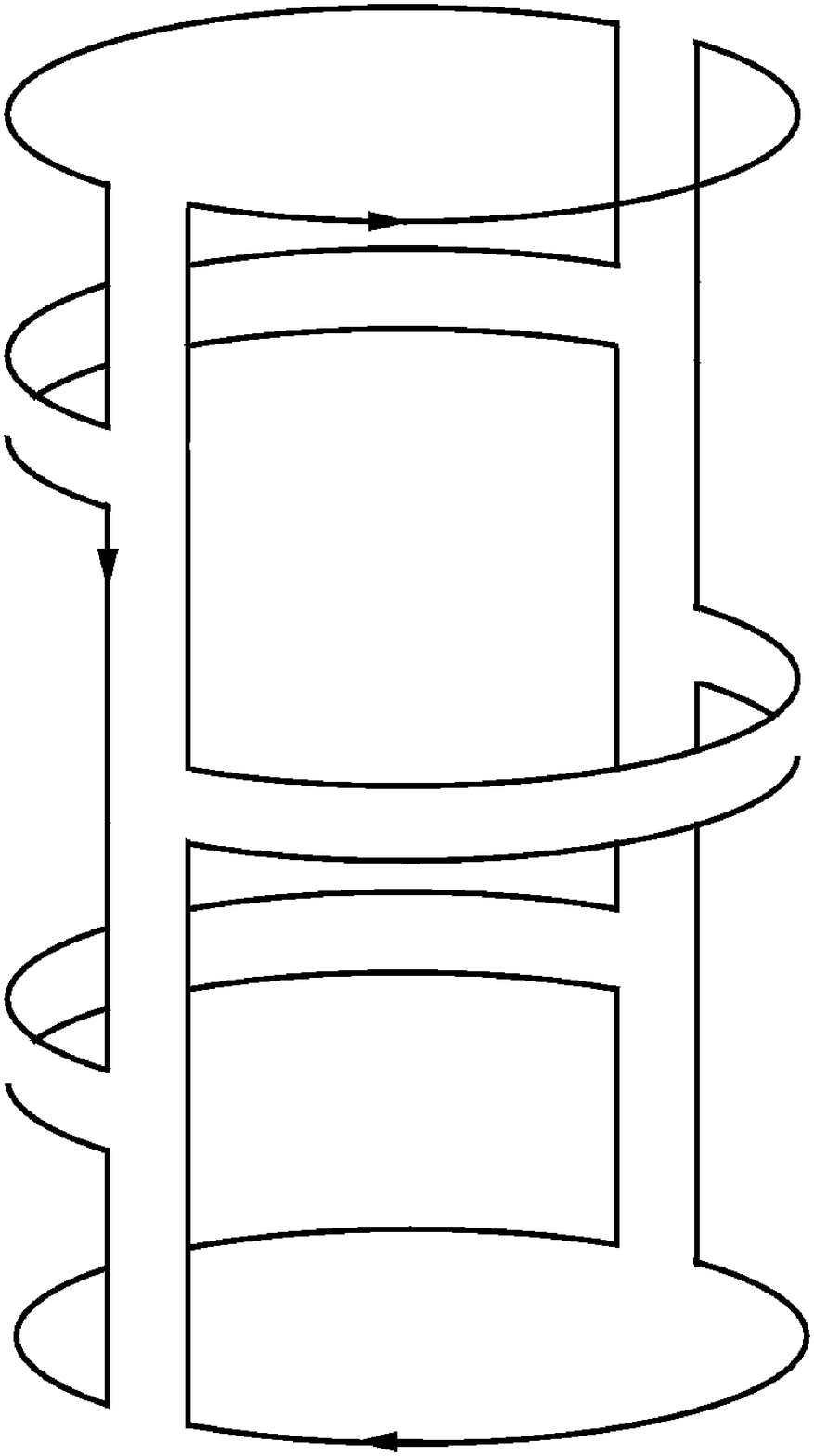}} 
\\
   \centering 
   \parbox{4.5cm}{\center (a)} 
   \parbox{4.5cm}{\center  (b)} 
   \parbox{4.5cm}{\center  (c)}
\caption{ \small Multi-gluon emission within the MRK. (a) The cut Feynman diagram (b) Combination of relevant color diagrams of the production vertex.   (c) The combination of (b) on the cylinder.  }
  \label{fig:fore_back}
\end{figure}
For each rung, we have two possibilities of connecting the $t$-channel
lines of the ladder, either on the forefront and or on the backside of
the cylinder \footnote{ The fact that, on the cylinder, each emission
  has two possibilities, one on the forefront and another one on the
  backside of the cylinder, has been realized also in
  \cite{DelDuca:1993pp,DelDuca:1995zy}.  However, the approach pursued in this paper
  is quite different form ours: it starts from Park-Taylor
  amplitudes.}  Similar to the planar case discussed before, the
summation over all production processes is done by formulating the
BFKL equation on the cylinder.  This integral equation coincides with
Eq.(\ref{eq:bfkl-eq}), except for the factor $2$ in front of
$K_{2\to2}$, which results from the cylinder topology. Technically, we
again split off the impact factor of the lower virtual photon and
consider the partial wave amplitude $ \mathcal{D}_{2} (\omega|{\bf
  k}_1, {\bf k}_2) $ for the scattering of a virtual photon on two
reggeized gluons, which as before we define to contain the Reggeon
propagator of the external reggeized gluons. We find
\begin{align}
  \label{eq:bfkl-singlet}
(\omega - \sum_i^2\beta({\bf{k}}_i)) \mathcal{D}_{2} (\omega|{\bf k}_1, {\bf k}_2) &= 
  \mathcal{D}_{2,0} ({\bf k}_1, {\bf k}_2)
+2\bar{\lambda}
  K_{2 \to 2}\otimes \mathcal{D}_{2} (\omega|{\bf k}_1, {\bf k}_2) .
\end{align}
The kernel coincides with the BFKL kernel. Unlike the planar BFKL-equation Eq.(\ref{eq:bfkl-eq}),
Eq.(\ref{eq:bfkl-singlet}) has no pole solution, but the solution is
known to have a cut in the complex $\omega$-plane. The solution is known
both for the forward $(t=0)$ \cite{Lipatov:1976zz,Kuraev:1976ge,Fadin:1975cb,
  Kuraev:1977fs,Balitsky:1978ic}  and the non-forward $(t \neq
0)$ case \cite{Lipatov:1985uk}. Furthermore, the BFKL-Green's
function, which is obtained from $\mathcal{D}_2(\omega)$ by splitting up
the impact factor of the above virtual photon, is invariant under
M\"obius transformations.

\section{Six-point amplitudes and the pair-of-pants -topology}
\label{sec:sixpoint}
\begin{figure}[htbp]
  \centering \parbox{4cm}{\center
    \includegraphics[height=3cm]{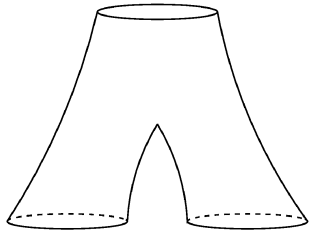}}
  \caption{ \small The sphere with three boundaries   $b=3$  and no handles $h = 0$ which yields the pair-of-pants.}
  \label{fig:pop}
\end{figure}

Returning to the expansion Eq.(\ref{eq:vacuumgraph}), we finally address
the term with three boundaries and zero handles $b=3 , h=0$.  It is
proportional to $1/N_c$ and leads to color factors that fit on the
pair-of-pants, as illustrated in Fig.\ref{fig:pop}.  As we did for
scattering amplitudes on the plane and on the cylinder, also for the
pair-of-pants we identify a QCD-amplitude which, using the LLA, will
be determined in a certain kinematical high energy limit.
As it was the case for the elastic amplitude, within the LLA, quark
loops occur only inside the coupling to external states. In order to
arrive at a surface with three boundaries, we are thus lead to a
six-point amplitude, which we will study in the triple-Regge limit, to
be specified in the following.

\begin{figure}
\centering
 \parbox{7.5cm}{ \includegraphics[height = 3.8cm]{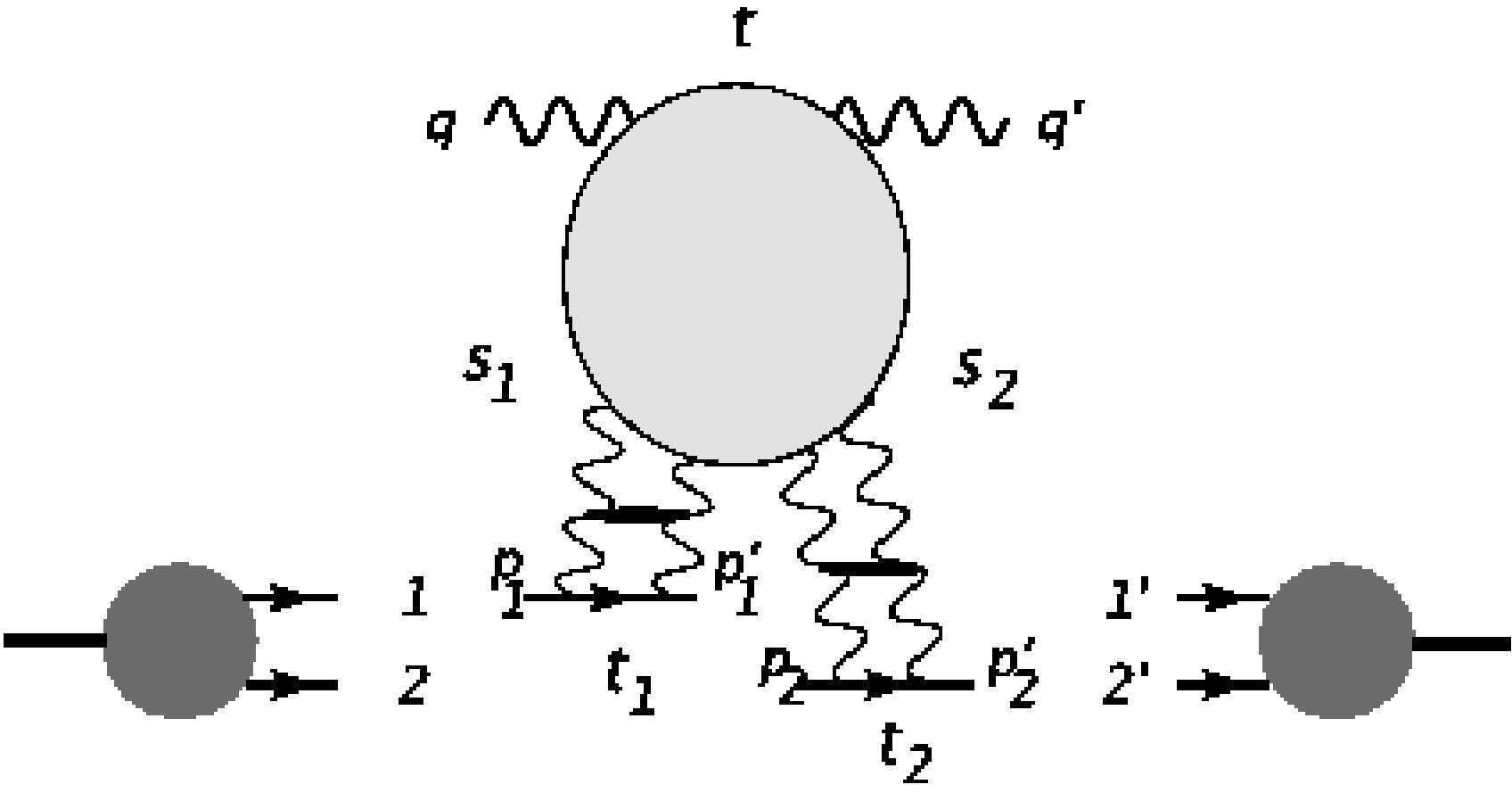}}
 \parbox{6.5cm}{ \includegraphics[height = 3.8cm]{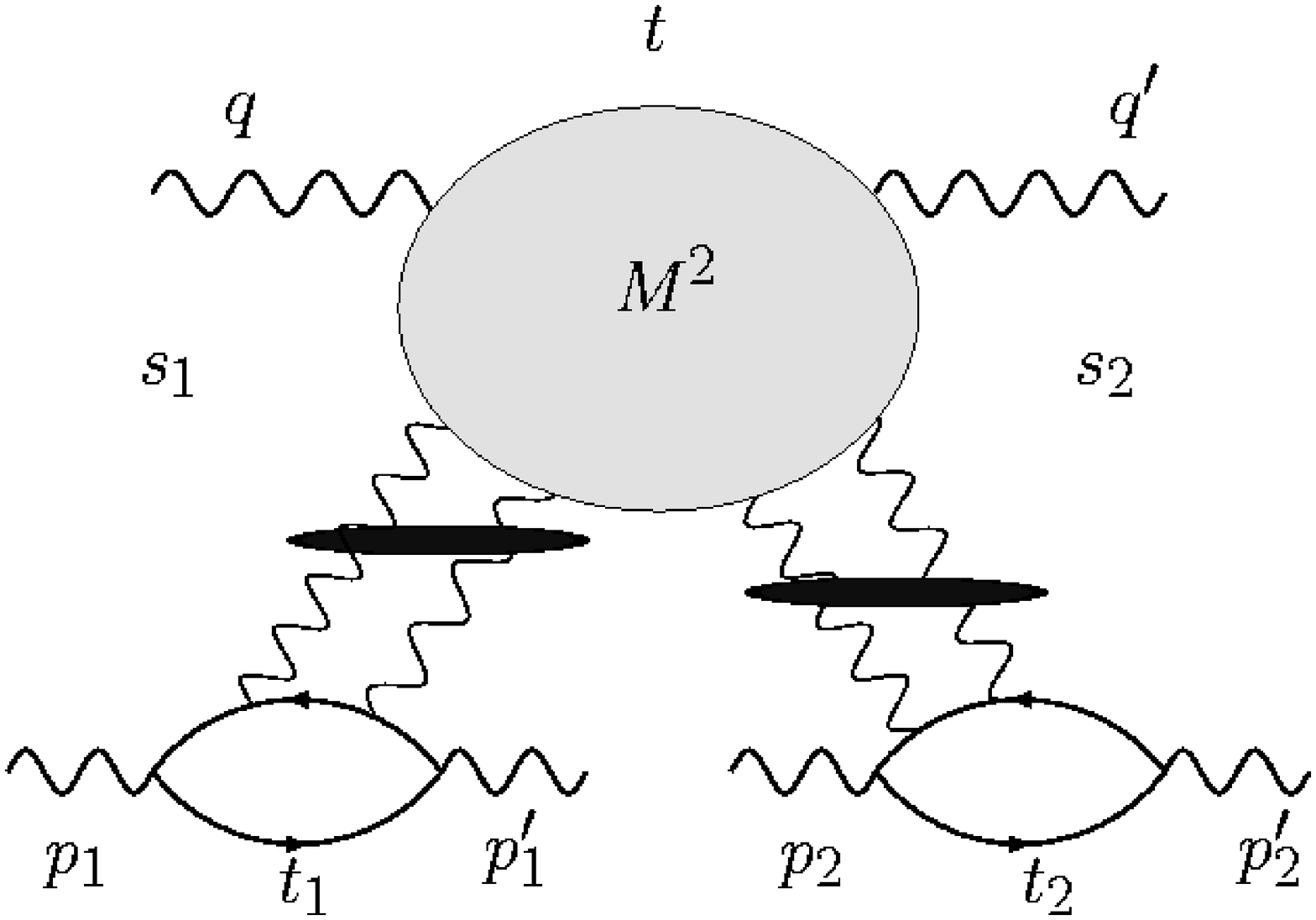}}
\caption{\small Scattering of a virtual photon on a weakly bound nucleus and the corresponding process where the lower nuclei are replaced by two highly virtual photons to the right.  }
\label{fig:nucl_phot}
\end{figure}

Within QCD, scattering amplitudes with more than $4$ external
particles arise naturally in the context of deep inelastic scattering
on a weakly bound nucleus, see  Fig.\ref{fig:nucl_phot}a.
To be definite, let us consider deep inelastic electron scattering on
deuterium. The total cross section of this scattering process is
obtained from the elastic scattering amplitude , $T_{\gamma^*(pn) \to
  \gamma ^* (pn)}$, which describes a $3 \to 3$ process,
\begin{equation}
\label{eq:sigmatot}
\sigma^{tot}_{\gamma^*(pn) \to \gamma ^* (pn)} = \frac{1}{s} \Im{\text m}\, 
T_{\gamma^*(pn) \to \gamma ^* (pn)},
\end{equation} 
where $s$ denotes the square of the total center of mass energy of the
scattering process.  To study the pair-of-pants topology, we replace
the two nucleons by virtual photons which furthermore provides us with
a clean perturbative environment for the study of the six-point
amplitude.  The kinematics is illustrated in Fig.\ref{fig:nucl_phot}.
Large energy variables are $s_1=(q+p_1)^2$, $s_2=(q'+p'_2)^2$, and $M^2
= (q+p_1 - p'_1)^2$ which gives the squared mass of the diffractive
system in which the upper virtual photon dissociates. The total energy
square is given by $s=(q+p_1+p_2)^2$.  Furthermore, we have the
momentum transfer variables $t=(q-q')^2$, $t_1=(p_1-p'_1)^2$, and
$t_2=(p_2-p'_2)^2$.  The triple Regge-limit is given by $ s_1 = s_2
\gg M^2 \gg t, t_1, t_2$.  The investigation of such a process, for
finite $N_c$, has been started in \cite{Bartels:1994jj}, in the
context of diffractive dissociation in deep inelastic scattering, and
in the following we will stay close to the methods used there.

Let us begin with finite $N_c$.  The $3 \to3 $ amplitude in the triple
Regge limit has the following analytic representation
\begin{align}
T_{3 \to 3}(s_1, s_2, M^2| t_1,t_2,t)=
\frac{s_1s_2}{M^2} \int  \frac{d\omega_1 d\omega_2 d\omega}{(2\pi i)^3}
&
 s_1^{\omega_1}{s_2}^{\omega_2} (M^2)^{\omega-\omega_1-\omega_2}
 \xi({\omega_1})  \xi({\omega_2})  \xi({\omega,\omega_1,\omega_2}) \notag \\
&\cdot
 F(\omega, \omega_1, \omega_2| t, t_1,t_2) .   
\label{tripleregge}
\end{align}
All three $t$-channels carry positive signature, and the signature factors
are given by
\begin{align}
  \label{eq:sig_facs}
            \xi(\omega) &= -\pi\frac{e^{-i\pi\omega}-1}{\sin(\pi\omega)}  
&\mbox{and}& &
 \xi({\omega,\omega_1,\omega_2}) &= -\pi\frac{e^{-i\pi(\omega - \omega_1 -\omega_2)} - 1}{\sin\pi(\omega - \omega_1-\omega_2)}.
\end{align}
The partial wave $F(\omega_1,\omega_2,\omega| t_1,t_2,t)$ has no phases and is
real valued. In analogy to the treatment of $T_{2 \to 2}$, we take
the triple energy discontinuity in $s_1$, $M^2$, and $s_2$,
\begin{equation}
\mathrm{disc}_{s_1} \mathrm{disc}_{s_2} \mathrm{disc}_{M^2} T_{3 \to 3} =
 \pi^3\frac{s_1s_2}{M^2} \int  \frac{d\omega_1 d\omega_2 d\omega}{(2\pi i)^3}
 s_1^{\omega_1}  {s_2}^{\omega_2}  (M^2)^{\omega-\omega_1-\omega_2}
\cdot
 F(\omega_1, \omega_2, \omega|t_1,t_2,t),
\label{tripledisc}
\end{equation}
which via a triple Mellin transform relates the partial wave
$F(\omega_1,\omega_2,\omega| t_1,t_2,t)$ to the real valued triple
energy discontinuity.

\begin{figure}[htbp]
  \centering
  \parbox{6cm}{\includegraphics[height = 3cm]{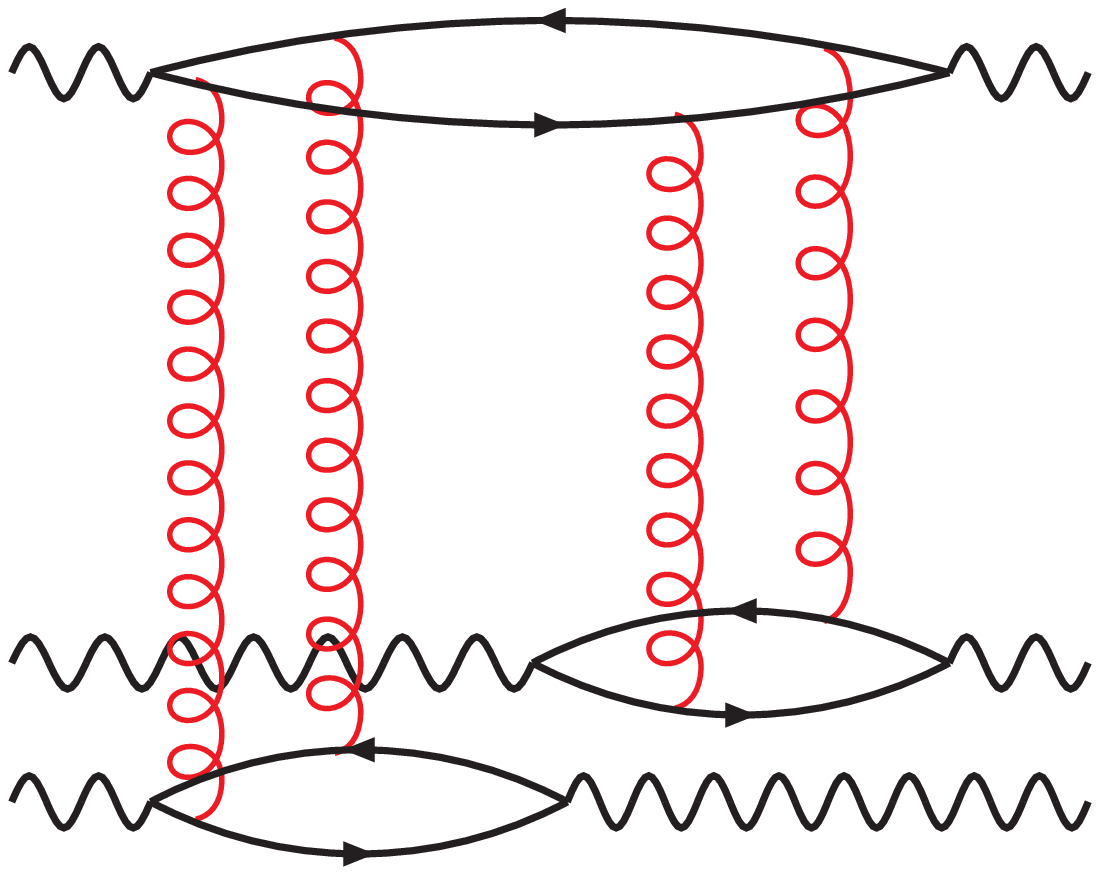}}
   \parbox{4cm}{\includegraphics[height = 3cm]{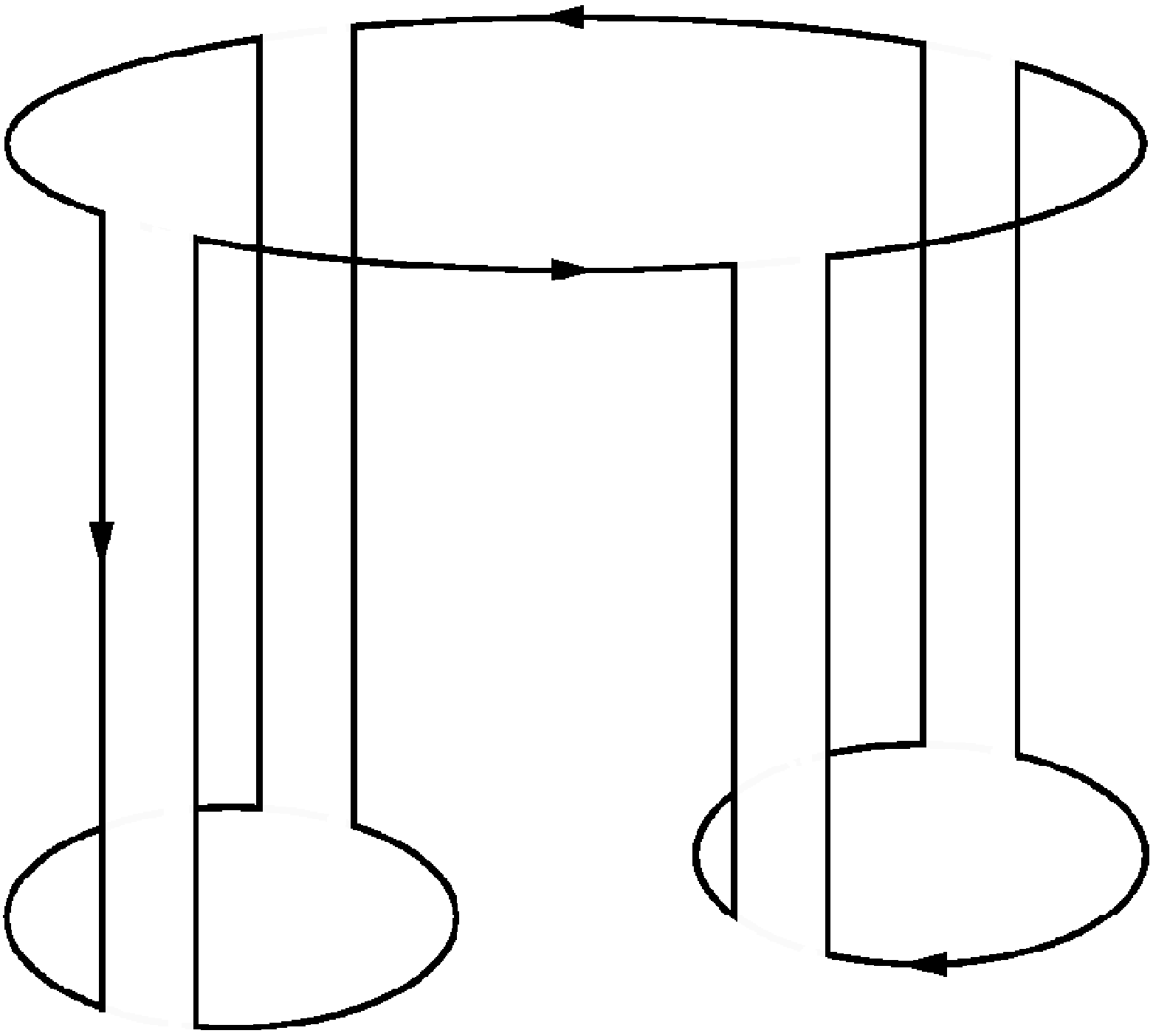}}
  \caption{ \small An example of a leading order diagram for the six-point amplitude with  its color factor depicted to the right. The diagram is proportional to $g^8N_c^3 = \lambda^4N_c^{-1}$ as required for the pair-of-pants.}
  \label{fig:trouser}
\end{figure}

\begin{figure}[ht]
\centering
\parbox{6cm}{\includegraphics[height = 5cm]{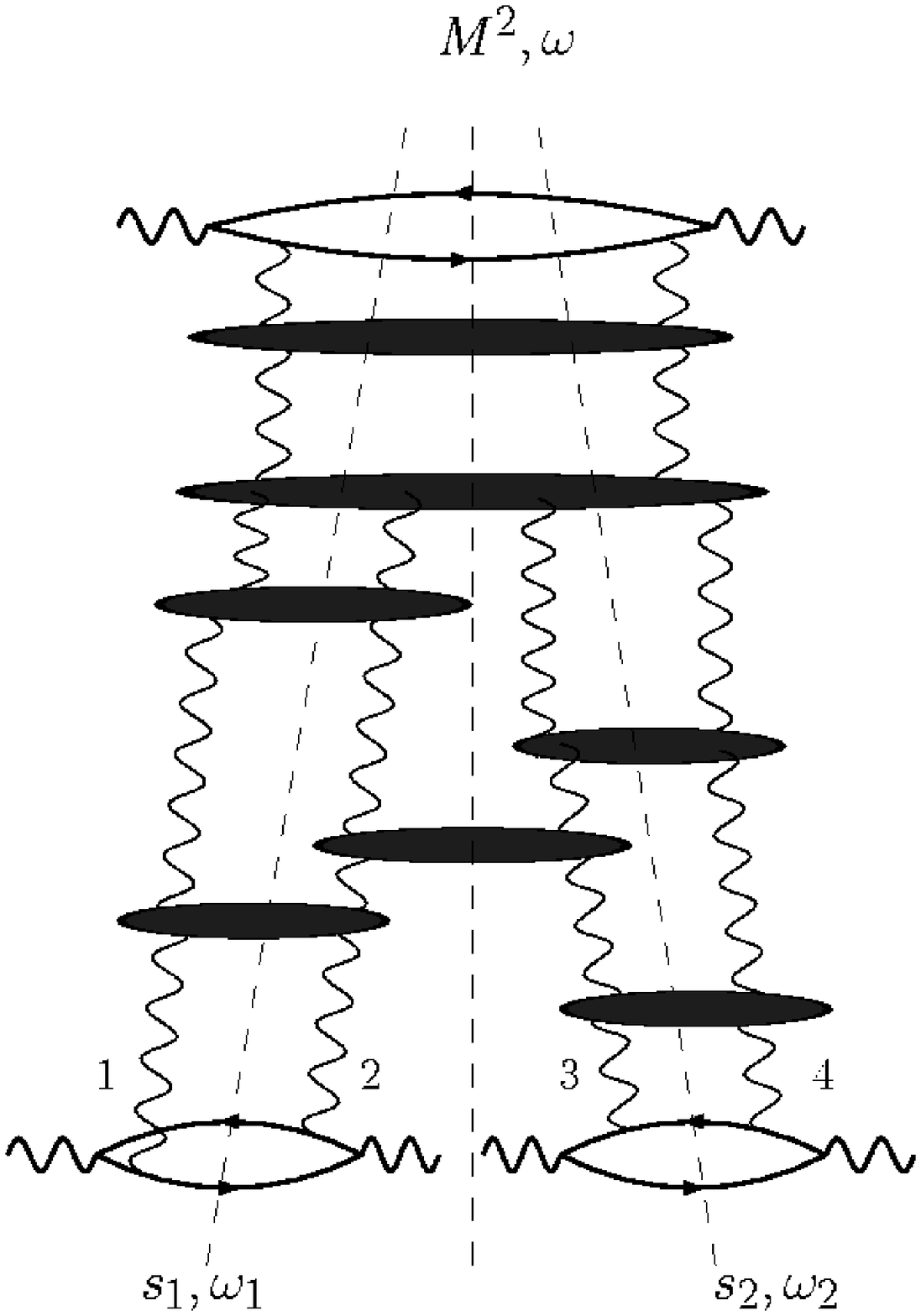}}
\parbox{6cm}{\includegraphics[height = 5cm]{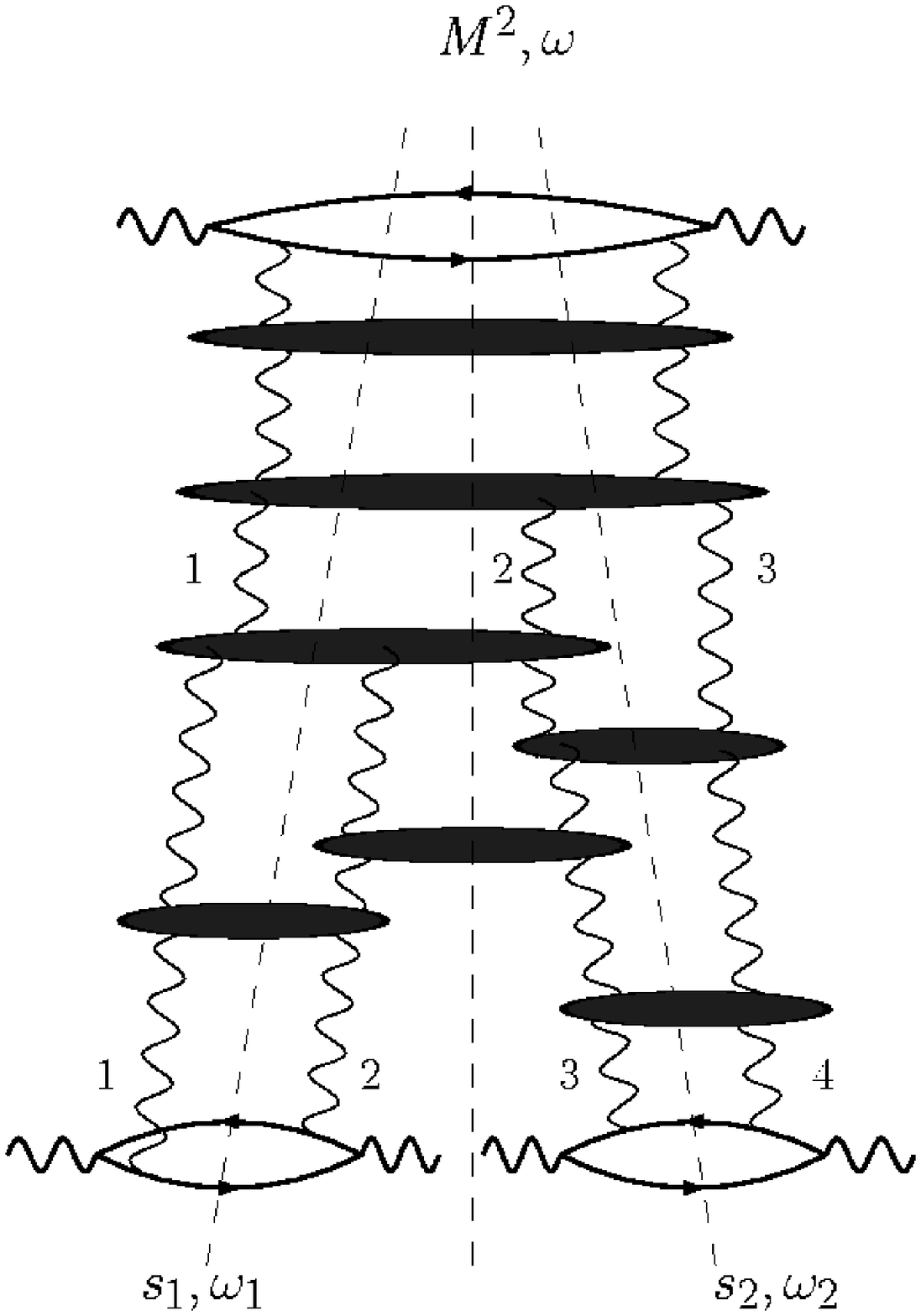}}
\caption{\small A few contributions to the triple energy discontinuity of Eq.(\ref{tripleregge}) }
\label{fig:triple_disc}
\end{figure}

Within the LLA, each of the three virtual photons couples to the
$t$-channel gluons by a quark-loop (which in the topological expansion
provides the three boundaries of the pair-of-pants). To leading order
in $g^2$, four $t$ channel gluons couple to the upper quark-loop, and
two gluons to each of the two lower quark-loops, which yields diagrams
like the one shown in Fig.\ref{fig:trouser}.
When taking the triple energy discontinuity, all intermediate states
between the $t$ channel gluons are on mass-shell.  For higher order
corrections to these diagrams, we make extensive use of unitarity and
compute sums over products of production processes.  Some examples are
shown in Fig.\ref{fig:triple_disc}.
Due to the triple discontinuity, $t$-channel gluons do not intersect,
and it is therefore useful to enumerate them from the left to right,
with the most left gluon carrying the index $1$. In order to form,
within the LLA, color singlet states which at the top couple to the
virtual photon and at the bottom to two virtual photons, we are lead
to QCD diagrams with four $t$-channel gluons at the lower and two,
three or four $t$-channel gluons at the upper end.  In contrast to our
previous discussion of scattering processes on the plane and on the
cylinder, the number of $t$-channel gluons can change, i.e. a $t$
channel gluon can emerge also from a $s$ channel produced gluon.
However, in LLA there is the restriction that, when moving downwards,
the number of reggeized $t$ channel gluons never decreases.

A special role in our analysis is taken by the diffractive mass $M^2$
as it defines the size of the upper 'cylinder' of the pair-of-pants.
In particular, the lowest intermediate state inside the
$M^2$-discontinuity (descending from the top to the bottom of the
diagram) defines the last interaction between the two 'legs' of the
pants, i.e. the branching point of the upper into the two lower
cylinders.  We use this branching point to factorize the partial wave
$F(\omega,\omega_1,\omega_2)$ into a convolution of three different
amplitudes. With $s_1, s_2 \gg M^2 \gg t, t_1, t_2$, the diagrams
below the branching point, within the LLA, do not depend on the
details of the dissociation of the upper virtual photon, and are
therefore described by two independent amplitudes
$\mathcal{D}_2(\omega_1)$ and $\mathcal{D}_2(\omega_2)$ which describe
scattering of two $t$-channel gluons and their coupling to the lower
photons.  The functions $\mathcal{D}_2(\omega_1)$ and
$\mathcal{D}_2(\omega_2)$ are given by BFKL-Pomeron Green's functions,
convoluted with the impact factors of the corresponding lower virtual
photon. In the following we will confirm that, in our topological
approach, on the pair-of-pants the required color factors work out
correctly.  The part above the the branching point is resummed by the
amplitude $\mathcal{D}_4(\omega)$, which describes the scattering of the upper
virtual photon on the four $t$-channel gluons.  In the following, our
interest will mainly concentrate on this part of our scattering
amplitude.

\section{Color factors with pair-of-pants topology}
\label{sec:prodtrous}

After our general outline of which QCD diagrams contribute to the
leading log approximation of the $3 \to 3$ process we have to select
now the contributions which belong to the large-$N_c$ limit on the
pair-of-pants. This will be done by attributing to each QCD diagram a
'color diagram' on the pair-of pants surface, where each gluon is
drawn by a pair of color lines with opposite directions, as done for
the plane and the cylinder in Sec.\ref{sec:elastic}.

\subsection{Color factors at Born-level}
\label{sec:pop_tree}
All diagrams that contribute to the Born approximation of the 
triple-discontinuity have the form of Fig.\ref{fig:trouser}, with $t$-channel
gluons coupling to the the quark-loop in all possible ways. 
Overall, one finds sixteen different diagrams. Because of the triple 
energy discontinuity $t$-channel gluons never 
intersect and hence can be labeled from the left to the right.
A closer look then shows that, inside the quark loop, we have 
the four different orderings of color matrices:
(1234), (2134), (1243), and (2143) (not distinguishing between cyclic permutations ). These four structures are illustrated in 
Fig.\ref{fig:born}
\begin{figure}[htbp]
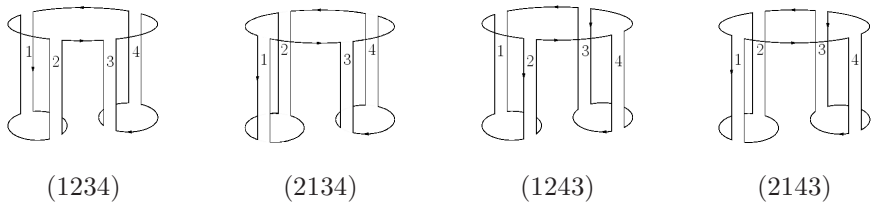

  \centering
    \begin{minipage}{.9\textwidth}
    \parbox{3cm}{\center   \includegraphics[width=2cm]{born1234.eps2}} 
    \parbox{3cm}{\center   \includegraphics[width=2cm]{born2134.eps2}}  
    \parbox{3cm}{\center   \includegraphics[width=2cm]{born1243.eps2}} 
    \parbox{3cm}{\center   \includegraphics[width=2cm]{born2143.eps2}} \\
    \parbox{3cm}{\center $(1234)  $}
    \parbox{3cm}{\center ${(2134)}  $} 
     \parbox{3cm}{\center ${(1243)} $}
    \parbox{3cm}{\center ${(2143)}  $}
  \end{minipage}
  \caption{\small The four different orderings of color factors of the Born-term.}
  \label{fig:born}
\end{figure}
and are all of the order
\begin{equation} 
g^8 N_c^3 = N_c^{-1} \lambda^4 .
\label{eq:borncounting}
\end{equation}

\subsection{Gluon production on the pair-of-pants surface}
\label{sec:prod_pop}

In order to study corrections to these Born-amplitudes, 
we consider real gluon production processes on the pair-of-pants surface 
($t$-channel gluons are always reggeized). 
As to the selection of diagrams, we are searching the maximal power 
of color factors $N_c$. With (\ref{eq:borncounting}) the diagrams 
we are going to collect will come with the weight 
\begin{align}
\label{eq:BBB}
g^8 N_c^3 (g^2 N_c)^k = N_c^{-1} \lambda^{4+k}, 
\end{align}
with $k$ being some positive integer number.
This leads to the requirement that, for each gluon, the pair of color 
lines has to be planar, i.e. it never intersects.   
\begin{figure}[htbp]
  \centering
 \parbox{7cm}{\center \includegraphics[height=4.5cm]{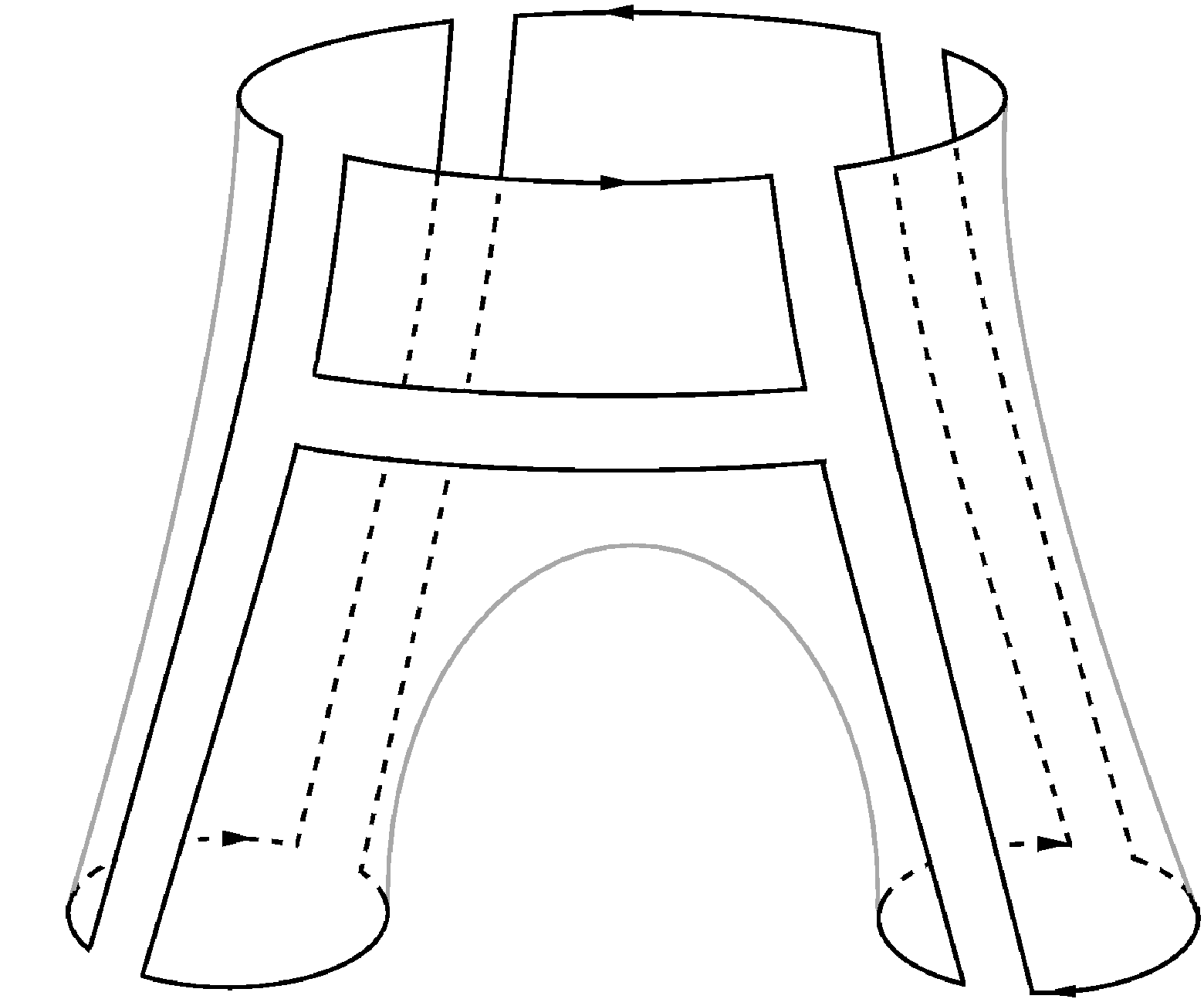}}
 \parbox{7cm}{\center \includegraphics[height=4.5cm]{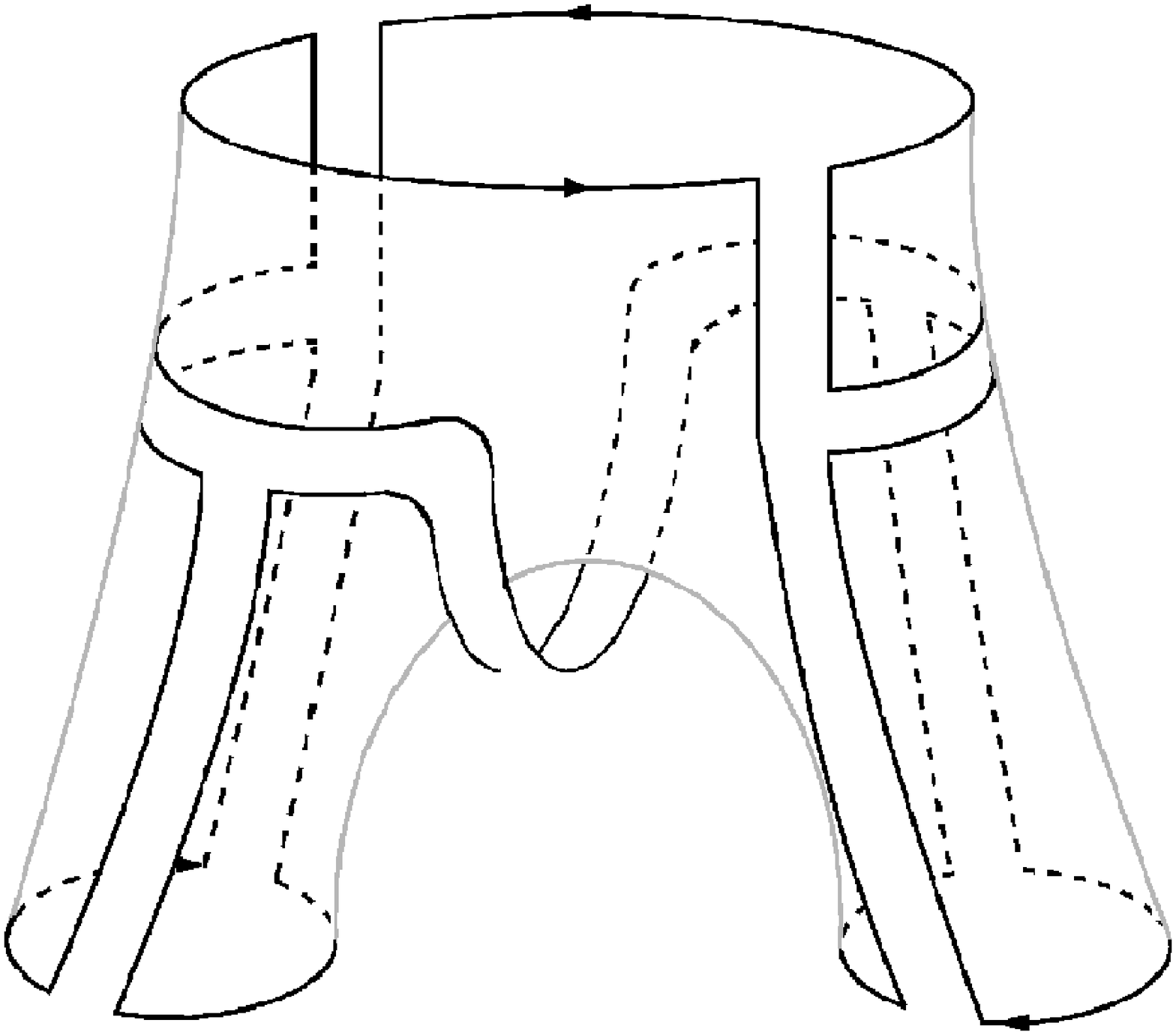}}\\
\parbox{7cm}{\center (a)}\parbox{7cm}{\center (b)}
  \caption{\small (a) A typical color factor of the planar class (A). (b) A typical color factor of the non-planar class (B) which has the interpretation of the Mandelstam diagram. }
  \label{fig:classes}
\end{figure}

As a result we find two classes of color
diagrams. The first class (A) consists of all diagrams which, by
contracting closed color loops, coincide with one of the lowest order
diagrams in Fig.\ref{fig:born}. An example for such a diagram is given
in Fig.\ref{fig:classes}.a. In the following we will refer to these
diagrams as 'planar' diagrams.  The second class (B) consists of those
diagrams where the 'last' vertex before the separation into the two
lower cylinders is of the form illustrated in Fig.\ref{fig:classes}.b:
one easily verifies that the power of color factors is $g^8 N_c^3 (g^2
N_c) = N_c^{-1} \lambda^{4+1}$, as required by condition
Eq.(\ref{eq:BBB}). Also, there is no intersection of color lines on
the pair-of-pants surface. These diagrams cannot be redrawn, by
contracting closed color loops in a straightforward way, such that they
coincide with one of the lowest order diagrams of Fig.\ref{fig:born}.
The structure shown in Fig.\ref{fig:classes}.b is reminiscent of the
Mandelstam diagram ('Mandelstam cross') which, when integrated over
the diffractive mass $M$, couples to the two Pomeron cut; we will
therefore call it 'non-planar' (although it fits onto the surface of
the pair-of-pants).  Note that this non-planar structure appears only
once, namely at the point where the upper cylinder splits into the two
lower ones.

In the remainder of this section, we describe these two sets of diagrams in 
more detail. In particular, we list the momentum space expressions which 
belong to the gluon transition kernels. In the following section, we derive 
integral equations which sum all diagrams of set (A). For set (B) we 
re-derive the triple Pomeron vertex.

\subsection{Two-to-two Reggeon transitions }
\label{sec:amplitudes-with-4}
\begin{figure}[htbp]
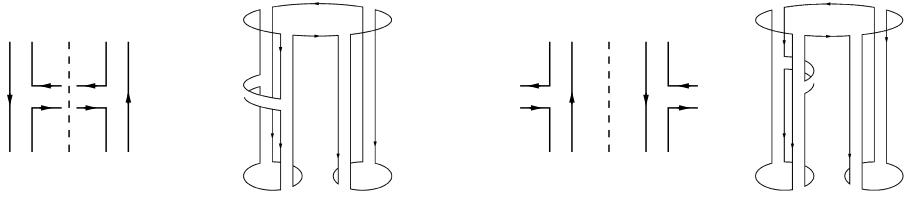

  \centering
  \parbox{3cm}{\includegraphics[height=1.5cm]{bfkl_1lopp_a.eps}}
  \parbox{3.5cm}{\includegraphics[height=2.5cm]{trouser_inta12_cyl_voll.eps2}} 
  \parbox{3cm}{\includegraphics[height=1.5cm]{bfkl_1lopp_b.eps}}
  \parbox{3.5cm}{\includegraphics[height=2.5cm]{trouser_inta.eps2}}
  \caption{\small Two combination of  color-factors for the interaction 
between gluons '1' and '2'.}
  \label{fig:same_ql}
\end{figure}

\begin{figure}[htbp]
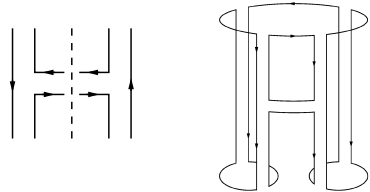

  \centering
   \parbox{2cm}{\includegraphics[height=1.5cm]{bfkl_1lopp_a.eps}}\parbox{3.5cm}{\center \includegraphics[height=2.5cm]{trouser_inta_cyl_voll12.eps2}} 
  \caption{\small Color-factors of an interacting between two neighboring $t$-channel gluons.}
  \label{fig:diff_ql_nearby}
\end{figure}

\begin{figure}[htbp]
  \centering
   \parbox{3cm}{\includegraphics[height=1.5cm]{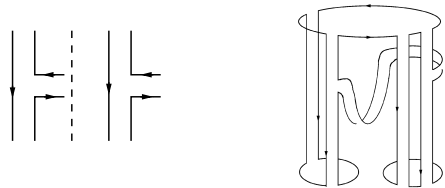}}\parbox{3.5cm}{\center \includegraphics[height=2.5cm]{trans22_alt2.eps2}} 
  \caption{\small Color factors of an interaction between two non-neighboring $t$-channel gluons.}
  \label{fig:diff_ql_no_nearby}
\end{figure}

We start with the case, where all four $t$-channel gluons couple to
the upper quark-loop. We therefore consider only insertions of the
two-to-two transition kernel,
\begin{align}
  \label{eq:22kernel_color}
\parbox{2cm}{\includegraphics[width=1.5cm]{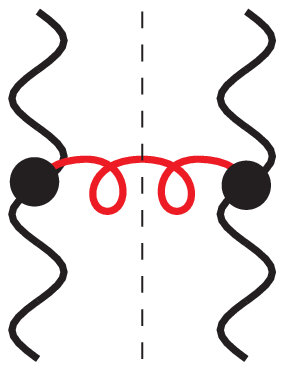}}  = 
\bar{g}^2\big(
     \parbox{.5cm}{\includegraphics[width=.5cm]{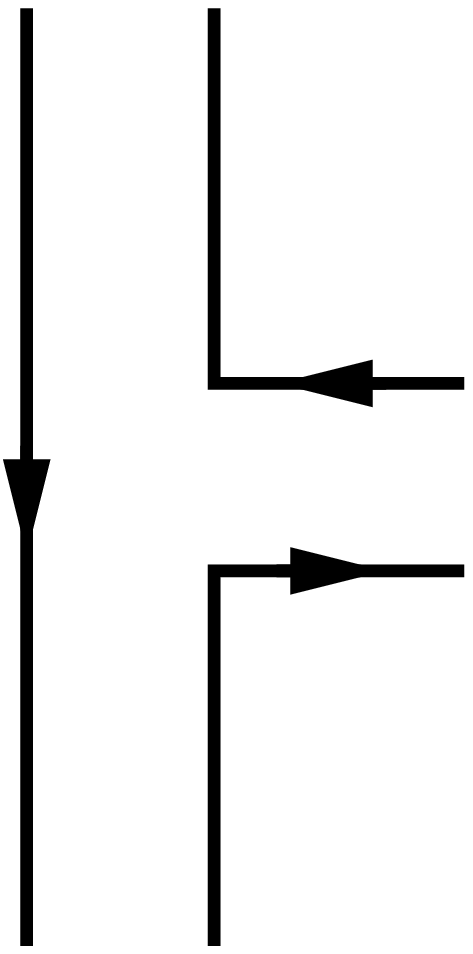}} 
     -
     \parbox{.5cm}{\includegraphics[width=.5cm]{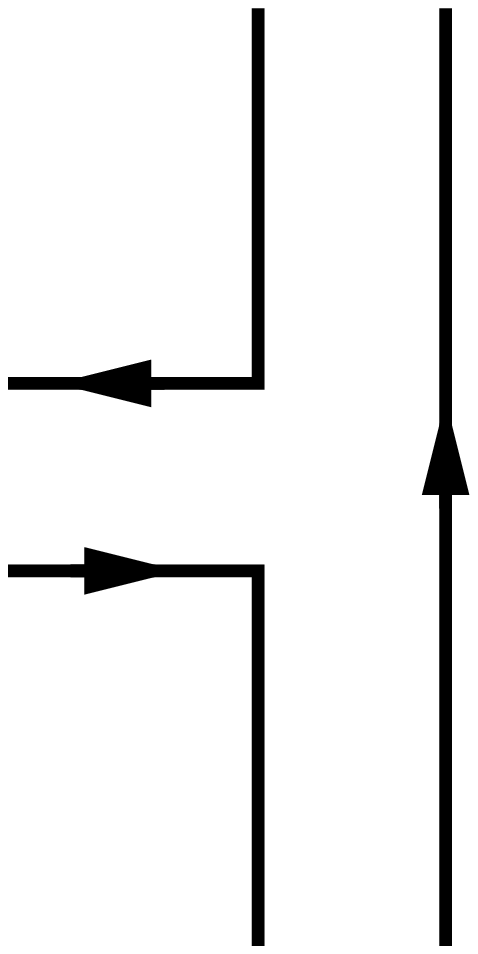}}
\big) \cdot
\big(
     \parbox{.5cm}{\includegraphics[width=.5cm]{structure2.eps}} 
     -
     \parbox{.5cm}{\includegraphics[width=.5cm]{structure1.eps}}
\big)
 K_{2 \to 2}({\bf l}_1, {\bf l }_2;{\bf k}_1, {\bf k}_2), 
\end{align}
into the Born diagrams of Fig.\ref{fig:born}, with $K_{2 \to 2}$ given
by Eq.(\ref{eq:22kernel_momis}). To be definite, we consider corrections 
to the first  diagram of Fig.\ref{fig:born}; the other Born diagrams are treated in the same way. 
We start with the case where
the interaction is between gluons which end up in the same lower
quark-loop, i.e. the interaction is inside the gluon pairs (12) or
(34). In the case of the gluon pair (12), the two combinations of
color factors of Fig.\ref{fig:same_ql} fit on the pair-of-pants,
similar to the cylinder discussed before. An analogous result holds for the 
gluon pair (34). If the
interacting $t$-channel gluons end up in different quark-loops we need
to distinguish between two different cases. In the first 
case, shown in Fig.\ref{fig:diff_ql_nearby}, the interaction is between 
gluons '2' and '3'; on the surface of the upper cylinder the two gluons are 
neighboring.
We call this interaction planar: by contracting the color loop above the rung between gluon '2' and '3',
we are back to the first Born diagram in Fig.\ref{fig:born}.  
In the second case, shown in Fig.\ref{fig:diff_ql_no_nearby}, the 
interaction is between gluon '2' and '4', These two gluons are  
not neighboring, and although the color diagram fits onto the surface of the 
pair-of-pants without any intersection of color lines, it will be referred to  
as 'non-planar'. It cannot be reduced to the Born diagram in Fig.\ref{fig:born}.
Counting closed color loops, it is of the same order as 
that of Fig.\ref{fig:diff_ql_nearby}. 
Note, however, that compared to the planar one it has a relative minus sign.  
The same discussion applies if the interaction is between gluon '1' and '3'.

\begin{figure}[htbp]
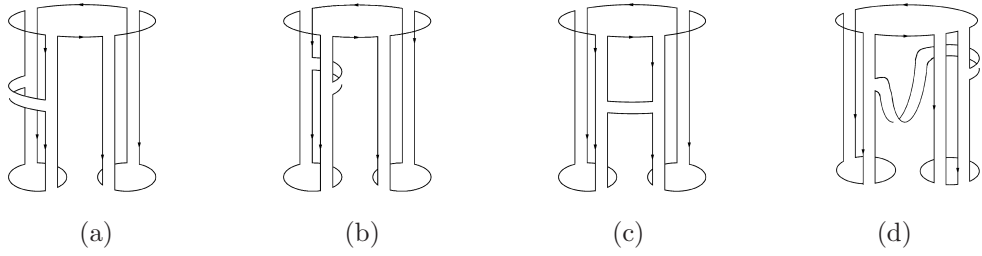

  \centering
     \parbox{3.5cm}{  \center \includegraphics[height=2.5cm]{trouser_inta12_cyl_voll.eps2}} 
     \parbox{3.5cm}{\center \includegraphics[height=2.5cm]{trouser_inta.eps2}} 
     \parbox{3.5cm}{\center \includegraphics[height=2.5cm]{trouser_inta_cyl_voll12.eps2}} 
  \parbox{3.5cm}{\center \includegraphics[height=2.5cm]{trans22_alt2.eps2}}   
\\
     \parbox{3.5cm}{\center (a)}\parbox{3.5cm}{\center (b)}\parbox{3.5cm}{\center (c)}\parbox{3.5cm}{\center (d)}
  \caption{\small  Different structures due to insertion of a two-to-two kernel on the pair-of-pants.}
  \label{fig:differentstructures}
\end{figure}

We summarize the four possibilities in Fig.\ref{fig:differentstructures}.
Apart form the last diagram, all diagrams belong to the class of
'planar' graphs.  Note, however, that while in the first graph we can
contract the closed color loop either above or below, in the second
graph we can contract only the lower loop on the lower cylinder: it is
planar w.r.t. the lower left cylinder.

When inserting more two-to-two interactions, it is useful to descend
from the top to the bottom of the diagram: we start by inserting
$s$-channel gluons between $t$-channel gluons which, on the upper
cylinder, are neighboring (Figs.\ref{fig:differentstructures}a and c).
This generates planar graphs.  Moving further down, these planar
insertions come to a stop as soon as one of the
following interactions is included:\\
(i) either an interaction of the type
Fig.\ref{fig:differentstructures}b which is still planar but belongs
to the lower left cylinder. Below such an interactions, further
interactions lie on the surface of one of the two lower cylinders,
i.e. they are inside the pairs $(12)$ or $(34)$. An interaction
between the two cylinders, e.g., between gluon '1' and '3',
loses a color factor $N_c$ and does not contribute on the pair-of-pants.\\
(ii) alternatively, a non-planar interaction of the type
Fig.\ref{fig:differentstructures}: This interaction occurs at most
once, and any further interaction below, again, lies on the surface of
one of the two lower cylinders and hence is inside the pairs $(12)$ or
$(34)$.

\subsection{Two-to-four Reggeon transition }
\label{sec:amplitudes-with-24}
\begin{figure}[htbp]
  \centering
  \parbox{6cm}{\includegraphics[width=4cm]{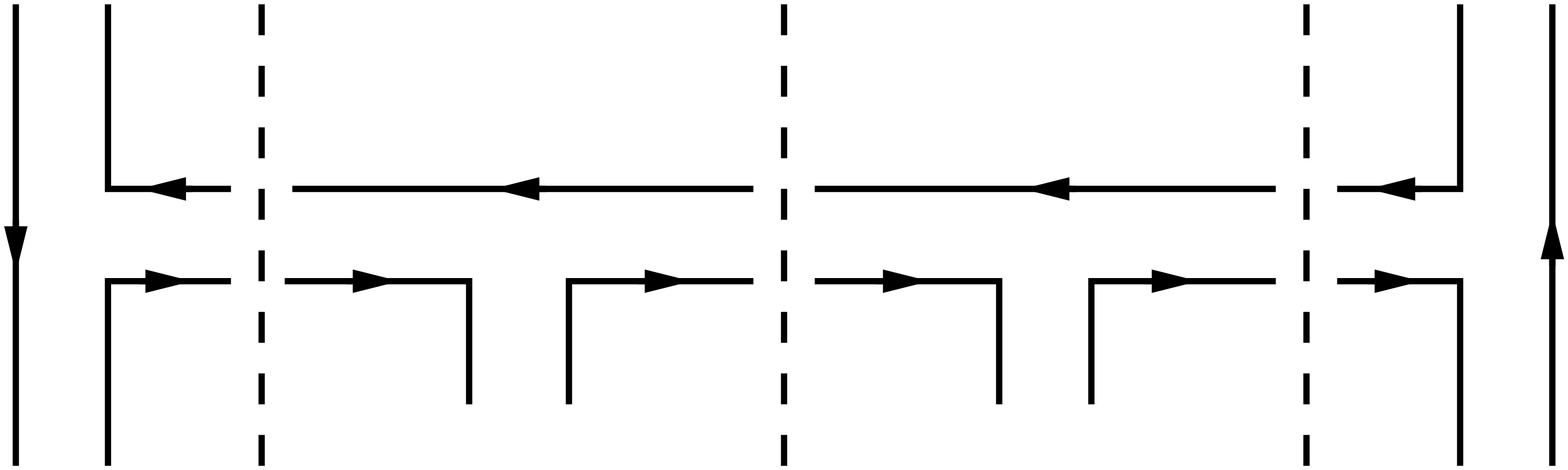}} \parbox{4cm}{    \includegraphics[height=2.2cm]{trouser_inta_cyl24.eps2}} \\
 \parbox{6cm}{\includegraphics[width=4cm]{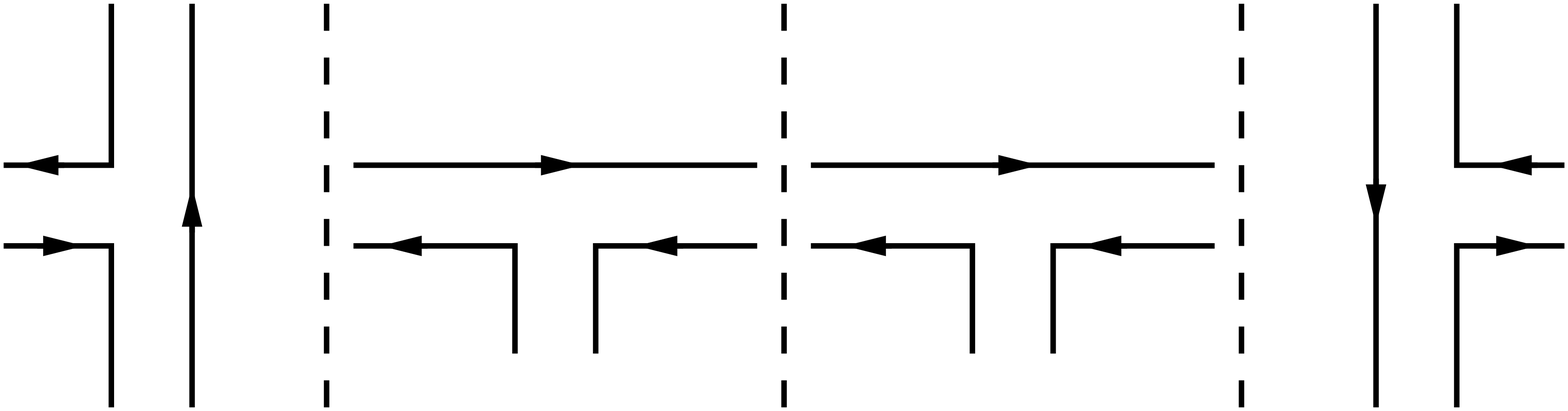}} \parbox{4cm}{    \includegraphics[height=2.2cm]{trouser_inta_cyl_24b.eps2}}
  \caption{\small The planar two-to-four Reggeon transition.}
  \label{fig:two4_1}
\end{figure}

\begin{figure}[htbp]
  \centering \parbox{6cm}{\includegraphics[width=4cm]{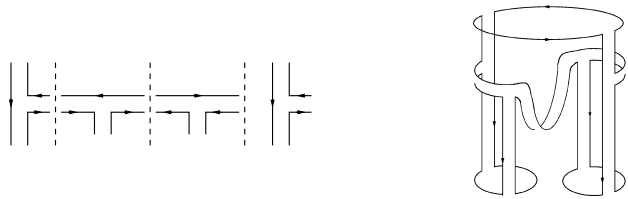}}
  \parbox{4cm}{
    \includegraphics[width=2cm]{trans24_alt2.eps2}}
  \caption{\small The non-planar two-to-four Reggeon transition.}
  \label{fig:two4_2}
\end{figure}

Apart from the above examples we have further the possibility that some of
the $t$-channel gluons do not start at the upper quark loop but emerge
from produced $s$-channel gluons. We start with the transition from
two to four $t$-channel gluons. In this case we start, at the upper
quark loop, with two $t$-channel gluons which end at a two-to-four
transition. Leaving for the moment details on the momentum structure
of the transition kernel $K_{2\to 4}$ aside, the transition from two
to four reggeized gluons is given by
\begin{align}
  \label{eq:kernel24}
\parbox{2cm}{\includegraphics[height=1cm]{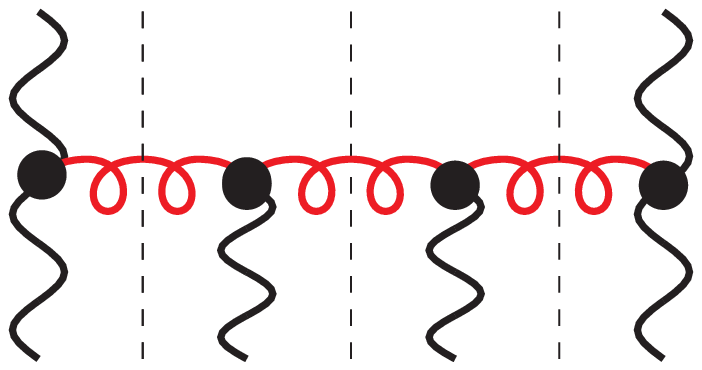}}  
 = {\bar g}^4
 \bigg(
       \parbox{.5cm}{\includegraphics[width=.5cm]{structure1.eps}}
       -
       \parbox{.5cm}{\includegraphics[width=.5cm]{structure2.eps}}
\bigg)
\bigg(
      \parbox{1.1cm}{\includegraphics[height=.5cm]{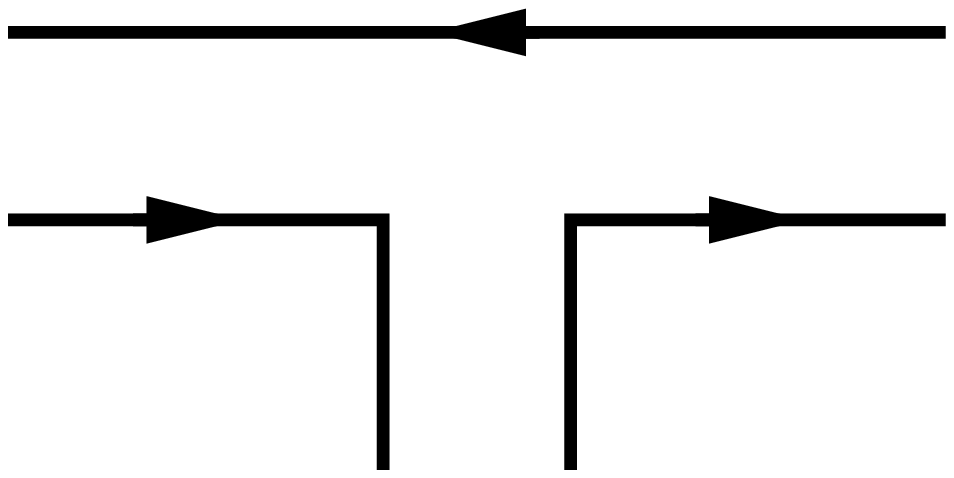}} 
       -
       \parbox{1.1cm}{\includegraphics[height=.5cm]{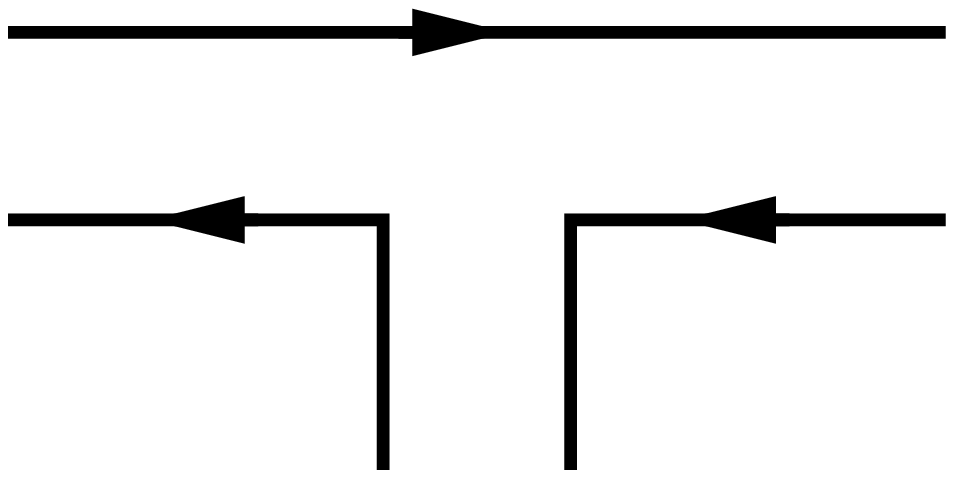}}
\bigg)
\bigg(
      \parbox{1.1cm}{\includegraphics[height=.5cm]{structure4.eps}} 
       -
       \parbox{1.1cm}{\includegraphics[height=.5cm]{structure3.eps}}
\bigg)
\bigg(
      \parbox{.5cm}{\includegraphics[width=.5cm]{structure2.eps}}
      -
      \parbox{.5cm}{\includegraphics[width=.5cm]{structure1.eps}}
\bigg)
K_{2 \to 4}.
\end{align}
Of the 16 possible combinations of color factors, only a subset fits
on the pair-of-pants, of which two examples are shown in
Figs.\ref{fig:two4_1} and \ref{fig:two4_2}.
Again, we have two distinct structures: Fig.\ref{fig:two4_1}, by
contracting the closed color loop on the upper cylinder, can be
reduced to one of the Born diagrams, and thus belongs to class of
planar diagrams. In contrast, Fig.\ref{fig:two4_2} cannot be
contracted and is non-planar.

Further rungs above the $2 \to 4$ transition vertex can always be
contracted to either Figs.\ref{fig:two4_1} or \ref{fig:two4_2}. As to
the interactions below the $2 \to 4$ transition, we have to
distinguish between the two classes.  For the planar class in
Figs.\ref{fig:two4_1}, we can continue by inserting $2\to2$
interactions as described in the previous subsection. For the
non-planar class in Fig.\ref{fig:two4_2} any further $2\to2$
interaction is either inside the pair $(12)$ or $(34)$ and can always
be reduced to Fig.\ref{fig:two4_2}.

For the momentum structure of the $2 \to 4$ transition vertex we need,
apart from the real gluon production vertex in
Eq.(\ref{eq:lipatov_factor}), the vertex which describes coupling of a
$t$-channel gluon to a real $s$-channel gluon, which is known as the
Reggeon-Particle-Particle (RPP)-vertex. Similar to the production
vertex, the RPP-vertex is an effective vertex. To lowest order, for
the scattering of a gluon on an antiquark at high center of mass
energies, this RPP-vertex is built up from the following diagrams:
\begin{align}
  \label{eq:1diag_rpr}
\parbox{1.5cm}{\includegraphics[width=1.5cm]{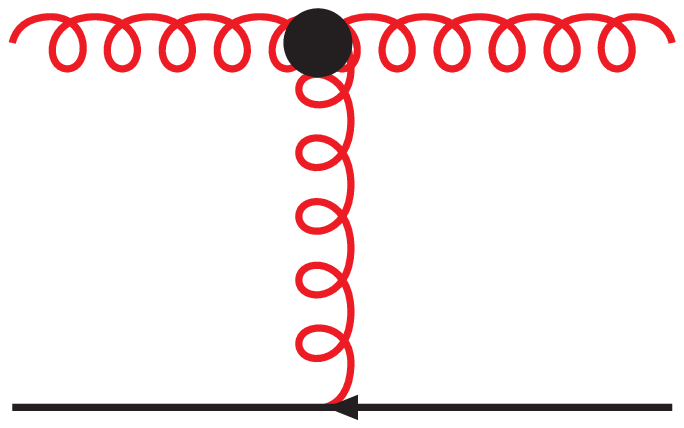}}
=
\parbox{1.5cm}{\includegraphics[width=1.5cm]{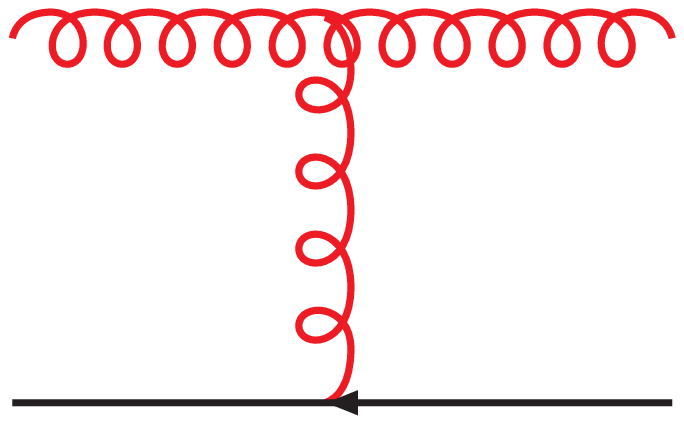}}
+
\parbox{1.5cm}{\includegraphics[width=1.5cm]{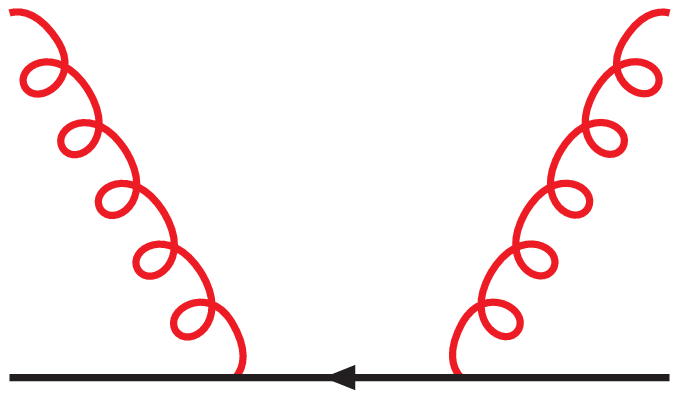}}
+
\parbox{1.5cm}{\includegraphics[width=1.5cm]{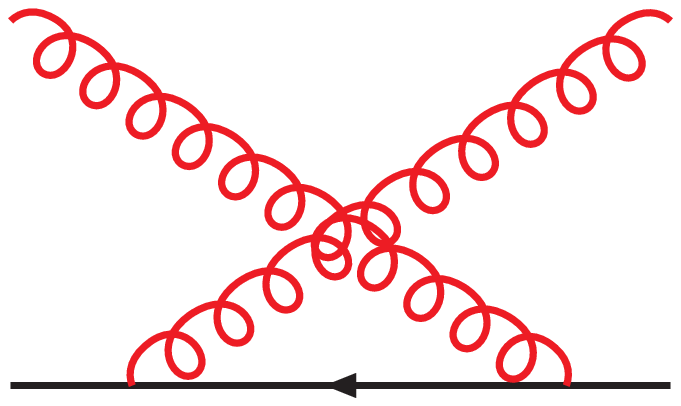}.}
\end{align}
Similar to the production vertex, at high center of mass energies the
last two diagrams coincide with each other up to a sign, and color
factor and momentum part of the RPP-vertex can be written in the 
factorized form
\begin{align}
  \label{eq:colorrpp_abstract}
-i{\bar{g}} \left[
\parbox{1.5cm}{\includegraphics[width=1.5cm]{structure4.eps}} 
-
 \parbox{1.5cm}{\includegraphics[width=1.5cm]{structure3.eps}}
\right] \epsilon^*_{(\lambda')} \!\!\cdot\!\Gamma\!\cdot\!\epsilon_{(\lambda)}.
\end{align}
Sandwiching one or two RPP vertices between two production vertices
then leads to two-to-three and two-to-four Reggeon transition kernels.
They have been constructed in \cite{Bartels:1980pe }  and for details we refer to
this reference.  The momentum part of the the $2\to 4$ transition is
then given by
\begin{align}
  \label{eq:def_k24}
K_{2 \to 4} =  {\bf q}^2
   &-
   \frac{{\bf l}_1^2 ({\bf q} - {\bf k}_1)^2}{ ({\bf l}_1 - {\bf k}_1)^2 } 
   -
   \frac{{\bf l}_2^2 ({\bf k}_1 + {\bf k}_2 + {\bf k}_3)^2}{ ({\bf l}_2 - {\bf k}_4)^2 }
   +
   \frac{{\bf l}_1^2 ({\bf q}  - {\bf k}_1)^2 {\bf l}^2_2 }{ ({\bf l}_1 - {\bf k}_1)^2 ({\bf l}_2 - {\bf k}_4)^2 }.
\end{align}

\subsection{Two-to-three Reggeon transition}
\label{sec:amplitudes-with-234}
Last we need to consider diagrams that contain, at least, one
transition from two to three gluons. Beginning with the momentum
structure, we remind that the coupling of an additional 'new'
$t$-channel to an $s$-channel gluon takes place by the RPP-vertex in
Eq.(\ref{eq:colorrpp_abstract}). The result for two-to-three Reggeon
transition is described by
\begin{align}
  \label{eq:kernel23}
\parbox{2cm}{\includegraphics[height=1cm]{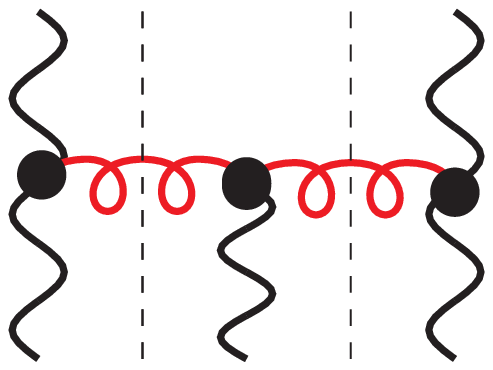}} 
 =
i\bar{g}^3 \bigg(
       \parbox{.5cm}{\includegraphics[width=.5cm]{structure1.eps}}
       -
       \parbox{.5cm}{\includegraphics[width=.5cm]{structure2.eps}}
\bigg)
\bigg(
      \parbox{1.1cm}{\includegraphics[height=.5cm]{structure4.eps}} 
       -
       \parbox{1.1cm}{\includegraphics[height=.5cm]{structure3.eps}}
\bigg)
\bigg(
      \parbox{.5cm}{\includegraphics[width=.5cm]{structure2.eps}}
      -
      \parbox{.5cm}{\includegraphics[width=.5cm]{structure1.eps}}
\bigg)
K_{2 \to 3}
\end{align}
with
\begin{align}
  \label{eq:k23_def}
 K_{2 \to 3}^{(\{12\} \to \{123\})} =   ( {\bf k}_1 +{\bf k}_2 + {\bf k}_3 )^2
         &-
         \frac{{\bf l}_1^2 ({\bf k}_2 + {\bf k}_3)^2}{ ({\bf l}_1 - {\bf k}_1)^2 } 
         -
         \frac{{\bf l}_2^2 ({\bf k}_1 + {\bf k}_2)^2}{ ({\bf l}_2 - {\bf k}_3)^2 }
         +
         \frac{{\bf l}_1^2 {\bf k}_2^2 {\bf l}^2_2 }{ ({\bf l}_1 - {\bf k}_1)^2 ({\bf l}_2 - {\bf k}_3)^2 } ,
\end{align}
where the superscripts on the kernel refer to the ingoing and outgoing
$t$-channel gluons respectively. As to the general structure of diagrams 
with $2 \to 3$ transitions, we have two possibilities: either we start, 
at the upper quark loop, 
with three gluons. In this case we need, somewhere further below, only one $2 \to 3$ transition.
Alternatively, we could start with two gluons and then need two $2 \to 3$ transition vertices.     

\begin{figure}[htbp]
  \centering
\parbox{5cm}{\includegraphics[width=3cm]{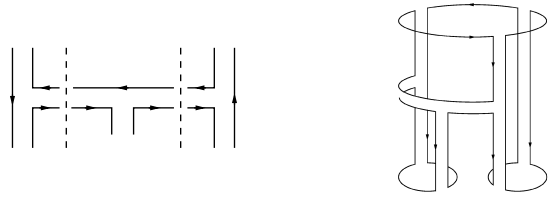}}  \parbox{4cm}{\includegraphics[height=2.5cm]{trouser_inta_cyl323.eps2}}\\
  \caption{\small Planar graph with one two-to-three transition.}
  \label{fig:two3_1}
\end{figure}

\begin{figure}[htbp]
  \centering
  \parbox{5cm}{\includegraphics[width=3cm]{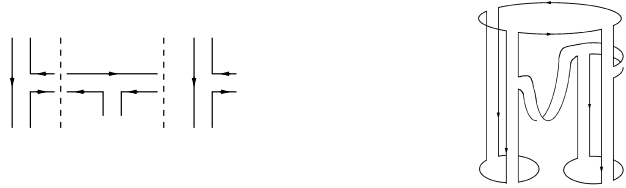}} 
  \parbox{4cm}{  \center \includegraphics[height=2.5cm]{trans23_alt2.eps2}}
  \caption{\small Non-planar graph with one two-to-three transition.}
  \label{fig:two3_2}
\end{figure}

For the discussion of the color diagrams we start with the former case. 
As before we encounter planar and non-planar graphs. 
An example of a planar graph is shown in Fig.\ref{fig:two3_1}, 
a non-planar graph in Fig.\ref{fig:two3_2}. In both cases, the planar class, Fig.\ref{fig:two3_1},
 and the non-planar class, Fig.\ref{fig:two3_2},
above the vertex any further $2\to2$ interaction is between lines which
are neighbored on the surface of the upper cylinder. Below the $2\to3$ vertex 
we have the same situation as for the $2\to4$ vertex: for the planar graph 
in  Fig.\ref{fig:two3_1} we proceed as described in Sec.\ref{sec:amplitudes-with-4}, whereas for 
the non-planar graph there is no further communication between the two lower 
cylinders.     

\begin{figure}[htbp]
  \centering
  \parbox{5cm}{\includegraphics[width=4cm]{two3_a.eps}}
  \parbox{5cm}{  \center \includegraphics[height=3cm]{trouser_inta_cyl323_corr2.eps2}} \\
  \parbox{5cm}{\includegraphics[width=4cm]{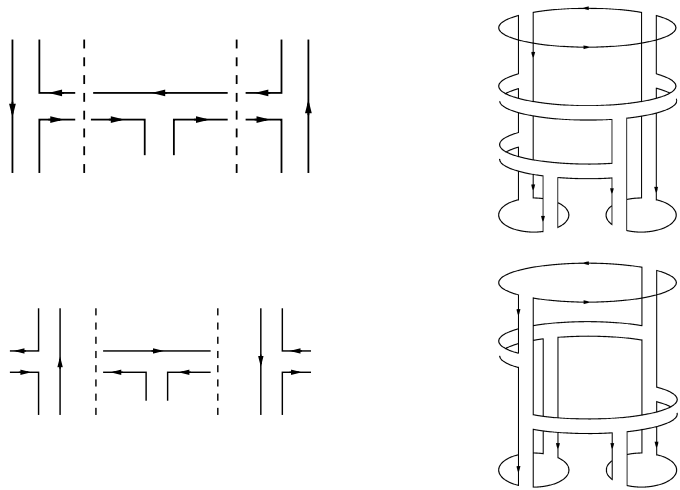}}
  \parbox{5cm}{  \center \includegraphics[height=3cm]{trouser_inta_cyl323_corr4.eps2}}
  \caption{\small Combination of two transitions of two to three $t$-channel gluons.     From Eq.(\ref{eq:kernel23}) one finds a relative minus sign for
   the second combination.}
  \label{fig:two3_4}
\end{figure}

For the graphs which, at the top, start with two $t$-channel gluons 
the two classes arise in the following way. Beginning at the top, the first 
$2\to3$ transition always lies on the surface of the upper cylinder.
Below this vertex we have the same situation which we have described 
a moment ago, i.e. the same as for the case where at the fermion quark loop 
we start with three gluons: the second $2\to3$ transition decides whether 
the graph is planar or non-planar. Planar examples are shown in 
Fig.\ref{fig:two3_4}.

\subsection{Planar and non-planar partial waves}
\label{sec:below}
In this final subsection we collect the results and prepare the 
summation of all diagrams on the pair-of-pants surface.
At the end of Sec.\ref{sec:sixpoint} we proposed to factorize
at the branching point, i.e. at the last interaction between the two 'legs' 
of the pair-of-pants,  the partial wave $F(\omega, \omega_1, \omega_2)$ 
into a convolution of three amplitudes $\mathcal{D}_4(\omega)$,
$\mathcal{D}_2(\omega_1)$, and $\mathcal{D}_2(\omega_2)$,
depending on $\omega$,
$\omega_1$, or  $\omega_2$, resp.. In order to find closed expressions 
for these amplitudes we shall, similar to our treatment of the diagrams in 
the plane and on the surface of the cylinder,
formulate integral equations which sum up the different classes of 
diagrams. 

The situation is easiest for the two
amplitudes that start from the two lower virtual photons,
$\mathcal{D}_2(\omega_1)$, and $\mathcal{D}_2(\omega_2)$. For these
amplitudes we need to resum, for both legs, contributions like those in
Fig.\ref{fig:same_ql}. Their resummation yields the BFKL-equation on the
cylinder Eq.(\ref{eq:bfkl-singlet}).  

The upper amplitude, $\mathcal{D}_4(\omega)$ is defined to include the
branching vertex, which by definition is the lowest interaction that
connects the gluon pair (12) with the pair (34). Below this vertex,
there are only interactions inside the pair $(12)$ or $(34)$. The
branching vertex can be either a $2\to 2$, a $2\to3$, or a $2\to4$
transition vertex.  From this definition it follows that the upper
amplitude always has four reggeized gluons at its lower end.

From now on it becomes important to distinguish between 'planar' and
'non-planar' graphs.
\begin{figure}[htbp]
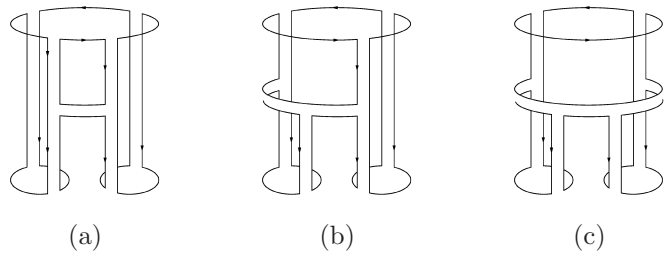

  \centering
  \parbox{3.2cm}{\center \includegraphics[height=2.5cm]{trouser_inta_cyl_voll12.eps2}} 
  \parbox{3.2cm}{\center \includegraphics[height=2.5cm]{trouser_inta_cyl323.eps2}}
  \parbox{3.2cm}{ \center   \includegraphics[height=2.5cm]{trouser_inta_cyl24.eps2}} \\
\parbox{3.2cm}{\center (a)}
\parbox{3.2cm}{\center (b)}
\parbox{3.2cm}{\center (c)}
  \caption{\small 'Planar' color factors on the pair-of-pants which reduce in an apparent way to the color structure of the Born-term by extracting a closed color loop with the \emph{upper} quark-loop.}
  \label{fig:planar_omega}
\end{figure}
\begin{figure}[htbp]
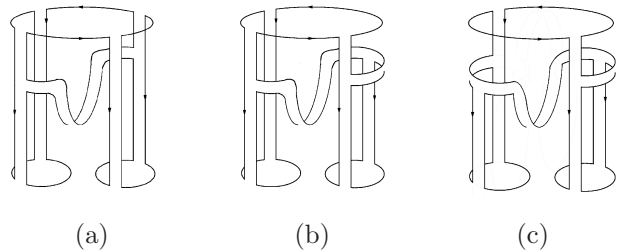

  \centering
  \parbox{2.9cm}{\center \includegraphics[height=2.5cm]{trans22_1pop_closed.eps2}}
  \parbox{2.9cm}{\center \includegraphics[height=2.5cm]{trans23_1pop_closed.eps2}}
  \parbox{2.9cm}{\center \includegraphics[height=2.5cm]{trans24_1pop_closed.eps2}}\\
\parbox{2.9cm}{\center (a)}\parbox{2.9cm}{\center (b)}\parbox{2.9cm}{\center (c)}
  \caption{\small 'Non-planar' color factors that cannot be reduced trivially to the color structure of the Born-term by extracting a closed color loop with the \emph{upper} quark-loop.}
  \label{fig:non_planar_omega}
\end{figure}
For convenience we list, once more, three examples of planar and non-planar 
graphs: Fig.\ref{fig:planar_omega} contains the planar example, and 
Fig.\ref{fig:non_planar_omega} the non-planar ones. 
They are of the order $g^8 N_c^3 (g^2 N_c)$.

Beginning with the planar diagrams in Fig.\ref{fig:planar_omega} 
one easily verifies that we always 
can contract one closed color loop to arrive at the first diagram of Fig.\ref{fig:born}. 
For the other diagrams in  Fig.\ref{fig:born} we have analogous sets of graphs.  
Next, the figures Fig.\ref{fig:planar_omega}a-c differ from each other in that, at the fermion loop at the top, 
in (a) we start with four gluons, in (b) with only three gluons, 
and in (c) with two gluons. 
Correspondingly, in (a) the branching vertex consists of a two-to-to kernel, 
in (b) a two-to-three kernel, and in (c)  a two-to-four kernel. 
At the branching vertex, all diagrams have four $t$-channel gluon 
lines. Higher order diagrams are obtained by inserting, above the
branching vertex, pairwise interactions between neighboring $t$-channel gluons. 
Also, in Fig.\ref{fig:planar_omega}b and d we could insert a $2\to2$ 
interaction between gluons '2' and '3' or '1' and '4': in this case, the branching vertex 
consists of a two-to-to kernel. In all cases,   
by drawing a horizontal cutting line just below the branching  
vertex, we arrive at amplitudes with four $t$-channel gluons.  
We will denote the sum of all graphs by $\mathcal{D}_4^{(1234)}(\omega)$,  
 $\mathcal{D}_4^{(2134)}(\omega)$, $\mathcal{D}_4^{(2143)}(\omega)$, 
and  $\mathcal{D}_4^{(2134)}(\omega)$ where 
the upper label refers to the four terms in Fig.\ref{fig:born}, 
i.e. it indicates to which of the four Born terms the diagrams can be 
contracted (a more precise definition will be given further below in section
\ref{sec:regg_ql}). 

For the task of summing of all diagrams it will be convenient to define also 
auxiliary amplitudes with two and three gluons in the $t$-channel. 
In Fig.\ref{fig:planar_omega}c
one sees that, starting at the upper quark loop, we begin with two 
gluons which, when higher order corrections are included, interact through 
two-to-two kernels. The sum of these 
diagrams leads to the BFKL cylinder and therefore coincides with 
$\mathcal{D}_2(\omega)$ of Sec.\ref{sec:cylinder}. 
\begin{figure}[htbp]
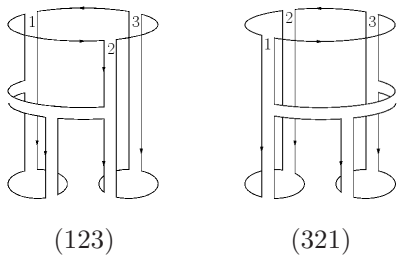

  \centering
  \parbox{3cm}{\center    \includegraphics[width=2cm]{born123.eps2} }
  \parbox{3cm}{\center    \includegraphics[width=2cm]{born321.eps2} }\\
  \parbox{3cm}{\center ${(123)}$ } 
  \parbox{3cm}{\center  ${(321)}$} 
  \caption{ \small The two possible orderings of three gluons along the quark-loop  which leads to the definition of $\mathcal{D}_{3}^{(123)}(\omega)$ and  $\mathcal{D}_{3}^{(321)}(\omega)$. }
  \label{fig:born_d3}
\end{figure}
For the three gluon 
amplitude above the branching vertex in Fig.\ref{fig:planar_omega}b 
we have two inequivalent couplings to the quark loops, as illustrated 
in Fig.\ref{fig:born_d3}.  
We denote them by $\mathcal{D}_3^{(123)}(\omega)$ and 
$\mathcal{D}_3^{(132)}(\omega)$.

When writing down integral equations for the amplitudes $\mathcal{D}_3$ and 
$\mathcal{D}_4$, we observe that they will be coupled. The three gluon state 
at the lower end of $\mathcal{D}_3$ can start as a two gluons at the quark 
loop, then undergo a two-to-three transition. Similarly, the four gluon state  
can start from two or three gluons and then make transitions to the final four 
gluon state. The formulation of these coupled integral equations 
will be carried out in the subsequent section. 

Next we turn to the non-planar diagrams of
Fig.\ref{fig:non_planar_omega}.  They cannot be reduced to the color
structure of the Born diagrams, and we therefore define an additional
partial wave that sums up these terms. They all have in common that the 
non-planar structure at the lower end of the upper cylinder is always 
the branching vertex: below this vertex, there is no further communication 
between the two lower cylinders. As an example, an interaction between 
gluon '2' and '3' would be subleading in powers $N_c$. Similar to the 
planar diagrams
in Fig.\ref{fig:planar_omega}, we see three different structures:
above the branching vertex we have four (Fig.\ref{fig:planar_omega}a),
three (Fig.\ref{fig:planar_omega}b), or two
(Fig.\ref{fig:planar_omega}c) $t$-channel gluons. This suggests that
above the branching vertex the structure is the same as for the planar
graphs, and we simply have to convolute the non-planar branching
vertex with the planar amplitudes $\mathcal{D}_2$, $\mathcal{D}_3$,
and $\mathcal{D}_4$.  The resulting expression will be denoted by
$\mathcal{D}_4^{(\text{NP})}(\omega)$.  Its derivation will be the
content of Sec.\ref{sec:trip_pom}.

\section{Integral equations: gluon amplitudes with planar color structure}
\label{sec:regg_ql}

In the following we will formulate integral equations for amplitudes
of the planar class with two, three and four $t$-channel gluons. The
amplitude with four $t$-channel gluons that belongs to the non-planar
class will be addressed in Sec.\ref{sec:trip_pom}.

\subsection{Integral equations for the three gluon amplitude}
\label{sec:inteq_d3}

The evolution of the three gluon state is described by the sum of
pairwise interactions, i.e. the two-to-two transition kernels acting
on the three gluon state
\begin{align}
  \label{eq:d3_22}
         \parbox{2cm}{\includegraphics[width=2cm]{ql3_k22_12.eps2}} 
         + 
         \parbox{2.3cm}{\includegraphics[width=2cm]{ql3_k22_23.eps2}} 
         +\,
         \parbox{2cm}{\includegraphics[width=2cm]{ql3_k22_31.eps2}} 
         \to 
          \bar{g}^2N_c \sum_{\substack{(12), (23) \\ (31)}} K_{2 \to 2}\otimes \mathcal{D}_3^{(123)}(\omega).
\end{align}
The transition from the two gluon state to the three gluon state is 
mediated by the two-to-three transition kernels acting on the two-gluon state,
\begin{align}
  \label{eq:d2_23}
 \parbox{2.5cm}{\includegraphics[width=2.5cm]{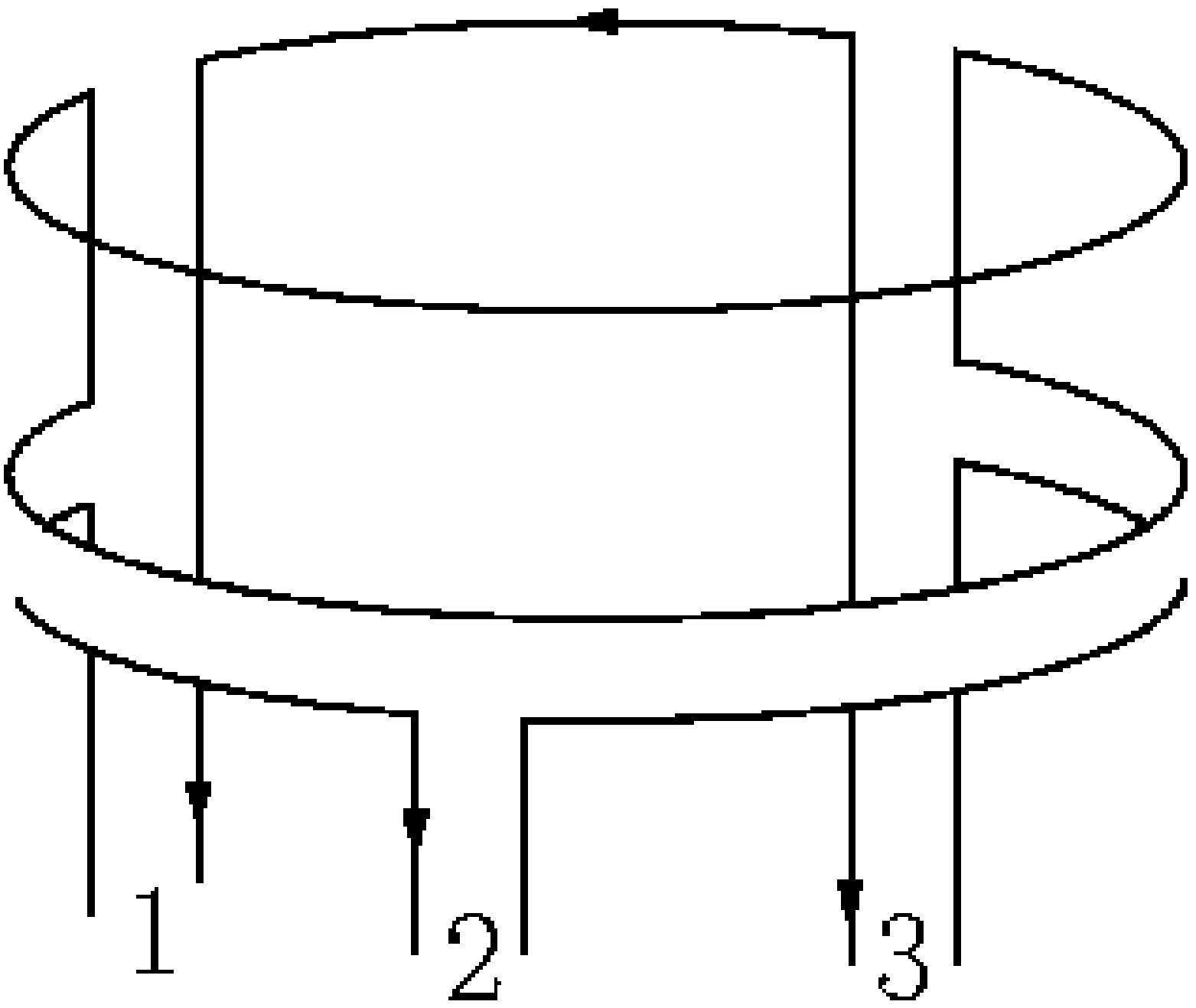}} \to  {\bar{g}^3}(-N_c) K_{2 \to 3}^{( \{12\} \to \{123\})} \otimes \mathcal{D}_2(\omega).
\end{align}
 Similar to $\mathcal{D}_2(\omega)$, also the
amplitude $\mathcal{D}_3(\omega)$ is defined to contain the Reggeon
propagator of the external reggeized gluons, $1/(\omega - \sum_i^3
\beta({\bf k}_i))$. With these ingredients, the integral equation for
$\mathcal{D}_{3}^{(123)}(\omega)$ is given by
\begin{align}
  \label{eq:bfkl-singlet3sincolor}
  \big(\omega - \sum_i^3 \beta({\bf{k}}_i)\big)
  \mathcal{D}_{3}^{(123)} (\omega| {\bf k}_1,{\bf k}_2,{\bf k}_3) =&
  \mathcal{D}^{(123)}_{(3,0)} + {\bar g}^3(-N_c) K^{( \{12\} \to
    \{123\})}_{2 \to 3}\otimes \mathcal{D}_2(\omega)
  \notag \\
  &+ \bar{g}^2N_c\sum K_{2 \to 2}\otimes
  \mathcal{D}^{(123)}_3(\omega).
\end{align}
In complete analogy the equation for
$\mathcal{D}_3^{(321)}(\omega)$ contains the two-to-three transition kernel
\begin{align}
  \label{eq:d2_23_backwards}
 \parbox{2.5cm}{\includegraphics[width=2.5cm]{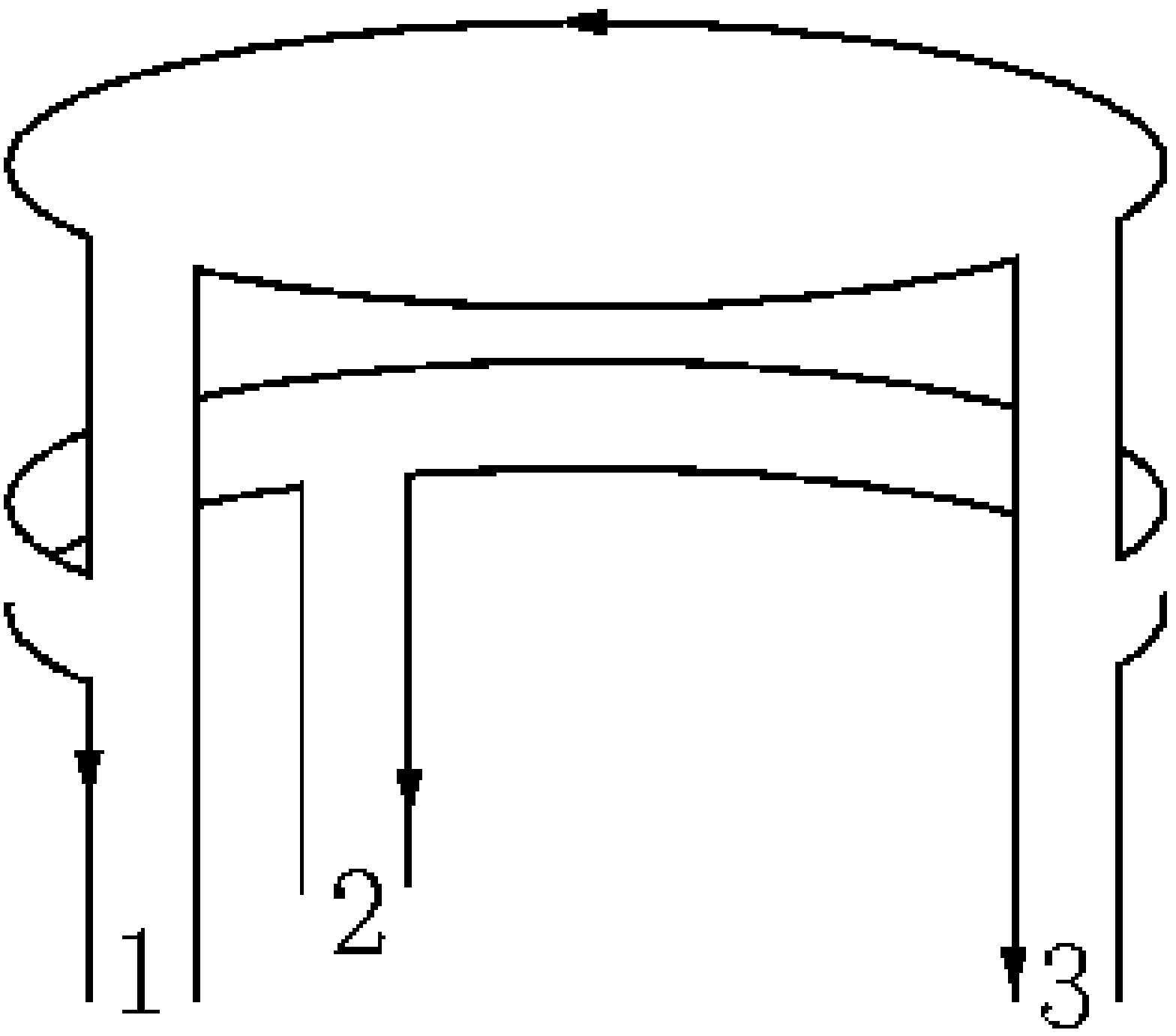}} \to {\bar g}^3N_c K_{2 \to 3}^{( \{12\} \to \{123\})} \otimes \mathcal{D}_2(\omega).
\end{align}
The integral equation is given by:
\begin{align}
  \label{eq:bfkl-singlet3sincolor_321}
\big((\omega - \sum_i^3 \beta({\bf{k}}_i)\big) \mathcal{D}_{3}^{(321)} (\omega| {\bf k}_1,{\bf k}_2,{\bf k}_3) =& 
 \mathcal{D}^{(321)}_{(3,0)}
+
 \bar{g}^3N_c K^{( \{12\} \to \{123\})}_{2 \to 3}\otimes \mathcal{D}_2(\omega)
\notag \\
&+
\bar{g}^2N_c \sum_{\substack{(13),(32) \\ (21)}}  K_{2 \to 2}\otimes \mathcal{D}^{(321)}_3(\omega)
.
\end{align}

\subsection{Reggeization of the three-gluon amplitude}
\label{sec:regg_d3}

To study the integral equations Eq.(\ref{eq:bfkl-singlet3sincolor})
and Eq.(\ref{eq:bfkl-singlet3sincolor_321}) we need the impact factors
$\mathcal{D}_{(3;0)}^{(123)}$ and $\mathcal{D}_{(3;0)}^{(321)}$.  Both
impact factors have the same properties as in the $N_c$ finite case
\cite{Bartels:1994jj}, and they can be written as a superposition of
two-gluon impact factors $\mathcal{D}_{(2;0)}$:
\begin{align}
  \label{eq:d30d20123}
\mathcal{D}_{(3;0)}^{(123)} ( {\bf k}_1,{\bf k}_2,{\bf k}_3) &= + \frac{{\bar g}}{2}  \big[\mathcal{D}_{(2;0)}(12,3) - \mathcal{D}_{(2;0)}(13,2) + \mathcal{D}_{(2;0)} (1,23) \big] \\
\label{eq:d30d20321}
D_{(3;0)}^{(321)} ( {\bf k}_1,{\bf k}_2,{\bf k}_3) &= -\frac{{\bar g}}{2}  \big[D_{(2;0)}(12,3) - D_{(2;0)}(13,2) + D_{(2;0)} (1,23) \big].
\end{align}
  On the right
hand side we introduced a short-hand notation: a string of numbers
stands for the sum of the corresponding transverse momenta, for
instance
\begin{align}
  \label{eq:shorthand1}
\mathcal{D}_{(2;0)}(12,3) \equiv \mathcal{D}_{(2;0)}( {\bf k}_1 +{\bf k}_2,{\bf k}_3).
\end{align}
We will make use of this notation in the following, whenever it proves
useful to clarify the underlying structure of complex expressions.  It
can be now demonstrated that the structure on the rhs of
Eqs.(\ref{eq:d30d20123}] and Eqs.(\ref{eq:d30d20321} also holds for
the solutions of the integral equations
Eq.(\ref{eq:bfkl-singlet3sincolor}) and
Eq.(\ref{eq:bfkl-singlet3sincolor_321}):
\begin{align}
  \label{eq:d3regge123}
  \mathcal{D}_{3}^{(123)} (\omega | {\bf k}_1,{\bf k}_2,{\bf k}_3) = 
+ \frac{{\bar g}}{2}   \big[\mathcal{D}_{2}(\omega|12,3) - \mathcal{D}_{2}(\omega|13,2) + \mathcal{D}_{2} (\omega| 1,23) \big] \\
\mathcal{D}_{3}^{(321)} (\omega| {\bf k}_1,{\bf k}_2,{\bf k}_3) =
 - \frac{{\bar g}}{2}  
\big[\mathcal{D}_{2}(\omega|12,3) - \mathcal{D}_{2}(\omega|13,2) + \mathcal{D}_{2} (\omega|1,23) \big].
\end{align}
Obviously this solution shares important properties with the
reggeization of the planar gluon discussed in
Sec.\ref{sec:reggeization}: in each term on the rhs, two gluons
'collapse' into a single gluon which, at the end, decays into two
gluons. This allows also to deform the color factor as shown in
Fig.\ref{fig:deform_D3}:
\begin{figure}[htbp]
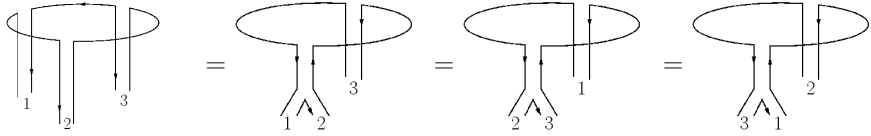

  \centering
     \parbox{2.5cm}{ \includegraphics[height=1.7cm]{ql3a_num.eps2}} = 
     \parbox{2.5cm}{ \includegraphics[height=1.7cm]{regg3_12.eps2}} =
     \parbox{2.5cm}{ \includegraphics[height=1.7cm]{regg3_23.eps2}} =
     \parbox{2.5cm}{ \includegraphics[height=1.7cm]{regg3b_13.eps2}} 
  \caption{ \small The color factor  associated with $\mathcal{D}_3^{(123)}(\omega)$  can be deformed such that the form of the color factor  coincides with the  momentum structure of the solution in terms of $\mathcal{D}_2$. A similar observation holds for  $\mathcal{D}_3^{(321)}(\omega)$. }
  \label{fig:deform_D3}
\end{figure}

\subsection{The integral equations for the four gluon amplitude with planar  color structure}
\label{sec:inteq_d4}

For the four gluon amplitude, $\mathcal{D}_4^{(1234)}(\omega)$, the
evolution of the four gluon state is describes by the sum of pairwise
interactions between neighboring gluons. For the amplitude
$\mathcal{D}_4^{(1234)}(\omega)$ the sum of the two-to-two transition
is given by
\begin{align}
  \label{eq:22_d4_1234}
\parbox{2cm}{ \includegraphics[width=2cm]{ql1234_k22_12.eps2}}
+
\parbox{2cm}{ \includegraphics[width=2cm]{ql1234_k22_23.eps2}}
+
\parbox{2cm}{ \includegraphics[width=2cm]{ql1234_k22_34.eps2}}
+
\parbox{2cm}{ \includegraphics[width=2cm]{ql1234_k22_41.eps2}}
\to  \bar{g}^2N_c \sum_{\substack{(12), (23) \\ (34), (41)}} K_{2 \to 2} 
\otimes \mathcal{D}_4^{(1234)}(\omega).
\end{align}
The corresponding expressions for $\mathcal{D}_4^{(2134)}(\omega)$,
$\mathcal{D}_4^{(2143)}(\omega)$ and $\mathcal{D}_4^{(1243)}(\omega)$
are easily obtained by simply exchanging the corresponding labels on
the $t$-channel gluons. For $\mathcal{D}_4^{(1234)}(\omega)$, the
two-to-three transitions are
\begin{align}
  \label{eq:23_d4_1234}
\parbox{2cm}{  \includegraphics[width=2cm]{trouser3a_inta1234.eps2}}  
+
\parbox{2cm}{  \includegraphics[width=2cm]{trouser3b_inta1234.eps2}}
    \quad \to
 \begin{array}{l}
    \bar{g}^3N_c  K_{2 \to 3}^{( \{12\} \to \{123\})} \otimes D_3^{(123)}(\omega) \\ \\
    + \quad \bar{g}^3 N_c 
    K_{2 \to 3}^{( \{23\} \to \{234\})} \otimes D_3^{(123)}(\omega),
\end{array}
\end{align}
while the two-to-four  transition is
\begin{align}
  \label{eq:24_d4_1234}
\parbox{2cm}{ \includegraphics[width=2cm]{two4_1_num.eps2}}
\quad \to
\bar{g}^4{N_c} K_{2 \to 4}^{( \{12\} \to \{1234\})} \otimes \mathcal{D}_2(\omega).
\end{align}
Also $\mathcal{D}_4^{(1234)}(\omega)$ is defined to contain the
Reggeon-propagator of the external reggeized gluons. Including this
contribution together with the four gluon impact factor,
$\mathcal{D}_{(4;0)}^{(1234)}$, the integral equation for
$\mathcal{D}_4^{(1234)}(\omega)$ is given by
\begin{align}
  \label{eq:int_eq_d4_1234}
 (\omega - \sum_i^4 \beta({\bf k}_i))  \mathcal{D}_4^{(1234)}(\omega) &= 
          \mathcal{D}^{(1234)}_{(4;0)}  
          +
          {\bar g}^3N_c\bigg[ K_{2 \to 3}^{( \{12\} \to \{123\})} \!\!\otimes\! \mathcal{D}_3^{(123)}(\omega) 
          +
          K_{2 \to 3}^{( \{23\} \to \{234\})}\!\! \otimes\! \mathcal{D}_3^{(123)}(\omega)\bigg]
          \notag \\
          &
          +            
          {\bar{g}^4}{N_c}^2  K_{2 \to 4}^{( \{12\} \to \{1234\})} \otimes \mathcal{D}_2(\omega)
          +
          \bar{g}^2N_c\sum_{\substack{(12), (23) \\ (34), (41)}} K_{2 \to 2} \otimes \mathcal{D}_4^{(1234)}(\omega).
\end{align}
For $\mathcal{D}_4^{(2143)}(\omega)$ the two-to-three transition arises from
\begin{align}
  \label{eq:23_d4_4321}
\parbox{2cm}{  \includegraphics[width=2cm]{trouser3a_inta2143.eps2}}  
+
\parbox{2cm}{  \includegraphics[width=2cm]{trouser3b_inta2143.eps2}}
  \quad  \to
    \begin{array}{l}
\\ {\bar g}^3 (-N_c)  K_{2 \to 3}^{( \{12\} \to \{123\})} \otimes \mathcal{D}_3^{(321)}(\omega) 
   \\ \\   + \quad 
    {\bar g}^3 (-N_c)      K_{2 \to 3}^{( \{23\} \to \{234\})} \otimes \mathcal{D}_3^{(321)}(\omega),
    \end{array}
   \end{align}
and the two-to-four transition comes from
\begin{align}
  \label{eq:24_d4_4321}
\parbox{2cm}{ \includegraphics[width=2cm]{two4_2_num.eps2}}
\quad \to
{ \bar{g}^4}{N_c} K_{2 \to 4}^{( \{12\} \to \{1234\})} \otimes \mathcal{D}_2(\omega).
\end{align}
We obtain the integral equation
\begin{align}
  \label{eq:int_eq_d4_4321}
(\omega - \sum_i^4 \beta({\bf k}_i))  \mathcal{D}_4^{(2143)}(\omega) &= 
          \mathcal{D}^{(2143)}_{(4;0)}  
          +
          {\bar g}^3 (\!-\!N_c) \bigg[ K_{2 \to 3}^{( \{12\}\! \to\! \{123\})} \!\!\otimes\! \mathcal{D}_3^{(321)}(\omega) 
          +
          K_{2 \to 3}^{( \{23\}\! \to\! \{234\})}\!\! \otimes\! \mathcal{D}_3^{(321)}(\omega)\bigg]
          \notag \\
          &
          +            
          {\bar{g}^4}{N_c}^2 K_{2 \to 4}^{( \{12\} \to \{1234\})} \otimes \mathcal{D}_2(\omega)
          +
          \bar{g}^2\sum_{\substack{(21), (14) \\ (43) , (32)}} K_{2 \to 2} \otimes \mathcal{D}_4^{(2143)}(\omega)
\end{align}
Finally, for $\mathcal{D}_4^{(2134)}(\omega)$ and
$\mathcal{D}_4^{(1243)}(\omega)$ the two-to-three transition are
\begin{align}
  \label{eq:23_d4_2134}
\parbox{2cm}{  \includegraphics[width=2cm]{trouser3b_inta2134.eps2}}  
+
\parbox{2cm}{  \includegraphics[width=2cm]{trouser3a_inta2134.eps2}}
    \quad  \to
\begin{array}{l} 
     \bar{g}^3(-N_c) 
      K_{2 \to 3}^{( \{13\} \to \{124\})} \otimes 
\mathcal{D}_3^{(123)}(\omega)  \\ \\    
+ \quad
   {\bar g}^3N_c
     K_{2 \to 3}^{( \{13\} \to \{134\})} \otimes \mathcal{D}_3^{(321)}(\omega)
\end{array}
\end{align}
and
\begin{align}
  \label{eq:23_d4_4312}
\parbox{2cm}{  \includegraphics[width=2cm]{three4_num124b.eps2}}  
+
\parbox{2cm}{  \includegraphics[width=2cm]{three4_num134a.eps2}}
   \quad \to
\begin{array}{l}
    {\bar g}^3N_c 
    K_{2 \to 3}^{( \{13\} \to \{124\})} \otimes \mathcal{D}_3^{(321)}(\omega) 
     \\ \\
     + \quad
    {\bar g}^3 (-N_c) 
    K_{2 \to 3}^{( \{13\} \to \{134\})} \otimes \mathcal{D}_3^{(123)}(\omega)
\end{array}
\end{align}
respectively, while two-to-four transition are absent in
that case. The integral equations are given by
\begin{align}
 (\omega - \sum_i^4 \beta({\bf k}_i))  \mathcal{D}_4^{(2134)}(\omega) &= 
          \mathcal{D}^{(2134)}_{(4;0)}  
          + {\bar g}^3 
         (-N_c) K_{2 \to 3}^{( \{13\} \to \{124\})} \!\!\otimes\! \mathcal{D}_3^{(123)}(\omega) 
         \notag \\
          & +{\bar g}^3 N_c
          K_{2 \to 3}^{( \{13\} \to \{134\})}\!\! \otimes\! \mathcal{D}_3^{(321)}(\omega)
          +
           \bar{g}^2N_c\sum_{\substack{(21),(13)\\(34),(42)}} K_{2 \to 2} \otimes \mathcal{D}_4^{(2134)}(\omega) 
  \label{eq:int_eq_d4_2134} \\
(\omega - \sum_i^4 \beta({\bf k}_i))  \mathcal{D}_4^{(1243)}(\omega) &= 
          \mathcal{D}^{(1243)}_{(4;0)}  
          +
          {\bar g}^3 
                         (-N_c)  K_{2 \to 3}^{( \{13\} \to \{134\})}\!\! \otimes\! \mathcal{D}_3^{(123)}(\omega)
          \notag \\
          &                
 +
                {\bar g}^3  N_c K_{2 \to 3}^{( \{13\} \to \{124\})} \!\!\otimes\! \mathcal{D}_3^{(321)}(\omega) 
                     +
           \bar{g}^2N_c\sum_{\substack{(12),(24) \\(43), (31)}} K_{2 \to 2} \otimes \mathcal{D}_4^{(1243)}(\omega) .  \label{eq:int_eq_d4_4312}
\end{align}

\subsection{Reggeization of four-gluon-amplitudes with planar color  structure}
\label{sec:regg_d4}
\begin{figure}[htbp]
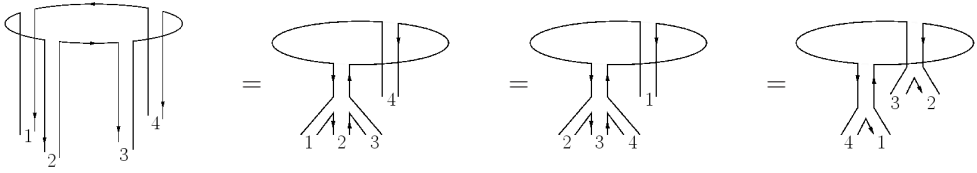

\centering   
  \parbox{3cm}{   \includegraphics[height=2.2cm]{cylinder_1234.eps2}}
=    \parbox{3cm}{ \includegraphics[height=1.7cm]{regg4_123.eps2}}
  =\parbox{3cm}{ \includegraphics[height=1.7cm]{regg4_234.eps2}}
= \parbox{3cm}{ \includegraphics[height=1.7cm]{regg4_14.eps2}}  
   \caption{ \small Color factors associated with the four gluon amplitude with planar color structure can be deformed to coincide with the momentum structure of the solution in terms of $\mathcal{D}_2$. Above we show the ordering (1234). }
  \label{fig:d4_color}
\end{figure}

Similar to the three-gluon amplitude, the four independent integral
equations for the four-gluon-amplitudes with planar color structure
are solved by a reggeization ansatz. Again the starting point is given by 
the virtual photon impact factors which, similar to the three-gluon
case, can be expressed in terms of the two-gluon impact factor:
\begin{align}
  \label{eq:d40_1234}
\mathcal{D}_{(4;0)}^{(1234)}({\bf k}_1,{\bf k}_2,{\bf k}_3,{\bf k}_4) 
    &=
    -\frac{\bar{\lambda}}{2N_c} \big[
\mathcal{D}_{(2;0)}(123,4) + \mathcal{D}_{(2;0)}(1,234) - \mathcal{D}_{(2;0)}(14,23)
\big] \\
  \label{eq:d40_4321}
\mathcal{D}_{(4;0)}^{(2143)}({\bf k}_1,{\bf k}_2,{\bf k}_3,{\bf k}_4) 
    &=
    - \frac{\bar{\lambda}}{2N_c} \big[
\mathcal{D}_{(2;0)}(123,4) + \mathcal{D}_{(2;0)}(1,234) - \mathcal{D}_{(2;0)}(14,23)
\big] \\
\label{eq:d40_2134}
\mathcal{D}_{(4;0)}^{(2134)}({\bf k}_1,{\bf k}_2,{\bf k}_3,{\bf k}_4) 
    &=
    -\frac{\bar{\lambda}}{2N_c} \big[
\mathcal{D}_{(2;0)}(134,2) + \mathcal{D}_{(2;0)}(124,3) - \mathcal{D}_{(2;0)}(12,34) - \mathcal{D}_{(2;0)}(13,24)
\big] \\
\label{eq:d40_4312}
\mathcal{D}_{(4;0)}^{(1243)}({\bf k}_1,{\bf k}_2,{\bf k}_3,{\bf k}_4) 
    &=
    - \frac{\bar{\lambda}}{2N_c} \big[
\mathcal{D}_{(2;0)}(134,2) + \mathcal{D}_{(2;0)}(124,3) - \mathcal{D}_{(2;0)}(12,34) - \mathcal{D}_{(2;0)}(13,24)
\big] .
\end{align}
This decomposition also holds for the solutions of the integral 
equations, Eq.(\ref{eq:int_eq_d4_1234}),
Eq.(\ref{eq:int_eq_d4_4321}), Eq.(\ref{eq:int_eq_d4_2134}) and
Eq.(\ref{eq:int_eq_d4_4312}):
\begin{align}
    \label{eq:d4_1234}
\mathcal{D}_{4}^{(1234)}(\omega|{\bf k}_1,{\bf k}_2,{\bf k}_3,{\bf k}_4) 
    &=
     -\frac{\bar{\lambda}}{2N_c} \big[
\mathcal{D}_{2}(\omega|123,4) + \mathcal{D}_{2}(\omega|1,234) - \mathcal{D}_{2}(\omega|14,23)
\big] \\
  \label{eq:d4_4321}
\mathcal{D}_{4}^{(2143)}(\omega|{\bf k}_1,{\bf k}_2,{\bf k}_3,{\bf k}_4) 
    &=
     -\frac{\bar{\lambda}}{2N_c} \big[
\mathcal{D}_{2}(\omega|123,4) + \mathcal{D}_{2}(\omega|1,234) - \mathcal{D}_{2}(\omega|14,23)
\big]\\ 
\label{eq:d4_2134}
\mathcal{D}_{4}^{(2134)}(\omega|{\bf k}_1,{\bf k}_2,{\bf k}_3,{\bf k}_4) 
    &=
     -\frac{\bar{\lambda}}{2N_c} \big[
\mathcal{D}_{2}(\omega|134,2) + \mathcal{D}_{2}(\omega|124,3) - \mathcal{D}_{2}(\omega|12,34) - \mathcal{D}_{2}(\omega|13,24)
\big] \\
\label{eq:d4_4312}
\mathcal{D}_{4}^{(1243)}(\omega|{\bf k}_1,{\bf k}_2,{\bf k}_3,{\bf k}_4) 
    &=
     -\frac{\bar{\lambda}}{2N_c} \big[
\mathcal{D}_{2}(\omega|134,2) + \mathcal{D}_{2}(\omega|124,3) - \mathcal{D}_{2}(\omega|12,34) - \mathcal{D}_{2}(\omega|13,24)
\big] .
\end{align}
Similar to the the three gluon case, each term on the rhs consists of
a two gluon state where at the end one (or both) gluons decay: this
again generalizes the reggeization discussed in the context of planar
amplitudes. It allows to redraw the color factors as shown in
Fig.\ref{fig:d4_color}.

For the sum of these planar four gluon amplitudes, it is convenient to
define, following \cite{Bartels:1994jj}, an effective $2 \to 4$ vertex
$V^{\text{R}}$, which sums the various decay-channels of the two
reggeized gluons,
\begin{align}
  \label{eq:V_R_def}
V^{\text{R}}({\bf l}_1, {\bf l}_2; {\bf k}_1,{\bf k}_2,{\bf k}_3, {\bf k}_4 ) = -{\bf l}_1^2 {\bf l}_2^2  
\bigg[&
     \delta^{(2)}({\bf l}_1 - {\bf k}_1- {\bf k}_2- {\bf k}_3 ) 
     +
     \delta^{(2)}({\bf l}_1 - {\bf k}_1- {\bf k}_2- {\bf k}_4 ) 
     \notag \\     
     &+
     \delta^{(2)}({\bf l}_1 - {\bf k}_1- {\bf k}_3- {\bf k}_4 ) 
     +  
     \delta^{(2)}({\bf l}_1 - {\bf k}_2- {\bf k}_3- {\bf k}_4 ) 
     \notag \\     
     -&
     \delta^{(2)}({\bf l}_1 - {\bf k}_1- {\bf k}_2 ) 
     -
     \delta^{(2)}({\bf l}_1 - {\bf k}_1- {\bf k}_3  )
     -
     \delta^{(2)}({\bf l}_1 - {\bf k}_1- {\bf k}_4)  \bigg],
\end{align}
such that 
\begin{align}
  \label{eq:d4_planar_vertex}
\sum_{ (\text{ijkl}) =\substack{(1234), (2143), \\(2134), (1243) }}  \!\!\!\!\!\!  \mathcal{D}_{4}^{(\text{ijkl})}(\omega|{\bf k}_1,{\bf k}_2,{\bf k}_3,{\bf k}_4)  = \frac{\bar{\lambda}}{N_c} V^{(\text{R})} \otimes \mathcal{D}_{2}(\omega).
\end{align}

It is instructive to compare the results of this section with the
finite $N_c$ results of \cite{Bartels:1994jj}. For finite $N_c$ it was
found that the four gluon amplitude $D_4$ can be written as a sum of
two terms, the reggeizing term and the triple Pomeron vertex:
\begin{equation}
D_4= D_4^R + D_4^I
\end{equation}
This decomposition was motivated by $t$-channel requirements: the
reggeizing pieces contained in $D_4^R$ are antisymmetric under the
exchange of the pair of gluons which forms the reggeized gluon,
whereas the remainder $D_4^I$ (which contains the triple Pomeron
vertex), is completely symmetric. In the present approach which is
based upon the expansion in topologies, this decomposition into
reggeizing and non-reggeizing pieces comes automatically, as a
consequence of different classes of color structures. This suggests
that the reggeization is deeply linked to the planar structure of
scattering amplitudes.

\section{The triple Pomeron vertex on the pair-of-pants}
\label{sec:trip_pom}

In this section we turn now to the partial wave amplitude associated
with the non-planar class, $\mathcal{D}_4^{(\text{NP})} (\omega)$.
As we have outlined at the end of Sec.\ref{sec:below},
we have to convolute, in Fig.\ref{fig:non_planar_omega}a - c,
the branching vertices with $\mathcal{D}_4$, $\mathcal{D}_3$, 
and $\mathcal{D}_2$. Let us go through these  
convolutions in more detail.  

We begin with those contributions where the branching vertex consists 
of a two-to-two transition (Fig.\ref{fig:non_planar_omega}a).
They are to be combined with of one of the four planar amplitudes 
discussed before. Contributions involving $\mathcal{D}_4^{(1234)}(\omega)$ are
\begin{align}
  \label{eq:d4pop_22_1234}
  \parbox{2cm}{ \includegraphics[width=2cm]{trans12_2pop1234.eps2}} 
  +
  \parbox{2cm}{\includegraphics[width=2cm]{trans13_2pop1234.eps2}}
  + 
  \parbox{2cm}{\includegraphics[width=2cm]{trans24_2pop1234.eps2}}
  +
  \parbox{2cm}{ \includegraphics[width=2cm]{trans34_2pop1234.eps2}} 
\to
 \bar{g}^2N_c\sum_{\substack{(12), (13), \\(24), (34) }}  K_{2 \to 2} \otimes \mathcal{D}_4^{(1234)}(\omega),
\end{align}
where, on the rhs, the subscripts under the sum indicate which pairings of 
gluons should be included. A comment on the first and the last 
terms is in place. These interactions are inside the 
gluon pairs (12) or (34) and thus do not belong to the production     
of an $s$-channel gluon inside the $M^2$ discontinuity. Hence, these 
interactions are not part of the branching vertex. It is only for our 
convenience that we, nevertheless, include them into our definition of 
$\mathcal{D}_4^{(\text{NP})} (\omega)$ (later on, when defining the partial 
waves, we will have to subtract them again).      
In (\ref{eq:d4pop_22_1234}), the second and third term 
(the interactions (13) and (24)) carry an additional minus sign, due  
to the color structure (see Sec.\ref{sec:amplitudes-with-4}).
Similarly one has 
\begin{align}
  \label{eq:d4pop_22_2134}
  \parbox{2cm}{ \includegraphics[width=2cm]{leg2134_12.eps2}} 
  + 
  \parbox{2cm}{\includegraphics[width=2cm]{pop23_2134.eps2}}
  + 
\parbox{2cm}{\includegraphics[width=2cm]{pop14_2134.eps2}}
  +
  \parbox{2cm}{ \includegraphics[width=2cm]{leg2134_34.eps2}} 
\to \bar{g}^2N_c\sum_{\substack{(12), (23), \\(14), (34) }}  K_{2 \to 2} \otimes \mathcal{D}_4^{(2134)}(\omega),
\end{align}
while the contributions containing $\mathcal{D}_4^{(2143)}(\omega) $ and $\mathcal{D}_4^{(1243)}(\omega)$ are 
\begin{align}
  \label{eq:d4pop_22_4312}
  \bar{g^2}N_c\sum_{\substack{(12), (13), \\(24), (34) }}  K_{2 \to 2} \otimes \mathcal{D}_4^{(2143)}(\omega)
+
\bar{g}^2N_c\sum_{\substack{(12), (23), \\(14), (34) }}  K_{2 \to 2} \otimes \mathcal{D}_4^{(1243)}(\omega).
\end{align}
In all  three sums, the second and third terms carry minus signs,
and the first and last terms are kept for convenience and 
will have to be removed later on. 

For the two-to-three transition (Fig.\ref{fig:non_planar_omega}b) 
we have two contributions: terms that originate from $\mathcal{D}_3^{(123)}(\omega) $ are given by
\begin{align}
  \label{eq:d4pop_23_123}
  \parbox{2cm}{ \includegraphics[width=2cm]{pop2_trans23.eps2}}
  +
  \parbox{2cm}{ \includegraphics[width=2cm]{trans124_3pop1234.eps2}}
  + 
  \parbox{2cm}{ \includegraphics[width=2cm]{trans134_3pop1234.eps2}}
  + 
  \parbox{2cm}{ \includegraphics[width=2cm]{trans234_3pop1234.eps2}}
\to  {\bar g}^3 N_c \sum K_{2 \to 3}  \otimes \mathcal{D}_3^{(123)}(\omega),
 \end{align}
where, on the rhs, the sum is over the four possible two-to-three 
transitions, allowed by the triple discontinuity. 
Again, in the second and the third terms  
a minus sign originating from the color factor has to be included.
 Contributions with $\mathcal{D}_3^{(321)}(\omega) $ are
\begin{align}
  \label{eq:d4pop_23_321}
  \parbox{2cm}{ \includegraphics[width=2cm]{trans123_3pop1234.eps2}}
  +
  \parbox{2cm}{ \includegraphics[width=2cm]{trans124_3popb.eps2}}
  + 
  \parbox{2cm}{ \includegraphics[width=2cm]{trans134_3popb.eps2}}
  + 
  \parbox{2cm}{ \includegraphics[width=2cm]{pop3_trans23.eps2}}
\to
 {\bar g}^3(-N_c)  \sum K_{2 \to 3}  \otimes \mathcal{D}_3^{(321)}(\omega),
 \end{align}
 where the sum is again over all four possible two-to-three
 transitions and for the second and third term a relative minus sign has
 to be included. Finally, from the two-to-four transition 
 (Fig.\ref{fig:non_planar_omega}c) we have two contributions.  They
 are given by
\begin{align}
  \label{eq:d4pop_24}
  \parbox{2cm}{\includegraphics[width=2cm]{trans_4popx.eps2}}
  +
  \parbox{2cm}{\includegraphics[width=2cm]{pop2_trans24.eps2}}
  \to
  2{\bar{g}^4}{N_c} K_{2 \to 4} \otimes \mathcal{D}_2(\omega).
\end{align}

Combining these three groups and including the Reggeon-propagators we obtain 
for the amplitude $\mathcal{D}^{\text{NP}}_{4}$:
\begin{align}
  \label{eq:d4pop_inteq}
 (\omega - \sum_i^4 \beta({\bf{k}}_i)  \mathcal{D}^{\text{NP}}_{4} (\omega|  {\bf k}_1,  {\bf k}_2, {\bf k}_3, {\bf k}_4)
&=
        {\bar g}^3 N_c   \sum K_{2 \to 3}  \otimes \mathcal{D}_3^{(123)}(\omega)
        +
        {\bar g}^3 (-N_c)  \sum K_{2 \to 3}  \otimes \mathcal{D}_3^{(321)}(\omega) 
        \notag \\
        +
        2{\bar{g}^4}{N_c}  K_{2 \to 4} \otimes \mathcal{D}_2(\omega)
        +
         \bar{g}^2N_c \bigg[ &
          \sum_{\substack{(12), (13), \\(24), (34) }}  K_{2 \to 2} \otimes \mathcal{D}_4^{(1234)}(\omega) 
           + 
           \sum_{\substack{(12), (13), \\(24), (34) }}  K_{2 \to 2} \otimes \mathcal{D}_4^{(2143)}(\omega)
          \notag \\
           &+
           \sum_{\substack{(12), (23), \\(14), (34) }}  K_{2 \to 2} \otimes \mathcal{D}_4^{(2134)}(\omega)
           +
           \sum_{\substack{(12), (23), \\(14), (34) }}  K_{2 \to 2} \otimes \mathcal{D}_4^{(1243)}(\omega)\bigg].
\end{align}
In the next step we make use of the fact that, according to the results of
Sec.\ref{sec:regg_d3} and Sec.\ref{sec:regg_d4}, all planar three and four
gluon partial amplitudes on the rhs, $\mathcal{D}_3$  and $\mathcal{D}_4$, 
can be written as superpositions of two gluon amplitudes
$\mathcal{D}_2(\omega)$. As a result, the right hand side of
Eq.(\ref{eq:d4pop_inteq}) depends only on
$\mathcal{D}_2(\omega)$, convoluted with the sum of transition kernels:
\begin{align}
  \label{eq:d4pop_inteq_vertex}
(\omega - \sum_{i = 1}^4 \beta({\bf k}_i)) \mathcal{D}_4^{\text{NP}}(\omega|{\bf k}_1,{\bf k}_2,{\bf k}_3,{\bf k}_4)
=&
\frac{\bar{\lambda}^2}{N_c} V_{\text{TPV}} \otimes \mathcal{D}_2(\omega) .
\end{align}
The right hand side of Eq.(\ref{eq:d4pop_inteq_vertex}) defines 
the Triple-Pomeron-Vertex $ V_{\text{TPV}} $ on the pairs-of-pants.
It describes the coupling of the upper BFKL Pomeron, $\mathcal{D}_2(\omega)$, 
to the lower ones, $\mathcal{D}_2(\omega_1)$ and $\mathcal{D}_2(\omega_2)$.

The function $V_{\text{TPV}}$ 
coincides with the $V$ function found in \cite{Bartels:1994jj}.  
In this paper, a finite $N_c$ calculation with $N_c = 3$ has been carried out,
and the following result for the $2 \to 4$ gluon vertex has been obtained:
\begin{align}
  \label{eq:v24_bartelswuesthoff}
V_{2 \to 4}^{a_1a_2a_3a_4}  ({\bf l}_1, {\bf l}_2 | {\bf k}_1,{\bf k}_2,{\bf k}_3,{\bf k}_4) )
 & = 
\delta^{a_1a_2}\delta^{a_3a_4} V ({\bf l}_1, {\bf l}_2 |{\bf k}_1,{\bf k}_2;{\bf k}_3,{\bf k}_4) )  \notag \\
+\, \delta^{a_1a_3}\delta^{a_2a_4}&  V ({\bf l}_1, {\bf l}_2 |{\bf k}_1,{\bf k}_3;{\bf k}_2,{\bf k}_4) ) + \delta^{a_1a_4}\delta^{a_2a_3} V ({\bf l}_1, {\bf l}_2 |{\bf k}_1,{\bf k}_4;{\bf k}_2,{\bf k}_3) .
\end{align}
Here $a_i$, $i = 1, \ldots 4$ denote color indices in the adjoint
representation of the $t$-channel gluons. To compare with our result,
we use the finite $N_c$ version of Eq.(\ref{eq:v24_bartelswuesthoff})
in \cite{Bartels:1999aw} which has been obtained for  arbitrary $N_c$ and we find
\begin{align}
  \label{eq:pp_barwue}
\frac{\bar{\lambda}^2}{N_c} V_{\text{TPV}}({\bf l}_1, {\bf l}_2 |{\bf k}_1,{\bf k}_2;{\bf k}_3,{\bf k}_4) ) = N_c V ({\bf l}_1, {\bf l}_2 |{\bf k}_1,{\bf k}_2;{\bf k}_3,{\bf k}_4) ).
\end{align}
It is easy to verify that this result is in complete agreement with
the large large-$N_c$ limit of \cite{Bartels:1994jj,Bartels:1999aw}.
Namely, if we project, in Eq.(\ref{eq:v24_bartelswuesthoff}), on the
color single state of the gluon-pairs (12)and (34) and then consider
the limit of large $N_c$, only the first term on the right hand side
of Eq.(\ref{eq:v24_bartelswuesthoff}) contributes, and we indeed
obtain agreement of the two results  as required. In \cite{Bartels:1995kf} it has been
shown that the vertex $V$ is invariant under M\"obius transformations.
Furthermore, $V$ can be expressed in terms of another function $G({\bf
  k}_1,{\bf k}_2,{\bf k}_3)$:
\begin{align}
  \label{eq:VbyG}
  V  ( {\bf k}_1,{\bf k}_2,{\bf k}_3,{\bf k}_4) =& \bar{\lambda} \big[
G(1,23,4) + G(2,13,4) + G(1, 24, 3) + G(2,14,3) - G(12,3,4)
\notag \\
&
 - G(12,4,3) - G(1,2,34) - G(2,1,34) + G(12, -,34)
\big],
\end{align}
where we suppressed the dependence on the momenta ${\bf l}_1$ and
${\bf l}_2$. This function $G$ has been first introduced for the
forward case and $N_c=3$ in \cite{Bartels:1994jj}, whereas the above
version was introduced in \cite{Braun:1997nu}. It has the nice property
that it is not only infra-red finite but also by itself invariant
under M\"obius transformations \cite{Braun:1997nu}.

\section{The six-point amplitude on the pair-of-pants surface}
\label{sec:6point_res}
\begin{figure}[htbp]
  \centering
  \parbox{6cm}{\includegraphics[height=4.5cm]{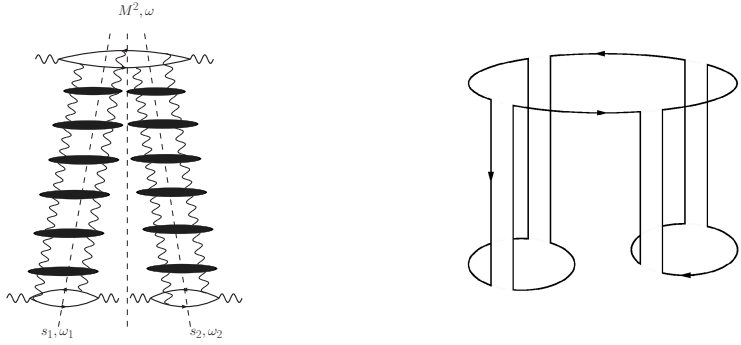}}
 \parbox{4cm}{\includegraphics[height = 3.2cm]{trouser_born.eps}}
  \caption{\small Diagrams where the  two Pomerons couple directly to the  upper quark-loop. 
They contain the lowest order diagram in Fig.\ref{fig:trouser}. The color factors  
can be reduced to the color factor to the right and therefore belong to the planar class.}
  \label{fig:quark_branching}
\end{figure}

With the results of the previous sections, we have almost all
constituents that are needed to build the triple Regge limit of the
six-point amplitude with the pair-of-pants topology.  However before
we add together the amplitudes of the foregoing sections, we need to
take care of a peculiarity inside the planar piece of the upper amplitude. 
To leading order in $\lambda$, the two lower Pomerons couple
directly to the upper quark-loop, as depicted in
Fig.\ref{fig:quark_branching} and the branching vertex is not given by
a real gluon, but by the quark-loop itself. As long as we have, in addition 
to the $q\bar{q}$ pair, one (or more) $s$-channel gluons that contribute to 
the $M^2$ discontinuity, the mass of the $q\bar{q}$-system is integrated over,
and this integration enters into the definition of the impact factors 
$\mathcal{D}_{4;0}^{(\text{ijkl})}$. Without such gluons, the mass 
of the $q\bar{q}$ pair coincides with the diffractive mass $M$ which
is a fixed external parameter. In this case, the coupling of the $t$-channel four gluon system
to the quark-loop is hence given by the triple discontinuity of the
quark loop without the integration over the diffractive mass of the 
$q\bar{q}$-system, and defines an 'unintegrated' four
gluon impact factor, $\bar{\mathcal{D}}_{(4;0)}(M^2)$. Its precise analytic form 
can be obtained from Eq.(2.8) of \cite{Bartels:1994jj}.
Correspondingly, its Mellin transform, which enters the partial wave $F(\omega, \omega_1,
\omega_2)$ will be denoted by $\bar{\mathcal{D}}_{(4;0)}(\omega)$.
When $\bar{\mathcal{D}}_{(4;0)}(M^2)$ is integrated over the squared 
diffractive mass $M^2$, it coincides 
with sum over the impact factors $\mathcal{D}_{4;0}^{(\text{ijkl)}}$. 
 
\begin{figure}[htbp]
  \centering
 \parbox{7cm}{\center \includegraphics[height=4.5cm]{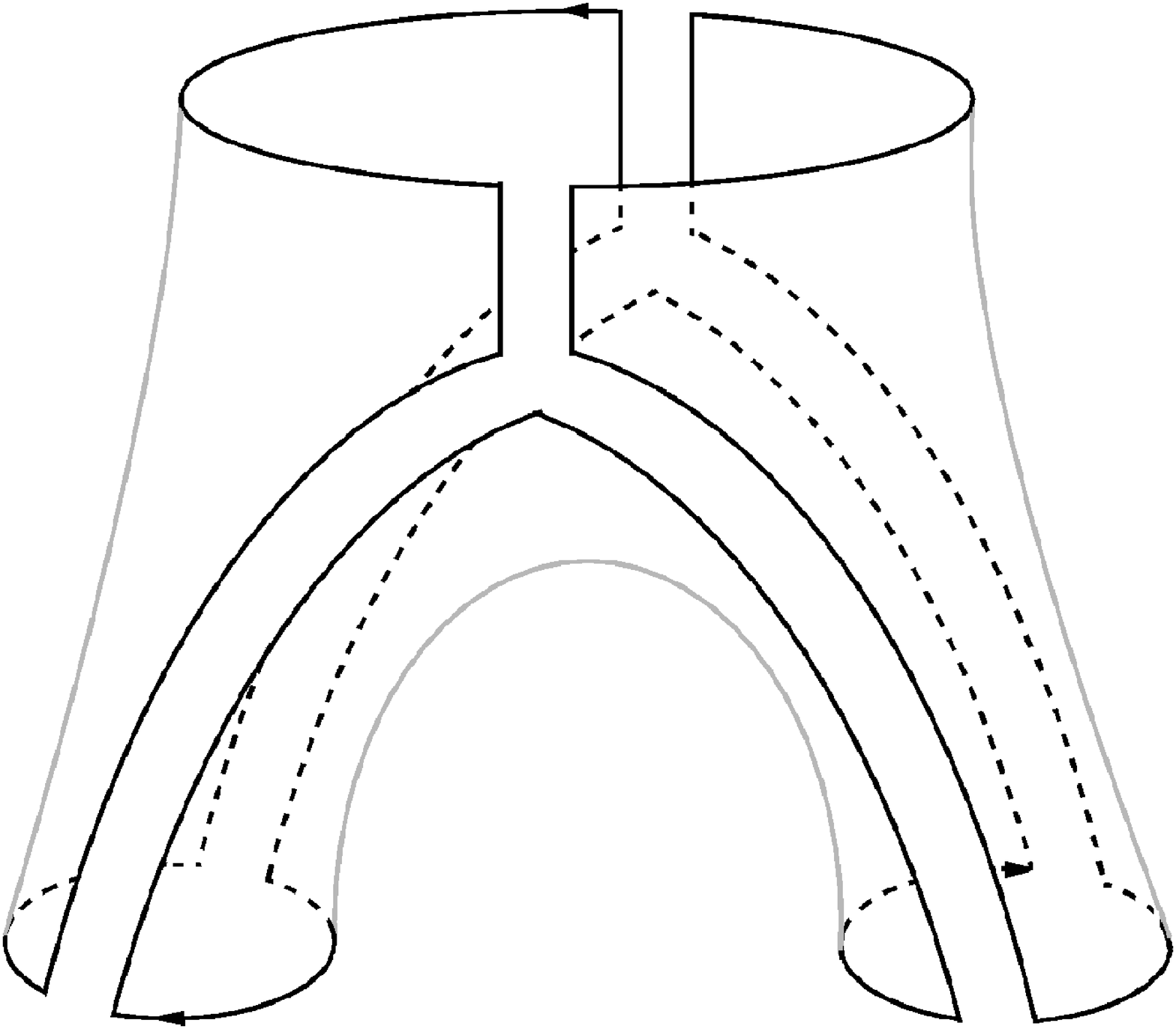}}
 \parbox{7cm}{\center \includegraphics[height=4.5cm]{vpp.eps}}\\
\parbox{7cm}{\center (a)}\parbox{7cm}{\center (b)}
  \caption{\small (a) Planar color graphs  associated with the decay of two reggeized gluons  and  (b) non-planar color graphs which lead to the Triple-Pomeron-Vertex. In both cases a two gluon state decays into a four gluon state.}
  \label{fig:vpp}
\end{figure}

After this remark we write the final result as the sum of two terms:
\begin{align}
  \label{eq:F_sum}
F(\omega, \omega_1, \omega_2)  = F^{(\text{P})}(\omega, \omega_1, \omega_2) 
+  F^{(\text{NP})}(\omega, \omega_1, \omega_2),
\end{align}
where the first term sums the planar diagrams (including  the 
'unintegrated' impact factors), the second term the
non-planar ones. When doing the convolution of the three amplitudes,
special care has to be taken of the counting of interactions inside
the gluon pairs (12) and (34) just below the branching point where the
upper cylinder splits into two cylinders (c.f. the discussion given in
Sec.\ref{sec:below}).  This issue has been addressed, for the
'$N_c$-finite' case, in Sec.4 of \cite{Bartels:1994jj} (see, in
particular Eqs.(4.14) and Eq.(4.15)), and in the following we shall
apply the same line of arguments. We identify, for the $M^2$
discontinuity, the lowest $s$-channel intermediate state which leads
to the 'last' interaction between the two gluon pairs (12) and (34):
below this interaction, the upper cylinder has split into two separate
cylinders.  In particular, this last interaction cannot consist of a
two-to-two kernel acting on the gluon pairs (12) or (34). Since, in
Secs.\ref{sec:inteq_d4} and \ref{sec:trip_pom}, we have defined the
functions $\mathcal{D}_{4}^{(\text{ijkl})}$ and
$\mathcal{D}_{4}^{(\text{NP})}$ in such a way that the last
interaction includes these contributions, we first have to remove
them. Furthermore, as explained above, the coupling of the
$\mathcal{D}^{(12)}_2(\omega_1) $ and $
\mathcal{D}^{(34)}_2(\omega_2)$ directly to the upper quark-loop is
described by $\bar{\mathcal{D}}_{4;0}(\omega)$, and we write this term separately. We therefore arrive at  the full
partial wave in the following form:
\begin{align}
  \label{eq:F_factor}
F(\omega, \omega_1, \omega_2) 
&=
4  \mathcal{D}^{(12)}_2(\omega_1) \otimes_{12} \mathcal{D}^{(34)}_2(\omega_2)\otimes_{34}\notag \\
 &\bigg(  \bar{\mathcal{D}}_{(4;0)}(\omega)
+
\big[\omega - \sum_{i = 1}^4 \beta({\bf k}_i)\big]  \!\!\!\!  
\sum_{(\text{ijkl}) = \substack{(1234), (2143), \\(2134), (1243) }}  \!\!\!\! 
\mathcal{D}_{4}^{(\text{ijkl})}(\omega)
    + \frac{{\bar{\lambda}}^2}{N_c} V_{\text{TPV}} \otimes \mathcal{D}_2(\omega)
\notag \\ 
& - 2{\bar{\lambda}}(  K^{(12)}_{2 \to 2} +  K^{(34)}_{2 \to 2} ) 
\otimes  \!\!\!\!  
\sum_{(\text{ijkl}) = \substack{(1234), (2143), \\(2134), (1243) }}  \!\!\!\! 
\mathcal{D}_{4}^{(\text{ijkl})}(\omega) 
 \bigg) 
\end{align}
In the first line, the indices of the convolution symbols indicate that
$\mathcal{D}^{(12)}_2(\omega_1)$ and $ \mathcal{D}^{(34)}_2(\omega_2)$
are to be contracted with the gluons (12) and (34), resp. 
The last line  subtracts the two-to-two kernels acting on gluon
pairs (12) and (34).
 Using the BFKL-equation 
$$ 2\bar{\lambda}{K}_{2 \to 2}^{(12)}
\mathcal{D}^{(12)}_2(\omega_1) = (\omega_1 - \beta({\bf k}_1) -
\beta({\bf k}_2) ) \mathcal{D}^{(12)}_2(\omega_1) -
\mathcal{D}^{(12)}_{(2;0)} $$ 
(and a similar expression for the gluon pair (34)),
and making further use of Eq.(\ref{eq:d4_planar_vertex}), we
arrive at
\begin{align}
  \label{eq:F_factor_2}
& F(\omega, \omega_1, \omega_2) =
\notag \\
& \quad
 4 \bigg\{ \mathcal{D}^{(12)}_2(\omega_1) \otimes_{12} \!\mathcal{D}^{(34)}_2(\omega_2)\otimes_{34}\!
 \bigg( 
 \bar{\mathcal{D}}_{(4;0)}(\omega)
+
\big[\omega \!-\! \omega_1 \!-\!\omega_2 \big]  V^{\text{R}}\!\otimes\! \mathcal{D}_2(\omega)
   + \frac{\bar{\lambda}^2}{N_c} V_{\text{TPV}} \! \otimes \!\mathcal{D}_2(\omega) \bigg) \notag \\ 
& 
\qquad
 + \mathcal{D}^{(12)}_2(\omega_1) \otimes_{12} \mathcal{D}^{(34)}_{(2;0)}\otimes_{34}
V^{\text{R}}\otimes \mathcal{D}_2(\omega)
+
\mathcal{D}^{(12)}_{(2;0)} \otimes_{12} \mathcal{D}^{(34)}_{2}(\omega_2)\otimes_{34}
V^{\text{R}}\otimes \mathcal{D}_2(\omega)
\bigg\}.
\end{align}
The terms in the last  line are either independent of $\omega_2$ or
$\omega_1$ and the integrals over  $\omega_2$
or $\omega_1$ vanish in Eq.(\ref{tripleregge}).  We therefore drop
these terms and obtain for the two partial waves $F^{(\text{P})}(\omega,
\omega_1, \omega_2)$ and $F^{(\text{NP})}(\omega, \omega_1, \omega_2)$ the
following results:
\begin{align}
  \label{eq:F_factor_3}
F^{(\text{P})}(\omega, \omega_1, \omega_2) 
&= 4 
 \mathcal{D}^{(12)}_2(\omega_1) \otimes_{12} \mathcal{D}^{(34)}_2(\omega_2)\otimes_{34}\bigg( \bar{\mathcal{D}}_{(4;0)}(\omega) +
   \big[\omega - \omega_1 -\omega_2 \big] \frac{\bar{\lambda}}{N_c} V^R \otimes
 \mathcal{D}_{2}(\omega)  \bigg) 
\end{align}
and
\begin{align}
  \label{eq:F_nonplanar}
F^{(\text{NP})}(\omega, \omega_1, \omega_2) 
&=
4  \mathcal{D}^{(12)}_2(\omega_1) \otimes_{12} \mathcal{D}^{(34)}_2(\omega_2)\otimes_{34}
   \frac{\bar{\lambda}^2}{N_c} V_{(\text{TPV})}  \otimes \mathcal{D}_2(\omega).
\end{align}
Both terms are illustrated in Figs.\ref{fig:vpp}a and \ref{fig:vpp}b. 

\section{Conclusion}
\label{sec:concl}
In the present analysis we have considered, in the large-$N_c$ limit,
the triple Regge-limit of the scattering of three virtual photons. Our
emphasis has been on the topology of color factors: we have summed, in
the generalized leading-log approximation, only those diagrams which
fit onto the pair-of-pants surfaces. These diagrams group themselves
naturally into two classes (Fig.\ref{fig:vpp}). The first class
consists of all diagrams which, by contracting closed color loops,
coincide with one of the lowest order graphs (of order $g^8 N_c^3$)
illustrated in Fig.13. Making use of the bootstrap conditions, the sum
of these graphs is shown to reduce to reggeized gluons with a simple
splitting vertex. In addition, starting at the order $g^8 N_c^3 (g^2
N_c)$, a second class of diagrams appears which, by contracting closed
color loops, cannot be drawn as one of the lowest order graphs shown
in Fig.13.  The sum of these graphs can be written as a convolution of
three BFKL amplitudes, connected by the triple Pomeron vertex found in
earlier papers.  This triple Pomeron vertex has been shown to be
invariant under M\"obius transformation \cite{Bartels:1995kf}.

The analytic expression for the six-point amplitude derived in this
chapter coincides with the large-$N_c$ limit of the QCD result in
\cite{Bartels:1994jj}. However, the analysis of the present chapter
provides another interpretation of gluon reggeization and the
appearance of the M\"obius invariant triple Pomeron vertex: on the
pair-of-pant surface, the reggeizing pieces are completely planar,
whereas the the triple Pomeron vertex belongs to a distinct class of
color diagrams which reflect the non-planar Mandelstam cross.

We believe that the topological approach pursued in this chapter is
well-suited for studying the Regge limit of N=4 Super Yang Mills
Theory within the AdS/CFT correspondence.  On the string side,
amplitudes are naturally expanded in terms of topologies. In
particular, the pair-of-pants topology studied in this chapter is the
same as that of the vertex which describes the coupling of three
closed strings.

For a systematic study of the AdS/CFT correspondence it is convenient to
make use of $R$-currents: they allow to formulate current correlators
which are well-defined both on the gauge theory side and on the string
side. As a first step of investigating, in the Regge limit, the
duality between N=4 Super Yang Mills Theory and $AdS_5$ string theory,
the 4-point function for such currents has been studied on the gauge
theory side in \cite{Bartels:2008zy}.  This study  confirms that the interaction between two $R$-currents in the
Regge-limit is, indeed, mediated by a BFKL-Pomeron, and it contains the 
two gluon impact factors in N=4 SYM. On the string
side, the scattering of two R-currents has been shown to involve, in
lowest order, the one-graviton exchange \cite{bkms}.  In the present
chapter we have addressed, on the gauge theory side, the next term in
the topological expansion. As a start, we have restricted ourselves to
nonsupersymmetric QCD($N_c$). The generalization to the supersymmetric
case where, inside the impact factors, fermions and scalar particles
in the adjoint representation have to be considered, will be presented
elsewhere \cite{behm}.  On the string side, the analogous 6-point
R-current correlator is expected to contain the triple graviton
vertex. This project is currently being studied.

\chapter{The topology of the triple Pomeron vertex in $\mathcal{N}=4$ SYM}
\label{cha:sym}

Within the AdS/CFT correspondence which relates ${\cal N}=4$ supersymmetric Yang Mills  
quantum field theory (SYM) in four dimensions to a string theory in a 
$AdS_5 \otimes S_5$ space correlators of $R$-currents provide a useful tool 
for investigating the Regge limit, in particular the correspondence between 
the BFKL Pomeron in ${\cal N}=4$ SYM and the graviton on the string side. 

In the simplest case, the elastic scattering of two $R$-currents, both 
ends of the correspondence have been investigated in leading order.
On the gauge theory side, the supersymmetric impact factors consisting of 
the sum of a fermion and of a scalar loop in the adjoint representation of the gauge group 
have been computed, and it has been verified that the high energy behavior 
is described by the BFKL Pomeron \cite{Bartels:2008zy}. On the string side, the leading contribution in the 
zero slope limit is given by the Witten diagram with graviton exchange \cite{bkms}. Beyond the zero slope limit,
the graviton is believed to reggeize.  

A next step along this line is the six point function of $R$-currents in the 
triple Regge limit. In this kinematic limit one expects to see the triple Pomeron vertex which 
represents, besides the BFKL kernel, another fundamental element of high energy QCD: it describes 
the splitting of a BFKL Pomeron into two BFKL Pomerons. On the string
side, one expects the graviton self interaction to play the analogous role.     
For QCD - with fermions in the fundamental 
representation and with the electromagnetic current in place of the $R$-current - 
the six point function at finite $N_c$ has first been studied in ~\cite{Bartels:1994jj}. As the main result, the 
triple Pomeron vertex has been calculated, which by now has been derived in several other approaches.       
In ~\cite{behm} the case of the six point function of $R$-currents in ${\cal N}=4$ SYM has been studied for 
finite $N_c$: whereas the triple Pomeron vertex remains the same as in the nonsupersymmetric case, 
a new contribution to the six point correlator appears which results from the adjoint color representation 
of the particles and has no counterpart in QCD.

Another line of interest is the integrability of the BKP equation: since in the leading logarithmic 
approximation there is no difference between QCD and its supersymmetric generalization, ${\cal N}=4$ SYM, one 
expects that the integrability which has been discovered for the large-$N_c$ limit of QCD in fact is 
'inherited' from ${\cal N}=4$ SYM. The environment where the large-$N_c$ limit of BKP states can be studied 
are higher order current correlators, e.g. the eight point function in a suitable Multi-Regge limit.  
Within the AdS/CFT correspondence, the counterpart of the BKP states has not been 
addressed at all.        

Recently, an attempt has been started to investigate these high energy limits within a topological 
approach: in the large-$N_c$ limit, the color structure of scattering amplitudes can be attributed to 
surfaces: spheres, planes, cylinders, pairs-of-pants, etc. The simplest examples include, in QCD with 
fermions in the fundamental representation, multigluon scattering amplitudes in the plane and the 
BFKL Pomeron on the cylinder. Recently \cite{Bartels:2009zc} also the six point correlator of electromagnetic currents has been 
studied in the large-$N_c$ limit by summing diagrams whose color structure lies on the surface 
of a pair-of-pants. The study of these color diagrams provides a new view on the reggeization of the 
gluon and on the triple Pomeron vertex: whereas the reggeization can be understood as a feature of 
planar QCD, the triple Pomeron vertex requires a non-planar structure, reminiscent of the non-planar 
Mandelstam cross diagram.
     
When turning, from QCD with fundamental fermions, to ${\cal N}=4$ SYM with fermions and scalars in the 
adjoint representation, one encounters changes in the topology of the surfaces and in the structure of color graphs.The first example is the BFKL Pomeron which lies on the surface of a sphere with zero boundaries.
Apart from this the form of the impact factor stays the same.     
In the present paper we address in the large-$N_c$ limit a six point correlator of $R$-currents in ${\cal N}=4$ SYM in the topological approach.  
We sum graphs whose color structure belongs to a specific deformation 
of a sphere, corresponding to the pair-of-pants investigated in ~\cite{Bartels:2009zc}. 

We first review the color structure of Feynman diagrams in the large-$N_c$ limit and define the 
classes of graphs which we are going to sum. We then formulate integral equations which sum these 
graphs and discuss their solutions. Our final result for the large-$N_c$ limit consists of three 
terms which represent three distinct classes of color diagrams on the surface of the deformed 
sphere.
     
\section{Topology of graphs in ${\cal N}=4$ SYM}

It may be useful to recapitulate the large-$N_c$ limit of QCD with fermions in the fundamental 
representation, and to recall a few features of the classical 
paper of 't Hooft \cite{'tHooft:1973jz}. Starting from the Fierz identity
\begin{align}
  \label{eq:fierz}
    (g^a)^i_j(g^a)^k_l  =  \delta^i_l \delta^k_j - \frac{1}{N_c} \delta^i_j\delta^k_l,
\end{align}
where $(g^a)^i_j$ denoted the $SU(N_c)$ generators in the fundamental representation 
(with the normalization $\tr(g^ag^b) = \delta^{ab}$), and from the identity
\begin{align}
  \label{eq:structure_const}
f^{abc}  
&=
\frac{1}{i\sqrt{2}}\big[\tr(g^ag^bg^c) - \tr(g^cg^bg^a) \big],
\end{align}  
one is lead to draw, in the large-$N_c$ limit, color diagrams with the 
following elements:\\
(i) for each quark in the fundamental representation,  a single line with an arrow, indicating the flow from
the upper to the lower index,
\begin{align}
  \label{eq:kronecker_delta}
\delta^i_j = \parbox{2cm}{\includegraphics[width=2cm]{delta.ps}};
\end{align}
(ii) for each inner gluon a double line
\begin{align}
  \label{eq:double_line}
\delta^i_l \delta^k_j= \parbox{2cm}{\includegraphics[width=2cm]{double.ps}};
\end{align}
(iii) for each triple gluon vertex
\begin{align}
  \label{eq:structure_const2}
f^{abc}  
&=
 \frac{1}{i\sqrt{2}}\big[\tr(g^ag^bg^c) - \tr(g^cg^bg^a) \big]  .
\end{align}  
As a result, each graph turns into a network of double and single lines. The resulting diagrams represent only the color factors. A usual Feynman diagram represents both the color factors and the momentum part. In the double line notation the momentum part has to be written separately.
   
The double line diagrams can be
drawn on a two-dimensional surface with Euler number $\chi = 2 -2h -b
$, where $h$ is the number of handles of the surface, and $b$ the
number of boundaries or holes.  A closed color-loop always delivers a
factor $N_c$, and a closed quark-loop, compared to a corresponding gluon-loop, is $1/N_c$
suppressed and leads to a boundary.  For an arbitrary vacuum
graph $T$ one arrives at the following expansion in $N_c$
\begin{align}
  \label{eq:vacuumgraph}
T = \sum_{h,b}^\infty N_c^{2 - 2h -b} T_{h,b}(\lambda)
\end{align}
where
\begin{align}
  \label{eq:thooft}
\lambda &= g^2N_c 
\end{align}
is the 't Hooft-coupling which is held fixed, while $N_c$ is taken to
infinity.

In the expansion Eq.(\ref{eq:vacuumgraph}) which matches the loop
expansion of a closed string theory with the string coupling $1/N_c$,
the leading-$N_c$ diagrams are those that have the
topology of a sphere: zero handles and zero boundaries, $h=b=0$. If
quarks are included, the leading diagrams have the topology of a disk,
i.e. the surface with zero handle and one boundary, $h=0, b=1$, 
fits on the plane, with the boundary as the outermost edge.
Diagrams with two boundaries and zero handles can be drawn on the
surface of a cylinder, those with three boundaries on the surface of a
pair-of-pants. Boundaries are also be obtained by removing, from the
sphere, one or more points. Removing one point, one obtains the disk,
which can be drawn on the plane, and by identifying the removed point
with infinity, the graphs can be drawn on the (infinite) plane.
Removing two points we obtain the cylinder and so on. By definition,
the expansion Eq.(\ref{eq:vacuumgraph}) is defined for vacuum graphs.
However, from the earliest days on \cite{'tHooft:1974hx}, the large-$N_c$ limit has been also applied to the scattering of colored
objects.  In order to consider the topological expansion of an
amplitude with colored external legs, one needs to embed it into a
vacuum graph which then defines the topological expansion of an
amplitude with colored external legs.

\begin{figure}[tbp]
\centering
\includegraphics[height=5cm]{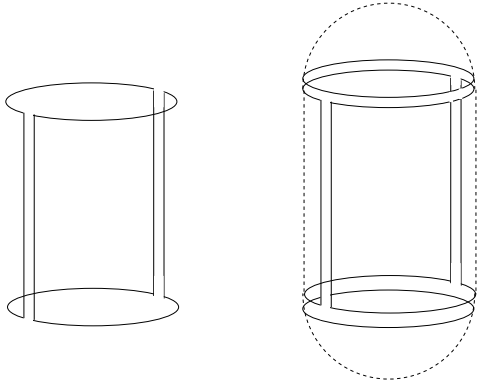}
\caption{Cylinder topology for the $ 2 \to 2$ scattering in QCD (left) and in 
${\cal N}=4$ SYM (right)}
\label{cylinder}
\end{figure}

\begin{figure}
\centering
\includegraphics[width=4cm]{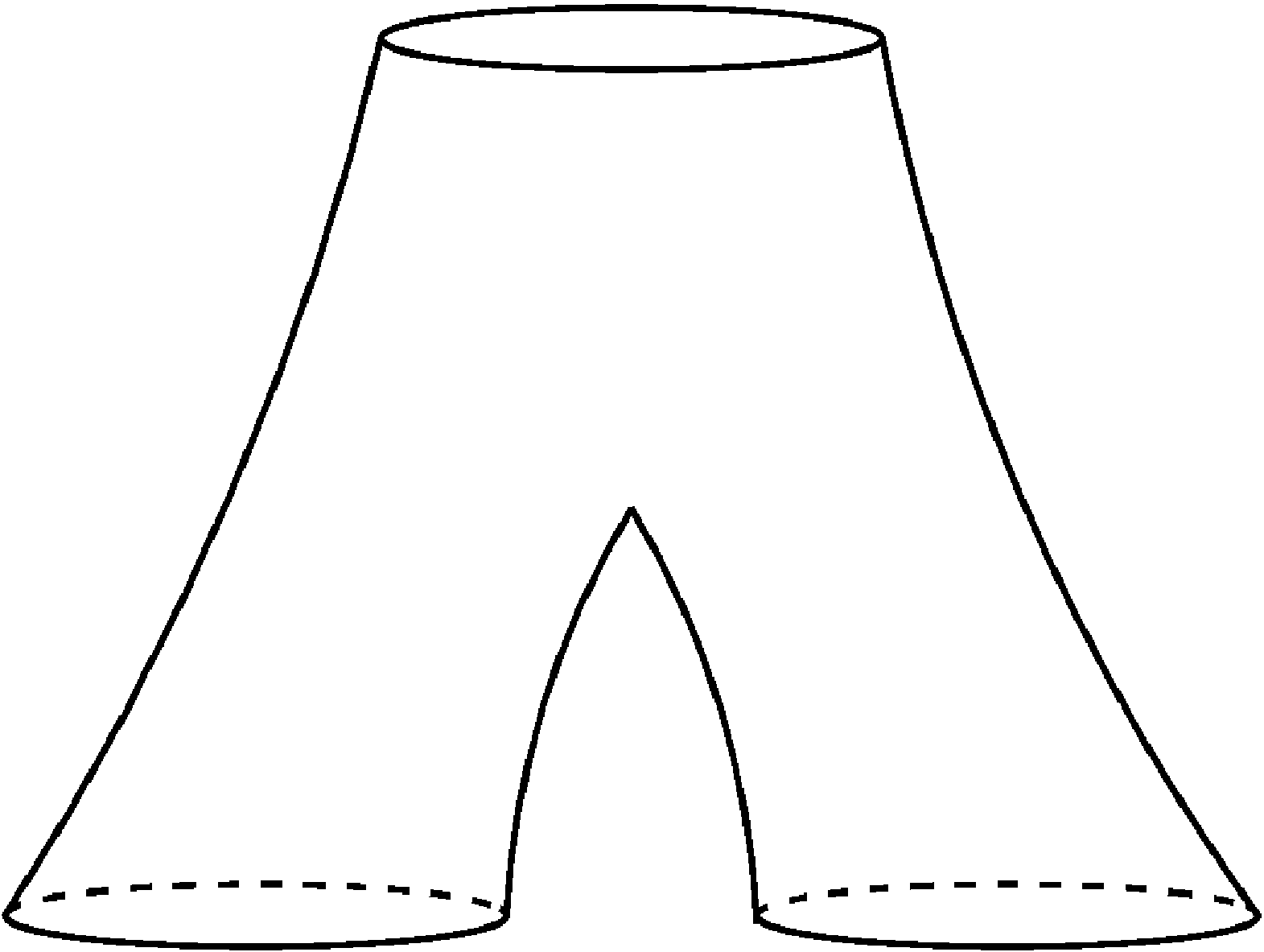}\\
\caption{Pair-of-pants topology for the $3 \to 3$ amplitude in the triple Regge limit}
\label{trousers}
\end{figure}

In the large-$N_c$ limit of high energy QCD, quark scattering
amplitudes are drawn on a plane; in particular, one can show that the
BFKL bootstrap condition is satisfied on the plane (zero handles, one
boundary). Next, for the BFKL Pomeron the color diagrams lie on the
surface of a cylinder (zero handles, two boundaries),
(Fig.\ref{cylinder}, left), and the triple Pomeron vertex is obtained
from diagrams which fit on the pair-of-pants surface (zero handles,
three boundaries), (Fig.\ref{trousers}).

Let us now turn to ${\cal N}=4$ SYM where fermions and scalars belong
to the adjoint representation and therefore are represented by double
lines, in the same way as the gauge bosons.  Now fermion loops no
longer define sections and therefore cannot be used define separate
topologies.  As an example, in QCD with fundamental quarks the
diagrams of the BFKL Pomeron fit onto the surface of a cylinder
(Fig.\ref{cylinder}, left): the closed quark loops at the top and at
the bottom define the two sections. In ${\cal N}=4$ SYM with adjoint
fermions and scalars, the top and the bottom obtain double lines
(Fig.\ref{cylinder}, right) and turn into caps, as a results of which
the cylinder turns into a (stretched) sphere (zero handles, zero
boundaries).

\begin{figure}
\centering
\includegraphics[height=5cm]{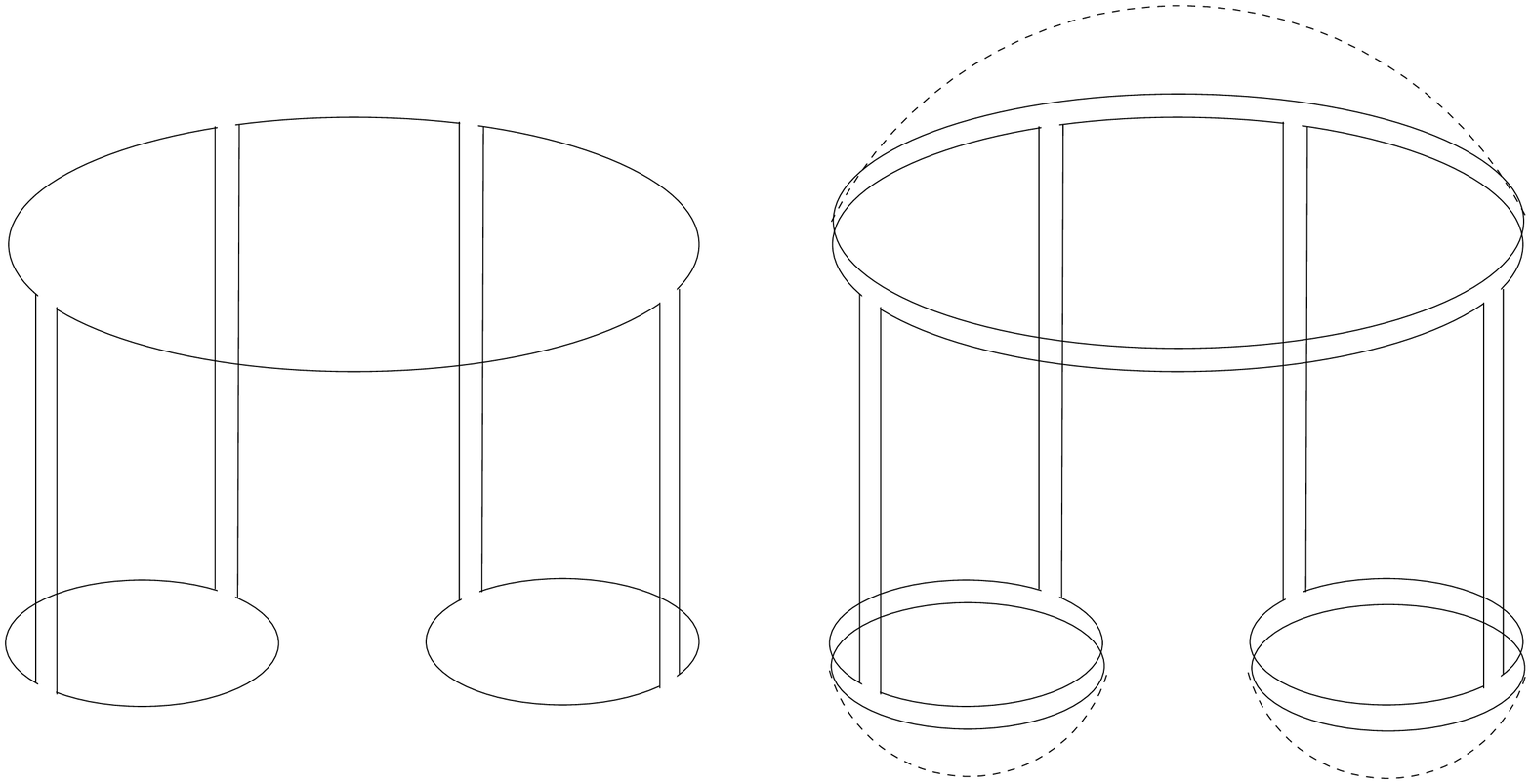}
\caption{Pair-of-pants topology for the $3 \to 3$ amplitude in QCD (left) 
and in ${\cal N}=4$ SYM (right)}
\label{pantsQCDSYM}
\end{figure}

An analogous result holds for the six point function of external currents in the triple Regge limit.
For QCD with fundamental quarks the sphere has three sections (for each impact factor, the closed 
fermion loop defines a section), (Fig.\ref{pantsQCDSYM}, left) and it has been shown in \cite{Bartels:2009zc} that the color diagrams fit on the surface   
of a pair-of-pants (Fig.\ref{trousers}). 
When switching to  ${\cal N}=4$ SYM where fermions and scalars belong to the 
adjoint representation, the closed lines of the upper and lower impact factors turn into double lines, 
and the sections of the pair-of-pants are replaced by caps, (Fig.\ref{pantsQCDSYM}, right).

As a result, we again arrive at a sphere, shaped as a 
pair of pants , and we are asked to sum, for the triple Regge limit, diagrams which lie 
on the surface of this body. As we shall see in the following, there are three distinct classes of 
diagrams: two of them are the same as in (nonsuperymmetric) QCD, whereas the third one 
is new and has no analogue in QCD.\\      

\section{Selection of diagrams}
 
\begin{figure}
\centering
\includegraphics[width=8.5cm]{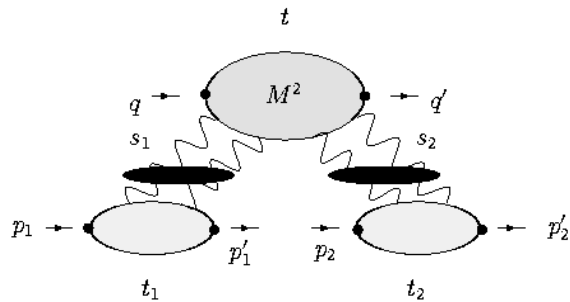}
\caption{Kinematics of a six point amplitude} 
\label{6point}
\end{figure}

Six point amplitudes depend on three energy variables, $s_1=(q+p_1)^2$, $s_2=(q'+p_2')^2$ and $M^2=(q+p_1-p_1')^2$. The momentum transfer variables are $t=(q-q')^2$, $t_1=(p_1-p_1')^2$ and $t_2=(p_2-p_2')^2$. The kinematics of a six point amplitude with external $R$-currents are illustrated in Fig.\ref{6point}.
Then the triple Regge limit is given by
\begin{equation}
\label{tripleRegge} 
s_1=s_2\gg M^2 \gg t,t_1,t_2.
\end{equation}
 In the Regge limit there is an easy way of computing six point amplitudes of $R$-currents, namely we take the triple energy discontinuity in $s_1$, $s_2$ and $M^2$. In lowest order the main ingredients of a diagram are three impact factors and four $t$-channel gluons. The impact factors consisting of a fermion and a scalar loop represent the coupling of the $t$-channel gluons to the external $R$-currents. The insertion of the $R$-currents is symbolized by the small black dots in Fig.\ref{6point}. Higher order corrections are taken into account by the production of real gluons in the Regge kinematics.

We now discuss in the double line notation the relevant diagrams which contribute to the triple Regge limit of the six point 
correlators of $R$-currents. We will closely follow the discussion of ~\cite{Bartels:2009zc}, and we will use 
the same notation. In the following double line diagrams we do not show the attached $R$-currents, we only consider the gluons and adjoint particle loops. 

We begin with a comment on the cylinder in QCD, (Fig.\ref{cylinder}, left). When replacing, at the top of the cylinder, 
the fermion in the fundamental representation by an adjoint fermion, we simply draw, above  
the already existing color line of the fermion, an additional closed color loop which generates an additional 
factor $N_c$. The double line notation now also includes scalar loops. The two $t$-channel gluons are attached to the same color line of the closed loop, either to the lower line or to the upper one as shown in Fig.\ref{cylinder}, right. Diagrams where the two $t$-channel lines are 
attached to different lines lose this additional factor $N_c$ and are suppressed. As a result of 
this simple observation, the contribution of the adjoint fermions to the impact factors is 
proportional to $N_c$  times that of a fundamental one\footnote{This is nothing else but the consequence of the different 
normalizations  of generators. In the fundamental representation we use $\mbox{tr}(t^a t^b)=\delta_{ab}$, 
whereas in the 
adjoint representation we have  $\mbox{tr}(T^a T^b)= N_c \delta_{ab}$. Note that our normalization of the fundamental generators deviates by a factor 2 from the standard normalization $\mbox{tr} (\tau^a\tau^b)=\delta^{ab}/2$.}. In ${\cal N}=4$ SYM, the 
four point correlator is of the form $N_c^2 A_{2 \to 2}(\lambda)$, whereas in nonsupersymmetric QCD it 
is of the form $N_c^0 A_{2 \to 2}(\lambda)$. 
                
Let us now turn to the six point function. In QCD in lowest order the four $t$-channel gluons couple to the upper loop in all possible ways, all together there a 16 different diagrams. A closer look shows that we have inside the 16 diagrams four different orderings of color matrices. For the lowest order diagrams in the nonsupersymmtric QCD case, 
the four different structures are illustrated in Fig.\ref{fig:born}.\begin{figure}[htbp]
  \centering
    \begin{minipage}{.9\textwidth}
    \parbox{3cm}{\center   \includegraphics[width=2cm]{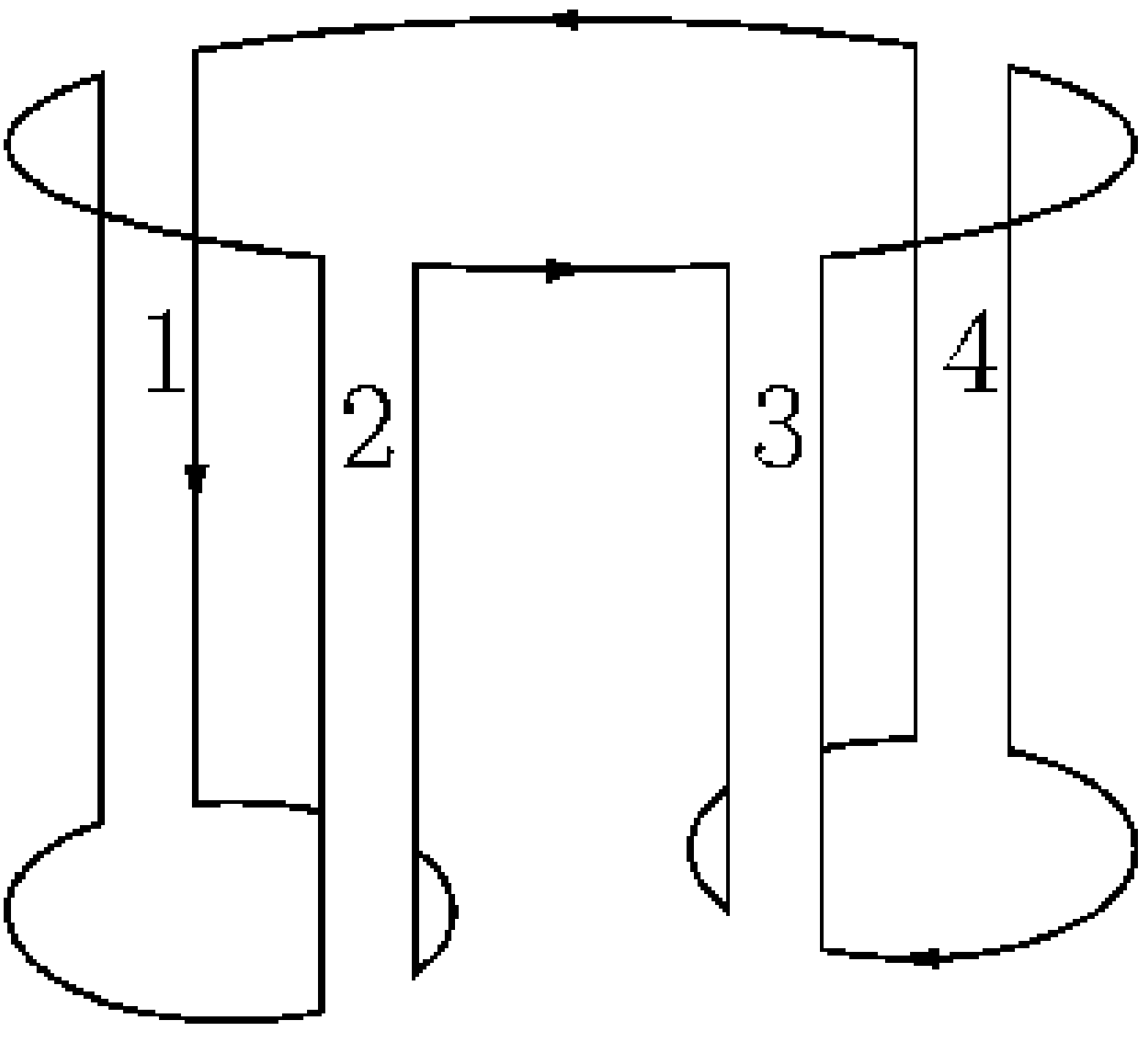}} 
    \parbox{3cm}{\center   \includegraphics[width=2cm]{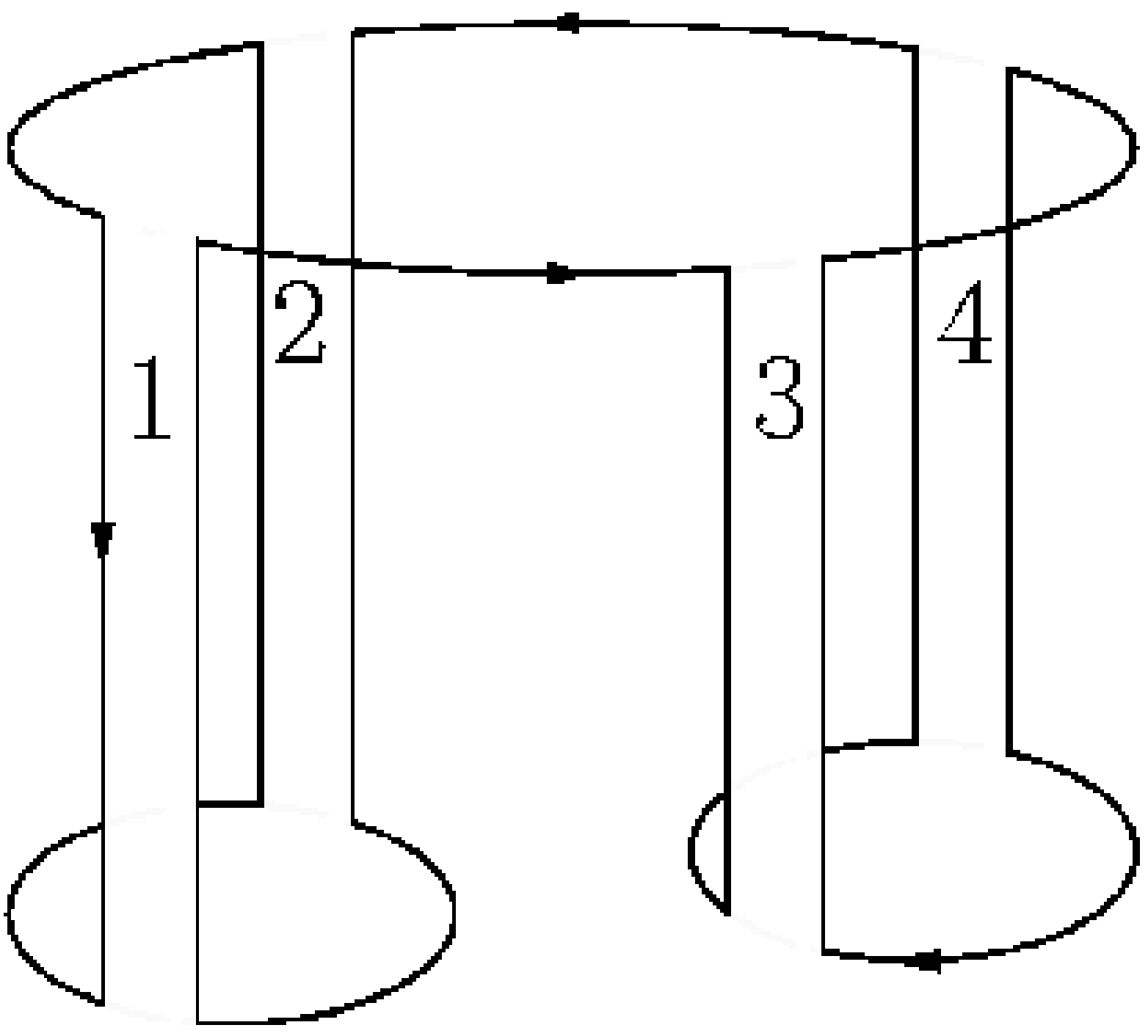}}  
    \parbox{3cm}{\center   \includegraphics[width=2cm]{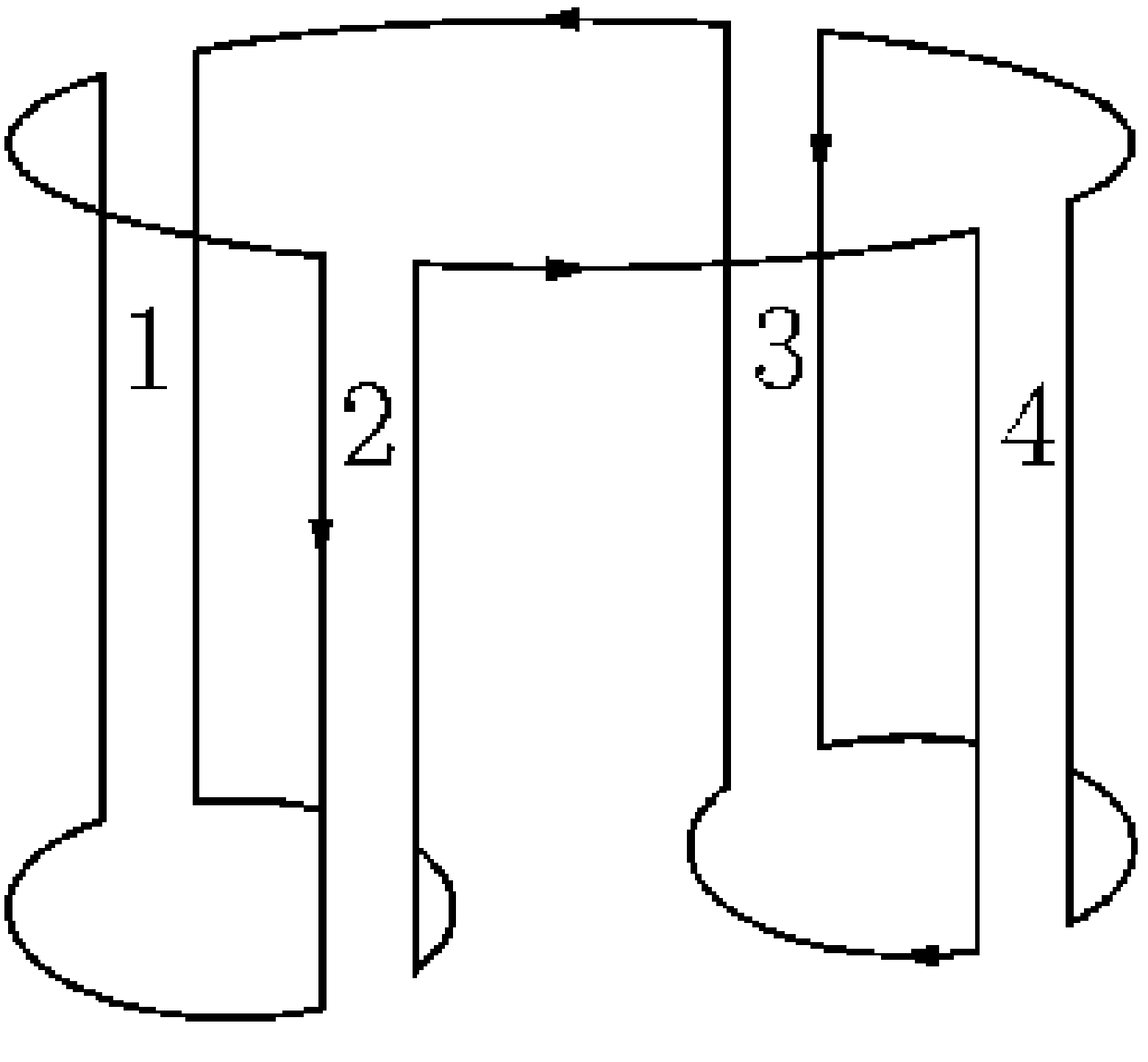}} 
    \parbox{3cm}{\center   \includegraphics[width=2cm]{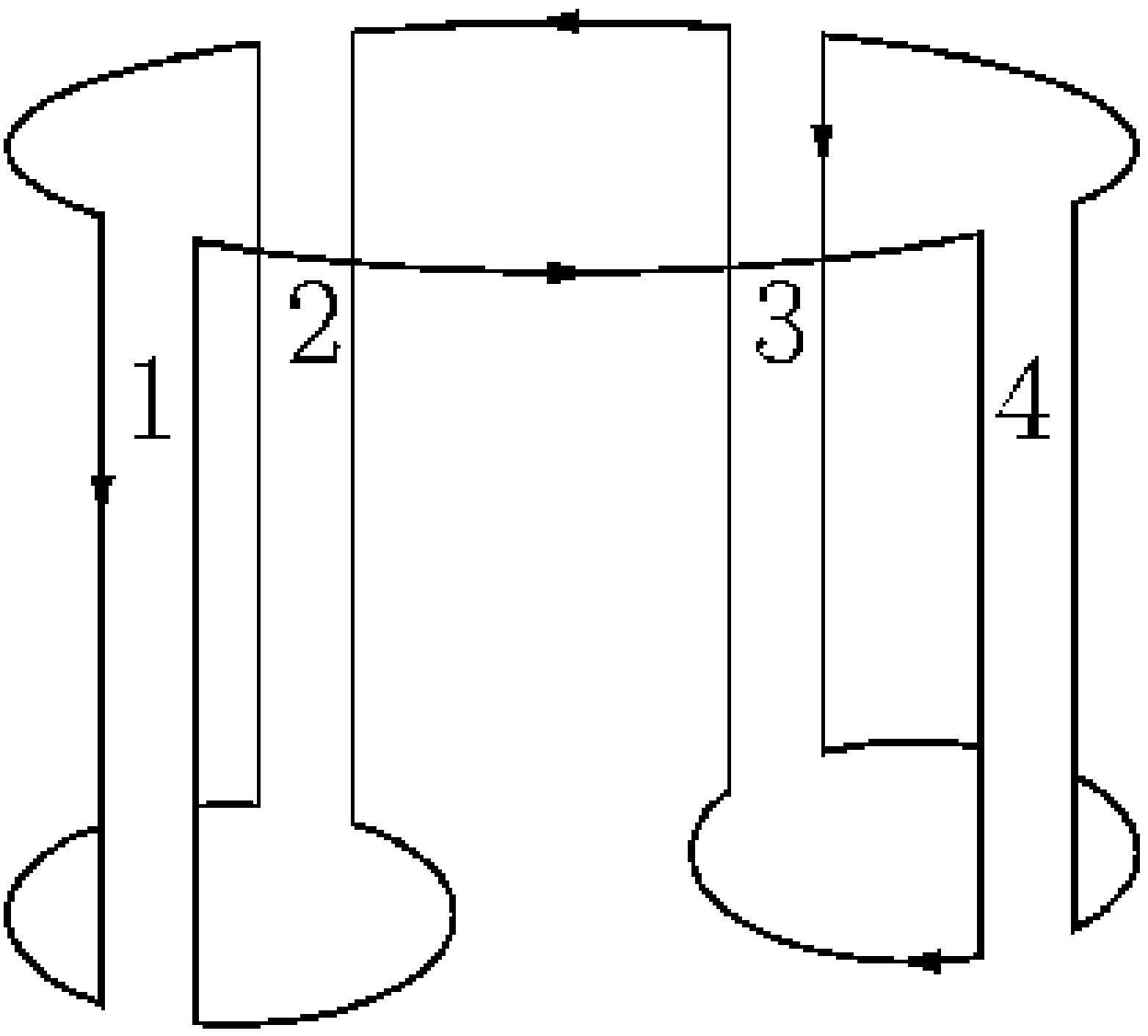}} \\
    \parbox{3cm}{\center $(1234)  $}
    \parbox{3cm}{\center ${(2134)}  $} 
     \parbox{3cm}{\center ${(1243)} $}
    \parbox{3cm}{\center ${(2143)}  $}
  \end{minipage}
  \caption{\small The four different orderings of color factors of the Born-term
}
  \label{fig:born}
\end{figure}
 Switching to ${\cal N}=4$ SYM,
we simply perform, for each of the three impact factors, the substitution we have just described for the BFKL 
cylinder, and we obtain the additional factor $N_c^3$, leading to a result of the order  
$N_c^2 \lambda^4$. Whereas in QCD the analogous lowest order graphs fit on the surface of a pair-of-pants in Fig.\ref{trousers}, 
the diagrams now have the shape of a deformed sphere as shown once again in Fig.\ref{Rpantsold}. Here both gluons cylinders are coupled to the same line of the upper loop.  
\begin{figure}
\centering
\subfigure[]{{\label{Rpantsold}}\includegraphics[height=5cm]{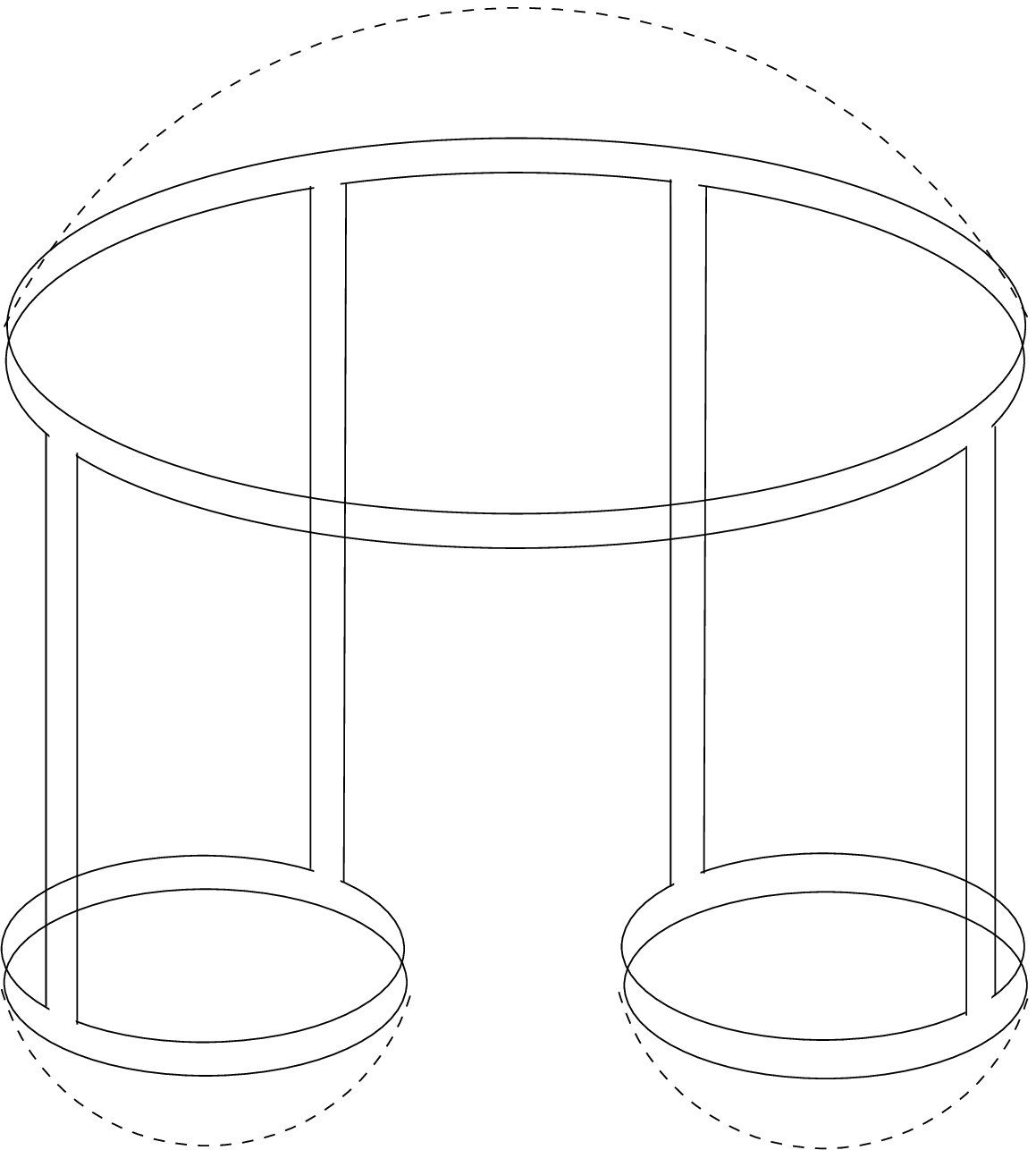}}
\hspace{1cm}
\subfigure[]{{\label{Rpantsnew}}\includegraphics[height=5cm]{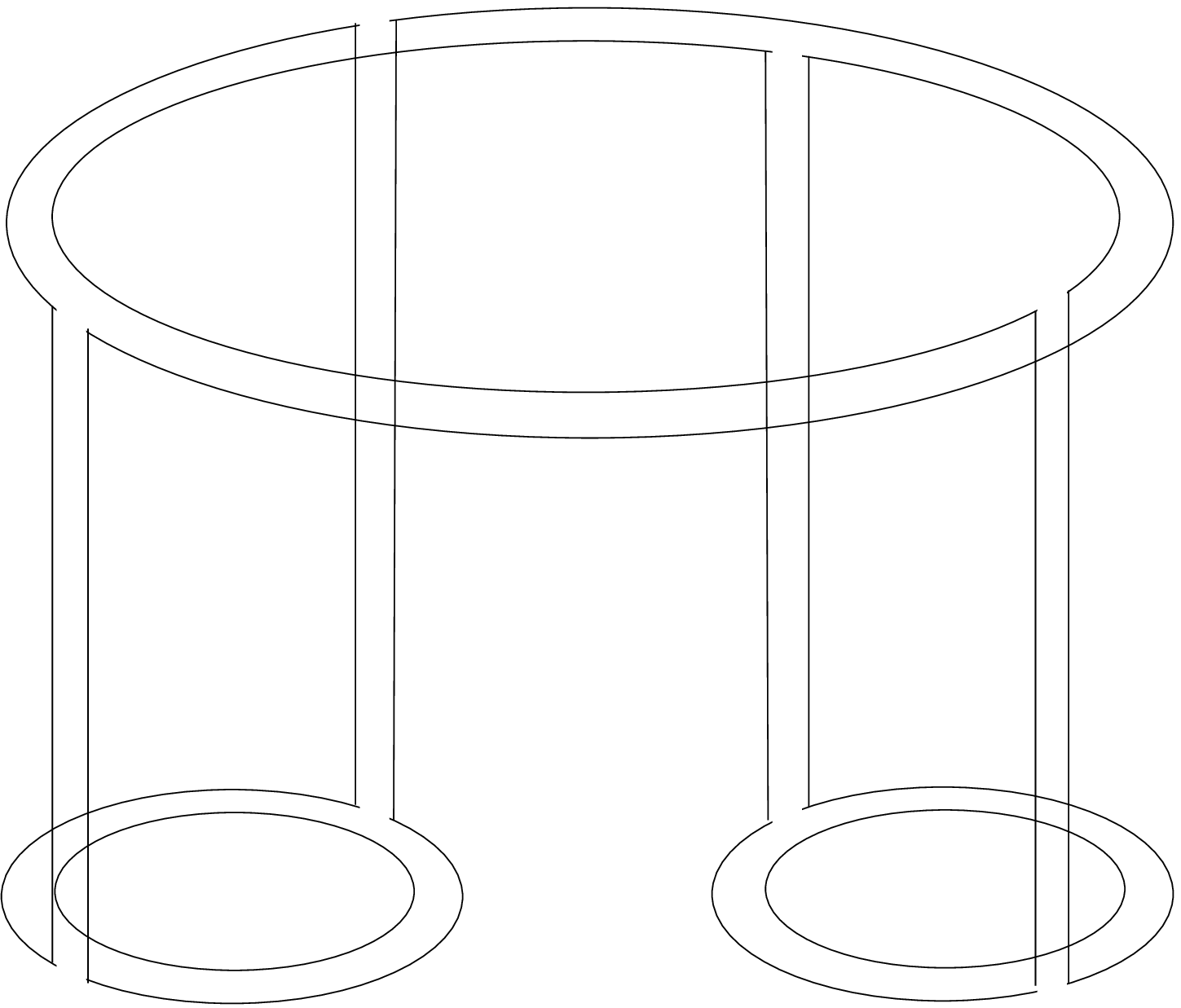}}
\caption{Pair-of-pants topology for the $3 \to 3$ amplitude in ${\cal N}=4$ SYM: (a) a color configuration already present in QCD, and  (b) a new one which exists only for adjoint particles}
\label{Rpants}
\end{figure} 

\begin{figure}
\centering
\includegraphics[height=4cm]{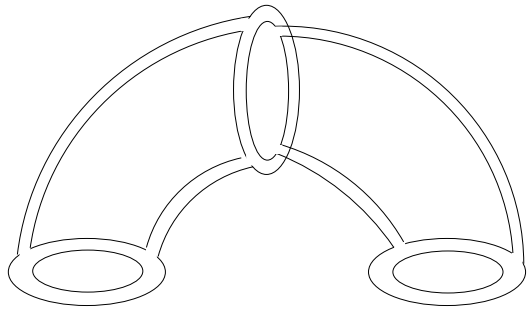}
\caption{An alternative way of drawing Fig.\ref{Rpantsnew}} 
\label{Sausage}
\end{figure}

A closer inspection shows that, in addition to Fig.\ref{Rpantsold}, another  
configuration is possible: without losing a factor $N_c$, we can attach one of the cylinders 
to the outer loop, the other one to the inner loop (Fig.\ref{Rpantsnew}). 
This 
additional piece in the four gluon impact factor which has no counterpart in the 
fundamental representation, has first been found in ~\cite{behm}. 
An alternative way of drawing this graph is shown in Fig.\ref{Sausage}. 

\begin{figure}
\centering
\includegraphics[height=5cm]{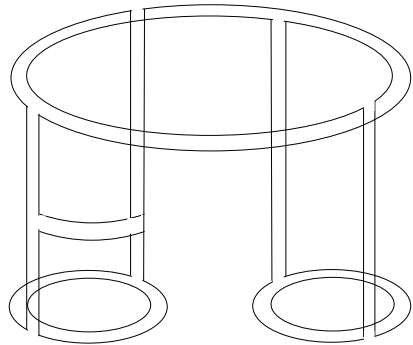}
\caption{Example of a next-to-leading order diagram of Fig.\ref{Rpantsnew}} 
\label{NLO}
\end{figure}

Moving on to the next order, we first note that for the diagram in Fig.\ref{Rpantsnew},  one can only start  
to 'dress' the two cylinders: an example is shown in Fig.\ref{NLO}.
In particular, any rung connecting 
the two cylinders loses a power of $N_c$. As a result, this class of diagrams simply consists 
of two BFKL Pomerons coupled to the four gluon impact factor, and the resulting amplitude 
is of the form $N_c^2 A_{3 \to 3}(\lambda)$. This class of diagrams will be named 'direct': 
the two BFKL Pomerons couple directly to the impact factor. 
As discussed in ~\cite{Bartels:2009zc}, on the cylinder each gluon rung 
comes in two different ways, one in front of the cylinder, the other one on the backside. This observation
also applies to our ${\cal N}=4$ SYM case.     

\begin{figure}[tb]
\centering
\includegraphics[height=4cm]{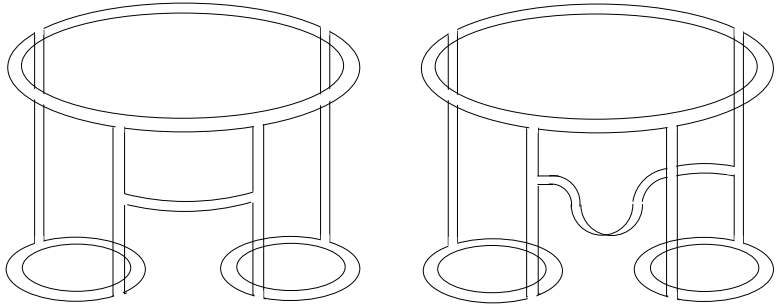}
\caption{Two classes of diagrams: planar graphs (left) and non-planar graphs (right)}   
\label{planarandnonplanar}
\end{figure}

Returning to the other diagrams in Fig.\ref{Rpants}, that are already present in nonsupersymmetric QCD, insertion of one more rung opens two distinct classes of graphs:
examples are given in Fig.\ref{planarandnonplanar}, and it is suggestive to name them as 'planar' and 'nonplanar', respectively.      
By definition, planar graphs have the property that, by contracting closed color loops, they can be 
reduced to the ${\cal N}=4$ SYM version of the graphs in Fig.\ref{fig:born}. For the non-planar ones, this is not possible, beginning with 
the graph shown in Fig.\ref{planarandnonplanar}, right, there is a new class of diagrams which cannot be deformed into planar graphs.  

In higher order $\lambda$, several possibilities arise. We briefly summarize the 
discussion given in ~\cite{behm}. The general structure of the diagrams is the following: At the upper impact factor we start with a $t$-channel state with two, three or four gluons. The propagation of the $t$-channel gluons is described by the BKP equations and transition between different states by vertices. We can have  $2\rightarrow 2$,  $2\rightarrow 3$, or  $2\rightarrow 4$ vertices. There is always a lowest interaction between the gluons defined by the $M^2$-discontinuity below which the upper cylinder breaks up into two disconnected ones. After this branching vertex the gluons interact only pairwise according to the BFKL equation and are coupled to the two impact factors of the $R$-currents at the bottom. 

We have to distinguish three different types of diagrams: the direct, the planar, and the nonplanar diagrams.
In the first case, the direct diagrams, the lowest interaction between the two cylinders is the upper impact factor itself. The four gluons couple directly the upper loop without interaction between the two disconnected BFKL Pomerons.
  
Planar diagrams, the second type of diagrams, are reggeizing pieces. At the upper loop they start with two, three, or four  $t$-channel gluons. These gluons undergo transitions by $2\rightarrow 2$, $2\rightarrow 3$, or $2\rightarrow 4$ vertices, respectively. One of these transitions is the branching vertex below which we always have four gluons but each two gluons only interact pairwise after the branching vertex. They form once again the two noninteracting BFKL Pomerons.

The last possibility are nonplanar diagrams. At the upper loop they can also start with two, three, or four  $t$-channel gluons and the structure above the branching vertex is the same as for planar diagrams. But the branching vertex itself now provides a nonplanar structure. Below this nonplanar vertex the known disconnected BFKL cylinders show up.

\section{Analytic expressions}

Let us now turn to analytic expressions. It is convenient to use the 
analytic representations of multiparticle amplitudes. A detailed discussion can be found 
in \cite{Brower:1974yv}. We restrict ourselves to those contributions which have a nonvanishing 
discontinuity in $M^2$. In the triple Regge 
limit, Eq.(\ref{tripleRegge}), we have for the $3\rightarrow 3$ amplitude:
\begin{align}  
T_{3 \to 3}(s_1, s_2, M^2| t_1,t_2,t)=
\frac{s_1s_2}{M^2} \int  \frac{d\omega_1 d\omega_2 d\omega}{(2\pi i)^3}
&
 s_1^{\omega_1}{s_2}^{\omega_2} (M^2)^{\omega-\omega_1-\omega_2}
 \xi({\omega_1})  \xi({\omega_2})  \xi({\omega,\omega_1,\omega_2}) \notag \\
&\cdot
 F(\omega, \omega_1, \omega_2| t, t_1,t_2) +\dots   
\label{eq:tripleregge}
\end{align}
The dots represent three further terms that appear in the triple Regge limit: they do not contribute to the 
$M^2$-discontinuity. The signature factors are given by
\begin{align}
  \label{eq:sig_facs}
            \xi(\omega) &= -\pi\frac{e^{-i\pi\omega}-1}{\sin(\pi\omega)}  
&\mbox{and}& &
 \xi({\omega,\omega_1,\omega_2}) &= -\pi\frac{e^{-i\pi(\omega - \omega_1 -\omega_2)} - 1}{\sin\pi(\omega - \omega_1-\omega_2)}.
\end{align} 
As discussed in the previous section, for the computation we have taken the triple energy discontinuity 
of the amplitude in $s_1$, $s_2$, and  $M^2$:
\begin{equation}
\mathrm{disc}_{s_1} \mathrm{disc}_{s_2} \mathrm{disc}_{M^2} T_{3 \to 3} =
 \pi^3\frac{s_1s_2}{M^2} \int  \frac{d\omega_1 d\omega_2 d\omega}{(2\pi i)^3}
 s_1^{\omega_1}  {s_2}^{\omega_2}  (M^2)^{\omega-\omega_1-\omega_2}
\cdot
 F(\omega,\omega_1, \omega_2 |t_1,t_2,t),
\label{tripledisc}
\end{equation}
which, via a triple Mellin transform, is related to the partial wave 
$F(\omega,\omega_1, \omega_2 |t_1,t_2,t)$. For the calculation of the 
triple discontinuity we have used unitarity integrals, and the summation 
of the diagrams has been performed by means of integral equations.
Details can be found in ~\cite{Bartels:2009zc}, and we only describe the 
results.
\begin{figure}
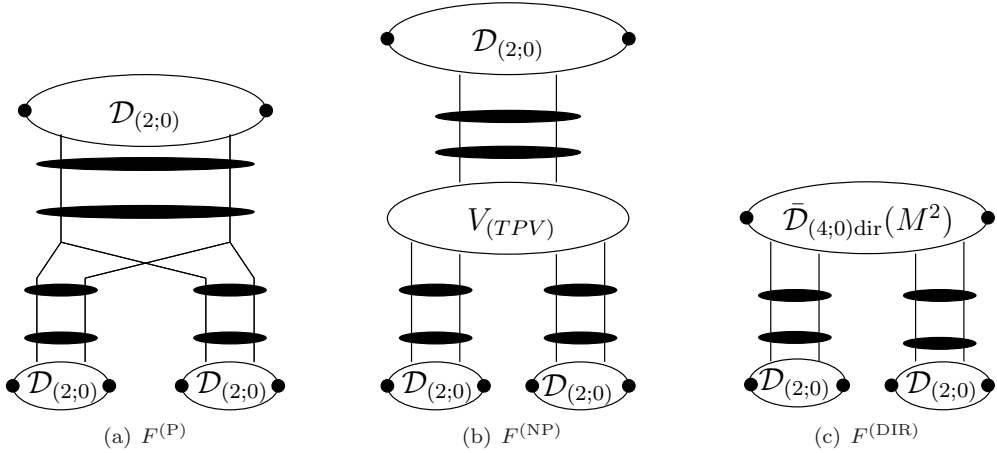

\centering 
\subfigure[$F^{(\text{P})}$]{{\label{Fplanar}}\input{Fplanar.pstex_t}}
\hspace{1cm}
\subfigure[$F^{(\text{NP})}$]{{\label{Fnonplanar}}\input{Fnonplanar.pstex_t}}
\hspace{1cm}
\subfigure[$F^{(\text{DIR})}$]{{\label{Fdirect}}\input{Fdirect.pstex_t}}
\caption{The three different parts of the six point function $T_{3\to3}$}
\label{partialwaves}
\end{figure}

Our result for the sum of the diagrams which fit on the surface of the deformed sphere (Fig.\ref{Rpants}) is 
given by the sum of three terms which sum different classes of diagrams: planar diagrams, non-planar diagrams, and direct diagrams:
\begin{equation}
T_{3 \to 3}= T_{3 \to 3}^{(\text{P})}+ T_{3 \to 3}^{(\text{NP})} + T_{3 \to 3}^{(\text{DIR})}
\end{equation}
The three different parts are illustrated in Fig.\ref{partialwaves}. In the triple Regge limit, Eq.(\ref{tripleRegge}), 
the partial waves factorize and consist of several building blocks. In our case we encounter two-gluon impact factors, 
BFKL Green's functions, and a triple vertex which connects them. The two-gluon impact factors, $D_{2;0}$, describe the coupling to the 
external $R$-currents. For ${\cal N}=4$ SYM they have been computed in \cite{Bartels:2008zy}. They contain both fermions
and scalar in the adjoint representation. 

Let us go into some detail. For the sum of the first two terms we use the representation (\ref{eq:tripleregge}) and write  
\begin{equation}
F(\omega, \omega_1,\omega_2)=F^{(\text{P})}(\omega, \omega_1,\omega_2)+F^{(\text{NP})}(\omega, \omega_1,\omega_2).
\end{equation}
As seen in Fig.\ref{partialwaves}, the impact factors appear at the three different ends 
of the diagrams, and they are connected by BFKL Green's functions and a triple vertex in the center.
The two terms differ from each other by the form of the triple vertex: in the second term, the vertex is due to the non-planar 
diagrams and coincides with the triple Pomeron vertex found in QCD. The first term which results from the planar 
diagrams is a direct consequence of the reggeization of the gauge boson. The partial wave has the form:
\begin{align}
  \label{eq:FP}
F^{(\text{P})}(\omega, \omega_1, \omega_2) 
&= 4 
 \mathcal{D}^{(12)}_2(\omega_1) \otimes_{12} \mathcal{D}^{(34)}_2(\omega_2)\otimes_{34}
   \big[\omega - \omega_1 -\omega_2 \big] \frac{\bar{\lambda}}{N_c} V^R \otimes
 \mathcal{D}_{2}(\omega)  
\end{align}
with 
\begin{equation}
\bar\lambda=\bar g^2 N_c=\frac{g^2N_c}{2}.
\end{equation}
The convolution symbol is defined as
\begin{equation}
\otimes_{12}=\int\frac{d^2{\bf k}_1}{(2\pi)^3{\bf k}_1^2{\bf k}_2^2},
\end{equation}
where ${\bf k}_1$ and ${\bf k}_2$ are the transverse momenta of the gluons 1 and 2. 
In (\ref{eq:FP}) we have introduced the three functions
$\mathcal{D}_2(\omega)$, $\mathcal{D}_2(\omega_1)$, $\mathcal{D}_2(\omega_2)$ which 
combine the three impact factors $D_{2;0}$ with their adjacent BFKL Green's functions.
The subscript $12$ at the convolution symbol indicates that the two gluon amplitude 
$\mathcal{D}^{(12)}_2(\omega_1)$ has to be contracted with the 
$t$-channel gluons 1 and 2. Similarly, $\otimes_{34}$ belongs 
to the gluons $3$ and $4$.  
Analytic expressions for the triple vertex $V^R$ can be found in ~\cite{Bartels:2009zc}.

The second part of the partial wave takes the form
\begin{align}
  \label{eq:F_nonplanar}
F^{(\text{NP})}(\omega, \omega_1, \omega_2) 
&=
4  \mathcal{D}^{(12)}_2(\omega_1) \otimes_{12} \mathcal{D}^{(34)}_2(\omega_2)\otimes_{34}
   \frac{\bar{\lambda}^2}{N_c} V_{(\text{TPV})}  \otimes \mathcal{D}_2(\omega).
\end{align}
The new ingredient here is the triple Pomeron vertex $V_{(\text{TPV})}$, described in ~\cite{Bartels:2009zc}, Eq.(87). 

Interesting enough, the first term, $F^{(\text{P})}$, is present only in the triple Regge limit with fixed $M^2$. As explained 
in ~\cite{Bartels:2009zc,behm}, after integration over $M^2$ and 
$t_1$ and $t_2$, our six-point function can be viewed as a part of the scattering of the upper 
$R$ current on a losely bound state of the two lower $R$ currents: in this case this part of the triple vertex 
disappears and turns into a special contribution to the initial conditions of the evolution of a 
BFKL Green's function, and only the triple Pomeron vertex remains (for a detailed discussion see ~\cite{Bartels:1994jj}). 

Finally, we have the third part in Fig.\ref{partialwaves}, $T_{3 \to 3}^{\text{dir}}$, where the two BFKL Green's functions 
couple directly to the upper 'unintegrated' impact factor, $D_{(4;0)\text{DIR}}(M^2)$. Here the dependence upon $M^2$ is contained 
inside $D_{(4;0)\text{DIR}}$, and instead of (\ref{eq:tripleregge}) we use 
\begin{align}  
T_{3 \to 3}^{\text{DIR}}(s_1, s_2, M^2| t_1,t_2,t)=
s_1s_2 \int  \frac{d\omega_1 d\omega_2}{(2\pi i)^2}
&
 \left(\frac{s_1}{M^2}\right)^{\omega_1}\left(\frac{s_2}{M^2} \right)^{\omega_2} 
 \xi({\omega_1})  \xi({\omega_2}) \;
 F^{(\text{DIR})}(M^2, \omega_1, \omega_2| t, t_1,t_2).  
\label{eq:direct}
\end{align}
The partial wave is given by
\begin{align}
\label{Fdirfinal}
F^{(\text{DIR})}(M^2,\omega_1,\omega_2)&=
4\, \mathcal{D}^{(12)}_2 (\omega_1) \otimes_{12} 
                           \mathcal{D}^{(34)}_2 (\omega_2)  \otimes_{34} \bar{\mathcal{D}}_{(4;0)\text{DIR}}(M^2).
\end{align}
The direct coupling of the amplitudes $\mathcal{D}^{(12)}_2(\omega_1)$ and $\mathcal{D}^{(34)}_2(\omega_2)$ to the upper loop is 
described by  the 'unintegrated' impact factor $\bar{\mathcal{D}}_{(4,0){\text{DIR}}}(M^2)$ which has both right hand and left hand cuts in $M^2$. 
In contrast to the 'normal' impact 
factors which are integrated over the mass $M^2$, in this case the coupling to the upper $R$-current is with fixed $M^2$. 
Restricting ourselves to the case of zero momentum transfers $t=t_1=t_2=0$, we have the expressions 
(see also ~\cite{behm}):
\begin{eqnarray}
\label{eq:unintIFhh}
\bar{\mathcal{D}}_{(4;0){\text{DIR}}}^{hh'}(\bk^2,{\bk'}^2,M^2)= \frac{g^4}{32 }\frac{1}{M^2} \delta_{hh'} \int_0^1 d \alpha\, I_v(\alpha, \bk^2,M^2) I_v(\alpha, {\bk'}^2,M^2)
\end{eqnarray}
and 
\begin{eqnarray}
\label{eq:unintIFLL}
\bar{\mathcal{D}}_{(4;0){\text{DIR}}}^{LL}(\bk^2,{\bk'}^2,M^2) = \frac{g^4}{32}Q^2 \int_0^1 d \alpha\,\alpha (1-\alpha) 
I_s(\alpha, \bk^2,M^2) I_s(\alpha, {\bk'}^2,M^2).
\end{eqnarray}
Here the dependence on $M^2$ is explicit. $L$ and $h$ denote the different polarizations of the incoming and outgoing $R$-currents. After integration over the angle $\varphi$ of the vector $\bk$, the  use of  $\delta$-functions to set the upper loop momentum $\bl^2 = \alpha(1-\alpha) Q^2$, we have defined
\begin{eqnarray}
\frac{\bl}{\bl^2} I_v(\alpha,\bk^2,M^2)
= \frac{\bl}{\bl^2} \left( \frac{Q^2 - M^2}{Q^2 + M^2} - 
\frac{\bk^2 +\alpha (1-\alpha) (Q^2 - M^2)}
{\sqrt{(\bk^2+\alpha(1-\alpha) (Q^2 - M^2))^2 + 4 \alpha^2(1-\alpha)^2 M^2 Q^2}}\right)
\end{eqnarray}   
and 
\begin{eqnarray}
I_s(\alpha,\bk^2,M^2)
= 2 \left( \frac{1}{\sqrt{(\bk^2+\alpha(1-\alpha) (Q^2 - M^2))^2 + 4 \alpha^2(1-\alpha)^2 M^2 Q^2}} 
- \frac{1}{\alpha(1-\alpha)(Q^2 + M^2)}\right).
\end{eqnarray}
Compared to the first two terms, this term is sub-leading for large $M^2$. Also, this term is 
present only in supersymmetric theories where the fermions are in the adjoint representation.    

The expressions listed in this section can directly be compared with results obtained from the analysis of Witten 
diagrams in the strong coupling regime ~\cite{BKMS2}.

\section{Conclusions}
In this chapter we have presented results for the six point $R$-current
correlator in ${\cal N}=4$ SYM in the triple Regge limit. In view of
the AdS/CFT correspondence, which relates the large-$N_c$ limit
of${\cal N}=4$ SYM to a string theory in $AdS_5$, we have concentrated
on those diagrams which fit on the surface of a sphere. This
generalizes an earlier calculation in nonsupersymmetric QCD, where the
leading diagrams fit on the surface of a pair-of-pants.

Our result consists of three terms which correspond to different
classes of color diagrams. Each term is composed of several buildings
blocks: impact factors, triple vertex, and BFKL Green's functions.
They should be compared with corresponding parts of the Witten
diagrams in the strong coupling region. Further work along these lines
is in progress ~\cite{BKMS2}.

\chapter{Summary and outlook}
\label{cha:concl}

In this last chapter we summarize our results and
present possible lines  which future research may follow.  Our 
goal has been the study of the high energy behavior of
QCD.  In this attempt  we followed   two different lines:

The object of the first part was given by the gauge invariant
effective action of high-energy QCD. The reformulation of QCD at high
energies in terms of an effective field theory provides a systematic
approach for the determination of high-energy QCD amplitudes.  It
factorizes high-energy QCD amplitudes into sub-amplitudes that are local in
rapidity, while non-local interaction between them is mediated
by the exchange of reggeized gluons. To separate local from non-local
interactions, a factorization parameter $\eta$ was furthermore
introduced.  However, even though the need for such a regularizing
parameter is outlined in the original formulation of the effective
action, a precise implementation that can be used by a practitioner is
missing. In the first part of this thesis a proposal has been given
for such a regularization that implements the requirements posed by
the effective action. It is stated in form of a Mellin-integral and
resembles in its precise formulation commonly used expressions in the
analysis of high-energy QCD. However its role is special in the
present context: We make  use of its property to implicitly impose a
lower bound on the squared center-of-mass energy that can be
associated with the exchange of a single reggeized gluon. At least
within the Leading-Logarithmic-Approximation (LLA) this constraint
provides the required locality in rapidity of the interaction of usual
gluons and quarks.  An extension of the method to calculations within
the Next-to-Leading-Logarithmic-Approximation (NLLA) has been proposed
as well, and it could be successfully applied to extract particularly
needed parts of these corrections. However no complete NLLA-result has
been derived so far and at present it is not clear whether the
proposed regularization is in that case adequate or not.  In
particular, at NLLA the necessity to consider renormalization of the
effective action arises which has not been addressed at all so far.

Apart from above regularization, the study of states of $n \geq 2$
reggeized gluon revealed the need to introduce so-called subtraction
diagrams.  At first a certain term, which is contained both in the
induced vertex and in the coupling of two reggeized gluons to a
particle, needs to be removed from the latter to avoid an over-counting,
which is achieved by the subtraction terms.  From a practical point of
view, subtraction terms make the integral over the longitudinal loop
momenta of loops containing reggeized gluons finite, and allow for
their evaluation.  The subtraction diagrams can be systematically
obtained by supplementing the Lagrangian of the effective action by a
subtraction term, which yields the required subtraction diagrams. This
method is particularly suitable, if entangled loops with $n > 2$
reggeized gluons are considered.

The proposed regularization methods allows to reproduce a number of
various well-known results, ranging from reggeization of the gluon and
the BFKL-equation, over production amplitudes to the analysis of the
states of the three and four reggeized gluons state. The most
non-trivial one is certainly the transition vertex from two-to-four
reggeized gluons, originally derived by Bartels and W\"usthoff. In
that case, the obtained result further allows for the study of the
two-to-four transition in the color octet, which has not been
determined so far. The results are however not yet in a suitable form
for an analysis of the properties of such a transition in the color
octet, and doing so remains as a task for the future.

In a second part, an element of Reggeon-field-theory has been studied
from the perspective of the large $N_c$ expansion in terms of
topologies of the color factor.  Within the AdS/CFT-correspondence,
the string coupling on the gravity side is proportional to $1/N_c$ and
such analysis seems to provide a suitable basis for the search of a
gravity dual of Reggeon-field-theory on the string side of the
correspondence.  Starting from a QCD-analysis of the scattering of
three virtual photons in the triple-Regge-limit, only diagrams with
the topology of the pair-of-pants surface have been resummed. Color
graphs have then be found to group themselves naturally into two
classes. While planar graphs reggeize, non-planar graphs yield the
triple-Pomeron-vertex, where the non-planar structure, which reminds
of the non-planar Mandelstam cross-diagram. Extending the study to the
triple-Regge-limit of the six-point-correlator of $R$-currents in
$\mathcal{N}=4$ SYM, the color graphs have at first no longer the
topology of a pair-of-pants, but of a sphere, as all particles are now
in the adjoint representation of $SU(N_c)$. Nevertheless it is
possible to group color factors into planar and non-planar graphs,
which appear as pair-of-pants-like deformations of the sphere. Apart
from these two contributions, a new group of color factors arises,
which is not present in QCD. In contrast to the pair-of-pants-like
structures, reggeizing contributions and triple-Pomeron-vertex, the
new piece particularly contributes for low diffractive masses.

For future work it seems suggestive to combine the results obtained
from the the two parts: The expansion of color factors in terms of
two-dimensional surfaces provides a new view on reggeization. For the
pair-of-pants, color graphs group diagrams into planar and non-planar
sets, which yield reggeization and the triple-Pomeron-vertex,
respectively. The question arises if this classification can be
extended to the general case at finite $N_c$. If so, it would be
natural to use such a classification in the formulation of the
effective action to separate reggeizing and not-reggeizing parts.

\cleardoublepage
\appendix
\setcounter{chapter}{0}
\renewcommand{\chaptername}{Appendix}

\chapter{The Rapidity scheme}
\label{cha:y_scheme}

In most of the calculations of the effective action in this thesis, a certain regularization scheme has been
used, that imposed a certain lower cut-offs $\Lambda_a$ and
$\Lambda_b$ on the squared center-of-mass energies of the
corresponding quark-gluon or gluon-gluon sub-amplitudes.

 This regularization scheme
seems very plausible, as it exactly imposes the requirements that are
needed to obtain the effective theory diagram out of the underlying
QCD-diagram. Furthermore, also imaginary parts, leading to negative
signature of the reggeized gluon are mapped one-to-one from the
QCD-diagram to the effective theory amplitude. Nevertheless the method
has some back-draws: To obtain the correct gluon trajectory function for instance,
it has been necessary to drop a certain part (see the discussion after
Eq.~(\ref{eq:cr_bar_1int})). The dropped part does not contribute
within the LLA and therefore everything is consistent within the
desired accuracy. However at the present stage it is not clear, whether
this dropped term can be  combined in a meaningful way with the other
corrections at NLLA or not. There exist also an alternative
possibility to regularize the above expression, which imposes the
lower bound on the rapidity of the gluon loop, rather than the
(sub)-center-of-mass-energies.

Also this 'rapidity scheme' relies on the Mellin transform, but instead of Eq.~(\ref{eq:theta_Mellin}) we  use 
\begin{align}
    \label{eq:theta_Melliny}
     \lim_{\nu \to 0}    
     \int_{0 - i \infty}^{0+   i \infty} \frac{d \omega}{2\pi i} 
     \frac{1}{\omega + \nu} \,
        e^{\omega Y_{ak} - \eta_a}
           = \left\{ 
             \begin{array}[h]{ll}
1 &  Y_{ak} = \frac{1}{2} \ln \frac{p_A^+k^-}{p_A^-k^+} > \eta_a = \ln \lambda_a \\ 
 & \\
0 & \text{otherwise}
             \end{array}
\right.
 \end{align}
and the central-rapidity diagram CR takes the following form
\begin{align}
  \label{eq:CR_komplettY}
 \text{CR}^{(Y)} =&      i\mathcal{M}_{2 \to 2}^{\text{tree}} (s,t)
     \int  \frac{d \omega_1}{4\pi i}  \int  \frac{d \omega_2}{4\pi i}   \frac{1}{\omega_1 + \nu}   \frac{1}{\omega_2 + \nu} 
\notag \\
&  
     \left[\left( \frac{-p_A^+}{m_A \lambda_a}\right)^{\omega_1}   + \left( \frac{p_A^+}{m_A\lambda_a}\right)^{\omega_1}   \right]      
 A^{(Y)} (\omega_1, \omega_2) 
 \left[\left( \frac{-p_B^-}{m_B \lambda_b}\right)^{\omega_2}   + \left( \frac{p_B^-}{m_B \lambda_b}\right)^{\omega_2}   \right]             \end{align}
with
\begin{align}
  \label{eq:a_y}
& A^{(Y)} (\omega_1, \omega_2) =  \int \frac{dk^+ dk^-}{2\pi i}
 \frac{1}{2}\left[
\frac{(k^- +i\epsilon)^{(\omega_1 - \omega_2)/2}}{k^- +i\epsilon}
 +
 \frac{(k^- -i\epsilon)^{(\omega_1 - \omega_2)/2}}{k^- -i\epsilon}
\right]
\notag \\
 & \frac{1}{2}\left[
\frac{(k^+ +i\epsilon)^{(\omega_2 - \omega_1)/2}}{k^+ +i\epsilon}
 +
 \frac{(k^+ -i\epsilon)^{(\omega_2 - \omega_1)/2}}{k^+ -i\epsilon}
\right]
 \frac{(-g^2N_c) }{2}  \int \frac{d^2 {\bm k}}{(2\pi)^3}      \frac{{\bm{q}}^2 }{ (k^2 + i \epsilon)((q -k)^2 + i \epsilon)}.  
  \end{align}
Substituting $k^- \to \mu = k^+k^-$ we obtain
\begin{align}
  \label{eq:a_y2}
A^{(Y)} (\omega_1, \omega_2) &=  \frac{1}{2} \int \frac{dk^+}{|k^+|}
(k^+)^{\omega_2 - \omega_1} \int \frac{d\mu}{2\pi i}
\left[ \frac{(\mu +i\epsilon)^{(\omega_1 - \omega_2)/2}}{\mu +i\epsilon}
 +
 \frac{(\mu -i\epsilon)^{(\omega_1 - \omega_2)/2}}{\mu -i\epsilon}
\right]
\notag \\
 & \qquad \qquad 
 \frac{(-g^2N_c) }{2}  \int \frac{d^2 {\bm k}}{(2\pi)^3}      \frac{{\bm{q}}^2 }{ (\mu - {\bm k}^2 + i \epsilon)(\mu - ({\bm q} -\bm{k})^2 + i \epsilon)}
\notag \\
&=
2\pi i \delta(\omega_1 -\omega_2) \beta({\bm q})
.  
  \end{align}
without the need to dropping any sub-leading term. Also for the calculation of reggeized gluon transition vertices, this kind of regularization can be used and prevents any dependence on transverse logarithms, which are then disregarded at LLA. On the other hand it is at the present stage not clear how to apply this method to loop-corrections to reggeized gluon-particle vertices. The proposed alternative scheme used throughout this thesis allows for such a treatment, while to verify whether it leads also to correct results at NLLA, it is necessary to carry out a full NLLA analysis, which remains as a task for the future.




\chapter{Integrals}
\label{cha:integrals}

In the following appendix we present the explicit evaluation of two
integrals that occored in Sec.\ref{sec:born+possig}, in the discussion
of one-loop corrections to the production vertex.

\section{The integral $I_1$}
\label{sec:integral-i_1}

 We consider 
\begin{align}
  \label{eq:1doublecut}
I_1 =&
\frac{i}{2} \int \frac{d k^+}{k^+} \frac{dk^-}{k^-} \frac{1}{k^2 + i\epsilon} \frac{1}{(q_1 - k)^2 + i\epsilon} \frac{1}{(q_2 - k)^2 + i\epsilon} 
\notag \\
&
\left[\left( \frac{-p_A^+k^- -i\epsilon}{s_1/\lambda_{12}}\right)^{\omega_1} 
+
\left( \frac{p_A^+k^--i\epsilon}{s_1/\lambda_{12}}\right)^{\omega_1} \right]
\left[\left( \frac{- p_B^-k^+-i\epsilon}{s_2/\lambda_{21}}\right)^{\omega_2} 
+
\left( \frac{p_B^-k^+-i\epsilon}{s_2/\lambda_{21}}\right)^{\omega_2} \right] 
&
\notag \\
=&
\frac{i}{2}\int \frac{d k^+}{k^+} \frac{dk^-}{k^-} \frac{1}{k^+k^- - {\bm k}^2 + i\epsilon}
\frac{1}{(k^+ - q_1^+)k^- -({\bm q}_1 - {\bm{k}})^2 + i \epsilon } \,
\frac{1}{k^+(k^- - q_2^-) - ({\bm{q}}_2 - {\bm{k}}^2) + i\epsilon}
\notag \\
&
\left[\left( \frac{-p_A^+k^- -i\epsilon}{s_1/\lambda_{12}}\right)^{\omega_1} 
+
\left( \frac{p_A^+k^--i\epsilon}{s_1/\lambda_{12}}\right)^{\omega_1} \right]
\left[\left( \frac{- p_B^-k^+-i\epsilon}{s_2/\lambda_{21}}\right)^{\omega_2} 
+
\left( \frac{p_B^-k^+-i\epsilon}{s_2/\lambda_{21}}\right)^{\omega_2} \right].
\end{align}

\subsubsection{The evaluation of the integral}

The following calculation follows closely \cite{Drummond:1969ft}.
Using  a  Schwinger parametrisation  of the propagators
\begin{align}
  \label{eq:2schwinger}
\frac{1}{A + i\epsilon} = -i \int d\lambda \exp(i\lambda (A+i\epsilon),
\end{align}
extracting phase factors, substituting $k^- \to -k^-$ and defining
rescaled integration variables $x = k^-q_1^+/p_A^+$ and $y =
-k^+q_2^-/p_B^-$, we arrive at
\begin{align}
  \label{eq:3dc}
I_1 = \frac{i}{2} \kappa^{-\omega_1- \omega_2}     \lambda_{12}^{\omega_1} \lambda_{21}^{\omega_2} 
 (-i)
\int_0^\infty d \lambda_1 d\lambda_2 d\lambda_3 &\exp[{-i \lambda_1 {\bm{k}}^2 + i\epsilon}] \exp[{-i\lambda_2({\bm{q}}_1 - \bm{k})^2 + i\epsilon}]
\notag \\
& \exp[{-i\lambda_3({\bm{q}}_2 - \bm{k})^2 + i\epsilon} ]
\notag \\
\int dx dy \big[e^{-i\pi\omega_1} (x + i\epsilon)^{\omega_1 -1 } +  (x - i\epsilon)^{\omega_1 -1 }\big] &
\big[ e^{-i\pi \omega_2}  (y +i\epsilon)^{\omega_2 -1} +  (y -i\epsilon)^{\omega_2 -1}\big]
\notag \\
&
\exp[i (- {\lambda x y}/{\kappa} + \lambda_2 x + \lambda_3 y)]
\notag \\
= \frac{-1}{2} \kappa^{-\omega_1- \omega_2}     \lambda_{12}^{\omega_1} \lambda_{21}^{\omega_2}    
 \int_0^\infty d \lambda_1 d\lambda_2 d\lambda_3 &\exp[{-i \lambda_1 {\bm{k}}^2 + i\epsilon}] \exp[{-i\lambda_2({\bm{q}}_1 - \bm{k})^2 + i\epsilon}]
\notag \\
& \exp[{-i\lambda_3({\bm{q}}_2 - \bm{k})^2 + i\epsilon} ]
\notag \\
\big[e^{-i\pi\omega_1} e^{-i\pi\omega_2 }J_1(+,+) +  e^{-i\pi\omega_2} J_1(-,+) &+
 e^{-i\pi\omega_1}J_1(+,-) + J_1(-,-)
\big],
\end{align}
where $\lambda = \lambda_1 + \lambda_2 + \lambda_3$ and $\kappa =
-q_1^+q_2^-$.  To carry out the integration over longitudinal
variables, it is therefore needed to determine
\begin{align}
  \label{eq:1long_intdc}
J_1(\pm, \pm) =\int dx dy  (x \pm i\epsilon)^{\omega_1 -1 } 
 (y \pm i\epsilon)^{\omega_2 -1} 
\exp[i (- {\lambda x y}/{\kappa} + \lambda_2 x + \lambda_3 y)].
\end{align}
Let us start with the case, where one of the $i\epsilon$ comes with a
plus sign, for definitness we chose the one associated with the
$x$-variable.  We begin the evaluation by a rescaling $x \to x
\lambda_2$ and $y \to y \lambda_3$.  Furthermore we introduce
$\alpha_i = \omega_i -1$ $i = 1,2$ and define $G := \kappa
\lambda_2\lambda_3/\lambda >0$. The
integral is then given by
\begin{align}
  \label{eq:1xandy}
J_1(+, \pm) &=\lambda_2^{-\alpha_1 -1}\lambda_3^{-\alpha_2-1} \exp[i G] \int_{-\infty}^\infty dx dy (x + i\epsilon)^{\alpha_1} (y\pm i\epsilon)^{\alpha_2}\exp[-i(x/G -1)(y-G)]
\notag \\
&=
\lambda_2^{-\alpha_1 -1}\lambda_3^{-\alpha_2-1} \exp[i G] \int_{-\infty}^\infty dx dy (x + i\epsilon)^{\alpha_1} (y + G \pm i\epsilon)^{\alpha_2}e^{-ixy/G} e^{ iy}.
\end{align}
The contour of the $x$-integration lies above the cut on the negative
$x$-axis. For $y < 0$, the exponential function guarentees convergence
in the upper semi-plane and the $x$-integral leads to a zero result.
For $y > 0$ convergence is given for the lower semi-plane and we can
enclose the contour around the cut. We obtain
\begin{align}
  \label{eq:3x+}
J_1(+, \pm) =
 e^{iG} \lambda_2^{-\alpha_1 -1}\lambda_3^{-\alpha_2 -1} 
\int_0^{-\infty} dx  \big[
              (x-i\epsilon)^{\alpha_1} - (x + i\epsilon)^{\alpha_1} 
              \big] 
\int_0^\infty dy
 (y + G \pm i\epsilon)^{\alpha_2} e^{-ixy/G} e^{iy}.
\end{align}
We are therefore lead to evaluate the discontinuity of the function $(x)^{\alpha_1}$ over the cut on the negative real axis. We find:
\begin{align}
  \label{eq:disc_cut}
 (x - i\epsilon)^{\alpha_1} - (x + i\epsilon)^{\theta_1}
& =
 \big( 
     e^{-i\pi \alpha_1} 
     -
     e^{i\pi\alpha_1}
\big) (-x)^{\alpha_1}
\notag \\
&=
-2i \sin(\pi \alpha_1) (-x)^{\alpha_1} = \frac{2\pi i}{\Gamma(-\alpha_1)\Gamma(1 + \alpha_1 )}
(-x)^{\alpha_1}.
\end{align}
We then further substitute $x \to -x$ and obtain
\begin{align}
  \label{eq:4x+}
J_1(+,\pm) &= e^{iG} \lambda_2^{-\alpha_1 -1}\lambda_3^{-\alpha_2 -1} \frac{-2\pi i}{\Gamma(-\alpha_1)\Gamma(1 - \alpha_1)} 
\int_o^\infty dy (y + G \pm i\epsilon)^{\alpha_2} e^{iy}
\int_0^\infty dx 
               x^{\alpha_1} e^{ixy/G}.
\end{align}
We rotate now the $x$-integral to the positive imaginary axis and
substitute $x \to is$. We  perform the integral which yields (toghether with the Jacobian factor from changing from $x$ to $s$) $\Gamma(1 + \alpha_1) (iG/y)^{\alpha_1 + 1}$. The complete expression then reads:
\begin{align}
  \label{eq:5x+}
J_1(+,\pm) &= 
e^{iG} \lambda_2^{-\alpha_1 -1}  \lambda_3^{-\alpha_2 - 1} \frac{2\pi i^{\alpha_1} G^{1 + \alpha_1 + \alpha_2}}{\Gamma(-\alpha_1)}
\int_0^\infty \frac{dy}{y} y^{-\alpha_1} \big(\frac{y}{G} + 1 \big)^{\alpha_2} e^{iy}.
\end{align}
Note that we are allowed to drop the $i\epsilon$ prescription for the $y$ integral as the integrad is in the considered region single valued.
Next we rotate the $y$-contour to the positive, imaginary axis and substitute $y = e^{i\pi/2}u$ which leads us to
\begin{align}
  \label{eq:22-xandy}
J_1(+,\pm) &= 
e^{iG} \lambda_2^{-\alpha_1 -1}  \lambda_3^{-\alpha_2 - 1} \frac{2\pi  G^{1 + \alpha_1 + \alpha_2}}{\Gamma(-\alpha_1)}
\int_0^\infty \frac{du}{u} u^{-\alpha_1} \big(\frac{u}{e^{-i\pi/2} G} + 1 \big)^{\alpha_2} e^{-u},
\end{align}
with \cite{Vaughn::2007ft, Watson::1962ft}
\begin{align}
  \label{eq:2confluent}
\int_0^\infty du (1 + \frac{u}{z})^{\alpha_2} u ^ {-\alpha_1 -1} e^u 
=
\Gamma(-\alpha_1) &\big[
z^{-\alpha_1} \frac{\Gamma(\alpha_1 -\alpha_2)}{\Gamma(-\alpha_2)} M(-\alpha_1 | \alpha_2 -\alpha_1 + 1| z) 
\notag \\
&+
 z^{-\alpha_2} \frac{\Gamma(\alpha_2 -\alpha_1)}{\Gamma(-\alpha_1)} M(-\alpha_2 | \alpha_1 -\alpha_2 + 1| z)
\big],
\end{align}
where $M(a|b|z) = {}_1F_1(a|b|z)$ is a confluent hypergeometric
function (Kummer's function \cite{Abramowitz::1949ft}). We  obtain
\begin{align}
  \label{eq:5x+andy}
J_1(+, \pm) =2\pi  G^{\alpha_1 + \alpha_2+ 1}  \lambda_2^{-\alpha_1 -1}\lambda_3^{-\alpha_2-1}
& 
e^{i G}  
\big[
(e^{-i\pi/2} G)^{-\alpha_1} \frac{\Gamma(\alpha_1 -\alpha_2)}{\Gamma(-\alpha_2)} M(-\alpha_1 | \alpha_2 -\alpha_1 + 1| e^{-i\pi/2} G) 
\notag \\
& +
(e^{-i\pi/2} G)^{-\alpha_2} \frac{\Gamma(\alpha_2 -\alpha_1)}{\Gamma(-\alpha_1)} M(-\alpha_2 | \alpha_1 -\alpha_2 + 1| e^{-i\pi/2} G).
\end{align}
 Reinserting $G = \kappa\lambda_2 \lambda_3/\lambda$ and $\alpha_i = \omega_i -1 $ we have for (\ref{eq:1long_intdc}) in the case where at least one of the $i\epsilon$ comes with a plus sign
\cite{Drummond:1969ft}
\begin{align}
  \label{eq:2long_int+eps}
J_1(+, \pm) &=2\pi i \lambda_2^{\omega_2 -1}\lambda_3^{\omega_1 -1} \lambda^{1 - \omega_1 -\omega_2} \kappa^{\omega_1 + \omega_2 -1} 
e^{i\lambda_2\lambda_3 \kappa/\lambda} 
\notag \\
&
\big[
 \bigg(e^{i\pi}\frac{\lambda}{\lambda_2\lambda_3 \kappa} \bigg)^{\omega_1 -1} e^{-i\pi \omega_1/2} \frac{\Gamma(\omega_1 -\omega_2)}{\Gamma(1 -\omega_2)}M(1-\omega_1 |\omega_2 - \omega_1 + 1 | -i \kappa \lambda_2 \lambda_3/\lambda) + (\omega_1 \leftrightarrow \omega_2)
\big],
\end{align}
which, applying a Kummer
transformation $M(a|b|z) = \exp(z)M(b-a|b|-z)$,  can be rewritten as
\begin{align}
  \label{eq:2long_int+eps_rew}
J_1(+, \pm) &=
2\pi i  \lambda_2^{\omega_2 -1}\lambda_3^{\omega_1 -1} \lambda^{1 - \omega_1 -\omega_2} \kappa^{\omega_1 + \omega_2 -1} 
\notag \\
&
\big[
 \bigg(e^{i\pi}\frac{\lambda}{\lambda_2\lambda_3 \kappa} \bigg)^{\omega_1 -1} e^{-i\pi \omega_1/2} \frac{\Gamma(\omega_1 -\omega_2)}{\Gamma(1 -\omega_2)}M(\omega_2 |\omega_2 - \omega_1 + 1 | i \kappa \lambda_2 \lambda_3/\lambda) + (\omega_1 \leftrightarrow \omega_2)
\big]
\notag \\
&= e^{i\omega_1\pi}  \kappa^{ \omega_2 } V(\omega_1, \omega_2, \kappa, \lambda_1, \lambda_2, \lambda_3) +  e^{i\omega_2\pi}  \kappa^{ \omega_1 }V(\omega_2, \omega_1, \kappa, \lambda_1, \lambda_3, \lambda_2),
\end{align}
where we introduced the function $V(\omega_1, \omega_2, \kappa, \lambda_1,
\lambda_2, \lambda_3) $ in order to compactify our expressions 
\begin{align}
  \label{eq:1F_left_right}
V(\omega_1, \omega_2, \kappa, \lambda_1, \lambda_2, \lambda_3) &:= -2\pi i \frac{ \lambda_2^{\omega_2 -\omega_1}  e^{-i\pi \omega_1/2}}{\lambda^{ \omega_2}} 
  \frac{\Gamma(\omega_1 -\omega_2)}{\Gamma(1 -\omega_2)}M(\omega_2 |\omega_2 - \omega_1 + 1 | i \kappa \lambda_2 \lambda_3/\lambda).
\end{align}
We come now to the case where both $i\epsilon$ occur with a minus
sign. After a rescaling and a shift $y \to y - G $, the integral is given by
\begin{align}
  \label{eq:2x-andy}
J_1(-,-) &= e^{iG} \lambda_2^{\alpha_1 -1}\lambda_3^{\alpha_2 -1} \int d x \int d y (x -i\epsilon)^{\alpha_1} (y + G -i\epsilon)^{\alpha_2} e^{-ixy/G}e^{iy}.
\end{align}
In contrast to the above case, the contour of the $x$-integration is now enclosed round the cut for negative $y$:
\begin{align}
  \label{eq:2x-andyneu}
J_1(-,-) &= e^{iG} \lambda_2^{\alpha_1 -1}\lambda_3^{\alpha_2 -1} \int_0^{-\infty} d x \big[ (x -i\epsilon)^{\alpha_1} - (x + i\epsilon)^{\alpha_1}\big] \int_0^{-\infty} {d y}(y + G -i\epsilon)^\alpha_2 e^{-ixy/G}e^{iy}.
\end{align}
We make now use again of Eq.(\ref{eq:disc_cut}) and substitute $x \ to -x$ and $y \to -y$, which gives us
\begin{align}
  \label{eq:3x-andy}
J_1(-,-) =  e^{iG} \lambda_2^{\alpha_1 -1}\lambda_3^{\alpha_2 -1}  \frac{2 i\pi}{\Gamma(-\alpha_1)\Gamma(1 + \alpha_1)}
\int_0^\infty dy (-y + G -i\epsilon)^{\alpha_2} e^{-iy}
\int_0^{\infty} dx
x^{\alpha_1} e^{-ixy/G}.
\end{align}
We rotate now the $x$-integral to the negative imaginary axis and substitute $x = -is$ and we obtain the for the integral over $x$ a factor $\Gamma(1 + \alpha_1) (-iG/y)^{1 + \alpha_1}$. The complete expression reads:
\begin{align}
  \label{eq:4x-andy}
J_1(-,-) =  e^{iG} \lambda_2^{\alpha_1 -1}\lambda_3^{\alpha_2 -1}  \frac{2 \pi (-i)^{\alpha_1} G^{1 + \alpha_1}}{\Gamma(-\alpha_1)}
\int_0^\infty \frac{dy}{y}   y^{-\alpha_1} (-y + G -i\epsilon)^{\alpha_2} e^{-iy}.
\end{align}
We use now
\begin{align}
  \label{eq:3_phase_raus}
  (-y + G -i\epsilon)^{\alpha_2} & = G^{\alpha_2} (\frac{-y}{G}  + 1 - i\epsilon)^{\alpha_2} =  G^{\alpha_2} (\frac{-y - i\epsilon}{G}  + 1 )^{\alpha_2} \notag \\
&=  G^{\alpha_2} (\frac{e^{-i\pi }y }{G}  + 1 )^{\alpha_2} = G^{\alpha_2} (\frac{y }{e^{i\pi}G}  + 1 )^{\alpha_2} ,
\end{align}
and rotate the $y-$contour to the negative, imaginary axis and substitute $y = e^{-i\pi/2} u$ which leads to
\begin{align}
  \label{eq:5x-andy}
J_1(-,-) =&  e^{iG} \lambda_2^{\alpha_1 -1}\lambda_3^{\alpha_2 -1}  \frac{2 \pi  G^{1 + \alpha_1 + \alpha_2}}{\Gamma(-\alpha_1)}
\int_0^\infty \frac{du}{u}   u^{-\alpha_1}  (\frac{u }{e^{i\pi3/2}G}  + 1 )^{\alpha_2}  e^{-u}
\notag \\
 =&
2\pi G^{1 + \alpha_1 + \alpha_2}  e^{iG} \lambda_2^{\alpha_1 -1}\lambda_3^{\alpha_2 -1} 
\big[
(e^{i\pi 3/2} G)^{-\alpha_1} \frac{\Gamma(\alpha_1 -\alpha_2)}{\Gamma(-\alpha_2)} M(-\alpha_1 | \alpha_2 -\alpha_1 + 1| e^{i\pi 3/2} G) 
\notag \\
&+
(e^{i\pi 3/2} G)^{-\alpha_2} \frac{\Gamma(\alpha_2 -\alpha_1)}{\Gamma(-\alpha_1)} M(-\alpha_2 | \alpha_1 -\alpha_2 + 1| e^{i\pi 3/2} G)
\big].
\end{align}
In the case where both $i\epsilon$ come with a minus sign,
(\ref{eq:1long_intdc}) is then given by
\begin{align}
  \label{eq:2long_int-eps_rew}
J_1(-,-) = &2\pi i \lambda_2^{\omega_2 -1}\lambda_3^{\omega_1 -1} \lambda^{1 - \omega_1 -\omega_2} \kappa^{\omega_1 + \omega_2 -1} 
\notag \\
&
\big[
 \bigg(e^{-\pi i}\frac{\lambda}{\lambda_2\lambda_3 \kappa} \bigg)^{\omega_1 -1} e^{-i\pi \omega_1/2} \frac{\Gamma(\omega_1 -\omega_2)}{\Gamma(1 -\omega_2)}M(\omega_2 |\omega_2 - \omega_1 + 1 | i \kappa \lambda_2 \lambda_3/\lambda) + (\omega_1 \leftrightarrow \omega_2)
\big]
\notag \\
&= e^{-i\omega_1\pi} \kappa^{ \omega_2 } V(\omega_1, \omega_2, \kappa, \lambda_1, \lambda_2, \lambda_3) +  e^{-i\omega_2\pi} \kappa^{ \omega_1 } V(\omega_2, \omega_1, \kappa, \lambda_1, \lambda_2, \lambda_3),
\end{align}
where we used that $M(a|b|z)$ is an entire function of $z$.
We obtain therefore for the last two lines of (\ref{eq:3dc}) 
\begin{align}
  \label{eq:3dc_lasttowlines}
&\int dx dy \big[e^{-i\pi\omega_1} (x + i\epsilon)^{\omega_1 -1 } +  (x - i\epsilon)^{\omega_1 -1 }\big] 
\big[ e^{-i\pi \omega_2}  (y +i\epsilon)^{\omega_2 -1} +  (y -i\epsilon)^{\omega_2 -1}\big]
\notag \\
&\exp[i (- {\lambda x y}/{\kappa} + \lambda_2 x + \lambda_3 y)]
\notag \\
&= \xi_{\omega_1}^{(-)}  \xi_{\omega_2}^{(-)} \big[ \phi^{\omega_1}  \kappa^{ \omega_2 } V(\omega_1, \omega_2, \kappa, \lambda_1, \lambda_2, \lambda_3)
 +
 \phi^{\omega_2}  \kappa^{ \omega_1 }V(\omega_2, \omega_1, \kappa, \lambda_1, \lambda_2, \lambda_3),
\end{align}
where we introduced, following closely \cite{Bartels:1974tj}, factors 
\begin{align}
  \label{eq:1phifactor_xis}
\phi^{\omega_i}_{\omega_1\omega_2} 
&=
e^{i\pi\omega_i} - \frac{1}{ \xi_{\omega_1}^{(-)}  \xi_{\omega_2}^{(-)}} (e^{-i\pi\omega_i} - e^{i\pi\omega_i}),
\end{align}
with
\begin{align}
  \label{eq:2phifactor_xis}
  \phi^{\omega_1}_{\omega_1\omega_2} &= \frac{1}{ \xi_{\omega_1}^{(-)}
    \xi_{\omega_2}^{(-)}}  \xi_{\omega_1}^{(-)} \xi_{\omega_2 \omega_1}^{(-,-)}
&
 \phi^{\omega_2}_{\omega_1\omega_2} &= \frac{1}{ \xi_{\omega_1}^{(-)}
    \xi_{\omega_2}^{(-)}} \xi_{\omega_2}^{(-)} \xi_{\omega_1 \omega_2}^{(-,-)}
\end{align}
and signature factors
\begin{align}
  \label{eq:2sig_facs}
 \xi_{\omega}^{(-)} &= e^{-i\pi\omega} + 1 
&
 \xi_{\omega_1 \omega_2}^{(-,-)} &= e^{-i\pi(\omega_1 -\omega_2)} + 1,
\end{align}
which results for (\ref{eq:3dc_lasttowlines}) into
\begin{align}
\label{eq:4dc_lasttowlines}
 ( \ref{eq:3dc_lasttowlines}) 
&=  \big[    \xi_{\omega_1}^{(-)} \xi_{\omega_2 \omega_1}^{(-,-)}  \kappa^{ \omega_2 } 
V(\omega_1, \omega_2, \kappa, \lambda_1, \lambda_2, \lambda_3) 
+
  \xi_{\omega_2}^{(-)} \xi_{\omega_1 \omega_2}^{(-,-)}  \kappa^{ \omega_1 }
  V(\omega_2, \omega_1, \kappa, \lambda_1, \lambda_3, \lambda_2). 
\end{align}
The complete integral (\ref{eq:1doublecut}) is then given by
\begin{align}
  \label{eq:4dc}
I_1&= - 
\lambda_{12}^{\omega_1} \lambda_{21}^{\omega_2} 
\int_0^\infty d \lambda_1 d\lambda_2 d\lambda_3 
e^{-i \lambda_1 {\bm{k}}^2 + i\epsilon}e^{-i\lambda_2({\bm{q}}_1 - \bm{k})^2 + i\epsilon}
 e^{-i\lambda_3({\bm{q}}_2 - \bm{k})^2 + i\epsilon} 
 \big[    \xi_{\omega_1}^{(-)} \xi_{\omega_2 \omega_1}^{(-,-)}  
\kappa^{ -\omega_1}  
\notag \\
& \qquad \qquad \times  V(\omega_1, \omega_2, \kappa, \lambda_1, \lambda_2, \lambda_3) 
+
  \xi_{\omega_2}^{(-)} \xi_{\omega_1 \omega_2}^{(-,-)}  \kappa^{ -\omega_2}
  V(\omega_2, \omega_1, \kappa, \lambda_1, \lambda_3, \lambda_2). 
\end{align}
In a next step we switch from Schwinger to Feynman-parameters
$\lambda_i = x_i \lambda$ with $i = 1,2,3$ and $x_i \in [0,1]$ and
obtain 
\begin{align}
  \label{eq:finalfortext}
I_1 = \lambda_{12}^{\omega_1'}\lambda_{21}^{\omega_2'} 
       \left[ \kappa_1^{-\omega_1}
    \xi_{\omega_1'}^{(-)} \xi_{\omega_2' \omega_1'}^{(-,-)}  \frac{F_1(\omega_1', \omega_2', t_1, t_2, \kappa_1)}{\sin \pi(\omega_{2}' - \omega_1')}   
+ 
\kappa_1^{-\omega_2'}
 \xi_{\omega_2'}^{(-)} \xi_{\omega_1' \omega_2'}^{(-,-)}  \frac{F_1(\omega_2', \omega_1', t_1, t_2, \kappa_1)}{\sin \pi(\omega_1' - \omega_2')} 
 \right],
\end{align}
with signature factors defined as in Eqs.~(\ref{eq:2sigi}) and (\ref{eq:doule_sig}) and
\begin{align}
  \label{eq:F11}
   F_1(\omega_1, \omega_2, t_1, t_2, \kappa) =&
 \frac{-i \pi}{2} \int \frac{d^2 {\bm k}}{(2\pi)^3}    \int_0^\infty d\lambda \lambda^{2 -\omega_1} e^{-i\pi \omega_1/2} \int \prod_{i=1}^3 
 dx_i \,\delta(1 - \sum_{i=1}^3 x_i) x_2^{\omega_2- \omega_1}
\notag \\ &
 e^{-i\lambda(x_1 {\bm{k}}^2 + x_2 ({\bm{q}}_1 -{\bm{k}}^2) + x_3 ({\bm{q}}_2 - \bm{k})^2 )} \frac{M(\omega_2| \omega_2 - \omega_1 + 1| i\kappa_1 \lambda x_2x_3)}{\Gamma(\omega_2 - \omega_1 + 1) \Gamma(1 - \omega_2)}.
\end{align}

\section{The integral $I_2$}
\label{sec:integral-i_2}

 We consider 
\begin{align}
  \label{eq:1doublecut}
I_2 =&
\frac{i}{2}\int \frac{d k^+}{k^+} \frac{dk^-}{k^-} \frac{1}{k^2 + i\epsilon}  \frac{1}{(q_2 - k)^2 + i\epsilon} 
\notag \\
&
\left[\left( \frac{-p_A^+k^- -i\epsilon}{s_1/\lambda_{12}}\right)^{\omega_1} 
+
\left( \frac{p_A^+k^--i\epsilon}{s_1/\lambda_{12}}\right)^{\omega_1} \right]
\left[\left( \frac{- p_B^-k^+-i\epsilon}{s_2/\lambda_{21}}\right)^{\omega_2} 
+
\left( \frac{p_B^-k^+-i\epsilon}{s_2/\lambda_{21}}\right)^{\omega_2} \right] 
&
\notag \\
=&\frac{i}{2}
\int \frac{d k^+}{k^+} \frac{dk^-}{k^-} \frac{1}{k^+k^- - {\bm k}^2 + i\epsilon}
\frac{1}{k^+(k^- - q_2^-) - ({\bm{q}}_2 - {\bm{k}}^2) + i\epsilon}
\notag \\
&
\left[\left( \frac{-p_A^+k^- -i\epsilon}{s_1/\lambda_{12}}\right)^{\omega_1} 
+
\left( \frac{p_A^+k^--i\epsilon}{s_1/\lambda_{12}}\right)^{\omega_1} \right]
\left[\left( \frac{- p_B^-k^+-i\epsilon}{s_2/\lambda_{21}}\right)^{\omega_2} 
+
\left( \frac{p_B^-k^+-i\epsilon}{s_2/\lambda_{21}}\right)^{\omega_2} \right].
\end{align}
Using Schwinger parametrisation we arrive at 
\begin{align}
  \label{eq:5I_6}
 I_2 =&
\frac{i}{2} \kappa^{-\omega_1 -\omega_2}\lambda_{12}^{\omega_1} \lambda_{21}^{\omega_2}  (-i)^2 
\int_0^\infty d\lambda_1 d\lambda_2 e^{-i\lambda_1 {\bm{k}}^2} e^{-i \lambda_2 ({\bm{q}}_2 -{\bm{k}})^2} 
\notag \\
&
\int dx dy \big[
e^{-i\pi\omega_1}(x + i\epsilon)^{\omega_1 -1} + (x -i\epsilon)^{\omega_1 -1}
\big]
\notag \\
&
\big[
e^{-i\pi\omega_2}(y + i\epsilon)^{\omega_2 -1} + (y-i\epsilon)^{\omega_2 -1}
\big]
e^{i(-\lambda xy/\kappa + \lambda_2 y)},
\end{align}
and we define
\begin{align}
  \label{eq:6J6}
J_2(\pm, \pm) &= \int dx dy 
(x \pm i\epsilon)^{\omega_1 -1} 
(y \pm i\epsilon)^{\omega_2 -1} 
e^{i(-\lambda xy/\kappa + \lambda_2 y)}.
\end{align}
We then obtain
\begin{align}
  \label{eq:7J6+}
J_2(+, \pm) &=
e^{i\pi (\omega_1 + 1)}\lambda_2^{-\omega_2}   2\sin(\pi \omega_1)      \int_0^\infty dx dy x^{\omega_1 -1} (y \pm i\epsilon)^{\omega_2 - 1}
e^{i(-\lambda xy/\kappa +  y)}
\notag \\
&=
\lambda_2^{-\omega_2}2 \sin(\pi \omega_1)e^{i\pi (\omega_1 + 1)}(e^{-i\pi/2} \frac{\lambda_2 \kappa}{\lambda})^{\omega_1}\Gamma(\omega_1)\int_0^\infty \frac{d y}{y}y^{\omega_2 -\omega_1}e^{iy}
\notag \\
&=
-2i\pi e^{i\pi\omega_2/2}\kappa^{\omega_1}\lambda_2^{\omega_1 -\omega_2}\lambda^{-\omega_1}\frac{\Gamma(\omega_2 -\omega_1)}{\Gamma(1-\omega_1)}.
\end{align}
For the  the last integral to be convergent it is necessary that  $\omega_2 > \omega_1$, according to the definition of the  Gamma-function \cite{Abramowitz::1949ft}.
Similar
\begin{align}
  \label{eq:8J6-}
J_2(-,\pm) &=
-e^{i\pi\omega_1}\lambda_2^{-\omega_2}\int_0^\infty dx \int_{-\infty}^0 dy (-2\sin(\pi\omega_1)) x^{\omega_1 -1}(y\pm i\epsilon)^{\omega_2 -1}e^{i(-\lambda xy/\kappa + \lambda_2 y)}
\notag \\
&=
\frac{2i\pi}{\Gamma(\omega_1)\Gamma(1-\omega_1)} \kappa^{\omega_1}e^{-i\pi\omega_1/2}\lambda_2^{\omega_1-\omega_2}\lambda^{-\omega_1}\Gamma(\omega_1)\int_{-\infty}^0 {dy} (y \pm i\epsilon)^{\omega_2 -1} y^{-\omega_1} e^{iy}.
\end{align}
We have to distinguish now different signs of the $i\epsilon$. We find
\begin{align}
  \label{eq:9J6-+}
  J_2(-,+)&=-2i\pi
  e^{i\pi\omega_2/2}\kappa^{\omega_1}\lambda_2^{\omega_1
    -\omega_2}\lambda^{-\omega_1}\frac{\Gamma(\omega_2
    -\omega_1)}{\Gamma(1-\omega_1)} 
= J_2(+, \pm),
\end{align}
and with 
\begin{align}
  \label{eq:1trick_phase}
(y -i\epsilon)^{\omega_2} = (y + i\epsilon)^{\omega_2}e^{-2\pi i \omega_2},
\end{align}
one has
\begin{align}
  \label{eq:9J6-}
J_2(-,-)&= 
 e^{-i2\pi\omega_2} J_3(+, \pm) .
\end{align}
We therefore find
\begin{align}
  \label{eq:2doublecutww}
I_2 =&   \lambda_{12}^{\omega_1}
\lambda_{21}^{\omega_2} 
      \kappa_1^{-\omega_2}
 \xi_{\omega_2}^{(-)} \xi_{\omega_1 \omega_2}^{(-,-)}  \frac{F_2(\omega_2, \omega_1,  t_2, \kappa_1)}{\sin \pi(\omega_{1} - \omega_2)}\bigg|_{\Re\text{e}\omega_2 > \Re\text{e}\omega_1}.
 \end{align}
and  $F_2$ is given by
\begin{align}
  \label{eq:F11}
   F_2(\omega_1, \omega_2,  t_2, \kappa) =&
 \frac{- \pi}{2} \int \frac{d^2 {\bm k}}{(2\pi)^3}    \int_0^\infty d\lambda \lambda^{1 -\omega_1} e^{-i\pi \omega_2/2} \int \prod_{i=1}^2 
 dx_i \,\delta(1 - \sum_{i=1}^2 x_i) x_2^{\omega_1- \omega_2}
\notag \\ &
 e^{-i\lambda(x_1 {\bm{k}}^2 + x_2 ({\bm{q}}_2 -{\bm{k}}^2) } \frac{1}{\Gamma(\omega_2 - \omega_1 + 1) \Gamma(1 - \omega_2)},
\end{align}


\cleardoublepage


\bibliography{thesis_notes}
\bibliographystyle{JHEP-2.bst} 
\cleardoublepage
\chapter*{Acknowledgements}
\addcontentsline{toc}{chapter}{\numberline{}Acknowledgements}
  
\begin{minipage}{13cm}
  First of all I want to thank my supervisor Prof. Jochen Bartels for
  his advice and guidance, for support and interest in my work.  I am
  also particularly grateful to Prof. Lev N. Lipatov for very helpful
  discussions and explanations that deepened my understanding of
  high-energy QCD.  I also wish to thank Prof. Mikhail Braun and Carlo
  Ewerz for useful conversations.
  \\
  
  Diverse conversations broadened and deepened my physical knowledge.
  For discussions on various topics concerned with high-energy-limit
  of QCD, I want to thank particularly Florian Schwennsen, Leszek
  Motyka, Agustin Sabio-Vera, Salvadore Michele, Gian Paolo Vacca,
  Anna-Maria Mischler, Alexander Prygarin, Krzysztof Kutak and Jan
  Kotanski.  For discussion on more general grounds I am grateful to
  Falk Neugebohrn, Michael Olschewsky, Mathias Butensch\"on, Thomas
  Danckaert, Hagen Triendl, Christian Hambrock, Sebastian Mendizabal,
  Marco Drewes, Torben Kneesch, Frank Fugel and Seyed Mohammad Moosavi
  Nejad.  Moreover I want to thank to the members of the II.  Institut
  f\"ur Theoretische Physik and the DESY Theory group for creating a
  very pleasant and stimulating working atmosphere.

  Financial support from the Graduiertenkolleg "Zuk\"unftige
  Entwicklungen in der Teilchenphysik" and from DESY is gratefully
  acknowledged.
  \\

  I want to thank my family, my parents Rosmarie and Gottfried, my
  brother Stefan and my sister Monika for their emotional support
  during the preparation of this thesis.
  \\

  Finally  I want to express my gratitude to my wife Mary
  for her continuous support, understanding and love during the preparation
  of this thesis.

\end{minipage}








\end{document}